\documentclass[a4paper,twoside,american,portuges,english,fontsize=11pt,paper=a4,oneside,openright,titlepage,numbers=noenddot,headinclude,BCOR=5mm,footinclude=true,cleardoublepage=empty,table]{scrreprt}

\usepackage[T1]{fontenc}
\usepackage[latin9]{inputenc}
\usepackage{babel}
\usepackage{array}
\usepackage{longtable}
\usepackage{booktabs}
\usepackage{units}
\usepackage{mathrsfs}
\usepackage{multirow}
\usepackage{amsmath}
\usepackage{amssymb}
\usepackage{graphicx}
\usepackage{setspace}
\usepackage{nomencl}
\providecommand{\printnomenclature}{\printglossary}
\providecommand{\makenomenclature}{\makeglossary}
\makenomenclature
\usepackage[unicode=true,pdfusetitle,
 bookmarks=true,bookmarksnumbered=false,bookmarksopen=false,
 breaklinks=false,pdfborder={0 0 1},backref=false,colorlinks=false]
 {hyperref}
\hypersetup{
 breaklinks}

\makeatletter

\pdfpageheight\paperheight
\pdfpagewidth\paperwidth

\providecommand{\tabularnewline}{\\}

\makeatother

\PassOptionsToPackage{eulerchapternumbers,listings,
				 pdfspacing,
				 subfig,beramono,parts}{classicthesis}

\usepackage{ifthen}
\newboolean{enable-backrefs} 
\setboolean{enable-backrefs}{false} 

\newcommand{\myTitle}{A Classic Thesis Style\xspace}

\newcommand{\myName}{Andr\'e Miede\xspace}

\newcommand{\myFaculty}{Put data here\xspace}

\newcommand{\myUni}{Put data here\xspace}

\providecommand{\mLyX}{L\kern-.1667em\lower.25em\hbox{Y}\kern-.125emX\@}

\PassOptionsToPackage{latin9}{inputenc}	
 \usepackage{inputenc}

 \usepackage{babel}

\PassOptionsToPackage{square,numbers,sort&compress}{natbib}
 \usepackage{natbib}

\PassOptionsToPackage{fleqn}{amsmath}		
 \usepackage{amsmath}

\PassOptionsToPackage{T1}{fontenc} 
	\usepackage{fontenc}
\usepackage{textcomp} 
\usepackage{scrhack} 
\usepackage{xspace} 
\usepackage{mparhack} 
\usepackage{fixltx2e} 
\PassOptionsToPackage{printonlyused,smaller}{acronym}
	\usepackage{acronym} 

\usepackage{tabularx} 
\usepackage{caption}
\usepackage{subfig}

\usepackage{listings}
\PassOptionsToPackage{hyperfootnotes=false,pdfpagelabels}{hyperref}
	\usepackage{hyperref}  
	\usepackage{graphicx}

\newcommand{\backrefnotcitedstring}{\relax}
\newcommand{\backrefcitedsinglestring}[1]{(Cited on page~#1.)}
\newcommand{\backrefcitedmultistring}[1]{(Cited on pages~#1.)}
\ifthenelse{\boolean{enable-backrefs}}%
{%
		\PassOptionsToPackage{hyperpageref}{backref}
		\usepackage{backref} 
		   \renewcommand*{\backref}[1]{}  
		   \renewcommand*{\backrefalt}[4]{
		      \ifcase #1 %
		         \backrefnotcitedstring%
		      \or%
		         \backrefcitedsinglestring{#2}%
		      \else%
		         \backrefcitedmultistring{#2}%
		      \fi}%
}{\relax}

\hypersetup{%
    colorlinks=true, linktocpage=true, pdfstartpage=3, pdfstartview=FitV,%
    breaklinks=true, pdfpagemode=UseNone, pageanchor=true, pdfpagemode=UseOutlines,%
    plainpages=false, bookmarksnumbered, bookmarksopen=true, bookmarksopenlevel=1,%
    hypertexnames=true, pdfhighlight=/O,
    urlcolor=webbrown, linkcolor=RoyalBlue, citecolor=webgreen, 
    pdftitle={\myTitle},%
    pdfauthor={\textcopyright\ \myName, \myUni, \myFaculty},%
    pdfsubject={},%
    pdfkeywords={},%
    pdfcreator={pdfLaTeX},%
    pdfproducer={LaTeX with hyperref and classicthesis}%
}

\makeatletter
\@ifpackageloaded{babel}%
    {%
       \addto\extrasamerican{%
				}%
       \addto\extrasngerman{%
				}%
    }{\relax}
\makeatother

\PreventPackageFromLoading{mathpazo}			
\usepackage{classicthesis}

\usepackage{lmodern} 

\makeatletter

\usepackage[all] {xy}
\usepackage{gensymb}
\usepackage{dsfont}
\usepackage{mathrsfs}
\hypersetup{pdfstartpage=1}
\usepackage[figure,table]{hypcap}
\usepackage{pifont}
\usepackage{bbold}
\usepackage{mathrsfs}
\usepackage{slashed}
\usepackage{notoccite}

\usepackage[square,numbers]{natbib}

\areaset[current]{390pt}{740pt}

\titleformat{\chapter}[display]%
{\relax}{\mbox{}\oldmarginpar{\vspace*{-3\baselineskip}\hspace*{-5mm}\color{halfgray}\chapterNumber\thechapter}}{0pt}%
{\raggedright\spacedallcaps}[\normalsize\vspace*{.8\baselineskip}\titlerule]

\newcommand{\cmark}{\ding{51}}%
\newcommand{\xmark}{\ding{55}}%

\definecolor{red}{rgb}{0.718,0,0.167}
\definecolor{green}{rgb}{0.196,0.490,0.2}
\definecolor{blue}{rgb}{0,0,0.608}
\definecolor{yellow}{rgb}{0.855,0.733,0}

\newcommand{\ab}{\allowbreak}

\newcommand\cleartooddpage{\clearpage
  \ifodd\value{page}\else\null\thispagestyle{empty}\clearpage\fi}

\makeatother

\begin{document}

\frenchspacing
\raggedbottom
\pagenumbering{roman}
\pagestyle{plain}
\sloppy

\thispagestyle {empty}

~\vspace{0.8cm}

~

~\hspace{-1.6cm}\includegraphics[bb=1.8600000000000001cm 0cm 10cm 3.1600000000000001cm,scale=0.3]{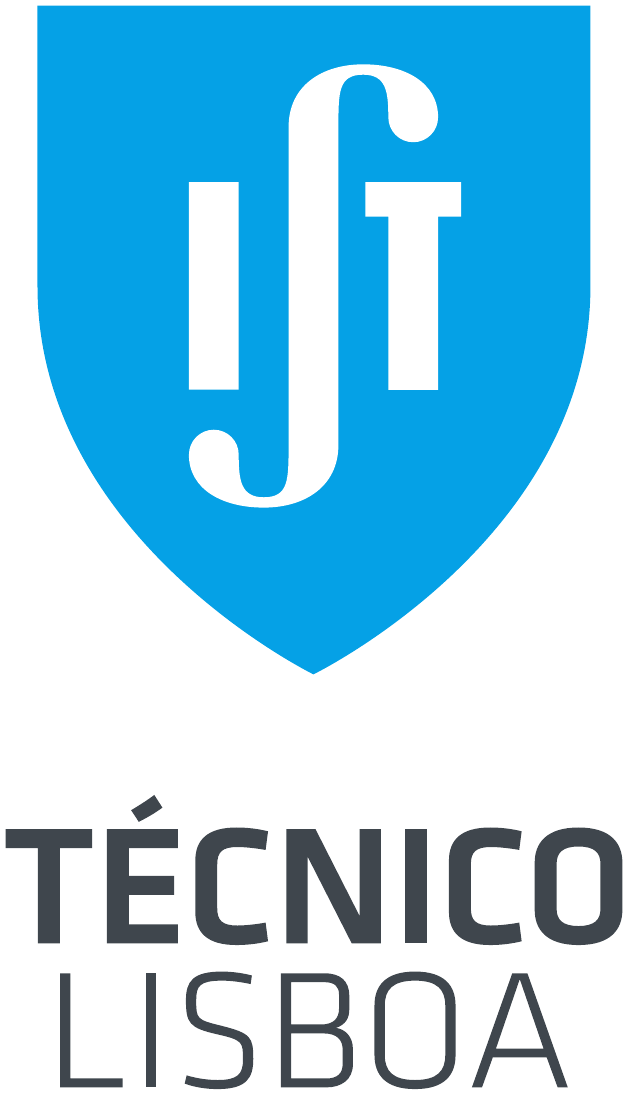}

\begin{center}
\vspace{-2.4cm}\textsc{\Large Universidade de Lisboa}
\par\end{center}{\Large \par}

\begin{center}
\textsc{\Large Instituto Superior Técnico}
\par\end{center}{\Large \par}

~

~

~

~

~

~

~

~

~

~\vspace{1.4cm}

\begin{center}
\textbf{\Large Renormalization in supersymmetric models}
\par\end{center}{\Large \par}

\textbf{\large ~}{\large \par}

\begin{center}
{\large \vspace{-0.5cm}}\textbf{\large Renato Miguel Sousa da Fonseca}
\par\end{center}{\large \par}

~

~

\begin{flushleft}
{\large \vspace{-0.5cm}Supervisor: Doctor Jorge Manuel Rodrigues
Crispim Romão}
\par\end{flushleft}{\large \par}

\begin{flushleft}
{\large \vspace{-0.3cm}Co-Supervisor: Doctor Ana Margarida Domingues
Teixeira}
\par\end{flushleft}{\large \par}

{\large ~}{\large \par}

~

\begin{center}
{\large \vspace{-0.5cm}Thesis approved in public session to obtain
the PhD Degree in}
\par\end{center}{\large \par}

\begin{center}
{\large \vspace{-0.3cm}}\textbf{\large Physics}
\par\end{center}{\large \par}

\begin{center}
{\large Jury final classification: Pass With Merit}{\Large }
\par\end{center}{\Large \par}

\begin{center}
{\large ~}
\par\end{center}{\large \par}

\begin{center}
{\large ~}
\par\end{center}{\large \par}

\begin{center}
{\large \vspace{-0.5cm}}\textbf{\large Jury}
\par\end{center}{\large \par}

\begin{flushleft}
{\large \vspace{-0.2cm}Chairperson: Chairman of the IST Scientific
Board}
\par\end{flushleft}{\large \par}

\begin{flushleft}
{\large \vspace{-0.3cm}Members of the Committee:}
\par\end{flushleft}{\large \par}

\begin{flushleft}
{\large \vspace{-0.3cm}\hspace{0.8cm}Doctor Jean Orloff}
\par\end{flushleft}{\large \par}

\begin{flushleft}
{\large \vspace{-0.3cm}\hspace{0.8cm}Doctor José Wagner Furtado Valle}
\par\end{flushleft}{\large \par}

\begin{flushleft}
{\large \vspace{-0.3cm}\hspace{0.8cm}Doctor Jorge Manuel Rodrigues
Crispim Romão}
\par\end{flushleft}{\large \par}

\begin{flushleft}
{\large \vspace{-0.3cm}\hspace{0.8cm}Doctor Filipe Rafael Joaquim}
\par\end{flushleft}{\large \par}

\begin{flushleft}
{\large \vspace{-0.3cm}\hspace{0.8cm}Doctor David Emmanuel da Costa}
\par\end{flushleft}{\large \par}

\begin{flushleft}
{\large \vspace{-0.3cm}\hspace{0.8cm}Doctor Ana Margarida Domingues
Teixeira}
\par\end{flushleft}{\large \par}

~

\begin{center}
\vspace{0.3cm}\textbf{\large 2013}\vspace{-2.0cm}
\par\end{center}

\newpage{}\thispagestyle {empty}~\newpage{}

\thispagestyle {empty}

~\vspace{0.8cm}

~

~\hspace{-1.6cm}\includegraphics[bb=1.8600000000000001cm 0cm 10cm 3.1600000000000001cm,scale=0.3]{Images/Logo_mod}

\begin{center}
\vspace{-2.3cm}\textsc{\Large Universidade de Lisboa}
\par\end{center}{\Large \par}

\begin{center}
\textsc{\Large Instituto Superior Técnico}
\par\end{center}{\Large \par}

~\vspace{1.4cm}

\begin{center}
\textbf{\Large Renormalization in supersymmetric models}
\par\end{center}{\Large \par}

\textbf{\large ~}{\large \par}

\begin{center}
{\large \vspace{-0.5cm}}\textbf{\large Renato Miguel Sousa da Fonseca}
\par\end{center}{\large \par}

~

~

\begin{flushleft}
{\large \vspace{-0.5cm}Supervisor: Doctor Jorge Manuel Rodrigues
Crispim Romão}
\par\end{flushleft}{\large \par}

\begin{flushleft}
{\large \vspace{-0.3cm}Co-Supervisor: Doctor Ana Margarida Domingues
Teixeira}
\par\end{flushleft}{\large \par}

{\large ~}{\large \par}

~

\begin{center}
{\large \vspace{-0.5cm}Thesis approved in public session to obtain
the PhD Degree in}
\par\end{center}{\large \par}

\begin{center}
{\large \vspace{-0.3cm}}\textbf{\large Physics}
\par\end{center}{\large \par}

\begin{center}
{\large Jury final classification: Pass With Merit}{\Large }
\par\end{center}{\Large \par}

\begin{center}
{\large ~}
\par\end{center}{\large \par}

\begin{center}
{\large ~}
\par\end{center}{\large \par}

\begin{center}
{\large \vspace{-0.5cm}}\textbf{\large Jury}
\par\end{center}{\large \par}

\begin{flushleft}
{\large \vspace{-0.2cm}Chairperson: Chairman of the IST Scientific
Board}
\par\end{flushleft}{\large \par}

\begin{flushleft}
{\large \vspace{-0.3cm}Members of the Committee:}
\par\end{flushleft}{\large \par}

\begin{flushleft}
{\large \vspace{-0.3cm}\hspace{0.8cm}Doctor Jean Orloff, Full Professor,
Laboratoire de Physique}
\par\end{flushleft}{\large \par}

\begin{flushleft}
{\large \vspace{-0.3cm}\hspace{0.8cm}Corpusculaire Clermont-Ferrand,
Université Blaise Pascal, France}
\par\end{flushleft}{\large \par}

\begin{flushleft}
{\large \vspace{-0.1cm}\hspace{0.8cm}Doctor José Wagner Furtado Valle,
Full Professor, CSIC, Instituto de}
\par\end{flushleft}{\large \par}

\begin{flushleft}
{\large \vspace{-0.3cm}\hspace{0.8cm}Física Corpuscular, Universitat
de València, Spain}
\par\end{flushleft}{\large \par}

\begin{flushleft}
{\large \vspace{-0.1cm}\hspace{0.8cm}Doctor Jorge Manuel Rodrigues
Crispim Romão, Full Professor,}
\par\end{flushleft}{\large \par}

\begin{flushleft}
{\large \vspace{-0.3cm}\hspace{0.8cm}Instituto Superior Técnico,
Universidade de Lisboa, Portugal}
\par\end{flushleft}{\large \par}

\begin{flushleft}
{\large \vspace{-0.1cm}\hspace{0.8cm}Doctor Filipe Rafael Joaquim,
Assistant  Professor, Instituto Superior}
\par\end{flushleft}{\large \par}

\begin{flushleft}
{\large \vspace{-0.3cm}\hspace{0.8cm}Técnico, Universidade de Lisboa,
Portugal}
\par\end{flushleft}{\large \par}

\begin{flushleft}
{\large \vspace{-0.1cm}\hspace{0.8cm}Doctor David Emmanuel da Costa,
Assistant  Researcher, Instituto}
\par\end{flushleft}{\large \par}

\begin{flushleft}
{\large \vspace{-0.3cm}\hspace{0.8cm}Superior Técnico, Universidade
de Lisboa, Portugal}
\par\end{flushleft}{\large \par}

\begin{flushleft}
{\large \vspace{-0.1cm}\hspace{0.8cm}Doctor Ana Margarida Domingues
Teixeira, Chargé de Recherche,}
\par\end{flushleft}{\large \par}

\begin{flushleft}
{\large \vspace{-0.3cm}\hspace{0.8cm}Laboratoire de Physique Corpusculaire
Clermont-Ferrand, Université}
\par\end{flushleft}{\large \par}

\begin{flushleft}
{\large \vspace{-0.3cm}\hspace{0.8cm}Blaise Pascal, France}
\par\end{flushleft}{\large \par}

\begin{flushleft}
{\large ~}
\par\end{flushleft}{\large \par}

\begin{center}
{\large ~}
\par\end{center}{\large \par}

\begin{center}
{\large \vspace{-0.5cm}}\textbf{\large Funding Institutions}
\par\end{center}{\large \par}

\begin{center}
{\large \vspace{-0.2cm}Fundação para a Ciência e a Tecnologia}
\par\end{center}{\large \par}

~

\begin{center}
\vspace{0.3cm}\textbf{\large 2013}\vspace{-2.0cm}
\par\end{center}

\thispagestyle {empty}

\newpage{}
\cleartooddpage

\section*{Acknowledgments}

\begin{onehalfspace}
\addcontentsline{toc}{section}{Acknowledgements}
\end{onehalfspace}

I would like to thank the following people:
\begin{itemize}
\item My supervisors, Jorge Romão and Ana Teixeira, for the trouble I have
given them these last years.
\item Those whom I have worked with in different research projects: Carolina
Arbeláez, Martin Hirsch, Michal Malinský, Werner Porod, Florian Staub,
and my supervisors as well.
\item Members of CFTP, where this doctoral program was carried out.
\item Members of the LPT in Orsay, the AHEP group in Valencia, and the LPC
in Clermont-Ferrand, where I have spent some time. In particular,
I thank Asmaa Abada, Martin Hirsch, and Ana Teixeira for having invited
me there.
\item Friends and colleagues which, knowingly or unknowingly, have been
very supportive. In order not to fill the whole page and still risk
forgetting someone, I will just mention here by name colleagues with
whom I have shared office with: Marco Cardoso, Nuno Cardoso, João
Esteves, António Figueiredo, David Forero, Mao Jing, Bruno Mera, Siavash
Neshatpour, Tharnier Oliveira, Laslo Reichert, Hugo Serôdio and Catarina
Simões. My gratitude goes also to Leonardo Pedro, for the interesting
discussions and for reading part of this thesis.
\item My family, to whom I dedicate this thesis.
\end{itemize}
Finally, I would like to acknowledge the support of the Portuguese
State, provided through the grant SFRH/BD/47795/2008 from the \textit{Fundação
para a Ciência e a Tecnologia}.
\cleartooddpage
\selectlanguage{portuges}%
\begin{onehalfspace}

\end{onehalfspace}

\begin{onehalfspace}

\section*{Resumo}
\end{onehalfspace}

\begin{onehalfspace}
\addcontentsline{toc}{section}{Resumo}
\end{onehalfspace}

Há motivos para acreditar que o Modelo Padrão é apenas uma teoria
efetiva, havendo nova Física para além deste. Extensões supersimétricas
são uma possibilidade: elas atenuam algumas das deficiências do Modelo
Padrão, como por exemplo a instabilidade  da massa do bosão de Higgs
sob correções radiativas. Nesta tese são analisados alguns temas
relacionados com a renormalização de modelos supersimétricos. Um deles
é a automatização do cálculo do Lagrangiano e das equações do grupo
de renormalização destes modelos --- feito à mão, este é um processo
complicado e onde podem ser introduzidos erros. As próprias equações
genéricas do grupo de renormalização são estendidas de forma a abranger
modelos que possuem um grupo de gauge com mais de um fator abeliano.
Casos deste tipo surgem, por exemplo, em teorias de grande unificação.
Para um vasto número de modelos inspirados no grupo $SO(10)$, é
igualmente mostrado que o grupo de renormalização imprime na massa
das spartículas alguma da informação sobre o comportamento destes
a altas energias. Finalmente, em alguns casos estas teorias introduzem
interações violadoras do sabor de leptões carregados, que podem levar
a alterações no rácio $\Gamma\left(K\rightarrow e\nu\right)/\Gamma\left(K\rightarrow\mu\nu\right)$.
Tendo em conta os limites experimentais noutras observáveis, a nossa
análise mostra que qualquer alteração à previsão do Modelo Padrão
será menor que a sensibilidade atual a esta observável.

~

~

~

~

~

~

~

~

~

~

~

~

~

~

~

~

~

~

~

~

~

\noindent \textsc{Palavras-chave:} Supersimetria, Grande unificação,
Equações do grupo de renormalização, Mistura de $U(1)$s, Teoria de
grupos, Susyno, Modelos inspirados em $SO(10)$, Violação de sabor
leptónico, Universalidade de sabor leptónico, Física de kaões\selectlanguage{english}

\cleartooddpage
\begin{onehalfspace}

\end{onehalfspace}

\begin{onehalfspace}

\section*{Abstract}
\end{onehalfspace}

\begin{onehalfspace}
\addcontentsline{toc}{section}{Abstract}
\end{onehalfspace}

There are reasons to believe that the Standard Model is only an effective
theory, with new Physics lying  beyond it. Supersymmetric extensions
are one possibility: they address some of the Standard Model's shortcomings,
such as the instability of the Higgs boson mass under radiative corrections.
In this thesis, some topics related to the renormalization of supersymmetric
models are analyzed. One of them is the automatic computation of the
Lagrangian and the renormalization group equations of these models,
which is a hard and error-prone process if carried out by hand. The
generic renormalization group equations themselves are extended so
as to include those models which have more than a single abelian gauge
factor group. Such situations can occur in grand unified theories,
for example. For a wide range of $SO(10)$-inspired supersymmetric
models, we also show that the renormalization group imprints on sparticle
masses some information on the higher energies behavior of the models.
Finally, in some cases these theories introduce charged lepton flavor
violating interactions, which can change the ratio $\Gamma\left(K\rightarrow e\nu\right)/\Gamma\left(K\rightarrow\mu\nu\right)$.
In light of experimental bounds on other observables, our analysis
shows that any change over the Standard Model prediction must be smaller
than the current experimental sensitivity on this observable.

~

~

~

~

~

~

~

~

~

~

~

~

~

~

~

~

~

~

~

~

~

~

~

\selectlanguage{portuges}%
\noindent \textsc{Keywords:}\foreignlanguage{english}{ Supersymmetry,
Grand unification, Renormalization group equations, $U(1)$ mixing,
Group theory, Susyno, $SO(10)$-inspired models, Lepton flavor violation,
Lepton flavor universality, Kaon physics}\selectlanguage{english}

\cleartooddpage

\tableofcontents{}\cleartooddpage
\listoftables\cleartooddpage

\addcontentsline{toc}{section}{List of Tables}

\listoffigures
\addcontentsline{toc}{section}{List of Figures}\cleartooddpage

\foreignlanguage{english}{\renewcommand{\nomname}{Abbreviations}
\renewcommand{\nomlabel}[1]{\textbf{#1}}}

\printnomenclature[2.5cm]{}\cleartooddpage

\addcontentsline{toc}{section}{Abbreviations}
\nomenclature{ATLAS}{A toroidal LHC apparatus}\nomenclature{BAO}{Baryon acoustic oscillations}\nomenclature{BBN}{Big Bang nucleosyntesis}\nomenclature{BSM}{Beyond the Standard Model}\nomenclature{CL}{Confidence Level}\nomenclature{cLFV}{Charged lepton flavor violation}\nomenclature{CLIC}{Compact linear collider}\nomenclature{CMB}{Cosmic microwave background}\nomenclature{CMS}{Compact muon solenoid}\nomenclature{cMSSM}{Constrained Minimal Supersymmetric Standard Model}\nomenclature{COMET}{Coherent muon to electron transition}\nomenclature{CP}{Charge-parity}\nomenclature{CPT}{Charge-parity-time}\nomenclature{DM}{Darkmatter}\nomenclature{EDM}{Electric dipole moment}\nomenclature{EW}{Electroweak}\nomenclature{EWSB}{Electroweak symmetry breaking}\nomenclature{FV}{Flavor violation}\nomenclature{GCU}{Gauge couplings unification}\nomenclature{GIM}{Glashow\textendash{}Iliopoulos\textendash{}Maiani}\nomenclature{GUT}{Grand unified theory}\nomenclature{HRS}{Hall-Rattazzi-Sarid}\nomenclature{ILC}{International linear collider}\nomenclature{KATRIN}{Karlsruhe tritium neutrino}\nomenclature{LEP}{Large electron\textendash{}positron collider}\nomenclature{LFC}{Lepton flavor conservation}\nomenclature{LFV}{Lepton flavor violation}\nomenclature{LHC}{Large hadron collider}\nomenclature{LSP}{Lightest supersymmetric particle}\nomenclature{MFV}{Minimal flavour violation}\nomenclature{MIA}{Mass insertion approximation}\nomenclature{MRV}{Malinský-Romão-Valle}\nomenclature{MSSM}{Minimal Supersymmetry Standard Model}\nomenclature{mSUGRA}{Minimal supergravity}\nomenclature{NMSSM}{Next-to-Minimal Supersymmetric Standard Model}\nomenclature{NUHM}{Non-universal Higgs mass}\nomenclature{PS}{Pati-Salam}\nomenclature{QCD}{Quantum chromodynamics }\nomenclature{RG}{Renormalization group}\nomenclature{RGE}{Renormalization group equation}\nomenclature{SM}{Standard Model}\nomenclature{SUSY}{Supersymmetry, Supersymmetric}\nomenclature{WMAP}{Wilkinson microwave anisotropy probe}\nomenclature{VEV}{Vacuum expectation value}\nomenclature{MEG}{Muon to electron and gamma}\nomenclature{NP}{New physics}\nomenclature{NP}{New physics}\nomenclature{NH}{Normal hierarchy}\nomenclature{IH}{Inverted hierarchy}\nomenclature{CKM}{Cabibbo\textendash{}Kobayashi\textendash{}Maskawa}\nomenclature{PMNS}{Pontecorvo-Maki-Nakagawa-Sakata}\nomenclature{LEP}{Large electron\textendash{}positron collider}\nomenclature{RPV}{R-parity violation}
\cleartooddpage

\setcounter{page}{1}
\pagenumbering{arabic}

\part{The Standard Model and beyond}\cleartooddpage

\begin{onehalfspace}

\chapter{\label{chap:Introduction}Introduction}
\end{onehalfspace}

Our understanding of the fundamental laws of Physics at the microscopic
level is encoded in a relativistic quantum field theory---the Standard
Model of Particle Physics (SM). It is a gauge theory based on the
group $U\left(1\right)_{Y}\times SU\left(2\right)_{L}\times SU\left(3\right)_{c}$
whose last missing piece, the Higgs particle, appears to have been
finally discovered at CERN. However, despite its many successes at
explaining observations made by different experiments, the SM has
several shortcomings.

One of them is the so called hierarchy problem. In principle, the
mass of a fundamental scalar, such as the Higgs doublet in the SM,
is subject to large radiative corrections, making it very sensitive
to physics at high energies. Therefore, it seems unnatural that there
is a scalar with a mass of the order of the electroweak scale, much
lower than the Planck scale at which standard quantum field theory
is expected to break down.

On the other hand, there is no explanation for the structure and parameter
values of the SM. Is there any justification for the measured values
of the electron mass and charge, for instance? It might just be that
the Universe turns out to be this way, without an underlying reason.
However, the idea that the fundamental laws or Physics are simpler
and more predictive than they appear, as in Grand Unified Theories
(GUTs), should not be dismissed.

The Standard Model also fails to explain some important experimental
and observational data. Massive neutrinos are one example. Another
one is the presence of non-luminous, weakly interacting matter in
the Universe, which is known to exist due to its gravitational effects
at galactic and cosmological scales. Through gravity, it is also known
that there is something---a dark energy---which permeates the Universe
and accelerates its expansion. Its density does not change significantly
with time and appears to be fairly homogeneous in space, behaving
as a vacuum energy. Yet, the observed dark energy density is much
lower than naive predictions made from the SM's energy scale. Another
important cosmological puzzle is the fact that, even though the Standard
Model predicts similar amounts of matter and anti-matter, astronomical
observations reveal very little anti-matter in the Universe.

It is worth mentioning that the force of gravity, from which the presence
of dark matter and dark energy is inferred, is not described by the
SM. This is due to the fact that the theories of General Relativity
and Quantum Mechanics are conceptually very different from one another,
so much so that almost a century of theoretical research has failed
to merge the two.

\medskip{}

Supersymmetric (SUSY) Grand Unified Theories described in this thesis
try to address some of these issues. Two distinct research lines are
examined. The first one is the computation of the renormalization
group equations (RGEs) which are needed for the study of the phenomenology
of these models at the energy scales accessible to experiments. A
second topic concerns the phenomenology of SUSY GUTs, in particular
relations between particle masses and the enhancement of the $\Gamma\left(K\rightarrow e\nu\right)/\Gamma\left(K\rightarrow\mu\nu\right)$
ratio through charged lepton flavor violation (cLFV) interactions.
Both of these research lines involve radiative corrections and renormalization
in SUSY models, hence the title of the thesis.

\medskip{}

This work is organized as follows. The next chapter contains a review
of the shortcomings of the Standard Model, explaining also how supersymmetry
(SUSY) can overcome some of them. As such, chapter \ref{chap:The-SM's-shortcomings}
also presents a description of supersymmetric (SUSY) models in general,
and the Minimal Supersymmetric Standard Model (MSSM) in particular.

The experimental evidence for massive neutrinos, as well as possible
ways of accommodating them by extending the SM, is presented separately
in chapter \ref{chap:Lepton-flavour-violation}. In it, the possibility
of having charged lepton flavor violation in supersymmetric models
(in addition to neutral lepton flavor violation in neutrino oscillations)
is also reviewed. These processes are interesting because any observation
of cLFV clearly signals the presence of new Physics.

Chapter \ref{chap:Symmetry} is dedicated to the use of symmetry in
Particle Physics, and its aim is twofold. On the one hand, it complements
chapter \ref{chap:The-SM's-shortcomings} by motivating the introduction
of supersymmetry from a theoretical perspective: it is the only possible
non-trivial extension of the space-time symmetry group. On the other
hand, it reviews the main concepts needed for a systematic analysis
of Lie algebras used in gauge theories.

This generic treatment of Lie algebras is essential to the Mathematica
program \texttt{Susyno}, which is described in the chapter that follows
(chapter \ref{chap:Susyno}) and which is based on \citep{Fonseca:2011sy}.
With the defining elements of a SUSY model, the program computes its
Lagrangian and provides the 2-loop RGEs.

Chapter \ref{chap:U1_mixing_paper}, adapted from \citep{Fonseca:2011vn},
presents the 2-loop RGEs of softly broken SUSY models with more than
one $U(1)$ gauge factor group. In these models, the $F_{\mu\nu}$
tensor associated to a $U(1)$ factor mixes with the other ones, and
this is a feature requiring special care.

Models with $U(1)$-mixing are not rare: the intermediate stages in
GUTs such as some of the ones analyzed in chapter \ref{chap:SlidingScale_models}
do have this feature. Details of the high energy structure of SUSY
GUTs are imprinted in the soft scalar masses at lower energies, and
in this chapter we consider $SO(10)$-inspired models in this context
\citep{Arbelaez:2013hr}.

Chapter \ref{chap:Revisiting_RK} revisits the ratio $\Gamma\left(K\rightarrow e\nu\right)/\Gamma\left(K\rightarrow\mu\nu\right)$
in constrained and unconstrained supersymmetric models. In principle,
in such models this ratio can be significantly different from the
SM prediction, due to cLFV interactions. However, one should look
carefully at this observable, taking into consideration also the bounds
on $\textrm{BR}\left(B_{u}\rightarrow\tau\nu\right)$, $\textrm{BR}\left(\tau\rightarrow e\gamma\right)$
and $\textrm{BR}\left(B_{s}\rightarrow\mu\mu\right)$ \citep{Fonseca:2012kr}.

Finally, chapter \ref{chap:Conclusion} presents some concluding remarks.

\chapter{\label{chap:The-SM's-shortcomings}The Standard Model and supersymmetry}

\begin{figure}[tbph]
\begin{centering}
\includegraphics{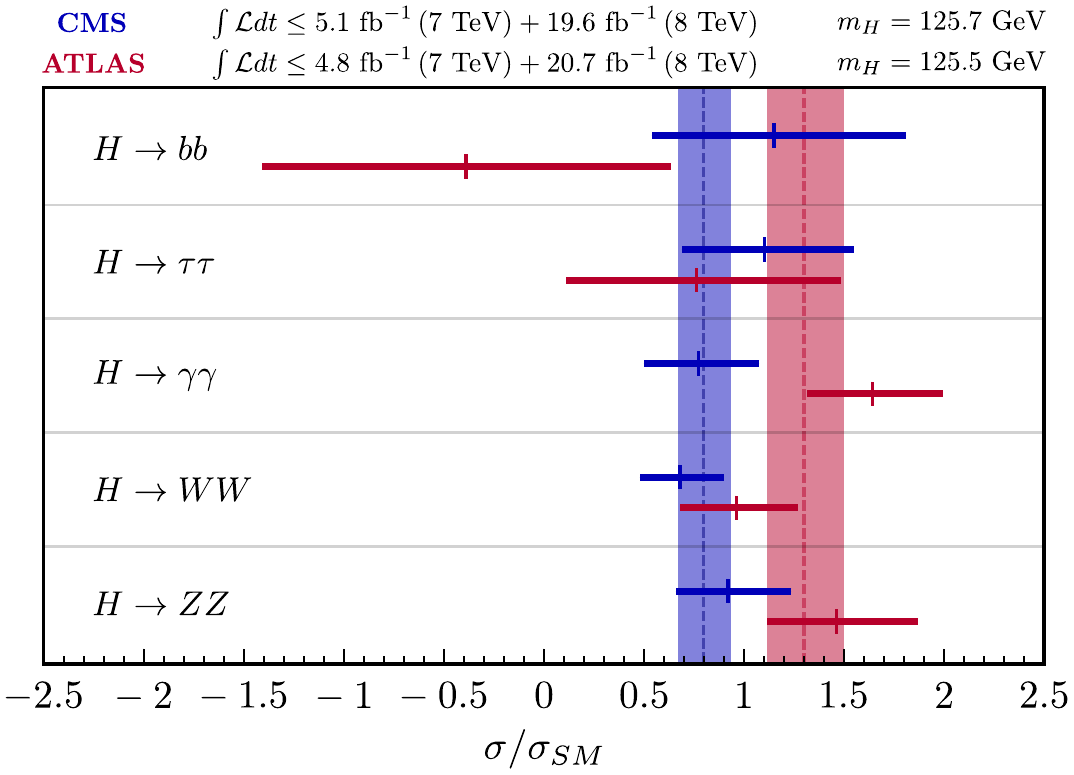}
\par\end{centering}

\caption{\label{fig:ALTAS-and-CMS-Higgs-couplings}ATLAS and CMS preliminary
data on the measured signal strength in five different channels, where
the horizontal bars represent the 68\% confidence level intervals.
The vertical lines show the result of combining all channels, together
with the 68\% confidence level bands (picture edited from \citep{ATLAS-CONF-2013-034}
and \citep{CMS-PAS-HIG-13-005}). Notably, the new 8 TeV data pulls
CMS's $H\rightarrow\gamma\gamma$ cross section below the one predicted
by the SM.}
\end{figure}

\begin{figure}[tbph]
\begin{centering}
\includegraphics{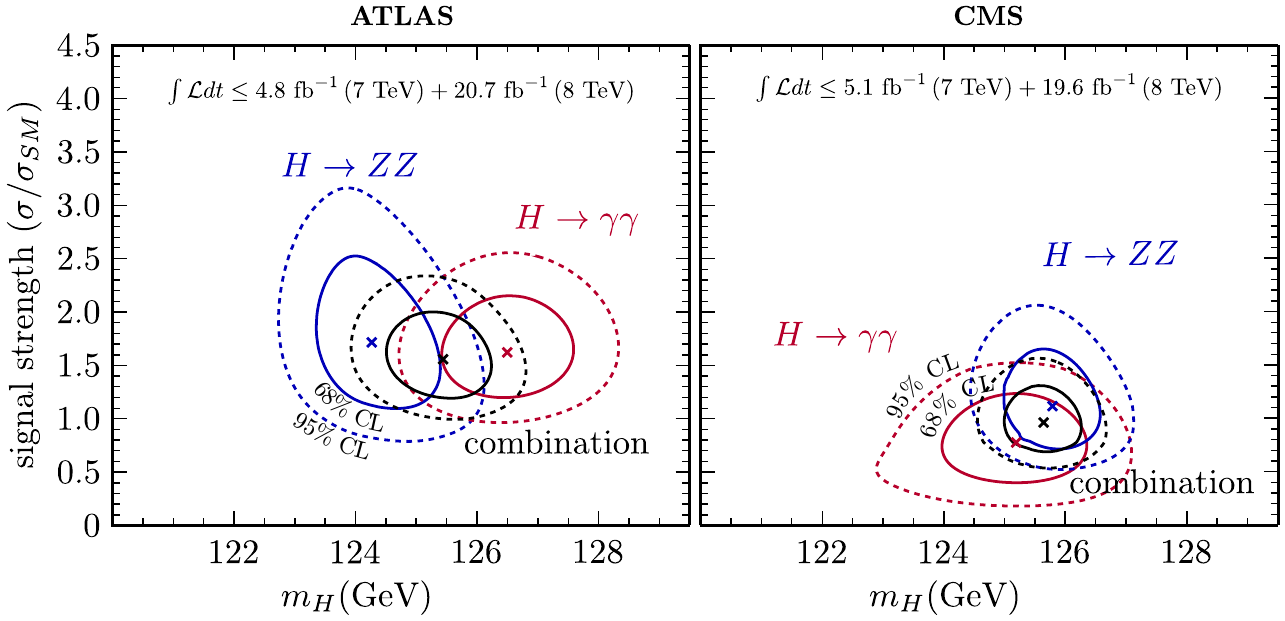}
\par\end{centering}

\caption{\label{fig:ALTAS-and-CMS-HiggsMass}ATLAS and CMS preliminary data
on the Higgs mass, in the two channels with the best energy resolution
($H\rightarrow\gamma\gamma$, $H\rightarrow ZZ$), and also the combined
results. Image edited from \citep{ATLAS-CONF-2013-014} and \citep{CMS-PAS-HIG-13-005}.}
\end{figure}

The Standard Model was developed in the 1960s and 1970s, incorporating
the electroweak (EW) theory \citep{Glashow:1961tr,Weinberg:1967tq,Salam:1968rm}
and the theory of strong interactions \citep{GellMann:1964nj,Zweig:1981pd,Zweig:1964jf,Bjorken:1968dy,Feynman:1969ej,Gross:1973id,Politzer:1973fx,Greenberg:1964pe,Han:1965pf,Georgi:1951sr}.
The discovery of neutral currents \citep{Hasert:1973ff} and the $W$/$Z$
bosons \citep{Arnison:1983rp}, the evidence that hadrons are composite
\citep{Breidenbach:1969kd,Bloom:1969kc,Friedman:1972sy}, and more
recently the discovery of a Higgs particle%
\footnote{The mechanism responsible for the breakdown of the EW symmetry was
suggested by Englert, Brout, Higgs, Guralnik, Hagen, and Kibble \citep{Englert:1964et,Higgs:1964ia,Higgs:1964pj,Guralnik:1964eu}.
For simplicity, we refer to the associated scalar particle(s) as \textit{Higgs
boson(s)}.%
} \citep{ATLAS_Higgs_discovery:2012gk,CMS_Higgs_discovery:2012gu}
(see figures \eqref{fig:ALTAS-and-CMS-Higgs-couplings} and \eqref{fig:ALTAS-and-CMS-HiggsMass}),
were important milestones in the confirmation of the Standard Model.

Being a Yang\textendash{}Mills theory \citep{Yang:1954ek}, the model
is defined by its group, $U\left(1\right)_{Y}\times SU\left(2\right)_{L}\times SU\left(3\right)_{c}$,
and its particle content, presented in table \eqref{tab:Representations-of-the-SM}.
The gauge symmetry, together with the space-time one, severely constrains
the possible Lagrangian terms (see appendix \ref{chap:SM_appendix}).
The strength of these terms is controlled by free parameters, which
have been measured over decades with ever increasing precision. Electroweak
precision tests at LEP indicate a remarkable agreement between theory
and experiment \citep{Schael:2013ita}. One noteworthy example is
the excellent agreement (better than 1 part in $10^{9}$) between
the measured anomalous magnetic moment of the electron, and its predicted
value \citep{Giudice:2012ms}. In the strong sector however, due to
difficulties in making precise non-perturbative computations, presently
the agreement between theory and experiment can only be checked to
the percent level.
\begin{table}[tbph]
\begin{centering}
\begin{tabular}{cccc}
\toprule 
\addlinespace[0.1cm]
Representation & $U(1)_{Y}\times SU(2)_{L}\times SU(3)_{C}$ & Spin & Flavors\tabularnewline\addlinespace[0.1cm]
\midrule
\addlinespace[0.1cm]
\selectlanguage{portuges}%
$Q=\left(u_{L},d_{L}\right)^{T}$\selectlanguage{english}
 & $\left(+\frac{1}{6},\mathbf{2},\mathbf{3}\right)$ & $\frac{1}{2}$ & 3\tabularnewline\addlinespace[0.1cm]
\addlinespace[0.1cm]
$u_{R}$ & $\left(+\frac{2}{3},\mathbf{1},\mathbf{3}\right)$ & $\frac{1}{2}$ & 3\tabularnewline\addlinespace[0.1cm]
\addlinespace[0.1cm]
$d_{R}$ & $\left(-\frac{1}{3},\mathbf{1},\mathbf{3}\right)$ & $\frac{1}{2}$ & 3\tabularnewline\addlinespace[0.1cm]
\addlinespace[0.1cm]
$L=\left(\nu_{L},e_{L}\right)^{T}$ & $\left(-\frac{1}{2},\mathbf{2},\mathbf{1}\right)$ & $\frac{1}{2}$ & 3\tabularnewline\addlinespace[0.1cm]
\addlinespace[0.1cm]
$e_{R}$ & $\left(-1,\mathbf{1},\mathbf{1}\right)$ & $\frac{1}{2}$ & 3\tabularnewline\addlinespace[0.1cm]
\addlinespace[0.1cm]
$H=\left(H^{+},H^{0}\right)^{T}$ & $\left(+\frac{1}{2},\mathbf{2},\mathbf{1}\right)$ & $0$ & 1\tabularnewline\addlinespace[0.1cm]
\bottomrule
\end{tabular}
\par\end{centering}

\caption{\label{tab:Representations-of-the-SM}Representations of the SM gauge
group. All listed fermions are left-handed.}

\end{table}

Despite these successes, the SM is widely regarded as just an effective
theory of a more fundamental one. In the next section, we review some
of the reasons why it is thought that there must be some new physics
beyond the Standard Model.

\section{Shortcomings of the Standard Model}

\subsection{The hierarchy problem}

There has been a constant quest to explore ever smaller distances,
or equivalently, ever higher energy regimes, in order to find a more
fundamental understanding of the laws of Nature. It is therefore curious
that the current Standard Model of Particle Physics is very sensitive
to the cut-off scale $\Lambda$ at which it ceases to be a valid effective
theory \citep{Susskind:1978ms,'tHooft:1979bh,Veltman:1980mj}. This
is because radiative corrections make the Higgs boson mass $\mu$
quadratically sensitive to this energy scale. Then, the measured value
of $\mu$, which is of the order of $\left(\textrm{Higgs vacuum expectation value}\right)^{2}$,
must be the sum of a bare squared mass and a self energy of the order
of $\Lambda^{2}$. As such, unless we are willing to accept that there
is a fine-tuned cancellation of these last two terms, the SM cut-off
scale $\Lambda$ must not be substantially larger than the electroweak
scale. This is why it is widely believed that some new Physics must
be present at the TeV energy range.

Following \citep{Murayama:1994kt,Murayama:2000dw}, we recall that
there is a precedent for this. In order to assemble a uniform, electrically
charged sphere of radius $R$, it is necessary to spend some energy
$E_{\textrm{self}}\sim\frac{1}{4\pi\varepsilon_{0}}\nicefrac{Q^{2}}{R}$
to overcome the repulsion between its charged components. Therefore,
in the case of an electron with charge $e$, its total energy $m_{e}c^{2}$
is given by the sum of some bare mass term and this self energy $E_{\textrm{\textrm{self}}}$.
Assuming that there is no accidental cancellation between these two
terms, and since $\nicefrac{\frac{1}{4\pi\varepsilon_{0}}e^{2}}{m_{e}c^{2}}$
is of the order of femtometers, there are two possibilities: either
the electron size is bigger than this, or classical electrodynamics
ceases to be valid at distances smaller than $\sim10^{-15}$ m. In
either case, something previously unaccounted for becomes relevant
at this scale. It turns out that the latter possibility is correct:
quantum effects become significant, and classical electrodynamics
is no longer a reliable theory at small distances.

In fact, with relativistic quantum electrodynamics, the divergence
of the electron mass, which was linear in the cut-off scale $\Lambda\propto R^{-1}$,
is reduced to a logarithmic one. The electron mass is said to be protected
by the chiral symmetry, which relates the electron with its antiparticle
(the newly introduced positron). Even though it is broken, this symmetry
ensures that radiative corrections to $m_{e}$ are proportional to
$m_{e}$ itself, implying that they can only depended on $\Lambda$
through logarithms.

Similarly, in Yang\textendash{}Mills theories such as the Standard
Model, gauge bosons are massless due to the gauge symmetry. If this
symmetry is spontaneously broken, the mass of gauge bosons becomes
proportional to the Higgs mass, but the Higgs mass itself is not protected
by any fundamental principle in the Standard Model. However, noting
that bosons and fermions contribute radiatively to it with different
signs, it is possible to build a symmetry that relates these two types
of fields in such a way that the quadratic dependence of the Higgs
mass on the cut off scale is canceled. Since it would mix different
irreducible representations of the space-time symmetry group (bosons
and fermions), this would therefore be a supersymmetry (see chapter
\ref{chap:Symmetry} for a detailed discussion).

\subsection{Unexplained structure and parameter values}

The origin of the structure of the Standard Model, if there is one,
remains a mystery. It is a Yang\textendash{}Mills theory, but we do
not known why there are three forces, with three distinct coupling
strengths. The matter fields are scattered through small irreducible
representations of the gauge group (there are only singlets, doublets
of $SU\left(2\right)_{L}$ and triplets of $SU\left(3\right)_{c}$),
but surprisingly the representations are such that the SM is an anomaly
free theory, even though separately each matter field would generate
anomalies (suggesting a connection between quarks and leptons). Left-
and right-handed fermions are treated differently by the gauge symmetry,
leading to a chiral theory with both charge-conjugation and parity
symmetry violation. There is also the flavor puzzle, which relates
to the existence of three copies of all fermionic representations
even though, from a theoretical point of view, there seems to be no
good reason for this. 

In any case, once the symmetries and content of the SM have been established,
building the most general Lagrangian consistent with them reveals
a total of 19 physical parameters which must be measured:
\begin{itemize}
\item 3 gauge coupling constants;
\item 9 fermion masses, 3 mixing angles and 1 phase;
\item 2 parameters in the Higgs potential ($\mu$ and $\lambda$);
\item 1 parameter related to topological effects and the strong CP problem
($\theta$).
\end{itemize}
In the Standard Model, these are all free parameters. Yet, their values,
obtained by different experiments, suggest that they are not completely
arbitrary. For example, $\theta$ is very close to zero and also,
in relation to the flavor structure of the SM, mixing angles are small
in the quark sector and large in lepton sector.%
\footnote{Here we are assuming implicitly that the SM is extended in order to
accommodate neutrinos with mass. Depending on whether or not neutrinos
are Majorana particles, at least 9(7) new parameters are introduced
in the theory.%
} On the other hand, fermion masses are spread over more than twelve
orders of magnitude, exhibiting a strong hierarchy in flavor space
(with the possible exception of neutrinos).

It is conceivable that our Universe is just the way it is, with no
underlying reason. However, it is widely believed that this not the
case, and that the Standard Model is just an effective model of a
more fundamental and predictive one. A particular hypothesis that
has received much attention over time \citep{Georgi:1974sy,Pati:1974yyX,Georgi:1974yf,Fritzsch:1974nn,Buras:1977yy,Georgi:1979df,Barr:1981qv,Dimopoulos:1981zb,Hewett:1988xc}
is the unification of all three forces of the SM in a grand unified
theory, at high energies. In these theories, the fundamental gauge
group is a simple one%
\footnote{It might also be a direct product of equal simple factor groups, together
with some discrete symmetry relating them.%
} which spontaneously breaks into the smaller $U\left(1\right)_{Y}\times SU\left(2\right)_{L}\times SU\left(3\right)_{c}$
at low energies. Since the gauge group of a GUT is larger than the
SM one, such a fundamental theory is more predictive than the effective
one at lower energies. This implies that SM parameters, such as gauge
and Yukawa couplings, are related amongst themselves in these theories.
Also, the SM fields are assembled  into larger representation of the
fundamental symmetry group, explaining (partially at least) why the
matter fields are in the observed representations of the SM gauge
group. It is indeed suggestive that all SM fermion representations
fit in just two representations of the $SU(5)$ group (the $\overline{\boldsymbol{5}}$
and $\boldsymbol{10}$). In turn, together with a singlet representation
(the right-handed neutrinos), these form an $SO(10)$ spinor representation
($\boldsymbol{16}$). We also note that, since non-$SU(n)$ simple
Lie groups are known to be anomaly free \citep{Georgi:1972bb,Okubo:1977sc},
this bundling of the SM fermions into a complete representation of
$SO(10)$  ensures that the Standard Model is anomaly free.

\begin{figure}[tbph]
\begin{centering}
\includegraphics[scale=0.82]{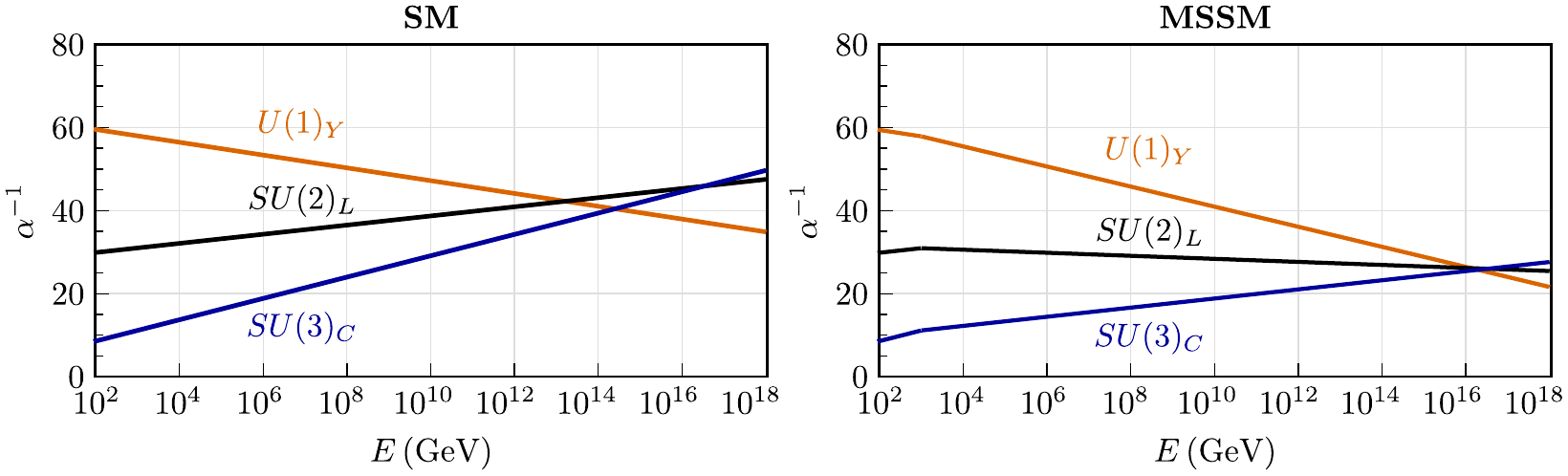}
\par\end{centering}

\caption{\label{fig:GCU_SM_vs_MSSM}Comparison of the 1-loop evolution of the
gauge coupling constants ($\alpha_{i}=\nicefrac{g_{i}^{2}}{4\pi}$)
with the energy scale $E$, for the Standard Model and the Minimal
Supersymmetric Standard Model. By changing the slope of the lines
in this plot, the extra fields in the MSSM allow the gauge coupling
constants to unify.}
\end{figure}

Unification of the three forces can only occur if their strengths
converge to a common value, as energy is increased and the full gauge
symmetry is restored. The analysis of the renormalization group (RG)
of the Standard Model does show that the differences between the three
gauge coupling constants shrink as energy is increased. However, they
never unify completely (figure \eqref{fig:GCU_SM_vs_MSSM}), but surprisingly,
if the SM is supersymmetrized in a minimal way, the additional fields
change the running of the gauge coupling constants in such a way that
unification is indeed achieved.%
\footnote{Note that SUSY is by no means the only way to achieve unification;
adding other combinations of fields to the SM works equally well.%
} Even though the sensitivity to the SUSY scale is only logarithmic,
we note that unification is consistent with $m_{SUSY}\sim1$ TeV.

Generally, GUTs lead to an unstable proton. The enlarged gauge symmetry
transforms quarks into leptons and vice versa, which means that gauge
interactions violate both baryon and lepton number. Indeed, through
the exchange of heavy vector bosons (leptoquarks) with mass $M_{X}$,
a dimension 6 effective operator allows the proton to decay into $e^{+}\pi^{0}$,
with a partial lifetime $\tau\left(p\rightarrow e^{+}\pi^{0}\right)\thicksim\nicefrac{M_{X}^{4}}{\alpha_{G}^{2}m_{p}^{5}}$
where $\alpha_{G}$ is the unified coupling strength. In the non-supersymmetric
SM, rough unification of the gauge coupling constants happens at $M_{X}\sim10^{14-15}$
GeV, which implies that the proton lifetime is of the order of $10^{30-32}$
years \citep{Langacker:1980js,Langacker:1994vf}. Yet experimental
evidence shows that the proton's lifetime is much higher than this;
Super-Kamiokande in particular has recently reported that $\tau\left(p\rightarrow e^{+}\pi^{0}\right)$
is larger than $1.29\times10^{34}$ years at 90\% confidence level
\citep{Nishino:2012ipa}. In SUSY GUTs though, unification happens
at higher energies, $M_{X}\sim10^{16}$ GeV, and consequently $\tau\left(p\rightarrow e^{+}\pi^{0}\right)$
is predicted to be of the order of $10^{34-38}$ years (see \citep{EmmanuelCosta:2003pu,Raby:2008gh}
and references therein). On the other hand, exchanges of sparticles
(the supersymmetric partners of known particles) can lead to dangerous
dimension 5 operators \citep{Farrar:1978xj,Sakai:1981pk,Weinberg:1981wj}
inducing the decay $p\rightarrow K^{+}\overline{\nu}$, which is experimentally
constrained to be very rare as well ($\tau\left(p\rightarrow K^{+}\overline{\nu}\right)>3.3\times10^{33}$
years at 90\% confidence level \citep{Miura:2010zz}). Viable SUSY
GUTs must therefore suppress or forbid altogether these operators.
Also, in SUSY theories there are often renormalizable operators violating
both baryon and lepton number which are gauge invariant. They are
usually forbidden by the introduction of a special discrete symmetry,
known as an R-symmetry. The MSSM is one such case, as we shall see
latter on.

\subsection{Massive neutrinos}

A few years after the existence of neutrinos was postulated to explain
the continuous beta decay spectrum, it was realized that their interaction
cross section with matter is extremely small \citep{Bethe:1934qn}.
With a mean free path of several light-years in water (for typical
beta decay energies), detection of such an elusive particle seemed
all but impossible at the time. Yet neutrino physics has evolved remarkably,
and nowadays it is possible to not only detect them but also measure
some of their properties.

One of the most significant recent developments was the discovery
that neutrinos have mass. In 1998 the Super-Kamiokande provided evidence
that neutrinos oscillate \citep{Fukuda:1998mi}, explaining the lower
than expected solar neutrino flux, which had been both predicted and
measured three decades earlier \citep{Bahcall:1964gx,Davis:1964hf,Davis:1968cp}.
Since then, other solar, atmospheric, reactor and accelerator experiments
have confirmed this phenomenon. Oscillations imply that neutrinos
are massive and that leptons mix, but the Standard Model cannot accommodate
this experimental evidence: neutrinos cannot have a Dirac mass (there
are no right-handed neutrinos) nor a Majorana mass ($B-L$ is preserved
even in non-perturbative processes \citep{Belavin:1975fg,'tHooft:1976fv,'tHooft:1976up,Klinkhamer:1984di}),
so they are strictly massless in the Standard Model and therefore
no leptonic mixing occurs.

This is an important and far reaching topic, which will be discussed
in some detail latter on, in chapter \ref{chap:Lepton-flavour-violation}.
There, we review the known neutrino properties, as well as some of
the theoretical frameworks that may account for them.

\subsection{Dark matter, dark energy and gravity}

For many decades, it has been known from the observation of the rotation
speed of galaxies \citep{Zwicky:1933gu} that either something is
wrong with the laws of gravity, or with the assumed sources of gravity.
Over time, strong evidence has appeared in support of the latter hypothesis:
it seems that there is more mass producing gravity beyond the ordinary
one in stars and gas clouds. For example, in the Bullet Cluster \citep{Clowe:2006eq},
two colliding galaxy clusters leave behind interstellar gas (containing
most of the ordinary baryonic matter), while most of the mass in the
clusters goes right through one another, as measured via the weak
gravitational lensing effect. This indicates that gravitational anomalies
are localized and unrelated to normal matter.

At bigger, cosmological scales there is also strong evidence for the
existence of non-luminous and non-baryonic cold dark matter (DM).
From the precise measurement of the small anisotropies in the Cosmic
Microwave Background (CMB), of the order of 1 part in 100000, it is
estimated that the energy density of dark matter $\rho_{cdm}$ in
the Universe is \citep{Ade:2013lta}
\begin{alignat}{1}
\Omega_{cdm} & =\frac{\rho_{cdm}}{\rho_{c}}=0.263\pm0.013\,.
\end{alignat}
Here, $\rho_{c}=\frac{3H^{2}}{8\pi G}$ is the critical density of
the Universe, $H$ is the Hubble parameter valued today at $H_{0}=67.4\pm1.4$
km/s/Mpc by the Planck space observatory, and $G$ is the gravitational
constant. This value of $\rho_{cdm}$ is much bigger than the energy
density of baryons $\rho_{B}$, but lower than the one of the mysterious
dark energy $\rho_{\Lambda}$, both of which are also obtainable from
the CMB \citep{Ade:2013lta}:
\begin{alignat}{1}
\Omega_{b} & =\frac{\rho_{b}}{\rho_{c}}=0.049\pm0.002\,,\\
\Omega_{\Lambda} & =\frac{\rho_{\Lambda}}{\rho_{c}}=0.686\pm0.020\,.\label{eq:DarkEnergy_Omega}
\end{alignat}

Most of the dark matter must be made of some particle or particles
yet to be discovered, and in order for its relic density to match
observations, it must be completely stable or very long lived. It
is also non-luminous and weakly interacting, having no electric charge
nor color. Also, observations of the large scale structure of the
Universe suggest that this dark matter must be mostly cold; in other
words, it must be made of non-relativistic massive particles. This
last property rules out an early dark matter candidate---the neutrinos
in the SM or the MSSM \citep{Marx:1972aa,*Cowsik:1972gh,*Cowsik:1973yj}
(it is now known that $\Omega_{\nu}\lesssim0.01$\citep{Ade:2013lta}).
However, right-handed or sterile neutrinos are still viable candidates.

Given the observational evidence, dark matter is most likely predominantly
made of particles with no electric charge nor color, with low velocities,
and which are stable or very long lived. This last requirement is
necessary in order to explain the observed dark matter relic abundance.
There are in fact many dark matter candidates; the following are some
of the best motivated ones (see for example \citep{Bertone:2004pz}):
\begin{itemize}
\item Axions introduced as part of a solution to the strong CP problem,
and their superpartners axinos.
\item Kaluza-Klein excitations of SM fields, introduced in extra dimensions
theories.
\item Gravitinos, the superpartners of gravitons in theories where supersymmetry
is imposed locally.
\item Superpartners of SM particles, in particular the lightest supersymmetric
particle (LSP) which often is a neutralino.
\end{itemize}
Concerning the last candidate in the above list, we note that with
the introduction of R-parity in the MSSM to prevent the fast decay
of the proton, the model gains automatically a convincing DM candidate.
The reason is simple: the R-charge assignment is such that SM fields
get a multiplicatively conserved quantum number $+1$, while their
yet to be discovered superpartners get $-1$ charges under this symmetry.
Every term in the Lagrangian is invariant under this charge assignment,
therefore perturbative processes must involve an even number of sparticles,
and so the LSP cannot decay.

There is currently an ongoing effort by many collaborations to detect
dark matter through its non-gravitational interactions. Direct detection
experiments are buried deep underground in order to reduce background
events and discern the rare scattering of dark matter particles off
atomic nuclei. Indirect detection experiments, on the other hand,
aim at observing the annihilation or decay products of dark matter,
such as photons, electrons, positrons and neutrinos.

Direct detection experiments have produced conflicting results so
far. The DAMA/NaI experiment and its successor DAMA/LIBRA have measured
a signal with an annual modulation \citep{Bernabei:2003za,Bernabei:2010mq}
consistent with the varying relative speed of dark matter particles
(with a mass of $\sim10$ or $\sim70$ GeV) as Earth orbits around
the Sun. Such a modulation was also seen by the CoGeNT collaboration
\citep{Aalseth:2011wp}, in this case pointing to a $\sim10$ GeV
dark matter particle. Also, the CRESST experiment reported \citep{Angloher:2011uu}
a statistical significant excess of events. However, this set of results
appears to be in contradiction \citep{Kopp:2011yr} with the non-observation
of an excess of events and/or an annual modulation by the CDMS \citep{Ahmed:2010wy,Ahmed:2012vq}
and XENON collaborations \citep{Aprile:2012nq}. Seemingly complicating
matters further, in 2008 the CDMS observed two DM candidate events
with germanium detectors (with low statistical significance) and recently,
the same collaborations claims to have seen three more such events
with its silicon detectors, consistent with a $6-15$ GeV dark matter
particle \citep{Agnese:2013rvf}. The CDMS signal appears to be compatible
with CoGeNT data, but not with the DAMA and CRESST results \citep{Agnese:2013rvf}.

Indirect detection experiments have also produced some positive results
(for a review, see \citep{Cirelli:2012tf}). The PAMELA \citep{Adriani:2008zr},
FERMI \citep{FermiLAT:2011ab} and AMS-02 \citep{Aguilar:2013qda}
experiments have measured an excess of positrons over the expected
astrophysical background. Also, ATIC-2 \citep{Chang:2008aa}, PPB-BETS
\citep{Torii:2008xu}, FERMI \citep{Abdo:2009zk} and HESS \citep{Aharonian:2008aa,Aharonian:2009ah}
see an excess of electrons plus positrons with energies of hundreds
of GeV. However, no such excess of anti-protons was seen by PAMELA
\citep{Adriani:2008zq,Adriani:2010rc}. If these signals are due to
the annihilation of dark matter particles, taken together, they imply
that dark matter is leptofilic, heavy (with a mass of $\thicksim$TeV)
and with an annihilation cross section two orders of magnitude bigger
than the one needed to explain its relic abundance. Finally, we note
that there has been some claims of dark matter signals also in photon
data; a notable one is that there is a gamma-ray line at around 130
GeV in FERMI's data \citep{Bringmann:2012vr,Weniger:2012tx}. \\
~

From the measurements of the anisotropies of the CMB, baryonic and
dark matter account for only 30\% of the energy density of the Universe.
The remaining 70\% are due to some dark energy (see equation \eqref{eq:DarkEnergy_Omega}),
which does not change much with time and appears to be roughly constant
throughout space. It behaves like a cosmological constant in Einstein's
equations, explaining therefore the observed accelerated expansion
of the Universe \citep{Perlmutter:1998np,Riess:1998cb}. The simplest,
microscopical explanation for the cosmological constant is that it
is due to the vacuum energy of some field(s). Even though these calculations
have not been carried out rigorously, barring some cancellations,
one expects that this energy density is roughly
\begin{align}
\rho_{DE}^{\textrm{th}} & \thicksim\int_{0}^{\Lambda}k^{2}\sqrt{k^{2}+m^{2}}dk\thicksim\Lambda^{4}\,,
\end{align}
for some cut-off energy $\Lambda$. Taking this $\Lambda$ to be equal
to the Planck mass ($\thicksim10^{18}$ GeV) or even the lower EW
scale ($\thicksim100$ GeV) yields a dark energy density dozens of
orders of magnitude above the observed one, $\rho_{DE}^{\textrm{obs}}\thicksim\left(10^{-3}\textrm{ eV}\right)^{4}$,
and this the so-called cosmological constant problem (see for example
\citep{Carroll:2001xs}). Related to this, it is also puzzling that
the energy density of the vacuum turns out to be, at the present time,
of the same order of magnitude as the matter energy density (the coincidence
problem).

~

To conclude this brief review of gravity related shortcomings of the
SM, we also point out that the gravitational force itself is not described
by the model. This is not a deficiency of the Standard Model in particular
though, as it is well known that Quantum Mechanics in general is incompatible
with General Relativity. Quantum Mechanics is very successful at describing
small scale physics, while General Relativity has been shown to accurately
explain the large scale dynamics of very massive systems. A back of
the envelope calculation shows that at energies of the order of the
Planck scale one would expect that quantum as well as gravitational
effects become relevant, therefore a fundamental understanding of
the laws of Physics appears to require a quantum theory of gravity.
Nevertheless, it has been difficult to combine the two theories because
they are very different in nature: in quantum field theory the metric
is a background entity where fields propagate, while in General Relativity
it is a classical but dynamical entity. Highlighting the challenging
nature of uniting these two theories, almost a century of research
has yielded many competing and unproven theories of quantum gravity,
such a string theory, quantum loop gravity, supergravity, noncommutative
geometry and twistor theory, just to name a few (for a recent review
of these research programs, see the introductory section of \citep{Ashtekar:2012np}).
We note in this regard that the Planck scale is much higher than the
energies we can currently probe, so without dramatic experimental
progress, it seems unlikely that this state of affairs will change.

\subsection{Baryogenesis and leptogenesis}

From the discussion above, one concludes that 95\% of the content
of the Universe is currently unknown. The remaining 5\% are known
to be mostly baryons, but even here there is a mystery (for reviews
on this topic, see \citep{Riotto:1999yt,Davidson:2008bu,turner1994early}).
For temperatures lower than the proton mass, but before freeze-out,
the baryon and anti-baryon number densities, $n_{b}$ and $n_{\overline{b}}$,
divided by the photon number density $n_{\gamma}$ was
\begin{alignat}{1}
\frac{n_{b}}{n_{\gamma}},\frac{n_{\overline{b}}}{n_{\gamma}} & \thicksim\left(\frac{m_{p}}{T}\right)^{\frac{3}{2}}\exp\left(-\frac{m_{p}}{T}\right)\,.\label{eq:baryogenesis_density_standard}
\end{alignat}
Using the known annihilation cross section for baryons, $\left\langle \sigma v\right\rangle \sim m_{\pi}^{-2}\sim\left(\unit[100]{MeV}\right)^{-2}$,
we conclude that freeze-out happens for a temperature of about $\unit[20]{MeV}$.
However, this number is too low; it is 50 times smaller than the proton
mass, meaning that the exponential factor in equation \eqref{eq:baryogenesis_density_standard}
heavily suppresses the number of baryons and anti-baryons, leading
to ratios $\nicefrac{n_{b}}{n_{\gamma}},\nicefrac{n_{\overline{b}}}{n_{\gamma}}$
of the order of $10^{-18}$ which would persist to this day. The observed
baryon abundance (making up 5\% of the Universe) implies a much larger
value of $\nicefrac{n_{b}}{n_{\gamma}}$ though. 

Yet another puzzle is that there seems to be very little anti-matter,
despite the well know $CPT$ symmetry relating matter and anti-matter.
One possibility is that the Universe is made of regions with matter/anti-matter
only, and that we happen to live in the middle of one of them. Since
protons and anti-protons annihilate into detectable photons, these
regions would need to be separated from one another by big gaps, possibly
of the order of megaparsecs. This leads us to a second possibility,
which is that those domains do not really exist, and the average baryon
density $n_{b}$ of the Universe is indeed bigger than its anti-baryon
density $n_{\overline{b}}$. From the anisotropies of the CMB we know
that \citep{Ade:2013lta}
\begin{alignat}{1}
\eta\equiv\frac{n_{b}-n_{\overline{b}}}{n_{\gamma}} & =\left(6.05\pm0.09\right)\times10^{-10}\,.
\end{alignat}
Big Bang nucleosynthesis (BBN) provides another, independent measurement
of $\eta$, because the production rate of some light nuclei, in particular
$^{2}\textrm{H}$, $^{3}\textrm{He}$, $^{4}\textrm{He}$ and $^{7}\textrm{Li}$,
depends on this parameter. The value obtained in this way \citep{Beringer_mod:1900zz},
\begin{alignat}{1}
\eta & =\left(5.8\pm0.7\right)\times10^{-10}\,,
\end{alignat}
agrees with the CMB value, although the abundance of $^{7}\textrm{Li}$
(and $^{6}\textrm{Li}$) seems to point to a smaller $\eta$.

It is conceivable, in principle, that the Universe started with a
baryon asymmetry which explains $\eta$, but this is unlikely because
the Universe's inflationary period would have washed out this small
initial asymmetry. So it seems more likely that at the beginning $n_{b}=n_{\overline{b}}$,
and over time a non-zero $\eta$ was dynamically generated. It turns
out that the SM contains all the necessary ingredients to generate
a baryonic asymmetry, even though it is a small one. These necessary
ingredients were written down long ago \citep{Sakharov:1967dj}: in
an out-of-equilibrium setting, there must be violation of baryon number,
as well as violation of the $C$ and $CP$ symmetries. The SM fulfills
all these conditions:
\begin{enumerate}
\item Due to its gauge invariance, in the SM it is possible to assign a
baryon and a lepton number ($B$ and $L$) to the various fields,
which are conserved quantities in perturbative processes. However,
these symmetries of the classical action are not symmetries of the
quantum field theory, with conserves only $B-L$. The source of $B+L$
violation is the following. Not all gauge field configurations can
be smoothly transformed into one another, and in particular there
are infinite inequivalent configurations that minimize the energy.
Separated by a potential barrier with a height $\nicefrac{4\pi v}{g}\sim\unit[5]{TeV}$,
these vacua solutions correspond to different values of $B+L$ and
so there are solutions to the field equations (instantons \citep{Belavin:1975fg,'tHooft:1976fv,'tHooft:1976up}
and sphalerons \citep{Klinkhamer:1984di}) related to transitions
between these vacua which violate $B+L$. At low temperatures, the
corresponding tunneling rate is negligible, but at the high temperatures
of the early Universe this suppression can be overcome and the production
rate of sphalerons can be significant \citep{Kuzmin:1985mm}.
\item The V-A structure of weak interactions violates the charge conjugation
symmetry $C$ maximally, and through the Cabibbo\textendash{}Kobayashi\textendash{}Maskawa
(CKM) matrix $V$ \citep{Kobayashi:1973fv} they also violate the
$CP$ symmetry: $\textrm{Im}\left(V_{ij}V_{kl}V_{il}^{*}V_{kj}^{*}\right)\equiv J\sum_{m,n}\varepsilon_{ikm}\varepsilon_{jln}$,
where $J=2.96_{-0.16}^{+0.20}\times10^{-5}$ is the Jarlskog invariant
\citep{Jarlskog:1985cw,Jarlskog:1985ht}. Nevertheless, this value
appears to be too small to generate the observed baryon asymmetry
of the Universe \citep{Gavela:1993ts,Gavela:1994dt,Huet:1994jb}.%
\footnote{The quantum chromodynamics  (QCD) $\theta$ parameter is also a source
of CP violation, but it too is very small (maybe even null). %
}
\item The SM provides out-of-equilibrium dynamics near its electroweak phase
transition. When the Universe's temperature dropped below the electroweak
energy scale, in the middle of a EW symmetric plasma, bubbles started
to form where the Higgs field had a (non-null) vacuum expectation
value (VEV). These bubbles would have grown until they filled all
space. For this to have happened though, this phase transition must
have precise properties, and in particular the Higgs mass would have
to be smaller than 70 GeV \citep{Bochkarev:1987wf,Kajantie:1995kf}
(for a review of this topic, see \citep{Morrissey:2012db}).
\end{enumerate}
So, generating the observed amount of baryons through the electroweak
phase transition is not possible in the SM, but nevertheless it is
an interesting idea which may work, for example, within the MSSM \citep{Delepine:1996vn,Cline:1998hy,Laine:1998qk,Carena:2008rt}
although recent LHC data seem to disfavor it \citep{Carena:2008vj,Curtin:2012aa,Chung:2012vg,Huang:2012wn,Cohen:2012zza}.

Another possibility is that the baryon asymmetry of the Universe
was created from a leptonic one \citep{Fukugita:1986hr}. As we shall
review in chapter \ref{chap:Lepton-flavour-violation}, the SM needs
to be extended in order to give mass to neutrinos, and if neutrinos
are Majorana particles their mass term violates lepton number by two
units. In the most conventional case, there are 3 heavy right-handed
neutrinos $N_{i}$, with masses $M_{i}$ and couplings $Y_{i}^{\nu}N_{i}LH$
to left-handed leptons and the Higgs doublet, which give a small mass
$m_{i}^{\nu}=\nicefrac{\left(Y_{i}^{\nu}v\right)^{2}}{M_{i}}$ to
the observed left-handed neutrinos through the seesaw mechanism. Produced
in the early Universe, these heavy states would have generated an
asymmetry of $L$'s through CP violating decays, which in turn would
have been transformed in a baryonic asymmetry by $B-L$ preserving
sphalerons. This amounts to baryogenesis through leptogenesis.

\section{Supersymmetry and the MSSM}

Supersymmetry \citep{Golfand:1971iw,Wess:1973kz,Wess:1974tw} addresses
some of the SM shortcomings discussed previously. It consists of a
symmetry which extends in a non-trivial way the one of Special Relativity
and, for this reason, its irreducible representations---the supermultiplets---contain
different irreducible representations of the Poincaré group (each
labeled with a helicity/spin, and a mass). In the following we review
the main features of the Minimal Supersymmetric Standard Model (MSSM)
which are mentioned latter on, throughout this thesis. Nevertheless,
a more abstract discussion of SUSY can be read in chapter \ref{chap:Symmetry},
which is dedicated exclusively to the use of symmetry in Particle
Physics.

Some models are more supersymmetric than others. Yet, even a minimal
amount of supersymmetry, sometimes called simple or $N=1$ supersymmetry,
is enough to severely restrict the couplings of a gauge theory. In
order to write down the renormalizable Lagrangian of such a theory,
a superpotential $W$ is built,
\begin{alignat}{1}
W & =\frac{1}{6}Y^{ijk}\Phi_{i}\Phi_{j}\Phi_{k}+\frac{1}{2}\mu^{ij}\Phi_{i}\Phi_{j}+L^{i}\Phi_{i}\,,\label{eq:Introduction_superpotential}
\end{alignat}
which is a cubic function of the chiral supermultiplets $\Phi_{i}$,
each containing a Weyl fermion $\psi_{i}$ and a scalar $\phi_{i}$.
Alternatively, the $\Phi_{i}$ can be viewed as the scalar component
$\phi_{i}$ of the chiral supermultiplets, in which case the Lagrangian
density of a gauge theory associated to $W$ can be written as
\begin{alignat}{1}
\mathscr{L}_{\textrm{SUSY}}= & -D^{\mu}\phi^{i*}D_{\mu}\phi_{i}+i\psi^{i\dagger}\overline{\sigma}^{\mu}D_{\mu}\psi_{i}-\frac{1}{2}\left(\frac{\delta^{2}W}{\delta\phi_{i}\delta\phi_{j}}\psi_{i}\psi_{j}+\textrm{h.c.}\right)-\frac{\delta W}{\delta\phi_{i}}\left(\frac{\delta W}{\delta\phi_{i}}\right)^{*}\nonumber \\
 & -\frac{1}{4}F_{\mu\nu}^{a}F^{a\mu\nu}+i\lambda^{a\dagger}\overline{\sigma}^{\mu}D_{\mu}\lambda_{a}-\frac{1}{2}g^{2}\left[\left(T^{a}\right)_{ij}\phi^{i*}\phi_{j}\right]^{2}-g\kappa^{a}\left(T^{a}\right)_{ij}\phi^{i*}\phi_{j}\nonumber \\
 & -\left[\sqrt{2}g\left(T^{a}\right)_{ij}\phi^{i*}\psi_{j}\lambda^{a}+\textrm{h.c.}\right]\,,\label{eq:LSusy}
\end{alignat}
with the field strength tensor and covariant derivatives as follows:
\begin{alignat}{1}
F_{\mu\nu}^{a} & =\partial_{\mu}A_{\nu}^{a}-\partial_{\nu}A_{\mu}^{a}+gf^{abc}A_{\mu}^{b}A_{\nu}^{c}\,,\\
D_{\mu}\lambda^{a} & =\partial_{\mu}\lambda^{a}+gf^{abc}A_{\mu}^{b}\lambda^{c}\,,\\
D_{\mu}\phi_{i} & =\partial_{\mu}\phi_{i}-igA_{\mu}^{a}\left(T^{a}\right)_{ij}\phi_{j}\,,\\
D_{\mu}\psi_{i} & =\partial_{\mu}\psi_{i}-igA_{\mu}^{a}\left(T^{a}\right)_{ij}\psi_{j}\,.
\end{alignat}
Here, $g$ is the gauge coupling constant (the gauge group is assumed
to be simple), $f^{abc}$ are the group structure constants, and $T^{a}$
are the representation matrices for each chiral supermultiplet. Note
that the Fayet-Iliopoulos term with the $\kappa$ parameter is only
allowed for $U(1)$ gauge factors.

Supersymmetry pairs each fermion with a boson of equal mass. Since
this is not seen experimentally, if SUSY is a symmetry of Nature,
it must be a broken one at low energies. However, breaking it carelessly
leads to a generic, non-supersymmetric theory with the scalar mass
stability problem discussed previously. Therefore, in order to keep
SUSY's good properties, namely the cancellation of quadratic divergences,
the breaking terms must be of dimension lower than 4. In other words,
SUSY breaking parameters must be dimensionful. If this is the case,
supersymmetry is said to be only softly broken \citep{Girardello:1981wz}.
The most generic form usually considered for such terms is given by
the following Lagrangian density,%
\footnote{In some conditions \citep{Martin:1997ns}, it is also possible to
add Dirac masses to gauginos, $\psi_{i}\lambda_{a}$, and trilinear
couplings $\phi^{i*}\phi_{j}\phi_{k}$.%
} $\mathscr{L}_{\textrm{soft}}$, which should be added to $\mathscr{L}_{\textrm{SUSY}}$
in equation \eqref{eq:LSusy}:
\begin{alignat}{1}
-\mathscr{L}_{\textrm{soft}} & =\left(\frac{1}{2}M_{a}\lambda^{a}\lambda^{a}+\frac{1}{6}h^{ijk}\phi_{i}\phi_{j}\phi_{k}+\frac{1}{2}b^{ij}\phi_{i}\phi_{j}+s^{i}\phi_{i}+\textrm{h.c.}\right)+\left(m^{2}\right)_{j}^{i}\phi_{i}\phi^{j*}\,.\label{eq:Introduction_Lsoft}
\end{alignat}
Even though quadratic dependencies on the cutoff scale are gone, scalar
masses in general, and the Higgs one in particular, still depend quadratically
on other particle masses. As such, there is a clear motivation for
an electroweak scale supersymmetry, where the soft masses are perhaps
an order of magnitude, at most, above the Higgs vacuum expectation
value.

If the SM is supersymmetrized, adding as little new fields as possible,
we end up with the MSSM---the Minimal Supersymmetric Standard Model.
Two of its features stand out. First, one Higgs doublet is no longer
enough%
\footnote{There are two reasons for this. The first one is that the superpotential
must be a holomorphic function of the superfields, which means that
the SM trick of using the doublets $H$ and $H^{*}$ to give mass
to down and up quarks/leptons cannot be used in supersymmetric theories.
The second one is that simply doubling the number of SM fields would
give rise to an anomalous theory, because of the introduction of a
fermionic partner of the Higgs doublet.%
} so the MSSM contains two, $H_{u}$ and $H_{d}$, with different hypercharges.
One of them ($H_{d}$) is in the same representation of the gauge
group as left-handed leptons therefore, unless some differentiation
is introduced, $H_{d}$ is a fourth generation of left-handed leptons.
Also, as discussed already in relation to the proton decay in GUTs,
it is necessary to suppress or restrict altogether some baryon number
violating couplings which are allowed by the gauge symmetry. To this
end, in the MSSM there is a $Z_{2}$-symmetry called R-parity, under
which the down Higgs doublet and left-handed leptons are assigned
different charges. The full chiral content of the MSSM is given in
table \eqref{tab:MSSMChiralContent}, while gauge bosons and gauginos
are presented in table \eqref{tab:MSSMGaugeSupermultiplets}.

\begin{table}[h]
\begin{centering}
\setlength{\tabcolsep}{2.9pt}
\thinmuskip=1mu
\medmuskip=1mu
\thickmuskip=1mu%
\begin{tabular}{cccccc}
\toprule 
\addlinespace[0.1cm]
\begin{tabular}{@{}r@{}}
Super-~~~\tabularnewline
multiplet\tabularnewline
\end{tabular} & Boson & Fermion & $\begin{array}{c}
U(1)_{Y}\times SU(2)_{L}\\
\times SU(3)_{C}
\end{array}$ & $Z_{2}\left(R\right)$ & Flavors\tabularnewline\addlinespace[0.1cm]
\midrule
\addlinespace[0.1cm]
$\widehat{Q}$ & \selectlanguage{portuges}%
$\widetilde{Q}=\left(\widetilde{u}_{L},\widetilde{d}_{L}\right)^{T}$\selectlanguage{english}
 & \selectlanguage{portuges}%
$Q=\left(u_{L},d_{L}\right)^{T}$\selectlanguage{english}
 & $\left(+\frac{1}{6},\mathbf{2},\mathbf{3}\right)$ & -1 & 3\tabularnewline\addlinespace[0.1cm]
\addlinespace[0.1cm]
$\widehat{U}^{c}$ & \selectlanguage{portuges}%
$\widetilde{u}_{R}^{*}$\selectlanguage{english}
 & \selectlanguage{portuges}%
$u_{R}^{\dagger}$\selectlanguage{english}
 & $\left(-\frac{2}{3},\mathbf{1},\overline{\mathbf{3}}\right)$ & -1 & 3\tabularnewline\addlinespace[0.1cm]
\addlinespace[0.1cm]
$\widehat{D}^{c}$ & \selectlanguage{portuges}%
$\widetilde{d}_{R}^{*}$\selectlanguage{english}
 & \selectlanguage{portuges}%
$d_{R}^{\dagger}$\selectlanguage{english}
 & $\left(+\frac{1}{3},\mathbf{1},\overline{\mathbf{3}}\right)$ & -1 & 3\tabularnewline\addlinespace[0.1cm]
\addlinespace[0.1cm]
$\widehat{L}$ & \selectlanguage{portuges}%
$\widetilde{L}=\left(\widetilde{\nu}_{L},\widetilde{e}_{L}\right)^{T}$\selectlanguage{english}
 & \selectlanguage{portuges}%
$L=\left(\nu_{L},e_{L}\right)^{T}$\selectlanguage{english}
 & $\left(-\frac{1}{2},\mathbf{2},\mathbf{1}\right)$ & -1 & 3\tabularnewline\addlinespace[0.1cm]
\addlinespace[0.1cm]
$\widehat{E}^{c}$ & \selectlanguage{portuges}%
$\widetilde{e}_{R}^{*}$\selectlanguage{english}
 & \selectlanguage{portuges}%
$e_{R}^{\dagger}$\selectlanguage{english}
 & $\left(+1,\mathbf{1},\mathbf{1}\right)$ & -1 & 3\tabularnewline\addlinespace[0.1cm]
\addlinespace[0.1cm]
$\widehat{H}_{u}$ & \selectlanguage{portuges}%
$H_{u}=\left(H_{u}^{+},H_{u}^{0}\right)^{T}$\selectlanguage{english}
 & \selectlanguage{portuges}%
$\widetilde{H}_{u}=\left(\widetilde{H}_{u}^{+},\widetilde{H}_{u}^{0}\right)^{T}$\selectlanguage{english}
 & $\left(+\frac{1}{2},\mathbf{2},\mathbf{1}\right)$ & +1 & 1\tabularnewline\addlinespace[0.1cm]
\addlinespace[0.1cm]
$\widehat{H}_{d}$ & \selectlanguage{portuges}%
$H_{d}=\left(H_{d}^{0},H_{d}^{-}\right)^{T}$\selectlanguage{english}
 & \selectlanguage{portuges}%
$\widetilde{H}_{d}=\left(\widetilde{H}_{d}^{0},\widetilde{H}_{d}^{-}\right)^{T}$\selectlanguage{english}
 & $\left(-\frac{1}{2},\mathbf{2},\mathbf{1}\right)$ & +1 & 1\tabularnewline\addlinespace[0.1cm]
\bottomrule
\end{tabular}
\par\end{centering}

\setlength{\tabcolsep}{6pt}

\caption{\label{tab:MSSMChiralContent}Chiral superfields in the MSSM}

\end{table}

\begin{table}[h]
\begin{centering}
\begin{tabular}{cccc}
\toprule 
\addlinespace[0.1cm]
Fermion & Boson & $U(1)_{Y}\times SU(2)_{L}\times SU(3)_{C}$ & $Z_{2}\left(R\right)$\tabularnewline\addlinespace[0.1cm]
\midrule
\addlinespace[0.1cm]
\selectlanguage{portuges}%
$\widetilde{B}$\selectlanguage{english}
 & \selectlanguage{portuges}%
$B$\selectlanguage{english}
 & $\left(0,\mathbf{1},\mathbf{1}\right)$ & +1\tabularnewline\addlinespace[0.1cm]
\addlinespace[0.1cm]
\selectlanguage{portuges}%
$\widetilde{W}^{a}$\selectlanguage{english}
 & $W^{a}$ & $\left(0,\mathbf{3},\mathbf{1}\right)$ & +1\tabularnewline\addlinespace[0.1cm]
\addlinespace[0.1cm]
\selectlanguage{portuges}%
$\widetilde{g}^{a}$\selectlanguage{english}
 & $g^{a}$ & $\left(0,\mathbf{1},\mathbf{8}\right)$ & +1\tabularnewline\addlinespace[0.1cm]
\bottomrule
\end{tabular}
\par\end{centering}

\thinmuskip=3mu
\medmuskip=4.0mu plus 2.0mu minus 4.0mu
\thickmuskip=5.0mu plus 5.0mu

\caption{\label{tab:MSSMGaugeSupermultiplets}Vector superfields in the MSSM}
\end{table}

The MSSM is completely specified by the chiral supermultiplet representations
under the gauge group, and the $Z_{2}\left(R\right)$ charge assignments.
All that remains is to write down the superpotential and soft SUSY
breaking Lagrangian, using some notation for the model parameters:
\begin{alignat}{1}
W & =Y_{ij}^{u}\widehat{U}_{i}^{c}\widehat{Q}_{j}\cdot\widehat{H}_{u}+Y_{ij}^{d}\widehat{D}_{i}^{c}\widehat{Q}_{j}\cdot\widehat{H}_{d}+Y_{ij}^{\ell}\widehat{E}_{i}^{c}\widehat{L}_{j}\cdot\widehat{H}_{d}+\mu\widehat{H}_{u}\cdot\widehat{H}_{d}\,,\label{eq:Introduction_MSSM_W}\\
-\mathscr{L}_{\textrm{soft}} & =\left[\frac{1}{2}M_{1}\widetilde{B}\widetilde{B}+\frac{1}{2}M_{2}\widetilde{W}^{a}\widetilde{W}^{a}+\frac{1}{2}M_{3}\widetilde{g}^{a}\widetilde{g}^{a}+h_{ij}^{u}\widetilde{u}_{Ri}^{*}\widetilde{Q}_{j}\cdot H_{u}\right.\nonumber \\
 & \left.+h_{ij}^{d}\widetilde{d}_{Ri}^{*}\widetilde{Q}_{j}\cdot H_{d}+h_{ij}^{\ell}\widetilde{e}_{Ri}^{*}\widetilde{L}_{j}\cdot H_{d}+bH_{u}\cdot H_{d}+\textrm{h.c.}\vphantom{\frac{1}{2}}\right]\nonumber \\
 & +\left(m_{\widetilde{Q}}^{2}\right)_{ij}\widetilde{Q}_{i}^{*}\widetilde{Q}_{j}+\left(m_{\widetilde{u}}^{2}\right)_{ji}\widetilde{u}_{Ri}^{*}\widetilde{u}_{Rj}+\left(m_{\widetilde{d}}^{2}\right)_{ji}\widetilde{d}_{Ri}^{*}\widetilde{d}_{Rj}\nonumber \\
 & +\left(m_{\widetilde{L}}^{2}\right)_{ij}\widetilde{L}_{i}^{*}\widetilde{L}_{j}+\left(m_{\widetilde{e}}^{2}\right)_{ji}\widetilde{e}_{Ri}^{*}\widetilde{e}_{Rj}+m_{H_{u}}^{2}H_{u}^{*}H_{u}+m_{H_{d}}^{2}H_{d}^{*}H_{d}\,.\label{eq:Introduction_MSSM_Lsoft}
\end{alignat}
The R-charges of the superfields are in tables \eqref{tab:MSSMChiralContent}
and \eqref{tab:MSSMGaugeSupermultiplets}, but frequently it is more
helpful to view things in terms of the bosons and fermions forming
a superfield. In this respect, we note that R-symmetries are special
since they express a symmetry of the SUSY algebra, which may or may
not be respected in a given model (see chapter \ref{chap:Symmetry}).
It is however important to note that they do not commute with supersymmetries
and so, it turns out that in the particular case of the MSSM or any
other $N=1$ supersymmetric model, the fermion and the boson in a
superfield are oppositely charged under R-parity. We can write this
charge as a function of the spin $s$ and the familiar baryon and
lepton numbers \citep{Farrar:1978xj}:
\begin{align}
R & =\left(-1\right)^{-2s+3B+L}\,.
\end{align}
It is tacitly assumed that the superfields $\widehat{Q}$ ($B=\nicefrac{1}{3}$)
and $\widehat{U}^{c}$, $\widehat{D}^{c}$ ($B=-\nicefrac{1}{3}$)
are the only ones with a non-null baryon number, while $\widehat{L}$
($L=1$) and $\widehat{E}^{c}$ ($L=-1$) are the only ones with a
non-null lepton number. Without R-parity, lepton number violating
terms $\widehat{L}\widehat{L}\widehat{E}^{c}$, $\widehat{L}\widehat{D}^{c}\widehat{Q}$,
$\widehat{L}\widehat{H}_{u}$ as well as the baryon number violation
term $\widehat{U}^{c}\widehat{D}^{c}\widehat{D}^{c}$ would be allowed
in the superpotential. Yet the stability of the proton requires that
the coupling of one of these two sets of terms must be very small
or null \citep{Dreiner:1997uz,Bhattacharyya:1997vv,Diaz:1997xc,Allanach:1999ic,Hirsch:2000ef,Barbier:2004ez}.

The EW symmetry breaking and the mass eigenstates of this supersymmetrized
version of the SM are the following. From the two Higgs VEVs, $\left\langle H_{u}^{0}\right\rangle $
and $\left\langle H_{d}^{0}\right\rangle $, we may compute a SM-like
one, $v=\sqrt{\left\langle H_{u}^{0}\right\rangle ^{2}+\left\langle H_{d}^{0}\right\rangle ^{2}}$,
and parametrize the ratio $\left\langle H_{u}^{0}\right\rangle /\left\langle H_{d}^{0}\right\rangle $
with an angle $\beta$: 
\begin{alignat}{1}
\left\langle H_{u}^{0}\right\rangle  & \equiv v_{u}=v\sin\beta\,,\quad\quad\left\langle H_{d}^{0}\right\rangle \equiv v_{d}=v\cos\beta\,.
\end{alignat}
At tree level, in addition to the $U\left(1\right)_{Y}$ and $SU\left(2\right)_{L}$
coupling constants, the neutral Higgs potential depends only on the
$\mu$, $b$, $m_{H_{u}}^{2}$ and $m_{H_{d}}^{2}$ parameters. As
such, there must to be a connection between these parameters and $\left\{ \beta,v\right\} $,
or alternatively $\left\{ \beta,m_{Z}\right\} $. This connection
is provided by the minimization conditions of the scalar potential:
\begin{alignat}{1}
\sin2\beta & =\frac{2b}{2\left|\mu\right|^{2}+m_{H_{u}}^{2}+m_{H_{d}}^{2}}\,,\label{eq:Introduction_switchMuB1}\\
m_{Z}^{2} & =\frac{\left|m_{H_{d}}^{2}-m_{H_{u}}^{2}\right|}{\sqrt{1-\sin^{2}2\beta}}-m_{H_{u}}^{2}-m_{H_{d}}^{2}-2\left|\mu\right|^{2}\,.\label{eq:Introduction_switchMuB2}
\end{alignat}
Using these relations, the variables $\left\{ \left|\mu\right|,b\right\} $
can then be swapped by $\left\{ m_{Z},\tan\beta\right\} $, which
is common practice. Note however that by opting to use the latter
ones, the sign of $\mu$ must still be provided.

Doubling the number of Higgs fields in the MSSM yields a total of
5 massive Higgs bosons after EW symmetry breaking. Three are neutral
(the CP-even $h^{0}$ and $H^{0}$, and the CP-odd $A^{0}$) and two
are charged ($H^{\pm}$):
\begin{alignat}{1}
\left(\begin{array}{c}
H_{u}^{0}\\
H_{d}^{0}
\end{array}\right) & =\left(\begin{array}{c}
v_{u}\\
v_{d}
\end{array}\right)+\frac{1}{\sqrt{2}}\left(\begin{array}{cc}
\cos\alpha & \sin\alpha\\
-\sin\alpha & \cos\alpha
\end{array}\right)\left(\begin{array}{c}
h^{0}\\
H^{0}
\end{array}\right)\nonumber \\
 & \qquad\qquad\qquad\qquad\qquad\qquad+\frac{i}{\sqrt{2}}\left(\begin{array}{cc}
\sin\beta & \cos\beta\\
-\cos\beta & \sin\beta
\end{array}\right)\left(\begin{array}{c}
G^{0}\\
A^{0}
\end{array}\right)\,,\\
\left(\begin{array}{c}
H_{u}^{+}\\
H_{d}^{-*}
\end{array}\right) & =\left(\begin{array}{cc}
\sin\beta & \cos\beta\\
-\cos\beta & \sin\beta
\end{array}\right)\left(\begin{array}{c}
G^{+}\\
H^{+}
\end{array}\right)\,.
\end{alignat}
The three $G$ fields in these expressions are the pseudo-Nambu-Goldstone
bosons \citep{Nambu:1960tm,Goldstone:1961eq}, which become the longitudinal
components of the massive EW bosons. The mixing angle $\alpha$ is
given by
\begin{alignat}{1}
\frac{\tan2\alpha}{\tan2\beta} & =\frac{2\left|\mu\right|^{2}+m_{H_{u}}^{2}+m_{H_{d}}^{2}+m_{Z}^{2}}{2\left|\mu\right|^{2}+m_{H_{u}}^{2}+m_{H_{d}}^{2}-m_{Z}^{2}}\,.
\end{alignat}

\begin{figure}[tbph]
\begin{centering}
\includegraphics[width=0.95\textwidth]{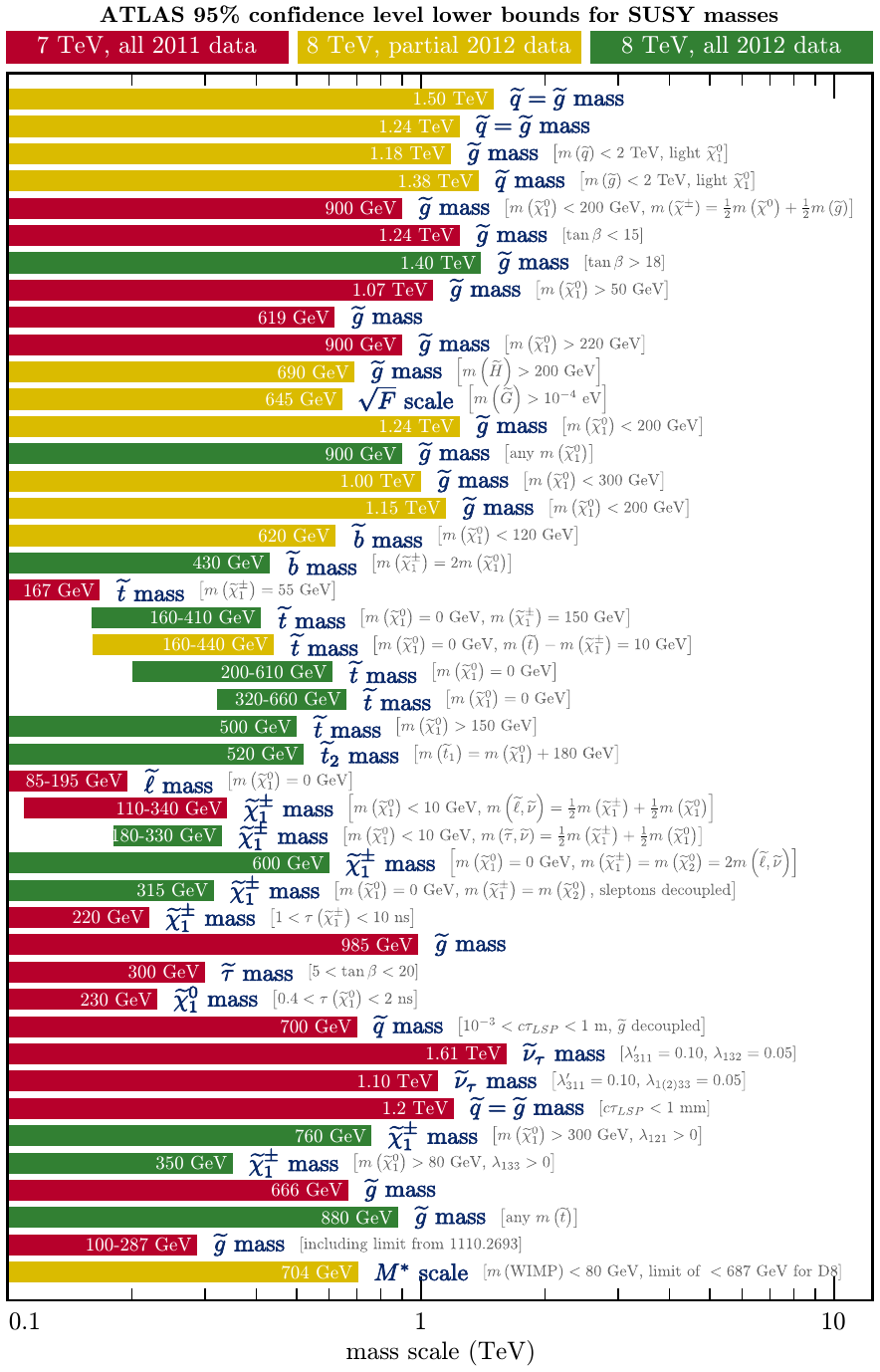}
\par\end{centering}

\caption{\label{fig:Introduction_ATLAS-95=000025-confidence}ATLAS 95\% confidence
level lower bounds for some SUSY masses and scales, obtained through
the analysis of different signals. Edited from \citep{website_ATLAS_SUSY_results}
(see this reference for details). The CMS collaboration has produced
comparable results \citep{Chatrchyan:2013sza}.}
\end{figure}

At tree level, the mass of the Higgs particles are the following:
\begin{alignat}{1}
m_{A^{0}}^{2} & =2\left|\mu\right|^{2}+m_{H_{u}}^{2}+m_{H_{d}}^{2}\,,\\
m_{h^{0},H^{0}}^{2} & =\frac{1}{2}\left(m_{A^{0}}^{2}+m_{Z}^{2}\mp\sqrt{\left(m_{A^{0}}^{2}-m_{Z}^{2}\right)^{2}+4m_{Z}^{2}m_{A^{0}}^{2}\sin^{2}2\beta}\right)\,,\\
m_{H^{\pm}}^{2} & =m_{A^{0}}^{2}+m_{W}^{2}\,.
\end{alignat}
In this approximation, the mass of $h^{0}$ (the lightest Higgs) increases
with $m_{A^{0}}^{2}$, so there is an upper bound for it which is
reached in the limit $m_{A^{0}}\gg m_{Z}$ \citep{Inoue:1982ej,Flores:1982pr}:
\begin{alignat}{1}
m_{h^{0}} & <m_{Z}\left|\cos2\beta\right|<m_{Z}\,.\label{eq:HiggsLowerMassLimit}
\end{alignat}
This stringent, electroweak related bound on the lightest Higgs mass
follows from the fact that $h^{0}$'s quartic coupling $\lambda$
is fixed and proportional to $g^{2}+g'^{2}$, unlike in the SM where
it is a free parameter. The mass value in equation \eqref{eq:HiggsLowerMassLimit}
was nevertheless excluded by LEP2, which set a 114.4 GeV lower limit
on the Higgs mass, at 95\% confidence level (CL) \citep{Barate:2003sz}.
This result seemingly precludes SUSY at the EW scale, or at least
a very big set of SUSY models, such as the MSSM. However, the Higgs
mass is known to be very sensitive to radiative corrections; the quadratic
dependence on the cutoff scale has been eliminated, but there are
still quadratic dependencies on the splitting between particle and
sparticle masses introduced by $\mathscr{L}_{\textrm{soft}}$. Therefore,
in theory, the lightest Higgs in the MSSM can have an arbitrarily
large mass, but in order not to reintroduce a fine tuning problem
in the theory, these radiative corrections should not be too large.
There is no unique and objective criterion to evaluate this, but there
is a general belief that the sparticles masses should not be much
heavier than 1 TeV. Yet, to this date, LHC searches for gluino and
squarks have found nothing (see figure \eqref{fig:Introduction_ATLAS-95=000025-confidence}).
However, the recent discovery of a 125 GeV Higgs particle is compatible
with a 1 TeV scale supersymmetry, because the limit on $m_{h^{0}}$
is raised to 135 GeV by radiative corrections (see \citep{Martin:1997ns}
and references therein). The main contribution usually comes from
an incomplete cancellation between top and stop loops \citep{Okada:1990vk,Ellis:1990nz,Haber:1990aw},
which can be written in an approximate way as follows (see also \citep{Carena:2002es,*Djouadi:2005gj}):
\begin{align}
m_{h^{0}}^{2} & <m_{Z}^{2}+\frac{3g^{2}m_{t}^{4}}{8\pi^{2}m_{W}^{2}}\left[\log\left(\frac{m_{S}^{2}}{m_{t}^{2}}\right)+\frac{X_{t}^{2}}{m_{S}^{2}}\left(1-\frac{X_{t}^{2}}{12m_{S}^{2}}\right)\right]\,,\label{eq:Introduction_mh0_radiative_correction}
\end{align}
where $X_{t}\equiv A_{t}-\mu\cot\beta$ ($h^{x}\equiv A_{x}Y^{x}$)
is the stop mixing parameter and $m_{S}\equiv\sqrt{m_{\widetilde{t}_{1}}m_{\widetilde{t}_{2}}}$
is their geometric mean mass.
 The stops $\widetilde{t}_{1}$ and $\widetilde{t}_{2}$ are mainly
a mixture of the superpartners of the right and left handed stops.
In fact, while the effect is more pronounced in the third generation,
all squarks and all sleptons mix amongst themselves. Consider the
following squark and slepton mass terms,\arraycolsep=2.7pt
\begin{multline}
-\mathscr{L}=\cdots+\left(\begin{array}{cc}
\widetilde{u}_{L}^{*} & \widetilde{u}_{R}^{*}\end{array}\right)M_{\widetilde{u}}^{2}\left(\begin{array}{c}
\widetilde{u}_{L}\\
\widetilde{u}_{R}
\end{array}\right)+\left(\begin{array}{cc}
\widetilde{d}_{L}^{*} & \widetilde{d}_{R}^{*}\end{array}\right)M_{\widetilde{d}}^{2}\left(\begin{array}{c}
\widetilde{d}_{L}\\
\widetilde{d}_{R}
\end{array}\right)\\
+\left(\begin{array}{cc}
\widetilde{e}_{L}^{*} & \widetilde{e}_{R}^{*}\end{array}\right)M_{\widetilde{\ell}}^{2}\left(\begin{array}{c}
\widetilde{e}_{L}\\
\widetilde{e}_{R}
\end{array}\right)+\widetilde{\nu}_{L}^{*}M_{\widetilde{\nu}}^{2}\widetilde{\nu}_{L}+\cdots\,.\label{eq:Introduction_sparticle_masses_Init}
\end{multline}
\arraycolsep=5.0ptSince there are three generations of chiral superfields,
$M_{\widetilde{u}}^{2}$, $M_{\widetilde{d}}^{2}$, $M_{\widetilde{\ell}}^{2}$
are $6\times6$ matrices and $M_{\widetilde{\nu}}^{2}$ is a $3\times3$
mass matrix. They have the following block form:
\begin{alignat}{1}
M_{\widetilde{u}}^{2} & =\left(\begin{array}{cc}
m_{\widetilde{Q}}^{2}+Y^{u\dagger}Y^{u}\left|v_{u}\right|^{2}+\Delta_{\widetilde{u}_{L}}\mathbb{1} & \left(\textrm{h.c.}\right)\\
h^{u}v_{u}-\mu^{*}Y^{u}v_{d}^{*} & m_{\widetilde{u}}^{2}+Y^{u}Y^{u\dagger}\left|v_{u}\right|^{2}+\Delta_{\widetilde{u}_{R}}\mathbb{1}
\end{array}\right)\,,\label{eq:Introduction_sparticle_masses_1}\\
M_{\widetilde{d}}^{2} & =\left(\begin{array}{cc}
m_{\widetilde{Q}}^{2}+Y^{d\dagger}Y^{d}\left|v_{d}\right|^{2}+\Delta_{\widetilde{d}_{L}}\mathbb{1} & \left(\textrm{h.c.}\right)\\
h^{d}v_{d}-\mu^{*}Y^{d}v_{u}^{*} & m_{\widetilde{d}}^{2}+Y^{d}Y^{d\dagger}\left|v_{d}\right|^{2}+\Delta_{\widetilde{d}_{R}}\mathbb{1}
\end{array}\right)\,,\label{eq:Introduction_sparticle_masses_2}\\
M_{\widetilde{\ell}}^{2} & =\left(\begin{array}{cc}
m_{\widetilde{L}}^{2}+Y^{\ell\dagger}Y^{\ell}\left|v_{d}\right|^{2}+\Delta_{\widetilde{e}_{L}}\mathbb{1} & \left(\textrm{h.c.}\right)\\
h^{\ell}v_{d}-\mu^{*}Y^{\ell}v_{u}^{*} & m_{\widetilde{e}}^{2}+Y^{\ell}Y^{\ell\dagger}\left|v_{d}\right|^{2}+\Delta_{\widetilde{e}_{R}}\mathbb{1}
\end{array}\right)\,,\\
M_{\widetilde{\nu}}^{2} & =\left(m_{\widetilde{L}}^{2}+\Delta_{\widetilde{\nu}_{L}}\mathbb{1}\right)\,,\label{eq:Introduction_sparticle_masses_3}
\end{alignat}
where the $\Delta$'s depend on the electric charge $Q$ and the $SU\left(2\right)_{L}$
isospin $T_{3}$ of each field, 
\begin{alignat}{1}
\Delta_{\phi} & \equiv\left[\left(T_{3\phi}-Q_{\phi}\right)m_{Z}^{2}+Q_{\phi}m_{W}^{2}\right]\cos2\beta\,.
\end{alignat}
The off-diagonal blocks in these mass matrices mix left and right
sparticles, but in some situations they can be ignored for the first
two generations. On the other hand, mixing between stops, sbottoms
and staus is significant, therefore $\widetilde{t}_{\nicefrac{L}{R}}$,
$\widetilde{b}_{\nicefrac{L}{R}}$ and $\widetilde{\tau}_{\nicefrac{L}{R}}$
give rise to mass eigenstates $\widetilde{t}_{\nicefrac{1}{2}}$,
$\widetilde{b}_{\nicefrac{1}{2}}$ and $\widetilde{\tau}_{\nicefrac{1}{2}}$.
For stops in particular, this mixing is enhanced with a big $X_{t}$
factor, and so a 125 GeV Higgs boson seems to imply a large top trilinear
coupling $A_{t}$ and/or a large $\tan\beta$.%
\footnote{The radiative corrections to $m_{h^{0}}^{2}$ given by equation \eqref{eq:Introduction_mh0_radiative_correction}
are maximal for $X_{t}=\sqrt{6}M_{S}$ and minimal when $X_{t}=0$. %
} However, these two parameters also have implications for low-energy
Physics; in particular many processes violating charged lepton number
depend on the sixth power of $\tan\beta$, while $\textrm{BR}\left(B_{s}\rightarrow\mu\mu\right)$
is sensitive to $A_{t}$. We shall mention this again in chapter \ref{chap:Lepton-flavour-violation}.

The fermionic superpartners of the Higgs and electroweak vector bosons---the
Higgsinos and electroweak gauginos---also mix, forming mass eigenstates
known as neutralinos $\chi_{1,2,3,4}^{0}$ (neutral) and charginos
$\chi_{1,2}^{\pm}$ (charged):
\begin{align}
-\mathscr{L} & =\frac{1}{2}\widetilde{N}^{T}M_{\widetilde{N}}\widetilde{N}+\frac{1}{2}\widetilde{C}^{T}M_{\widetilde{C}}\widetilde{C}+\textrm{h.c.}+\cdots\,,
\end{align}
with $\widetilde{N}=\left(\widetilde{B},\widetilde{W}^{0},\widetilde{H}_{d}^{0},\widetilde{H}_{u}^{0}\right)^{T}$,
$\widetilde{C}=\left(\widetilde{W}^{+},\widetilde{H}_{u}^{+},\widetilde{W}^{-},\widetilde{H}_{d}^{-}\right)^{T}$
and the matrices $M_{\widetilde{N}}$, $M_{\widetilde{C}}$ are given
by
\begin{align}
M_{\widetilde{N}} & =\left(\begin{array}{cccc}
M_{1} & . & . & (\textrm{sym.})\\
0 & M_{2} & . & .\\
-\frac{g'v_{d}}{\sqrt{2}} & \frac{gv_{d}}{\sqrt{2}} & 0 & .\\
\frac{g'v_{u}}{\sqrt{2}} & -\frac{gv_{u}}{\sqrt{2}} & -\mu & 0
\end{array}\right)\,,\label{eq:Introduction_sparticle_masses_4}\\
M_{\widetilde{C}} & =\left(\begin{array}{cccc}
0 & . & . & (\textrm{sym.})\\
0 & 0 & . & .\\
M_{2} & gv_{u} & 0 & .\\
gv_{d} & \mu & 0 & 0
\end{array}\right)\,.\label{eq:Introduction_sparticle_masses_End}
\end{align}
\medskip{}

The MSSM as described above contains 124 physical real degrees of
freedom \citep{Dimopoulos:1995ju,Haber:1997if}, which hardly makes
it a minimal model in terms of number of parameters. The SM, with
just 19 degrees of freedom, is able to describe successfully most
of the experimental data, so it is not surprising that a large portion
of the MSSM's bigger parameter space is already excluded. Unsuppressed
couplings and mass terms can lead to large charged lepton flavor violation,
flavor changing neutral currents and/or new sources of CP violation
which lead to big electric dipole moments, all of which are experimentally
ruled out (see chapter \ref{chap:Lepton-flavour-violation}). For
this reason, analyzing the phenomenology of the MSSM is complicated
unless its parameter space is reduced by taking into account low energy
experimental data. The phenomenological MSSM (pMSSM) \citep{Djouadi:1998di}
is one commonly used constrained model, where there are only 19 input
parameters (although this model is often generalized): it assumes
that there are no new complex phases in $\mathscr{L}_{\textrm{soft}}$,
that the soft SUSY breaking scalar masses are diagonal and the same
for the first/second generations, and also that trilinear scalar couplings
are null for the first/second generations.

From a theoretical point of view, the large parameter space of the
MSSM is also a problem because this model is more fundamental than
the SM, and yet it is less predictive. The source of this problem
is our lack of knowledge of the SUSY breaking mechanism, which forces
the introduction of the most general soft SUSY breaking Lagrangian.
Knowing this, several \textit{ansätze}  for $\mathscr{L}_{\textrm{soft}}$
 were proposed over the years by different authors, inspired on different
SUSY breaking mechanisms. We shall mention here only the minimal supergravity
inspired MSSM (mSUGRA) \citep{Chamseddine:1982jx,Barbieri:1982eh,Hall:1983iz},
also known as the constrained MSSM (cMSSM) \citep{Kane:1993td}, which
hypothesizes that SUSY is broken in some hidden sector which only
communicates with the visible one through the gravitational interaction.
The assumptions on the parameters at the GUT scale $m_{G}$ (which
is the scale at which the gauge couplings unify) are the following:
\begin{enumerate}
\item Gaugino mass unification: $M_{1}\left(m_{G}\right)=M_{2}\left(m_{G}\right)=M_{3}\left(m_{G}\right)=M_{1/2}$.
\item Universal scalar SUSY breaking masses: $m_{\widetilde{Q}}^{2}=m_{\widetilde{L}}^{2}=m_{\widetilde{u}}^{2}=m_{\widetilde{d}}^{2}=m_{\widetilde{e}}^{2}\equiv m_{0}^{2}\mathbb{1}$
and $m_{H_{u}}^{2}=m_{H_{d}}^{2}=m_{0}^{2}$.
\item Universal trilinear SUSY breaking terms: $h^{x}\equiv A_{0}Y^{x}$,
$x=u,d,\ell$.
\end{enumerate}
As for the parameters $\mu$ and $b$, through the equations \eqref{eq:Introduction_switchMuB1}
and \eqref{eq:Introduction_switchMuB2} they can be replaced by $\tan\beta$
and $\textrm{sign}\mu$. Therefore, mSUGRA requires only 5 input parameters:
$M_{1/2}$ (the common gaugino mass), $m_{0}$ (the common scalar
mass), $A_{0}$ (the common trilinear coupling factor), $\tan\beta$
(the ratio of neutral Higgs VEVs at the electroweak scale), and $\textrm{sign}\mu$.
The limited parameter space, as well as the smallness of the induced
CP and flavor violations, makes the mSUGRA interesting. However, the
discovery at CERN of a heavy Higgs implies that SUSY masses must be
heavier than originally thought, and this is particularly true for
the mSUGRA since there are less ways to raise $m_{h^{0}}$ without
raising the whole sparticle spectrum \citep{Arbey:2011ab,Carena:2011aa,Heinemeyer:2011aa,Baer:2011ab,Draper:2011aa,Kadastik:2011aa,Gunion:2012zd,Aparicio:2012iw,Ellis:2012aa,Baer:2012uya,Hirsch:2012ti,Baer:2012mv,Mayes:2013qmc}.
In chapter \ref{chap:Revisiting_RK} we shall encounter a generalization
of the mSUGRA, where the Higgs scalar masses $m_{H_{u}}^{2}$ and
$m_{H_{d}}^{2}$ are free parameters. This Non-Universal Higgs Masses
(NUHM) model \citep{Berezinsky:1995cj,Nath:1997qm} in interesting
because it decouples the Higgs masses from the other scalar masses,
potentially enhancing the amplitude of some Higgs mediated processes.
\cleartooddpage

\chapter{\label{chap:Lepton-flavour-violation}Lepton flavor violation}

\section{\label{sec:Massive-neutrinos}Massive neutrinos}

\subsection{A brief history of neutrino experiments}

Pontecorvo was the first to suggest that detection of neutrinos was
feasible, using big amounts of liquid chlorine \citep{Pontecorvo:1946mv}.
With this method, in 1956 Cowan and Reines detected reactor anti-neutrinos
$\overline{\nu}_{e}$ \citep{Cowan:1992xc} (their initial idea was
to use nuclear explosions \citep{Close:2010zz}) and years latter,
in 1962 and 2000, neutrinos associated with the muon and the tau were
also directly observed \citep{Danby:1962nd,Kodama:2000mp}. In the
meantime, several experiments were built to detect neutrinos of extraterrestrial
origin. Nuclear reactions in the Sun's core make it by far the biggest
available source of such neutrinos in the detectable energy range
($\thicksim\textrm{MeV}$). When the flux of these neutrinos was first
measured \citep{Davis:1968cp}, it did not match the theoretical prediction
which was derived with remarkable precision from nuclear cross sections
and life-times \citep{Bahcall:1964gx}. It was realized \citep{Pontecorvo:1957cp,Maki:1962mu,Pontecorvo:1967fh}
that neutrino oscillations/conversion could be the reason for the
solar neutrino deficit \citep{Wolfenstein:1977ue,Mikheev:1986gs,Mikheev:1986wj}:
most of the Sun's energy is produced by the proton-proton chain reaction,
where protons combine to produce heavier nuclei (such as deuterium,
$^{2}\textrm{He}$, $^{3}\textrm{He}$, $^{4}\textrm{He}$, $^{7}\textrm{Li}$,
$^{7}\textrm{Be}$, $^{8}\textrm{Be}$ and $^{8}\textrm{B}$) as well
as positrons and electron neutrinos $\nu_{e}$. If these neutrinos
change their nature between production and detection, the measured
flux can be lower than the calculated value. After mounting evidence
from the Homestake \citep{Davis:1968cp}, Kamiokande \citep{Hirata:1989zj},
SAGE \citep{Abazov:1991rx}, and GALLEX \citep{Anselmann:1992um}
experiments, in 2002 the SNO collaboration confirmed that this is
indeed the explanation for the solar neutrino problem \citep{Ahmad:2001an}.

A few years earlier though, the heavy water Cherenkov detector Super-Kamiokande,
built to detect proton decay, had already found evidence of the oscillation
of neutrinos with higher energies ($\thicksim\textrm{GeV}$) produced
by the collisions of cosmic rays with nuclei in the upper atmosphere
\citep{Fukuda:1998mi}. These collisions produce pions which then
decay into electron and muon neutrinos, and data showed a deficit
of the latter which could however be explained by the oscillation
of $\nu_{\mu}$ into another neutrino type, possibly $\nu_{\tau}$.
The fact that this deficit depended on the neutrino flight distance
(or equivalently the zenith angle) made this evidence even more compelling.

Other solar, atmospheric, reactor and beam neutrino experiments have
since then provided increasingly accurate values of the neutrino oscillation
parameters (see for example \citep{Balantekin:2013tqa} for an up-to-date
review of the experimental and theoretical progress).

\subsection{Neutrino oscillation parameters}

Assuming that neutrinos are massive and that leptons mix, the mass
eigenstates $\nu_{i}$ of non-degenerate neutrinos are a linear combination
of the neutrino states $\nu_{\alpha}$ produced in weak interactions.
These two bases are related by the leptonic mixing matrix $U$---the
so called Pontecorvo-Maki-Nakagawa-Sakata (PMNS) matrix:
\begin{alignat}{1}
\nu_{\alpha} & =\sum_{i}U_{\alpha i}\nu_{i}\,.
\end{alignat}

In the plane wave formalism, ultrarelativistic neutrinos $\nu_{\alpha}$
with an energy $E$ have a probability of oscillating to a different
flavor $\nu_{\beta}$, after traveling a distance $L$ in vacuum,
which is given by \citep{Bilenky:1987ty}
\begin{alignat}{1}
P\left(\nu_{\alpha}\rightarrow\nu_{\beta}\right) & =\delta_{\alpha\beta}-4\sum_{i<j}\textrm{Re}\left(J_{\alpha\beta}^{ij}\right)\sin^{2}\varphi_{ij}+2\sum_{i<j}\textrm{Im}\left(J_{\alpha\beta}^{ij}\right)\sin2\varphi_{ij}\,,\label{eq:GenericNuOscillationProbability}
\end{alignat}
where
\begin{alignat}{1}
J_{\alpha\beta}^{ij}\equiv U_{\alpha i}^{*}U_{\beta i}U_{\alpha j}U_{\beta j}^{*} & \,,\quad\varphi_{ij}\equiv\frac{\Delta m_{ij}^{2}L}{4E}\,,\quad\Delta m_{ij}^{2}=m_{i}^{2}-m_{j}^{2}\,.\label{eq:Jij_and_varphiij}
\end{alignat}
This expression implies that oscillation experiments are only sensitive
to squared mass differences. In other words, shifting all squared
masses by a constant term, $m_{i}^{2}\rightarrow m_{i}^{2}+M^{2}$,
does not change the oscillation probabilities.

It is instructive to consider just two neutrino flavors. In such case,
there is only one mass squared difference $\Delta m^{2}$ and the
mixing matrix is described by a single angle $\theta$: $U_{11}=U_{22}=\cos\theta$
and $U_{12}=-U_{21}=\sin\theta$. In this simplified scenario, equation
\eqref{eq:GenericNuOscillationProbability} is given by
\begin{alignat}{1}
P\left(\nu_{\alpha}\rightarrow\nu_{\beta}\right) & =\sin^{2}2\theta\sin^{2}\frac{\Delta m^{2}L}{4E}\,,\,\alpha\neq\beta\,.
\end{alignat}
There are two important quantities here: the oscillation depth $\sin^{2}2\theta$
which corresponds to the maximum oscillation probability ($\theta=\pm\nicefrac{\pi}{4}$
maximizes it), and the oscillation length $l_{\nu}$ given by
\begin{alignat}{1}
l_{\nu} & \equiv\frac{4\pi E}{\Delta m^{2}}\approx2.48\frac{E}{\unit[1]{GeV}}\frac{\unit[1]{eV^{2}}}{\Delta m^{2}}\unit{km}\,.\label{eq:oscillation_length}
\end{alignat}

The picture that emerges from various experiments is that the three
neutrinos weakly interacting with charged leptons mix into three mass
eigenstates. The two mass squared differences and the mixing angles
obtained from a global analysis are given in table \eqref{tab:Neutrino_parameter_values}.
From matter effects in the Sun (see below), it is known that the mass
eigenstate denoted by $\nu_{1}$ is lighter than $\nu_{2}$, and since
$\left|\Delta m_{31}^{2}\right|\gg\Delta m_{21}^{2}$, there are two
possible orderings of the three neutrino masses, depending on the
sign of $\Delta m_{31}^{2}$: the normal hierarchy (NH) $m_{1}<m_{2}<m_{3}$
and the inverted hierarchy (IH) $m_{3}<m_{1}<m_{2}$.

\begin{table}[tbph]
\begin{centering}
\begin{tabular}{ll}
\toprule 
\addlinespace[0.1cm]
Parameter & Best fit $\pm$$1\sigma$ errors\tabularnewline\addlinespace[0.1cm]
\midrule
\addlinespace[0.1cm]
$\Delta m_{21}^{2}$$(\unit[10^{-5}]{eV^{2}})$ & $7.62_{-0.19}^{+0.19}$\tabularnewline\addlinespace[0.1cm]
\addlinespace[0.1cm]
$\left|\Delta m_{31}^{2}\right|$$(\unit[10^{-3}]{eV^{2}})$ & $\begin{cases}
2.55_{-0.09}^{+0.06} & \textrm{NH}\\
2.43_{-0.06}^{+0.07} & \textrm{IH}
\end{cases}$\tabularnewline\addlinespace[0.1cm]
\addlinespace[0.1cm]
$\sin^{2}\theta_{12}$ & $0.320_{-0.017}^{+0.016}$\tabularnewline\addlinespace[0.1cm]
\addlinespace[0.1cm]
$\sin^{2}\theta_{23}$ & $\begin{cases}
0.613_{-0.04}^{+0.022}\left(0.427_{-0.027}^{+0.034}\right) & \textrm{NH}\\
0.600_{-0.031}^{+0.026} & \textrm{IH}
\end{cases}$\tabularnewline\addlinespace[0.1cm]
\addlinespace[0.1cm]
$\sin^{2}\theta_{13}$ & $\begin{cases}
0.0246_{-0.0028}^{+0.0029} & \textrm{NH}\\
0.0250_{-0.0027}^{+0.0026} & \textrm{IH}
\end{cases}$\tabularnewline\addlinespace[0.1cm]
\bottomrule
\end{tabular}
\par\end{centering}

\caption{\label{tab:Neutrino_parameter_values}Best fit and $1\sigma$ ranges
for the neutrino parameters obtained from a global three neutrino
oscillation analysis \citep{Tortola:2012te}. The value of $\sin^{2}\theta_{23}$
in parenthesis is also compatible with experimental data. See also
\citep{Fogli:2012ua,GonzalezGarcia:2012sz,NuFit} for comparable values.}
\end{table}

The three angles in table \eqref{tab:Neutrino_parameter_values} are
a reference to the usual parametrization of the lepton mixing matrix,\arraycolsep=2.7pt
\begin{align}
U & =\left(\begin{array}{ccc}
1 & 0 & 0\\
0 & c_{23} & s_{23}\\
0 & -s_{23} & c_{23}
\end{array}\right)\left(\begin{array}{ccc}
c_{13} & 0 & s_{13}e^{-i\delta}\\
0 & 1 & 0\\
-s_{13}e^{i\delta} & 0 & c_{13}
\end{array}\right)\left(\begin{array}{ccc}
c_{12} & s_{12} & 0\\
-s_{12} & c_{12} & 0\\
0 & 0 & 1
\end{array}\right)\,,\label{eq:LFV_Umatrix}
\end{align}
\arraycolsep=5.0ptwhere $s_{ij},c_{ij}\equiv\sin\theta_{ij},\cos\theta_{ij}$,
in analogy to the quark sector (see equation \eqref{eq:Appendix_SM_Vckm}).
This assumes that neutrinos are Dirac particles, with a mass term
$\overline{\nu}_{L}m_{\nu}\nu_{R}+\textrm{h.c.}$, where $\nu_{L}$
are the active neutrinos and $\nu_{R}$ are singlets under the gauge
group (right-handed neutrinos). However, neutrinos can be Majorana
particles, in which case the effective light neutrino mass term is
of the form $\nu_{L}^{T}m_{\nu}C\nu_{L}+\textrm{h.c.}$. If so, neutrinos
can be created and annihilated in pairs, violating lepton number.
In this latter instance, in general it is not possible to remove from
the mixing matrix as many complex phases as in the Dirac case, so
a diagonal phase matrix $\textrm{diag}\left(1,\exp\nicefrac{i\alpha_{21}}{2},\exp\nicefrac{i\alpha_{31}}{2}\right)$
must be added to $U$ in equation \eqref{eq:LFV_Umatrix}. These Majorana
phases ($\alpha_{21}$ and $\alpha_{31}$) have not yet been measured,
and in fact it is still an open question whether or not neutrinos
are Majorana particles. Neutrinoless double beta decay experiments,
described latter on, are and will be attempting to answer this question,
although the determination of the Majorana phases is hampered by uncertainties
in nuclear matrix elements \citep{Pascoli:2001by,Pascoli:2002qm,Barger:2002vy,deGouvea:2002gf,Pascoli:2005zb,Faessler:2008xj}.

Another unanswered question is whether or not $CP$ is violated in
the leptonic sector (see \citep{Branco:2011zb} for a review of this
topic). $CP$ invariance implies that $P\left(\overline{\nu}_{\alpha}\rightarrow\overline{\nu}_{\beta}\right)=P\left(\nu_{\alpha}\rightarrow\nu_{\beta}\right)$,
and since $P\left(\overline{\nu}_{\alpha}\rightarrow\overline{\nu}_{\beta}\right)=P\left(\nu_{\alpha}\rightarrow\nu_{\beta}\right)^{*}$,
this relation can be rewritten as $\sum_{i<j}\textrm{Im}\left(J_{\alpha\beta}^{ij}\right)\sin2\varphi_{ij}=0$.
From this expression, similarly to the quark sector, one concludes
that $CP$ violation requires three neutrinos (or more), all mass
differences and mixing angles must be non-null, and in addition $\delta$
must be different from $0,\pi$, otherwise the mixing matrix $U$
is real. Currently, this Dirac phase $\delta$ is poorly constrained
from data, so much so that at $1\sigma,2\sigma$ CL it can take any
value \citep{Tortola:2012te,Fogli:2012ua}. As such, it is not yet
known if leptons conserve $CP$, but given that $\theta_{13}$ was
recently measured to be quite large, this might be determined in the
near future through a combination of reactor and superbeam data \citep{Branco:2011zb}.

It should be noted that there are some anomalies in data which cannot
be explained within the 3 active neutrino oscillations paradigm. There
is a well known claim concerning the observation of neutrinoless double
beta decay \citep{KlapdorKleingrothaus:2001ke} (see below) and also
some signals that could be pointing towards the existence of sterile
neutrinos: the appearance of electron neutrinos and anti-neutrinos
in the LSND and MiniBoone experiments ($\nu_{\mu}\rightarrow\nu_{e}$,
$\overline{\nu}_{\mu}\rightarrow\overline{\nu}_{e}$) \citep{Aguilar:2001ty,AguilarArevalo:2007it,AguilarArevalo:2010wv,Aguilar-Arevalo:2012eua,Aguilar-Arevalo:2013pmq},
as well as the disappearance of $\nu_{e}$'s in gallium experiments,
and of $\overline{\nu}_{e}$'s in reactor experiments \citep{Acero:2007su,Kaether:2010ag,Giunti:2011gz,Mention:2011rk,Giunti:2012tn}.

\subsection{Matter effects and solar neutrinos}

When neutrinos travel through large bodies, such as the Earth or the
Sun, there are important matter effects that must be taken into account
\citep{Wolfenstein:1977ue,Mikheev:1986gs,Mikheev:1986wj}. Matter
contains electrons but no muon nor taus, so there is a coherent scattering
of $\nu_{e}$'s with electrons only. The effective neutrino Hamiltonian
is changed by this charged current interaction,
\begin{alignat}{1}
H=\frac{1}{2E}U^{\dagger}m^{2}U & \rightarrow H'=H+\sqrt{2}G_{F}n_{e}\textrm{diag}\left(1,0,0\right)\,,
\end{alignat}
where $m^{2}$ may be taken to be $\textrm{diag}\left(0,\Delta m_{21}^{2},\Delta m_{31}^{2}\right)$,
$G_{F}$ is the Fermi constant, and $n_{e}$ is the electron density
in the medium. Importantly, the eigenstates $\nu_{i}^{(m)}$ of $H'$
are not the same as the vacuum ones $\nu_{i}$ (the eigenstates of
$H$). In addition, the three energy eigenvalues also change from
$\nicefrac{m_{i}^{2}}{2E}$ (the eigenvalues of $H$) to some $H_{i}'$
(the eigenvalues of $H'$), so the phases $\varphi_{ij}$ in equation
\eqref{eq:Jij_and_varphiij} must be replaced by $\left(H_{i}'-H_{j}'\right)L$.
This is best seen with only two neutrino flavors: $\nu_{e}$ (electron
neutrinos) and $\nu_{a}$ (muon and tau neutrinos taken together),
with a squared mass splitting $\Delta m^{2}$ and one mixing angle
$\theta$. There are now two distinct and competing scales (see \citep{Smirnov:2003da}
and references contained therein): the oscillation length $l_{\nu}$
previously introduced in equation \eqref{eq:oscillation_length},
and the neutrino refraction length
\begin{alignat}{1}
l_{m} & \equiv\frac{\sqrt{2}\pi}{G_{F}n_{e}}\,.
\end{alignat}
When the two are roughly similar ($l_{\nu}=l_{m}\cos2\theta$ to be
precise) there is a resonance and the mixing angle in matter $\theta^{(m)}$
is maximal ($\sin^{2}2\theta^{(m)}=1$). As such, for neutrinos with
energy $E$ there is a layer in the Sun where the matter density is
close to the one satisfying this resonance condition: $n_{e}^{R}=\nicefrac{\Delta m^{2}\cos2\theta}{2\sqrt{2}EG_{F}}$.
Depending on whether or not the neutrino originates from a region
with $n_{e}$ larger than $n_{e}^{R}$, the neutrino state that emerges
from the Sun can be very different. In this brief discussion, we mention
only the extreme but important case when at the neutrino production
point one has $n_{e}\gg n_{e}^{R}$. In this case, mixing is small
at production ($\theta^{\left(m\right)}\approx\nicefrac{\pi}{2}$),
so the electron neutrino $\nu_{e}$ is essentially an energy eigenstate
$\nu_{2}^{(m)}$. As the neutrino travels through different Solar
layers, there is an electron density gradient which in principle allows
transitions between energy eigenstates $\nu_{2}^{(m)}$ and $\nu_{1}^{(m)}$;
in practice this density variation is small, so this transition can
be neglected (adiabatic approximation). On the other hand, the flavor
decomposition of these two mass eigenstates changes with density,
so a $\nu_{e}=\nu_{2}^{(m)}$ neutrino produced in the Sun's core
emerges from from its surface as a $\nu_{2}^{(m)}=\nu_{2}\neq\nu_{e}$
neutrino. The probability that the electron neutrino survives this
journey is given by $\left|\left\langle \nu_{2}|\nu_{e}\right\rangle \right|^{2}=\sin^{2}\theta$.
We note however that for lower energy neutrinos the conversion probability
is different.

Solar neutrinos then travel in the vacuum until they reach the Earth.
Even though the velocity difference between the ultra relativistic
neutrinos produced in the Sun is small, $\Delta v\approx\nicefrac{\Delta m^{2}}{2E}$,
the Earth-Sun distance is big enough to destroy any coherence that
existed at the source. As such, high energy solar neutrinos ($E\sim\unit[10]{MeV}$)
reach the Earth as an incoherent flux of $\nu_{2}$'s, which still
suffer minor matter effects as they transverse the Earth, giving rise
to a small day/night variation in the detection rate of electron neutrinos,
of the order of a few percent. In summary then, the measured flux
of solar neutrinos is significantly smaller than originally expected
essentially because electron neutrinos are adiabatically converted
in the Sun into other neutrino species.

\subsection{Neutrinoless double beta decay experiments}

Despite the major contribution of oscillation experiments to our understanding
of neutrinos, they are unable to measure all neutrino properties.
In particular, the nature and absolute value of neutrino masses are
not yet known, and one promising way to probe and determine them is
through neutrinoless double beta decay experiments. If neutrinos are
Majorana particles it should be possible for a nuclide $\left(A,Z\right)$
to decay into $\left(A,Z+2\right)+2e$ without the emission of neutrinos,
thus violating lepton number: a neutrinoless double beta decay ($0\nu\beta\beta$).
In fact, if neutrinoless double beta decays are detected, the 6-point
$\overline{u}\overline{u}dd\overline{e}\overline{e}$ effective vertex
responsible for it gives rise to a small neutrino Majorana mass via
a four loop diagram, therefore the two concepts are inseparably linked
\citep{Schechter:1981bd,Nieves:1984sn,Takasugi:1984xr,Hirsch:2006yk}.

To observe such rare decays, normal beta decays must be suppressed.
Some nuclides with an even atomic number $Z$ and an even mass number
$A$ ($^{48}$Ca, $^{76}$Ge, $^{82}$Se, $^{100}$Mo, $^{116}$Cd,$^{130}$Te,
$^{136}$Xe, $^{150}$Nd) are well suited for these experiments, as
their energy is higher than the one of $\left(A,Z+2\right)$ but lower
than the one of $\left(A,Z\pm1\right)$, so decays into these last
states are kinematically forbidden. Even so, the second order lepton
flavor conserving decay $\left(A,Z\right)\rightarrow\left(A,Z+2\right)+2e+2\overline{\nu}_{e}$
($2\nu\beta\beta$) must be taken into account, and because both $0\nu\beta\beta$
and $2\nu\beta\beta$ are rare processes, measuring them is challenging.
Several collaborations have either already attempted to do it or will
try to do so in the future (for a recent review, see \citep{Schwingenheuer:2012zs}).

Neutrinoless double beta decay can arise due to various mechanisms,
for example in SUSY theories \citep{Hirsch:1995zi,Hirsch:1997dm,Mohapatra:1986su}.
Assuming however that this process is driven by the exchange of light
Majorana neutrinos, the half life $T_{\nicefrac{1}{2}}^{0\nu}$ can
be written in a model independent way as
\begin{alignat}{1}
\frac{1}{T_{\nicefrac{1}{2}}^{0\nu}} & =G^{0\nu}\left|ME\right|^{2}m_{ee}^{2}\,,
\end{alignat}
where $G^{0\nu}$ is a phase-space factor \citep{Doi:1985dx}, $ME$
is the nuclear matrix element and
\begin{alignat}{1}
m_{ee} & \equiv\sum_{i}\left|U_{ei}^{2}m_{i}\right|\,.
\end{alignat}
Even though there are sizable theoretical uncertainties in the calculation
of nuclear matrix elements \citep{Menendez:2008jp,Barea:2009zza,Rodriguez:2010mn,Rath:2010zz,Suhonen:2010zzc,Menendez:2011zza,Iachello:2011zzc},
by measuring the $0\nu\beta\beta$ decay rate it is possible to shed
light on the absolute neutrino mass scale and, at least in theory,
on the Majorana phases (see however \citep{Pascoli:2001by,Branco:2002ie,Pascoli:2002qm,Barger:2002vy,deGouvea:2002gf,Pascoli:2005zb,Faessler:2008xj}).
Table \eqref{tab:LFV_UpperBounds_0Nu2BetaDecay} summarizes current
upper bounds on $m_{ee}$ from these experiments. It should be mentioned
here that there is also a well known claim by two members of the Heidelberg-Moscow
group that $0\nu\beta\beta$ decays have already been recorded, yielding
$m_{ee}=0.11-0.56$ at 95\% CL \citep{KlapdorKleingrothaus:2001ke}.
\begin{table}
\begin{centering}
\begin{tabular}{lll}
\toprule 
Experiment & Isotope & $m_{ee}$ upper limit (eV; 90\% CL)\tabularnewline
\midrule 
Heidelberg-Moscow \citep{KlapdorKleingrothaus:2000sn} & $^{76}$Ge & 0.35\tabularnewline
IGEX \citep{Aalseth:2002rf} & $^{76}$Ge & 0.33 -- 1.35\tabularnewline
CUORICINO \citep{Andreotti:2010vj} & $^{130}$Te & 0.30 -- 0.70\tabularnewline
KamLAND-Zen \citep{KamLANDZen:2012aa} & $^{136}$Xe & 0.3 -- 0.6 \tabularnewline
EXO-200 \citep{Auger:2012ar} & $^{136}$Xe & 0.14 -- 0.38\tabularnewline
NEMO-3 \citep{Simard:2012gb} & $^{100}$Mo & 0.31 -- 0.96\tabularnewline
NEMO-3 \citep{Simard:2012gb} & $^{82}$Se & 0.94 -- 2.6\tabularnewline
\bottomrule
\end{tabular}
\par\end{centering}

\caption{\label{tab:LFV_UpperBounds_0Nu2BetaDecay}Upper limits on $m_{ee}$
from different $0\nu\beta\beta$ experiments, obtained using the isotopes
$^{76}$Ge, $^{82}$Se, $^{100}$Mo, $^{130}$Te, $^{136}$Xe. }

\end{table}

\subsection{Beta decay experiments}

A more obvious way to measure the neutrino absolute mass scale, suggested
by Pauli upon postulating its existence, is by measuring the end-point
of the electron spectrum in normal beta decay reactions, which depends
on
\begin{alignat}{1}
m_{\beta} & \equiv\sqrt{\sum_{i}\left|U_{ei}^{2}\right|m_{i}^{2}}\,.
\end{alignat}
Unlike $m_{ee}$ measured by $0\nu\beta\beta$ experiments which can
be zero even for non-null neutrino masses, in $m_{\beta}$ no cancellations
can occur between the different terms involving the neutrino masses
$m_{i}$. By measuring tritium beta decays, the Mainz \citep{Kraus:2004zw}
and Troitzk \citep{Lobashev:2001uu,Aseev:2011dq} experiments have
established that $m_{\beta}<$2.3 eV, 2.2 eV (95\% CL), respectively,
and in the future the KATRIN collaboration is expected to have a $5\sigma$
discovery potential for $m_{\beta}=0.35$ eV. Tritium is an isotope
particularly well suited for these experiments because it possesses
a low electron spectrum end-point (18.6 keV), and at the same time
its short half-life makes it very active. This is important because
it increases the number of events, which is crucial for the measurement
of the electron's spectrum near the endpoint, as it is expected that
only 1 in $5\times10^{12}$ beta decays will produce an electron in
the last eV of the spectrum \citep{Drexlin:2013lha}.

\subsection{Cosmological bounds on neutrino masses}

Neutrino masses influence several astrophysical observables, therefore
it is possible make inferences about them by looking at the Cosmos
(for a pre-Planck review, see \citep{Lesgourgues:2012uu}). In particular,
by combining CMB data from Planck \citep{Ade:2013lta}, WMAP \citep{Komatsu:2010fb}
and high resolution experiments \citep{Keisler:2011aw,Reichardt:2011yv,Story:2012wx,Das:2013zf}
the Planck collaboration obtained the 95\% confidence level limit
\begin{align}
\sum_{\nu} & m_{\nu}<0.66\textrm{ eV}\quad\textrm{[Planck+WMAP+high res.]}
\end{align}
on the sum of neutrino masses. Even though baryon acoustic oscillation
(BAO) \citep{Percival:2009xn,Blake:2011en,Beutler:2011hx,Padmanabhan:2012hf}
are not sensitive to neutrino masses, they break the degeneracy between
some of the other parameters in the standard cosmological model.
As such, using this additional input, the limit on neutrinos masses
is significantly reduced:
\begin{align}
\sum_{\nu} & m_{\nu}<0.23\textrm{ eV}\quad\textrm{[Planck+WMAP+high res.+BAO]}\,.
\end{align}
Matter power spectrum data, on the other hand, is sensitive to $\sum_{\nu}m_{\nu}$.
As such, using data from the WiggleZ Dark Energy Survey \citep{Parkinson:2012vd},
the authors of \citep{Riemer-Sorensen:2013jsa} go even further and
set the limit
\begin{align}
\sum_{\nu} & m_{\nu}<0.15\textrm{ eV}\quad\textrm{[Planck+BAO+WiggleZ]}\,.
\end{align}
Noting that for normal (inverted) hierarchy neutrinos one has $0.05(0.1)\leq\sum_{\nu}m_{\nu}$,
it is conceivable that in the near future the two hierarchies might
be discernible through cosmological observations.

\subsection{The origin and smallness of neutrino masses}

In analogy to quarks and charged leptons, the simplest way to give
mass to neutrinos in the SM is to introduce right-handed neutrinos
$\nu_{R}$:%
\footnote{In the remainder of this section, repeated indices are to be summed
over.%
}
\begin{alignat}{1}
-\mathscr{L}^{I}=\cdots+ & Y_{ij}^{\nu}\overline{\nu}_{Ri}L_{j}\cdot H+\textrm{h.c.}\,.\label{eq:neutrinosYukawaCouplings}
\end{alignat}
One problem with such a lepton number conserving Dirac mass term is
that neutrinos have masses smaller than the electronvolt, implying
that the entries of the Yukawa matrix $Y^{\nu}$ must be of the order
of $10^{-12}$. This is a very small number and it should be compared
to the top's Yukawa coupling, which is close to 1 (such strong fermion
mass hierarchy is connected to the flavor problem which was mentioned
in chapter \ref{chap:The-SM's-shortcomings}). In addition, if we
introduce right-handed neutrinos in the theory, we are bound to include
all terms allowed by the symmetries and not just the Yukawa coupling
in equation \eqref{eq:neutrinosYukawaCouplings}. Since the $\nu_{R}$'s
are fermions and singlets under the SM gauge group, there is only
one extra renormalizable term allowed:
\begin{align}
-\mathscr{L}^{I}=\cdots+ & \frac{1}{2}\nu_{R}^{T}m_{R}^{*}C\nu_{R}+\textrm{h.c.},\quad C\equiv i\gamma^{2}\gamma^{0}\,.
\end{align}
Together with the Dirac mass term shown in equation \eqref{eq:neutrinosYukawaCouplings},
this Majorana mass term for the right-handed neutrinos generates a
Majorana mass for the light, mostly left-handed neutrinos, at tree
level:
\begin{align}
-\mathscr{L}^{I}=\cdots+ & \frac{1}{2}\nu_{L}^{T}m_{\nu}^{I}C\nu_{L}+\textrm{h.c.},\quad m_{\nu}^{I}=-Y^{\nu T}m_{R}^{-1}Y^{\nu}\left\langle H^{0}\right\rangle ^{2}\,.\label{eq:neutrinoSeesawImassMatrix}
\end{align}
To derive this expression, one assumes that right-handed Majorana
neutrino masses are much heavier than Dirac masses, in such a way
that for practical purposes the states $\nu_{R}$ become non-dynamical
and can be integrated out. As such, equation \eqref{eq:neutrinoSeesawImassMatrix}
is to be seen as the $\nu_{L}$ mass generated at tree level by the
exchange of heavy $\nu_{R}$ states, after electroweak symmetry breaking
(EWSB). This scenario, where the heavy mediators are fermions which
are singlets under the Standard Model gauge group, is known as seesaw
type-I. In the basis where both $m_{R}$ and the light neutrino mass
matrix $m_{\nu}^{I}$ are diagonal, the Yukawa matrix $Y^{\nu}$ has
the form \citep{Casas:2001sr}
\begin{align}
Y^{\nu} & =\frac{i}{\left\langle H^{0}\right\rangle }\sqrt{m_{R}}\mathcal{O}\sqrt{m_{\nu}^{I}}U^{\dagger}\,,
\end{align}
where $\mathcal{O}$ is some orthogonal matrix which accounts for
the mixing involving the heavy neutrino states.

We note however that this is just one of many possibilities of generating
a Majorana mass. Gauge invariance does not forbid neutrino masses;
it is rather the accidental global $U(1)_{L}$ symmetry of the renormalizable
SM Lagrangian which does, so once lepton number violating interactions
are introduced, the dimension 5 Weinberg effective operator \citep{Weinberg:1979sa}
\begin{equation}
\frac{c_{ij}}{2}\left(\varepsilon_{\alpha\gamma}\varepsilon_{\beta\delta}L_{\alpha i}^{T}CL_{\beta j}H_{\gamma}H_{\delta}\right)+\textrm{h.c.}\label{eq:LFV_WeinbergOperator}
\end{equation}
is generated (Greek and Roman indices denote $SU(2)_{L}$ and flavor
components, respectively). Consequently, after EWSB neutrinos get
a mass term $\frac{c_{ij}}{2}\left\langle H^{0}\right\rangle ^{2}\nu_{Li}^{T}C\nu_{Lj}+\textrm{h.c.}$
with coefficients $c_{ij}$ of the order $M^{-1}$, where $M$ is
the mass scale of the mechanism which generates the effective Weinberg
operator. Unlike Dirac neutrino masses which require exceedingly small
Yukawa interactions, Majorana neutrino masses are automatically small
even if the Yukawa interactions are large, provided that $M\gg\left\langle H^{0}\right\rangle $.
\begin{figure}[h]
\begin{centering}
\includegraphics[scale=0.8]{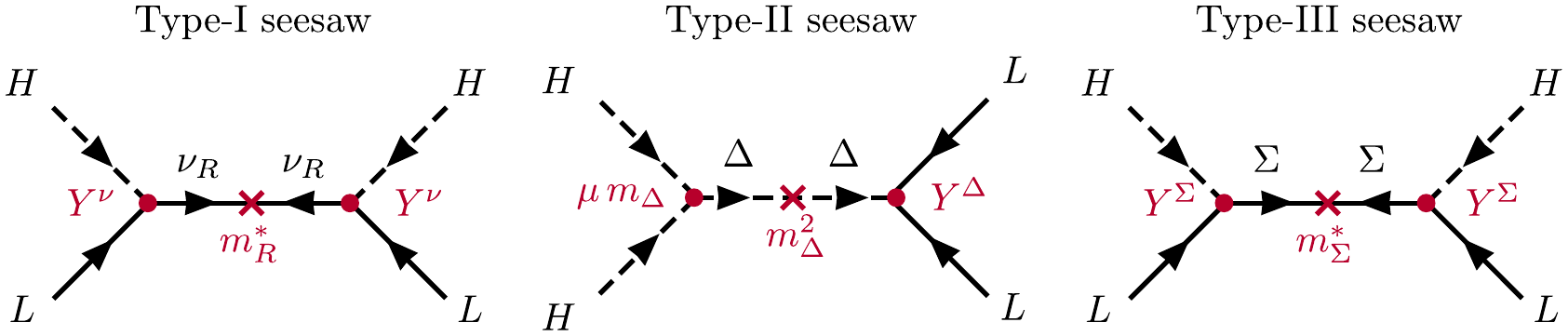}
\par\end{centering}

\caption{Diagrams which generate the different seesaw mechanisms. The mediator
field might be a fermionic singlet (type-I seesaw), a scalar triplet
(type-II seesaw), or a fermionic triplet (type-III seesaw).}
\end{figure}

There are in fact other ways to generate the above operator at tree
level (it can also be generated radiatively \citep{Zee:1980ai,Zee:1985id,Babu:1988ki,Branco:1988ex}).
There ought to be two vertices with $L$'s, $H$'s and some mediator
field. If the two $H$'s are in the same vertex, $HH$ must be in
a triplet representation of $SU(2)_{L}$ (because the singlet combination
is antisymmetric), so the mediator must be a scalar triplet $\Delta=\left(\Delta^{++},\Delta^{+},\Delta^{0}\right)$
(type-II seesaw \citep{Barbieri:1979ag,*Marshak:1980yc,*Cheng:1980qt,*Magg:1980ut,*Lazarides:1980nt,*Schechter:1980gr,*Mohapatra:1980yp}).
If, on the other hand, each vertex contains both an $L$ and an $H$,
then the $LH$ combination can be either in an invariant or in a triplet
representation of $SU\left(2\right)_{L}$, so the mediator field must
be a fermion singlet $\nu_{R}$ (type-I seesaw \citep{Minkowski:1977sc,GellMann:1980vs,*Yanagida:1979as,*Glashow:1979nm,*Mohapatra:1979ia})
or a fermion triplet $\Sigma=\left(\Sigma^{+},\Sigma^{0},\Sigma^{-}\right)$
(type-III seesaw \citep{Foot:1988aq}).

The type-I seesaw mechanism has been described above, in equations
\eqref{eq:neutrinosYukawaCouplings}--\eqref{eq:neutrinoSeesawImassMatrix}.
In type-II seesaw, the scalar triplet $\Delta$ has the following
mass and interaction terms with leptons and the SM Higgs doublet $H$:
\begin{align}
-\mathscr{L}^{II}= & Y_{ij}^{\Delta}\left[\Delta^{++}e_{Li}^{T}Ce_{Lj}-\nicefrac{1}{\sqrt{2}}\Delta^{+}\left(e_{Li}^{T}C\nu_{Lj}+i\leftrightarrow j\right)+\Delta^{0}\nu_{Li}^{T}C\nu_{Lj}\right]\nonumber \\
 & -\mu^{*}m_{\Delta}^{*}\left[\Delta^{++}\left(H^{+*}\right)^{2}+\sqrt{2}\Delta^{+}H^{+*}H^{0*}+\Delta^{0}\left(H^{0*}\right)^{2}\right]\nonumber \\
 & +\frac{m_{\Delta}^{2}}{2}\left[\left|\Delta^{++}\right|^{2}+\left|\Delta^{+}\right|^{2}+\left|\Delta^{0}\right|^{2}\right]+\textrm{h.c.}\,,
\end{align}
leading to an effective neutrino mass matrix of the form
\begin{align}
m_{\nu}^{II} & =\frac{\mu\left\langle H^{0}\right\rangle ^{2}}{m_{\Delta}}Y^{\Delta}\,.
\end{align}
In type-III seesaw,  two or more fermionic triplets $\Sigma_{i}=\left(\Sigma_{i}^{+},\Sigma_{i}^{0},\Sigma_{i}^{-}\right)$
are necessary to reproduce neutrino oscillation data: 
\begin{align}
-\mathscr{L}^{III}= & Y_{ij}^{\Sigma}\left(\sqrt{2}H^{+}\overline{\Sigma}_{i}^{+}\nu_{Lj}+H^{+}\overline{\Sigma}_{i}^{0}e_{Lj}+H^{0}\overline{\Sigma}_{i}^{0}\nu_{Lj}-\sqrt{2}H^{0}\overline{\Sigma}_{i}^{-}e_{Lj}\right)\nonumber \\
 & +\frac{1}{2}\left(m_{\Sigma}^{*}\right)_{ij}\left(\Sigma_{i}^{+T}C\Sigma_{j}^{-}+\Sigma_{i}^{-T}C\Sigma_{j}^{+}+\Sigma_{i}^{0T}C\Sigma_{j}^{0}\right)+\textrm{h.c.}\,.
\end{align}
Here, we used $\overline{\Sigma}_{i}^{\pm}\equiv\overline{\left(\Sigma_{i}^{\pm}\right)}$.
The neutral component of triplets plays an analogous role to the one
of $\nu_{R}$ in a type-I seesaw. As such, the effective neutrino
mass matrix is given by 
\begin{align}
m_{\nu}^{III} & =-Y^{\Sigma T}m_{\Sigma}^{-1}Y^{\Sigma}\left\langle H^{0}\right\rangle ^{2}\,.
\end{align}

There are also other, more complex tree level seesaw realizations.
We shall mention here the inverse \citep{Mohapatra:1986bd} and linear
\citep{Malinsky:2005bi} seesaws, both of which can be seen as particular
cases of type-I seesaw since there is the introduction of fermion
singlets $\nu_{R},S$ in the theory. However, in these models the
couplings of $S$ with $\nu_{R},\nu_{L}$ is constrained: in general,
one can write a neutrino mass matrix
\begin{equation}
m_{\nu}=\left(\begin{array}{ccc}
0 & m_{D} & m_{LS}\\
m_{D}^{T} & m_{R} & m_{RS}\\
m_{LS}^{T} & m_{RS}^{T} & m_{S}
\end{array}\right)
\end{equation}
in the basis $\left(\nu_{L},\nu_{R}^{c},S\right)$. If $m_{LS}=m_{R}=0$
and $m_{RS}\gg m_{S}$ there is a double or inverse seesaw, yielding
an effective light neutrino effective mass
\begin{equation}
m_{\nu}^{ISS}=m_{D}\left(m_{RS}^{T}\right)^{-1}m_{S}m_{RS}^{-1}m_{D}^{T}\,,
\end{equation}
which is the same as $m_{D}M^{-1}m_{D}^{T}$ with $M\equiv m_{RS}m_{S}^{-1}m_{RS}^{T}$.
Note that lepton number is conserved when $m_{S}\rightarrow0$, so
the assumption $m_{RS}\gg m_{S}$ is natural \citep{'tHooft:1979bh}.
As such, the smallness of neutrino masses can be achieved in this
framework by a small $m_{S}$ instead of a big $m_{RS}$. Alternatively,
setting $m_{S}=0$ and having a small lepton number violating $m_{LS}$
yields the linear seesaw mechanism:
\begin{equation}
m_{\nu}^{LSS}=m_{LS}m_{RS}^{-1}m_{D}^{T}+\textrm{(transpose)}\,.
\end{equation}

It is possible as well to incorporate the different seesaw mechanisms
in extended frameworks, as in the case of SUSY models. Consider the
MSSM, with a superpotential given by equation \eqref{eq:Introduction_MSSM_W}.
A type-I seesaw is obtained by introducing invariant superfields $\widehat{N}_{i}^{c}$
(usually three), each containing a right-handed neutrino $\nu_{Ri}$
and its scalar superpartner $\widetilde{\nu}_{Ri}$, and by adding
to the superpotential the following terms:
\begin{align}
W^{I} & =Y_{ij}^{\nu}\widehat{N}_{i}^{c}\widehat{L}_{j}\cdot\widehat{H}_{u}+\frac{1}{2}\left(m_{R}\right)_{ij}\widehat{N}_{i}^{c}\widehat{N}_{j}^{c}\,.
\end{align}
There are also additional soft SUSY breaking terms: $h_{ij}^{\nu}\widetilde{\nu}_{Ri}^{*}\widetilde{L}_{j}\cdot H_{u}$,
$(m_{\nu}^{2})_{ij}\widetilde{\nu}_{Ri}\widetilde{\nu}_{Rj}^{*}$,
$b^{\nu}\tilde{\nu}_{R}^{*}\tilde{\nu}_{R}^{*}$, and $s^{\nu}\tilde{\nu}_{R}^{*}$.
Should this model be embedded in a cMSSM framework, the parameters
$m_{\nu}^{2}$ and $h^{\nu}$ will obey universality conditions ($m_{\nu}^{2}=m_{0}^{2}\mathbb{1}$
and $h^{\nu}=A_{0}Y^{\nu}$). The light neutrino mass matrix is the
same as in equation \eqref{eq:neutrinoSeesawImassMatrix}.

The implementation of a type II SUSY seesaw model requires the addition
of at least two $SU(2)_{L}$ triplet superfields \citep{Rossi:2002zb}.
However, if gauge coupling unification is to be preserved, complete
$SU(5)$ multiplets must be added to the MSSM content: the $\boldsymbol{15}$
and $\overline{\boldsymbol{15}}$. Under the SM gauge group, the $\boldsymbol{15}$
decomposes as $\widehat{S}\oplus\widehat{T}\oplus\widehat{Z}$, with
$\widehat{S}=(\boldsymbol{6},\boldsymbol{1},-\nicefrac{2}{3})$, $\widehat{T}=(\boldsymbol{1},\boldsymbol{3},1)$
and $\widehat{Z}=(\boldsymbol{3},\boldsymbol{2},\nicefrac{1}{6})$.
On the other hand, $\overline{\boldsymbol{15}}=\widehat{\overline{S}}\oplus\widehat{\overline{T}}\oplus\widehat{\overline{Z}}$,
with $\widehat{\overline{S}}=(\overline{\boldsymbol{6}},\boldsymbol{1},\nicefrac{2}{3})$,
$\widehat{\overline{T}}=(\boldsymbol{1},\boldsymbol{3},-1)$ and $\widehat{\overline{Z}}=(\overline{\boldsymbol{3}},\boldsymbol{2},-\nicefrac{1}{6})$.
In the $SU(5)$ broken phase, below the GUT scale, the superpotential
contains the following terms:
\begin{align}
W^{II} & =\frac{1}{\sqrt{2}}\left(Y^{T}\widehat{L}\widehat{T}\widehat{L}+Y^{S}\widehat{D}^{c}\widehat{S}\widehat{D}^{c}\right)+Y^{Z}\widehat{D}^{c}\widehat{Z}\widehat{L}+\frac{1}{\sqrt{2}}\left(\lambda_{1}\widehat{H}_{d}\widehat{T}\widehat{H}_{d}+\lambda_{2}\widehat{H}_{u}\widehat{\overline{T}}\widehat{H}_{u}\right)\nonumber \\
 & +m_{T}\widehat{T}\widehat{\overline{T}}+m_{Z}\widehat{Z}\widehat{\overline{Z}}+m_{S}\widehat{S}\widehat{\overline{S}}\,,\label{eq:LFV_SUSY_seesawII}
\end{align}
where gauge and flavor indices we omitted for simplicity (the soft
breaking Lagrangian can be found in \citep{Rossi:2002zb}). After
having integrated out the heavy fields, the effective neutrino mass
matrix reads
\begin{equation}
m_{\nu}^{II}=\frac{\lambda_{2}\left\langle H_{u}^{0}\right\rangle ^{2}}{m_{T}}Y^{T}\,.\label{eq:LFV_SUSY_seesawII:light}
\end{equation}

In order to embed a type III seesaw in SUSY models, the $\boldsymbol{24}$
representation of $SU(5)$ is used \citep{Ma:1998dn}. It contains
the fermionic triplet $\Sigma$ mentioned above: $\boldsymbol{24}=\widehat{B}+\widehat{G}+\widehat{W}+\widehat{X}+\widehat{\overline{X}}$,
with $\widehat{B}=\left(\boldsymbol{1},\boldsymbol{1},0\right)$,
$\widehat{G}=\left(\boldsymbol{8},\boldsymbol{1},0\right)$, $\widehat{W}=\left(\boldsymbol{0},\boldsymbol{3},0\right)$,
$\widehat{X}=\left(\boldsymbol{3},\boldsymbol{2},-\nicefrac{5}{6}\right)$
and $\widehat{\overline{X}}=\left(\overline{\boldsymbol{3}},\boldsymbol{2},\nicefrac{5}{6}\right)$.
The extra superpotential terms are the following:
\begin{align}
W^{III} & =Y^{W}\widehat{H}_{u}\widehat{L}\widehat{W}-\sqrt{\frac{3}{10}}Y^{B}\widehat{H}_{u}\widehat{L}\widehat{B}+Y^{X}\widehat{H}_{u}\widehat{\overline{X}}\widehat{D}^{c}\nonumber \\
 & +\frac{1}{2}m_{B}\widehat{B}\widehat{B}+\frac{1}{2}m_{G}\widehat{G}\widehat{G}+\frac{1}{2}m_{W}\widehat{W}\widehat{W}+\frac{1}{2}m_{X}\widehat{X}\widehat{\overline{X}}\,.
\end{align}
The fermionic component of $\widehat{W}$ is the $\Sigma$, but there
is also a gauge invariant $\widehat{B}$ superfield, whose fermionic
component behaves like a right-handed neutrino $\nu_{R}^{c}$. As
such, after integration of the heavy fields there is a mixture of
type-I and type-III seesaw contributions to the effective light neutrino
mass:
\begin{align}
m_{\nu}^{III} & =-\left\langle H_{u}^{0}\right\rangle ^{2}\left[\frac{3}{10}Y^{B}m_{B}^{-1}\left(Y^{B}\right)^{T}+\frac{1}{2}Y^{W}m_{W}\left(Y^{W}\right)^{T}\right]\,.
\end{align}

Alternatively, neutrino masses can be generated without changing the
MSSM field content, violating R-parity instead \citep{Aulakh:1982yn,Ross:1984yg,Ellis:1984gi,Abada:2001zh,Grossman:1999hc,Bednyakov:1998cx,Joshipura:2002fc,Chang:1999nu}---for
a review, see \citep{Grossman:2003gq,Hirsch:2004he,VicenteMontesinos:2011pf}.
We have mentioned in the previous chapter that if R-parity is not
imposed as a symmetry of the Lagrangian, the following baryon and
lepton number terms are allowed in the superpotential:
\begin{align}
W^{\slashed{R}_{p}} & =\frac{1}{2}\lambda^{ijk}\widehat{L}_{i}\cdot\widehat{L}_{j}\widehat{E}_{k}^{c}+\lambda'^{ijk}\widehat{L}_{i}\cdot\widehat{Q}_{j}\widehat{D}_{k}^{c}+\frac{1}{2}\lambda''^{ijk}\widehat{U}_{i}^{c}\widehat{D}_{j}^{c}\widehat{D}_{k}^{c}+\epsilon^{i}\widehat{L}_{i}\cdot\widehat{H}_{u}\,.
\end{align}
Some neutrinos get a tree level mass due to mixing with the neutral
electroweak gauginos and Higgsinos, while other neutrino masses are
generated radiatively. Crucially, unlike in the seesaw mechanism where
neutrino masses are generated at high-energy scales, in SUSY models
with broken R-parity only electroweak scale physics is at play. These
models are particularly interesting because neutrino parameters (masses,
mixing angles, phases) can be related with accelerator observables,
such as the decay properties of the LSP \citep{Porod:2000hv}.

Nevertheless, generic R-parity breaking models have an obvious down-side:
they introduce a large number of extra parameters ($\lambda^{ijk}$,
$\lambda'^{ijk}$, $\lambda''^{ijk}$, $\epsilon^{i}$, plus the ones
associated with soft SUSY breaking couplings). The bilinear R-parity
violating model avoids this problem by introducing only the 6 bilinear
R-parity breaking terms: $\epsilon^{i}\widehat{L}_{i}\cdot\widehat{H}_{u}$
in the superpotential and $-b^{i}\epsilon^{i}\widetilde{L}_{i}\cdot\widetilde{H}_{u}$
in $-\mathscr{L}_{\textrm{soft}}$ \citep{Hall:1983id}. This minimal
extension of the MSSM shares some of the features of more general
models \citep{Mukhopadhyaya:1998xj,Allanach:1999bf,Choi:1999tq,Romao:1999up,Bartl:2000yh,Hirsch:2000ef,Porod:2000hv,Restrepo:2001me,Hirsch:2002ys,Diaz:2003as,Bartl:2003uq,Hirsch:2003fe}.
The neutrino mass matrix is given by
\begin{align}
m_{\nu} & =\frac{M_{1}g^{2}+M_{2}g'^{2}}{\mu v_{u}v_{d}\left(M_{1}g^{2}+M_{2}g'^{2}\right)-2\mu^{2}M_{1}M_{2}}\Lambda\Lambda^{T}\,,\label{eq:LFV_BRpV_neutrino_masses}
\end{align}
where $\Lambda$ is a vector with components
\begin{align}
\Lambda_{i} & =\mu\left\langle \widetilde{\nu}_{i}\right\rangle +v_{d}\epsilon^{i}\,,\; i=e,\mu,\tau\,.
\end{align}
The sneutrino VEVs appearing in this last equation are non-zero once
$\epsilon^{i}\neq0$; they obey the following equations:
\begin{align}
0 & =-bv_{u}+\left(m_{H_{d}}^{2}+\mu^{2}\right)v_{d}+v_{d}D-\mu\left\langle \widetilde{\nu}_{i}\right\rangle \epsilon^{i}\,,\\
0 & =-bv_{d}+\left(m_{H_{u}}^{2}+\mu^{2}\right)v_{u}-v_{u}D+\left\langle \widetilde{\nu}_{i}\right\rangle b^{i}\epsilon^{i}+v_{u}\epsilon^{i}\epsilon^{i}\,,\\
0 & =\left\langle \widetilde{\nu}_{j}\right\rangle D+\epsilon^{j}\left(-\mu v_{d}+b^{j}v_{u}+\left\langle \widetilde{\nu}_{i}\right\rangle \epsilon^{i}\right)+\left\langle \widetilde{\nu}_{i}\right\rangle \textrm{Re}\left(m_{\widetilde{L}}^{2}\right)_{ij}\,,
\end{align}
where
\begin{align}
D & \equiv\frac{1}{8}\left(g^{2}+g'^{2}\right)\left(v_{d}^{2}-v_{u}^{2}+\left\langle \widetilde{\nu}_{i}\right\rangle ^{2}\right)\,.
\end{align}
In these expressions the index $i$ is to be summed over, while $j$
is not. The neutrino mass matrix $m_{\nu}$ in equation \eqref{eq:LFV_BRpV_neutrino_masses}
has a single eigenvalue different from zero (given by the trace of
$m_{\nu}$), which is associated with the atmospheric mass scale.
On the other hand, the smaller solar mass scale is generated radiatively
by bottom-sbottom, tau-stau and neutrino-sneutrino pairs in loops.
At tree level, since two masses are degenerate, one mixing angle can
be rotated away (it reappears nevertheless once radiative corrections
are taken into account). The two remaining ones are given by
\begin{align}
\tan\theta_{13} & =-\frac{\Lambda_{e}}{\sqrt{\Lambda_{\mu}^{2}+\Lambda_{\tau}^{2}}}\,,\\
\tan\theta_{23} & =-\frac{\Lambda_{\mu}}{\Lambda_{\tau}}\,.
\end{align}

\section{\label{sec:Charged-lepton-flavour}Charged lepton flavor violation}

Neutrino oscillations do not conserve the flavor of neutral leptons.
As such, one is lead to consider the possibility that in the charged
sector there are analogue processes which also violate lepton flavor.
In the following, we discuss this possibility.

In the Standard Model one can assign a lepton quantum number $L$
to all fields such that all terms in the Lagrangian preserve the associated
$U(1)_{L}$ symmetry. In fact, for massless neutrinos, it is possible
to do so for each lepton flavor: three separate quantum numbers $L_{e,\mu,\tau}$
are consequently preserved in perturbative processes. However, once
the Standard Model is minimally extended to accommodate massive neutrinos,
the $U(1)_{L_{e,\mu,\tau}}$ flavor symmetries are broken, even though
$L$ is still preserved if neutrinos are Dirac particles. The observed
neutrino oscillations imply that this lepton flavor violation (LFV)
is sizable in the neutral sector, yet the resulting effect in charged
leptons is small. For definiteness, consider the branching ratio of
$\mu\rightarrow e\gamma$ which is GIM suppressed; it is kept small
by the unitarity of the leptonic mixing matrix and by the smallness
of the (Dirac) neutrino squared mass differences \citep{Petcov:1976ff,Bilenky:1977du,Lee:1977qz,Marciano:1977wx,Cheng:1980tp}:
\begin{alignat}{1}
\textrm{Br}\left(\mu\rightarrow e\gamma\right) & =\frac{3\alpha}{32\pi m_{W}^{4}}\left|\sum_{i}U_{ei}U_{\mu i}^{*}m_{\nu_{i}}^{2}\right|^{2}\sim\frac{3\alpha}{32\pi}c_{13}^{2}s_{13}^{2}s_{23}^{2}\left(\frac{\Delta m_{31}^{2}}{m_{W}^{2}}\right)^{2}\sim10^{-55}\,.\label{eq:LFV_small_MuEGamma}
\end{alignat}
Charged lepton flavor violation (cLFV) processes with such small branching
ratios are not measurable. This turns out to be an interesting feature
of these processes: since the Standard Model and its trivial extensions
predict no charged lepton flavor violation, its observation would
be a clear signal of new Physics (for example, low scale seesaw models,
extra dimensions, little Higgs models, etc.).

In some softly broken SUSY models the situation changes dramatically;
generic soft SUSY breaking terms introduce large sources of cLFV,
to the extent that such terms must be constrained to avoid conflict
with current experimental bounds. For example, in the constrained
MSSM the trilinear terms are proportional to the Yukawa couplings
and the soft SUSY breaking masses are assumed to be diagonal in flavor
space at the GUT scale (chapter \ref{chap:The-SM's-shortcomings}).
As such, the Yukawa couplings alone control the flavor structure of
the model and for this reason there is no LFV, even accounting for
the effect of the renormalization group evolution of the parameters
down to low scales.

This principle of having only SM sources of flavor violation in an
extended theory is the main idea behind the concept of minimal flavor
violation (MFV) \citep{Buras:2000dm}. With the exception of Yukawa
interactions, all terms in the SM's Lagrangian are invariant under
a global $SU(3)_{Q}\times SU(3)_{u}\times SU(3)_{d}\times SU(3)_{L}\times SU(3)_{e}$
symmetry. As such, to control its breaking, the Yukawa couplings are
usually promoted to constant spurion fields which transform under
this symmetry in the necessary way to preserve it. In the quark sector,
this means that the Yukawa matrices transform as $\left(Y_{u}\right)_{ij}\thicksim\left(\boldsymbol{3},\overline{\boldsymbol{3}},\boldsymbol{1},\boldsymbol{1},\boldsymbol{1}\right)$
and $\left(Y_{d}\right)_{ij}\thicksim\left(\boldsymbol{3},\boldsymbol{1},\overline{\boldsymbol{3}},\boldsymbol{1},\boldsymbol{1}\right)$.
Therefore, in a MFV MSSM the squark masses and trilinear couplings
in the soft SUSY breaking sector of the MSSM must have the following
form (see \citep{Antonelli:2009ws}):
\begin{alignat}{1}
m_{\widetilde{Q}}^{2} & =\widetilde{m}^{2}\left(a_{1}\mathbb{1}+a_{2}Y^{u\dagger}Y^{u}+a_{3}Y^{d\dagger}Y^{d}+\cdots\right)\,,\\
m_{\widetilde{x}}^{2} & =\widetilde{m}^{2}\left(b_{1}^{x}\mathbb{1}+b_{2}^{x}Y^{x*}Y^{xT}+\cdots\right),\; x=u,d\,,\\
h^{x} & =A_{0}Y^{x}\left(c_{1}^{x}\mathbb{1}+c_{2}^{x}Y^{u\dagger}Y^{u}+c_{3}^{x}Y^{d\dagger}Y^{d}+\cdots\right),\; x=u,d\,,
\end{alignat}
where the $a_{i}$, $b_{i}^{u,d}$ and $c_{i}^{u,d}$ are \textit{a
priori} free dimensionless parameters. They are not constrained in
any way unless the MSSM is embedded in some more fundamental, higher
energy theory. The cMSSM, as mentioned above, is such an example:
at the energy scale where the three gauge couplings unify, all coefficients
are zero except for $a_{1}=b_{1}^{u,d}=c_{1}^{u,d}=1$. Even so, we
must keep in mind that the renormalization group flow will generate
non-zero contributions to the coefficients shown explicitly above
of the order of $\nicefrac{1}{\left(4\pi\right)^{2}}\log\nicefrac{m_{GUT}}{m_{SUSY}}$.

Unlike the quark sector, in the lepton sector the situation is not
as straightforward. To establish a MFV framework, a flavor symmetry
as well as a `minimal' set of parameters that are allowed to violate
it must be defined. Crucially, in the lepton sector this depends on
the unknown nature of neutrinos. We shall not consider all possibilities
here, but it is instructive to see what happens in the case where
no new fields are added to the MSSM (R-parity must be a broken symmetry).
In this case, the sources of LFV are the charged lepton Yukawa matrix,
$Y^{\ell}\thicksim\left(\boldsymbol{1},\boldsymbol{1},\boldsymbol{1},\boldsymbol{3},\overline{\boldsymbol{3}}\right)$,
and the neutrino mass matrix $m_{\nu}\thicksim\left(\boldsymbol{1},\boldsymbol{1},\boldsymbol{1},\overline{\boldsymbol{6}},\boldsymbol{1}\right)$
generated from the Weinberg operator in equation \eqref{eq:LFV_WeinbergOperator}.
In analogy to the quark sector in the MSSM, the slepton masses and
the leptonic trilinear couplings are restricted to the following form:
\begin{alignat}{1}
m_{\widetilde{L}}^{2} & =\widetilde{m}^{2}\left(a_{1}\mathbb{1}+a_{2}Y^{\ell\dagger}Y^{\ell}+a_{3}m_{\nu}^{*}m_{\nu}+\cdots\right)\,,\\
m_{\widetilde{e}}^{2} & =\widetilde{m}^{2}\left(b_{1}\mathbb{1}+b_{2}Y^{\ell*}Y^{\ell T}+b_{3}Y^{\ell*}Y^{\ell T}Y^{\ell*}Y^{\ell T}+b_{4}Y^{\ell*}m_{\nu}m_{\nu}^{*}Y^{\ell T}+\cdots\right)\,,\\
h^{\ell} & =A_{0}Y^{\ell}\left(c_{1}\mathbb{1}+c_{2}Y^{\ell\dagger}Y^{\ell}+c_{3}m_{\nu}^{*}m_{\nu}+\cdots\right)\,.
\end{alignat}
Once again $a_{i}$, $b_{i}$ and $c_{i}$ should be seen as free
parameters in an effective theory framework. Note that, unlike $m_{\widetilde{L}}^{2}$
and $h^{\ell}$, the right handed slepton mass matrix $m_{\widetilde{e}}^{2}$
only depends on $m_{\nu}$ through a term $m_{\nu}^{2}(Y^{\ell})^{2}$,
whose leading coefficient is suppressed by a two loop factor $\nicefrac{1}{\left(4\pi\right)^{4}}$
(see next subsection). The predictive power of the MFV hypothesis
can be seen in these equations, as they connect the amplitudes of
charged lepton flavor violating processes with neutrino oscillations
parameters. Even though the coefficients of each term in the equations
above are not predicted, their order of magnitude can be estimated.
Furthermore, a systematic suppression of otherwise dangerous cLFV
interactions is achieved.

\subsection{Charged lepton flavor violation and SUSY}

Stringent upper bounds on cLFV observables (see next subsection) imply
that the misalignment between the soft SUSY breaking mass matrices
$m_{\widetilde{e}}^{2}$, $m_{\widetilde{L}}^{2}$ and the lepton
Yukawa matrix $Y^{\ell}$ is small. For various applications, it is
useful to make an expansion on these misalignment. Indeed, in the
basis where the Yukawa couplings are diagonal,%
\footnote{The same rotation is performed on leptons and sleptons, in such a
way that the lepton-slepton-gaugino and lepton-slepton-Higgsino interactions
are kept diagonal in flavor space.%
} rather than trying to fully diagonalize the $6\times6$ sleptons
mass matrix
\begin{alignat}{1}
M_{\widetilde{\ell}}^{2} & \equiv\begin{pmatrix}\widetilde{m}_{LL}^{2} & \widetilde{m}_{LR}^{2}\\
\widetilde{m}_{LR}^{2\,\dagger} & \widetilde{m}_{RR}^{2}
\end{pmatrix}\,,
\end{alignat}
it is often more convenient not to do so and instead separate each
block $\widetilde{m}_{XY}^{2}$ into a diagonal and an off-diagonal
part. This last part is usually written with a $\delta$:
\begin{alignat}{1}
\left(\widetilde{m}_{XY}^{2}\right)_{ij} & \equiv\delta_{ij}^{XY}\sqrt{\left(\widetilde{m}_{XX}^{2}\right)_{ii}\left(\widetilde{m}_{YY}^{2}\right)_{jj}}\;\textrm{ for }\left(i,X\right)\neq\left(j,Y\right)\,.
\end{alignat}
The diagonal masses $\left(\widetilde{m}_{LL}^{2}\right)_{ii}$ and
$\left(\widetilde{m}_{RR}^{2}\right)_{ii}$ are Gaussian integrated
in the path integrals together with the kinetic terms, appearing therefore
in the slepton propagators. On the other hand, the off-diagonal $\delta_{ij}^{XY}$
terms are treated as interactions (two-point vertices). This approach
is known as the mass insertion approximation (MIA) \citep{Gabbiani:1988rb,Gabbiani:1996hi,Masina:2002mv,Hall:1985dx}.
The effect of actually performing the rotation to the slepton mass
basis is achieved in this scheme in the limit where all diagrams with
an arbitrarily large number of mass insertions is considered. As such,
it is to be expected that this approximation is good as long as the
$\delta$'s are small (see for instance \citep{Paradisi:2005fk} for
a quantitative analysis). The advantage of the MIA is that it provides
simpler analytical expressions for the amplitude of lepton flavor
violating processes, making their dependence on the Lagrangian parameters
more transparent, as there are no rotation matrices other than the
CKM one (for hadronic processes) and possibly the PMNS one. 

Alternatively, one can compute the amplitude of the desired processes
in the mass eigenbasis, and then make a polynomial expansion of the
loop functions which depend on the masses of virtual particles. Doing
so makes it possible to eliminate the rotation matrices appearing
in the expressions and convert them into the Lagrangian parameters
that contribute to the sparticles mass matrices. In this approach,
the validity of the MIA is directly tied to the smallness of the splitting
between the physical masses of sparticles with the same quantum numbers,
which in turn can be related to the smallness of the off-diagonal
entries of the mass matrices in the gauge basis.

We have mentioned above that the introduction of neutrino masses in
the Standard Model generates negligible charged lepton flavor violation.
However, in SUSY models this is no longer true. For definiteness,
we shall consider the cMSSM which, at the GUT scale $m_{G}$ contains
universal and diagonal soft mass terms. Since the flavor structure
is controlled uniquely by the Yukawa matrix $Y^{\ell}$, there is
no cLFV. Once a seesaw mechanism is introduced, the situation changes
as there are then two or more flavored matrices in the lepton sector.
Equation \eqref{eq:LFV_small_MuEGamma} would suggest that the resulting
cLFV is small yet, due to the effect of the renormalization group
in the charged left-slepton mass matrix and trilinear soft couplings,
this turns out not to be the case \citep{Borzumati:1986qx}. In type
I seesaw
\begin{alignat}{1}
\left(m_{\widetilde{L}}^{2}\right)_{ij} & \approx-\frac{1}{8\pi^{2}}\left(3m_{0}^{2}+A_{0}^{2}\right)\left(Y^{\nu\dagger}LY^{\nu}\right)_{ij}\,,i\neq j\,,\\
h_{ij}^{\ell} & \approx-\frac{3}{16\pi^{2}}A_{0}\left(Y^{\ell}Y^{\nu\dagger}LY^{\nu}\right)_{ij}\,,i\neq j\,,
\end{alignat}
with $L_{ij}=\log\nicefrac{m_{G}}{m_{R_{i}}}\delta_{ij}$. Since right
handed charged leptons do not couple directly with $Y^{\nu}$, at
one loop order 
\begin{alignat}{1}
\left(m_{\widetilde{e}}^{2}\right)_{ij} & \approx0\,,i\neq j\,.
\end{alignat}
With type II seesaw, the expressions for the renormalization group
induced LFV are similar \citep{Rossi:2002zb}. In particular, the
off-diagonalities in $m_{\widetilde{L}}^{2}$ and $h^{\ell}$ are
proportional to the combination $Y^{T\dagger}LY^{T}$, where $Y^{T}$
was introduced in equation \eqref{eq:LFV_SUSY_seesawII}.

Processes violating the flavor of charged leptons are sensitive to
these off-diagonalities. For example, the decays $\ell_{i}\rightarrow\ell_{j}\gamma$
are induced by neutralino-slepton and chargino-sneutrino loops, yielding
a branching ratio (using $\delta_{ij}^{RR}\approx0$) \citep{Hisano:1995cp,Hisano:1995nq,Hisano:1996qq,Gabbiani:1996hi}
\begin{align}
\textrm{BR}\left(\ell_{i}\rightarrow\ell_{j}\gamma\right) & \propto\frac{\alpha^{3}}{G_{F}^{2}}\frac{\left(\delta_{ij}^{LL}\right)^{2}}{m_{SUSY}^{4}}\tan^{2}\beta\,,
\end{align}
where $m_{SUSY}$ is the mass scale of the virtual sparticles in the
loops. As such, in SUSY GUTs the problem of neutrino mass generation
is directly related to potentially large cLFV effects, which may be
measurable in low \citep{Hisano:1995cp,Hisano:1995nq,Hisano:1998fj,Buchmuller:1999gd,Kuno:1999jp,Ellis:1999uq,Casas:2001sr,Lavignac:2001vp,Bi:2001tb,Ellis:2002fe,Deppisch:2002vz,Fukuyama:2003hn,Brignole:2004ah,Masiero:2004js,Fukuyama:2005bh,Petcov:2005jh,Arganda:2005ji,Deppisch:2005rv,Yaguna:2005qn,Calibbi:2006nq,Antusch:2006vw,Arganda:2007jw,Arganda:2008jj,Esteves:2010ff,Buras:2010pm,Biggio:2010me,Calibbi:2012gr,Chowdhury:2013jta}
and high energy experiments \citep{ArkaniHamed:1996au,Krasnikov:1996np,Bityukov:1998va,Hisano:1998wn,Agashe:1999bm,Baer:2000cb,Hinchliffe:2000np,Guchait:2001us,Deppisch:2003wt,Deppisch:2004pc,Cannoni:2005gy,Carvalho:2002jg,Buckley:2006nv,Cannoni:2008bg,Hirsch:2008dy,Carquin:2008gv,Buras:2009sg,Herrero:2009tm,Esteves:2009vg,Abada:2010kj,Calibbi:2011dn,Calibbi:2011vi,Galon:2011wh,Arbelaez:2011bb,Hirsch:2011cw,Carquin:2011rg,Abada:2011mg,Abada:2012re,Hirsch:2012yv}.

\subsection{cLFV observables and limits on effective couplings}

Charged lepton flavor violation can be analyzed in a model independent,
effective field theory framework, where heavy fields are integrated
out (see \citep{Abada:2007ux,Raidal:2008jk}). With the SM field content,
it is not possible to build renormalizable lepton flavor violating
operators, therefore these operators must be of dimension $n=$5 or
higher. Observables can then be expressed as a function of the coefficients
of such operators, which can be calculated for specific models. From
a dimensional analysis alone, these LFV operators are expected to
be suppressed by a factor $\nicefrac{1}{m_{NP}^{n-4}}$ where $m_{NP}$
is the new Physics mass scale ($\gtrsim$ TeV), so usually it is enough
to consider operators up to dimension 6.

The only dimension 5 LFV operator is the one mentioned previously
in equation \eqref{eq:LFV_WeinbergOperator}, which gives mass to
neutrinos after EWSB (it is common to all seesaw mechanisms). On the
other hand, there are various dimension 6 operators, which depend
on the high energy model: they may contain two leptons and a gauge
boson, four leptons, or two leptons and two quarks. 
\begin{alignat}{1}
\mathcal{O}_{ij}^{\ell B\left(W\right)} & =\overline{L}_{i}\left[\gamma^{\mu},\gamma^{\nu}\right]e_{Rj}HB_{\mu\nu}\left(W_{\mu\nu}\right)\,,\label{eq:LFV_EffectiveOperators1}\\
\mathcal{O}_{ijkl}^{LL} & =\left(\overline{L}_{i}\gamma^{\mu}L_{j}\right)\left(\overline{L}_{k}\gamma_{\mu}L_{l}\right)\,,\\
\mathcal{O}_{ijkl}^{ee} & =\left(\overline{e}_{Ri}\gamma^{\mu}e_{Rj}\right)\left(\overline{e}_{Rk}\gamma_{\mu}e_{Rl}\right)\,,\\
\mathcal{O}_{ijkl}^{\ell\ell} & =\left(\overline{L}_{i}e_{Rj}\right)\left(\overline{e}_{Rk}L_{l}\right)\,,\\
\mathcal{O}_{ijkl}^{LQ} & =\left(\overline{L}_{i}\gamma^{\mu}L_{j}\right)\left(\overline{Q}_{k}\gamma_{\mu}Q_{l}\right)\,,\\
\mathcal{O}_{ijkl}^{Lu\left(d\right)} & =\left(\overline{e}_{Ri}\gamma^{\mu}e_{Rj}\right)\left(\overline{u\left(d\right)}_{Rk}\gamma_{\mu}u\left(d\right)_{Rl}\right)\,,\\
\mathcal{O}_{ijkl}^{\ell q} & =\left(\overline{L}_{i}e_{Rj}\right)\left(\overline{d}_{Rk}Q_{l}\right)\,,\label{eq:LFV_EffectiveOperators2}
\end{alignat}
For simplicity, $SU(2)$ indices were omitted in these expressions.
After EWSB, the operators $\mathcal{O}_{ij}^{\ell B\left(W\right)}$
give rise to electric and magnetic dipole moments, as well as $\ell_{i}\rightarrow\ell_{j}\gamma$
transitions. The four lepton operators $\mathcal{O}_{ijkl}^{LL}$,
$\mathcal{O}_{ijkl}^{ee}$ and $\mathcal{O}_{ijkl}^{\ell\ell}$ contribute
to the processes $\ell_{i}\rightarrow\ell_{j}\ell_{k}\ell_{l}$, $\ell_{i}\ell_{j}\rightarrow\ell_{k}\ell_{l}$,
and they can also be probed in $Z$ decays into a pair of leptons.
Finally, $\mathcal{O}_{ijkl}^{LQ}$, $\mathcal{O}_{ijkl}^{Lu\left(d\right)}$
and $\mathcal{O}_{ijkl}^{\ell q}$ contribute to leptonic and semileptonic
decays of mesons.

Bounds can be placed on the coefficients that multiply these operators.
For example, consider the charged part of $\mathcal{O}_{ij}^{\ell\gamma}$---a
mixture of $\mathcal{O}_{ij}^{\ell B}$ and $\mathcal{O}_{ij}^{\ell W}$---after
EWSB:
\begin{alignat}{1}
\mathscr{L} & =\frac{em_{\ell_{i}}}{8}A_{ij}\overline{e}_{Ri}\left[\gamma^{\mu},\gamma^{\nu}\right]e_{Lj}F_{\mu\nu}^{\textrm{em}}+\textrm{h.c.}\,,\label{eq:llGamma_effective_lagrangian}
\end{alignat}
where $A_{ij}$ are some coefficients. When $i\neq j$, this operator
contributes to the dipole transitions $\ell_{i}\rightarrow\ell_{j}\gamma$,
while diagonal entries $i=j$ generate leptonic anomalous magnetic
moments, $a_{i}\equiv\nicefrac{g_{i}}{2}-1$, and electric dipole
moments $d_{i}$ (see for example \citep{Giudice:2012ms}):

\begin{alignat}{1}
\frac{\textrm{BR}\left(\ell_{i}\rightarrow\ell_{j}\gamma\right)}{\textrm{BR}\left(\ell_{i}\rightarrow\ell_{j}\nu_{i}\overline{\nu}_{j}\right)} & =\frac{48\pi^{3}\alpha}{G_{F}^{2}}\left(\left|A_{ij}\right|^{2}+\left|A_{ji}\right|^{2}\right)\,,\\
\Delta a_{i} & =2m_{\ell_{i}}^{2}\textrm{Re}\left(A_{ii}\right)\quad\textrm{(no sum in }i\textrm{)\,,}\\
d_{i} & =em_{\ell_{i}}\textrm{Im}\left(A_{ii}\right)\quad\textrm{(no sum in }i\textrm{)\,.}
\end{alignat}
The anomalous magnetic moments of the two lightest charged leptons
have been measured to a good accuracy, so much so that $a_{e}$ is
used to determine the fine structure constant $\alpha$, and $\Delta a_{\mu}$
provides a precision test for the EW theory at the quantum level.
Interestingly, the present experimental value of $\Delta a_{\mu}$
differs from the SM prediction by more than 3$\sigma$ (see table
\eqref{tab:anomalous_magnetic_moment}). It is also possible to compute
a discrepancy between $a_{e}^{exp}$ and $a_{e}^{SM}$ as long as
$\alpha$ is obtained from another observable; this was done in \citep{Giudice:2012ms},
where the fine structure constant was extracted from the measurement
of the atomic recoil frequency shift of photons absorbed or emitted
by $^{133}\textrm{Cs}$ atoms using atom interferometry. On the other
hand, due to its short lifetime, the anomalous magnetic moment of
the $\tau$ is poorly measured.

\begin{table}
\begin{centering}
\begin{tabular}{cccc}
\toprule 
$i$ & $a_{i}^{exp}$ & $a_{i}^{SM}$ & $\Delta a_{i}$\tabularnewline
\midrule
$e$ $\left(\times10^{14}\right)$ & $11\;596\;521\;8076\left(27\right)$ & $11\;596\;521\;8178\left(76\right)$ & $-102\left(81\right)$\tabularnewline
$\mu$ $\left(\times10^{11}\right)$ & $116\;592\;089\left(63\right)$ & $116\;591\;828\left(49\right)$ & $261\left(80\right)$\tabularnewline
$\tau$ $\left(\times10^{3}\right)$ & $-18\left(17\right)$ & $1.177\;21\left(5\right)$ & $-19\left(17\right)$\tabularnewline
\bottomrule
\end{tabular}
\par\end{centering}

\caption{\label{tab:anomalous_magnetic_moment}Values of $a_{i}\equiv\nicefrac{\left(g_{i}-2\right)}{2},\, i=e,\mu,\tau$
taken from \citep{Giudice:2012ms,Hagiwara:2011af,Passera:2007fk}
and \citep{Mohr:2012tt,Bennett:2006fi,Abdallah:2003xd} ($1\sigma$
uncertainties are in parentheses). Note that $a_{e}^{SM}$ differs
from $a_{e}^{exp}$ because $\alpha$ is not being extracted from
the anomalous magnetic moment of the electron---see \citep{Giudice:2012ms}.}
\end{table}

To the magnetic moment $\mu$ measuring the strength of the coupling
between a particle\textquoteright{}s spin $\overrightarrow{S}$ and
an external magnetic field $\overrightarrow{B}$, there is an associated
electric dipole moment (EDM) $d$ which measures the coupling strength
between $\overrightarrow{S}$ and an external electric field $\overrightarrow{E}$.
A $P$, $T$ and $CP$ (by the $CPT$ theorem) preserving theory yields
$d=0$ for the various particles, because a term $\overrightarrow{S}\cdot\overrightarrow{E}$
changes sign under these symmetries. The Standard Model does violate
these symmetries through the complex phase in the CKM matrix, but
even so the predicted electric dipole moments are minute, assuming
that the QCD $\theta$ parameter is zero. Experimentally, it has been
confirmed that electric dipole moments are small (see table \eqref{tab:electric_dipole_moments}
for the leptonic limits); current measurements are in fact compatible
with null electric dipole moments, but future improvements in the
experimental sensibilities can change this \citep{Fukuyama:2012np}.

\begin{table}
\begin{centering}
\begin{tabular}{cc}
\toprule 
$\ell$ & $d_{\ell}^{exp}$($e\cdot\textrm{cm}$)\tabularnewline
\midrule
$e$ & $\left(-2.4\pm5.9\right)\times10^{-28}$ \citep{Hudson:2011zz}\tabularnewline
$\mu$ & $\left(-1\pm9\right)\times10^{-20}$ \citep{Bennett:2008dy}\tabularnewline
$\tau$ & $\begin{array}{c}
\left(1.15\pm1.70\right)\times10^{-17}\textrm{ (Re)}\\
\left(-0.83\pm0.86\right)\times10^{-17}\textrm{ (Im)}
\end{array}$\citep{Inami:2002ah}\tabularnewline
\bottomrule
\end{tabular}
\par\end{centering}

\caption{\label{tab:electric_dipole_moments}Experimental bounds on leptonic
electric dipole moments.}
\end{table}

On the other hand, the off-diagonal entries of the $A$ matrix in
equation \eqref{eq:llGamma_effective_lagrangian} are also constrained
by strict bounds on the branching ratios of $\ell_{i}\rightarrow\ell_{j}\gamma$
decays, as well as $\mu-e$ coherent conversion in the vicinity of
an atomic nucleus (table \eqref{tab:transition_moments}). Some collaborations
plan to decrease significantly these limits in the future; in particular,
the $\mu-e$ coherent conversion bound is expected to be improved
by 4 orders of magnitude in the next decade by the Mu2e and COMET
collaborations.

\begin{table}
\begin{centering}
\begin{tabular}{ccc}
\toprule 
 & Current bound & Future sensitivity\tabularnewline
\midrule
$\textrm{BR}\left(\mu\rightarrow e\gamma\right)$ & $5.7\times10^{-13}$ \citep{Adam:2013mnn} & $6\times10^{-14}$ \citep{Baldini:2013ke}\tabularnewline
$\textrm{BR}\left(\tau\rightarrow e\gamma\right)$ & $3.3\times10^{-8}$ \citep{Aubert:2009ag} & $3\times10^{-9}$ \citep{O'Leary:2010af}\tabularnewline
$\textrm{BR}\left(\tau\rightarrow\mu\gamma\right)$ & $4.4\times10^{-8}$ \citep{Aubert:2009ag} & $\left(5-10\right)\times10^{-9}$ \citep{Abe:2010sj}, $2.4\times10^{-9}$
\citep{O'Leary:2010af}\tabularnewline
$\frac{\sigma\left(\mu N\rightarrow eN\right)}{\sigma\left(\mu N\rightarrow\textrm{capture}\right)}$ & %
\begin{tabular}{c}
$7\times10^{-13}$ (Au) \citep{Bertl:2006up}\tabularnewline
$4.3\times10^{-12}$ (Ti) \citep{Dohmen:1993mp}\tabularnewline
\end{tabular} & %
\begin{tabular}{c}
$10^{-16}$ (Al) \citep{Cui:2009zz}, $2\times10^{-17}$ (Al) \citep{Glenzinski:2010zz}\tabularnewline
$10^{-18}$ (Ti) \citep{Barlow:2011zza}\tabularnewline
\end{tabular} \tabularnewline
\bottomrule
\end{tabular}
\par\end{centering}

\caption{\label{tab:transition_moments}Experimental bounds on the branching
ratio of decays $\ell_{i}\rightarrow\ell_{j}\gamma$ (90\% CL) as
well as expected future sensitivities.}
\end{table}

A thorough listing of limits of the other effective operators in equations
\eqref{eq:LFV_EffectiveOperators1}--\eqref{eq:LFV_EffectiveOperators2}
can be found in \citep{Raidal:2008jk}. Table \eqref{tab:4l_and_2l2q_limits}
contains some of the most important observables which constraint the
dimension 6 operators with four leptons and two leptons plus two quarks.
\begin{table}
\begin{centering}
\begin{tabular}{ccc}
\toprule 
 & Current bound/value & Future sensitivity\tabularnewline
\midrule
$\textrm{BR}\left(\mu\rightarrow eee\right)$ & $<1.0\times10^{-12}$ \citep{Bellgardt:1987du} & $\sim10^{-16}$ \citep{Blondel:2013ia}\tabularnewline
$\textrm{BR}\left(\tau\rightarrow eee\right)$ & $<2.7\times10^{-8}$ \citep{Hayasaka:2010np} & $\sim10^{-10}$ \citep{O'Leary:2010af}\tabularnewline
$\textrm{BR}\left(\tau\rightarrow\mu\mu\mu\right)$ & $<2.1\times10^{-8}$ \citep{Hayasaka:2010np} & $\left(1-3\right)\times10^{-9}$ \citep{Abe:2010sj}, $\sim10^{-10}$
\citep{O'Leary:2010af}\tabularnewline
$\textrm{BR}\left(B_{s}\rightarrow\mu\mu\right)$ & $3.2_{-1.2}^{+1.5}\times10^{-9}$ \citep{Aaij:2012nna} & $0.15\times10^{-9}$ \citep{Schune:2013wi}\tabularnewline
$\textrm{BR}\left(B\rightarrow\tau\nu\right)$ & $\left(1.65\pm0.34\right)\times10^{-4}$ \citep{Beringer_mod:1900zz} & 3\% -- 4\% \citep{Buchalla:2008jp}\tabularnewline
\bottomrule
\end{tabular}
\par\end{centering}

\caption{\label{tab:4l_and_2l2q_limits}Currents bounds (90\% CL) or values
of some important observables which depend on the $4\ell$ and $2\ell2q$
cLFV effective operators in equations in \eqref{eq:LFV_EffectiveOperators1}--\eqref{eq:LFV_EffectiveOperators2}.
Future sensitivities are also shown. The Standard Model predicts $\textrm{BR}\left(B_{s}\rightarrow\mu\mu\right)=\left(3.23\pm0.27\right)\times10^{-9}$
\citep{Buras:2012ru} and $\textrm{BR}\left(B\rightarrow\tau\nu\right)=\left(1.11\pm0.27\right)\times10^{-4}$
(using $f_{B}=190.6\pm4.7$ MeV \citep{latticeaverages_website,Laiho:2009eu,Bazavov:2011aa,Na:2012kp},
$V_{ub}=\left(4.15\pm0.49\right)\times10^{-3}$ \citep{Beringer_mod:1900zz}).}
\end{table}

\cleartooddpage

\chapter{\label{chap:Symmetry}Group theory in particle physics}

\section{The role of symmetry}

Symmetry seems to be an essential feature of the fundamental laws
of Physics. It is not a question of aesthetics though: symmetry makes
predictions by restricting the set of theories which can describe
Nature. In this chapter we discuss two cases of relevance in High
Energy Physics, namely space-time and gauge symmetries.

Consider the Theory of Relativity, where a system is described by
a metric $g$ and a manifold $\mathcal{M}$. Under a change of coordinates
$x\rightarrow x'$ the metric transforms as $g_{ab}\rightarrow g'_{ab}=\frac{\partial x^{c}}{\partial x'^{a}}\frac{\partial x^{d}}{\partial x'^{b}}g_{cd}$,
and we can ask what are the transformations that preserve the metric,
$g=g'$. These are called \textit{isometries}, and in the case of
the Minkowski metric $\eta=\textrm{diag}\left(1,-1,-1,-1\right)$
of Special Relativity they form the Poincaré group. If translations
are ignored, we are left with the group of homogeneous isometries---the
Lorentz group. We shall look into these two groups latter on in this
chapter, and also at how SUSY enlarges this space-time symmetry in
a unique, non-trivial way. Knowing how symmetry is of such importance
in fundamental Physics, the fact that the space-time symmetry group
can be extended, is in itself a powerful theoretical motivation for
considering supersymmetric theories, as alluded already in chapter
\ref{chap:The-SM's-shortcomings}.

On the other hand, modern Particle Physics models are Yang\textendash{}Mills
theories which possess some continuous gauge symmetry. This is a Lie
group, just like the Poincaré and Lorentz groups, but there is a significant
difference between them: unlike the gauge symmetry group, the Lorentz
and Poincaré groups are not compact because of the metric's signature,
and as a consequence the irreducible representations of these latter
groups cannot be simultaneously finite dimensional and unitary. We
shall see this in some detail latter on.

The SM, as well as the MSSM, are based on the $U\left(1\right)_{Y}\times SU\left(2\right)_{L}\times SU\left(3\right)_{c}$
group which at energies below the EW scale breaks into $U\left(1\right)_{em}\times SU\left(3\right)_{c}$
due to the Higgs mechanism (see the appendix \ref{chap:SM_appendix}).
Early on, it was realized that the SM gauge group itself could be
the remnant of a larger symmetry that is broken at low energies. The
prospect of having theories with a single gauge coupling constant
is particularly interesting; this happens only if the gauge group
is simple or the direct product of equal simple factors, together
with some discrete symmetry that permutes these factors \citep{Georgi:1974sy,Pati:1974yyX,Georgi:1974yf,Fritzsch:1974nn,Buras:1977yy,Georgi:1979df,Barr:1981qv,Dimopoulos:1981zb,Hewett:1988xc}.
Such models would unify all forces, thereby providing a simpler description
of the laws of Physics. Ideally, something analogous would happen
to the particle content of the theory: there would be only one representation
of this fundamental gauge group containing all the known matter particles.
This would dramatically reduce the number of parameters and thus yield
a much more predictive model.

At this point however, it seems difficult to formulate a completely
unified model. Nevertheless, the quest for a bigger gauge symmetry
group is an interesting and actively pursued one. In order to have
a global view of the possible ways of extending the SM in this way,
we shall be exploring some theoretical features of Lie algebras (see
also \citep{Cahn:1985wk,Fuchs:2003aa,Slansky:1981yr,Kerf:1990aa,Jacobson:1979aa,Tung:1985na}).
Some of the aspects discussed here were used on the Mathematica program
\texttt{Susyno}, described latter on in chapter \ref{chap:Susyno}.
One important result is the Serre-Chevalley relations in equations
\eqref{eq:Serre_Chevalley_relations1}--\eqref{eq:Serre_Chevalley_relations3}.
In words, they state that a simple Lie algebra which allows the simultaneous
diagonalization of $n$ of its generators can be seen as being made
of $n$ copies of the $SU(2)$ group. The way that the 3 generators
of each $SU(2)$ interact/commute with the generators of the other
$SU(2)$'s defines the structure of the algebra and, surprisingly,
these commutation relations are very simple in a particular basis;
they are completely determined by the so-called Cartan matrix of the
algebra. Therefore, with equations \eqref{eq:Serre_Chevalley_relations1}--\eqref{eq:Serre_Chevalley_relations3}
it is possible to generalize the procedure used to build the explicit
representation matrices of $SU(2)$ to any simple group! In turn,
with the explicit matrices of any representation of any simple group,
it is possible to write the Lagrangian invariant under such group
(see chapter \ref{chap:Susyno} and also appendix \ref{chap:Implementation-details-of}).

Finally, let us mention in passing the role of a few discrete symmetries
in High Energy Physics. In connection with the relativistic nature
of field theories, there are the charge, parity and time reversal
operations ($C$, $P$ and $T$) which are of great importance in
our understanding of these theories. Then there are the abelian discrete
symmetries associated with baryon ($B$) and lepton ($L$) number
conservation, which are related to the R-parity in the MSSM. Also,
as mentioned in chapter \ref{chap:The-SM's-shortcomings}, our lack
of understanding of the flavor structure of the SM has lead to the
use of discrete non-abelian flavor symmetries as a means to predict,
or at least constrain some of the mixing angles and fermion masses.
It is worth noting however that, unlike continuous symmetries, the
discrete symmetries that we know of appear to be violated by Nature:
$C$, $P$ and $T$ are broken symmetries, even though the last two
are part of the Lorentz group; baryon and lepton number are violated
in non-perturbative processes \citep{Klinkhamer:1984di,Belavin:1975fg,'tHooft:1976fv,'tHooft:1976up}
and their conservation at the perturbative level can be seen as a
consequence of the Lagrangian gauge invariance (a continuous symmetry).
Nevertheless, broken or not, discrete symmetries are of great relevance
in fundamental Physics.

\section{\label{sec:Lie-groups-and-Lie-algebras}Lie groups and their connection
to Lie algebras}

Particle Physics deals almost invariably with Lie algebras instead
of Lie groups, to the point that sometimes both these expressions
are used to denote the group's algebra. Therefore, before addressing
Lie algebras, we begin by briefly reviewing the connection between
these two concepts.

A set $G$ forms a group if there is an operation $\cdot$ (group
multiplication) such that for any $a,b\in G$, $a\cdot b$ is also
an element of $G$ and, in addition, the following holds:
\begin{itemize}
\item The group multiplication is associative, meaning that $\left(a\cdot b\right)\cdot c=a\cdot\left(b\cdot c\right)$
for any $a,b,c\in G$.
\item $G$ contains an identity element $e$ such that $a\cdot e=a$ for
any $a\in G$.
\item For all $a\in G$ there is an $a^{-1}\in G$ (the inverse of $a$)
such that $a\cdot a^{-1}=e$.
\end{itemize}
The number of elements of $G$ may be finite in number or not. This
latter case is the one which interests us, in particular when the
group elements can be labeled by continuous parameters ($G$ is said
to be a \textbf{continuous group}). In addition, the set $G$ can
be a manifold differentiable to all orders, with a group multiplication
function $G\times G\rightarrow G$ also differentiable to all orders.
This is the main feature of a Lie group, even though the exact definition
of such a group varies across the literature. The requirement that
$G$ is a group and an infinitely differentiable  manifold at the
same time gives rise to an interesting object which combines these
two mathematical structures in a non-trivial way.

At this point, it is worth mentioning that Yang\textendash{}Mills
theories deal with invertible linear transformations that act on some
$n$ complex fields.%
\footnote{The fact that some fields are strictly real is of no consequence to
the present discussion.%
} These transformations forms a group, the \textbf{complex general
linear group} $GL\left(n,\mathbb{C}\right)$, which consists of all
$n\times n$ complex matrices with a non-null determinant, together
with the operation of matrix multiplication. Any subgroup of $GL\left(n,\mathbb{C}\right)$
is called a \textbf{linear group} and we may therefore restrict our
analysis to these ones.

We shall now move on to consider algebras. An \textbf{algebra} $\mathfrak{g}$
over a field $K$ is a vector space $\mathfrak{g}$ over a field $K$
together with a bilinear operation $\times$, such that $a\times b\in\mathfrak{g}$
for any $a,b\in\mathfrak{g}$. This means that for any $a,b,c\in\mathfrak{g}$
and $k\in K$ the following holds:
\begin{itemize}
\item $\left(a+b\right)\times c=a\times c+b\times c$ and $a\times\left(b+c\right)=a\times b+a\times c$;
\item $\left(ka\right)\times b=a\times\left(kb\right)=k\left(a\times b\right)$.
\end{itemize}
This definition is very general and as such the resulting object does
not possess much structure. A Lie algebra is one that satisfies two
additional conditions:
\begin{itemize}
\item $a\times a=0$;
\item $\left(a\times b\right)\times c+\left(c\times a\right)\times b+\left(b\times c\right)\times a=0$.
\end{itemize}
In most cases, we consider Lie algebras of linear transformations,
so a particular notation is used for this bilinear operation $\times$,
which is a commutator $\left[,\right]$. The reason for this is simple:
consider two linear transformations represented in a given basis by
two matrices $A$ and $B$. Then the matrix commutator $[A,B]=AB-BA$
defines a bilinear relation with the properties of the $\times$ just
described. Therefore the set of linear transformations on a given
vector space forms a Lie algebra. This is the case we are interested
in, so we shall use this notation henceforth.

We have so far described the concepts of Lie group and Lie algebra
separately without connecting the two. Intuitively, there is a Lie
algebra associated to a Lie group that describes its local structure.
It turns out that this Lie algebra contains almost everything there
is to know about the underlying Lie group and for this reason, in
many situations we only consider the Lie algebras, or equivalently,
elements of the Lie group infinitesimally close to the identity element
$e$.

Recall that for a point $g$ of a differentiable manifold $G$, the
tangent space at $g$ ($\equiv T_{g}G$) is the vector space consisting
of all $\gamma'\left(0\right)$ where $\gamma\left(t\right)$ is any
path in $G$ such that $\gamma\left(0\right)=g$. This vector space
has the same dimension as the manifold $G$ and it forms a Lie algebra
under the so-called Lie bracket. Elaborating a little on this point,
note that a vector $X$ in $T_{e}G$ (the tangent space at the group's
identity element $e$) can be used to create a vector field in all
of $G$ by using the group's multiplication to transport it everywhere.
The resulting vector field is said to be left or right invariant depending
on how this translation is done. Suppose then that there are two such
vector fields $X$ and $Y$---their Lie bracket is defined to be the
vector field 
\begin{align}
\left[X,Y\right] & =\left(X^{b}\partial_{b}Y^{a}-Y^{b}\partial_{b}X^{a}\right)\partial_{a}\,,\label{eq:Symmetry_Lie_bracket}
\end{align}
which is also left/right invariant. Under this operation, invariant
vector fields (or equivalently $T_{e}G$ since we can move vectors
on $G$) forms a Lie algebra. The right side of equation \eqref{eq:Symmetry_Lie_bracket}
can be understood as the commutator of the vectors $X$ and $Y$ if
we use as a coordinate basis the vectors $\nicefrac{\partial}{\partial x^{a}}=\partial_{a}$
such that $X=X^{a}\partial_{a}$ and $Y=Y^{a}\partial_{a}$.

One useful tool in making the connection between the Lie group $G$
and its Lie algebra $T_{e}G$ is the so-called \textbf{exponential
map}. As we will see, a Lie algebra admits a pseudo-metric (the Killing
form) so we may define geodesics in $G$ \citep{Fuchs:2003aa}. It
turns out that these geodesic curves correspond to one-parameter subgroups
of $G$. If $\gamma$ is one such geodesic with $\gamma\left(0\right)=e$
then the exponential map is defined by the relation
\begin{align}
\textrm{Exp}\left[\gamma'\left(0\right)\right] & \equiv\gamma\left(1\right)\,.\label{eq:Symmetry_exponentialmap}
\end{align}
Following \citep{Fuchs:2003aa}, we used a capital {}``E'' here
because this rather abstract definition generalizes the usual one
of the exponential function, as applied to matrices and complex numbers.
Nevertheless, henceforth we will safely consider $\textrm{Exp}=\exp$.
The exponential map in equation \eqref{eq:Symmetry_exponentialmap}
relates a member of the group, $\gamma\left(1\right)$, with a member
of its algebra, $\gamma'\left(0\right)$, and it turns out that for
finite dimensional Lie groups, any element of $G$ close enough to
the identity can be written in this way. In fact, this is valid for
all elements connected to the identity $e$ of a compact group. On
the other hand, since the exponential map is a continuous function,
elements of $G$ disconnected from $e$ cannot be given in this form.
Also, if $G$ is not compact, the exponential map may not be surjective.
As an example, consider $\mathbb{R}$ which is a Lie group under addition;
its Lie algebra is also $\mathbb{R}$ but $\exp\left(\mathbb{R}\right)=\mathbb{R}^{+}\neq\mathbb{R}$.

From this discussion, it follows that different groups can have the
same algebra. We shall see this with two examples. The first pair
of groups we shall consider is $G=\mathbb{R}$, the real numbers with
the sum operation, and $G=U\left(1\right)$, the group of complex
numbers of unit modulus, with the multiplication operation. In both
cases the Lie algebra is isomorphic to $\mathbb{R}$, even though
the groups are different.

Consider next $SU(2)$, the group of $2\times2$ unitary complex matrices
with determinant 1, and $SO(3)$, the group of $3\times3$ real symmetry
matrices with unit determinant, or alternatively the group of linear
transformations that preserves the scalar product $\boldsymbol{x}\cdot\boldsymbol{y}=x^{i}y_{i}$
of two $\mathbb{R}^{3}$ vectors, $\boldsymbol{x}$ and $\boldsymbol{y}$.
Defining $\Sigma\left(\boldsymbol{x}\right)\equiv x^{i}\sigma_{i}$
where $\sigma_{i}$ are the three Pauli matrices, it is easy to check
that
\begin{align}
\textrm{Tr}\left[\Sigma\left(\boldsymbol{x}\right)\Sigma\left(\boldsymbol{x}\right)\right] & =2\boldsymbol{x}\cdot\boldsymbol{y}\,.\label{eq:SU2SO3Relation}
\end{align}
On the other hand, notice that if $U$ is a unitary matrix, then $U\Sigma\left(\boldsymbol{x}\right)U^{-1}$
can also be written as $\Sigma\left(\boldsymbol{x'}\right)$ for some
vector $\boldsymbol{x'}$. This vector is a rotation of the original
one, $\boldsymbol{x'}=R_{U}\boldsymbol{x}$ since $\boldsymbol{x'}\cdot\boldsymbol{y'}=\boldsymbol{x}\cdot\boldsymbol{y}$
by relation \eqref{eq:SU2SO3Relation}. This map $U\rightarrow R_{U}$
from $SU(2)$ to $SO(3)$ preserves the group structure of $SO(3)$
(meaning that $UV\rightarrow R_{U}R_{V}$) and in fact we can get
all $\mathbb{R}^{3}$ rotations in this way. However, $R_{U}=R_{-U}$
for any $U\in SU(2)$ which means that different elements of $SU(2)$,
$U$ and $-U$, give the same rotation so these two groups are not
the same ($SU(2)$ covers twice the $SO(3)$ group). Nonetheless,
these two groups share the same algebra and this helps to explain
why the $SU(2)$ group can play the role of the rotation group in
quantum mechanics. Latter on, in connection to the relativistic Lorentz
group, we shall encounter a similar situation.

\section{Lie Algebras}

\subsection{\label{sub:Symmetry_Basic-concepts}Basic concepts}

Since it is a vector space, we can choose a basis for a Lie algebra
$\mathfrak{g}$. We call the elements of such basis the \textbf{generators
of the algebra}, $t_{a}$, with $a=1,\cdots,n$, where $n$ is the
\textbf{dimension of the algebra}. As such, any $x\in\mathfrak{g}$
can be written as a linear combination $c^{a}t_{a}$. In a \textbf{real
algebra} these coefficients $c_{a}$ are real, while in a\textbf{
complex algebra} they are complex. In subsection \ref{sub:Symmetry_PhysicalConventions},
we shall see that despite the proliferation of complex quantities
in Yang\textendash{}Mills theories, the gauge symmetry must have a
real Lie algebra. Nevertheless, unless otherwise stated, in the following
we consider the algebras to be complex.

The concepts of abelian, simple, semi-simple and reductive Lie algebra
are equally important. An \textbf{abelian Lie algebra} $\mathfrak{g}$
is one for which the Lie bracket is always zero, $\left[\mathfrak{g},\mathfrak{g}\right]=0$.%
\footnote{This notation means the following: the set generated by taking the
commutator of all combinations of $x_{1},x_{2}\in\mathfrak{g}$ contains
only one element---the zero vector.%
} Note that every abelian Lie algebra can be broken down into several
$\mathfrak{u}(1)$'s.

To proceed we need the concept of an ideal of an algebra: a set $\mathfrak{s}\subset\mathfrak{g}$
is a \textbf{subalgebra} of $\mathfrak{g}$ if $\mathfrak{s}$ closes
under the Lie bracket operation: $\left[\mathfrak{s},\mathfrak{s}\right]\subset\mathfrak{s}$.
If $\mathfrak{s}\subset\mathfrak{g}$ meets the more demanding condition
that $\left[\mathfrak{g},\mathfrak{s}\right]\subset\mathfrak{s}$
then $\mathfrak{s}$ is an \textbf{ideal} of $L$ (and clearly a subalgebra
too). In a way, subalgebras and ideals stand for Lie algebras in the
same way as subgroups and invariant subgroups stand for groups, respectively.
Note that for every Lie algebra $\mathfrak{g}$ there are two trivial
ideals, the $0$ algebra of null dimension, and $\mathfrak{g}$ itself;
the other non-trivial ones are called \textbf{proper ideals} of $\mathfrak{g}$.

A Lie algebra is \textbf{simple} if it is non-abelian and has no proper
ideals. On the other hand, a \textbf{semi-simple Lie algebra} is a
non-null Lie algebra with no proper abelian ideals \citep{Cahn:1985wk}.
The connection between these two last concepts is made more clear
with the notion of \textbf{direct sum} $\mathfrak{g}=\mathfrak{g}_{1}\oplus\cdots\oplus\mathfrak{g}_{n}$
of Lie algebras $\mathfrak{g}_{1},\cdots,\mathfrak{g}_{n}$. Any element
of this $\mathfrak{g}$ can be written as a linear combination of
elements of the $\mathfrak{g}_{i}$, and in addition each $\mathfrak{g}_{i}$
must be an ideal of $\mathfrak{g}$. As a consequence of this last
requirement, $\left[\mathfrak{g}_{i},\mathfrak{g}_{j}\right]=0$ for
$i\neq j$. Then, the important result is that a semi-simple Lie algebra
is a direct sum of simple Lie algebras. On the other hand, a direct
sum of simple and abelian Lie algebras is called a \textbf{reductive
Lie algebra} \citep{Fuchs:2003aa}.

In subsection \ref{sub:Symmetry_PhysicalConventions} we will argue
that gauge symmetries must be given by a reductive Lie algebra. Since
abelian Lie algebras are just direct sums of trivial $\mathfrak{u}(1)$
algebras, the complexity of reductive Lie algebras is completely encoded
in the structure of simple groups. Therefore in the following we focus
on simple Lie algebras.

Another important concept is that of a \textbf{representation} (of
a Lie algebra): it is a linear map $\rho$ which associates to every
$x\in\mathfrak{g}$ a linear operator over some vector space, typically
$\mathbb{C}^{n}$. In other words, $\rho\left(x\right)$ can be seen
as a matrix and the linearity of $\rho$ implies that $\rho\left(\alpha x+\beta y\right)=\alpha\rho\left(x\right)+\beta\rho\left(y\right)$
for some $x,y\in\mathfrak{g}$ and $\alpha,\beta\in\mathbb{R}$ or
$\mathbb{C}$. In addition, to be a representation this linear map
must preserve the structure of $\mathfrak{g}$, which means that $\rho\left(\left[x,y\right]\right)=\left[\rho\left(x\right),\rho\left(y\right)\right]$.
Note that we are only considering here the cases when the Lie algebra
$\mathfrak{g}$ itself is a set of matrices, so there is a blurring
between $\mathfrak{g}$ and its trivial representation $\rho=\textrm{Identity}$.
As such, the set of matrices that defines $\mathfrak{g}$ is sometimes
called its \textbf{fundamental }or\textbf{ defining representation}.

While usually $\rho\left(x\right)$ is a linear operator over $\mathbb{C}^{n}$,
there is an important case where the vector space is $\mathfrak{g}$
itself. Consider again $x,y\in\mathfrak{g}$; then for every $x$
we can associate a linear transformation over $\mathfrak{g}$, $\textrm{ad }x$,
such that $\textrm{ad }x\left(y\right)\equiv\left[x,y\right]$. Note
that this map is linear and also $\textrm{ad }\left(\left[x,y\right]\right)=\left[\textrm{ad }\left(x\right),\textrm{ad }\left(y\right)\right]$,
so it is a representation. We call $\textrm{ad }x$ the \textbf{adjoint
representation} of $x$. Furthermore, if $t_{1},\cdots,t_{n}$ are
some generators of the Lie algebra, we can use them as a basis and
get the adjoint representation as a set of $n\times n$ matrices.
To do so, we compute $\textrm{ad }t_{i}\left(t_{j}\right)=\left[t_{i},t_{j}\right]\equiv c_{ij}^{k}t_{k}$
so $\textrm{ad }t_{i}$ can be seen as a matrix with entries $\left(\textrm{ad }t_{i}\right)_{kj}=c_{ij}^{k}$.
These $c_{ij}^{k}$ coefficients are known as the \textbf{structure
constants} of $\mathfrak{g}$ and are often viewed as carrying fundamental
information of the underlying Lie algebra. However, note that they
depend on the choice of generators, and are therefore basis dependent
numbers.

\subsection{\label{sub:Symmetry_Roots-of-simple}Roots of simple Lie algebras}

A useful tool in the study of simple Lie algebras is the so called
root space decomposition, where a very particular basis for the algebra
is used. The starting point consists of finding the largest set $\mathfrak{h}\in\mathfrak{g}$
such that $\left[\mathfrak{g},\mathfrak{h}\right]=\mathfrak{h}$ (so
that $\mathfrak{h}$ is an ideal of $\mathfrak{g}$) and $\left[\mathfrak{h},\mathfrak{h}\right]=0$.
Such $\mathfrak{h}$ is called a \textbf{maximal abelian subalgebra},
\textbf{maximal toral subalgebra} or \textbf{Cartan subalgebra} of
$\mathfrak{g}$. Note that there can be different Cartan subalgebras
for a given $\mathfrak{g}$, but it can be shown that all choices
are equivalent. The dimension of any of these Cartan subalgebras of
$\mathfrak{g}$ is known as the \textbf{rank} of the Lie algebra.

In Physics, this subalgebra can be directly related to the quantum
number of fields since its elements are a maximal set of matrices
that can be simultaneously diagonalizable. For example, in $SU(2)$
we have the 1-dimensional subalgebra generated by $\sigma_{3}=\textrm{diag}\left(1,-1\right)$
and in $SU(3)$ we may consider the 2-dimensional space generated
by the Gell-Mann matrices $\lambda_{3}=\textrm{diag}\left(1,-1,0\right)$
and $\lambda_{8}=\frac{1}{\sqrt{3}}\textrm{diag}\left(1,1,-2\right)$. 

Once we have chosen a particular Cartan subalgebra $\mathfrak{h}$
of $\mathfrak{g}$, there are bases of $\mathfrak{g}$, known as \textbf{Cartan-Weyl
bases}, such that the adjoint representation of all $h\in\mathfrak{h}$
is diagonal. Thus, if we write these basis elements as $e_{\alpha}$,
$\left[h,e_{\alpha}\right]$ is proportional to $e_{\alpha}$ and
we may denote the proportionality constant as $\alpha\left(h\right)$:
\begin{alignat}{1}
\left[h,e_{\alpha}\right] & \equiv\alpha\left(h\right)e_{\alpha}\,.\label{eq:IntroducingRoots}
\end{alignat}
The $\alpha\left(h\right)$ appearing in this equation is a number
which depends linearly on $h$, so for each $e_{\alpha}$ it can be
viewed as function $\alpha$ that converts elements of $\mathfrak{h}$
into numbers (a functional). Therefore these $\alpha$ belong to the
dual vector space of $\mathfrak{h}$, which is denoted as $\mathfrak{h}^{*}$,
and they are known as \textbf{roots} (the eigenvalues of $\textrm{ad}\left(h\right)$)
while the $e_{\alpha}$ are called \textbf{root vectors} (the eigenvectors
of $\textrm{ad}\left(h\right)$). We stress again that even though
$\alpha$ is an eigenvalue of $\textrm{ad}\left(h\right)$, it is
not to be seen as a number because we want to consider an arbitrary
$h\in\mathfrak{h}$ and not just a specific one, so $\alpha$ is a
function of $h$. The root vector $e_{\alpha}$ on the other hand
must be an eigenvector of $\textrm{ad}\left(h\right)$, for all $h$.
In practice, since every $h$ is a linear combination of some basis
elements $h_{1},\cdots,h_{n}$ of $\mathfrak{h}$, for every $e_{\alpha}$
the corresponding root $\alpha$ can be seen as the list of numbers
$\alpha\left(h_{1}\right),\cdots,\alpha\left(h_{n}\right)$. There
is however a detail: if a given $\alpha$ is zero, or in other words
$\alpha\left(h_{1}\right)=\cdots=\alpha\left(h_{n}\right)=0$, then
it is not considered a root and this happens only when $e_{\alpha}$
itself is in the Cartan subalgebra $\mathfrak{h}$. The\textbf{ root
system} $\Delta$ is the name given to the set of all roots of a Lie
algebra.

We have just achieved a \textbf{root space decomposition} of $\mathfrak{g}$:
\begin{alignat}{1}
\mathfrak{g}= & \underset{\alpha}{\oplus}\mathfrak{g}_{\alpha}=\mathfrak{h}+\underset{\alpha\neq0}{\oplus}\mathfrak{g}_{\alpha}\,,
\end{alignat}
where $\mathfrak{g}_{\alpha}$ is the subspace of the Lie algebra
$\mathfrak{g}$ consisting of all elements $x\in\mathfrak{g}$ such
that $\left[h,x\right]=\alpha\left(h\right)x$ for an arbitrary $h\in\mathfrak{h}$
($\mathfrak{g}_{\alpha}$ is called a \textbf{root space}). In other
words, $\mathfrak{g}_{\alpha}$ is the eigenspace of $\textrm{ad}\left(h\right)$
with eigenvalue $\alpha$ and it can be shown that $\mathfrak{g}_{\alpha\neq0}$
is always a 1-dimensional space: it is made up of $e_{\alpha}$ and
multiples of it---confer with equation \eqref{eq:IntroducingRoots}.
Note that this is usually not true for $\mathfrak{g}_{0}$, which
coincides with the Cartan subalgebra $\mathfrak{h}$, and is therefore
an $n$-dimensional space.

In order to make this discussion less abstract, consider $\mathfrak{su}_{\mathbb{C}}(2)$,
which is the \textbf{complexified} algebra of $SU(2)$ (linear combinations
of the generators can be complex). We do this complexification because
we will be assuming that the algebras are complex, and with this understanding
we will drop the $\mathbb{C}$ subscript. The Pauli matrices are commonly
chosen as generators,
\begin{align}
\sigma_{1} & =\left(\begin{array}{cc}
0 & 1\\
1 & 0
\end{array}\right)\,,\;\sigma_{2}=\left(\begin{array}{cc}
0 & -i\\
i & 0
\end{array}\right)\,,\;\sigma_{3}=\left(\begin{array}{cc}
1 & 0\\
0 & -1
\end{array}\right)\,,
\end{align}
whose commutators are
\begin{alignat}{1}
\left[\sigma_{i},\sigma_{j}\right] & =2i\varepsilon_{ijk}\sigma_{k}\,.\label{eq:Symmetry_PauliMatrices_commutator}
\end{alignat}
The third Pauli matrix can be taken as the generator of the Cartan
subalgebra, but $\left[\sigma_{3},\sigma_{1(2)}\right]$ is not proportional
to $\sigma_{1(2)}$ (see equation \eqref{eq:IntroducingRoots}), so
we change to a Cartan-Weyl basis:
\begin{gather}
e=\frac{1}{2}\left(\sigma_{1}+i\sigma_{2}\right)=\left(\begin{array}{cc}
0 & 1\\
0 & 0
\end{array}\right)\,,\; f=\frac{1}{2}\left(\sigma_{1}-i\sigma_{2}\right)=\left(\begin{array}{cc}
0 & 0\\
1 & 0
\end{array}\right)\,,\nonumber \\
h=\sigma_{3}=\left(\begin{array}{cc}
1 & 0\\
0 & -1
\end{array}\right)\,.\label{eq:EFH_SU2}
\end{gather}
This $e$ and $f$ are known in Physics as raising and lowering operators.
According to equation \eqref{eq:IntroducingRoots}, $e$ and $f$
are root vectors with roots
\begin{alignat}{1}
\alpha_{e}\left(h\right)= & 2\,,\quad\alpha_{f}\left(h\right)=-2\,,
\end{alignat}
so the root system of $\mathfrak{su}(2)$ is $\Delta=\left\{ \alpha_{e},\alpha_{f}\right\} =\left\{ \pm\alpha_{e}\right\} $.
Similarly, for $\mathfrak{su}(3)$ we may use\setlength{\tabcolsep}{4pt}
\thinmuskip=1.5mu
\medmuskip=1mu
\thickmuskip=2mu
\begin{alignat}{1}
\left\{ x_{1},\cdots,x_{8}\right\} = & \frac{1}{2}\left\{ \lambda_{1}+\lambda_{2},\lambda_{1}-\lambda_{2},\lambda_{4}+\lambda_{5},\lambda_{4}-\lambda_{5},\lambda_{6}+\lambda_{7},\lambda_{6}-\lambda_{7},\lambda_{3},\frac{2}{\sqrt{3}}\lambda_{8}\right\} \,,
\end{alignat}
\thinmuskip=3mu
\medmuskip=4.0mu plus 2.0mu minus 4.0mu
\thickmuskip=5.0mu plus 5.0muas a basis of the algebra ($\lambda_{i}$ are the eight Gell-Mann
matrices). The Cartan subalgebra is generated by $x_{7}$ and $x_{8}$
and
\begin{alignat}{1}
\alpha_{1\left(2\right)}\left(ax_{7}+bx_{8}\right) & =\pm a\,,\\
\alpha_{3\left(4\right)}\left(ax_{7}+bx_{8}\right) & =\pm\left(-\frac{1}{2}a+b\right)\,,\\
\alpha_{5\left(6\right)}\left(ax_{7}+bx_{8}\right) & =\pm\left(\frac{1}{2}a+b\right)\,,\\
\alpha_{7\left(8\right)}\left(ax_{7}+bx_{8}\right) & =0\,,
\end{alignat}
so the root system is $\left\{ \alpha_{1},\cdots,\alpha_{6}\right\} =\left\{ \pm\alpha_{1},\pm\alpha_{3},\pm\left(\alpha_{1}+\alpha_{3}\right)\right\} $.
Two features emerge here which turn out to be true for any simple
algebra: the first one is that for every root $\alpha$ there is an
opposite one, $-\alpha$. The second is that some of the roots are
linear combinations of other roots, which is to be expected, as roots
are objects in $\mathfrak{h}^{*}$ (the dual vector space of the Cartan
subalgebra $\mathfrak{h}$), and therefore there can be at most $n$
independent $\alpha$'s, where $n$ is the algebra rank. In fact,
it turns out there are always exactly $n$ linearly independent roots.
For latter use, we would like now to define negative and positive
roots and this can be done as follows. First we choose $n$ linearly
independent roots $\left\{ \alpha_{1},\cdots,\alpha_{n}\right\} $,
with some arbitrary but fixed ordering. Every root $\alpha$ can then
be written as a linear combination $\alpha=c^{i}\alpha_{i}$ and $\alpha$
is said to be a \textbf{positive root} if the first non-zero $c^{i}$
is a positive number. Any positive root that is not the sum of two
positive roots is called a \textbf{simple root} and there are always
$n$ such roots. We shall denote by $\Delta_{+}$ the set of all positive
roots of a simple Lie algebra, and $\Pi$ will represent the set of
simple roots.

Simple roots are very important because, as we shall see, they carry
all the information on the structure of the Lie algebra, and for this
reason they are used in the classification of complex simple Lie algebras.
Before discussing this, we need to introduce a pseudo-inner product
called the Killing form. As for the notation, for convenience we shall
indicate summations explicitly when there is one, in the following
three subsections.

\subsection{\label{sub:Symmetry_The-Killing-form}The Killing form}

The \textbf{Killing form} $\left(\cdot,\cdot\right)$ provides something
similar to a scalar product in a Lie algebra $\mathfrak{g}$:
\begin{align}
\left(y,z\right) & \equiv\textrm{Tr}\left[\textrm{ad}\left(y\right)\textrm{ad}\left(z\right)\right]\label{eq:KillingFormDefinition}
\end{align}
for $y,z\in\mathfrak{g}$. With a basis $x_{1},\cdots,x_{n}$ of $\mathfrak{g}$,
this is the same as $\sum_{i}\left[y,\left[z,x_{i}\right]\right]_{\textrm{coefficient in }x_{i}}$.
Notice that while the Killing form is symmetric, $\left(y,z\right)=\left(z,y\right)$,
in general it is degenerate, meaning that for some non-null $y$ we
have $\left(y,z\right)=0$ for any $z$. However, there is a theorem
due to Cartan which states that a Lie algebra is semi-simple if and
only if the Killing form is non-degenerate. Even so, if the algebra
is complex then $\left(iy,iy\right)=-\left(y,y\right)$, which means
that $\left(\cdot,\cdot\right)$ is neither positive nor negative
definite. On the other hand, the particular real Lie algebras used
in gauge theories (see subsection \ref{sub:Symmetry_PhysicalConventions})
are such that, as long as $y\neq0$, $\left(y,y\right)$ is always
negative (this is a theorem due to Weyl).

It can be shown that if another representation is used in equation
\eqref{eq:KillingFormDefinition} instead of the adjoint one, the
resulting bilinear function $\left(\cdot,\cdot\right)'$ differs from
$\left(\cdot,\cdot\right)$ by just a multiplicative factor. As such,
we note that the Dynkin index $S\left(R\right)$ used in Physics must
be related to the Killing form by some multiplicative factor (subsection
\ref{sub:Symmetry_PhysicalConventions}).

The Killing form, being a bilinear non-degenerate form on the simple
Lie algebra $\mathfrak{g}$ and in particular on its Cartan subalgebra
$\mathfrak{h}$, can be used to identify the dual space $\mathfrak{h}^{*}$
(the space to which roots belong) with $\mathfrak{h}$ itself. Indeed,
for every root $\alpha\in\mathfrak{h}^{*}$ there is an $h_{\alpha}\in\mathfrak{h}$
such that
\begin{alignat}{1}
\alpha\left(k\right) & =\left(h_{\alpha},k\right)\,,\label{eq:RelationBetweenRootsAndCartanSubalgebraElements}
\end{alignat}
with $k\in\mathfrak{h}$. Consider then $\mathfrak{h}_{0}^{*}$, the
space of real combinations of the roots such that $\left(h_{\alpha},h_{\beta}\right)$
is positive or zero for any $\alpha,\beta\in\mathfrak{h}_{0}^{*}$.
With the Killing form it is possible to define a genuine inner product
\begin{alignat}{1}
\left\langle \alpha,\beta\right\rangle  & \equiv\left(h_{\alpha},h_{\beta}\right)
\end{alignat}
on this space, which means that we can talk about norms and angles
between roots. We note in particular that equation \eqref{eq:IntroducingRoots}
can be written as $\left[h_{\beta},e_{\alpha}\right]=\alpha\left(h_{\beta}\right)e_{\alpha}=\left\langle \alpha,\beta\right\rangle e_{\alpha}$.
Normalizing both root vectors $e_{\alpha}$ and Cartan subalgebra
elements $h_{\beta}$ will lead us in the next subsection to the Chevalley-Serre
basis.

\subsection{\label{sub:Symmetry_The-Cartan-matrix}The Cartan matrix and the
classification of all complex simple Lie algebras}

If $\left\{ \alpha_{1},\cdots,\alpha_{n}\right\} $ are the simple
roots of a complex simple Lie algebra $\mathfrak{g}$ then the \textbf{Cartan
matrix} $A$ of $\mathfrak{g}$, defined as
\begin{alignat}{1}
A_{ij} & \equiv\frac{2\left\langle \alpha_{i},\alpha_{j}\right\rangle }{\left\langle \alpha_{j},\alpha_{j}\right\rangle }\,,\label{eq:CartanMatrixDefinition}
\end{alignat}
can be shown to encode all the information about $\mathfrak{g}$,
and therefore it provides a way to classify all such algebras. This
is a $n\times n$ matrix with peculiar properties which are a consequence
of restrictions in the angles and norms of simple roots. Without proof,
some of the more important ones are the following:
\begin{enumerate}
\item $A_{ii}=2$;
\item $A_{ij}=0,-1,-2$ or $-3$ for $i\neq j$;
\item $A_{ij}A_{ji}=0,1,2$ or $3$ for $i\neq j$, and in the first case
$A_{ij}=A_{ji}=0$;
\item There is at most one entry in $A$ with a value smaller than -1;
\item The sum of the negative entries of each column or row of $A$ is never
smaller than -3;
\item $\det A\neq0$.
\end{enumerate}
Notice that the Cartan matrix is not symmetric in general. The reason
is that different $\alpha_{i}$ may have different squared norms $\left\langle \alpha_{i},\alpha_{i}\right\rangle $.
In any case, from properties 2, 3 and 4 of the Cartan matrix, we can
infer that a simple Lie algebra $\mathfrak{g}$ contains at most simple
roots of two different squared norms, differing by a factor of 2 or
3. The absolute normalization of the $\alpha_{i}$ is irrelevant though.

Since $A$ is such a special matrix, it is very often translated into
a \textbf{Dynkin diagram}. This is done by representing each $\alpha_{i}$
by a dot and connecting the dots of $\alpha_{i}$ and $\alpha_{j}\neq\alpha_{i}$
by $A_{ij}A_{ji}=\max\left(\left|A_{ij}\right|,\left|A_{ji}\right|\right)$
lines. If all roots have the same norm, $A_{ij}A_{ji}$ will always
be 0 or 1, so all connections are with a single line, and the algebra
is called \textbf{simply laced}. If this is not the case, the simple
roots with double or triple line connections must be distinguished
in the Dynkin diagram, and a common convention is to use a white dot
for the one with the bigger norm and a black dot for the simple root
with the smaller norm.

\begin{figure}[tbph]
\begin{centering}
\includegraphics[scale=0.67]{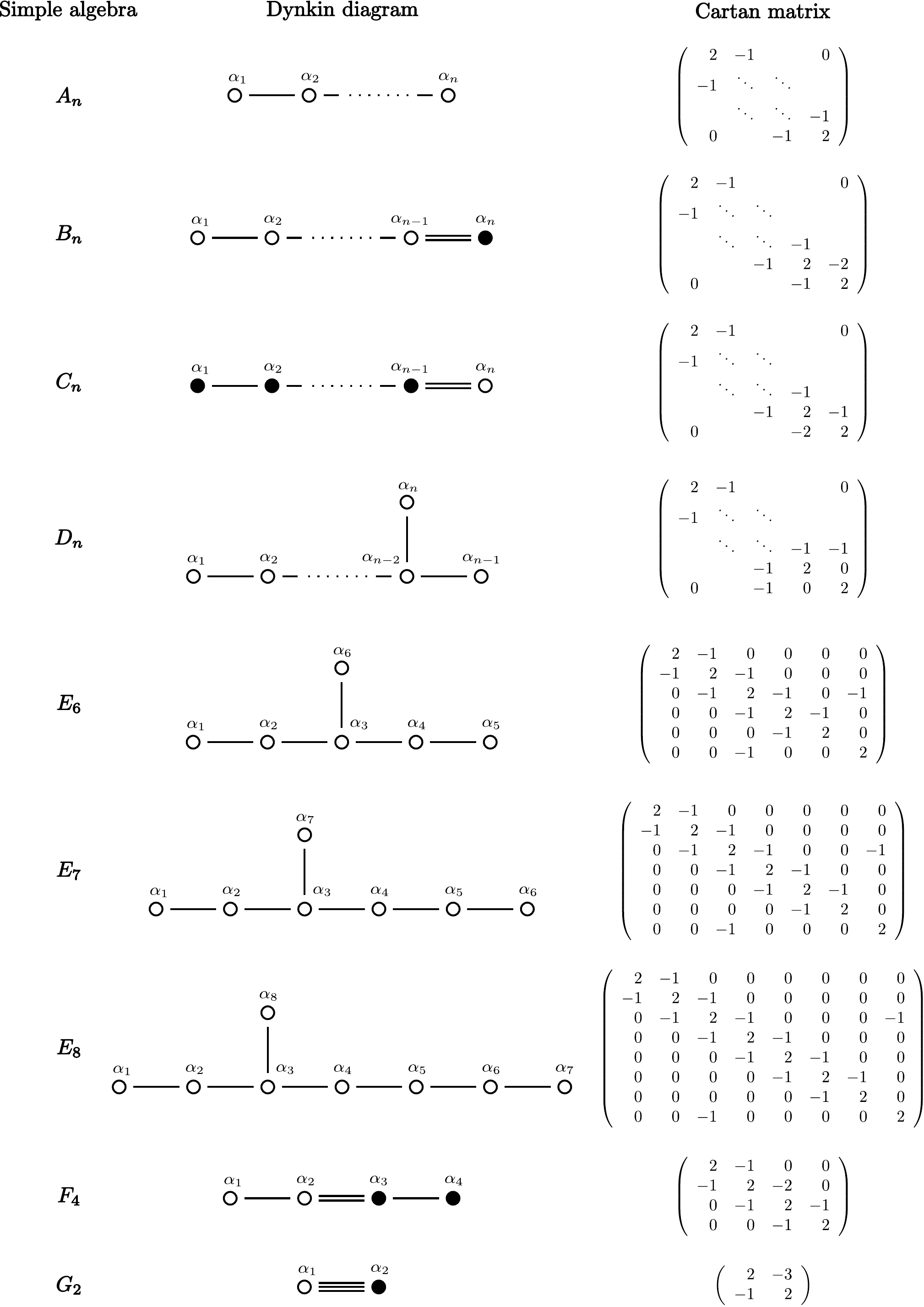}
\par\end{centering}

\caption{\label{fig:Symmetry_ListOfComplexSimpleLieAlgebras}List of all the
complex simple Lie algebras. In the Dynkin diagrams, we have added
a label to each dot indicating the simple root it represents. A permutation
of these labels leads to a Cartan matrix with a different arrangement
of rows and columns, but the underlying Lie algebra is the same.}

\end{figure}

As a consequence of the peculiarities of the Cartan matrix, it is
easy to list all possibilities.There are four infinite families of
Lie algebras, also known as the \textbf{classical Lie algebras},
\begin{align}
A_{n} & =\mathfrak{su}\left(n+1\right) & n\geq1 & \;, & B_{n} & =\mathfrak{so}\left(2n+1\right) & n\geq2 & \;,\label{eq:AnBn}\\
C_{n} & =\mathfrak{sp}\left(2n\right) & n\geq3 & \;, & D_{n} & =\mathfrak{so}\left(2n\right) & n\geq4 & \;,\label{eq:CnDn}
\end{align}
and five singular or \textbf{exceptional Lie algebras} designated
as
\begin{equation}
G_{2}\;,\; F_{4}\;,\; E_{6}\;,\; E_{7}\;,\; E_{8}\;.
\end{equation}
Their Cartan matrices and Dynkin diagrams are shown in figure \eqref{fig:Symmetry_ListOfComplexSimpleLieAlgebras}.
From the diagrams, it is clear that the constraints in equations \eqref{eq:AnBn}
and \eqref{eq:CnDn} on $n$ can be relaxed a little: for example
there is a $C_{2}$, but it is the same as $B_{2}$. Also, $D_{3}=A_{3}$
and $A_{1}=B_{1}=C_{1}=D_{1}$. As for $D_{2}$, its Dynkin diagram
consists of two disconnected dots, meaning that this algebra consists
of two $A_{1}$'s that are independent of each other ($D_{2}=A_{1}\oplus A_{1}$),
so it is not a simple Lie algebra.

The Cartan matrix is related to the commutator of elements of the
algebra. For a set of simple roots $\alpha_{1},\cdots,\alpha_{n}$
we define the following $3n$ elements of the Lie algebra (the \textbf{Chevalley-Serre
basis}): 
\begin{alignat}{1}
e_{i} & \equiv e_{\alpha_{i}}\,,\label{eq:Chevalley_Serre_E}\\
f_{i} & \equiv\frac{2}{\left(e_{\alpha_{i}},e_{-\alpha_{i}}\right)\left\langle \alpha_{i},\alpha_{i}\right\rangle }e_{-\alpha_{i}}\,,\label{eq:Chevalley_Serre_F}\\
h_{i} & \equiv\frac{2}{\left\langle \alpha_{i},\alpha_{i}\right\rangle }h_{\alpha_{i}}\,.\label{eq:Chevalley_Serre_H}
\end{alignat}
Recall that $e_{\alpha}$($e_{-\alpha}$) are the root vectors associated
to the root/eigenvalue $\alpha$($-\alpha$) and $h_{\alpha}$ is
the member of the Cartan subalgebra $\mathfrak{h}$ such that $\alpha\left(k\right)=\left(h_{\alpha},k\right)$
for a $k\in\mathfrak{h}$ (equation \eqref{eq:RelationBetweenRootsAndCartanSubalgebraElements}).
Then, it can be shown that the commutator between these elements of
the algebra is given by the \textbf{Chevalley-Serre relations}: 
\begin{alignat}{1}
\left[e_{i},f_{j}\right] & =\delta_{ij}h_{j}\,,\label{eq:Serre_Chevalley_relations1}\\
\left[h_{i},e_{j}\right] & =A_{ji}e_{j}\,,\\
\left[h_{i},f_{j}\right] & =-A_{ji}f_{j}\,,\label{eq:Serre_Chevalley_relations3}
\end{alignat}
where $A$ is the Cartan matrix. Except for $\mathfrak{su}(2)$, these
$3n$ elements do not generate the algebra, which is larger. For example,
$\mathfrak{su}(3)$ has rank 2 ($n=2$) but its dimension is $8$,
which is bigger than $3\times2$. However, the missing generators
can be obtained by successive commutations of the lowering and raising
operators $e_{i}$ and $f_{i}$, until no new elements are created:
$\left[e_{i},e_{j}\right],\left[e_{i},\left[e_{j},e_{k}\right]\right],\cdots,\left[f_{i},f_{j}\right],\left[f_{i},\left[f_{j},f_{k}\right]\right],\cdots$.
It is also worth noting the similarity between $e_{i}$, $f_{i}$
and $h_{i}$ in equations \eqref{eq:Serre_Chevalley_relations1}--\eqref{eq:Serre_Chevalley_relations3}
and the raising ($e$), lowering ($f$) and diagonal ($h$) operators
of $\mathfrak{su}(2)$ in equation \eqref{eq:EFH_SU2}. In a transparent
way, the Chevalley-Serre relations tell us that a Lie algebra of rank
$n$ can be viewed as being made up of $n$ copies of $\mathfrak{su}(2)$,
one for each dot on the Dynkin diagram, and that the interactions
between these $\mathfrak{su}(2)$ copies are encoded by the Cartan
matrix. Despite the apparent simplicity of these relations, they endow
simple Lie algebras with intricate properties, all of which are calculable
from the Cartan matrix.

\subsection{\label{sub:Symmetry_Representations,-weights-and}Representations,
weights and the Casimir operator}

In $\mathfrak{su}(2)$ every irreducible representation is identifiable
by an non-negative integer $2s$, where $s$ is the half-integer spin
in Particle Physics, and this representation contains $2s+1$ different
isospin components $2t_{3}=2s,2s-2,\cdots,-2s+2,-2s$. Noting that
$A_{1}=\mathfrak{su}(2)$ is the most basic of simple groups, we shall
now see how this generalizes for other simple groups with a rank $n>1$.
The generalization of twice the isospin components $t_{3}$ are called
the weights of a representation, and in practice they are a list of
$n$ integers each (for $\mathfrak{su}(2)$, $n=1$ so each weight
is a number). The weights of an irreducible representation can be
sorted, and the representation itself is labeled by its highest weight,
which in practice is a list of $n$ non-negative integers---the so-called
Dynkin coefficients of the representation. 

To see this in some detail, consider a $\Phi$ which transforms under
some representation of a Lie algebra $\mathfrak{g}$, in a basis where
all the representation matrices $\mathcal{H}$ of the Cartan subalgebra
elements $h$ are diagonal. Then for each component $w$ of $\Phi^{w}$
\begin{alignat}{1}
\mathcal{H}\Phi^{w} & \equiv M^{w}\left(h\right)\Phi^{w}\,.
\end{alignat}
The $M^{w}$ are called \textbf{weights}, and they are functions that
transform elements of the Cartan subalgebra into plain numbers, just
like roots. So, as with roots, we can view them as a list of $n$
numbers by establishing a basis $\left\{ h_{i}\right\} $ for the
Cartan subalgebra. In particular, if we use the Chevalley-Serre basis
such that $\mathcal{H}_{i}$ is the representation matrix of $h_{i}$
then, by equation \eqref{eq:Chevalley_Serre_H}, we have 
\begin{alignat}{1}
\mathcal{H}_{i}\Phi^{w} & =\frac{2\left\langle M^{w},\alpha_{i}\right\rangle }{\left\langle \alpha_{i},\alpha_{i}\right\rangle }\Phi^{w}\equiv M_{i}^{w}\Phi^{w}\,,
\end{alignat}
so the weight $M^{w}$ is reduced to a list of plain numbers $M_{i}^{w}$,
$i=1,\cdots,n$. It turns out that these numbers are always integers,
just like $2t_{3}$ in $\mathfrak{su}(2)$. Once the weights $M^{w}$
for the different $w$ are sorted, the biggest one $\Lambda\equiv\max\left\{ M^{w}\right\} $
can be used to label the representation, as mentioned above. The $n$
numbers
\begin{alignat}{1}
\Lambda_{i} & =\frac{2\left\langle \Lambda,\alpha_{i}\right\rangle }{\left\langle \alpha_{i},\alpha_{i}\right\rangle }
\end{alignat}
are non-negative and are called the\textbf{ Dynkin coefficients} of
a representation. With (a) the Cartan matrix of the algebra and (b)
the $\Lambda_{i}$ Dynkin coefficients of a representation, all properties
of the representation can be computed.

At this point, we should mention that, unlike in $\mathfrak{su}(2)$,
two different components $\Phi^{w}$ and $\Phi^{w'}$ of the vector
$\Phi$ may have the same weight ($M^{w}=M^{w'}$). In other words,
the eigenspaces of the representation matrix $\mathcal{H}$ of an
arbitrary element of the Cartan subalgebra (the \textbf{weight spaces})
are in general degenerate/multi-dimensional.%
\footnote{The adjoint representation of an $n$-rank algebra is an excellent
example: since the $n$ generators of the Cartan subalgebra commute
between themselves, the weight space associated to $M=0$ is $n$-dimensional. %
} This weight multiplicity can be computed with \textbf{Freudenthal's
formula}, and summing together the multiplicities of all weights yields
the dimension of the representation, which is given by \textbf{Weyl's
dimension formula} (see \citep{Cahn:1985wk,Fuchs:2003aa,Jacobson:1979aa}
for more details).

These weights have a number of interesting properties. For example,
they come in $\alpha$-strings: given a weight $M$ and some root
$\alpha$ of the algebra, in general there is a sequence of $M-m\alpha,M-\left(m-1\right)\alpha,\cdots,M,\cdots,M+p\alpha$
where the limits $m$ and $p$ are easily calculable: it can be shown
for example that $m-p=\frac{2\left\langle M,\alpha\right\rangle }{\left\langle \alpha,\alpha\right\rangle }$.
Consider now the following. Each string has a virtual middle at $M_{0}\equiv M+\frac{\left(p-m\right)}{2}\alpha=M-\frac{\left\langle M,\alpha\right\rangle }{\left\langle \alpha,\alpha\right\rangle }\alpha$,
which may or may not be a real weight. So an inversion of an $\alpha$-string
will transform the generic weight $M$ into $M_{0}-\left(M-M_{0}\right)$$=M-2\frac{\left\langle M,\alpha\right\rangle }{\left\langle \alpha,\alpha\right\rangle }\alpha$.
This is a symmetry of the weight system:
\begin{alignat}{1}
S_{\alpha}:\; & M\rightarrow M-2\frac{\left\langle M,\alpha\right\rangle }{\left\langle \alpha,\alpha\right\rangle }\alpha\,.
\end{alignat}
All such symmetries, when taking into consideration different $\alpha$'s,
generate the \textbf{Weyl group}. Indeed, the set of reflections induced
by the $n$ simple roots is enough to generate the whole Weyl group.
However, this does not mean that every element of the Weyl group is
of the form $S_{\alpha}$ for some root $\alpha$ (the Weyl group
is often much larger than the set $\left\{ S_{\alpha}\right\} $).
We shall not discuss it any further, but this symmetry of the weight
system has many application, in particular it is often used to speed
up computations.

As a final topic concerning representations, we note that it is possible
to build an operator, quadratic in the generators, that commutes with
the whole algebra. This is the well know \textbf{Casimir operator}
$C$. It turns out that a  Cartan-Weyl basis is more suitable to build
such an operator: if for every root $\alpha$ of the algebra the root
vectors are normalized such that $\left[e_{\alpha},e_{-\alpha}\right]=h_{\alpha}$,
then
\begin{align}
C & =\sum_{\alpha_{i},\alpha_{j}\in\Pi}\frac{2\left(A^{-1}\right)_{ij}}{\left\langle \alpha_{i},\alpha_{i}\right\rangle }h_{\alpha_{i}}h_{\alpha_{j}}+2\sum_{\alpha\in\Delta^{+}}e_{\alpha}e_{-\alpha}\,,\label{eq:Symmetry_Casimir_CartanWeyl}
\end{align}
where $A$ is the Cartan matrix. We recall here that $\Pi$ and $\Delta^{+}$
are the sets of simple and positive roots, respectively (see subsection
\ref{sub:Symmetry_Roots-of-simple}). Since it commutes with the algebra,
$C$ is proportional to the identity operator, and after some calculations
it can be shown that when applied to an irreducible representation
with highest weight $\Lambda$, the Casimir operator is given by
\begin{align}
C & =\left\langle \lambda,\lambda+\sum_{\alpha\in\Delta^{+}}\alpha\right\rangle \mathbb{1}=\sum_{i,j}\frac{1}{2}\Lambda_{i}\left(A^{-1}\right)_{ij}\left\langle \alpha_{j},\alpha_{j}\right\rangle \left(\Lambda_{j}+2\right)\mathbb{1}\,,
\end{align}
which is rather easy to calculate from the Cartan matrix $A$ and
Dynkin coefficients $\Lambda_{i}$ of the representation. If the smallest
root is taken to be of norm 1, $\min\left(\left\langle \alpha_{i},\alpha_{i}\right\rangle \right)=1$,
then it turns out that $C=\left(n^{2}-1\right)/2n\,\mathbb{1}$ for
the \textbf{fundamental representation} of $\mathfrak{su}(n)$, whose
Dynkin coefficients are $\Lambda_{i}=\delta_{i1}$. This matches the
normalization used in Particle Physics.

\subsection{Subalgebras and branching rules}

We shall not deal at length with the issue of finding the subalgebras
of a given Lie algebra, even though this is of great importance in
Particle Physics and GUTs in particular. Sometimes the vacuum state
in a quantum field theory breaks the gauge symmetry, and in such a
case, one first tries to find which part of the original symmetry
is still preserved. Once this is known, it is then necessary to study
how the representations of the original Lie algebra behave under the
new one, which is preserved by the vacuum state (the \textbf{branching
rules}). To address the first part of the problem in a systematic
way, the concept of \textbf{maximal subalgebra} is needed: $\mathfrak{g}'$
is said to be a maximal subalgebra of $\mathfrak{g}$ if there is
no other subalgebra $\mathfrak{g}''$ of $\mathfrak{g}$ such that
$\mathfrak{g}\subset\mathfrak{g}''\subset\mathfrak{g}'$ (other than
the trivial cases $\mathfrak{g}''=\mathfrak{g}$ or $\mathfrak{g}'$).
By studying maximal subalgebras only, there is no need to say, for
instance, that $\mathfrak{su}(3)$ may break into $\mathfrak{u}\left(1\right)^{3}$
as this becomes obvious once it is known that $\mathfrak{su}(3)$
can break into its maximal subalgebra $\mathfrak{su}\left(2\right)\oplus\mathfrak{u}\left(1\right)$,
and in turn $\mathfrak{su}(2)$ can break into its maximal subalgebra
$\mathfrak{u}\left(1\right)^{2}$.

A subalgebra is classified as regular or special depending on how
its Cartan subalgebra is related to the one of its parent algebra.
A \textbf{regular subalgebra} $\mathfrak{g}'$ of $\mathfrak{g}$
is one whose Cartan subalgebra $\mathfrak{h}'$ is contained in the
Cartan subalgebra $\mathfrak{h}$ of $\mathfrak{g}$ and the set of
roots $\Delta'$ of $\mathfrak{g}'$ is contained in the set of roots
$\Delta$ of $\mathfrak{g}$. If this is not the case, $\mathfrak{g}'$
is said to be a \textbf{special subalgebra} of $\mathfrak{g}$. Grand
Unified Theories deal almost invariably with regular subalgebras.
This turns out to be very convenient, because the maximal regular
subalgebras with an $\mathfrak{u}(1)$ ideal of a simple Lie algebra
are easy to derive: deleting a dot in the Dynkin diagram of $\mathfrak{g}$
yields a semi-simple algebra $\mathfrak{m}$, and $\mathfrak{u}(1)\oplus\mathfrak{m}$
is shown to be a maximal regular subalgebra of $\mathfrak{g}$. As
an example, deleting the appropriate dots, we immediately conclude
that $E_{8}\rightarrow E_{7}\oplus\mathfrak{u}(1)$, $E_{7}\rightarrow E_{6}\oplus\mathfrak{u}(1)$,
$E_{6}\rightarrow\mathfrak{so}\left(10\right)\oplus\mathfrak{u}(1)$,
$\mathfrak{so}\left(10\right)\rightarrow\mathfrak{su}\left(5\right)\oplus\mathfrak{u}(1)$,
$\mathfrak{su}\left(5\right)\rightarrow\mathfrak{su}\left(3\right)\oplus\mathfrak{su}\left(2\right)\oplus\mathfrak{u}(1)$,
which is a symmetry breaking chain  potentially applicable in High
Energy Physics.

\subsection{\label{sub:Symmetry_PhysicalConventions}The Lie algebra of gauge
symmetries}

Consider now the use of Lie algebras in Yang\textendash{}Mills theories.
To preserve the kinetic term of the fields in the Lagrangian, these
must be in a unitary representation of the Lie algebra. Resuming the
use of Einstein's summation convention for repeated indices, usually
a gauge transformation is written as 
\begin{align}
U & =\exp\left[i\varepsilon^{a}R\left(t_{a}\right)\right]\,,\label{eq:Symmetry_gaugetransformation}
\end{align}
with an explicit $i$. Therefore, the parameters $\varepsilon^{a}$
of the transformation must be real and the representation matrices
$R\left(t_{a}\right)$ of $t_{a}$ must be hermitian, otherwise $U$
is not unitary. This is an important observation: even though complex
numbers appear often in these theories (for example in the Pauli matrices),
the gauge symmetry must be given by a real Lie algebra. As such, note
that strictly speaking the generators of the algebra are $it_{a}$
instead of $t_{a}$. For example, consider the Pauli matrices which
obey the relation $\left[\sigma_{i},\sigma_{j}\right]=2i\varepsilon_{ijk}\sigma_{k}$:
these cannot be the generators of the real algebra $\mathfrak{su}_{\mathbb{R}}\left(2\right)$
because they do not close under the Lie bracket operation. Note also
that the raising and lower operators in equation \eqref{eq:EFH_SU2}
do not generate the same real Lie algebra as $i\sigma_{i}$, because
the two basis are related by complex coefficients.

At this point, it might seem odd that in this section, as well as
in several textbooks \citep{Cahn:1985wk,Slansky:1981yr}, it is assumed
that the Lie algebras are complex. In particular, figure \eqref{fig:Symmetry_ListOfComplexSimpleLieAlgebras}
contains the classification of all complex simple Lie algebras. There
is a good reason for this though. In order to see it, we need to say
a few words about the connection between real and complex algebras.
Suppose that  $\mathfrak{g}$ is a real Lie algebra: instead of taking
just real linear combinations of its generators, if we take complex
linear combinations as well, the resulting set will always close under
commutations, so $\mathfrak{g}_{\mathbb{C}}\equiv\mathfrak{g}\oplus i\mathfrak{g}$
is a complex Lie algebra. This $\mathfrak{g}_{\mathbb{C}}$ is called
the \textbf{complexification} of $\mathfrak{g}$, and $\mathfrak{g}$
is said to be a \textbf{real form} of $\mathfrak{g}_{\mathbb{C}}$.
There is a {}``many to one'' relation then: the complexification
of multiple real Lie algebras can be the same complex Lie algebra,
or equivalently, a complex Lie algebra can have many real forms. For
example, the real Lie algebras generated by $\left\{ i\sigma_{i}\right\} $
and $\left\{ e,g,h\right\} $ are two real forms of $\mathfrak{su}_{\mathbb{C}}\left(2\right)$:
they are called the \textbf{compact real form }and the \textbf{normal
}or \textbf{split real form}, respectively, and they exist for any
complex simple Lie algebra.

It turns out that the Lie algebra of gauge theories must be the (unique)
compact real form of some complex simple Lie algebra, if $U$ in equation
\eqref{eq:Symmetry_gaugetransformation} is to be a unitary matrix.
This is a consequence of the following considerations:
\begin{enumerate}
\item The generators $\left\{ it_{a}\right\} $ of any real simple Lie algebra
can be rotated and normalized such that $\left(it_{a},it_{b}\right)=\textrm{sign}\left(a\right)\delta_{ab}$;
in other words, the Killing form can be diagonalized, but its signature
cannot be changed as the algebra is real. The compact real form is
the unique real form with $\left(it_{a},it_{b}\right)=-\delta_{ab}$,
or in other words the signature of the Killing form is $\left(-,-,\cdots,-\right)$.
Recall that the Killing form $\left(it_{a},it_{b}\right)$ is proportional
to $-\textrm{Tr}\left[R\left(t_{a}\right)R\left(t_{a}\right)\right]$
for any non-trivial representation $R$, with a positive proportionality
factor.
\item The hermitian matrices $R\left(t_{a}\right)$ in equation \eqref{eq:Symmetry_gaugetransformation}
have real eigenvalues, so $\textrm{Tr}\left[R\left(t_{a}\right)R\left(t_{b}\right)\right]$
must always be non-negative for any representation $R$.
\end{enumerate}
In other words, the hermiticity of the matrices $R\left(t_{a}\right)$
requires that $\textrm{Tr}\left[R\left(t_{a}\right)R\left(t_{b}\right)\right]$
is positive, and this is only true for the compact real form. In relation
to this, note that in Particle Physics one has the trace condition:
\begin{align}
\textrm{Tr}\left[R\left(t_{a}\right)R\left(t_{b}\right)\right] & =S\left(R\right)\delta_{ab}\,,
\end{align}
where the positive number $S\left(R\right)$ is the \textbf{Dynkin
index} of the representation $R$. Therefore, in summary, for every
semi-simple Lie algebra (see figure \eqref{fig:Symmetry_ListOfComplexSimpleLieAlgebras})
there is a unique compact Lie algebra, with similar properties, whose
generators can be chosen to satisfy this last equation. The expression
for the Casimir operator,
\begin{align}
C & =R\left(t_{a}\right)R\left(t_{a}\right)=C\left(R\right)\mathbb{1}\,,\label{eq:Symmetry_CasimirPhysics}
\end{align}
provides a simple way to convert a Cartan-Weyl basis into the one
used in Physics, by comparing equations \eqref{eq:Symmetry_Casimir_CartanWeyl}
and \eqref{eq:Symmetry_CasimirPhysics}.

To conclude the analysis of the relation between the physical and
mathematical canonical approach to Lie algebras, we must consider
one final issue. It is often said that the gauge symmetry must be
given by a direct sum of a semi-simple Lie algebra and $\mathfrak{u}\left(1\right)$'s
(i.e., a reductive Lie algebra), but the reason for it is usually
omitted. There are many other non-reductive Lie algebras, for example
the one generated by $x_{1},x_{2}$ such that $\left[x_{1},x_{2}\right]=x_{2}$.
We have previously seen why a gauge symmetry must be associated to
a Lie algebra, but why should it be a reductive one? To answer this
question we start by noting that the gauge bosons are in the adjoint
representation of the gauge group:
\begin{align}
A_{\mu} & \rightarrow\exp\left[i\varepsilon^{a}R_{\textrm{ad}}\left(t_{a}\right)\right]A_{\mu}\label{eq:Symmetry_gaugebosons_transformation}
\end{align}
for a space-time independent transformation. The adjoint representation
matrices are connected to the structure constants, $\left[R_{\textrm{ad}}\left(t^{a}\right)\right]_{bc}=ic_{ac}^{b}$,
and their hermiticity implies that the structure constants $c_{ac}^{b}$
must be antisymmetric in all three indices. If the Lie algebra $\mathfrak{g}$
is a direct sum of two vector spaces, $\mathfrak{m}$ and $\mathfrak{m}^{T}$,
and if $\mathfrak{m}$ is an ideal of $L$ ($\left[\mathfrak{g},\mathfrak{m}\right]\subset\mathfrak{m}$)
then the hermiticity of the adjoint representation implies that the
orthogonal vector space $\mathfrak{m}^{T}$ is also an ideal of $\mathfrak{g}$:
$\left[\mathfrak{g},\mathfrak{m}^{T}\right]\subset\mathfrak{m}^{T}$.
Therefore, $\mathfrak{g}$ is the direct sum of the subalgebras $\mathfrak{m}$
and $\mathfrak{m}^{T}$, and if we pick $\mathfrak{m}$ to be the
biggest abelian ideal of $\mathfrak{g}$, then $\mathfrak{m}^{T}$
is semi-simple. In conclusion, the unitarity of the transformation
\eqref{eq:Symmetry_gaugebosons_transformation} implies that $\mathfrak{g}=\left(\textrm{abelian algebra}\right)\oplus\left(\textrm{semi-simple algebra}\right)=\oplus\left(\mathfrak{u}(1)\textrm{ or simple algebras}\right)$.

\section{Space-time symmetries}

\subsection{The Lorentz and Poincaré groups}

The way the laws of Physics are written in a given coordinate frame
depends on the space-time metric. As mentioned at the beginning of
this chapter, the Poincaré group is the group of space-time transformations
which leaves the Minkowski metric $\eta$ invariant, so it is the
space-time symmetry group of the laws of Physics in flat space-time.
It is easy to verify that such transformations must be of the form
\citep{Tung:1985na}
\begin{alignat}{1}
x^{\mu} & \rightarrow x'^{\mu}=\Lambda_{\nu}^{\mu}x^{\nu}+b^{\mu}\,,\label{eq:Symmetry_Poincare_group}
\end{alignat}
for some matrix $\Lambda$ and a vector $b$. The inhomogeneous part
of these transformations, given by the $b$ vector, can take any value,
and it corresponds to translations in the four space-time directions.
Ignoring these, we are left with the homogeneous part of the Poincaré
group---the Lorentz group. Each transformation of the $x^{\mu}$ under
this group is given by a $\Lambda$ matrix and, in order for equation
\eqref{eq:Symmetry_Poincare_group} to be an isometry of the flat
space-time metric, we must ensure that
\begin{align}
\eta & =\Lambda^{T}\eta\Lambda\,.\label{eq:Symmetry_Lorentz_group}
\end{align}
The group of $\Lambda$'s which satisfy this equation is sometimes
denoted by $O\left(1,3\right)$, as the above equation matches the
definition of the 4-dimensional orthogonal group, except that the
signature of $\eta$ is $\left(+---\right)$ instead of $\left(++++\right)$.
It is well known that this group, the Lorentz group, is made up of
rotations between the last three coordinates (the spacial ones), and
pseudo-rotations/boosts between the first coordinate (time) and the
other ones. We shall come shortly to this, when we review the algebra
of the Lorentz group. But before going into the topic of infinitesimal
transformations, it is worth mentioning that the matrices
\begin{align}
\Lambda_{T} & \equiv\textrm{diag}\left(-1,+1,+1,+1\right)\,,\\
\Lambda_{P} & \equiv\textrm{diag}\left(+1,-1,-1,-1\right)
\end{align}
also satisfy equation \eqref{eq:Symmetry_Lorentz_group}, although
they represent neither boosts nor rotations; they represent time ($\Lambda_{T}$)
and space ($\Lambda_{P}$) reversal operations. Because of the existence
of these transformations, topologically $O\left(1,3\right)$ is not
a connected set. But if we remove them (appropriately), the resulting
group, which is named the restricted or proper Lorentz group and denoted
by $SO\left(1,3\right)^{+}$, is indeed connected. To summarize this
relation, we can write%
\footnote{The symbol $\rtimes$ stands for a semi-direct product of two groups.
If $G=N\rtimes H$, it means that each element $g\in G$ can be written
as the product of an element $n\in N$ and $h\in H$. The product
of $g_{1}=\left(n_{1},h_{1}\right)$ with $g_{2}=\left(n_{2},h_{2}\right)$
is given by $\left(n_{3},h_{1}h_{2}\right)$ with $n_{3}=n_{1}h_{1}n_{2}h_{1}^{-1}$,
which is different from the relation $n_{3}=n_{1}n_{2}$ in a direct
product. Clearly both $N$ and $H$ are automatically subgroups of
$G$, but in addition it is necessary for $N$ to be an invariant
subgroup of $G$ for this construction to make sense.

Another way to relate the two groups is the following: $SO\left(1,3\right)^{+}$
is an invariant subgroup of $O\left(1,3\right)$, so it divides $O\left(1,3\right)$
in cosets (4 in this case). The cosets form a group (generically called
the factor group) denoted by $O\left(1,3\right)/SO\left(1,3\right)^{+}$,
which turns out to be $\left\{ \mathbb{1},\Lambda_{T},\Lambda_{P},\Lambda_{T}\Lambda_{P}\right\} $.%
} 
\begin{align}
O\left(1,3\right) & =SO\left(1,3\right)^{+}\rtimes\left\{ \mathbb{1},\Lambda_{T},\Lambda_{P},\Lambda_{T}\Lambda_{P}\right\} \,.
\end{align}
Note that by removing in this way the $\Lambda_{T}$ and $\Lambda_{P}$
transformations from $O\left(1,3\right)$, all the remaining $\Lambda$
have unit determinant, yet we cannot call the resulting group $SO\left(1,3\right)$
because we also removed $\Lambda_{T}\Lambda_{P}=-\mathbb{1}\in SO\left(1,3\right)$.

It turns out that, at the microscopic level, the laws of Physics are
not invariant under $\Lambda_{T}$ and $\Lambda_{P}$, at least at
the energies probed so far,%
\footnote{It may be that time-reversal $T$ and parity $P$ are fundamental
symmetries of Nature which are broken at the energies we can probe
experimentally. %
} so at this point $SO\left(1,3\right)^{+}$, the proper Lorentz group,
would seem to be the true space-time symmetry group. If there were
only scalar and vector quantities such as the Higgs fields ($H^{0},\, H^{+}$),
the electromagnetic field ($A^{\mu}$), or coordinates ($x^{\mu}$)
this would be true. However, the study of the electron essentially
reveals that a 360\degree$\:$ rotation adds a minus sign to its
wave function, instead of leaving it invariant. The implication of
this experimental result is that the space-time symmetry group cannot
be just $SO\left(1,3\right)^{+}$; it must be the bigger $SL\left(2,\mathbb{C}\right)$,
which is the group of $2\times2$ matrices with complex entries and
unit determinant (it is named the two dimensional special linear group
over $\mathbb{C}$). Much like the case of $SO(3)$ and $SU(2)$ discussed
previously, there is a 1:2 relation between $SO\left(1,3\right)^{+}$
and $SL\left(2,\mathbb{C}\right)$, and this accounts for the minus
sign gained by the electron wavefunction under a 360\degree$\:$rotation.
To see this\textit{ double covering} of $SO\left(1,3\right)^{+}$
by the $SL\left(2,\mathbb{C}\right)$ group%
\footnote{\label{fn:Symmetry_SpinGroup_footnote}The group $SL\left(2,\mathbb{C}\right)$
is sometimes called $\textrm{Spin}(1,3)^{+}$. In this nomenclature,
$\textrm{Pin}(1,3)$ and $\textrm{Spin}(1,3)$ are the double covers
of $O\left(1,3\right)$ and $SO\left(1,3\right)$, respectively. See
for instance \citep{Berg:2000ne}. %
} we can use the $\sigma^{\mu}=\left(1,-\boldsymbol{\sigma}\right)$
matrices. First note that any $\Lambda\in SO\left(1,3\right)^{+}$
transforms a 4-vector $x^{\mu}$ while preserving the pseudo-norm
$x^{T}\eta x$. Then we identify any of these 4-vectors $x^{\mu}$
with the $2\times2$ matrix $x^{\mu}\sigma_{\mu}$ whose determinant
is precisely $x^{T}\eta x$, so under a transformation $\Lambda\in SO\left(1,3\right)^{+}$
of $x^{\mu}\rightarrow x'^{\mu}=\Lambda_{\nu}^{\mu}x^{\nu}$ we can
associate the following change of $x^{\mu}\sigma_{\mu}$:
\begin{align}
x^{\mu}\sigma_{\mu} & \rightarrow\lambda\left(x^{\mu}\sigma_{\mu}\right)\lambda^{\dagger}=x'^{\mu}\sigma_{\mu}\,,
\end{align}
for some unknown $\lambda$ matrix with determinant $\pm1$. The equality
in this expression follows from the fact that any hermitian $2\times2$
matrix is a linear combination of the four $\sigma_{\mu}$ matrices.
Taking only the cases with $\det\lambda=1$, we can therefore relate
a $\Lambda\in SO\left(1,3\right)^{+}$ with a $\lambda\in SL\left(2,\mathbb{C}\right)$
but crucially we note that both $\lambda,-\lambda\in SL\left(2,\mathbb{C}\right)$
are associated to the same $\Lambda$: 
\begin{align}
\Lambda_{\nu}^{\mu}x^{\nu}\sigma_{\mu}=x'^{\mu}\sigma_{\mu} & =\lambda\left(x^{\mu}\sigma_{\mu}\right)\lambda^{\dagger}=\left(-\lambda\right)\left(x^{\mu}\sigma_{\mu}\right)\left(-\lambda\right)^{\dagger}\,.
\end{align}

We now briefly describe how this is related to relativistic fermions.
We first note that the well-known $\gamma$ matrices form a $2^{4}$-dimensional
Clifford algebra generated by the matrices $\mathbb{1},\,\gamma^{\mu},\,\gamma^{\mu}\gamma^{\nu},\,\gamma^{\mu}\gamma^{\nu}\gamma^{\sigma},\,\gamma^{\mu}\gamma^{\nu}\gamma^{\sigma}\gamma^{\rho}$
($0\leq\mu<\nu<\sigma<\rho\leq3$) with each $\gamma$ obeying the
relation
\begin{align*}
\gamma^{\mu}\gamma^{\nu}+\gamma^{\nu}\gamma^{\mu} & =2\eta^{\mu\nu}\,.
\end{align*}
When $x\rightarrow x'$, the Dirac equation for a spin $\nicefrac{1}{2}$
field $\Psi$ is known to be invariant if $\Psi\left(x\right)\rightarrow\exp\left(\sum_{ij}c_{ij}\left[\gamma_{i},\gamma_{j}\right]\right)\Psi\left(x\right)$
where the $c_{ij}$ are some real numbers, and it can be shown that
these transformations with an even number of $\gamma$ matrices and
unit determinant form the group $SL\left(2,\mathbb{C}\right)$. More
details on this connection between Clifford algebras and the Spin
groups ($SL\left(2,\mathbb{C}\right)=\textrm{Spin}(1,3)^{+}$; see
footnote \ref{fn:Symmetry_SpinGroup_footnote}) for an arbitrary number
of spacial and temporal dimensions can be found in \citep{RauschdeTraubenberg:2005aa}
and references contained therein.

\subsection{Lie algebras and representations of the Poincaré and Lorentz groups}

The previous discussion concerned mainly the global properties of
the space-time symmetry group. However, according to the discussion
in section \ref{sec:Lie-groups-and-Lie-algebras} many of important
features of the a Lie group are encoded in its local structure. Therefore,
without worrying too much about the details of the last subsection,
we shall now briefly review the relevant aspects of infinitesimal
Lorentz and Poincaré transformations, which lead directly to their
algebras.

Consider first an infinitesimal translation given by $\delta b^{\mu}$
\begin{align}
x^{\mu} & \rightarrow T\left(\delta b\right)x^{\mu}=x^{\mu}+\delta b^{\mu}\,.
\end{align}
The transformation $T\left(\delta b\right)$, valid not just for a
vector such as coordinates $x$, can be obtained with the usual trick
of considering $\partial_{\mu}$ to be the basis vectors in which
we are taking the coordinates $x^{\mu}$ and $\delta b^{\mu}$. In
this way, we have $x\equiv x^{\mu}\partial_{\mu}$ and $\delta b\equiv\delta b^{\mu}\partial_{\mu}$
vectors and the transformation we seek is given by
\begin{align}
T\left(\delta b\right) & =\mathbb{1}+\delta b\equiv\mathbb{1}-i\delta b^{\mu}P_{\mu}\,,
\end{align}
or
\begin{align}
T\left(b\right) & =\exp\left(-ib^{\mu}P_{\mu}\right)
\end{align}
for finite translations. Here $P_{\mu}=i\partial_{\mu}$ is the conserved
4-momentum vector, the generator of translations. Similarly, $J_{\mu\nu}=i\left(x_{\mu}\partial_{\nu}-x_{\nu}\partial_{\mu}\right)$
generates boosts and rotations of the proper Lorentz group:
\begin{align}
\Lambda\left(\omega\right) & =\exp\left(-\frac{i}{2}\omega^{\mu\nu}J_{\mu\nu}\right)\,,
\end{align}
where $\omega^{\mu\nu}$ are real parameters which are taken to be
antisymmetric in $\left(\mu\nu\right)$ since $J_{\mu\nu}=-J_{\nu\mu}$,
so there are 6 independent real parameters (3 boosts and 3 rotations).

The computation of the Lie algebra of the Poincaré group is straightforward,
and it yields the following:
\begin{align}
\left[P_{\mu},P_{\nu}\right] & =0\,,\label{eq:Symmetry_Poincare_algebra1}\\
\left[P_{\mu},J_{\nu\rho}\right] & =i\left(\eta_{\mu\nu}P_{\rho}-\eta_{\mu\rho}P_{\nu}\right)\,,\\
\left[J_{\mu\nu},J_{\rho\sigma}\right] & =i\left(\eta_{\mu\rho}J_{\sigma\nu}-\eta_{\nu\sigma}J_{\mu\rho}+\eta_{\nu\rho}J_{\mu\sigma}-\eta_{\mu\sigma}J_{\rho\nu}\right)\,.\label{eq:Symmetry_Poincare_algebra3}
\end{align}

We shall now use the algebra of the Poincaré and Lorentz groups to
derive their irreducible representations. Starting with the latter
one, we shall see that Lorentz group algebra is similar to the one
of $SU(2)\times SU(2)$. To reach such conclusion, first separate
$J_{\mu\nu}$ into the 3 generators of rotations $J_{i}$ and the
3 generators of boosts $K_{i}$:
\begin{alignat}{2}
J_{i} & \equiv\frac{1}{2}\varepsilon_{ijk}J^{jk}\,;\, & K_{i} & \equiv J_{i0}\,,\qquad i,j,k\in\left\{ 1,2,3\right\} \,.
\end{alignat}
We can then define
\begin{align}
A_{i}^{\nicefrac{R}{L}} & \equiv\frac{1}{2}\left(J_{i}\pm iK_{i}\right)\,,\qquad i=1,2,3\,,\label{eq:Symmetry_A+-}
\end{align}
and the interesting result is that the three $A_{i}^{R}$ as well
as the three $A_{i}^{L}$ obey the $SU(2)$ algebra, and in addition
the generators of one kind commute with those of the other:
\begin{align}
\left[A_{i}^{\nicefrac{R}{L}},A_{j}^{\nicefrac{R}{L}}\right] & =i\varepsilon_{ijk}A_{k}^{\nicefrac{R}{L}}\,,\\
\left[A_{i}^{\nicefrac{R}{L}},A_{j}^{\nicefrac{L}{R}}\right] & =0\,.
\end{align}
However, there is an important detail here. The algebra of the proper
Lorentz group algebra is not exactly the same as the one of $SU(2)\times SU(2)$
because we used a complex combination of $J_{i}$'s and $K_{i}$'s
in equation \eqref{eq:Symmetry_A+-}, even though we are working with
real algebras. In other words, the proper Lorentz group is given by
the exponentiation of $\left(\textrm{real coeficients}\right)\times iJ_{i},\, iK_{i}$,
while $SU(2)\times SU(2)$ is given by the exponentiation of $\left(\textrm{real coeficients}\right)\times iA_{i}^{L},\, iA_{i}^{R}$
and the two are not the same. This is directly related to the fact
that $SU(2)\times SU(2)$ is a compact Lie group, while $SL\left(2,\mathbb{C}\right)$
is not. As a consequence, the (finite) representations of the proper
Lorentz group are not unitary.

Just like $SU(2)_{L}\times SU(2)_{R}$ generated by $A_{i}^{\nicefrac{R}{L}}$,
each of the irreducible representations of the proper Lorentz group
is given by two non-negative half-integers $\left(j_{L},j_{R}\right)$.
Since $A_{i}^{R*}=-A_{i}^{L}$, we have the relation $\left(j_{L},j_{R}\right)=\left(j_{R},j_{L}\right)^{*}$.
Also, the basis vectors $\left|m_{L},m_{R}\right\rangle $ of such
a representation take the values $m_{L}=-j_{L},-j_{L}+1,\cdots,j_{L}$
and $m_{R}=-j_{R},-j_{R}+1,\cdots,j_{R}$ so $\left(j_{L},j_{R}\right)$
is a $\left(2j_{L}+1\right)\left(2j_{R}+1\right)$-dimensional representation.
Note also that since the $J_{3}$ generator of rotations is given
by $A_{3}^{R}+A_{3}^{L}$, the angular quantum number $m$ is equal
to $m_{R}+m_{L}$, which means that a representation $\left(j_{L},j_{R}\right)$
of the proper Lorentz groups is composed of $j=j_{L}+j_{R},\, j_{L}+j_{R}-1,\cdots,\,\left|j_{L}-j_{R}\right|$
irreducible representations of the rotation group. Consider the following
examples:
\begin{itemize}
\item $\left(j_{L},j_{R}\right)=\left(0,0\right)$ is a 1-dimensional representation
with $j=0$. Such a field $\phi$ is called a scalar.
\item $\left(j_{L},j_{R}\right)=\left(\nicefrac{1}{2},0\right)$ is a 2-dimensional
representation with $j=\nicefrac{1}{2}$. Such a field $\psi_{L}$
is a left-handed Weyl spinor.
\item $\left(j_{L},j_{R}\right)=\left(0,\nicefrac{1}{2}\right)$ is a 2-dimensional
representation with $j=\nicefrac{1}{2}$. Such a field $\psi_{R}$
is a right-handed Weyl spinor.
\item $\left(j_{L},j_{R}\right)=\left(\nicefrac{1}{2},\nicefrac{1}{2}\right)$
is a 4-dimensional representation with a $j=0$ part and another one
with $j=1$. Such a field $A_{\mu}$ is called a 4-vector (its first
component $A_{\mu}^{0}$ is a scalar under rotations and the other
three components form a vector).
\end{itemize}
On the other hand, a Dirac spinor $\Psi=\left(\nicefrac{1}{2},0\right)\oplus\left(0,\nicefrac{1}{2}\right)$
does not form an irreducible representation of the proper Lorentz
group, as it is made of right- and left-handed Weyl spinors.

We now consider the unitary representations of the Poincaré group,
which are infinite dimensional. We note in passing that, while the
true symmetry of space-time is given by the Poincaré group, the wave
functions used to write Lagrangian densities are Lorentz representations
(see \citep{Tung:1985na,Streater:1989vi} for details on this connection).
From equations \eqref{eq:Symmetry_Poincare_algebra1}--\eqref{eq:Symmetry_Poincare_algebra3}
we see that the Poincaré group is not just the product of the translations
group with the Lorentz group, and for that reason its representations
are markedly different from the ones of the proper Lorentz group.
We can start by diagonalizing the $P_{\mu}$ operator such that for
an eigenstate state $\left|p\right\rangle $ we have
\begin{alignat}{1}
P_{\mu}\left|p\right\rangle  & \equiv p_{\mu}\left|p\right\rangle \,.
\end{alignat}
Without entering into details, if $p_{\mu}p^{\mu}>0$ the irreducible
representations of the Poincaré group are labeled with the continuous
parameter $m$ (interpreted physically as a mass) and a non-negative
half-integer number $s$ (the spin). In particular, these quantities
are related to the two group Casimir operators
\begin{align}
P_{\mu}P^{\mu} & =m^{2}\,,\\
W_{\mu}W^{\mu} & =-m^{2}s\left(s+1\right)\,,
\end{align}
where $W^{\lambda\mu\nu\sigma}=-\nicefrac{1}{2}\varepsilon^{\lambda\mu\nu\sigma}J_{\mu\nu}p_{\sigma}$
is the so-called Pauli-Luba\'{n}ski pseudo-vector \citep{Lubanski:1942},
which is given by $W^{0}=0$ and $W^{i}=mJ^{i}$, $i=1,2,3$ in the
frame where $P_{\mu}=\left(m,\boldsymbol{0}\right)$. The (infinite)
set of states $\left\{ \left|p_{\mu},\lambda\right\rangle \right\} $
such that $p_{\mu}p^{\mu}=m^{2}$ and $\lambda=-s,-s+1,\cdots,s$
is the eigenvalue of the $J_{3}$ generator of rotations forms a basis
for the vector space of the irreducible representation $\left(m,s\right)$.

If $p_{\mu}p^{\mu}=0$, special care is needed. If $p_{\mu}=0$ then
this 4-vector is an invariant and the irreducible representations
of the Poincaré group can be labeled as $\left(j_{L},j_{R}\right)$,
in analogy to the representations of the Lorentz group. Physically
however, the interesting situation is when $p_{\mu}\neq0$, corresponding
to physical particles with no mass. The irreducible representations
in this case can be labeled with a single half-integer $\lambda$
(the helicity) which is the eigenvalue of the $J_{3}$ generator of
rotations. Again, the infinite set of states $\left\{ \left|p_{\mu},\lambda\right\rangle \right\} $
such that $p_{\mu}p^{\mu}=0$ constitutes a basis for the vector space
of the $\lambda-$representation, but notice that $\lambda$ is fixed
here, unlike in the $p_{\mu}p^{\mu}>0$ case. In other words, Poincaré
transformations do not change the helicity of a particle. An implication
of this is that the photon, with two polarizations, is actually a
reducible representation of the Poincaré group, $-1\oplus+1$ (CPT
invariance requires the simultaneous presence of positive and negative
helicities).

There is one final case, when $p_{\mu}p^{\mu}<0$, which corresponds
to tachyons. We shall not deal with it here and instead point to \citep{Tung:1985na}
for details.

\section{Supersymmetry as a super-Poincaré group}

Can the symmetry of Nature be non-trivially larger than the Poincaré
group? Under the assumptions of the 1967 Coleman-Mandula theorem \citep{Coleman:1967ad},
the answer is negative: the symmetry group $G$ of the scattering
matrix $S$ must be a direct product of the Poincaré group and some
other internal symmetries such as the gauged ones.%
\footnote{This also defines an internal symmetry: it consists of any symmetry
commuting with the Poincaré group.%
} This celebrated theorem assumes the following (ignoring some technical
details):
\begin{enumerate}
\item $G$ contains a subgroup locally isomorphic to the Poincaré group;
\item There is a finite number of one-particle states with finite mass,
and their energy is always positive;
\item Elastic-scattering amplitudes are analytic functions of the $s$ and
$t$ Mandelstam variables;
\item Any two plane waves scatter at almost all energies (i.e., the scattering
matrix $S$ is non-trivial);
\item $G$ is a connected symmetry group which can be built from the generators
of infinitesimal symmetry transformations.
\end{enumerate}
The Coleman-Mandula theorem therefore does not allow symmetries to
change simultaneously space-time coordinates and internal quantum
numbers of fields. As a consequence, particles with a given mass $m$
and spin $s$ (or just helicity $\lambda$), which are irreducible
representations of the Poincaré group, cannot be related/grouped together
in bigger representations of a bigger group.

The assumptions presented above are the list given in \citep{Coleman:1967ad}.
However, implicitly the Coleman-Mandula theorem also assumes that
$G$ transforms bosons into bosons and fermions into fermions and
it turns out \citep{Golfand:1971iw,Wess:1973kz,Wess:1974tw} that
a super-Poincaré symmetry (supersymmetry) relating bosons to fermions
is actually possible. The Coleman-Mandula theorem was eventually extended
by Haag, {\L}opusza\'{n}ski and Sohnius \citep{Haag:1974qh} to
include this possibility, and it became clear that the structure of
such supersymmetries is very constrained, making these extensions
of the Poincaré symmetry almost unique.

Let us then review some of the theoretical aspects of supersymmetry,
following \citep{Weinberg:2000cr,Sohnius:1985qm,Lykken:1996xt,Martin:1997ns,Aitchison:2007fn,Bilal:2001nv,Argyres:1998vt,Wess:1992cp,West:1990tg,Nilles:1983ge}.
As a first step, we shall try to motivate the existence of commutator
and anti-commutator relations in a supersymmetric algebra. An infinitesimal
supersymmetric transformation can be written as
\begin{align}
S\left(\delta\alpha\right) & =\mathbb{1}-i\delta\alpha^{a}G_{a}\,,
\end{align}
where the $\delta\alpha^{a}$ are the transformation parameters and
the $G_{a}$ the generator operators. Here the $\delta\alpha^{a}$
are assumed to be $\mathbb{Z}_{2}$-graded parameters, meaning that
they may commute or anticommute between themselves, depending on some
numbers $\eta_{A}=\pm1$ associated with them (the grading):
\begin{align}
\delta\alpha^{a}\delta\alpha^{b} & =\left(-1\right)^{\eta_{a}\eta_{b}}\delta\alpha^{b}\delta\alpha^{a}\,.
\end{align}
If we require that the transformation $S\left(\delta\alpha\right)$
commutes with these graded parameters, then the generators $G_{a}$
themselves must behave like graded parameters:
\begin{align}
G_{a}G_{b} & =\left(-1\right)^{\eta_{a}\eta_{b}}G_{b}G_{a}\,.
\end{align}
We now subtract to $S\left(\delta\alpha\right)S\left(\delta\beta\right)$
the product of these two supersymmetric transformations applied in
the reverse order. The result must itself be a supersymmetric transformation:
\begin{align}
S\left(\delta\alpha\right)S\left(\delta\beta\right)-S\left(\delta\beta\right)S\left(\delta a\right) & =S\left(\delta\gamma\right)\quad\textrm{for some }\delta\gamma\,.
\end{align}
With a few computations this $\delta\gamma$ is shown to be
\begin{align}
\delta\gamma^{c}G_{c} & =-i\delta\beta^{b}\delta\alpha^{a}\left[G_{a},G_{b}\right\} \,,\label{eq:Symmetry_SuperAlgebra1}
\end{align}
where, for two operators $A$ and $B$, $\left[\cdot,\cdot\right\} $
is defined to be the following generalization of the commutator and
anti-commutator:
\begin{align}
\left[A,B\right\}  & \equiv AB-\left(-1\right)^{\eta_{A}\eta_{B}}BA\,.\label{eq:Symmetry_CommutatorGeneral}
\end{align}
Equation \eqref{eq:Symmetry_SuperAlgebra1} implies that $\left[G_{a},G_{b}\right\} $
must be a linear combination of the generators $G_{c}$, so in this
way we are lead to the concept of structure constants $c_{ab}^{c}$
of a $\mathbb{Z}_{2}$-graded Lie algebra:
\begin{align}
\left[G_{a},G_{b}\right\}  & \equiv ic_{ab}^{c}G_{c}\,.
\end{align}
The generators are either bosonic ($\eta_{A}=0$) or fermionic ($\eta_{A}=1$),
and according to equation \eqref{eq:Symmetry_CommutatorGeneral},
apart from relations between fermionic generators, which are anticommutating,
all other relations between the $\mathbb{Z}_{2}$-graded Lie algebra
are given by commutators.

Returning now to the extension of the Poincaré symmetry, we note that
the generators $P_{\mu}$ and $J_{\mu\nu}$ are bosonic operators,
transforming under the proper Lorentz group as $\left(\nicefrac{1}{2},\nicefrac{1}{2}\right)$
and $\left(1,0\right)\oplus\left(0,1\right)$, respectively. The Haag-{\L}opusza\'{n}ski-Sohnius
theorem states that the fermionic generators $Q^{I}$, $I=1,\cdots,N$
of supersymmetries must be either in the $\left(\nicefrac{1}{2},0\right)$
representation of the proper Lorentz group or its conjugate $\left(0,\nicefrac{1}{2}\right)$.
Therefore, without loss of generality, such $Q^{I}$ can be taken
to be left-handed Weyl spinors, while their hermitian conjugate operators
$Q^{I\,\dagger}$ are right-handed Weyl spinors. Both have therefore
two components: $Q_{\alpha}^{I}$, $Q_{\alpha}^{I\,\dagger}$ with
$\alpha=1\,,2$. It can be shown that the (anti)commutation relations
between these fermionic charges and the generators $P_{\mu}$ and
$J_{\mu\nu}$ of the Poincaré symmetry are the following:
\begin{align}
\left[P_{\mu},Q_{\alpha}^{I}\right] & =0\,, & \left[P_{\mu},Q_{\alpha}^{\dagger I}\right] & =0\,,\label{eq:Symmetry_SuperAlgebraRelations1}\\
\left[J_{\mu\nu},Q_{\alpha}^{I}\right] & =i\left(\sigma_{\mu\nu}\right)_{\alpha\beta}Q_{\beta}^{I}\,, & \left[J_{\mu\nu},Q_{\alpha}^{\dagger I}\right] & =i\left(\overline{\sigma}_{\mu\nu}\right)_{\alpha\beta}Q_{\beta}^{\dagger I}\,,\\
\left\{ Q_{\alpha}^{I},Q_{\beta}^{J}\right\}  & =\varepsilon_{\alpha\beta}Z^{IJ}\,, & \left\{ Q_{\alpha}^{\dagger I},Q_{\beta}^{\dagger J}\right\}  & =\varepsilon_{\alpha\beta}Z^{IJ\,*}\,,\label{eq:Symmetry_SuperAlgebraRelations3}\\
\left\{ Q_{\alpha}^{I},Q_{\beta}^{\dagger J}\right\}  & =2\left(\sigma^{\mu}\right)_{\alpha\beta}P_{\mu}\delta^{IJ}\,.\label{eq:Symmetry_SuperAlgebraRelations4}
\end{align}
These relations complement the ones in equations \eqref{eq:Symmetry_Poincare_algebra1}--\eqref{eq:Symmetry_Poincare_algebra3}.
In terms of notation, we used $\sigma^{\mu\nu}=-\overline{\sigma}_{\mu\nu}^{\dagger}\equiv\nicefrac{1}{4}\left(\sigma^{\mu}\overline{\sigma}^{\nu}-\sigma^{\nu}\overline{\sigma}^{\mu}\right)$,
where $\sigma^{0}=\overline{\sigma}^{0}\equiv\mathbb{1}$ and $\overline{\sigma}^{1,2,3}=-\sigma^{1,2,3}\equiv\sigma_{1,2,3}$
are the usual Pauli matrices. As for the $Z^{IJ}$ in equations \eqref{eq:Symmetry_SuperAlgebraRelations3},
they are bosonic symmetry generators which commute will all generators
(including themselves) and for that reason they are called \textit{central
charges}. Note that in the important case where there is just a single
pair of fermionic generators $\left\{ Q,Q^{\dagger}\right\} $ ($N=1$)
then, since $Z^{IJ}=-Z^{JI}$, we conclude that there are no central
charges. The $N=1$ case is known as \textit{simple supersymmetry},
while $N>1$ is sometimes called \textit{$N$-extended supersymmetry}.
With the exception of the present section, this thesis discusses only
$N=1$ supersymmetric models; more fermionic charges lead to a bigger
symmetry group and also to bigger irreducible representations, which
in turn means that such theories are very restrictive.

In general, when $Z^{IJ}=0$, one can perform any unitary transformation
$U^{(R)}$ on the charges,
\begin{align}
Q^{I}\rightarrow Q'^{I}=U_{IJ}^{(R)}Q^{J}\,, & Q^{\dagger I}\rightarrow Q'^{\dagger I}=U_{IJ}^{(R)*}Q^{\dagger J}\,,
\end{align}
and the new $\left\{ Q'^{I},Q'^{\dagger I}\right\} $ will generate
the same supersymmetry as before, since relations \eqref{eq:Symmetry_SuperAlgebraRelations1}--\eqref{eq:Symmetry_SuperAlgebraRelations4}
are preserved. If $N=1$ then, this so-called \textit{R-symmetry}
is a $U(1)$ global symmetry of the super-Poincaré algebra, which
nonetheless does not need to be a symmetry of the action. Nevertheless,
as explained in chapter \ref{chap:The-SM's-shortcomings}, in physically
interesting models where supersymmetry is softly broken, the Lagrangian
and the action are made to be invariant under a discrete $\mathbb{Z}_{2}$
subgroup of this continuous R-symmetry (\textit{R-parity}) in order
to avoid dangerous couplings which would lead to rapid proton decay.

We note that the generators of internal symmetries are missing from
equations \eqref{eq:Symmetry_Poincare_algebra1}--\eqref{eq:Symmetry_Poincare_algebra3}
and \eqref{eq:Symmetry_SuperAlgebraRelations1}--\eqref{eq:Symmetry_SuperAlgebraRelations4}.
By definition, the generators $T_{i}$ of such symmetries commute
with $P_{\mu}$, $J_{\mu\nu}$ (and also with the central charges
$Z^{IJ}$), but they do not necessarily commute with the fermionic
generators:
\begin{align}
\left[T_{i},T_{j}\right] & =ic_{ij}^{k}T_{k}\,,\\
\left[T_{i},Q_{\alpha}^{I}\right] & =s_{iJ}^{I}Q_{\alpha}^{J}\,,\quad\left[T_{i},Q_{\alpha}^{\dagger I}\right]=-s_{iJ}^{I*}Q_{\alpha}^{\dagger J}.\label{eq:Symmetry_QT_commutator}
\end{align}
The $c_{ij}^{k}$ are the structure constants of the internal symmetry
algebra. The $T_{i}$ are Lorentz scalars (by definition of internal
symmetry) so their commutator with $Q^{I}$ must be a linear combination
of these fermionic charges, and this is precisely the statement made
by equation \eqref{eq:Symmetry_QT_commutator}, for some coefficients
$s_{iJ}^{I}=\left(s_{iI}^{J}\right)^{*}$. Since there are $N$ charges
$Q^{I}$ and because the representations of the compact internal symmetry
are unitary, we deduce that the $Q^{I}$ transform under an internal
symmetry transformation as a representation of a subgroup of $U(N)$.
In simple supersymmetry then, where $N=1$, the single $Q$ is in
a 1-dimentional representation of the internal symmetry group and
since non-abelian Lie algebras do not have non-trivial 1-dimensional
representations, $\left[T_{i},Q\right]$ must be 0 for non-abelian
internal symmetry subgroups. One practical consequence is that all
generators of the SM gauge group must commute with the fermionic charges
$Q$, $Q^{\dagger}$ in $N=1$ supersymmetric theories, with the possible
exception of the hypercharge generator.%
\footnote{In the MSSM even the hypercharge generator commutes with $Q$, $Q^{\dagger}$
though.%
}

To conclude this discussion about the theoretical aspects of supersymmetry,
we shall make some remarks about the irreducible representations of
the supersymmetry group---the supermultiplets.
\begin{enumerate}
\item As in the Poincaré subgroup, $P_{\mu}$ still commutes with everything,
so $P_{\mu}P^{\mu}\equiv m^{2}$ is also a Casimir of the supersymmetry
algebra. As a consequence, the mass of all components of a supermultiplet
is the same \citep{O'Raifeartaigh:1965jg}. Interestingly, this mass
cannot be negative as it can be shown that there is the following
lower bound:
\begin{align}
m & \geq\frac{1}{2N}\textrm{Tr}\sqrt{Z^{\dagger}Z}\,,
\end{align}
where $\sqrt{Z^{\dagger}Z}$ is the unique hermitian matrix satisfying
$\left(\sqrt{Z^{\dagger}Z}\right)_{IJ}\left(\sqrt{Z^{\dagger}Z}\right)_{JK}=Z_{JI}^{*}Z_{JK}$.
In simple supersymmetry, where no central charges $Z_{IJ}$ exist,
the bound is $m\geq0$.
\item The contraction $W_{\mu}W^{\mu}$ of the Pauli-Luba\'{n}ski pseudo-vector
does not commute with the fermionic charges, which means that it is
not a Casimir of the supersymmetry algebra. This is somewhat obvious,
since a supermultiplet will group different irreducible representations
of the Poincaré group, with different spins, so it does not have a
single spin associated to it. However, for $N=1$ supersymmetry (see
also \citep{Galperin:1982fx}) there is a new Casimir operator $C_{\mu\nu}C^{\mu\nu}$,
with
\begin{align}
C_{\mu\nu} & \equiv B_{\mu}P_{\nu}-B_{\nu}P_{\mu}\,,\quad B_{\mu}\equiv W_{\mu}-\frac{1}{4}\left(\overline{\sigma}_{\mu}\right)_{\alpha\beta}Q_{\alpha}^{\dagger}Q_{\beta}\,,
\end{align}
which generalizes $W_{\mu}W^{\mu}$. For a massive particle, $C_{\mu\nu}C^{\mu\nu}=2j\left(j+1\right)m^{4}$
where $j\left(j+1\right)$ is the eigenvalue of the operator $\sum_{i=1,2,3}\widetilde{J}_{i}^{2}$
with
\begin{align}
\widetilde{J}_{i} & \equiv J_{i}-\frac{1}{4m}Q_{\alpha}^{\dagger}Q_{\beta}\left(\overline{\sigma}_{\mu}\right)_{\alpha\beta}\,.
\end{align}
These $\widetilde{J}_{i}$ obey the same $SU(2)$ algebras as the
rotation generators: $\left[\widetilde{J}_{i},\widetilde{J}_{k}\right]=i\varepsilon_{ijk}\widetilde{J}_{k}$.
\item Simple calculations reveal that a massless supermultiplet will contain
particles with helicities $\lambda_{0},\lambda_{0}+\nicefrac{1}{2},\cdots,\lambda_{0}+\nicefrac{N}{2}$
for some half-integer $\lambda_{0}$, and the number of states with
helicity $\lambda_{0}+\nicefrac{i}{2}$ is $\nicefrac{N!}{i!\left(N-i\right)!}$.
On the other hand, a massive supermultiplet is composed of states
with spins $\max\left(0,s_{0}-\nicefrac{N}{2}\right),\max\left(0,s_{0}-\nicefrac{N}{2}\right)+\nicefrac{1}{2},\cdots,s_{0}+\nicefrac{N}{2}$.
In either case, the supermultiplets contain states with spins/helicities
differing by as much as $\nicefrac{N}{2}$ and this means that particles
with spin/helicity modulus equal or bigger than $\nicefrac{N}{4}$
will be present. Since renormalizable field theories without(with)
gravity%
\footnote{Gravity can be incorporated in these theories by promoting supersymmetry
to a local symmetry, instead of leaving it as global one, as we have
tacitly been assuming. Supergravity \citep{Nath:1975nj,Arnowitt:1975xg,Freedman:1976xh,Deser:1976eh,Freedman:1976py,Cremmer:1978hn,Bagger:1982ab,Cremmer:1982en}
however will not be addressed in this thesis.%
} cannot describe particles with spins or helicities higher than $1$($2$)
we can have at most $N=4$(8).
\item It can be shown that the trace of the operator $\left(-1\right)^{2s}$
($s$ is the spin or helicity) over each supermultiplet, times $P_{\mu}$,
is null. This means that for the physically known cases where the
4-momentum vector is non-null, the number of fermionic and bosonic
degrees of freedom in a supersymmetric theory is the same.\end{enumerate}

\cleartooddpage

\part{Topics on the renormalization of SUSY models}\cleartooddpage

\chapter{\label{chap:Susyno}Calculating the renormalization group equations
of a SUSY model with \texttt{Susyno}}

\section{Introduction}

The analysis of the theoretical and phenomenological implications
of SUSY GUT models requires a careful study of the evolution of the
fundamental parameters from the high-energy scale down to the electroweak
one, at which observables are computed and constraints applied. As
such, knowledge of the renormalization group equations is necessary.
Although the RGEs of several models (for example the MSSM and the
NMSSM) are already known \cite{Martin:1993zk,Ellwanger:2009dp}, for
other SUSY extensions of the SM complicated general equations must
be used \cite{Martin:1993zk,Yamada:1994id}.

In this chapter we describe \texttt{Susyno}, a Mathematica-based package
that addresses this issue. The program takes as input the gauge group,
the representations (i.e., the chiral superfield content), the number
of flavors/copies of each representation, and any abelian discrete
symmetries (e.g., R-parity). \texttt{Susyno} then computes the form
of the most general superpotential and soft SUSY breaking Lagrangian
consistent with the field content and symmetries imposed. Once these
elements have been derived, \texttt{Susyno} calculates the 2-loop
$\beta$-functions of all the parameters of the model, which is its
main output. The program also contains a variety of group theoretical
functions which may be of interest on their own (see also the Mathematica
application \texttt{LieART} \cite{Feger:2012bs}).

There is another Mathematica package, \texttt{SARAH} \cite{Staub:2009bi,Staub:2010jh,Staub:2012pb,Staub:2013tta},
which provides an extensive list of functions which can be used to
automate many of the computations necessary to build and analyze supersymmetric
models (including the RGEs).%
\footnote{Also, see \cite{Lyonnet:2013dna} for non-SUSY models.%
} Given that it worked originally for models based on $SU(n)$ gauge
factor only and that \texttt{Susyno} is prepared to accept any gauge
group as working input, the two programs were linked as of \texttt{SARAH4}. 

This chapter is organized as follows. Section \ref{sec:Installation-and-quick}
explains how to install version 2 of the program and run a first,
simple example (MSSM based). Sections \ref{sec:Input} and \ref{sec:Output}
explain how to prepare the input and how to read and interpret the
output, also using as examples the MSSM case. Section \ref{sec:Tests-made}
summarizes the tests conducted to validate the code, and finally section
\ref{sec:List-of-available} lists some of the functions available
to the end user.

Note that the theoretical concepts related to Lie algebras which are
detailed in chapter \ref{chap:Symmetry} are fundamental for the functioning
of the program. Because of their technical nature, these implementation
details have been placed separately in appendix \ref{chap:Implementation-details-of}.

\section{\label{sec:Installation-and-quick}Installation and quick start}

\texttt{Susyno} works on Windows, Linux and Mac OS provided that Mathematica
7 (or a latter version) is installed. The program is obtainable from\\

\url{http://web.ist.utl.pt/renato.fonseca/susyno.html}\\

The files \textit{IO.m}, \textit{LieGroups.m}, \textit{ModelBuilding.m},
\textit{Models.m}, \textit{SimplifyEinsteinNotation.m}, and \textit{SusyRGEs.m}
are the core of the program. These and other auxiliary files can be
found inside the folder \textit{Susyno}, which must be extracted from
the downloaded \textit{Susyno-2.0.zip} file to a location that is
visible to Mathematica. Typing \texttt{\$Path} in Mathematica will
show a complete list of acceptable locations. One possibility is to
place the whole folder (not just its contents) in\\

\noindent \textit{(Mathematica base directory)/AddOns/Applications}\\

\noindent (note that in a Windows system the slashes {}``/'' must
be replaced by backslashes {}``\textbackslash{}''). The package
can be loaded by typing\\

\noindent \texttt{<\textcompwordmark{}< Susyno$\grave{\,\textrm{\,}}$}~\\

\noindent in Mathematica's front end. A text message is returned,
informing that a built-in help system provides a detailed description
of the program and its functions (see also section \ref{sec:List-of-available}).
A tutorial is also included.

The \texttt{Susyno} lines below allow a simple and easy first run:
the example consists in a possible way of writing the MSSM input (we
shall call this model \texttt{myMSSM} because \texttt{MSSM} is already
defined in the program by default). \\

\noindent \texttt{group{[}myMSSM{]} \textasciicircum{}= \{U1, SU2,
SU3\};}~\\

\noindent \texttt{fieldNames{[}myMSSM{]} \textasciicircum{}= \{u,
d, Q, e, L, Hu, Hd\};}

\noindent \texttt{normalization = Sqrt{[}3/5{]};}

\noindent \texttt{reps{[}myMSSM{]} \textasciicircum{}= \{\{-2/3 normalization,
\{0\}, \{0, 1\}\},}

\noindent \texttt{\{1/3 normalization, \{0\}, \{0, 1\}\}, \{1/6 normalization,
\{1\}, \{1, 0\}\},}

\noindent \texttt{\{normalization, \{0\}, \{0, 0\}\}, \{-1/2 normalization,
\{1\}, \{0, 0\}\},}

\noindent \texttt{\{1/2 normalization, \{1\}, \{0, 0\}\}, \{-1/2 normalization,
\{1\}, \{0, 0\}\}\};}~\\
\texttt{}~\\
\texttt{nFlavs{[}myMSSM{]} \textasciicircum{}= \{3, 3, 3, 3, 3, 1,
1\};}~\\
\texttt{discreteSym{[}myMSSM{]} \textasciicircum{}= \{-1, -1, -1,
-1, -1, 1, 1\};}~\\
\texttt{}~\\
\texttt{GenerateModel{[}myMSSM{]}}~\\
\texttt{}~\\
Evaluation of this simple code generates the 2-loop $\beta$-functions
of the model (MSSM in this case). Notice that no external input or
output files are used---everything happens on Mathematica's front
end.

\section{\label{sec:Input}The input of \texttt{Susyno}: defining a model}

A SUSY model contains two building blocks: a superpotential and a
soft SUSY breaking Lagrangian (see equations \eqref{eq:Introduction_superpotential}
and \eqref{eq:Introduction_Lsoft}). \texttt{Susyno} works as follows:
it requires as input the gauge group, the representations/fields,
the number of flavors of each representation/field and the discrete
abelian symmetries (if there are any) of the model. With this information
the program then internally builds the superpotential and the soft
SUSY breaking Lagrangian using an algorithm to automatically name
the parameters of the model (see the next section). Once this information
has been assigned to a \texttt{model} variable, the user must then
call the function \texttt{GenerateModel} as follows:\\
 \\
 \texttt{GenerateModel{[}model{]}}~\\
 ~\\
 We shall focus now on each of the elements that characterize a model.
We will take the Minimal Supersymmetric Standard Model as an example.

\subsection{Gauge group}

The program needs a complete list of all the abelian and simple Lie
groups of the model.%
\footnote{We emphasize here that we are actually dealing with algebras, not
groups (see chapter \ref{chap:Symmetry}). Nevertheless, we will adopt
the common practice in high-energy physics of using the word \textit{group}
for both these concepts.%
} For the MSSM this would correspond to ($U(1)$ factors must come
first) \\

\noindent \texttt{group{[}myMSSM{]} \textasciicircum{}= \{U1, SU2,
SU3\};}\\

\noindent Note that this code assigns to the variable \texttt{myMSSM}
(instead of \texttt{group}) the information on the right, therefore
the use of \texttt{\textasciicircum{}=} is important.

Any simple group can be given as a factor: the simple gauge factor
groups, as well as their corresponding \texttt{Susyno} input are collected
in table \eqref{table:gaugefactor}---see also chapter \ref{chap:Symmetry}
for an explanation on how one arrives at this list of simple gauge
factor groups. 
\begin{table}[tbph]
\begin{centering}
\begin{tabular}{cc}
Simple gauge factor group  & \texttt{Susyno} input\tabularnewline
\hline 
$SU(n)$  & \texttt{SU2}, \texttt{SU3}, \texttt{SU4}, \texttt{SU5}, ...\tabularnewline
$SO(n)$  & \texttt{SO3}, \texttt{SO5}, \texttt{SO6}, \texttt{SO7}, ...\tabularnewline
$Sp(2n)$  & \texttt{SP2}, \texttt{SP4}, \texttt{SP6}, ...\tabularnewline
$G_{2}$ & \texttt{G2}\tabularnewline
$F_{4}$  & \texttt{F4}\tabularnewline
$E_{6}$, $E_{7}$, $E_{8}$  & \texttt{E6}, \texttt{E7}, \texttt{E8}\tabularnewline
\end{tabular}
\par\end{centering}

\caption{\label{table:gaugefactor}Simple gauge factor groups}
\end{table}

\subsection{Representations/fields: the content of the model}

As mentioned before, \texttt{Susyno} is designed to accept an arbitrary
field content. An input must be provided in the form of two lists\\

\noindent \texttt{fieldNames{[}model{]} \textasciicircum{}= \{fieldName1,
fieldName2,...\};}

\noindent \texttt{reps{[}model{]} \textasciicircum{}= \{rep1,rep1,...\};}\\

\noindent The first one should simply contain a list of names chosen
by the user for each field. It is important to note that the ordering
of the fields is arbitrary. However, the user must consistently adhere
to the chosen ordering when inputting lists composed of field attributes
(for example the number of flavors). In our example\\
\\
\texttt{fieldNames{[}myMSSM{]} \textasciicircum{}= \{u, d, Q, e, L,
Hu, Hd\};}\\
\\
The other list must contain all the gauge group irreducible representations
present in the model. Each of these \texttt{rep} should be a list
with representations of the gauge factor groups:\\

\noindent \texttt{rep=\{hChrg1,hChrg2,...,hChrgM,rep\_simplegroup1,rep\_simplegroup2,...\};}\\

\noindent The first entries correspond to the hypercharges of \texttt{rep}
(if any), which are just real numbers. After the hypercharges one
must declare the representations of \texttt{rep} under each of the
simple gauge factor groups mentioned above. These representations
must be specified by their Dynkin coefficients (see subsection \ref{sub:Symmetry_Representations,-weights-and}
of chapter \ref{chap:Symmetry} for details). In table \eqref{tab:Table_representations}
we list some of the representations of $SU(2)$, $SU(3)$, $SU(5)$
and $SO(10)$ (Dynkin coefficients and corresponding dimensions).
\begin{table}[h!]
\begin{centering}
\begin{tabular}{lccc}
\toprule 
\multirow{2}{*}{\textbf{Group} } & \multicolumn{3}{c}{\textbf{Representation}}\tabularnewline
 & \textit{Dynkin coefficients}  & \textit{Dimension} & \textit{Name}\tabularnewline
\midrule
 & \{0\}  & $\boldsymbol{1}$ & Trivial/Singlet\tabularnewline
$SU(2)$ & \{1\}  & $\boldsymbol{2}$ & Fundamental/Doublet\tabularnewline
 & \{2\}  & $\boldsymbol{3}$ & Adjoint/Triplet\tabularnewline
\midrule
\multirow{4}{*}{$SU(3)$} & \{0,0\}  & $\boldsymbol{1}$ & Trivial/Singlet\tabularnewline
 & \{1,0\}  & $\boldsymbol{3}$ & Fundamental\tabularnewline
 & \{0,1\}  & $\overline{\boldsymbol{3}}$ & Anti-fundamental\tabularnewline
 & \{1,1\}  & $\boldsymbol{8}$ & Adjoint\tabularnewline
\midrule
 & \{0,0,0,0\}  & $\boldsymbol{1}$ & Trivial/Singlet\tabularnewline
 & \{1,0,0,0\}  & $\boldsymbol{5}$ & Fundamental\tabularnewline
 & \{0,0,0,1\}  & $\overline{\boldsymbol{5}}$ & Anti-fundamental\tabularnewline
$SU(5)$ & \{0,1,0,0\}  & $\boldsymbol{10}$ & \tabularnewline
 & \{2,0,0,0\}  & $\boldsymbol{15}$ & \tabularnewline
 & \{0,0,0,2\}  & $\overline{\boldsymbol{15}}$ & \tabularnewline
 & \{1,0,0,1\}  & $\boldsymbol{24}$ & Adjoint\tabularnewline
\midrule
 & \{0,0,0,0,0\}  & $\boldsymbol{1}$ & Trivial/Singlet\tabularnewline
 & \{1,0,0,0,0\}  & $\boldsymbol{10}$ & Fundamental\tabularnewline
 & \{0,0,0,0,1\}  & $\boldsymbol{16}$ & Spinor\tabularnewline
 & \{0,0,0,1,0\}  & $\overline{\boldsymbol{16}}$ & Spinor's conjugate\tabularnewline
 & \{0,1,0,0,0\}  & $\boldsymbol{45}$ & Adjoint\tabularnewline
$SO(10)$ & \{2,0,0,0,0\}  & $\boldsymbol{54}$ & \tabularnewline
 & \{0,0,1,0,0\}  & $\boldsymbol{120}$ & \tabularnewline
 & \{0,0,0,0,2\}  & $\boldsymbol{126}$ & \tabularnewline
 & \{0,0,0,2,0\}  & $\overline{\boldsymbol{126}}$ & \tabularnewline
 & \{0,0,0,1,1\} & $\boldsymbol{210}$ & \tabularnewline
 & \{3,0,0,0,0\}  & $\boldsymbol{210'}$ & \tabularnewline
\bottomrule
\end{tabular}
\par\end{centering}

\caption{\label{tab:Table_representations}List of some frequently used representations
of $SU(2)$, $SU(3)$, $SU(5)$ and $SO(10)$}
\end{table}
There are functions in \texttt{Susyno} that compute properties of
the representations (e.g., \texttt{DimR} calculates the dimension
of a representation, \texttt{ReduceRepProduct} reduces products of
representations) and they are documented in the built-in help files
(see also section \ref{sec:List-of-available}). These should be enough
to identify a representation by its Dynkin coefficients; however,
should the user wish to consult lists of representations, these are
available in the literature (see for example \cite{Slansky:1981yr}).

To understand how the MSSM was specified in the example of section
\ref{sec:Installation-and-quick}, we just need the following information
from table \eqref{tab:Table_representations}: 
\begin{itemize}
\item the Dynkin coefficients of the trivial and fundamental representations
of $SU(2)$: \{0\} and \{1\}.
\item the Dynkin coefficients of the trivial, fundamental ($\boldsymbol{3}$)
and anti-fundamental ($\overline{\boldsymbol{3}}$) representations
of $SU(3)$: \{0,0\}, \{1,0\}, \{0,1\}.
\end{itemize}
\noindent For the MSSM each field must then be cast in the format\\

\noindent \texttt{rep=\{U1\_charge,SU(2)\_rep,SU(3)\_rep\};}~\\

\noindent Further normalizing the hypercharges with the usual $\sqrt{\frac{3}{5}}$
factor (from an embedding of the MSSM in an $SU(5)$ based model),%
\footnote{Notice however that \texttt{Susyno} accepts any choice for the normalization
of the hypercharges.%
} we can then write the following:\\

\noindent \texttt{normalization = Sqrt{[}3/5{]};}

\noindent \texttt{reps{[}myMSSM{]} \textasciicircum{}= \{\{-2/3 normalization,
\{0\}, \{0, 1\}\},}

\noindent \texttt{\{1/3 normalization, \{0\}, \{0, 1\}\}, \{1/6 normalization,
\{1\}, \{1, 0\}\},}

\noindent \texttt{\{normalization, \{0\}, \{0, 0\}\}, \{-1/2 normalization,
\{1\}, \{0, 0\}\},}

\noindent \texttt{\{1/2 normalization, \{1\}, \{0, 0\}\}, \{-1/2 normalization,
\{1\}, \{0, 0\}\}\};}~\\

We must emphasize here that although the user is free to choose the
ordering of the simple factor groups, $SU(2)$ and $SU(3)$, once
this is set (e.g., \texttt{group{[}myMSSM{]} \textasciicircum{}= \{U1,
SU2, SU3\}}) one must adhere to the (user-established) convention,
and define the representations of the fields accordingly:\\

\noindent \texttt{rep=\{U1\_charge,SU(2)\_rep,SU(3)\_rep\};}~\\

\subsection{Number of flavors and abelian discrete symmetries}

\texttt{Susyno} needs two more input lists: one containing the number
of flavors of each field and another defining its abelian discrete
symmetries. The ordering of both these lists must be consistent with
the representations list we have just discussed. In our \texttt{myMSSM}
example we used the ordering \texttt{\{u,d,Q,L,e,Hu,Hd\}}, so\\

\noindent \texttt{nFlavs{[}myMSSM{]} \textasciicircum{}= \{3, 3, 3,
3, 3, 1, 1\};}~\\
\texttt{discreteSym{[}myMSSM{]} \textasciicircum{}= \{-1, -1, -1,
-1, -1, 1, 1\};}~\\

\noindent In this particular case, it is clear that the discrete symmetry
imposed corresponds to R-parity. For the most general (R-parity violating)
MSSM we have\\

\noindent \texttt{discreteSym{[}RPVMSSM{]} \textasciicircum{}= \{1,
1, 1, 1, 1, 1, 1\};}~\\

\noindent Let us consider another example: for instance, if we were
to modify the MSSM to include $m$ copies of $\widehat{H}_{u}$ and
$\widehat{H}_{d}$, we would write\\

\noindent \texttt{nFlavs{[}myMSSMmod{]} \textasciicircum{}=\{3, 3,
3, 3, 3, m, m\};}~\\

\subsection{Calling the function \texttt{GenerateModel}}

Once the \texttt{model} variable has been defined (=\texttt{myMSSM}
in our case), the \texttt{GenerateModel} function can be invoked as
follows:\\

\noindent \texttt{GenerateModel{[}myMSSM{]}}~\\

\noindent There are two Boolean optional parameters which can be passed
to this function: \texttt{CalculateEverything->False,True} (default
value is \texttt{False}) and \texttt{Verbose->False,True} (default
value is \texttt{True}). The first one, \texttt{CalculateEverything},
can be used to force the program to compute explicitly the most general
superpotential and soft SUSY breaking Lagrangian consistent with the
definitions of the model. On the other hand, the option \texttt{Verbose}
can be used to suppress the printing of the results on the screen.
In any case, the RGEs are always saved to the variable \texttt{betaFunctions{[}myMSSM{]}},
and the model parameters are saved to \texttt{parameters{[}myMSSM{]}}:
for properly bounded indices $i$, $j$, the 1- and 2-loop $\beta$
functions of \texttt{parameters{[}myMSSM{]}{[}{[}i,j{]}{]}} are \texttt{betaFunctions{[}myMSSM{]}{[}{[}i,1,j{]}{]}}
and \texttt{betaFunctions{[}myMSSM{]}{[}{[}i,2,j{]}{]}} respectively.

\section{\label{sec:Output}The output of \texttt{Susyno}}

Once all the definitions have been provided, \texttt{Susyno} automatically
computes the form of the most general superpotential and soft SUSY
breaking Lagrangian consistent with them. In particular, parameter
names are generated by the program (they are not given by the user).
The advantage of this approach is that inputting a model becomes very
easy, since it is not even necessary to know the exact number of its
parameters.%
\footnote{The parameters considered throughout this chapter are the fundamental
degrees of freedom of a model, with no experimental input taken into
consideration (such as the requirement of EWSB, for example).%
} On the other hand, this notation renders the output harder to read
(and hence not particularly user-friendly), since the names of the
parameters are chosen by the program. There is nonetheless a built-in
function---\texttt{RenameParametersWithRule}---which provides a way
for the user to change \texttt{Susyno}'s default notation (see below).

We note that \texttt{Susyno} does not have custom built-in functions
to export the results. Users who wish to do so must do it manually
or with the help of Mathematica's built-in functions \texttt{CForm}
and \texttt{FortranForm}.

\subsection{Naming of parameters}

\texttt{Susyno} assigns names to the parameters of a model in such
a way that the user can identify which representations/fields they
are multiplying:\\

\texttt{y{[}\{field1,field2,field3\}, <InvIndex>, \{<flav1>,<flav2>,<flav3>\}{]}}

\texttt{$\mu${[}\{field1,field2\}, \{<flav1>,<flav2>\}{]}}

\texttt{l{[}\{field1\}, \{<flav1>\}{]}}

\texttt{h{[}\{field1,field2,field3\}, <InvIndex>, \{<flav1>,<flav2>,<flav3>\}{]}}

\texttt{b{[}\{field1,field2\}, \{<flav1>,<flav2>\}{]}}

\texttt{s{[}\{field1\}, \{<flav1>\}{]}}

\texttt{m2{[}\{field1,field2\}, \{<flav1>,<flav2>\}{]}}\\

A few comments concerning the above (output) tensors are in order: 
\begin{itemize}
\item \texttt{y}, \texttt{$\mu$}, \texttt{l}, \texttt{h}, \texttt{b}, \texttt{s}
and \texttt{m2} can easily be identified with the different types
of couplings and dimensionful parameters of the superpotential and
the soft SUSY breaking Lagrangian (see equations \eqref{eq:Introduction_superpotential}
and \eqref{eq:LSusy});
\item \texttt{field1}, \texttt{field2}, \texttt{field3} are the fields entering
a given coupling. In our example above, where we used \texttt{fieldNames{[}myMSSM{]}
\textasciicircum{}= \{u,d,Q,e,L,Hu,Hd\}}, the up-quark Yukawa couplings
would be \texttt{y{[}\{u,Q,Hu\},...{]}};
\item There is the possibility that the product of 3 representations, $R_{1}\otimes R_{2}\otimes R_{3}$,
contains more than one invariant. Therefore an addition label \texttt{InvIndex=1,2,...}
might be necessary to distinguish them. This is rare though, so in
most cases (e.g., the MSSM) this index is omitted. Notice that in
linear and bilinear terms, $R_{1}$ and $R_{1}\otimes R_{2}$, this
problem does not arise since there is at most one invariant;
\item \texttt{<flav1>}, \texttt{<flav2>} , \texttt{<flav3>} are the flavor
indices of \texttt{field1}, \texttt{field2}, \texttt{field3}. If any
of these fields has only one flavor, the corresponding index is omitted.
Consider again the example of the up-quark Yukawa couplings: we would
have \texttt{y{[}\{u,Q,Hu\},\{i,j,k\}{]}} where \texttt{i} = flavor
of $\widehat{u}$, \texttt{j} = flavor of $\widehat{Q}$, \texttt{k}
= flavor of $\widehat{H}_{u}$. Yet $\widehat{H}_{u}$ only has one
flavor so the correct parameter name is \texttt{y{[}\{u,Q,Hu\},\{i,j\}{]}}. 
\end{itemize}
Additionally, there are also the gauge coupling constants and the
gaugino masses:%
\footnote{With more than one $U(1)$ gauge factor group, according to the discussion
in chapter \ref{chap:U1_mixing_paper} there is $U(1)$-mixing and
both the gauge coupling constants and the gaugino masses should be
seen as matrices in $U(1)$ space. As a consequence, parameters \texttt{g{[}1,1{]}},
\texttt{g{[}1,2{]}}, ... , \texttt{M{[}1,1{]}}, \texttt{M{[}1,2{]}},
... are necessary.%
}\\

\noindent \texttt{g{[}1{]}, g{[}2{]}, ...}

\noindent \texttt{M{[}1{]}, M{[}2{]}, ...}\\

\subsection{\label{sub:Susyno_Normalization-of-the}Normalization of the parameters}

Consider for example the MSSM's $\mu$ parameter. According to the
discussion in the previous subsection, \texttt{Susyno}'s name for
$\mu$ will be \texttt{$\mu${[}\{Hu,Hd\}{]}}, but this identification
is only valid up to some multiplicative factor, since we do not know
how the doublet indices of $\widehat{H}_{u}$ and $\widehat{H}_{d}$
are being contracted. In principle the program could be assuming that
the $\mu$ term is \texttt{$\mu${[}\{Hu,Hd\}{]}}$\widehat{H}_{u}\cdot\widehat{H}_{d}$,
\texttt{$-\mu${[}\{Hu,Hd\}{]}}$\widehat{H}_{u}\cdot\widehat{H}_{d}$,
2\texttt{$\mu${[}\{Hu,Hd\}{]}}$\widehat{H}_{u}\cdot\widehat{H}_{d}$
or any other multiple of these expressions. Therefore with the generic
description that \texttt{$\mu${[}\{Hu,Hd\}{]}} is the parameter that
multiplies the contraction of $\widehat{H}_{u}$ and $\widehat{H}_{d}$
in the superpotential, we can only say that \texttt{$\mu\propto$$\mu${[}\{Hu,Hd\}{]}}.

Version 2 of the program no longer computes explicitly a Lagrangian
in order to get the RGEs (although the user can still ask the program
to compute it), but even with two explicit Lagrangians written with
different conventions and notations, it is not straightforward to
compare them, because they may differ by irrelevant/unphysical unitary
transformations of the gauge representations. Fortunately, there is
a simple way to compare the normalization of their parameters. First,
we describe the parameter normalization convention used by \texttt{Susyno}:
\begin{enumerate}
\item The trilinear superpotential couplings of a generic superpotential
can be encoded in a tensor $Y^{ijk}$ (see equation \eqref{eq:Introduction_superpotential}).
The program uses the normalization $Y^{ijk}Y_{ijk}=$$\sum_{y}\sqrt{\dim(R_{1})\dim(R_{2})\dim(R_{3})}y^{\alpha\beta\gamma}y_{\alpha\beta\gamma}$,
where the sum is over all trilinear superpotential parameters $y$,
and $R_{1}$, $R_{2}$ and $R_{3}$ are the participating representation/fields.
Note that the flavor indices $\alpha$, $\beta$, $\gamma$ contract
between $y^{\alpha\beta\gamma}$ and $y_{\alpha\beta\gamma}=\left(y^{\alpha\beta\gamma}\right)^{*}$.
Consider the MSSM's case, where there are three such parameters: \texttt{y{[}\{u,Q,Hu\},\{i,j\}{]}},
\texttt{y{[}\{d,Q,Hd\},\{i,j\}{]}} and \texttt{y{[}\{e,L,Hd\},\{i,j\}{]}}.
The dimensions of the $\widehat{u}$, $\widehat{d}$, $\widehat{Q}$,
$\widehat{e}$, $\widehat{L}$, $\widehat{H}_{u}$ and $\widehat{H}_{d}$
representations are 3, 3, 6, 1, 2, 2 and 2, respectively, therefore
the Yukawa parameters are normalized in such a way that $Y^{ijk}Y_{ijk}=$\texttt{6$\,$y{[}\{u,Q,Hu\},\{m,n\}{]}
y{[}\{u,Q,Hu\},\{m,n\}{]}$^{*}$ + 6$\,$y{[}\{d,Q,Hd\},\{m,n\}{]}
y{[}\{d,Q,Hd\},\{m,n\}{]}$^{*}$ + 3$\,$y{[}\{e,L,Hd\},\{m,n\}{]}
y{[}\{e,L,Hd\},\{m,n\}{]}$^{*}$};
\item If there is a singlet representation $\widehat{S}$, \texttt{Susyno}
assumes that a bilinear term $R_{1}\otimes R_{2}$ is written in the
same way as the trilinear one $R_{1}\otimes R_{2}\otimes\widehat{S}$,
the only difference being that the singlet field is eliminated and,
of course, a different parameter name must be given. In the NMSSM
for example, if there is a term $\left(\textrm{parameter}\right)\widehat{S}\widehat{H}_{u}\cdot\widehat{H}_{d}$
in the superpotential, then the bilinear one must be written as $\left(\textrm{parameter'}\right)\widehat{H}_{u}\cdot\widehat{H}_{d}$,
with no relative phases or factors. The same is true for a linear
term so, given the normalization in the condition 1, this means that
a linear term is of the form $\left(\textrm{parameter}\right)\widehat{S}$;
\item The trilinear, bilinear and linear terms in the soft SUSY breaking
Lagrangian ($-\mathscr{L}_{\textrm{soft}}$) are obtained by copying
the ones in the superpotential $W$ and simply renaming the parameters:
\texttt{y{[}...{]}} $\rightarrow$ \texttt{h{[}...{]}}, \texttt{$\mu${[}...{]}}
$\rightarrow$ \texttt{b{[}...{]}} and \texttt{l{[}...{]}} $\rightarrow$
\texttt{s{[}...{]}}. In particular, notice that there are no relative
phases or factors between the parameters in $W$ and the equivalent
ones in $-\mathscr{L}_{\textrm{soft}}$;
\item The soft scalar masses $m^{2}$ are assumed to be, as usual, of the
trivial form $\left(\textrm{mass parameter of }R_{i}\right)\left(R_{i}^{1}R_{i}^{1*}+R_{i}^{2}R_{i}^{2*}+\cdots\right)$
for a representation $R_{i}$ of the gauge group with components $R_{i}^{1},R_{i}^{2},\cdots$.
\end{enumerate}
The crucial statement is the following one: the RGEs of the parameters
of any other Lagrangian, possibly written in a different form and
with different parameter names, are the same as the ones provided
by \texttt{Susyno} as long as these conditions are obeyed. Note that
these conditions are necessary and sufficient. As an example, the
RGEs of the MSSM would not change even if the $\mu$ parameter was
doubled everywhere ($\mu\rightarrow2\mu$), as long as we also doubled
the $b$ parameter in the soft SUSY breaking Lagrangian (condition
3).

In conclusion, the user must see how his/her own way of writing the
model parameters compares with conditions 1-4 above and, according
to the result of such comparison, make adequate adaptations of \texttt{Susyno}'s
output (if necessary). Since conditions 2, 3 and 4 are reasonably
standard, the only non-trivial one is the first.

\subsection{Changing the default notation}

The user can change the default notation by providing a list of substitution
rules for the parameter names:\\
\\
\texttt{parameterRenamingRules{[}model{]}\textasciicircum{}=substitutionRules;}~\\
\texttt{}~\\
In our \texttt{myMSSM} example, from the previous subsections we know
what are the parameters names used by \texttt{Susyno}, and how they
are normalized. As such, we can derive table \eqref{tab:Appendix_ParametersMSSM},
which compares the program's notation with the more standard one in
equations \eqref{eq:Introduction_MSSM_W} and \eqref{eq:Introduction_MSSM_Lsoft}.
\begin{table}[tbph]
\begin{centering}
\begin{tabular}{cc}
Parameter  & \texttt{Susyno}'s default notation\tabularnewline
\hline 
$g_{1}$, $g_{2}$, $g_{3}$  & \texttt{g{[}1{]}, g{[}2{]}, g{[}3{]}}\tabularnewline
$M_{1}$, $M_{2}$, $M_{3}$  & \texttt{M{[}1{]}, M{[}2{]}, M{[}3{]}}\tabularnewline
$\left(Y_{u}\right)_{ij}$  & \texttt{y{[}\{u,Q,Hu\},\{i,j\}{]}}\tabularnewline
$\left(Y_{d}\right)_{ij}$  & \texttt{y{[}\{d,Q,Hd\},\{i,j\}{]}}\tabularnewline
$\left(Y_{e}\right)_{ij}$  & \texttt{y{[}\{e,L,Hd\},\{i,j\}{]}}\tabularnewline
$\mu$  & \texttt{mu{[}\{Hu,Hd\}{]}}\tabularnewline
$\left(h_{u}\right)_{ij}$  & \texttt{h{[}\{u,Q,Hu\},\{i,j\}{]}}\tabularnewline
$\left(h_{d}\right)_{ij}$  & \texttt{h{[}\{d,Q,Hd\},\{i,j\}{]}}\tabularnewline
$\left(h_{e}\right)_{ij}$  & \texttt{h{[}\{e,L,Hd\},\{i,j\}{]}}\tabularnewline
$b$  & \texttt{b{[}\{Hu,Hd\}{]}}\tabularnewline
$\left(m_{\widetilde{u}}^{2}\right)_{ij}$  & \texttt{m2{[}\{u,u\},\{i,j\}{]}}\tabularnewline
$\left(m_{\widetilde{d}}^{2}\right)_{ij}$  & \texttt{m2{[}\{d,d\},\{i,j\}{]}}\tabularnewline
$\left(m_{\widetilde{Q}}^{2}\right)_{ij}$  & \texttt{m2{[}\{Q,Q\},\{j,i\}{]}}\tabularnewline
$\left(m_{\widetilde{e}}^{2}\right)_{ij}$  & \texttt{m2{[}\{e,e\},\{i,j\}{]}}\tabularnewline
$\left(m_{\widetilde{L}}^{2}\right)_{ij}$  & \texttt{m2{[}\{L,L\},\{j,i\}{]}}\tabularnewline
$m_{H_{u}}^{2}$  & \texttt{m2{[}\{Hu,Hu\}{]}}\tabularnewline
$m_{H_{d}}^{2}$  & \texttt{m2{[}\{Hd,Hd\}{]}}\tabularnewline
\end{tabular}
\par\end{centering}

\caption{\label{tab:Appendix_ParametersMSSM}Parameters of the MSSM assuming
a field ordering \texttt{\{u,d,Q,e,L,Hu,Hd\}} and the gauge factor
group ordering \texttt{\{U1,SU2,SU3\}}.}
\end{table}
 Then, it is possible to match the two with the following code:\texttt{}~\\
\texttt{}~\\
\texttt{parameterRenamingRules{[}myMSSM{]}\textasciicircum{}=\{g{[}i\_{]}:>$\mathtt{g}_{\mathtt{i}}$,
M{[}i\_{]}:>$\mathtt{M}_{\mathtt{i}}$, y{[}\{x\_,\_\_\},\{i\_,j\_\}{]}:>
$\mathtt{Y}_{\mathtt{x}}${[}i,j{]}, $\mathtt{\mu}${[}\{\_\_\}{]}:>$\mathtt{\mu}$,
h{[}\{x\_,\_\_\},\{i\_,j\_\}{]}:>$\mathtt{h}_{\mathtt{x}}${[}i,j{]},
b{[}\{\_\_\}{]}:>b, m2{[}\{Q,Q\},\{i\_,j\_\}{]}:>$\mathtt{m}_{\mathtt{\tilde{Q}}}^{\mathtt{2}}${[}j,i{]},
m2{[}\{L,L\},\{i\_,j\_\}{]}:>$\mathtt{m}_{\mathtt{\tilde{L}}}^{\mathtt{2}}${[}j,i{]},
m2{[}\{x\_,\_\},\{i\_,j\_\}{]}:>$\mathtt{m}_{\mathtt{\tilde{x}}}^{\mathtt{2}}${[}i,j{]},
m2{[}\{x\_,\_\}{]}:>$\mathtt{m}_{\mathtt{x}}^{\mathtt{2}}$, f{[}i\_{]}:>FromCharacterCode{[}104+i{]},
Conjugate{[}x\_\_{]}:>$\mathtt{x}^{*}$\}{]};}~\\
\texttt{}~\\
\texttt{myMSSM}~\\
\\
The last two rules change the default flavor indices \texttt{f{[}1{]}},
\texttt{f{[}2{]}}, ... (\texttt{f{[}i\_{]}:>FromCharacterCode {[}104+i{]}}),
and compactify the notation of the conjugation operation (\texttt{Conjugate{[}x\_\_{]}:>$\mathtt{x}^{*}$}).
Note that by simply running the model's name in the console (\texttt{myMSSM}),
the program will detect that it is a \texttt{Susyno} model and print
all relevant information, in the new notation.

\section{\label{sec:Tests-made}Tests/validation of \texttt{Susyno}}

The output of \texttt{Susyno} was confronted with the analysis of
some models available in the literature. In particular, the RGEs generated
by \texttt{Susyno} were compared with the results of \cite{Martin:1993zk}
(MSSM), \cite{Allanach:1999mh} (\nomenclature{RPV}{R-parity violation}RPV-MSSM),
\cite{Ellwanger:2009dp} (general NMSSM) as well as \cite{Borzumati:2009hu}
($SU(5)$-based models). The program's RGEs are consistent with the
results collected in the latest version of these publications.

\section{\label{sec:List-of-available}List of available functions}

\texttt{Susyno}'s code is spread over many functions. Due to their
nature, some of these functions may be useful on their own, and they
were thus built in a user-friendly way, and are documented.

Below is a list of functions that can be called directly by the user
in Mathematica's front-end, followed by a brief description. The package's
built-in help system describes in detail how to use them. Extensive
use is made of the Lie algebra concepts mentioned in chapter \ref{chap:Symmetry}
(see also appendix \ref{chap:Implementation-details-of} for some
implementation details of some of these functions).
\begin{itemize}
\item \texttt{Adjoint}: Computes the Dynkin coefficients of the adjoint
representation of a group.
\item \texttt{CartanMatrix}: Computes the Cartan matrix of a group.
\item \texttt{Casimir}: Computes the quadratic Casimir of a representation.
\item \texttt{CMtoName}: Returns the name of the group with a given a Cartan
matrix.
\item \texttt{ConjugateIrrep}: Computes the Dynkin coefficients of the conjugate
of a representation.
\item \texttt{DecomposeSnProduct}: Decomposes the product of an arbitrary
number of representations of the discrete $S_{n}$ group in its irreducible
parts.
\item \texttt{DimR}: Computes the dimension of a representation.
\item \texttt{DynkinIndex}: Computes the Dynkin index of a representation.
\item \texttt{GenerateModel}: Computes the 1- and 2-loop RGEs of a SUSY
model, among other things.
\item \texttt{HookContentFormula}: Counts the number of semi-standard Young
tableaux of shape given by a partition $\lambda$ and with the cells
filled with the numbers $1,...,n$ \cite{Stanley:0928.05001}.
\item \texttt{Invariants}: Computes (in some basis) the invariant combination(s)
of an arbitrary number of representations. These are essentially generalized
Clebsch\textendash{}Gordan coefficients (see also the similar function
\texttt{IrrepInProduct}).
\item \texttt{IrrepInProduct}: Computes (in some basis) the combination(s)
of two representations which transforms according to a particular
irreducible representation of the group. For the $SU(2)$ group, these
are known as the Clebsch\textendash{}Gordan coefficients.
\item \texttt{PermutationSymmetryOfTensorProductParts}: Computes the transformation
properties of the irreducible parts of a product of fields/representations
(of the gauge group) under a permutation of the fields being multiplied.
The related function \texttt{PermutationSymmetryOfInvariants} only
returns the gauge invariant parts in these products of fields.
\item \texttt{Plethysms}: Computes the plethysms in a product of an arbitrary
number of representations of a group \cite{Leeuwen:1992aa}. The related
function \texttt{InvariantsPlethysms} only returns those phethysms
which are invariants under the (Lie) group. See also appendix \ref{chap:Implementation-details-of}
for a description of what are phethysms and why do they need to be
computed by the program.
\item \texttt{PositiveRoots}: Computes the positive roots of a group.
\item \texttt{ReduceRepProduct}: Decomposes a direct product representation
in its irreducible parts \cite{Snow:0883.17002,Snow:0885.22001}.
\item \texttt{RepMatrices}: Computes (in some basis) the explicit matrices
of any representation.
\item \texttt{RepMinimalMatrices}: Computes (in some basis) the explicit
representation matrices of the generators appearing in the Chevalley-Serre
relations \eqref{eq:Serre_Chevalley_relations1}--\eqref{eq:Serre_Chevalley_relations3}.
\item \texttt{RepsUpToDimN}: Computes all representations of a given group
up to some dimension.
\item \texttt{RepsUpToDimNNoConjugates}: Computes all representations of
a given group up some dimension, returning for each pair of conjugate
representations only one of them.
\item \texttt{SimplifyEinsteinNotation}: Simplifies an expression written
in Einstein's notation.
\item \texttt{SnClassCharacter}: Computes for a given representation of
the discrete $S_{n}$ group the character of a conjugacy class \cite{Bernstein:1125.20302}.
\item \texttt{SnClassOrder}: Computes the dimension of a conjugacy class
of the discrete $S_{n}$ group (see for example \cite{Stanley:1247.05003}). 
\item \texttt{SnIrrepDim}: Computes the dimension of a representation of
the discrete $S_{n}$ group.
\item \texttt{TriangularAnomalyValue}: Computes the contribution of a representation
for the triangular gauge anomalies \cite{Okubo:1977sc}.
\item \texttt{Weights}: Computes the weights of a representation, including
degeneracy.
\end{itemize}
Unless otherwise stated, in the above list, \textit{group} and \textit{representation}
refers to a simple Lie group and a representation of a simple Lie
group, respectively (not to be confused with the discrete $S_{n}$
group and its representations). We note that the functions \texttt{RepMatrices},
\texttt{Invariants}, \texttt{Plethysms}, \texttt{SimplifyEinsteinNotation}
and related functions are discussed in some detail in appendix \ref{chap:Implementation-details-of}.

\section{Summary}

In this chapter, the Mathematica package \texttt{Susyno} was described.
Given only the defining elements of a softly broken SUSY model---the
gauge group, the representations, the number of flavors/copies of
each representation, and any abelian discrete symmetries (such as
R-parity)---it calculates the 2-loop RGEs. For each model, this is
a long and complicated calculation which should be automated, otherwise
it is very likely that mistakes will be made.

The program also contains several group theoretical functions (related
to both Lie groups and to the discrete permutation group $S_{n}$)
which may be of interest on their own. In other words, even if there
is no intention of computing renormalization group equations, these
functions can still be used. It should be pointed out that there is
an almost complete absence of Mathematica packages with this kind
of functionality (a notable exception is \texttt{LieART} \cite{Feger:2012bs},
which has since been published precisely with the aim of filling this
gap). In fact, even beyond Mathematica, at a theoretical level, the
problem of calculating with all generality the representation matrices
and Clebsch-Gordon coefficients appearing in gauge theories had not
received much attention (see appendix \ref{chap:Implementation-details-of}).
\cleartooddpage

\chapter{\label{chap:U1_mixing_paper}Running soft parameters in SUSY models
with multiple $U(1)$ gauge factors}

\section{\label{sec:U1_mixing_Introduction}Introduction}

The two-loop RGEs for a generic softly broken SUSY model have been
known for quite some time \cite{Martin:1993zk,Yamada:1994id}. However,
these expressions are not completely general. For instance, in the
presence of one $U(1)$ factor group, it is possible to form a Fayet-Iliopoulos
term $\kappa D$ in the superpotential with the non-dynamical $D$
field of the $U(1)$ group, because it is gauge invariant. As such,
there is one extra free parameter $\kappa$ in the theory and the
corresponding RGEs were given in \cite{Jack:1999zs,Jack:2000jr,Jack:2000nm}.
Another issue is the potential presence of Dirac gaugino mass terms
$m_{D}^{iA}\psi_{i}\lambda_{A}$ if there are superfields in the adjoint
representation of one of the gauge factor groups (see \cite{Jack:1999ud,Jack:1999fa,Goodsell:2012fm}).
Yet another problem occurs when there are multiple $U(1)$ factor
groups, a situation that leads to something that is known as $U(1)$-mixing
\cite{Holdom:1985ag,Babu:1997st}. A two-loop renormalization group
analysis for non-SUSY theories with this feature is available in \cite{Luo:2002iq},
while the SUSY case was addressed in \cite{Fonseca:2011vn}. The discussion
summarized in this chapter, as well as the contents of appendix \ref{chap:Matching_conditions},
are mostly taken from this last work. In \cite{Martin:1993zk,Yamada:1994id}
the RGEs are given in a first stage for a single gauge group and then
a list of substitution rules of certain terms is provided, which can
be used to generalize the expressions for multiple gauge groups. In
this chapter we shall see how these rules can be changed to account
for the $U(1)$-mixing effects.

The practical applications of these results are extensive. For instance,
in SUSY GUTs featuring an extended intermediate $U(1)_{R}\times U(1)_{B-L}$
phase, see e.g. \cite{Malinsky:2005bi}, the $U(1)$-mixing effects
can shift the effective MSSM bino soft mass by several per cent with
respect to the naive estimate where such effects are neglected. In
principle, this can have non-negligible effects for the low-energy
phenomenology. In this respect, let us just mention that theories
with a gauged $U(1)_{B-L}$ surviving to the proximity of the soft
SUSY-breaking scale have become rather appealing recently due to their
interesting implications for R-parity and the mechanism of its spontaneous
violation \cite{Khalil:2007dr,FileviezPerez:2010ek,Barger:2008wn},
for leptogenesis \cite{Pelto:2010vq,Babu:2009pi}, etc.

\section{$U(1)$ mixing}

Consider then that the gauge group of a given model is $U(1)^{n}$
and that there are $m$ supermultiplets $\Phi_{i}$, $i=1,\cdots,m$.
At this point, it should be stressed that if the number of supermultiplets
$m$ is smaller than the number of $U(1)$ factors $n$, the $U(1)$'s
may be redefined such that $n-m$ of them (or more) are rotated away.
To see this, first define $Q_{i}^{a}$ as the charge of $\Phi_{i}$
under the $a$-th $U(1)$ group, which we can see as component $a$
for a vector $\boldsymbol{Q}_{i}$. We can then make a rotation%
\footnote{Strictly speaking this $\mathcal{O}_{1}$ matrix does not need to
be a rotation matrix; it is only necessary for it to be invertible.%
} $\mathcal{O}_{1}$ in $U(1)$ space such that $\boldsymbol{Q}_{i}\rightarrow\mathcal{O}_{1}\boldsymbol{Q}_{i}$
for every index $i$, in such a way that $\left(\mathcal{O}_{1}\boldsymbol{Q}_{i}\right)^{a=m+1,\cdots,n}=0$.
In other words, all the $\Phi_{i}$'s can be made to have vanishing
charges under at least $n-m$ of the new, rotated $U(1)$'s, making
them invisible.

Assuming henceforth that $m\geq n$, we now look for the most general
Lagrangian invariant under this $\textrm{U(1}\textrm{)}^{n}$ gauge
group. To do that we must introduce, as usual, $n$ gauge bosons $A_{\mu}^{a}$
which are to be seen also as components of a vector $\boldsymbol{A_{\mu}}$.
Under a gauge transformation with parameters $\alpha^{a}$ we have
the following:
\begin{alignat}{1}
\Phi_{i} & \rightarrow\exp\left(iQ_{i}^{a}\alpha^{a}\right)\Phi_{i}\equiv\exp\left(i\boldsymbol{Q}_{i}^{T}\boldsymbol{\alpha}\right)\Phi_{i}\,,\\
\boldsymbol{A_{\mu}} & \rightarrow\boldsymbol{A_{\mu}}+\boldsymbol{G}^{-1}\partial_{\mu}\boldsymbol{\alpha}\,,
\end{alignat}
where once more $\boldsymbol{\alpha}$ is defined as a vector in $U(1)$
space with $\alpha^{a}$ components ($a=1,\cdots,n$). In the last
equation, a $\boldsymbol{G}$ matrix shows up. In the spirit of making
the most general gauge transformation, $\boldsymbol{G}$ can be any
real $n\times n$ matrix. In particular, this means that the transformation
of $A_{\mu}^{a}$ may depend on some gauge transformation parameter
$\alpha^{b}$ with $b\neq a$. It is straightforward to see that the
Lagrangian will be invariant under this transformation if the covariant
derivative for the supermultiplet $\Phi_{i}$ has the form
\begin{alignat}{1}
D_{\mu}\Phi_{i} & =\left(\partial_{\mu}-i\boldsymbol{Q}_{i}^{T}\boldsymbol{G}\boldsymbol{A_{\mu}}\right)\Phi_{i}\,,
\end{alignat}
which supports the idea that $\boldsymbol{G}$ is a $U(1)$ gauge
couplings matrix. Notice that even though it is a square matrix in
$U(1)$ space, its left and right indices contract with different
vectors: on the left we have the vector with the hypercharges of $\Phi_{i}$,
while on the right there is the $U(1)$ gauge bosons vector. There
is a generic gauge kinetic term to be considered,%
\footnote{The same mixing parameter $\boldsymbol{\xi}$ appears in the gaugino
kinetic term and also in the $\nicefrac{1}{2}D^{a}D^{a}$ term.%
} 
\begin{equation}
-\frac{1}{4}\boldsymbol{F}_{\boldsymbol{\mu\nu}}^{T}\boldsymbol{\xi}\boldsymbol{F^{\mu\nu}}\,,\label{eq:xi}
\end{equation}
and also, in a softly broken supersymmetric theory, the $U(1)$ gaugino
mass term
\begin{equation}
-\frac{1}{2}\boldsymbol{\lambda}^{T}\boldsymbol{M}\boldsymbol{\lambda}+\textrm{h.c.}\,.
\end{equation}
Here we have introduced more vectors in $U(1)$ space: $\boldsymbol{F_{\mu\nu}}$
is a vector whose components are $F_{\mu\nu}^{a}\equiv\partial_{\mu}A_{\nu}^{a}-\partial_{\nu}A_{\mu}^{a}$,
while the $a$ component of $\boldsymbol{\lambda}$ is the gaugino
field $\lambda^{a}$ associated with $U(1)^{a}$. Therefore we have
new $n\times n$ matrices $\boldsymbol{\xi}$ and $\boldsymbol{M}$
to consider, which are free parameters of the theory, containing $\tfrac{1}{2}n\left(n-1\right)$
extra real degrees of freedom. The advantage of having $\boldsymbol{\xi}\neq\mathbb{1}$
is that the gauge coupling matrix can be made diagonal with a rotation
of the gauge boson and gaugino fields. As far as the renormalization
group analysis is concerned, it is now necessary to include both the
effect of $\boldsymbol{\xi}$ on the evolution of the other parameters
and also to describe the evolution of $\boldsymbol{\xi}$ itself.
This is indeed the method adopted in some of the first studies of
the subject (see for example \cite{Luo:2002iq}). But there is an
alternative: the fact that $\boldsymbol{\xi}\neq\mathbb{1}$ means
that the gauge boson fields (as well as the gaugino fields) are not
canonically normalized and we can therefore rotate and rescale these
fields such that in the new basis one always has $\boldsymbol{\xi}=\mathbb{1}$.
In this way, the $U(1)$-mixing in the kinetic term is completely
encoded in a matrix of gauge couplings $\boldsymbol{G}$ and in a
matrix of gaugino masses $\boldsymbol{M}$. This latter approach is
the one adopted in the rest of this chapter.

We have discussed above that the hypercharges $\boldsymbol{Q}_{i}$
can be rotated by a transformation $\mathcal{O}_{1}$ and that the
same can be done to the gauge bosons. Indeed, in general we may perform
two rotations, $\mathcal{O}_{1}$ and $\mathcal{O}_{2}$, which affect
the different parameters and fields in the following way:
\begin{alignat}{1}
\boldsymbol{Q_{i}} & \rightarrow\mathcal{O}_{1}\boldsymbol{Q_{i}}\,,\label{eq:U1_mixing_symmetries1}\\
\boldsymbol{A_{\mu}}\left(\boldsymbol{\lambda}\right) & \rightarrow\mathcal{O}_{2}\boldsymbol{A_{\mu}}\left(\boldsymbol{\lambda}\right)\,,\label{eq:U1_mixing_symmetries_O2}\\
\boldsymbol{G} & \rightarrow\mathcal{O}_{1}\boldsymbol{G}\mathcal{O}_{2}^{T}\,,\label{eq:U1_mixing_symmetriesG}\\
\boldsymbol{M} & \rightarrow\mathcal{O}_{2}\boldsymbol{M}\mathcal{O}_{2}^{T}\,.\label{eq:U1_mixing_symmetries2}
\end{alignat}
As a result of this freedom, it is not straightforward to count the
true number of degrees of freedom in the $\boldsymbol{G}$ and $\boldsymbol{M}$
matrices. We may proceed as follows: an $\mathcal{O}_{2}$ rotation
is used to diagonalize $\boldsymbol{M}$ and then, with the so-called
QR matrix decomposition \cite{golub1996matrix}, it is possible to
put $\boldsymbol{G}$ in an upper or a lower triangular form with
a particular choice of $\mathcal{O}_{1}$. Another possibility is
to use the so-called polar matrix decomposition to cast $\boldsymbol{G}$
in a symmetric form, while keeping $\boldsymbol{M}$ diagonal. In
both these situations, and up to some discrete transformations, we
exhaust the freedom to rotate $\boldsymbol{Q_{i}}$ and $\boldsymbol{A_{\mu}}$
so we can count in these particular bases the number of degrees of
freedom in $\boldsymbol{G}$ and $\boldsymbol{M}$ as being $\frac{1}{2}n\left(n+3\right)$.

Naturally, these symmetries must be reflected at the RGE level. Thus,
for instance, only those combinations $C$ of $\boldsymbol{G}$ and
$\boldsymbol{\gamma}\propto\sum_{i}\boldsymbol{Q_{i}}\boldsymbol{Q_{i}}^{T}$
that transform as $C\to\mathcal{O}_{1}C\mathcal{O}_{2}^{T}$ are allowed
to enter the right-hand side of the renormalization group equation
for $\boldsymbol{G}$. However, at the one-loop level, there is only
one structure involving a third power of $\boldsymbol{G}$ and one
power of $\boldsymbol{\gamma}$ that can arise from a matter-field
loop in the gauge propagator, namely $\boldsymbol{G}\boldsymbol{G}^{T}\boldsymbol{\gamma}\boldsymbol{G}$,
so one immediately concludes that 
\begin{equation}
\beta_{\boldsymbol{G}}^{\textrm{(one loop)}}\propto\boldsymbol{G}\boldsymbol{G}^{T}\boldsymbol{\gamma}\boldsymbol{G}\,.\label{eq:U1_mixing_1loop_example}
\end{equation}
The proportionality coefficient is trivially obtained by matching
this to the single $U(1)$ case. However, at the two loop level the
same exercise becomes more complicated because the increased complexity
of the underlying Feynman diagrams means that more matrices enter
each term. In particular, and unlike in equation \eqref{eq:U1_mixing_1loop_example},
it is possible that specific terms in \cite{Martin:1993zk,Yamada:1994id}
need to be expanded into multiple terms because now, instead of dealing
with numbers which always commute, we have to deal with matrices which
do not.

It might be tempting to think that without the gaugino mass matrix
$\boldsymbol{M}$ in non-SUSY theories, we can diagonalize $\boldsymbol{G}$
and therefore get rid of the $U(1)$-mixing effects. However this
is not the case, as radiative effects will reintroduce off-diagonalities
in the gauge couplings matrix. In fact, already at one loop level,
the anomalous dimension $\boldsymbol{\gamma}$ which controls the
RGE of $\boldsymbol{G}$ is in general a non-diagonal matrix in $U(1)$
space and as such, even if $\boldsymbol{G}$ is diagonal at a given
energy scale, non-diagonal entries will be radiatively generated.

It turns out that there are exceptions to this rule. For instance,
it can be that all the relevant $U(1)$ couplings originate from a
common gauge factor and thus, barring threshold effects, all of them
happen to be equal at a certain scale. In such a case, the charges
and the gauge fields can be simultaneously rotated at the one-loop
level so that no off-diagonalities appear in $\boldsymbol{\gamma}$
\cite{delAguila:1988jz,Martin:1993zk} and one can use the simple
form of the RGEs for individual gauge couplings. Note again that this
will only work in the non-SUSY case where only the gauge sector has
to be taken into account. Also, at two loops, Yukawa couplings and
trilinear soft SUSY breaking couplings appear in the RGEs, rendering
this approach useless.

\section{\label{sect-results}Two loop RGEs}

In this section, we describe the generic method of constructing the
fully general two-loop RGEs for softly-broken supersymmetric gauge
theories out of the results of \cite{Martin:1993zk,Yamada:1994id}
relevant to the case of (at most) a single abelian gauge-group factor.
For the sake of completeness, the relevant formulae for the cases
of (i) a simple gauge group and (ii) the product of several simple
factors with at most a single $U(1)$ are reproduced in appendix \ref{chap:Appendix_Two_loop_RGEs}.
The computation has been done using the $\overline{DR}'$ scheme defined
in \cite{Jack:1994rk}.

\subsection{Notation and conventions}

The gauge group is taken to be $G_{A}\times G_{B}\times\cdots\times U\left(1\right)^{n}$,
where the $G_{X}$'s are simple groups. We shall use uppercase indices
for simple group-factors only; lowercase indices are used either for
all groups or, in some specific cases, for $U(1)$'s only.%
\footnote{This will be evident from the context; we follow as closely as possible
\cite{Martin:1993zk} and when quoting results contained therein,
the $a$ and $b$ indices go over all groups (simple and $U(1)$ groups).
On other occasions, when referring to particular components of the
$U(1)$-related $\boldsymbol{G}$, $\boldsymbol{M}$ and $\boldsymbol{V}$
matrices and vectors, $a$ and $b$ stretch over the $U(1)$ groups
only. %
} As mentioned before, the $U(1)$ sector should be treated as a whole
and described in terms of a general real $n\times n$ gauge couplings
matrix $\boldsymbol{G}$, a $n\times n$ symmetric soft SUSY breaking
gaugino mass matrix $\boldsymbol{M}$ and a column vector of charges
$\boldsymbol{Q_{i}}$ for each chiral supermultiplet $\Phi_{i}$.
Notice, however, that $\boldsymbol{V_{i}}\equiv\boldsymbol{G}^{T}\boldsymbol{Q_{i}}$
for each $i$ are the only combinations of $\boldsymbol{Q_{i}}$ and
$\boldsymbol{G}$ which appear in the Lagrangian (recall the $\mathcal{O}_{1}$
rotation freedom discussed in the previous section) and thus, all
the general RGEs can be written in terms of $\boldsymbol{V}$'s and
$\boldsymbol{M}$ only. We shall follow this convention with a single
exception---the evolution equations for the gauge couplings---which
are traditionally written in terms of $\nicefrac{d\boldsymbol{G}}{dt}$
rather than $\nicefrac{d\boldsymbol{V}}{dt}$. Here, we shall adhere
to the usual practice and as a consequence we expect an isolated $\boldsymbol{G}$
in these equations.

Before proceeding any further we shall define some of the expressions
that are used in the RGEs: 
\begin{itemize}
\item $C_{a}\left(i\right)$: quadratic Casimir invariant of the representation
of superfield $\Phi_{i}$ under the group $G_{a}$.
\item $C\left(G_{a}\right)$: quadratic Casimir invariant of the adjoint
representation of group $G_{a}$.
\item $S_{a}\left(i\right)$: Dynkin index of the representation of superfield
$\Phi_{i}$ under the group $G_{a}$.
\item $d_{a}\left(i\right)$: Dimension of the representation of $\Phi_{i}$
under the group $G_{a}$.
\item $d\left(G_{a}\right)$: dimension of group $G_{a}$.
\item $S_{a}\left(R\right)$: Dynkin index of group $G_{a}$ summed over
all chiral supermultiplets---$S_{a}\left(R\right)=\sum_{i}\frac{S_{a}\left(i\right)}{d_{a}\left(i\right)}$.
\item $S_{a}\left(R\right)C_{b}\left(R\right)$: defined as $\sum_{i}\frac{S_{a}\left(i\right)C_{b}\left(i\right)}{d_{a}\left(i\right)}$.
\item $S_{a}\left(R\right)\boldsymbol{V}_{\boldsymbol{R}}^{T}\boldsymbol{V_{R}}$:
defined as $\sum_{i}\frac{S_{a}\left(i\right)\boldsymbol{V}_{\boldsymbol{i}}^{T}\boldsymbol{V_{i}}}{d_{a}\left(i\right)}$.
\item $S_{a}\left(R\right)\boldsymbol{V}_{\boldsymbol{R}}^{T}\boldsymbol{M}\boldsymbol{V_{R}}$:
defined as $\sum_{i}\frac{S_{a}\left(i\right)\boldsymbol{V}_{\boldsymbol{i}}^{T}\boldsymbol{M}\boldsymbol{V_{i}}}{d_{a}\left(i\right)}$.
\end{itemize}
In addition, sometimes one has to deal with the explicit representation
matrices of the gauge groups (denoted in \cite{Martin:1993zk} by
${\bf t}_{i}^{Aj}$). Notice that here $A$ is not a group index but
rather a coordinate in the adjoint representation of the corresponding
Lie algebra (for example, $A=1,..,3$ in $SU(2)$, and $A=1,..,8$
in $SU(3)$).

Naturally, whenever we refer to results of \cite{Martin:1993zk,Yamada:1994id}
for a simple gauge group (collected in section \ref{sect-rges-simple}
of appendix \ref{chap:Appendix_Two_loop_RGEs}), the $a$ and $b$
indices will be omitted. In all cases, repeated indices are not implicitly
summed over.

\subsection{Strategy for the constructing the general substitution rules}

Let us now describe in more detail the strategy for upgrading the
{}``product'' substitution rules of section III in reference \cite{Martin:1993zk}
to the most general case of an arbitrary gauge group. For the moment
we shall focus on a limited number of terms; later on, in section
\ref{sec:Obtaining_the_substitution_rules}, a more elaborate discussion
addresses all remaining situations.

Let us begin with the term $g^{2}C\left(r\right)$ appearing for instance
in equation \eqref{eq:U1_mixing_example1} and, subsequently, in the
substitution rules of \cite{Martin:1993zk} for product groups, equation
\eqref{eq:U1_mixing_example1productX}. It is clear that this has
to be replaced by $\sum_{A}g_{A}^{2}C_{A}\left(r\right)+\text{`}U(1)\text{ part'}$.
For a single $U(1)$, $g^{2}C\left(r\right)=g^{2}y_{r}^{2}\sim\boldsymbol{V_{r}}\boldsymbol{V_{r}}$
so this `$U(1)$ part' can only take the form%
\footnote{If there is a single abelian factor group, we denote by $y_{i}$ the
(hyper)charge of chiral superfield $\Phi_{i}$, which is just a number%
} $\boldsymbol{V}_{\boldsymbol{r}}^{T}\boldsymbol{V_{r}}=\boldsymbol{Q}_{\boldsymbol{r}}^{T}\boldsymbol{G}\boldsymbol{G}^{T}\boldsymbol{Q_{r}}$.
There is no other way to obtain a number from two vectors $\boldsymbol{V_{r}}$.
This expression automatically sums the contributions of all the $U(1)$'s.

Similarly, $Mg^{2}C\left(r\right)$ (in equation \eqref{eq:U1_mixing_example4}
for example) is replaced by $\sum_{A}M_{A}g_{A}^{2}C_{A}\left(r\right)+\text{`}U(1)\text{ part'}$;
the ingredients for the construction of the `$U(1)$ part' are two
vectors $\boldsymbol{V_{r}}$ and the gaugino mass matrix $\boldsymbol{M}$.
Only $\boldsymbol{V}_{\boldsymbol{r}}^{T}\boldsymbol{M}\boldsymbol{V_{r}}$
forms a number.

In fact, this simple procedure allows us to generalize many of the
terms in the RGEs of \cite{Martin:1993zk,Yamada:1994id}, section
II (reproduced in appendix \ref{chap:Appendix_Two_loop_RGEs}). As
a more involved example, consider for instance the $g^{4}{\bf t}_{i}^{Aj}\textrm{Tr}\left[{\bf t}^{A}C\left(r\right)m^{2}\right]$
structure appearing in equation \eqref{eq:U1_mixing_example3}. It
is not difficult to see that all terms where the representation matrices
${\bf t}^{A}$ appear explicitly are zero unless $A$ corresponds
to an abelian group. Hence, if for a single $U(1)$ one has%
\footnote{In this context, $g$ is the $U(1)$ gauge coupling.%
}
\begin{align}
g^{4}{\bf t}_{i}^{Aj}\textrm{Tr}\left[{\bf t}^{A}C\left(r\right)m^{2}\right] & =g^{2}\delta_{i}^{j}y_{i}\sum_{p}y_{p}\left[\sum_{B}g_{B}^{2}C_{B}\left(p\right)+g^{2}y_{p}^{2}\right]\left(m^{2}\right)_{p}^{p}\,,
\end{align}
it can be immediately deduced that, in the general case,
\begin{align}
g^{4}{\bf t}_{i}^{Aj}\textrm{Tr}\left[{\bf t}^{A}C\left(r\right)m^{2}\right] & \rightarrow\delta_{i}^{j}\sum_{p}\left(\boldsymbol{V}_{\boldsymbol{i}}^{T}\boldsymbol{V_{p}}\right)\left[\sum_{B}g_{B}^{2}C_{B}\left(p\right)+\left(\boldsymbol{V}_{\boldsymbol{p}}^{T}\boldsymbol{V_{p}}\right)\right]\left(m^{2}\right)_{p}^{p}\,.
\end{align}
The RGEs of $\boldsymbol{G}$ and $\boldsymbol{M}$ represent a bigger
challenge, because they are matrix equations (in other words, there
are uncontracted gauge indices). On the other hand, this should be
viewed as an advantage because all the relevant equations must then
respect the reparametrization symmetries \eqref{eq:U1_mixing_symmetries1}--\eqref{eq:U1_mixing_symmetries2}.
Notice that these equations imply that the $\boldsymbol{V}$'s transform
as $\boldsymbol{V_{i}}\rightarrow\mathcal{O}_{2}\boldsymbol{V_{i}}$.
These symmetries are especially powerful in the $\beta$-functions
for the gauge couplings which, due to equation \eqref{eq:U1_mixing_symmetriesG},
inevitably take the generic form $\boldsymbol{G}\boldsymbol{V_{i}}\left(\cdots\right)\boldsymbol{V}_{\boldsymbol{j}}^{T}$
for some chiral indices $i,\, j$. For example, $g^{3}S\left(R\right)\sim g^{3}\sum_{p}y_{p}^{2}$
can only take the form $\boldsymbol{G}\sum_{p}\boldsymbol{V_{p}}\boldsymbol{V}_{\boldsymbol{p}}^{T}$.

Concerning the gaugino soft masses $M$, let us for instance consider
the $2g^{2}S\left(R\right)M$ term appearing in equation \eqref{eq:U1_mixing_example5}.
Its generalized variant should be built out of a pair of $\boldsymbol{V_{p}}$
vectors and the $\boldsymbol{M}$ matrix. However, there are only
two combinations of these objects that transform correctly under $\mathcal{O}_{2}$,
namely, $\boldsymbol{M}\boldsymbol{V_{p}}\boldsymbol{V}_{\boldsymbol{p}}^{T}$
and $\boldsymbol{V_{p}}\boldsymbol{V}_{\boldsymbol{p}}^{T}\boldsymbol{M}$.
Thus, due to the symmetry of $\boldsymbol{M}$, one obtains $2g^{2}S\left(R\right)M\rightarrow\boldsymbol{M}\sum_{p}\boldsymbol{V_{p}}\boldsymbol{V}_{\boldsymbol{p}}^{T}+\sum_{p}\boldsymbol{V_{p}}\boldsymbol{V}_{\boldsymbol{p}}^{T}\boldsymbol{M}$.

Another important ingredient of the analysis is provided by the existing
substitution rules linking the case of a simple gauge group (section
II in \cite{Martin:1993zk} or section \ref{sect-rges-simple} of
appendix \ref{chap:Appendix_Two_loop_RGEs}) to the settings with
group products (section III in \cite{Martin:1993zk} or section \ref{sect-rges-products}
of appendix \ref{chap:Appendix_Two_loop_RGEs}). Consider, for example,
the $g^{5}S\left(R\right)C\left(R\right)$ term in equation \eqref{eq:U1_mixing_g5SRCR}
which, according to \cite{Martin:1993zk}, is replaced by $\sum_{b}g_{a}^{3}g_{b}^{2}S_{a}\left(R\right)C_{b}\left(R\right)$;
see formula \eqref{eq:U1_mixing_g5SRCRsubst} for the product groups.
Let us recall that the expression $S\left(R\right)C\left(R\right)$
has a very particular meaning: it is the sum of the Dynkin indices
weighted by the quadratic Casimir invariant, so $\sum_{b}g_{a}^{3}g_{b}^{2}S_{a}\left(R\right)C_{b}\left(R\right)=\sum_{b,p}g_{a}^{3}g_{b}^{2}\frac{S_{a}\left(p\right)C_{b}\left(p\right)}{d_{a}\left(p\right)}$.
With this in mind, whenever $a$ refers to the abelian part of the
gauge group, one should replace $g_{a}^{3}S_{a}\left(p\right)\rightarrow\boldsymbol{G}\boldsymbol{V_{p}}\boldsymbol{V}_{\boldsymbol{p}}^{T}$
($d_{a}\left(p\right)=1$), while $\sum_{b}g_{b}^{2}C_{b}\left(p\right)\rightarrow\sum_{B}g_{B}^{2}C_{B}\left(p\right)+\boldsymbol{V}_{\boldsymbol{p}}^{T}\boldsymbol{V_{p}}$.
Therefore, for the abelian sector, $g^{5}S\left(R\right)C\left(R\right)\rightarrow\sum_{p}\boldsymbol{GV_{p}}\boldsymbol{V}_{\boldsymbol{p}}^{T}\left[\sum_{B}g_{B}^{2}C_{B}\left(p\right)+\boldsymbol{V}_{\boldsymbol{p}}^{T}\boldsymbol{V_{p}}\right]$.

However, sometimes even a detailed inspection of the underlying expressions
does not allow an unambiguous identification of its generalized form.
When this happens, a careful analysis of the structure of the contributing
Feynman diagrams is necessary. However, remarkably, the number of
such cases is small, as shown in section \ref{sec:Obtaining_the_substitution_rules}.

\subsection{Substitution rules}

Depending on the group sector (abelian or simple), we get different
RGEs for the gauge couplings and the gaugino masses. The parameters
are then either the matrices $\boldsymbol{G}$, $\boldsymbol{M}$
or the numbers $g_{A}$, $M_{A}$. For the abelian sector, one obtains:
\begin{align}
C\left(G\right) & \rightarrow0\,,\label{eq:U1_mixing_rule31}\\
g^{3}S\left(R\right) & \rightarrow\boldsymbol{G}\sum_{p}\boldsymbol{V_{p}}\boldsymbol{V}_{\boldsymbol{p}}^{T}\,,\label{eq:U1_mixing_rule32}\\
g^{5}S\left(R\right)C\left(R\right) & \rightarrow\sum_{p}\boldsymbol{G}\boldsymbol{V_{p}}\boldsymbol{V}_{\boldsymbol{p}}^{T}\Bigl[\sum_{B}g_{B}^{2}C_{B}\left(p\right)+\boldsymbol{V_{p}}\boldsymbol{V}_{\boldsymbol{p}}^{T}\Bigr]\,,\label{eq:U1_mixing_rule33}\\
\frac{g^{3}C\left(k\right)}{d\left(G\right)} & \rightarrow\boldsymbol{G}\boldsymbol{V_{k}}\boldsymbol{V}_{\boldsymbol{k}}^{T}\,,\label{eq:U1_mixing_rule35}\\
2g^{2}S\left(R\right)M & \rightarrow\boldsymbol{M}\sum_{p}\boldsymbol{V_{p}}\boldsymbol{V}_{\boldsymbol{p}}^{T}+\sum_{p}\boldsymbol{V_{p}}\boldsymbol{V}_{\boldsymbol{p}}^{T}\boldsymbol{M}\,,\label{eq:U1_mixing_rule36}\\
g^{2}C\left(k\right) & \rightarrow\boldsymbol{V_{k}}\boldsymbol{V}_{\boldsymbol{k}}^{T}\,,\label{eq:U1_mixing_rule39}\\
2g^{2}C\left(k\right)M & \rightarrow\boldsymbol{M}\boldsymbol{V_{k}}\boldsymbol{V}_{\boldsymbol{k}}^{T}+\boldsymbol{V_{k}}\boldsymbol{V}_{\boldsymbol{k}}^{T}\boldsymbol{M}\,,\label{eq:U1_mixing_rule40}\\
16g^{4}S\left(R\right)C\left(R\right)M & \rightarrow\sum_{p}\left\{ 4\left(\boldsymbol{M}\boldsymbol{V_{p}}\boldsymbol{V}_{\boldsymbol{p}}^{T}+\boldsymbol{V_{p}}\boldsymbol{V}_{\boldsymbol{p}}^{T}\boldsymbol{M}\right)\left[\sum_{B}g_{B}^{2}C_{B}\left(p\right)+\boldsymbol{V_{p}}\boldsymbol{V}_{\boldsymbol{p}}^{T}\right]\right.\nonumber \\
 & \left.+8\boldsymbol{V_{p}}\boldsymbol{V}_{\boldsymbol{p}}^{T}\left[\sum_{B}M_{B}g_{B}^{2}C_{B}\left(p\right)+\boldsymbol{V}_{\boldsymbol{p}}^{T}\boldsymbol{M}\boldsymbol{V_{p}}\right]\right\} \,.\label{eq:U1_mixing_rule37}
\end{align}

For a simple group factor $G_{A}$, the substitution rules of \cite{Martin:1993zk}
do not need to be changed except for two cases:
\begin{align}
g^{5}S\left(R\right)C\left(R\right) & \rightarrow g_{A}^{3}S_{A}\left(R\right)\left[\sum_{B}g_{B}^{2}C_{B}\left(R\right)+\boldsymbol{V}_{\boldsymbol{R}}^{T}\boldsymbol{V_{R}}\right]\,,\label{eq:U1_mixing_rule34}\\
16g^{4}S\left(R\right)C\left(R\right)M & \rightarrow8g_{A}^{2}M_{A}S_{A}\left(R\right)\left[\sum_{B}g_{B}^{2}C_{B}\left(R\right)+\boldsymbol{V}_{\boldsymbol{R}}^{T}\boldsymbol{V_{R}}\right]\nonumber \\
 & +8g_{A}^{2}S_{A}\left(R\right)\left[\sum_{B}M_{B}g_{B}^{2}C_{B}\left(R\right)+\boldsymbol{V}_{\boldsymbol{R}}^{T}\boldsymbol{MV_{R}}\right]\,.\label{eq:U1_mixing_rule38}
\end{align}

As for the rest of the parameters in a SUSY model, the relevant substitution
rules read:{\allowdisplaybreaks
\begin{align}
g^{2}C\left(r\right) & \rightarrow\sum_{A}g_{A}^{2}C_{A}\left(r\right)+\boldsymbol{V}_{\boldsymbol{r}}^{T}\boldsymbol{V_{r}}\,,\label{eq:U1_mixing_rule14}\\
Mg^{2}C\left(r\right) & \rightarrow\sum_{A}M_{A}g_{A}^{2}C_{A}\left(r\right)+\boldsymbol{V}_{\boldsymbol{r}}^{T}\boldsymbol{MV_{r}}\,,\label{eq:U1_mixing_rule15}\\
M^{*}g^{2}C\left(r\right) & \rightarrow\sum_{A}M_{A}^{*}g_{A}^{2}C_{A}\left(r\right)+\boldsymbol{V}_{\boldsymbol{r}}^{T}\boldsymbol{M}^{\dagger}\boldsymbol{V_{r}}\,,\label{eq:U1_mixing_rule16}\\
MM^{*}g^{2}C\left(r\right) & \rightarrow\sum_{A}M_{A}M_{A}^{*}g_{A}^{2}C_{A}\left(r\right)+\boldsymbol{V}_{\boldsymbol{r}}^{T}\boldsymbol{M}\boldsymbol{M}^{\dagger}\boldsymbol{V_{r}}\,,\label{eq:U1_mixing_rule17}\\
g^{4}C\left(r\right)S\left(R\right) & \rightarrow\sum_{A}g_{A}^{4}C_{A}\left(r\right)S_{A}\left(R\right)+\sum_{p}\left(\boldsymbol{V}_{\boldsymbol{r}}^{T}\boldsymbol{V_{p}}\right)^{2}\,,\label{eq:U1_mixing_rule18}\\
Mg^{4}C\left(r\right)S\left(R\right) & \rightarrow\sum_{A}M_{A}g_{A}^{4}C_{A}\left(r\right)S_{A}\left(R\right)+\sum_{p}\left(\boldsymbol{V}_{\boldsymbol{r}}^{T}\boldsymbol{MV_{p}}\right)\left(\boldsymbol{V}_{\boldsymbol{r}}^{T}\boldsymbol{V_{p}}\right)\,,\label{eq:U1_mixing_rule19}\\
g^{4}C^{2}\left(r\right) & \rightarrow\left[\sum_{A}g_{A}^{2}C_{A}\left(r\right)+\boldsymbol{V}_{\boldsymbol{r}}^{T}\boldsymbol{V_{r}}\right]^{2}\,,\label{eq:U1_mixing_rule20}\\
Mg^{4}C^{2}\left(r\right) & \rightarrow\left[\sum_{A}M_{A}g_{A}^{2}C_{A}\left(r\right)+\boldsymbol{V}_{\boldsymbol{r}}^{T}\boldsymbol{MV_{r}}\right]\left[\sum_{A}g_{A}^{2}C_{A}\left(r\right)+\boldsymbol{V}_{\boldsymbol{r}}^{T}\boldsymbol{V_{r}}\right]\,,\label{eq:U1_mixing_rule21}\\
g^{4}C\left(G\right)C\left(r\right) & \rightarrow\sum_{A}g_{A}^{4}C\left(G_{A}\right)C_{A}\left(r\right)\,,\label{eq:U1_mixing_rule22}\\
Mg^{4}C\left(G\right)C\left(r\right) & \rightarrow\sum_{A}M_{A}g_{A}^{4}C\left(G_{A}\right)C_{A}\left(r\right)\,,\label{eq:U1_mixing_rule23}\\
MM^{*}g^{4}C\left(G\right)C\left(r\right) & \rightarrow\sum_{A}M_{A}M_{A}^{*}g_{A}^{4}C\left(G_{A}\right)C_{A}\left(r\right)\,,\label{eq:U1_mixing_rule24}\\
g^{2}{\bf t}_{i}^{Aj}\textrm{Tr}\left({\bf t}^{A}m^{2}\right) & \rightarrow\delta_{i}^{j}\sum_{p}\boldsymbol{V}_{\boldsymbol{i}}^{T}\boldsymbol{V_{p}}\left(m^{2}\right)_{p}^{p}\,,\label{eq:U1_mixing_rule25}\\
g^{2}{\bf t}_{i}^{Aj}\left({\bf t}^{A}m^{2}\right)_{r}^{l} & \rightarrow\delta_{i}^{j}\boldsymbol{V}_{\boldsymbol{l}}^{T}\boldsymbol{V_{i}}\left(m^{2}\right)_{r}^{l}\,,\label{eq:U1_mixing_rule26}\\
g^{4}{\bf t}_{i}^{Aj}\textrm{Tr}\left[{\bf t}^{A}C\left(r\right)m^{2}\right] & \rightarrow\delta_{i}^{j}\sum_{p}\boldsymbol{V}_{\boldsymbol{i}}^{T}\boldsymbol{V_{p}}\left[\sum_{B}g_{B}^{2}C_{B}\left(p\right)+\boldsymbol{V}_{\boldsymbol{p}}^{T}\boldsymbol{V_{p}}\right]\left(m^{2}\right)_{p}^{p}\,,\label{eq:U1_mixing_rule27}\\
g^{4}C\left(i\right)\textrm{Tr}\left[S\left(r\right)m^{2}\right] & \rightarrow\sum_{A}g_{A}^{4}C_{A}\left(i\right)\textrm{Tr}\left[S_{A}\left(r\right)m^{2}\right]+\sum_{p}\left(\boldsymbol{V}_{\boldsymbol{i}}^{T}\boldsymbol{V_{p}}\right)^{2}\left(m^{2}\right)_{p}^{p}\,,\label{eq:U1_mixing_rule28}\\
24g^{4}MM^{*}C\left(i\right)S\left(R\right) & \rightarrow24\sum_{A}g_{A}^{4}M_{A}M_{A}^{*}C_{A}\left(i\right)S_{A}\left(R\right)\nonumber \\
 & +8\sum_{p}\left[\left(\boldsymbol{V}_{\boldsymbol{i}}^{T}\boldsymbol{MV_{p}}\right)\left(\boldsymbol{V}_{\boldsymbol{i}}^{T}\boldsymbol{M}^{\dagger}\boldsymbol{V_{p}}\right)+\left(\boldsymbol{V}_{\boldsymbol{i}}^{T}\boldsymbol{M}\boldsymbol{M}^{\dagger}\boldsymbol{V_{p}}\right)\left(\boldsymbol{V}_{\boldsymbol{i}}^{T}\boldsymbol{V_{p}}\right)\right.\nonumber \\
 & +\left.\left(\boldsymbol{V}_{\boldsymbol{i}}^{T}\boldsymbol{M}^{\dagger}\boldsymbol{MV_{p}}\right)\left(\boldsymbol{V}_{\boldsymbol{i}}^{T}\boldsymbol{V_{p}}\right)\right]\,,\label{eq:U1_mixing_rule29}\\
48g^{4}MM^{*}C\left(r\right)^{2} & \rightarrow\sum_{A,B}g_{A}^{2}g_{B}^{2}C_{A}\left(r\right)C_{B}\left(r\right)\left(32M_{A}M_{A}^{*}+16M_{A}M_{B}^{*}\right)\nonumber \\
 & +\sum_{A}g_{A}^{2}C_{A}\left(r\right)\left(32M_{A}M_{A}^{*}\boldsymbol{V}_{\boldsymbol{r}}^{T}\boldsymbol{V_{r}}+16M_{A}\boldsymbol{V}_{\boldsymbol{r}}^{T}\boldsymbol{M}^{\dagger}\boldsymbol{V_{r}}\right.\nonumber \\
 & +\left.32\boldsymbol{V}_{\boldsymbol{r}}^{T}\boldsymbol{M}\boldsymbol{M}^{\dagger}\boldsymbol{V_{r}}+16M_{A}^{*}\boldsymbol{V}_{\boldsymbol{r}}^{T}\boldsymbol{MV_{r}}\right)\nonumber \\
 & +32\left(\boldsymbol{V}_{\boldsymbol{r}}^{T}\boldsymbol{M}\boldsymbol{M}^{\dagger}\boldsymbol{V_{r}}\right)\left(\boldsymbol{V}_{\boldsymbol{r}}^{T}\boldsymbol{V_{r}}\right)+16\left(\boldsymbol{V}_{\boldsymbol{r}}^{T}\boldsymbol{MV_{r}}\right)\left(\boldsymbol{V}_{\boldsymbol{r}}^{T}\boldsymbol{M}^{\dagger}\boldsymbol{V_{r}}\right)\,.\label{eq:U1_mixing_rule30}
\end{align}
}

\section{\label{sec:Obtaining_the_substitution_rules}Obtaining the substitution
rules}

We discuss now in more detail the methods used throughout the derivation
of the substitution rules given in section \ref{sect-results}, with
particular emphasis on the few cases where this is not straightforward.

\subsection{The role of the $\boldsymbol{V_{i}}$ vectors and the $\boldsymbol{M}$
matrix}

\begin{figure}[h]
\begin{centering}
\includegraphics[scale=0.8]{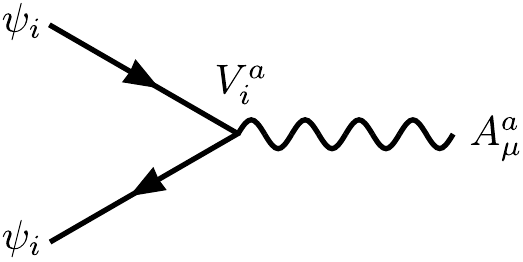}
\par\end{centering}

\caption{\label{fig:U1_mixing_PsiPsiA_vertex}The vertex $\psi^{i\dagger}\psi^{i}A_{\mu}^{a}$
is proportional to $V_{i}^{a}$.}
\end{figure}

As mentioned before, the $U(1)$ gauge coupling matrix $\boldsymbol{G}$
and the charge vectors $\boldsymbol{Q_{i}}$ of the chiral superfields
$\Phi_{i}$ appear always through the combination $\boldsymbol{V_{i}}=\boldsymbol{G}^{T}\boldsymbol{Q_{i}}$.
The only exception are the RGEs of $\boldsymbol{G}$, where there
should be a leading free $\boldsymbol{G}$. For example, the $\psi^{i\dagger}\psi^{i}A_{\mu}^{a}$
vertex is proportional to $V_{i}^{a}$ (component $a$ of the vector
$\boldsymbol{V_{i}}$)---figure \eqref{fig:U1_mixing_PsiPsiA_vertex}.

\begin{figure}[h]
\begin{centering}
\includegraphics[scale=0.8]{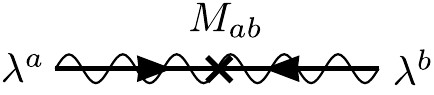}
\par\end{centering}

\caption{\label{fig:U1_mixing_gaugino_mass_diagram}The $U(1)$ gaugino mass
matrix $\boldsymbol{M}$ mixes gaugino fields. $M_{ab}$ is the $a,b$
component of $\boldsymbol{M}$.}
\end{figure}

Similarly the vertices $\phi^{*i}\phi^{i}A_{\mu}^{a}$, $\phi^{*i}\phi^{i}A_{\mu}^{a}A_{\nu}^{b}$,
$\phi^{i*}\psi^{i}\lambda^{a}$ and the Yukawa independent part of
$\phi^{i*}\phi^{i}\phi^{j*}\phi^{j}$ are proportional to $V_{i}^{a}$,
$V_{i}^{a}V_{i}^{b}$, $V_{i}^{a}$ and $\boldsymbol{V}_{\boldsymbol{i}}^{T}\boldsymbol{V_{j}}$
respectively. In addition, one must consider the $U(1)$ gaugino mass
matrix $\boldsymbol{M}$ (see figure \eqref{fig:U1_mixing_gaugino_mass_diagram}).

\subsection{RGEs with no $U(1)$ indices}

\begin{figure}[h]
\begin{centering}
\includegraphics[scale=0.8]{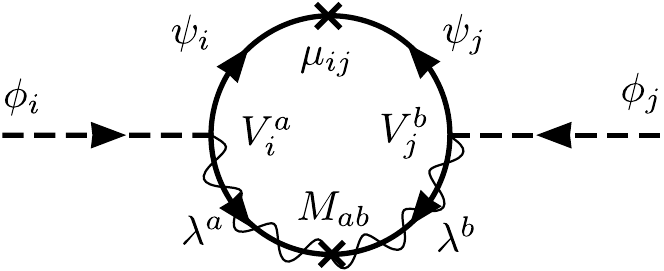}
\par\end{centering}

\caption{\label{fig:U1_mixing_D1}Diagram contributing to the one-loop RGE
of the bilinear scalar soft terms $b^{ij}$. Notice the contraction
of the $a$ and $b$ indices between the $\boldsymbol{V}$'s and $\boldsymbol{M}$.}
\end{figure}

Diagrams needed to compute the RGEs of $\boldsymbol{G}$ and $\boldsymbol{M}$
are the only ones with external $U(1)$ gauge bosons/gauginos. In
all other equations, while vectors $\boldsymbol{V_{i}}$ and the matrix
$\boldsymbol{M}$ may be present, they must form scalar combinations,
so no free $U(1)$ indices exist. Consider the $Mg^{2}C\left(i\right)$
appearing in the one-loop RGE of the bilinear scalar soft terms $b^{ij}$,
which is to be replaced by $\sum_{A}M_{A}g_{A}^{2}C_{A}\left(i\right)+\boldsymbol{V}_{\boldsymbol{i}}^{T}\boldsymbol{MV_{i}}$.
The simple groups contribution, $\sum_{A}M_{A}g_{A}^{2}C_{A}\left(i\right)$,
can safely be neglected in this discussion. We can see that $\boldsymbol{V}_{\boldsymbol{i}}^{T}\boldsymbol{MV_{i}}$
is the only structure that can generalize the expression $Mg^{2}C\left(i\right)=Mg^{2}y_{i}^{2}$
for one $U(1)$ group only. Notice also the contraction of the $U(1)$
indices in the expression---it comes from the possibility of having
any of the $U(1)$ gauginos in the internal lines of the contributing
diagram in figure \eqref{fig:U1_mixing_D1}.
The amplitude is proportional to $\sum_{a,b}V_{i}^{a}M_{ab}V_{j}^{b}\mu^{ij}=\mu^{ij}\boldsymbol{V}_{\boldsymbol{i}}^{T}\boldsymbol{MV_{j}}$.
Note that for any pair of values $i,\, j$ the gauge symmetry forces
$\mu^{ij}=0$ unless $\boldsymbol{V_{i}}+\boldsymbol{V_{j}}=0$, which
means that $\mu^{ij}\boldsymbol{V_{j}}=-\mu^{ij}\boldsymbol{V_{i}}$
so the amplitude of the diagram is indeed proportional to $\boldsymbol{V}_{\boldsymbol{i}}^{T}\boldsymbol{MV_{i}}$.

\begin{figure}[h]
\centering{}\includegraphics[scale=0.7]{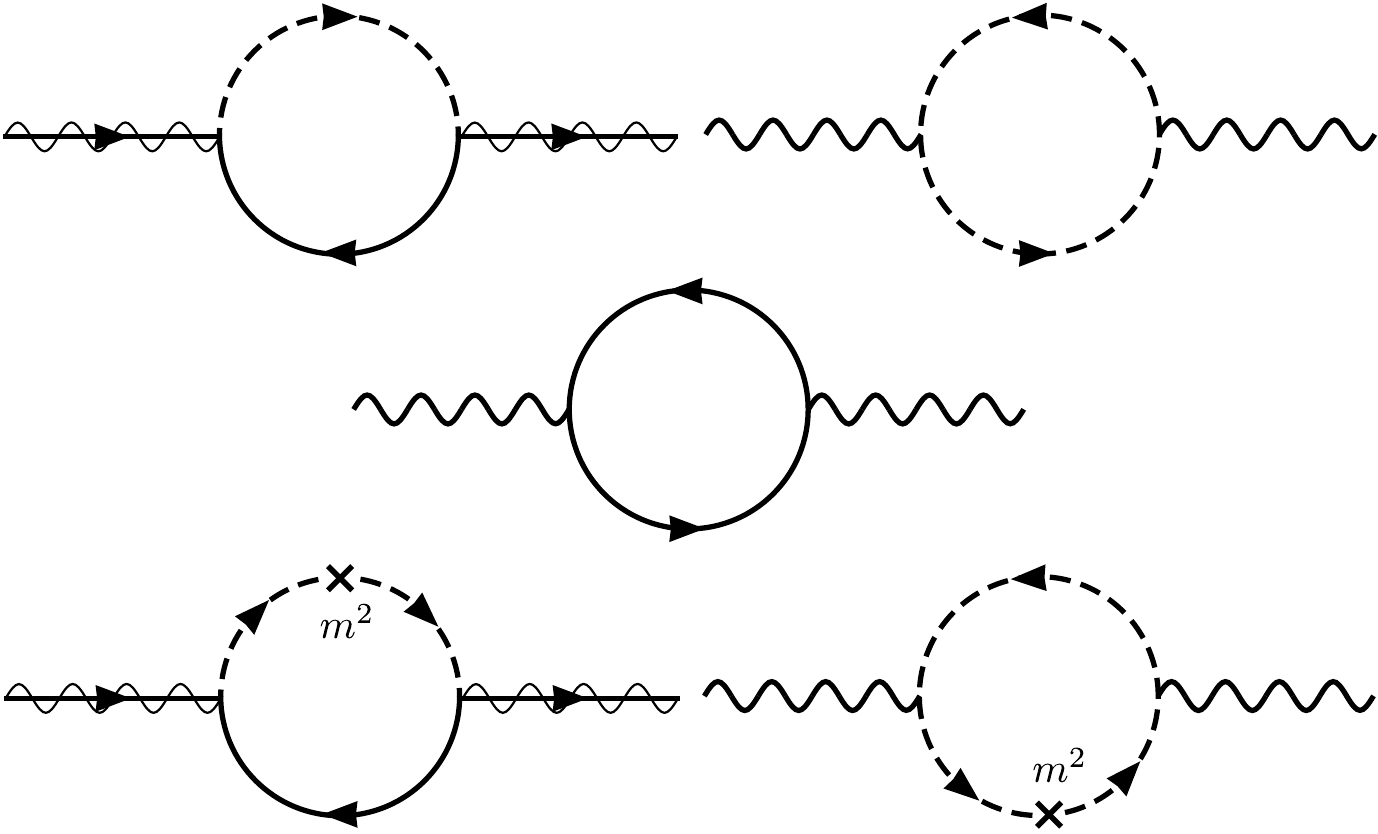}\caption{\label{fig:U1_mixing_D2}Sub-diagrams which give rise to $S\left(R\right)$
and $\textrm{Tr}\left[S\left(r\right)m^{2}\right]$ factors in the
RGEs.}
\end{figure}

\begin{figure}[h]
\begin{centering}
\includegraphics[scale=0.9]{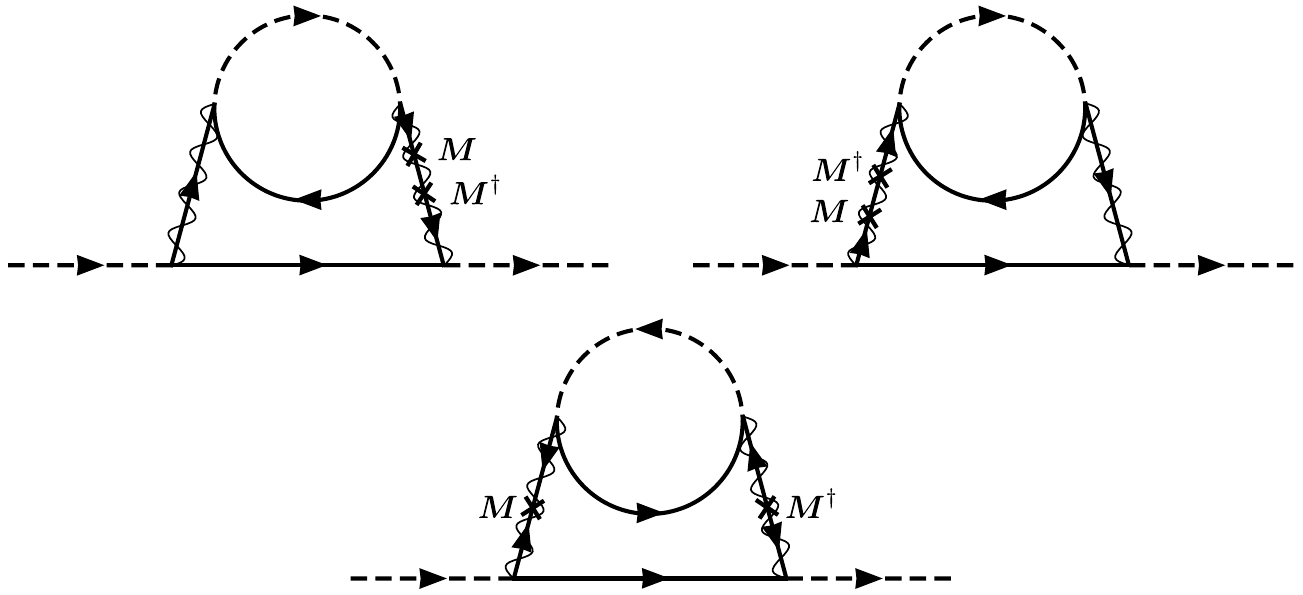}
\par\end{centering}

\caption{\label{fig:U1_mixing_D3}The three contribution from $U(1)$ groups
to the term $24g^{4}MM^{*}C\left(r\right)S\left(R\right)$.}
\end{figure}

This requirement that expressions with $\boldsymbol{V}$'s and $\boldsymbol{M}$'s
must form scalars is enough to derive equations \eqref{eq:U1_mixing_rule34}--\eqref{eq:U1_mixing_rule17},
\eqref{eq:U1_mixing_rule20}--\eqref{eq:U1_mixing_rule27} and \eqref{eq:U1_mixing_rule30}
from the existing substitution rules for gauge groups with multiple
factors. We are left with the terms $g^{4}C\left(r\right)S\left(R\right)$,
$Mg^{4}C\left(r\right)S\left(R\right)$, $g^{4}C\left(i\right)\textrm{Tr}\left[S\left(r\right)m^{2}\right]$
and $24g^{4}MM^{*}C\left(i\right)S\left(R\right)$. Note that one
can write $S\left(R\right)$ as $\textrm{Tr}\left[S\left(r\right)\right]$
in the notation of reference \cite{Martin:1993zk}, so in all four
cases there is a sum over field components of chiral superfields.
For diagrams with up to two-loops and with no external gauginos nor
gauge bosons, the factors $S\left(R\right)$ and $\textrm{Tr}\left[S\left(r\right)m^{2}\right]$
can only come from the sub-diagrams in figure \eqref{fig:U1_mixing_D2}.%
\footnote{It is conceivable that they could come also from diagrams with one
$\phi^{*}\phi^{*}\phi\phi$ vertex, but we may choose an appropriate
gauge, the Landau gauge, where these vanish because an external scalar
line always couples to a gauge bosons at a three point vertex.%
}
Take for example $g^{4}C\left(r\right)S\left(R\right)\sim\sum_{p}g^{4}y_{r}^{2}y_{p}^{2}$.
One cannot immediately generalize this expression to include $U(1)$-mixing
effects because in theory it could take the form $\sum_{p}\left(\boldsymbol{V}_{\boldsymbol{r}}^{T}\boldsymbol{V_{p}}\right)\left(\boldsymbol{V}_{\boldsymbol{r}}^{T}\boldsymbol{V_{p}}\right)$
or $\sum_{p}\left(\boldsymbol{V}_{\boldsymbol{r}}^{T}\boldsymbol{V_{r}}\right)\left(\boldsymbol{V}_{\boldsymbol{p}}^{T}\boldsymbol{V_{p}}\right)$.
But looking at the diagrams in figure \eqref{fig:U1_mixing_D2}, such
ambiguities disappear because in all cases the $\boldsymbol{V}$'s
which are summed over (the $\boldsymbol{V_{p}}$'s) do not contract
with each other; they contract with something else at the other end
of the gauge boson/gaugino lines. As such, there are no $\boldsymbol{V}_{\boldsymbol{p}}^{T}\boldsymbol{V_{p}}$'s
in these expressions; with this piece of information, combined with
the known rules for a gauge group with multiple factors, equations
\eqref{eq:U1_mixing_rule18}, \eqref{eq:U1_mixing_rule19} and \eqref{eq:U1_mixing_rule28}
follow. The substitution rule given in equation \eqref{eq:U1_mixing_rule29}
for $24g^{4}MM^{*}C\left(r\right)S\left(R\right)$ appearing in the
two-loop equation of the soft scalar masses is more complicated since
the placement of the $\boldsymbol{M}$, $\boldsymbol{M}^{\dagger}$
gaugino mass matrices between these $\boldsymbol{V}$'s is relevant.
Nevertheless, from the diagrams in figure \eqref{fig:U1_mixing_D3}
we can infer that the $U(1)$'s contribution to this term is $8\sum_{p}\left[\left(\boldsymbol{V}_{\boldsymbol{i}}^{T}\boldsymbol{MV_{p}}\right)\left(\boldsymbol{V}_{\boldsymbol{i}}^{T}\boldsymbol{M}^{\dagger}\boldsymbol{V_{p}}\right)+\left(\boldsymbol{V}_{\boldsymbol{i}}^{T}\boldsymbol{M}\boldsymbol{M}^{\dagger}\boldsymbol{V_{p}}\right)\left(\boldsymbol{V}_{\boldsymbol{i}}^{T}\boldsymbol{V_{p}}\right)\right.$
$\left.+\left(\boldsymbol{V}_{\boldsymbol{i}}^{T}\boldsymbol{M}^{\dagger}\boldsymbol{MV_{p}}\right)\left(\boldsymbol{V}_{\boldsymbol{i}}^{T}\boldsymbol{V_{p}}\right)\right]$.

\subsection{RGEs with $U(1)$ indices}

\begin{figure}[h]
\begin{centering}
\includegraphics[scale=0.81]{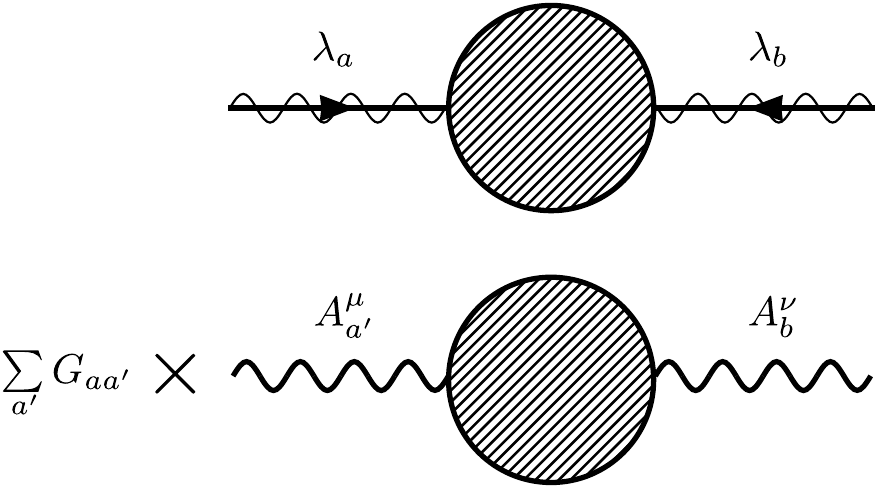}
\par\end{centering}

\caption{\label{fig:U1_mixing_D4}Diagrams from which the $\beta$ functions
of $U(1)$ gaugino masses and gauge couplings are calculated. Notice
the isolated $\boldsymbol{G}$ which appears in the RGEs of $\boldsymbol{G}$
itself.}
\end{figure}

The RGEs for $\boldsymbol{G}$ and $\boldsymbol{M}$ are the only
ones with free $U(1)$ indices. For the $\beta$ functions of the
gaugino masses, we will be interested in diagrams with two incoming
gauginos. As for the coupling constant, due to the Ward identities,
the contributing diagrams are those with two external gauge bosons.
From the amplitude of these diagrams we still have to add a $\boldsymbol{G}$
factor in order to obtain $\beta_{\boldsymbol{G}}$ (see figure \eqref{fig:U1_mixing_D4}).
Note that all the terms in $\beta_{\boldsymbol{G}}$ must be of the
form $\boldsymbol{G}\boldsymbol{V}_{\boldsymbol{i}}^{T}\left(\cdots\right)\boldsymbol{V_{j}}$
for some $i,\, j$ as mentioned before, and also
\begin{enumerate}
\item the RGEs are invariant under the set of transformations $\boldsymbol{G}\rightarrow\mathcal{O}_{1}\boldsymbol{G}\mathcal{O}_{2}^{T}$,
$\boldsymbol{V_{i}}\rightarrow\mathcal{O}_{2}\boldsymbol{V_{i}}$,
$\boldsymbol{M}\rightarrow\mathcal{O}_{2}\boldsymbol{M}\mathcal{O}_{2}^{T}$
for any orthogonal matrices $\mathcal{O}_{1}$, $\mathcal{O}_{2}$; 
\item $\boldsymbol{M}$ is a symmetric matrix, therefore $\frac{d\boldsymbol{M}}{dt}$
must be so as well. 
\end{enumerate}
Taken together, these considerations allow us to deduce equations
\eqref{eq:U1_mixing_rule32}--\eqref{eq:U1_mixing_rule40} (equation
\eqref{eq:U1_mixing_rule31} is trivial).

We shall exemplify this for the case of $16g^{4}S\left(R\right)C\left(R\right)M$
which for multiple factor groups is replaced by $8\sum_{b}g_{a}^{2}g_{b}^{2}S_{a}\left(R\right)C_{b}\left(R\right)\left(M_{a}+M_{b}\right)$
in the RGEs of $M_{a}$. This is the same as $8\sum_{p,b}g_{a}^{2}g_{b}^{2}\frac{S_{a}\left(p\right)C_{b}\left(p\right)}{d_{a}\left(p\right)}\left(M_{a}+M_{b}\right)$.
Groups $a$ and $b$ are independent so the expressions $\sum_{b}g_{b}^{2}C_{b}\left(p\right)$,
$\sum_{b}M_{b}g_{b}^{2}C_{b}\left(p\right)$ are decoupled from $g_{a}^{2}\frac{S_{a}\left(p\right)}{d_{a}\left(p\right)}$,
$M_{a}g_{a}^{2}\frac{S_{a}\left(p\right)}{d_{a}\left(p\right)}$.
Inclusion of $U(1)$ mixing effects in the first pair of expressions
is easy because there are no free $U(1)$ indices: $\sum_{b}g_{b}^{2}C_{b}\left(p\right)\rightarrow\sum_{B}g_{B}^{2}C_{B}\left(p\right)+\boldsymbol{V}_{\boldsymbol{p}}^{T}\boldsymbol{V_{p}}$
and $\sum_{b}M_{b}g_{b}^{2}C_{b}\left(p\right)\rightarrow\sum_{B}M_{B}g_{B}^{2}C_{B}\left(p\right)+\boldsymbol{V}_{\boldsymbol{p}}^{T}\boldsymbol{MV_{p}}$.
If the group $a$ is a $U(1)$, then in the single $U(1)$ case this
corresponds to $g_{a}^{2}\frac{S_{a}\left(p\right)}{d_{a}\left(p\right)}=g^{2}y_{p}^{2}$,
which generalizes to $g_{a}^{2}\frac{S_{a}\left(p\right)}{d_{a}\left(p\right)}\rightarrow\boldsymbol{V_{p}}\boldsymbol{V}_{\boldsymbol{p}}^{T}$.
Similarly, the only symmetric matrix expression which respects the
$\mathcal{O}_{2}$ symmetry that can generalize $M_{a}g_{a}^{2}\frac{S_{a}\left(p\right)}{d_{a}\left(p\right)}$
is $\frac{1}{2}(\boldsymbol{MV_{p}}\boldsymbol{V}_{\boldsymbol{p}}^{T}+\boldsymbol{V_{p}}\boldsymbol{V}_{\boldsymbol{p}}^{T}\boldsymbol{M})$.

\begin{figure}[h]
\begin{centering}
\includegraphics[scale=0.7]{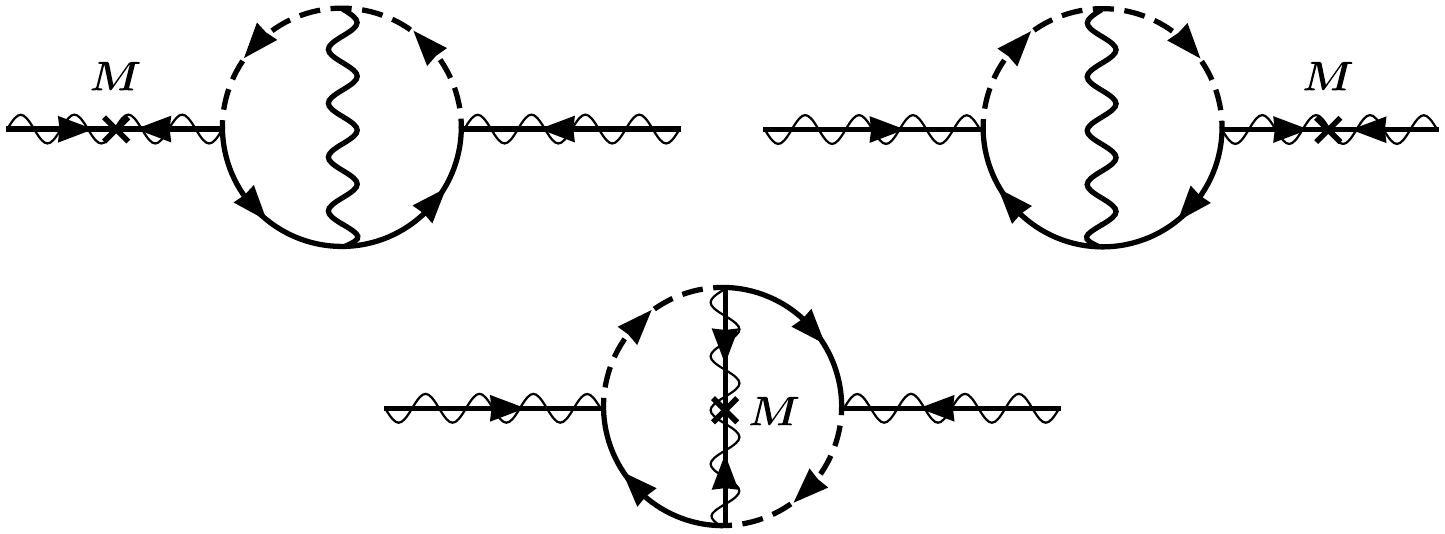}
\par\end{centering}

\caption{\label{fig:U1_mixing_D5}Diagrams contributing to the term $16g^{4}S\left(R\right)C\left(R\right)M$
in the RGEs.}
\end{figure}

Assembling these pieces gives equation \eqref{eq:U1_mixing_rule37}
for $16g^{4}S\left(R\right)C\left(R\right)M$. The structure of the
final expression is verifiable by considering the relevant diagrams
(figure \eqref{fig:U1_mixing_D5}).

\section{\label{sect:numerics}Comparison with other methods of including
$U(1)$-mixing effects}

\subsection{\label{sect:general}General discussion}

So far, several approaches to the SUSY $U(1)$-mixing problem have
been proposed in the literature. In the following, we discuss some
of them and comment on their limitations when compared to the complete
two-loop treatment presented in this chapter.

As we have already mentioned, one can attempt to choose a convenient
pair of bases in the $U(1)$-charge and gauge field spaces for which
the situation might simplify \cite{delAguila:1988jz,Martin:1993zk}.
For instance, it is always possible to diagonalize the one-loop anomalous
dimensions
\begin{equation}
\boldsymbol{\gamma}=\sum_{i}\boldsymbol{Q}_{i}\boldsymbol{Q}_{i}^{T}\label{eq:U1mixing_gamma}
\end{equation}
by means of a suitable $\mathcal{O}_{1}$ rotation $\boldsymbol{Q}_{i}\to\mathcal{O}_{1}\boldsymbol{Q}_{i}\equiv\boldsymbol{Q}'_{i}$
(cf. equation \eqref{eq:U1_mixing_symmetries1}), so that $\boldsymbol{\gamma'}=\mathcal{O}_{1}\boldsymbol{\gamma}\mathcal{O}_{1}^{T}$
is diagonal. This changes the gauge coupling matrix as well: $\boldsymbol{G}\to\mathcal{O}_{1}\boldsymbol{G}$.
However, if all the relevant $U(1)$ gauge couplings arise at a single
scale, or in other words $\boldsymbol{G}\propto\mathbb{1}$, this
$\mathcal{O}_{1}$ matrix commutes with $\boldsymbol{G}$ and can
be absorbed by a suitable redefinition of the gauge fields (equation
\eqref{eq:U1_mixing_symmetries_O2}) where now $\mathcal{O}_{2}=\mathcal{O}_{1}$.
In this way, the one-loop evolution of $\boldsymbol{G}$ is driven
by a diagonal $\boldsymbol{\gamma'}$ and the initial condition $\boldsymbol{G}\propto\mathbb{1}$
remains intact. Thus, no off-diagonalities emerge in this case and
it is consistent to work with the usual RGEs for individual gauge
couplings, one for each $U(1)$ factor.

This approach, however, is generally limited to situations where there
is a complete $U(1)$ unification. This is very often not the case
in practice, in particular for GUTs in which the hypercharge is a
non-trivial linear combination of diagonal generators of the higher
energy gauge group, such as in left-right models based on $SU(3)_{c}\times SU(2)_{L}\times SU(2)_{R}\times U(1)_{B-L}$
(see the next section and also chapter \ref{chap:SlidingScale_models}).
Moreover, we should take into consideration the $U(1)$ gaugino soft
masses, which should also be universal at the unification scale; otherwise
the method fails in the soft sector already at the one-loop level.
The crucial point is that only then the generalized one-loop relation
\begin{equation}
\boldsymbol{G}\boldsymbol{M}^{-1}\boldsymbol{G}^{T}=\text{constant}\label{generalizedgauginogaugecorrelation}
\end{equation}
between the gauge couplings and the gaugino masses ensures the gaugino
mass diagonality along the unification trajectory.

At the two-loop level more complicated structures such as higher powers
of charges, gauge couplings and Yukawa couplings enter the anomalous
dimensions and therefore, in general, there is no way to diagonalize
simultaneously all the evolution equations. Though there is still
a technique one can implement in the gauge sector if the $U(1)$ couplings
do not unify \cite{delAguila:1988jz}, in the supersymmetric case
there is no general way out for the gauginos, as also discussed in
\cite{Braam:2011xh}. Thus, a full-fledged two-loop approach as presented
in this work is necessary and, in fact, it turns out to be even technically
indispensable if there happen to be more than two abelian gauge groups
as, for instance, in \cite{delAguila:1987st}, \cite{Perez:2011dg}
and many string-inspired constructions.

\subsection{\label{sect:numericsMRV}Quantifying $U(1)$-mixing effects with
simple examples }

In this section, we shall see through some examples the importance
of the kinetic mixing effects in simple scenarios which exhibit all
the prominent features discussed above.

\subsubsection{One-loop effects}

\paragraph{Gauge coupling constants:}

We shall consider the one-loop evolution of the gauge couplings in
the SUSY $SO(10)$ model of \cite{Malinsky:2005bi}, in which the
unified gauge symmetry is broken down to the MSSM in three steps,
namely, $SO(10)\to SU(3)_{c}\times SU(2)_{L}\times SU(2)_{R}\times U(1)_{B-L}\to SU(3)_{c}\times SU(2)_{L}\times U(1)_{R}\times U(1)_{B-L}\to{\rm MSSM}$;
the corresponding breaking scales shall be denoted by $m_{G}$, $m_{R}$
and $m_{B-L}$, respectively. Further details including the field
contents at each of the symmetry breaking stages can be found in \cite{Malinsky:2005bi}
(see also {}``class-III models'' in chapter \ref{chap:SlidingScale_models}).

For our purpose, it is crucial that in this model the ratio $\nicefrac{m_{R}}{m_{B-L}}$
can be as large as $10^{10}$ and, hence, the $U(1)$-mixing effects
become important. Note that even a short $SU(3)_{c}\times SU(2)_{L}\times SU(2)_{R}\times U(1)_{B-L}$
stage is sufficient to split the $g_{R}$ and the $g_{B-L}$ gauge
couplings such that the extended gauge-coupling matrix $\boldsymbol{G}$
at the $m_{R}$ scale is somewhat far from being proportional to the
unit matrix. Thus, there is no way to choose the $\mathcal{O}_{1}$
and $\mathcal{O}_{2}$ rotation matrices such that both $\boldsymbol{G}$
and
\begin{equation}
\boldsymbol{\gamma}=\boldsymbol{N}\left(\begin{array}{cc}
\frac{15}{2} & -1\\
-1 & 18
\end{array}\right)\boldsymbol{N}\label{eq:U1mixing_Nmatrix}
\end{equation}
are simultaneously diagonalized. Here $\boldsymbol{N}=\textrm{diag}\left(1,\sqrt{\nicefrac{3}{8}}\right)$
ensures the canonical normalization of the $B-L$ charge within the
$SO(10)$ framework. Therefore, the one-loop evolution equation relevant
to the $U(1)_{R}\times U(1)_{B-L}$ stage has to be matrix-like. In
the abelian sector it reads
\begin{equation}
\frac{d}{dt}\boldsymbol{A}^{-1}=-\boldsymbol{\gamma}\,,\label{eq:U1mixing_Amatrix}
\end{equation}
where $\boldsymbol{A}^{-1}=4\pi(\boldsymbol{G}\boldsymbol{G}^{T})^{-1}$
and $t=\nicefrac{\log\left(\nicefrac{E}{E_{0}}\right)}{2\pi}$.

The reason why the $U(1)_{R}\times U(1)_{B-L}$ phase can be spread
over a broad energy range has to do with the fact that this gauge
symmetry is broken by neutral components of an $SU(2)_{R}$ doublet
pair, namely, $(1,1,+\tfrac{1}{2},-1)+(1,1,-\tfrac{1}{2},+1)=\chi_{R}^{0}+\overline{\chi}_{R}^{0}$
which are SM singlets and, as such, they do not affect the low-energy
value of $\alpha_{Y}^{-1}$. Indeed, the would-be change inflicted
on $\alpha_{Y}^{-1}$ by the presence or absence of $\chi_{R}^{0}+\overline{\chi}_{R}^{0}$
is given by
\begin{equation}
\Delta\alpha_{Y}^{-1}=p_{Y}^{T}\cdot\Delta\boldsymbol{A}^{-1}\left(m_{B-L}\right)\cdot p_{Y}\propto p_{Y}^{T}\cdot\Delta\boldsymbol{\gamma}\cdot p_{Y}=0\,,\label{Deltaalphainverse}
\end{equation}
where $p_{Y}^{T}=\left(\sqrt{\nicefrac{3}{5}},\sqrt{\nicefrac{2}{5}}\right)$
is the vector describing the combination of $U(1)_{R}\times U(1)_{B-L}$
charges which constitutes the MSSM hypercharge, and $\Delta\boldsymbol{\gamma}$
denotes the relevant change of the $\boldsymbol{\gamma}$ matrix.
Therefore, at the one-loop level, the $m_{B-L}$ scale is not constrained
by the low-energy data and hence, barring other phenomenological constraints,
it can be pushed as close to the MSSM scale ($m_{SUSY}$) as desired.
\begin{figure}[t]
\centering{} \includegraphics[width=8cm]{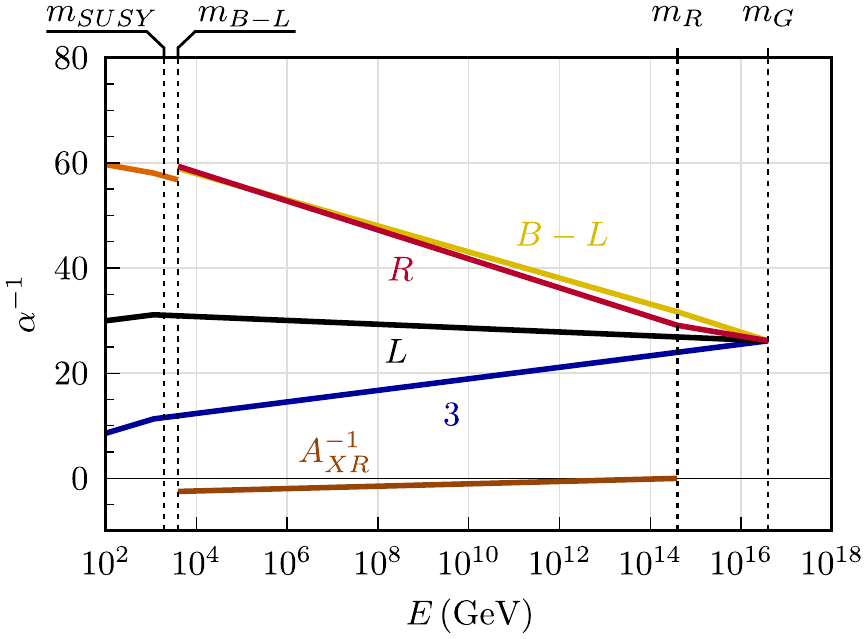} \caption{\label{MRVrunningOK}One-loop gauge-coupling evolution in the MRV
model \cite{Malinsky:2005bi}. The position of the GUT scale, the
unified gauge coupling and the intermediate symmetry-breaking scale
$m_{R}$ were chosen in such a way as to fit the electroweak data
with $\alpha_{Y}^{-1}(m_{Z})=59.73$. The close-to-zero brown line
in the $\left[m_{B-L},m_{R}\right]$ energy range depicts the evolution
of the off-diagonal entries of the $\boldsymbol{A}^{-1}=4\pi\left(\boldsymbol{G}\boldsymbol{G}^{T}\right)^{-1}$
matrix which, at the one-loop level, scales linearly with $\log E$.
Likewise, in this energy range the red and yellow lines are the $\left(1,1\right)$
and $\left(2,2\right)$ diagonal entries of this matrix. The apparent
discontinuity in $\alpha_{Y}^{-1}$ at the $m_{B-L}$ scale is due
to the generalized matching condition (see appendix \ref{chap:Matching_conditions}
for details).}
\end{figure}

\begin{figure}[t]
\centering{} \includegraphics[width=8cm]{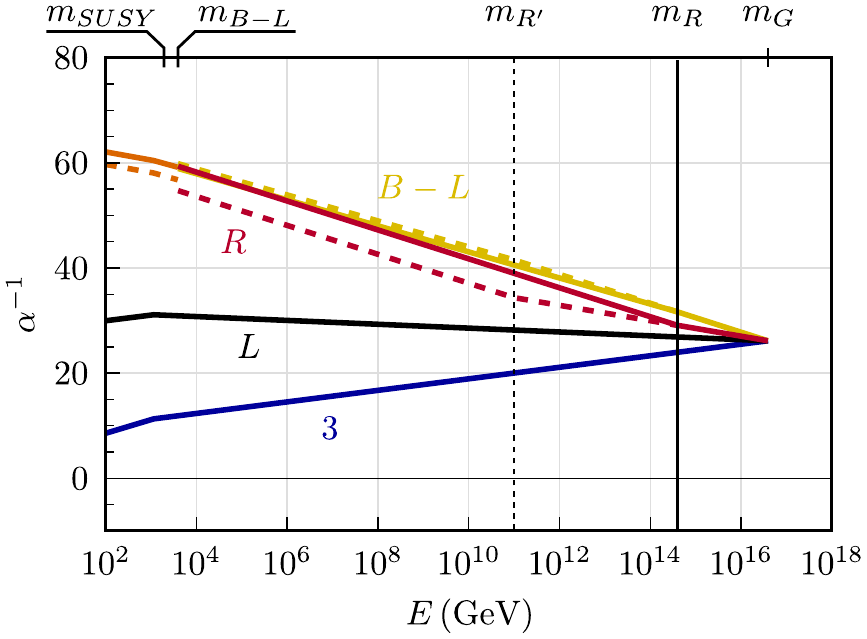} \caption{\label{MRVrunningWRONG}The same as in figure \eqref{MRVrunningOK}
but without the kinetic mixing effects taken into account. With the
same GUT-scale boundary condition and $m_{R}$, the low-energy value
of $\alpha_{Y}^{-1}$ is $62.51$ (orange solid line) and it differs
from the one obtained in the full calculation by as much as 4\%. Alternatively,
if one attempts to obtain the right value of $\alpha_{Y}^{-1}\left(m_{Z}\right)$
by adjusting the $SU(2)_{R}$ breaking scale, the new $m'_{R}$ scale
must be shifted with respect to the correct $m_{R}$ by as much as
4 orders of magnitude (vertical solid and dashed lines).}
\end{figure}

However, this simple argument only works if the $U(1)$-mixing effects
are properly taken into account. Remarkably, if they are neglected,
$\Delta\boldsymbol{\gamma}$ contains only diagonal non-null entries
and $\alpha_{Y}^{-1}\left(m_{Z}\right)$ becomes a function of $m_{B-L}$.
Moreover, by stretching the $m_{B-L}$--$m_{R}$ range to the maximum,
the value of $\alpha_{Y}^{-1}\left(m_{Z}\right)$ can be incorrectly
shifted by as much as 4\%, as can be seen by comparing figures \eqref{MRVrunningOK}
and \eqref{MRVrunningWRONG}. Alternatively, in order to retain the
desired value of $\alpha_{Y}^{-1}\left(m_{Z}\right)$, one would have
to re-adjust $m_{R}$ by several orders of magnitude. However, this
could have a large impact on the MSSM soft spectrum \cite{DeRomeri:2011ie,Arbelaez:2013hr},
and, in more general constructions, also on $m_{G}$ and $\alpha_{G}$,
with potential consequences for $d=6$ proton decay.

Finally, let us note that the \textit{rotated-basis} method discussed
previously is only partially successful because the $g_{R}$ and $g_{B-L}$
gauge couplings do not coincide at the $m_{R}$ scale. In fact, the
value of $\alpha_{Y}^{-1}(m_{Z})$ obtained in this way is $60.93$,
which is $\sim2\%$ off the correct value, but still this number is
closer to the correct value than the one obtained by considering no
mixing at all.

\paragraph{Gaugino masses:}

The full impact of the method presented in this chapter can be understood
by considering the interplay between the gauge and the soft sectors.
At one loop-level, we can use equation \eqref{generalizedgauginogaugecorrelation}
which relates the gauge couplings $\boldsymbol{G}$ with the gaugino
soft masses $\boldsymbol{M}$ in an invariant combination. As a consequence,
at low energies the bino mass is given by 
\begin{equation}
M_{Y}\left(m_{SUSY}\right)=\frac{\alpha_{Y}\left(m_{SUSY}\right)}{\alpha_{G}}p_{Y}^{T}\boldsymbol{M_{1/2}}p_{Y}\,,\label{binomass}
\end{equation}
where $\boldsymbol{M_{1/2}}$ is the GUT-scale gaugino soft mass matrix.
From equation \eqref{binomass} we see that the ratio $M_{Y}\left(m_{SUSY}\right)/\alpha_{Y}\left(m_{SUSY}\right)$
depends on whether the mixing effects are included or not, as was
already noticed in \cite{Kribs:1998rb}. Note that if $\boldsymbol{M_{1/2}}$
is not proportional to the unit matrix at the GUT scale, the $p_{Y}^{T}\boldsymbol{M_{1/2}}p_{Y}$
term will mix all entries of $\boldsymbol{M_{1/2}}$. Moreover, in
the special case in which the abelian gauge couplings unify, even
the one-loop gaugino sector evolution can be fully accounted for by
the \textit{rotated-basis} technique.

\subsubsection{Two-loop effects}

At the two-loop level this method becomes important in cases with
gauge coupling unification at a certain scale. We illustrate this
by taking as an example the model presented in reference \cite{FileviezPerez:2010ek}
where an intermediate $SU(3)_{c}\times SU(2)_{L}\times U(1)_{Y}\times U(1)_{B-L}$
gauge symmetry is assumed to originate from a grand unified framework.
We consider two cases: (i) full gauge coupling unification at $2\times10^{16}$GeV
and (ii) a small difference of 5\% between the two $U(1)$ couplings
caused by possible GUT-scale threshold effects. In the gaugino sector
we assume universal boundary conditions in both cases, but the effect
gets even stronger if in addition one considers threshold effects
in the gaugino sector as well.

The results are given in table \eqref{tab:comparison_running}. Interestingly,
besides the expected equivalence of the \textit{rotated-basis} method
and the full-fledged calculation at the one-loop level, the relevant
effective hypercharge gauge coupling turns out to be identical to
the one obtained even at two loop-level if exact gauge coupling unification
is assumed. The reason is that all additional states not present in
the MSSM are charged only with respect to $U(1)_{B-L}$ and are neutral
under the MSSM gauge group. In the gaugino sector, the first deviations
show up already in this case, even though they are only of the per-mile
order. If one includes also threshold corrections at the GUT-scale
the effects are at the percent level leading to shifts in sparticle
masses potentially measurable at the LHC.

Lastly, it should be kept in mind that the effects would be even larger
if the $U(1)_{Y}$ would result from the breaking of $U(1)_{R}\times U(1)_{B-L}$
as discussed in the previous example.

\begin{table*}[htb]
\scalebox{0.81}{
\setlength{\tabcolsep}{3pt}

\begin{tabular}{ccccccccc}
\toprule 
 & \multicolumn{3}{c}{One-loop results} &  & \multicolumn{4}{c}{Two-loop results}\tabularnewline
\cmidrule{2-4} \cmidrule{6-9} 
 &  %
\begin{tabular}{@{}r@{}}
No kinetic\tabularnewline
mixing~~~\tabularnewline
\end{tabular} & %
\begin{tabular}{@{}r@{}}
Rotated basis\tabularnewline
method~~~~~\tabularnewline
\end{tabular}  & %
\begin{tabular}{@{}r@{}}
Complete\tabularnewline
RGEs~~~\tabularnewline
\end{tabular}  &  & %
\begin{tabular}{@{}r@{}}
No kinetic\tabularnewline
mixing~~~\tabularnewline
\end{tabular}  & %
\begin{tabular}{@{}r@{}}
Complete\tabularnewline
RGEs~~~\tabularnewline
\end{tabular} & %
\begin{tabular}{@{}r@{}}
No kinetic\tabularnewline
mixing~~~\tabularnewline
\end{tabular} & %
\begin{tabular}{@{}r@{}}
Complete\tabularnewline
RGEs~~~\tabularnewline
\end{tabular}\tabularnewline
\midrule
\ensuremath{g_{YY}}
  & 0.4511  & 0.4700  & 0.4700  &  & 0.4487  & 0.4677 & 0.4487  & 0.4686\tabularnewline
\ensuremath{g_{BLBL}}
  & 0.4083  & 0.4243  & 0.4243  &  & 0.4070  & 0.4231 & 0.4131  & 0.4298\tabularnewline
\ensuremath{g_{BLY},g_{YBL}}
  & 0.0  & -0.0723  & -0.0723  &  & 0.0  & -0.0725 & 0.0  & -0.0725\tabularnewline
$g_{Y}$  & 0.4511  & 0.4511  & 0.4511  &  & 0.4487  & 0.4487 & 0.4487  & 0.4500\tabularnewline
\ensuremath{M_{YY}}
  & 196.34  & 218.13  & 218.13  &  & 185.82  & 207.96 & 185.80  & 208.71 \tabularnewline
\ensuremath{M_{BLBL}}
  & 160.83  & 178.67  & 178.67  &  & 154.88  & 173.19 & 144.26  & 161.97\tabularnewline
\ensuremath{M_{BLY},M_{YBL}}
  & 0.0  & - 62.39  & - 62.39  &  & 0.0  & -63.10  & 0.0  & -62.15\tabularnewline
$M_{Y}$ & 196.34  & 196.34  & 196.34  &  & 185.82  & 185.96 & 185.80  & 187.04 \tabularnewline
\midrule
 & \multicolumn{6}{c}{Exact unification} & \multicolumn{2}{c}{$g_{BL}^{{\rm GUT}}=1.05\, g_{Y}^{{\rm GUT}}$}\tabularnewline
\bottomrule
\end{tabular}

}
\setlength{\tabcolsep}{6pt}\caption{\label{tab:comparison_running}Low energy values of the entries of
the gauge coupling and gaugino mass matrices ($g_{ab}$ and $M_{ab}$
with $a,b=Y,\, BL$) and the correctly fitted MSSM parameters ($g_{Y}$,
$M_{Y}$)---see equations \eqref{eq:gauge_matching_condition_general}
and \eqref{eq:gaugino_matching_condition_general}. All gaugino mass
parameters are in GeV. We have set the GUT scale at $2\times10^{16}$
with $g_{G}=0.72$ and imposed an mSUGRA boundary condition taking
$\boldsymbol{M_{1/2}}=\mathbb{1}\times500$ GeV. At the one-loop level,
we compare the case with no kinetic mixing effects included, the \textit{rotated
basis} and the full-fledged calculation. At the two-loop level, we
include the case where $g_{Y}$ and $g_{BL}$ are split at the GUT
scale due to threshold corrections.}
\end{table*}

\section{Conclusions and outlook}

In this chapter we have discussed the structure of the renormalization
group equations in softly broken supersymmetric models with more than
a single abelian gauge factor group. In such models there are $U(1)$-mixing
effects which must be taken into consideration, as explained in section
\ref{sec:U1_mixing_Introduction}.

Although the evolution equations available in the literature do not
formally exhibit any obvious pathologies if such subtleties are not
taken into account, the calculations based on these formulas are in
general incomplete and thus, the results are internally inconsistent.
This is even more pronounced in the context of SUSY models because
it affects also the evolution of the soft SUSY parameters, in particular
the evolution of the gaugino mass parameters.

Remarkably enough, the issue of $U(1)$-mixing in softly broken supersymmetric
gauge theories has never been addressed in full generality, even at
one loop. The main aim of reference \cite{Fonseca:2011vn}, which
is reprinted here, was to fill this gap and provide a fully self-consistent
method of dealing with the renormalization group evolution of all
the parameters in such models, up to two loops. To this end, the existing
two-loop renormalization group equations valid for models with at
most a single abelian gauge factor were extended.

In particular, we have argued that all the $U(1)$-mixing effects
can be consistently included if the gauge couplings and the soft SUSY-breaking
gaugino masses associated to the individual abelian gauge-group factors
are generalized to matrices and these are then substituted into the
formulae in \cite{Martin:1993zk,Yamada:1994id} in a specific manner.
However, this is a non-trivial task because the new matrix-like structures
do not commute, and as a consequence the generalization of some expressions
can be ambiguous. In this respect, the residual reparametrization
invariance of the covariant derivative associated to the redefinition
of the abelian gauge fields turned out to be a very useful tool, yet
in many cases one had to resort to a detailed analysis of the relevant
Feynman diagrams.

The general method has been illustrated for two cases: at the one
loop level, for a model with different gauge coupling strengths due
to a breaking of the original simple group in two steps; and at the
two loop level, in a model where gauge coupling unification occurs
in a single step, but only with threshold corrections taken into consideration.
In both case we obtain effects in the percent range which none of
the previously proposed partial treatments can fully account for.

Lastly, let us stress again that our results are generic and, as such,
they do not require any specific assumptions about the charges of
the chiral multiplets in the theory and/or the boundary conditions
applied to the relevant gauge couplings. This makes the framework
suitable for implementation into computer algebraic codes calculating
two-loop renormalization group equations in softly-broken supersymmetric
gauge theories such as \texttt{SARAH} \cite{Staub:2008uz,Staub:2009bi,Staub:2010jh,Staub:2013tta}
and \texttt{Susyno} \cite{Fonseca:2011sy} (see chapter \ref{chap:Susyno}).
\cleartooddpage

\chapter{\label{chap:SlidingScale_models}Supersymmetric $SO(10)$-inspired
GUTs with sliding scales}

\section{Introduction}

In the MSSM gauge couplings unify at an energy scale of about $m_{G}\approx2\times10^{16}$
GeV (see chapter \ref{chap:The-SM's-shortcomings}). Arbitrarily adding
particles to the MSSM easily destroys this attractive feature. Thus,
relatively few SUSY models have been discussed in the literature which
have a particle content larger than MSSM at experimentally accessible
energies. However, neutrino oscillation experiments \cite{Fukuda:1998mi,Ahmad:2002jz,Eguchi:2002dm}
have shown that at least one neutrino must have a mass bigger than
$0.05$ eV (confer with $\Delta m^{2}$ on table \eqref{tab:Neutrino_parameter_values}).
A Majorana neutrino mass of this order hints at the existence of a
new energy scale below $m_{G}$. For models with renormalizable interactions
and perturbative couplings, as for example in the classical seesaw
models \cite{Minkowski:1977sc,*Yanagida:1979as,GellMann:1980vs,Mohapatra:1979ia},
this new scale should lie below $10^{15}$ GeV, approximately.

From the theoretical point of view, GUT models based on the group
$SO(10)$ \cite{Fritzsch:1974nn} offer a number of advantages compared
to the simpler models based on $SU(5)$. For example, several of the
chains through which $SO(10)$ can be broken to the SM gauge group
contain the left-right symmetric group $SU(3)_{c}\times SU(2)_{L}\times SU(2)_{R}\times U(1)_{B-L}$
as an intermediate step \cite{Mohapatra:1986uf} (see also chapter
\ref{chap:Symmetry}), thus potentially explaining the observed left-handedness
of the weak interactions. However, probably the most interesting aspect
of $SO(10)$ is that it automatically contains the necessary ingredients
to generate a seesaw mechanism \cite{Mohapatra:1979ia}: (i) the right-handed
neutrino is included in the ${\bf 16}$ which forms a fermion family;
and (ii) $(B-L)$ is one of the generators of $SO(10)$.

Left-right (LR) symmetric models usually break the LR symmetry at
a rather large energy scale, $m_{R}$. For example, if LR is broken
in the SUSY LR model by the VEV of $(B-L)=2$ triplets \cite{Cvetic:1983su,Kuchimanchi:1993jg}
or by a combination of $(B-L)=2$ and $(B-L)=0$ triplets \cite{Aulakh:1997ba,Aulakh:1997fq},
$m_{R}\approx10^{15}$ GeV is the typical scale consistent with gauge
coupling unification (GCU). The authors of \cite{Majee:2007uv} find
a lower limit of $m_{R}\gtrsim10^{9}$ GeV from GCU for models where
the LR symmetry is broken by triplets, even if one allows additional
non-renormalizable operators or sizable GUT-scale thresholds to be
present. On the other hand, in models with an extended gauge group
it is possible to formulate sets of conditions on the $\beta$-coefficients
for the gauge couplings, which allow to enforce GCU independently
of the energy scale at which the extended gauge group is broken. This
was called the \textit{sliding mechanism} in \cite{DeRomeri:2011ie}.%
\footnote{A different (but related) approach to enforcing GCU is taken by the
authors of \cite{Calibbi:2009cp} with what they call \textit{magic
fields}.%
} However, reference \cite{DeRomeri:2011ie} was not the first to present
examples of sliding scale models in the literature. In \cite{Malinsky:2005bi}
it was shown that, if the left-right group is broken to $SU(2)_{L}\times U(1)_{R}\times U(1)_{B-L}$
by the VEV of a scalar field $\Phi_{1,1,3,0}$ then%
\footnote{The indices denote the transformation properties under the LR group,
see next section and appendix \ref{chap:Lists_of_superfields_in_LR_models}
for notation.%
} the resulting $U(1)_{R}\times U(1)_{B-L}$ can be broken to the SM
$U(1)_{Y}$, in agreement with experimental data at any energy scale.
In \cite{Majee:2007uv} the authors demonstrated that in fact a complete
LR group can be lowered to the TeV-scale, if certain carefully chosen
fields are added and the LR-symmetry is broken by right doublets.
A particularly simple model of this kind was discussed in \cite{Dev:2009aw}.
Finally, the authors of \cite{DeRomeri:2011ie} also discussed an
alternative way of constructing a sliding LR scale by relating it
to an intermediate Pati-Salam stage. We note in passing that these
papers are not in contradiction with earlier works \cite{Cvetic:1983su,Kuchimanchi:1993jg,Aulakh:1997ba,Aulakh:1997fq},
all of which have a large $m_{R}$: as discussed briefly in the next
section, it is not possible to construct a sliding scale \textit{variant}
of an LR model including pairs of $\Phi_{1,1,3,-2}$ and $\Phi_{1,3,1,-2}$.

Three different constructions, based on different $SO(10)$ breaking
chains, were considered in \cite{DeRomeri:2011ie}:
\begin{itemize}
\item In chain-I $SO(10)$ is broken in exactly one intermediate (LR symmetric)
step to the Standard Model group;
\item In chain-II $SO(10)$ is broken first to the Pati-Salam group \cite{Pati:1974yyX}
and at lower energies to the LR group;
\item In chain-III there is a LR symmetric phase, which then breaks into
a phase where instead of the full $SU(2)_{R}$ group there is only
a $U(1)_{R}$ gauge symmetry.
\end{itemize}
In other words,\thinmuskip=3mu
\medmuskip=2mu
\thickmuskip=6mu
\begin{align}
 & \textrm{chain I:}\quad & SO(10) & \to SU(3)_{c}\times SU(2)_{L}\times SU(2)_{R}\times U(1)_{B-L}\to{\rm MSSM}\,,\label{eq:SlidingScale_chainI}\\
 & \textrm{chain II:}\quad & SO(10) & \to SU(4)\times SU(2)_{L}\times SU(2)_{R}\nonumber \\
 &  &  & \to SU(3)_{c}\times SU(2)_{L}\times SU(2)_{R}\times U(1)_{B-L}\to{\rm MSSM}\,,\label{eq:SlidingScale_chainII}\\
 & \textrm{chain III:}\quad & SO(10) & \to SU(3)_{c}\times SU(2)_{L}\times SU(2)_{R}\times U(1)_{B-L}\nonumber \\
 &  &  & \to SU(3)_{c}\times SU(2)_{L}\times U(1)_{R}\times U(1)_{B-L}\to{\rm MSSM}\,.\label{eq:SlidingScale_chainIII}
\end{align}
\thinmuskip=3mu
\medmuskip=4.0mu plus 2.0mu minus 4.0mu
\thickmuskip=5.0mu plus 5.0mu 

In all cases, the last symmetry breaking scale before reaching the
SM group can be as low as ${\cal O}(1\textrm{ TeV})$ maintaining
nevertheless GCU.%
\footnote{In fact, the \textit{sliding mechanism} would also work at even lower
energy scales. However, this possibility is excluded phenomenologically.%
} References \cite{Malinsky:2005bi,Majee:2007uv,Dev:2009aw,DeRomeri:2011ie}
mentioned above give at most one or two sample models for each chain;
in other words they present a {}``proof of principle'' that models
with the stipulated conditions can indeed be constructed in agreement
with experimental constraints. It is then perhaps natural to ask:
how unique are the models discussed in these papers? In reference
\cite{Arbelaez:2013hr} we set out to address this question, and the
discussion contained in this chapter (as well as in appendix \ref{chap:Lists_of_superfields_in_LR_models})
is taken from it.

Predictably perhaps, we found that there is a huge number of variants
in each class. Even in the simplest class, corresponding to the symmetry
breaking chain I, there is a total of 53 variants (up to 5324 configurations,
see next section) which can have perturbative GCU and a LR scale below
10 TeV, consistent with experimental data. For the two other classes,
chain-II and chain-III, we have found literally thousands of variants.

With such a huge number of variants corresponding essentially to equivalent
constructions, an immediate concern is whether there is any way of
experimentally distinguishing among all of these constructions. Tests
could be either direct or indirect. Direct tests are possible, because
of the sliding scale feature of the classes of models we discuss---see
section \ref{sec:SlidingScale_models}. Different variants predict
different additional (s)particles, some of which (being colored) could
give rise to spectacular resonances at the LHC. However, even if the
new gauge symmetry and all additional fields are outside the reach
of the LHC, all variants have different $\beta$-coefficients and
thus different running of MSSM parameters, both the gauge couplings
and the SUSY soft masses. Thus, if one assumes the validity of a certain
SUSY breaking scheme, such as for example mSUGRA, indirect traces
of the different variants remain in the SUSY spectrum, potentially
measurable at the LHC and a future linear collider (ILC/CLIC). This
was discussed earlier in the context of indirect tests for the SUSY
seesaw mechanism in \cite{Buckley:2006nv,Hirsch:2008gh,Esteves:2010ff}
and for extended gauge models in \cite{DeRomeri:2011ie}. We generalize
the discussion of \cite{DeRomeri:2011ie} and show how the \textit{invariants},
which are certain combinations of SUSY soft breaking parameters, can
themselves be organized into a few classes, which in principle allow
to distinguish class-II models from class-I or class-III and, if sufficient
precision could be reached experimentally, even select specific variants
within a class and provide indirect information about the new energy
scale(s).

In the rest of this chapter, we first lay out the general conditions
for the construction of the models we are interested in (next section),
and afterwards we discuss variants and examples of configurations
for all of the three classes we consider. Section \ref{sect:SlidingScale_invariants}
then addresses the \textit{invariants}, which are combinations of
SUSY soft parameters in the different model classes. A short summary
and discussion of the main results can also found at the end of this
chapter. Finally, appendix \ref{chap:Lists_of_superfields_in_LR_models}
contains the lists of chiral superfields considered, as well as a
quick discussion on the necessary ones to achieve a given symmetry
breaking sequence.

\section{\label{sec:SlidingScale_models}Models}

\subsection{\label{subsec:SlidingScale_so10models}Supersymmetric $SO(10)$ models:
General considerations}

Before entering into the details of the different model classes, we
first list some general requirements which we use in all constructions.
These requirements are the basic conditions any \textit{proto-model},
as we shall call them for now, has to fulfill to guarantee that a
phenomenologically realistic model based on it exists. We use the
following conditions:
\begin{itemize}
\item \textbf{Perturbative $\boldsymbol{SO(10)}$ unification}: gauge couplings
unify (at least) as successfully as in the MSSM and the value of $\alpha_{G}$
is in the perturbative regime.
\item \textbf{The GUT scale should lie above $\boldsymbol{\sim10^{16}}$
GeV}: this bound is motivated by the limit on the proton decay half-life. 
\item \textbf{Sliding mechanism}: this requirement translates into a set
of conditions (different conditions for the different classes of models)
on the allowed $\beta$-coefficients of the gauge couplings, which
ensures that the additional gauge group structure can be broken at
any energy scale, while still achieving GCU. 
\item \textbf{Renormalizable symmetry breaking}: at each intermediate step
we assume that there is a minimal number of scalar fields needed for
symmetry breaking. 
\item \textbf{Fermion masses and in particular neutrino masses}: the field
content of the extended gauge groups must be rich enough to account
for experimental data, although we will not attempt detailed fits
of all data. In particular, we require the presence of the fields
necessary to generate Majorana neutrino masses through seesaw, either
ordinary seesaw or inverse/linear seesaw.
\item \textbf{Anomaly cancellation}: we accept as valid proto-models only
field configurations which are anomaly free. 
\item \textbf{$\boldsymbol{SO(10)}$ completion}: all fields used in a lower
energy phase must be parts of a multiplet present at the next higher
symmetry phase. In particular, all fields should come from the decomposition
of one of the $SO(10)$ irreducible representations that we consider,
which are the ones up to ${\bf 126}$.
\item \textbf{Correct MSSM limit}: all proto-models must have a rich particle
content, so that at low energies the MSSM can emerge as an effective
theory. 
\end{itemize}
A few more words on our naming convention and notations is necessary.
We consider the three different $SO(10)$ breaking chains, equations
\eqref{eq:SlidingScale_chainI}--\eqref{eq:SlidingScale_chainIII},
and we will call these \textit{model classes}. In each class there
are fixed sets of $\beta$-coefficients, all leading to GCU, but with
different values of $\alpha_{G}$ and different values of $\alpha_{R}$
and $\alpha_{B-L}$ at low energies. These different sets are called
\textit{variants} in the following. Finally, (nearly) all of the variants
can be created by more than one possible set of superfields. We will
call such a set of superfields a \textit{configuration}. Configurations
are what usually is called \textit{model} by model builders, although
we prefer to think of these as \textit{proto-models}, in other words
constructions fulfilling all our basic requirements. These are only
proto-models (and not full-fledged models), since for each configuration
we do not check in a detailed calculation that all the fields required
in that configuration can remain light. We believe that for many (but
probably not all of the configurations) one can find conditions for
the required field combinations being {}``light'', following conditions
similar to those discussed in the prototype class-I model of \cite{Dev:2009aw}.
Having said this, and for the sake of simplicity, we will henceforth
call \textit{proto-models} just \textit{models}.

All superfields are named as $\Phi_{3_{c},2_{L},2_{R},1_{B-L}}$ (in
the left-right symmetric stage), $\Psi_{4,2_{L},2_{R}}$ (in the Pati-Salam
regime) and $\Phi_{3_{c},2_{L},1_{R},1_{B-L}}^{'}$ (in the $U(1)_{R}\times U(1)_{B-L}$
regime), with the indices giving the transformation properties under
the group. The conjugate of a field is distinguished by an overbar
($\overline{\Phi},\,\overline{\Psi},\,\overline{\Phi}^{'}$) as in
$\overline{\Phi}_{3_{c},2_{L},2_{R},1_{B-L}}$, for example, but note
that no overbar or minus sign is added to the indices. In appendix
\ref{chap:Lists_of_superfields_in_LR_models} we list all fields used,
together with their transformation properties and their $SO(10)$
origin, complete up to the ${\bf 126}$ representation of $SO(10)$.

\subsection{\label{subsect:SlidingScale_LRm}Model class-I: One intermediate
(left-right) scale}

We start our discussion with the simplest class of models with only
one new intermediate scale (LR):
\begin{equation}
SO(10)\rightarrow SU(3)_{c}\times SU(2)_{L}\times SU(2)_{R}\times U(1)_{B-L}\rightarrow\text{MSSM}\,.
\end{equation}
We do not discuss the first symmetry breaking step in detail, since
it is not relevant for the following discussion; we only mention that
$SO(10)$ can be broken to the LR group either via the interplay of
VEVs from a ${\bf 45}$ and a ${\bf 54}$, as done for example in
\cite{Dev:2009aw}, or via a ${\bf 45}$ and a ${\bf 210}$, an approach
followed in \cite{Malinsky:2005bi}. In the left-right symmetric stage
we consider all irreducible representations which can be constructed
from $SO(10)$ multiplets up to dimension ${\bf 126}$. This allows
for a total of 24 different representations (plus conjugates), whose
transformation properties under the LR group and their $SO(10)$ origin
are summarized in table \eqref{tab:List_of_LR_fields} of appendix
\ref{chap:Lists_of_superfields_in_LR_models}.

First, consider gauge coupling unification. If we take the MSSM particle
content as a starting point, the $\beta$-coefficients in the different
regimes are given as:%
\footnote{For $b_{1}^{SM'}$ and $b_{2}^{SM'}$ we use the SM particle content
plus one additional Higgs doublet.%
}
\begin{align}
\left(b_{3}^{SM'},b_{2}^{SM'},b_{Y}^{SM'}\right) & =\left(-7,-3,\frac{21}{5}\right)\,,\\
\left(b_{3}^{MSSM},b_{2}^{MSSM},b_{Y}^{MSSM}\right) & =\left(-3,1,\frac{33}{5}\right)\,,\\
\left(b_{3}^{LR},b_{2}^{LR},b_{R}^{LR},b_{B-L}^{LR}\right) & =\left(-3,1,1,6\right)+\left(\Delta b_{3}^{LR},\Delta b_{2}^{LR},\Delta b_{R}^{LR},\Delta b_{B-L}^{LR}\right)\,,
\end{align}
where we have used the canonical normalization for $B-L$, which is
related to the usual one%
\footnote{The canonical normalization comes from the requirement that all generators
$T^{a}$ of $SO(10)$, including therefore $B-L$, share the same
norm $\textrm{Tr}\left(T^{a}T^{a}\right)$, while the usual normalization
assumes that $B-L$ for the left quarks $Q$ and left leptons $L$
is $\nicefrac{1}{3}$ and $-1$ respectively.%
} as follows: $(B-L)^{c}=\sqrt{\frac{3}{8}}(B-L)$. Here, $\Delta b_{i}^{LR}$
stands for the contributions of chiral superfields which are not present
in the MSSM.

\begin{figure}[htb]
\centering{} \includegraphics[width=0.6\linewidth]{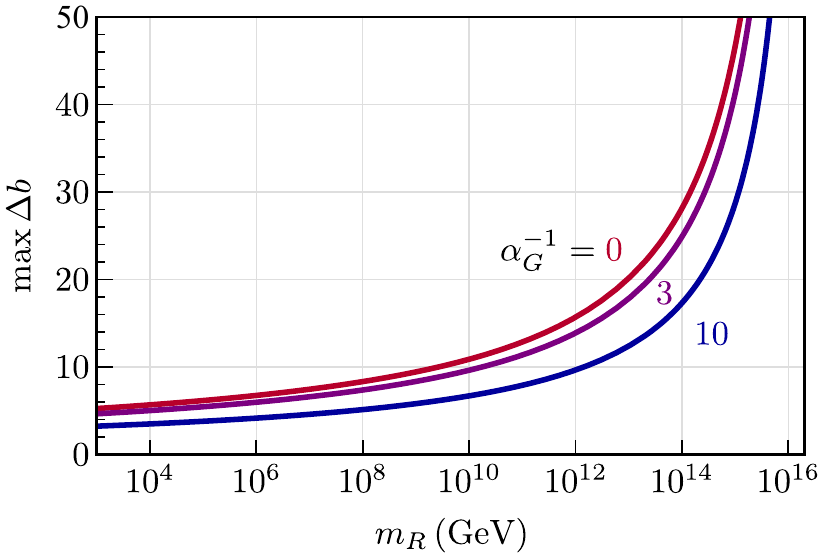}
\caption{\label{fig:MaxDb}Maximum value of $\Delta b$ allowed by perturbativity
as function of the scale $m_{R}$ (in GeV). The three different lines
have been calculated for three different values for the unified coupling
$\alpha_{G}^{-1}$, namely $\alpha_{G}^{-1}=0,\,3,\,10$. A LR scale
below 10 TeV (1 TeV) requires $\Delta b_{3}\lesssim5.7$ ($5.2$)
if the extreme value of $\alpha_{G}^{-1}=0$ is chosen, and $\Delta b_{3}\lesssim5.1$
($4.7$) for $\alpha_{G}^{-1}=3$.}
\end{figure}

As is well known, in contrast to the MSSM, putting an additional LR
scale below the GUT scale with all $\Delta b_{i}^{LR}$ equal to zero
destroys unification. Nevertheless GCU can be maintained if some simple
conditions on the $\Delta b_{i}^{LR}$ are fulfilled. First, since
in the MSSM $\alpha_{3}=\alpha_{2}$ at roughly $2\times10^{16}$
GeV one has that $\Delta b_{2}^{LR}=\Delta b_{3}^{LR}\equiv\Delta b$
in order to preserve this situation for an arbitrary LR scale (sliding
condition). Next, recall the matching condition
\begin{equation}
\alpha_{Y}^{-1}\left(m_{R}\right)=\frac{3}{5}\alpha_{R}^{-1}\left(m_{R}\right)+\frac{2}{5}\alpha_{B-L}^{-1}\left(m_{R}\right)\label{eq:SlidingScale_match}
\end{equation}
which, by substitution of the LR scale by an arbitrary one above $m_{R}$,
allows the definition of an artificial continuation of the hypercharge
coupling constant $\alpha_{Y}$ into the LR stage. The $\beta$-coefficient
of this dummy coupling constant for $E>m_{R}$ is $\frac{3}{5}b_{R}^{LR}+\frac{2}{5}b_{B-L}^{LR}$
and it should be compared with $b_{Y}^{MSSM}$ ($E<m_{R}$); the difference
is $\frac{3}{5}\Delta b_{R}^{LR}+\frac{2}{5}\Delta b_{B-L}^{LR}-\frac{18}{5}$,
which must be equal to $\Delta b$ in order for the difference between
this $\alpha_{Y}$ coupling and $\alpha_{3}=\alpha_{2}$ at the GUT
scale to be independent of the scale $m_{R}$. These are the two conditions
imposed by the sliding requirement of the LR scale on the $\beta$-coefficients---see
equations \eqref{eq:SlidingScale_variantsLR1} and \eqref{eq:SlidingScale_variantsLR2}.
Note, however, that we did not require (approximate) unification of
$\alpha_{R}$ and $\alpha_{B-L}$ with $\alpha_{3}$ and $\alpha_{2}$;
it is sufficient to require that $\alpha_{2}^{-1}=\alpha_{3}^{-1}\approx\frac{3}{5}\alpha_{R}^{-1}+\frac{2}{5}\alpha_{B-L}^{-1}$.
In any case, we can always achieve the desired unification because
the splitting between $\alpha_{R}$ and $\alpha_{B-L}$ at the $m_{R}$
scale is a free parameter, so it can be used to force $\alpha_{R}=\alpha_{B-L}$
at the scale where $\alpha_{3}$ and $\alpha_{2}$ unify, which leads
to an almost perfect unification of the four couplings. Also, we require
that unification is perturbative, i.e. the value of the common coupling
constant at the GUT scale is $\alpha_{G}^{-1}>0$. From the experimental
value of $\alpha_{3}(m_{Z})$ \cite{Beringer_mod:1900zz} one can
easily calculate the maximal allowed value of $\Delta b$ as a function
of the scale at which the LR group is broken to the SM group. This
is shown in figure \eqref{fig:MaxDb} for three different values of
$\alpha_{G}^{-1}$. The smallest value of $\max\Delta b$ is obtained
when $m_{R}$ is smallest as well (and $\alpha_{G}^{-1}$ is largest).
For $\alpha_{G}^{-1}$ in the interval $\left[0,3\right]$ one obtains
$\max\Delta b$ in the range $\left[4.7,5.7\right]$, which motivates
us to consider $\Delta b$ up to 5 (however, see the discussion below).

\begin{figure}[htb]
\centering{}\includegraphics[width=0.5\linewidth]{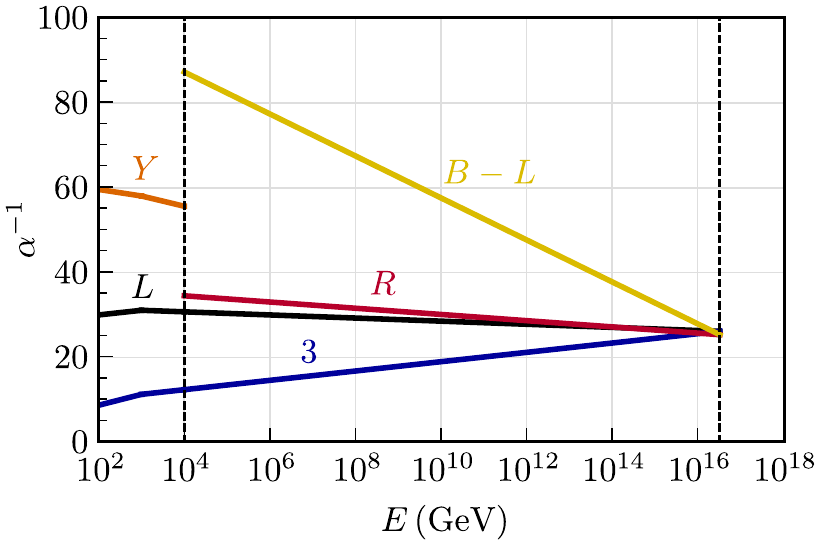}\includegraphics[width=0.5\linewidth]{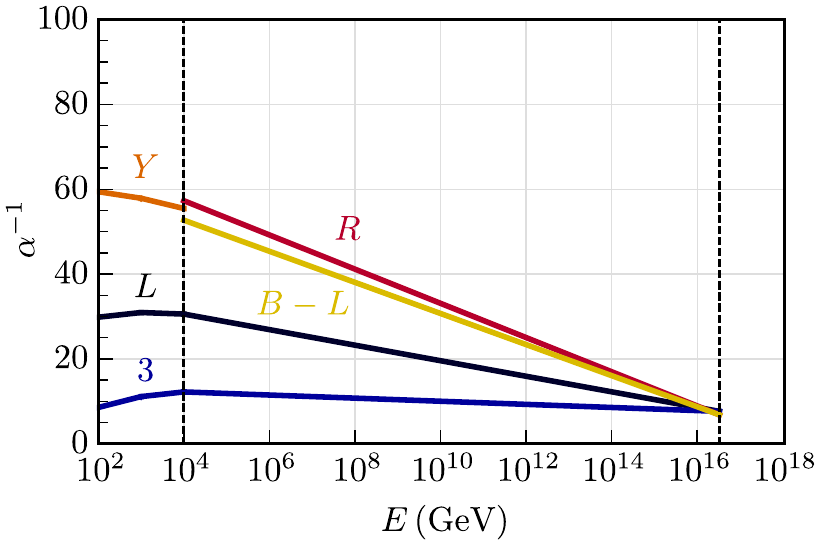}
\caption{\label{fig:SlidingScale_LR-Running}Gauge coupling unification in
LR models with $m_{R}=10^{4}$ GeV. The left panel is for ($\Delta b_{3}^{LR},\ab\Delta b_{2}^{LR},\ab\Delta b_{R}^{LR},\ab\Delta b_{B-L}^{LR}$)
$=(0,\ab0,\ab1,\ab\nicefrac{15}{2})$ while the right one is for $(4,\ab4,\ab10,\ab4)$.}
\end{figure}

Altogether, these considerations result in the following constraints
on the allowed values for the $\Delta b_{i}^{LR}$:
\begin{eqnarray}
\Delta b_{2}^{LR}=\Delta b_{3}^{LR}\equiv\Delta b & \le & 5\,,\label{eq:SlidingScale_variantsLR1}\\[2mm]
\Delta b_{B-L}^{LR}+\frac{3}{2}\Delta b_{R}^{LR}-9=\frac{5}{2}\Delta b & \le & \frac{25}{2}\,.\label{eq:SlidingScale_variantsLR2}
\end{eqnarray}
Given equations \eqref{eq:SlidingScale_variantsLR1} and \eqref{eq:SlidingScale_variantsLR2},
one can calculate all allowed variants of sets of $\Delta b_{i}^{LR}$,
guaranteed to give GCU. Two examples are shown in figure \eqref{fig:SlidingScale_LR-Running}.
The figure displays the running of the inverse gauge couplings as
a function of the energy scale, for an assumed value of $m_{R}=10$
TeV and a SUSY scale of 1 TeV and ($\Delta b_{3}^{LR},\Delta b_{2}^{LR},\Delta b_{R}^{LR},\Delta b_{B-L}^{LR}$)
$=(0,0,1,15/2)$ (left) or $(4,4,10,4)$ (right). The example on the
left has $\alpha_{G}^{-1}\approx25$ as in the MSSM, while the example
on the right has $\alpha_{G}^{-1}\approx6$. Note that while both
examples lead by construction to the same value of $\alpha_{Y}(m_{Z})$,
they have very different values for $\alpha_{R}(m_{R})$ and $\alpha_{B-L}(m_{R})$,
and thus predict different couplings for the gauge bosons $W_{R}$
and $Z'$ of the extended gauge group.

With the constraints in equations \eqref{eq:SlidingScale_variantsLR1}
and \eqref{eq:SlidingScale_variantsLR2}, we find that a total of
65 different variants can be built. However, after requiring that
at least one of the fields that breaks correctly the $SU(2)_{R}\times U(1)_{B-L}$
symmetry to $U(1)_{Y}$ is indeed present, either a $\Phi_{1,1,3,-2}$
or a $\Phi_{1,1,2,-1}$ (and/or their conjugates), the number of variants
is reduced to 53. We list them in table \eqref{tab:SlidingScale_LR_field_configuration},
together with one example of a field configuration for each variant.

\setlength{\tabcolsep}{3pt}

{
\thinmuskip=1mu
\medmuskip=1mu
\thickmuskip=1mu
\rowcolors{2}{white}{gray!20} %
\begin{longtable}{ll}
\caption{\label{tab:SlidingScale_LR_field_configuration}List of the 53 variants
with a single LR scale. In each case, the fields shown are the extra
ones, which are not part of the MSSM (the 2 Higgs doublets are assumed
to come from one bi-doublet $\Phi_{1,2,2,0}$). The $\Delta b_{3},\ab\,\Delta b_{2},\ab\,\Delta b_{R},\ab\,\Delta b_{B-L}$
values can be obtained from the first column through equations \eqref{eq:SlidingScale_variantsLR1}
and \eqref{eq:SlidingScale_variantsLR2}.}
\tabularnewline
\toprule 
\rowcolor{gray!0}\raisebox{2pt}{\scalebox{0.85}{$\boldsymbol{(\Delta b,\Delta b_{R})}$}} & \raisebox{2pt}{\scalebox{0.85}{\textbf{Sample field combination}}}\tabularnewline
\midrule
\endfirsthead
\toprule 
\rowcolor{gray!0}\raisebox{2pt}{\scalebox{0.85}{$\boldsymbol{(\Delta b,\Delta b_{R})}$}} & \raisebox{2pt}{\scalebox{0.85}{\textbf{Sample field combination}}}\tabularnewline
\midrule
\endhead
\raisebox{2pt}{\scalebox{0.85}{(0, 1)}} & \raisebox{2pt}{\scalebox{0.85}{$\overline{\Phi}_{1,1,2,-1}+2\overline{\Phi}_{1,1,1,2}+\Phi_{1,1,2,-1}+2\Phi_{1,1,1,2}$}}\tabularnewline
\raisebox{2pt}{\scalebox{0.85}{(0, 2)}} & \raisebox{2pt}{\scalebox{0.85}{$2\overline{\Phi}_{1,1,2,-1}+\overline{\Phi}_{1,1,1,2}+2\Phi_{1,1,2,-1}+\Phi_{1,1,1,2}$}}\tabularnewline
\raisebox{2pt}{\scalebox{0.85}{(0, 3)}} & \raisebox{2pt}{\scalebox{0.85}{$\overline{\Phi}_{1,1,2,-1}+\overline{\Phi}_{1,1,1,2}+\Phi_{1,1,2,-1}+\Phi_{1,1,3,0}+\Phi_{1,1,1,2}$}}\tabularnewline
\raisebox{2pt}{\scalebox{0.85}{(0, 4)}} & \raisebox{2pt}{\scalebox{0.85}{$2\overline{\Phi}_{1,1,2,-1}+2\Phi_{1,1,2,-1}+\Phi_{1,1,3,0}$}}\tabularnewline
\raisebox{2pt}{\scalebox{0.85}{(0, 5)}} & \raisebox{2pt}{\scalebox{0.85}{$\overline{\Phi}_{1,1,2,-1}+\Phi_{1,1,2,-1}+2\Phi_{1,1,3,0}$}}\tabularnewline
\raisebox{2pt}{\scalebox{0.85}{(1, 1)}} & \raisebox{2pt}{\scalebox{0.85}{$\overline{\Phi}_{1,2,1,1}+\overline{\Phi}_{1,1,2,-1}+2\overline{\Phi}_{1,1,1,2}+\overline{\Phi}_{3,1,1,-\frac{2}{3}}+\Phi_{1,2,1,1}+\Phi_{1,1,2,-1}+2\Phi_{1,1,1,2}+\Phi_{3,1,1,-\frac{2}{3}}$}}\tabularnewline
\raisebox{2pt}{\scalebox{0.85}{(1, 2)}} & \raisebox{2pt}{\scalebox{0.85}{$\overline{\Phi}_{1,1,2,-1}+2\overline{\Phi}_{1,1,1,2}+\overline{\Phi}_{3,1,1,-\frac{2}{3}}+\Phi_{1,1,2,-1}+\Phi_{1,2,2,0}+2\Phi_{1,1,1,2}+\Phi_{3,1,1,-\frac{2}{3}}$}}\tabularnewline
\raisebox{2pt}{\scalebox{0.85}{(1, 3)}} & \raisebox{2pt}{\scalebox{0.85}{$2\overline{\Phi}_{1,1,2,-1}+\overline{\Phi}_{1,1,1,2}+\overline{\Phi}_{3,1,1,-\frac{2}{3}}+2\Phi_{1,1,2,-1}+\Phi_{1,2,2,0}+\Phi_{1,1,1,2}+\Phi_{3,1,1,-\frac{2}{3}}$}}\tabularnewline
\raisebox{2pt}{\scalebox{0.85}{(1, 4)}} & \raisebox{2pt}{\scalebox{0.85}{$\overline{\Phi}_{1,1,2,-1}+\overline{\Phi}_{1,1,1,2}+\overline{\Phi}_{3,1,1,-\frac{2}{3}}+\Phi_{1,1,2,-1}+\Phi_{1,1,3,0}+\Phi_{1,2,2,0}+\Phi_{1,1,1,2}+\Phi_{3,1,1,-\frac{2}{3}}$}}\tabularnewline
\raisebox{2pt}{\scalebox{0.85}{(1, 5)}} & \raisebox{2pt}{\scalebox{0.85}{$2\overline{\Phi}_{1,1,2,-1}+\overline{\Phi}_{3,1,1,-\frac{2}{3}}+2\Phi_{1,1,2,-1}+\Phi_{1,1,3,0}+\Phi_{1,2,2,0}+\Phi_{3,1,1,-\frac{2}{3}}$}}\tabularnewline
\raisebox{2pt}{\scalebox{0.85}{(1, 6)}} & \raisebox{2pt}{\scalebox{0.85}{$\overline{\Phi}_{1,1,2,-1}+\overline{\Phi}_{3,1,1,-\frac{2}{3}}+\Phi_{1,1,2,-1}+2\Phi_{1,1,3,0}+\Phi_{1,2,2,0}+\Phi_{3,1,1,-\frac{2}{3}}$}}\tabularnewline
\raisebox{2pt}{\scalebox{0.85}{(2, 1)}} & \raisebox{2pt}{\scalebox{0.85}{$\overline{\Phi}_{1,1,2,-1}+3\overline{\Phi}_{1,1,1,2}+2\overline{\Phi}_{3,1,1,-\frac{2}{3}}+\Phi_{1,1,2,-1}+\Phi_{1,3,1,0}+3\Phi_{1,1,1,2}+2\Phi_{3,1,1,-\frac{2}{3}}$}}\tabularnewline
\raisebox{2pt}{\scalebox{0.85}{(2, 2)}} & \raisebox{2pt}{\scalebox{0.85}{$2\overline{\Phi}_{1,1,2,-1}+2\overline{\Phi}_{1,1,1,2}+2\overline{\Phi}_{3,1,1,-\frac{2}{3}}+2\Phi_{1,1,2,-1}+\Phi_{1,3,1,0}+2\Phi_{1,1,1,2}+2\Phi_{3,1,1,-\frac{2}{3}}$}}\tabularnewline
\raisebox{2pt}{\scalebox{0.85}{(2, 3)}} & \raisebox{2pt}{\scalebox{0.85}{$\overline{\Phi}_{1,1,2,-1}+2\overline{\Phi}_{1,1,1,2}+2\overline{\Phi}_{3,1,1,-\frac{2}{3}}+\Phi_{1,1,2,-1}+2\Phi_{1,2,2,0}+2\Phi_{1,1,1,2}+2\Phi_{3,1,1,-\frac{2}{3}}$}}\tabularnewline
\raisebox{2pt}{\scalebox{0.85}{(2, 4)}} & \raisebox{2pt}{\scalebox{0.85}{$2\overline{\Phi}_{1,1,2,-1}+\overline{\Phi}_{1,1,1,2}+2\overline{\Phi}_{3,1,1,-\frac{2}{3}}+2\Phi_{1,1,2,-1}+2\Phi_{1,2,2,0}+\Phi_{1,1,1,2}+2\Phi_{3,1,1,-\frac{2}{3}}$}}\tabularnewline
\raisebox{2pt}{\scalebox{0.85}{(2, 5)}} & \raisebox{2pt}{\scalebox{0.85}{$\overline{\Phi}_{1,1,2,-1}+\overline{\Phi}_{1,1,1-2}+2\overline{\Phi}_{3,1,1,-\frac{2}{3}}+\Phi_{1,1,2,-1}+\Phi_{1,1,3,0}+2\Phi_{1,2,2,0}+\Phi_{1,1,1,2}+2\Phi_{3,1,1,-\frac{2}{3}}$}}\tabularnewline
\raisebox{2pt}{\scalebox{0.85}{(2, 6)}} & \raisebox{2pt}{\scalebox{0.85}{$2\overline{\Phi}_{1,1,2,-1}+2\overline{\Phi}_{3,1,1,-\frac{2}{3}}+2\Phi_{1,1,2,-1}+\Phi_{1,1,3,0}+2\Phi_{1,2,2,0}+2\Phi_{3,1,1,-\frac{2}{3}}$}}\tabularnewline
\raisebox{2pt}{\scalebox{0.85}{(2, 7)}} & \raisebox{2pt}{\scalebox{0.85}{$\overline{\Phi}_{1,1,2,-1}+2\overline{\Phi}_{3,1,1,-\frac{2}{3}}+\Phi_{1,1,2,-1}+2\Phi_{1,1,3,0}+2\Phi_{1,2,2,0}+2\Phi_{3,1,1,-\frac{2}{3}}$}}\tabularnewline
\raisebox{2pt}{\scalebox{0.85}{(2, 8)}} & \raisebox{2pt}{\scalebox{0.85}{$\overline{\Phi}_{1,1,2,-1}+\overline{\Phi}_{3,1,2,\frac{1}{3}}+\Phi_{1,1,2,-1}+\Phi_{1,1,3,0}+2\Phi_{1,2,2,0}+\Phi_{3,1,2,\frac{1}{3}}$}}\tabularnewline
\raisebox{2pt}{\scalebox{0.85}{(3, 1)}} & \raisebox{2pt}{\scalebox{0.85}{$\overline{\Phi}_{1,2,1,1}+\overline{\Phi}_{1,1,2,-1}+4\overline{\Phi}_{1,1,1,2}+\Phi_{1,2,1,1}+\Phi_{1,1,2,-1}+\Phi_{1,3,1,0}+\Phi_{8,1,1,0}+4\Phi_{1,1,1,2}$}}\tabularnewline
\raisebox{2pt}{\scalebox{0.85}{(3, 2)}} & \raisebox{2pt}{\scalebox{0.85}{$\overline{\Phi}_{1,1,2,-1}+4\overline{\Phi}_{1,1,1,2}+\Phi_{1,1,2,-1}+\Phi_{1,3,1,0}+\Phi_{1,2,2,0}+\Phi_{8,1,1,0}+4\Phi_{1,1,1,2}$
}}\tabularnewline
\raisebox{2pt}{\scalebox{0.85}{(3, 3)}} & \raisebox{2pt}{\scalebox{0.85}{$2\overline{\Phi}_{1,1,2,-1}+3\overline{\Phi}_{1,1,1,2}+2\Phi_{1,1,2,-1}+\Phi_{1,3,1,0}+\Phi_{1,2,2,0}+\Phi_{8,1,1,0}+3\Phi_{1,1,1,2}$}}\tabularnewline
\raisebox{2pt}{\scalebox{0.85}{(3, 4)}} & \raisebox{2pt}{\scalebox{0.85}{$\overline{\Phi}_{1,2,1,1}+\overline{\Phi}_{1,1,3,-2}+\Phi_{1,2,1,1}+\Phi_{1,3,1,0}+\Phi_{8,1,1,0}+\Phi_{1,1,3,-2}$}}\tabularnewline
\raisebox{2pt}{\scalebox{0.85}{(3, 5)}} & \raisebox{2pt}{\scalebox{0.85}{$\overline{\Phi}_{1,1,3,-2}+\Phi_{1,3,1,0}+\Phi_{1,2,2,0}+\Phi_{8,1,1,0}+\Phi_{1,1,3,-2}$}}\tabularnewline
\raisebox{2pt}{\scalebox{0.85}{(3, 6)}} & \raisebox{2pt}{\scalebox{0.85}{$\overline{\Phi}_{1,1,2,-1}+2\overline{\Phi}_{1,1,1,2}+\Phi_{1,1,2,-1}+\Phi_{1,1,3,0}+3\Phi_{1,2,2,0}+\Phi_{8,1,1,0}+2\Phi_{1,1,1,2}$
}}\tabularnewline
\raisebox{2pt}{\scalebox{0.85}{(3, 7)}} & \raisebox{2pt}{\scalebox{0.85}{$2\overline{\Phi}_{1,1,2,-1}+\overline{\Phi}_{1,1,1-2}+2\Phi_{1,1,2,-1}+\Phi_{1,1,3,0}+3\Phi_{1,2,2,0}+\Phi_{8,1,1,0}+\Phi_{1,1,1,2}$}}\tabularnewline
\raisebox{2pt}{\scalebox{0.85}{(3, 8)}} & \raisebox{2pt}{\scalebox{0.85}{$\overline{\Phi}_{1,1,2,-1}+\overline{\Phi}_{1,1,1,2}+\Phi_{1,1,2,-1}+2\Phi_{1,1,3,0}+3\Phi_{1,2,2,0}+\Phi_{8,1,1,0}+\Phi_{1,1,1,2}$
}}\tabularnewline
\raisebox{2pt}{\scalebox{0.85}{(3, 9)}} & \raisebox{2pt}{\scalebox{0.85}{$2\overline{\Phi}_{1,1,2,-1}+2\Phi_{1,1,2,-1}+2\Phi_{1,1,3,0}+3\Phi_{1,2,2,0}+\Phi_{8,1,1,0}$}}\tabularnewline
\raisebox{2pt}{\scalebox{0.85}{(3, 10)}} & \raisebox{2pt}{\scalebox{0.85}{$\overline{\Phi}_{1,1,2,-1}+\Phi_{1,1,2,-1}+3\Phi_{1,1,3,0}+3\Phi_{1,2,2,0}+\Phi_{8,1,1,0}$}}\tabularnewline
\raisebox{2pt}{\scalebox{0.85}{ (4, 1)}} & \raisebox{2pt}{\scalebox{0.85}{$\overline{\Phi}_{1,1,2,-1}+5\overline{\Phi}_{1,1,1,2}+\overline{\Phi}_{3,1,1,-\frac{2}{3}}+\Phi_{1,1,2,-1}+2\Phi_{1,3,1,0}+\Phi_{8,1,1,0}+5\Phi_{1,1,1,2}+\Phi_{3,1,1,-\frac{2}{3}}$}}\tabularnewline
\raisebox{2pt}{\scalebox{0.85}{(4, 2)}} & \raisebox{2pt}{\scalebox{0.85}{$2\overline{\Phi}_{1,1,2,-1}+4\overline{\Phi}_{1,1,1,2}+\overline{\Phi}_{3,1,1,-\frac{2}{3}}+2\Phi_{1,1,2,-1}+2\Phi_{1,3,1,0}+\Phi_{8,1,1,0}+4\Phi_{1,1,1,2}+\Phi_{3,1,1,-\frac{2}{3}}$
}}\tabularnewline
\raisebox{2pt}{\scalebox{0.85}{(4, 3)}} & \raisebox{2pt}{\scalebox{0.85}{$\overline{\Phi}_{1,1,2,-1}+4\overline{\Phi}_{1,1,1,2}+\overline{\Phi}_{3,1,1,-\frac{2}{3}}+\Phi_{1,1,2,-1}+\Phi_{1,3,1,0}+2\Phi_{1,2,2,0}+\Phi_{8,1,1,0}$}}\tabularnewline
 & \raisebox{2pt}{\scalebox{0.85}{$+4\Phi_{1,1,1,2}+\Phi_{3,1,1,-\frac{2}{3}}$}}\tabularnewline
\raisebox{2pt}{\scalebox{0.85}{(4, 4)}} & \raisebox{2pt}{\scalebox{0.85}{$\overline{\Phi}_{1,1,1,2}+\overline{\Phi}_{3,1,1,-\frac{2}{3}}+\overline{\Phi}_{1,1,3,-2}+2\Phi_{1,3,1,0}+\Phi_{8,1,1,0}+\Phi_{1,1,1,2}+\Phi_{3,1,1,-\frac{2}{3}}+\Phi_{1,1,3,-2}$
}}\tabularnewline
\raisebox{2pt}{\scalebox{0.85}{(4, 5)}} & \raisebox{2pt}{\scalebox{0.85}{$\overline{\Phi}_{1,1,2,-1}+\overline{\Phi}_{3,1,1,-\frac{2}{3}}+\overline{\Phi}_{1,1,3,-2}+\Phi_{1,1,2,-1}+2\Phi_{1,3,1,0}+\Phi_{8,1,1,0}+\Phi_{3,1,1,-\frac{2}{3}}+\Phi_{1,1,3,-2}$}}\tabularnewline
\raisebox{2pt}{\scalebox{0.85}{(4, 6)}} & \raisebox{2pt}{\scalebox{0.85}{$\overline{\Phi}_{3,1,1,-\frac{2}{3}}+\overline{\Phi}_{1,1,3,-2}+\Phi_{1,3,1,0}+2\Phi_{1,2,2,0}+\Phi_{8,1,1,0}+\Phi_{3,1,1,-\frac{2}{3}}+\Phi_{1,1,3,-2}$}}\tabularnewline
\raisebox{2pt}{\scalebox{0.85}{(4, 7)}} & \raisebox{2pt}{\scalebox{0.85}{$\overline{\Phi}_{1,1,2,-1}+2\overline{\Phi}_{1,1,1,2}+\overline{\Phi}_{3,1,1,-\frac{2}{3}}+\Phi_{1,1,2,-1}+\Phi_{1,1,3,0}+4\Phi_{1,2,2,0}+\Phi_{8,1,1,0}$}}\tabularnewline
 & \raisebox{2pt}{\scalebox{0.85}{$+2\Phi_{1,1,1,2}+\Phi_{3,1,1,-\frac{2}{3}}$}}\tabularnewline
\raisebox{2pt}{\scalebox{0.85}{(4, 8)}} & \raisebox{2pt}{\scalebox{0.85}{$2\overline{\Phi}_{1,1,2,-1}+\overline{\Phi}_{1,1,1,2}+\overline{\Phi}_{3,1,1,-\frac{2}{3}}+2\Phi_{1,1,2,-1}+\Phi_{1,1,3,0}+4\Phi_{1,2,2,0}+\Phi_{8,1,1,0}$}}\tabularnewline
 & \raisebox{2pt}{\scalebox{0.85}{$+\Phi_{1,1,1,2}+\Phi_{3,1,1,-\frac{2}{3}}$}}\tabularnewline
\raisebox{2pt}{\scalebox{0.85}{(4, 9)}} & \raisebox{2pt}{\scalebox{0.85}{$\overline{\Phi}_{1,1,2,-1}+\overline{\Phi}_{1,1,1,2}+\overline{\Phi}_{3,1,1,-\frac{2}{3}}+\Phi_{1,1,2,-1}+2\Phi_{1,1,3,0}+4\Phi_{1,2,2,0}+\Phi_{8,1,1,0}$}}\tabularnewline
 & \raisebox{2pt}{\scalebox{0.85}{$+\Phi_{1,1,1,2}+\Phi_{3,1,1,-\frac{2}{3}}$}}\tabularnewline
\raisebox{2pt}{\scalebox{0.85}{(4, 10)}} & \raisebox{2pt}{\scalebox{0.85}{$2\overline{\Phi}_{1,1,2,-1}+\overline{\Phi}_{3,1,1,-\frac{2}{3}}+2\Phi_{1,1,2,-1}+2\Phi_{1,1,3,0}+4\Phi_{1,2,2,0}+\Phi_{8,1,1,0}+\Phi_{3,1,1,-\frac{2}{3}}$}}\tabularnewline
\raisebox{2pt}{\scalebox{0.85}{(4, 11)}} & \raisebox{2pt}{\scalebox{0.85}{$\overline{\Phi}_{1,1,2,-1}+\overline{\Phi}_{3,1,1,-\frac{2}{3}}+\Phi_{1,1,2,-1}+3\Phi_{1,1,3,0}+4\Phi_{1,2,2,0}+\Phi_{8,1,1,0}+\Phi_{3,1,1,-\frac{2}{3}}$}}\tabularnewline
\raisebox{2pt}{\scalebox{0.85}{(5, 1)}} & \raisebox{2pt}{\scalebox{0.85}{$\overline{\Phi}_{1,2,1,1}+\overline{\Phi}_{1,1,2,-1}+5\overline{\Phi}_{1,1,1,2}+2\overline{\Phi}_{3,1,1,-\frac{2}{3}}+\Phi_{1,2,1,1}+\Phi_{1,1,2,-1}+2\Phi_{1,3,1,0}+\Phi_{8,1,1,0}$}}\tabularnewline
 & \raisebox{2pt}{\scalebox{0.85}{$+5\Phi_{1,1,1,2}+2\Phi_{3,1,1,-\frac{2}{3}}$}}\tabularnewline
\raisebox{2pt}{\scalebox{0.85}{(5, 2)}} & \raisebox{2pt}{\scalebox{0.85}{$\overline{\Phi}_{1,1,2,-1}+5\overline{\Phi}_{1,1,1,2}+2\overline{\Phi}_{3,1,1,-\frac{2}{3}}+\Phi_{1,1,2,-1}+2\Phi_{1,3,1,0}+\Phi_{1,2,2,0}+\Phi_{8,1,1,0}+5\Phi_{1,1,1,2}$}}\tabularnewline
 & \raisebox{2pt}{\scalebox{0.85}{$+2\Phi_{3,1,1,-\frac{2}{3}}$}}\tabularnewline
\raisebox{2pt}{\scalebox{0.85}{(5, 3)}} & \raisebox{2pt}{\scalebox{0.85}{$2\overline{\Phi}_{1,1,2,-1}+4\overline{\Phi}_{1,1,1,2}+2\overline{\Phi}_{3,1,1,-\frac{2}{3}}+2\Phi_{1,1,2,-1}+2\Phi_{1,3,1,0}+\Phi_{1,2,2,0}+\Phi_{8,1,1,0}+4\Phi_{1,1,1,2}$}}\tabularnewline
 & \raisebox{2pt}{\scalebox{0.85}{$+2\Phi_{3,1,1,-\frac{2}{3}}$}}\tabularnewline
\raisebox{2pt}{\scalebox{0.85}{(5, 4)}} & \raisebox{2pt}{\scalebox{0.85}{$\overline{\Phi}_{1,2,1,1}+\overline{\Phi}_{1,1,1,2}+2\overline{\Phi}_{3,1,1,-\frac{2}{3}}+\overline{\Phi}_{1,1,3,-2}+\Phi_{1,2,1,1}+2\Phi_{1,3,1,0}+\Phi_{8,1,1,0}$}}\tabularnewline
 & \raisebox{2pt}{\scalebox{0.85}{$+\Phi_{1,1,1,2}+2\Phi_{3,1,1,-\frac{2}{3}}$}}\tabularnewline
\raisebox{2pt}{\scalebox{0.85}{(5, 5)}} & \raisebox{2pt}{\scalebox{0.85}{$\overline{\Phi}_{1,1,1,2}+2\overline{\Phi}_{3,1,1,-\frac{2}{3}}+\overline{\Phi}_{1,1,3,-2}+2\Phi_{1,3,1,0}+\Phi_{1,2,2,0}+\Phi_{8,1,1,0}+\Phi_{1,1,1,2}+2\Phi_{3,1,1,-\frac{2}{3}}$}}\tabularnewline
 & \raisebox{2pt}{\scalebox{0.85}{$+\Phi_{1,1,3,-2}$}}\tabularnewline
\raisebox{2pt}{\scalebox{0.85}{(5, 6)}} & \raisebox{2pt}{\scalebox{0.85}{$\overline{\Phi}_{1,1,2,-1}+2\overline{\Phi}_{3,1,1,-\frac{2}{3}}+\overline{\Phi}_{1,1,3,-2}+\Phi_{1,1,2,-1}+2\Phi_{1,3,1,0}+\Phi_{1,2,2,0}+\Phi_{8,1,1,0}+2\Phi_{3,1,1,-\frac{2}{3}}$}}\tabularnewline
 & \raisebox{2pt}{\scalebox{0.85}{$+\Phi_{1,1,3,-2}$}}\tabularnewline
\raisebox{2pt}{\scalebox{0.85}{(5, 7)}} & \raisebox{2pt}{\scalebox{0.85}{$2\overline{\Phi}_{3,1,1,-\frac{2}{3}}+\overline{\Phi}_{1,1,3,-2}+\Phi_{1,3,1,0}+3\Phi_{1,2,2,0}+\Phi_{8,1,1,0}+2\Phi_{3,1,1,-\frac{2}{3}}+\Phi_{1,1,3,-2}$}}\tabularnewline
\raisebox{2pt}{\scalebox{0.85}{(5, 8)}} & \raisebox{2pt}{\scalebox{0.85}{$\overline{\Phi}_{3,1,2,\frac{1}{3}}+\overline{\Phi}_{1,1,3,-2}+2\Phi_{1,3,1,0}+\Phi_{1,2,2,0}+\Phi_{8,1,1,0}+\Phi_{3,1,2,\frac{1}{3}}+\Phi_{1,1,3,-2}$}}\tabularnewline
\raisebox{2pt}{\scalebox{0.85}{(5, 9)}} & \raisebox{2pt}{\scalebox{0.85}{$2\overline{\Phi}_{1,1,2,-1}+\overline{\Phi}_{1,1,1,2}+2\overline{\Phi}_{3,1,1,-\frac{2}{3}}+2\Phi_{1,1,2,-1}+\Phi_{1,1,3,0}+5\Phi_{1,2,2,0}+\Phi_{8,1,1,0}+\Phi_{1,1,1,2}$
}}\tabularnewline
 & \raisebox{2pt}{\scalebox{0.85}{$+2\Phi_{3,1,1,-\frac{2}{3}}$ }}\tabularnewline
\raisebox{2pt}{\scalebox{0.85}{(5, 10)}} & \raisebox{2pt}{\scalebox{0.85}{$\overline{\Phi}_{1,1,2,-1}+\overline{\Phi}_{1,1,1,2}+2\overline{\Phi}_{3,1,1,-\frac{2}{3}}+\Phi_{1,1,2,-1}+2\Phi_{1,1,3,0}+5\Phi_{1,2,2,0}+\Phi_{8,1,1,0}+\Phi_{1,1,1,2}$}}\tabularnewline
 & \raisebox{2pt}{\scalebox{0.85}{$+2\Phi_{3,1,1,-\frac{2}{3}}$}}\tabularnewline
\raisebox{2pt}{\scalebox{0.85}{(5, 11)}} & \raisebox{2pt}{\scalebox{0.85}{$2\overline{\Phi}_{1,1,2,-1}+2\overline{\Phi}_{3,1,1,-\frac{2}{3}}+2\Phi_{1,1,2,-1}+2\Phi_{1,1,3,0}+5\Phi_{1,2,2,0}+\Phi_{8,1,1,0}+2\Phi_{3,1,1,-\frac{2}{3}}$
}}\tabularnewline
\raisebox{2pt}{\scalebox{0.85}{(5, 12)}} & \raisebox{2pt}{\scalebox{0.85}{$\overline{\Phi}_{1,1,2,-1}+2\overline{\Phi}_{3,1,1,-\frac{2}{3}}+\Phi_{1,1,2,-1}+3\Phi_{1,1,3,0}+5\Phi_{1,2,2,0}+\Phi_{8,1,1,0}+2\Phi_{3,1,1,-\frac{2}{3}}$}}\tabularnewline
\raisebox{2pt}{\scalebox{0.85}{(5, 13)}} & \raisebox{2pt}{\scalebox{0.85}{$\overline{\Phi}_{1,1,2,-1}+\overline{\Phi}_{3,1,2,\frac{1}{3}}+\Phi_{1,1,2,-1}+2\Phi_{1,1,3,0}+5\Phi_{1,2,2,0}+\Phi_{8,1,1,0}+\Phi_{3,1,2,\frac{1}{3}}$
}}\tabularnewline
\midrule
\end{longtable}} 

We give only one example for each configuration in table \eqref{tab:SlidingScale_LR_field_configuration},
although we went through the exercise of finding all possible configurations
for the 53 variants with the field content of table \eqref{tab:List_of_LR_fields}.
In total there are 5324 anomaly-free configurations \cite{website_sliding_models}.
The variants (0,1), (0,2), (0,4) and (0,5) are the only ones which
have a single configuration; for the other variants, in particular
those with larger values of $\Delta b_{3}^{LR}$, there are many configurations.

Not all the fields in table \eqref{tab:List_of_LR_fields} can lead
to valid configurations: the fields which never give an anomaly-free
configuration are $\Phi_{8,2,2,0}$, $\Phi_{3,2,2,\frac{4}{3}}$,
$\Phi_{3,3,1,-\frac{2}{3}}$, $\Phi_{3,1,3,-\frac{2}{3}}$, $\Phi_{6,3,1,\frac{2}{3}}$,
$\Phi_{6,1,3,\frac{2}{3}}$ and $\Phi_{1,3,3,0}$. Also, the field
$\Phi_{3,2,2,-\frac{2}{3}}$ appears exactly once, in the configuration
$4\Phi_{1,2,1,1}+\Phi_{3,1,1,-\frac{2}{3}}+\Phi_{3,2,2,-\frac{2}{3}}+4\Phi_{1,1,2,1}+2\Phi_{1,1,1,2}+5\Phi_{3,1,1,-\frac{2}{3}}$
which is a (5,5) variant. Note that the examples we give for variants
(1,3) and (1,4) are not the model-II and model-I discussed in \cite{DeRomeri:2011ie}.

Many of the 53 variants only have configurations with $\Phi_{1,1,2,-1}$
(and conjugate) for the breaking of the LR-symmetry. To generate neutrino
masses via a seesaw mechanism these variants need either the presence
of $\Phi_{1,3,1,0}$, as for example in the configuration shown for
variant (2,1), or $\Phi_{1,1,3,0}$ (see, for instance variant (1,4)),
or an additional singlet $\Phi_{1,1,1,0}$ (which is not shown in
table \eqref{tab:SlidingScale_LR_field_configuration} since it does
not affect the $\Delta b_{i}^{LR}$). Using the $\Phi_{1,1,1,0}$
one could construct either an inverse \cite{Mohapatra:1986bd} or
a linear \cite{Akhmedov:1995ip,Akhmedov:1995vm} seesaw mechanism,
while with $\Phi_{1,3,1,0}$ a type-III seesaw \cite{Foot:1988aq}
is a possibility, and finally a $\Phi_{1,1,3,0}$ allows for an inverse
type-III seesaw \cite{DeRomeri:2011ie}. The first example where a
valid configuration with $\Phi_{1,1,3,-2}$ appears is in the variant
(3,4). The simplest configuration is $\Phi_{1,2,1,1}+\Phi_{1,3,1,0}+\Phi_{8,1,1,0}+\Phi_{1,1,3,-2}+\overline{\Phi}_{1,2,1,1}+\overline{\Phi}_{1,1,3,-2}$
(which is not the example given in table \eqref{tab:SlidingScale_LR_field_configuration}).
The VEV of the $\Phi_{1,1,3,-2}$ does not only break the LR symmetry,
but it can also generate a Majorana mass term for the right-handed
neutrino fields, i.e. configurations with $\Phi_{1,1,3,-2}$ can generate
a type-I seesaw, in principle. Finally, the simplest possibility with
a valid configuration including $\Phi_{1,3,1,-2}$ is found in variant
(4,1) with $\Phi_{1,1,2,-1}+\Phi_{8,1,1,0}+\Phi_{1,1,1,2}+\Phi_{3,1,1,\frac{4}{3}}+\Phi_{1,3,1,-2}+\overline{\Phi}_{1,1,2,-1}+\overline{\Phi}_{1,1,1,2}+\overline{\Phi}_{3,1,1,\frac{4}{3}}+\overline{\Phi}_{1,3,1,-2}$.
The presence of $\Phi_{1,3,1,-2}$ allows the construction of a type-II
seesaw mechanism for the neutrinos.

As mentioned in the beginning of this chapter, it is not possible
to construct a sliding scale model in which the LR symmetry is broken
by two pairs of triplets: $\Phi_{1,3,1,-2}+\overline{\Phi}_{1,3,1,-2}+\Phi_{1,1,3,-2}+\overline{\Phi}_{1,1,3,-2}$.
The sum of the $\Delta b$'s for these fields adds up to $(\Delta b_{3}^{LR},b_{L}^{LR},\Delta b_{R}^{LR},\Delta b_{B-L}^{LR})=(0,4,4,18)$
and so, because $\Delta b_{B-L}^{LR}+\frac{3}{2}\Delta b_{R}^{LR}-9=15>\frac{25}{2}$
(confer with equation \eqref{eq:SlidingScale_variantsLR2}), there
are no configurations with this combination of fields. This observation
is consistent with the analysis done in \cite{Majee:2007uv}, where
the authors have shown that a supersymmetric LR-symmetric model, where
the LR symmetry is broken by two pairs of triplets, requires a minimal
LR scale of at least $10^{9}$ GeV (and, actually, a much larger scale
in minimal renormalizable models, if GUT scale thresholds are small).

Regarding the variants with $\Delta b_{2}^{LR}=\Delta b_{3}^{LR}=0$,
strictly speaking none of these variants is guaranteed to give a valid
model in the sense defined in subsection \ref{subsec:SlidingScale_so10models},
as they contain only one $\Phi_{1,2,2,0}\rightarrow(H_{u},H_{d})$
and no vector-like quarks ($\Phi_{3,1,1,\frac{4}{3}}$ or $\Phi_{3,1,1,-\frac{2}{3}}$).
With such a minimal configuration, the CKM matrix is trivial at the
energy scale where the LR symmetry is broken. We nevertheless list
these variants, since in principle a CKM matrix for quarks consistent
with experimental data could be generated at 1-loop level from flavor
violating soft terms, as discussed in \cite{Babu:1998tm}.

\subsection{\label{subsec:SlidingScale_PSmodels}Model class-II: Additional intermediate
Pati-Salam scale}

In the second class of supersymmetric $SO(10)$ models that we are
considering, $SO(10)$ is broken first to the Pati-Salam (PS) group.
The complete breaking chain is thus
\begin{eqnarray}
SO(10) & \to & SU(4)\times SU(2)_{L}\times SU(2)_{R}\nonumber \\
 & \to & SU(3)_{c}\times SU(2)_{L}\times SU(2)_{R}\times U(1)_{B-L}\to{\rm MSSM}\,.
\end{eqnarray}
The representations available from the decomposition of $SO(10)$
multiplets up to ${\bf 126}$ are listed in table \eqref{tab:List_of_PatiSalam_fields}
of appendix \ref{chap:Lists_of_superfields_in_LR_models}, together
with their possible $SO(10)$ origin. Breaking $SO(10)$ to the PS
group requires that $\Psi_{1,1,1}$ from the ${\bf 54}$ develops
a VEV. The subsequent breaking of the PS group to the LR group requires
that the LR singlet in $\Psi_{15,1,1}$, originally from the ${\bf 45}$
of $SO(10)$, acquires a VEV. And finally, as before in the LR-class,
the breaking of LR to $SU(3)_{c}\times SU(2)_{L}\times U(1)_{Y}$
can be done either with a $\Phi_{1,1,2,-1}$ or $\Phi_{1,1,3,-2}$
(and/or conjugates).

\begin{figure}[htb]
\centering{} \includegraphics[width=0.6\linewidth]{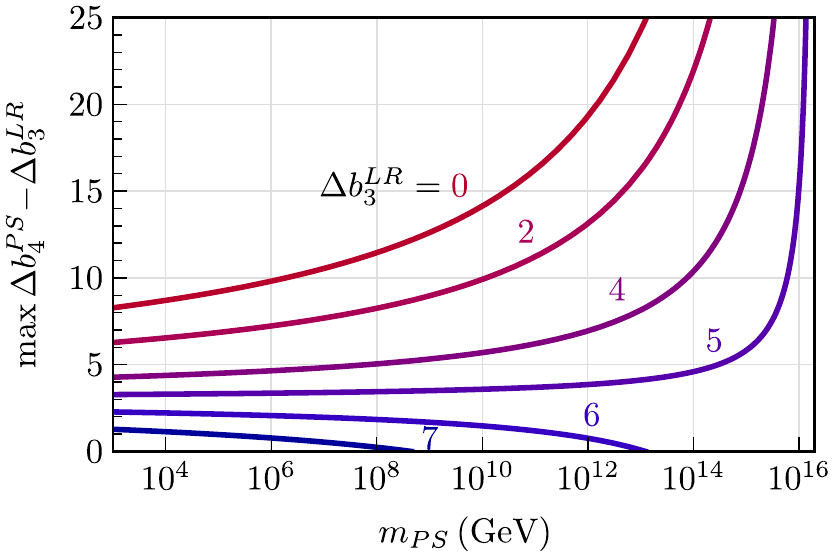}
\caption{\label{fig:SlidingScale_MaxDbPS}Maximum value of $\Delta b_{4}^{PS}-\Delta b_{3}^{LR}$
allowed by perturbativity as function of the scale $m_{PS}$ in GeV.
The different lines have been calculated for six different values
of $\Delta b_{3}^{LR}$. For this plot we assumed that $m_{R}=1$
TeV.}
\end{figure}

The additional $b_{i}$ coefficients for the regime $\left[m_{PS},m_{GUT}\right]$
are given by:
\begin{align}
\left(b_{4}^{PS},b_{2}^{PS},b_{R}^{PS}\right) & =\left(-6,1,1\right)+\left(\Delta b_{4}^{PS},\Delta b_{2}^{PS},\Delta b_{R}^{PS}\right)\label{eq:deltaBPS}
\end{align}
where, as before, the $\Delta b_{i}^{PS}$ include contributions from
superfields which are not part of the MSSM field content.

In this class of models, the unification scale is independent of the
LR one if the following condition is satisfied:{
\thinmuskip=1mu \medmuskip=1mu \thickmuskip=1mu
\begin{alignat}{1}
0= & \begin{pmatrix}\Delta b_{3}^{LR}-\Delta b_{2}^{LR}\\
\frac{3}{5}\Delta b_{R}^{LR}+\frac{2}{5}\Delta b_{B-L}^{LR}-\Delta b_{2}^{LR}-\frac{18}{5}
\end{pmatrix}^{T}.\begin{pmatrix}\begin{array}{rr}
2 & 3\\
-5 & 0
\end{array}\end{pmatrix}.\begin{pmatrix}\Delta b_{4}^{PS}-\Delta b_{2}^{PS}-3\\
\Delta b_{R}^{PS}-\Delta b_{2}^{PS}-12
\end{pmatrix}\,.
\end{alignat}
\thinmuskip=3mu
\medmuskip=4.0mu plus 2.0mu minus 4.0mu
\thickmuskip=5.0mu plus 5.0mu
}It is worth noting that also requiring $m_{PS}$ to be independent
of the LR scale would lead to the conditions \eqref{eq:SlidingScale_variantsLR1}--\eqref{eq:SlidingScale_variantsLR2},
which are the sliding conditions for LR models. This must be so, and
we can see it as follows: for some starting values of the three gauge
couplings at $m_{PS}$, the scales $m_{PS}$ and $m_{G}$ can be adjusted
such that the two splittings between the three gauge couplings are
reduced to zero at $m_{G}$. This fixes these scales, which must not
change even if $m_{R}$ is varied. As such $\alpha_{3}^{-1}\left(m_{PS}\right)-\alpha_{2}^{-1}\left(m_{PS}\right)$
and $\alpha_{3}^{-1}\left(m_{PS}\right)-\alpha_{R}^{-1}\left(m_{PS}\right)$
are also fixed, and they can be determined by running the MSSM up
to $m_{PS}$. The situation is therefore equal to the one that led
to the equalities in \eqref{eq:SlidingScale_variantsLR1}--\eqref{eq:SlidingScale_variantsLR2},
namely the splittings between the gauge couplings at some fixed scale
must be independent of $m_{R}$.

Since there are now two unknown scales in the problem, the maximum
$\Delta b_{i}^{X}$ allowed by perturbativity in one regime do not
depend only on the new scale $X$, but on the $\Delta b_{i}^{Y}$
in the other regime $Y$ as well. As an example, in figure \eqref{fig:SlidingScale_MaxDbPS}
we show the maximum $\Delta b_{4}^{PS}$ allowed by $\alpha_{G}^{-1}\ge0$
for different values of $\Delta b_{3}^{LR}$ and assuming that $m_{R}=10^{3}$
GeV and $m_{G}\geq10^{16}$ GeV. The dependence of max $\Delta b_{4}^{PS}$
on $m_{R}$ is rather weak, as long as $m_{R}$ does not approach
the GUT scale.

\begin{figure}[htb]
\centering{}\includegraphics{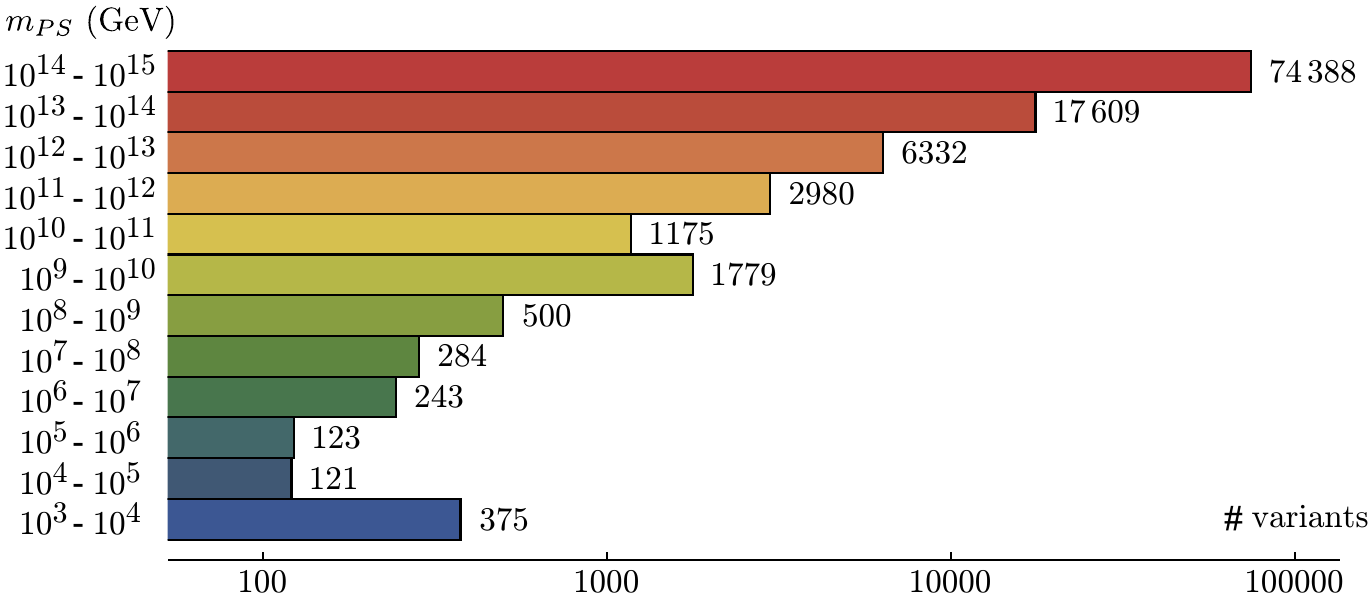} \caption{\label{fig:SlidingScale_ScanConfigsPS}The number of possible variants
of class-II models as a function of $m_{PS}$, assuming $m_{R}\approx1$
TeV and $10^{16}\textrm{ GeV}\lesssim m_{G}\lesssim2\times10^{18}$
GeV.}
\end{figure}

If all the $\Delta b$'s are to be bounded, an upper limit must be
placed on the PS scale. For example, if $m_{PS}\leq10^{6}$ GeV it
is possible to derive the following bounds:%
\footnote{In fact, the bounds shown here exclude a few variants with $m_{PS}<10^{6}$
GeV. This is because of the following: while in most cases the most
conservative assumption is to assume $m_{PS}$ as large as possible
in deriving these bounds ($=10^{6}$ GeV, leading to a smaller running
in the PS regime), there are some cases where this is not true. This
is a minor complication which nonetheless was taken into account in
our computation.%
}
\begin{align}
\Delta b_{2}^{PS}+\frac{3}{10}\Delta b_{2}^{LR} & <7.2\,,\\
\Delta b_{4}^{PS}+\frac{3}{10}\Delta b_{3}^{LR} & <10\,,\\
\frac{2}{5}\Delta b_{4}^{PS}+\frac{3}{5}\Delta b_{R}^{PS}+\frac{3}{10}\left(\frac{2}{5}\Delta b_{B-L}^{LR}+\frac{3}{5}\Delta b_{R}^{LR}\right) & <17\,.
\end{align}
The large values of $\max\Delta b^{LR}$ and $\max\Delta b^{PS}$
allow, in principle, a huge number of class-II variants to be constructed.
This is demonstrated in figure \eqref{fig:SlidingScale_ScanConfigsPS},
where we show the number of variants for an assumed $m_{R}\approx1$
TeV as a function of the scale $m_{PS}$ (up to $m_{PS}=10^{15}$
GeV). For larger values of the PS scale we have only scanned a finite
(though large) set of possible variants. Also note that these are
variants and not configurations. As in the case of class-I models,
almost all variants can be realized through several anomaly-free configurations.
The exhaustive list of variants (up to $m_{PS}=10^{15}$ GeV) containing
a total of 105909 possibilities can be found in \cite{website_sliding_models}.

With such a huge number of possible variants, we can discuss only
some general features. First of all, within the exhaustive set of
models with $m_{PS}=10^{15}$ GeV, there are a total of 1570 different
sets of $\Delta b_{i}^{LR}$, each of which can be completed by more
than one set of $\Delta b_{i}^{PS}$. Variants with the same set of
$\Delta b_{i}^{LR}$ but different completion of $\Delta b_{i}^{PS}$
have the same configuration in the LR-regime, but are associated to
a different value for $m_{PS}$ for fixed $m_{R}$. Thus, they have
in general different values for $\alpha_{B-L}$ and $\alpha_{R}$
at the LR scale and, as discussed in the following section, different
values of the invariants. For example, for the smallest possible values
of $\Delta b_{i}^{LR}$, $\Delta b_{i}^{LR}=(0,0,1,\nicefrac{3}{2})$,
there are 342 different completing sets of $\Delta b_{i}^{PS}$.

The simplest possible set of $\Delta b_{i}^{LR}$, $\Delta b_{i}^{LR}=(0,0,1,\nicefrac{3}{2})$,
corresponds to the configuration $\Phi_{1,1,2,-1}+\overline{\Phi}_{1,1,2,-1}$.
These fields are necessary to break $SU(2)_{R}\times U(1)_{B-L}\to U(1)_{Y}$.
Their presence in the LR regime requires that in the PS-regime we
have at least one set of copies of $\Psi_{4,1,2}+\overline{\Psi}_{4,1,2}$.
In addition, to break the PS group to the LR one, we need at least
one copy of $\Psi_{15,1,1}$. However, the combination $\Psi_{4,1,2}+\overline{\Psi}_{4,1,2}+\Psi_{15,1,1}$
is not sufficient to generate a sliding scale mechanism and the simplest
configuration that can do so is $3\Psi_{1,2,2}+4\Psi_{1,1,3}+\Psi_{4,1,2}+\overline{\Psi}_{4,1,2}+\Psi_{15,1,1}$,
leading to $\Delta b_{i}^{PS}=(6,3,15)$ and a very low possible value
of $m_{PS}=8.2$ TeV for $m_{R}=1$ TeV (see, however, the discussion
on leptoquarks below). The next possible completion for $\Phi_{1,1,2,-1}+\overline{\Phi}_{1,1,2,-1}$
is $3\Psi_{1,2,2}+5\Psi_{1,1,3}+\Psi_{4,1,2}+\overline{\Psi}_{4,1,2}+\Psi_{15,1,1}$,
with $\Delta b_{i}^{PS}=(6,3,17)$ and $m_{PS}=1.3\times10^{8}$ GeV
(for $m_{R}=1$ TeV), and so forth.

As noted already in subsection \ref{subsect:SlidingScale_LRm}, one
copy of $\Phi_{1,2,2,0}$ is not sufficient to produce a realistic
CKM matrix at tree-level. Thus, the minimal configuration of $\Phi_{1,1,2,-1}+\overline{\Phi}_{1,1,2,-1}$
relies on the possibility of generating all the departure of the CKM
matrix from unity with flavor violating soft masses \cite{Babu:1998tm}.
There are at least two possibilities to generate a non-trivial CKM
at tree-level, either by adding another $\Phi_{1,2,2,0}$ plus (at
least) one copy of $\Phi_{1,1,3,0}$, or with one copy of {}``vector-like
quarks'' $\Phi_{3,1,1,\frac{4}{3}}$ or $\Phi_{3,1,1,-\frac{2}{3}}$.
First consider the configuration $\Phi_{1,1,2,-1}+\overline{\Phi}_{1,1,2,-1}+\Phi_{1,2,2,0}+\Phi_{1,1,3,0}$,
which leads to $\Delta b_{i}^{LR}=(0,1,4,\nicefrac{3}{2})$. Since
$\Phi_{1,2,2,0}$ and $\Phi_{1,1,3,0}$ must come from $\Psi_{1,2,2}$
(or $\Psi_{15,2,2}$) and $\Psi_{1,1,3}$, respectively, the simplest
completion for this set of $\Delta b_{i}^{LR}$ is again $3\Psi_{1,2,2}+4\Psi_{1,1,3}+\Psi_{4,1,2}+\overline{\Psi}_{4,1,2}+\Psi_{15,1,1}$,
leading to $\Delta b_{i}^{PS}=(6,3,15)$ and $m_{PS}=5.4$ TeV, for
$m_{R}=1$ TeV. Again, many completions with different $\Delta b_{i}^{PS}$
exist for this set of $\Delta b_{i}^{LR}$.

The other possibility for generating CKM at tree-level, adding for
example a pair of $\Phi_{3,1,1,-\frac{2}{3}}+\overline{\Phi}_{3,1,1,-\frac{2}{3}}$,
corresponds to $\Delta b_{i}^{LR}=(1,0,1,\nicefrac{5}{2})$ and its
simplest PS-completion is $4\Psi_{1,2,2}+4\Psi_{1,1,3}+\Psi_{4,1,2}+\overline{\Psi}_{4,1,2}+\Psi_{6,1,1}+\Psi_{15,1,1}$,
with $\Delta b_{i}^{PS}=(7,4,16)$ and a $m_{PS}=4.6\times10^{6}$
TeV for $m_{R}=1$ TeV. In this case one can also find very low values
of $m_{PS}$. For example, adding a $\Phi_{1,2,2,0}$ to this LR-configuration
yields $\Delta b_{i}^{LR}=(1,1,2,\nicefrac{5}{2})$, and one finds
that, with the same $\Delta b_{i}^{PS}$, the PS scale is now $m_{PS}=8.3$
TeV, for $m_{R}=1$ TeV.

We note in passing that in our notation the original PS-class model
of \cite{DeRomeri:2011ie} corresponds to $\Delta b_{i}^{LR}=(1,2,10,4)$
and $\Phi_{1,1,2,-1}+\overline{\Phi}_{1,1,2,-1}+\Phi_{1,2,1,1}+\overline{\Phi}_{1,2,1,1}+\Phi_{1,2,2,0}+4\Phi_{1,1,3,0}+\Phi_{3,1,1,-\frac{2}{3}}+\overline{\Phi}_{3,1,1,-\frac{2}{3}}$,
completed by $\Delta b_{i}^{PS}=(9,5,13)$ with $\Psi_{4,1,2}+\overline{\Psi}_{4,1,2}+\Psi_{4,2,1}+\Psi_{4,2,1}+\Psi_{1,2,2}+4\Psi_{1,1,3}+\Psi_{6,1,1}+\Psi_{15,1,1}$.
The lowest possible $m_{PS}$, corresponding to $m_{R}=1$ TeV, is
$m_{PS}=2.4\times10^{8}$ GeV. Obviously, this example is not the
simplest construction in class-II. We also mention that, although
this would not have an impact in the $\beta$-coefficients, the superfield
$\Phi_{1,1,3,0}$ can be interpreted either as a {}``Higgs'' field
or as a {}``matter'' field, and in the original construction \cite{DeRomeri:2011ie}
the 4 copies of $\Phi_{1,1,3,0}$ were viewed as one $\Omega=\Phi_{1,1,3,0}$
({}``Higgs'') and three $\Sigma^{c}=\Phi_{1,1,3,0}$ ({}``matter'').
In this way, $\Omega^{c}$ can be used to generate the CKM matrix
at tree-level (together with the extra bi-doublet $\Phi_{1,2,2,0}$),
while the $\Sigma^{c}$ can be used to generate an inverse type-III
seesaw accounting for neutrino masses.

As figure \eqref{fig:SlidingScale_ScanConfigsPS} shows, there are
more than 600 variants in which $m_{PS}$ can in principle be lower
than $10^{6}$ GeV. However, such low PS scales are already constrained
by searches for rare decays, such as $B_{s}\to\mu^{+}\mu^{-}$. This
is because $\Psi_{15,1,1}$, which must be present in all our constructions
for the breaking of the PS group, contains two leptoquark states.
We will not study in detail leptoquark phenomenology \cite{Davidson:1993qk},
but it is worth mentioning that in a recent work \cite{Kuznetsov:2012ad}
an absolute lower bound $\approx40$ TeV on the mass of leptoquarks
within PS models was derived. There are 426 variants for which we
find $m_{PS}$ lower than this bound, provided that $m_{R}=$ 1 TeV.
Due to the sliding scale nature of our construction, this does not
mean that these models are ruled out by the lower limit found in \cite{Kuznetsov:2012ad}.
Instead, for these models one can calculate a lower limit on $m_{R}$
from the requirement that $m_{PS}=40$ TeV. It turns out that from
this requirement, and depending on the model, the minimum $m_{R}$
must be in the range $\left[1.3,27.7\right]$ TeV for these 426 variants.

\begin{figure}[htb]
\centering{}\includegraphics[width=0.5\linewidth]{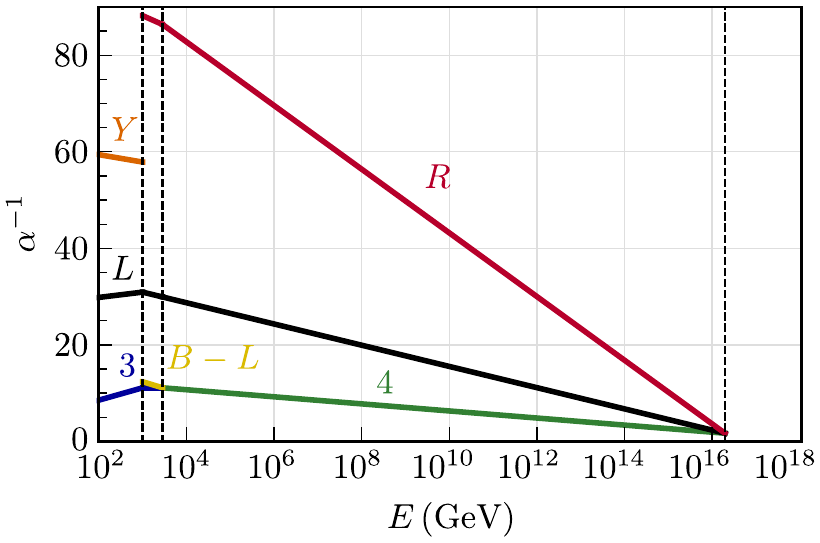}\includegraphics[width=0.5\linewidth]{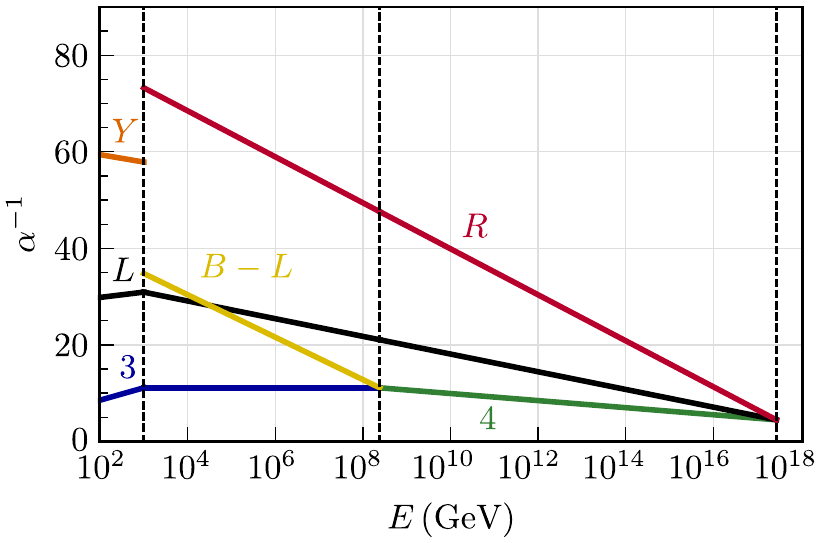}
\caption{\label{fig:LowPS_Scale}Gauge coupling unification for PS models with
$m_{R}=1$ TeV. In the plot to the left $\left(\Delta b_{3}^{LR},\ab\Delta b_{L}^{LR},\ab\Delta b_{R}^{LR},\ab\Delta b_{B-L}^{LR},\ab\Delta b_{4}^{PS},\ab\Delta b_{L}^{PS},\ab\Delta b_{R}^{PS}\right)=\left(3,\ab5,\ab10,\ab\nicefrac{3}{2},\ab8,\ab5,\ab17\right)$,
while the plot to the right corresponds to $\Delta b^{'}s=\left(3,\ab4,\ab12,\ab6,\ab8,\ab4,\ab12\right)$.
Note that in the left plot $m_{PS}$ is lower than 40 TeV therefore,
in order to respect leptoquark mass bounds, the $m_{R}$ scale must
be raised.}
\end{figure}

Two variants can be seen in figure \eqref{fig:LowPS_Scale}: we have
chosen one example with a very low $m_{PS}$ (left) and another with
an intermediate $m_{PS}$ (right). In both graphs $m_{R}$ was chosen
to be 1 TeV, and we note that in the example on the left this leads
to $m_{PS}<40$ TeV, therefore the scale $m_{R}$ must be higher.
Note also that, unlike in class-I models, in class-II models the GUT
scale is no longer fixed to the MSSM value $m_{G}\approx2\times10^{16}$
GeV.

\subsection{Models with an $U(1)_{R}\times U(1)_{B-L}$ intermediate scale}

Finally, we consider models where there is an additional intermediate
$U(1)_{R}\times U(1)_{B-L}$ phase that follows the $SU(2)_{R}\times U(1)_{B-L}$
stage. The field content relevant to this model is given in table
\eqref{tab:List_of_LR_fields_U1} of appendix \ref{chap:Lists_of_superfields_in_LR_models}.
In this case, the original $SO(10)$ is broken down to the MSSM in
three steps:
\begin{eqnarray}
SO(10) & \to & SU(3)_{c}\times SU(2)_{L}\times SU(2)_{R}\times U(1)_{B-L}\nonumber \\
 & \to & SU(3)_{c}\times SU(2)_{L}\times U(1)_{R}\times U(1)_{B-L}\to{\rm MSSM}\,\textrm{.}
\end{eqnarray}
The first step is achieved in the same way as in class-I models. The
subsequent breaking $SU(2)_{R}\times U(1)_{B-L}\rightarrow U(1)_{R}\times U(1)_{B-L}$
is triggered by $\Phi_{5}=\Phi_{1,1,3,0}$ and the last one requires
$\Phi_{4}^{'}=\Phi'_{1,1,\frac{1}{2},-1}$, $\Phi_{20}^{'}=\Phi'_{1,1,1,-2}$
or their conjugates.\\

\noindent As mentioned previously in chapter \ref{chap:U1_mixing_paper},
theories with more that one $U(1)$ gauge factor give rise to $U(1)$-mixing,
and to account for it the $U(1)$ gauge couplings should be seen as
a matrix. In the present case, 
\begin{align}
\boldsymbol{G} & =\left(\begin{array}{cc}
g_{RR} & g_{RX}\\
g_{XR} & g_{XX}
\end{array}\right)\,,
\end{align}
where an $X$ is used instead of $B-L$ because of the $\sqrt{\nicefrac{3}{8}}$
normalization factor mentioned previously. From section \ref{sect:numericsMRV},
we recall that we can build a generalization of $\alpha$, which is
$\boldsymbol{A}=\nicefrac{\boldsymbol{G}\boldsymbol{G}^{T}}{4\pi}$,
and whose evolution under the renormalization group is controlled
by the anomalous dimensions matrix $\boldsymbol{\gamma}$ (see equations
\eqref{eq:U1mixing_gamma} and \eqref{eq:U1mixing_Amatrix}, as well
as \cite{DeRomeri:2011ie}). Taking the MSSM's field content, we find
that
\begin{align}
\boldsymbol{\gamma} & =\left(\begin{array}{cc}
7 & 0\\
0 & 6
\end{array}\right)\,.
\end{align}
Note again that to ensure the canonical normalization of the $B-L$
charge within the $SO(10)$ framework, $\boldsymbol{\gamma}$ should
be normalized as $\boldsymbol{\gamma}^{\textrm{can}}=\boldsymbol{N}\boldsymbol{\gamma}^{\textrm{usual}}\boldsymbol{N}$,
where $\boldsymbol{N}=\text{diag}(1,\sqrt{\nicefrac{3}{8}})$---compare
with equation \eqref{eq:U1mixing_Nmatrix}. The gauge coupling $g_{Y}$
of the $U(1)_{Y}$ group of the MSSM is obtainable from the following
expression, which is valid at the $m_{B-L}$ energy scale:
\begin{align}
\alpha_{Y}^{-1} & =\boldsymbol{p}_{\boldsymbol{Y}}^{T}\cdot\boldsymbol{A}^{-1}\cdot\boldsymbol{p_{Y}}\,.
\end{align}
For completeness, we recall here that $\boldsymbol{p}_{\boldsymbol{Y}}^{T}=\left(\sqrt{\nicefrac{3}{5}},\sqrt{\nicefrac{2}{5}}\right)$;
for generic details of the matching procedure in models with multiple
$U(1)$'s, see appendix \ref{chap:Matching_conditions}.

\begin{figure}[htb]
\centering{} \includegraphics[width=0.5\linewidth]{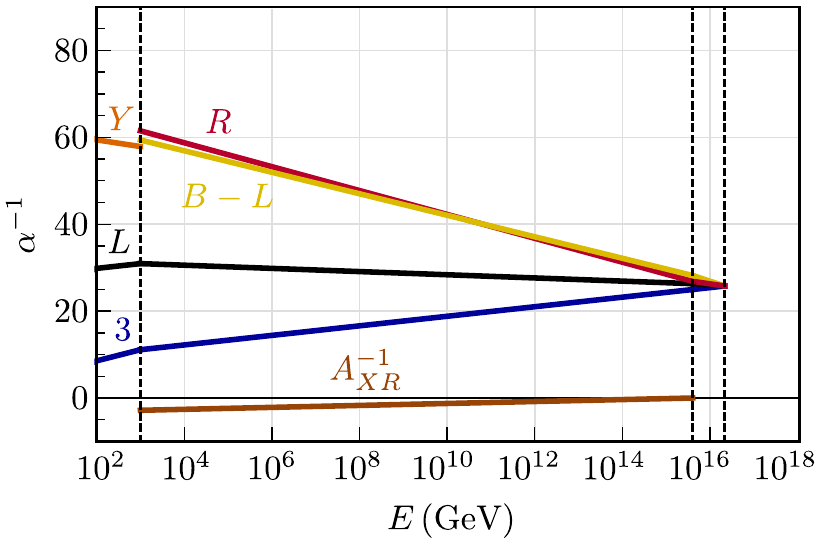}\includegraphics[width=0.5\linewidth]{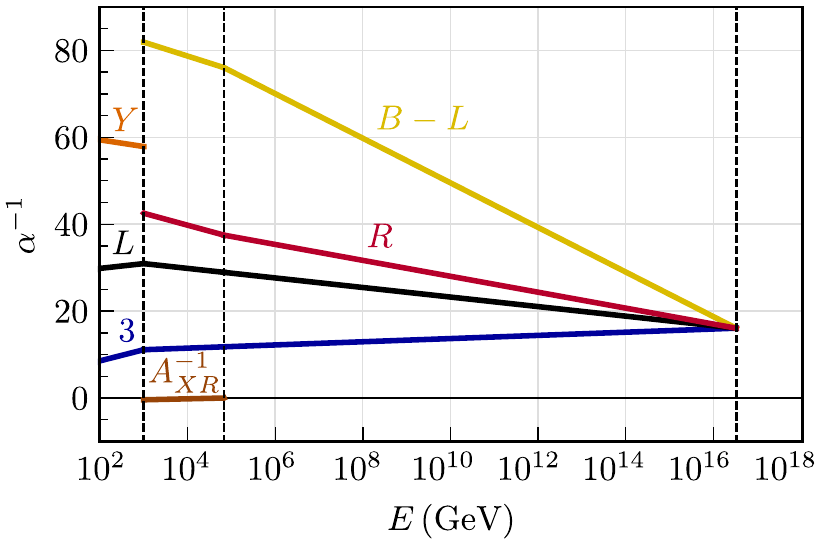}
\caption{\label{fig:U1MixingLowScale}Gauge coupling unification in models
with an $U(1)_{R}\times U(1)_{B-L}$ intermediate scale, for $m_{R}=10^{3}$
GeV. Left: $(\Delta b_{3}^{LR},\ab\Delta b_{L}^{LR},\ab\Delta b_{R}^{LR},\ab\Delta b_{B-L}^{LR},\ab\Delta b_{3}^{B-L},\ab\Delta b_{L}^{B-L},\ab\Delta\boldsymbol{\gamma}_{RR},\ab\Delta\boldsymbol{\gamma}_{XR},\ab\Delta\boldsymbol{\gamma}_{XX})=(0,\ab1,\ab3,\ab3,\ab0,\ab0,\ab1/2,\ab-\sqrt{\nicefrac{3}{8}},\ab\nicefrac{3}{4})$;
Right: ($2,\ab2,\ab4,\ab8,\ab2,\ab2,\ab1/2,\ab-\sqrt{\nicefrac{3}{8}},\ab\nicefrac{11}{4}$).
The brown line, which appears close to zero in the $U(1)_{R}\times U(1)_{B-L}$
regime, is the running of the off-diagonal element of the matrix $\boldsymbol{A}^{-1}$,
measuring the size of the $U(1)$-mixing in the model. The running
of the diagonal components $\left(1,1\right)$ and $\left(2,2\right)$
of this matrix in the $U(1)_{R}\times U(1)_{B-L}$ regime are given
by the red and yellow lines, respectively.}
\end{figure}

The additional $\beta$-coefficients for the running step $\left[m_{B-L},m_{R}\right]$
are given by{
\thinmuskip=1mu
\medmuskip=1mu
\thickmuskip=1mu
\begin{align}
\left(b_{3}^{B-L},b_{2}^{B-L},\boldsymbol{\gamma}_{RR}^{B-L},\boldsymbol{\gamma}_{XR}^{B-L},\boldsymbol{\gamma}_{XX}^{B-L}\right) & =\left(-3,1,6,0,7\right)\nonumber \\
 & +\left(\Delta b_{3}^{B-L},\Delta b_{2}^{B-L},\Delta\boldsymbol{\gamma}_{RR},\Delta\boldsymbol{\gamma}_{XR},\Delta\boldsymbol{\gamma}_{XX}\right)\,.
\end{align}
\thinmuskip=3mu
\medmuskip=4.0mu plus 2.0mu minus 4.0mu
\thickmuskip=5.0mu plus 5.0mu
}Similarly to what was done in the previous class of models, we consider
$m_{B-L}=10^{3}$ GeV, $m_{G}\ge10^{16}$ GeV, $m_{R}\leq10^{6}$
GeV, and extract bounds for the $\Delta b$:
\begin{align}
\Delta b_{2}^{LR} & +\frac{3}{10}\Delta b_{2}^{B-L}<7.1\,,\\
\Delta b_{3}^{LR} & +\frac{3}{10}\Delta b_{3}^{B-L}<6.9\,,\\
\frac{3}{5}\Delta b_{R}^{LR}+\frac{2}{5}\Delta b_{B-L}^{LR}+ & \frac{3}{10}\boldsymbol{p}_{\boldsymbol{Y}}^{T}\cdot\Delta\boldsymbol{\gamma}\cdot\boldsymbol{p_{Y}}<10.8\,.
\end{align}
Even with this restriction in the scales, we found 15610 solutions,
more than in the PS case with similar conditions, due to the fact
that there are more $\Delta b^{'}s$ that can be varied to obtain
solutions. The qualitative features of the running of the gauge couplings
are shown for two cases in figure \eqref{fig:U1MixingLowScale}. In
these two examples, $(\Delta b_{3}^{LR},\Delta b_{L}^{LR},\Delta b_{R}^{LR},\Delta b_{B-L}^{LR},\Delta b_{3}^{B-L},\Delta b_{L}^{B-L},\Delta\boldsymbol{\gamma}_{RR},\Delta\boldsymbol{\gamma}_{XR},\Delta\boldsymbol{\gamma}_{XX})$
is equal to $(0,1,3,3,0,0,1/2,-\sqrt{\nicefrac{3}{8}},\nicefrac{3}{4})$
(left) and ($2,2,4,8,2,2,1/2,-\sqrt{\nicefrac{3}{8}},\nicefrac{11}{4}$)
(right). The former corresponds to the minimal configuration $\Phi'_{1,1,\nicefrac{1}{2},-1}+\overline{\Phi}'_{1,1,\nicefrac{1}{2},-1}$,
in the lower energy regime, and $\Phi_{1,1,2,-1}+\overline{\Phi}_{1,1,2,-1}+\Phi_{1,1,3,0}+\Phi_{1,2,1,1}+\overline{\Phi}_{1,2,1,1}$
in the higher one (the LR-symmetric regime). The latter corresponds
to $\Phi'_{1,1,\nicefrac{1}{2},-1}+\overline{\Phi}'_{1,1,\nicefrac{1}{2},-1}+\Phi'_{1,3,0,0}+2\Phi'_{3,1,1,-\nicefrac{2}{3}}+2\overline{\Phi}'_{3,1,1,-\nicefrac{2}{3}}$
and $2\Phi_{1,1,2,-1}+2\overline{\Phi}_{1,1,2,-1}+\Phi_{1,1,3,0}+\Phi_{1,3,1,0}+\Phi_{1,1,1,2}+\overline{\Phi}_{1,1,1,2}+2\Phi_{3,1,1,-\nicefrac{2}{3}}+2\overline{\Phi}_{3,1,1,-\nicefrac{2}{3}}$,
respectively.

For models in this class, the sliding condition requires that the
unification scale is independent of $m_{B-L}$, and this happens only
if{
\thinmuskip=1mu
\medmuskip=1mu
\thickmuskip=1mu
\begin{alignat}{1}
0 & =\begin{pmatrix}\Delta b_{3}^{B-L}-\Delta b_{2}^{B-L}\\
\boldsymbol{p}_{\boldsymbol{Y}}^{T}\cdot\Delta\boldsymbol{\gamma}\cdot\boldsymbol{p_{Y}}-\Delta b_{2}^{B-L}
\end{pmatrix}^{T}.\begin{pmatrix}\begin{array}{rr}
0 & 1\\
-1 & 0
\end{array}\end{pmatrix}.\begin{pmatrix}\Delta b_{3}^{LR}-\Delta b_{2}^{LR}\\
\frac{3}{5}\Delta b_{R}^{LR}+\frac{2}{5}\Delta b_{B-L}^{LR}-\Delta b_{2}^{LR}-\frac{18}{5}
\end{pmatrix}\,.
\end{alignat}
\thinmuskip=3mu
\medmuskip=4.0mu plus 2.0mu minus 4.0mu
\thickmuskip=5.0mu plus 5.0mu
}Similarly to PS models, in this class of models the higher intermediate
scale ($m_{R}$) depends in general on the lower one ($m_{B-L}$).
However, there is also a special condition in the present case which
makes both $m_{R}$ and $m_{G}$ simultaneously independent of $m_{B-L}$:
\begin{align}
\Delta b_{3}^{LR} & =\Delta b_{2}^{LR}=\boldsymbol{p}_{\boldsymbol{Y}}^{T}\cdot\Delta\boldsymbol{\gamma}\cdot\boldsymbol{p_{Y}}\,.
\end{align}
Models of this kind are, for example, those with $\Delta b_{3}=0$
and large $m_{R}$ ($\gtrsim10^{13}$ GeV). One such case is given
in \cite{DeRomeri:2011ie}, where $m_{R}\approx4\times10^{15}$ GeV.

\section{\label{sect:SlidingScale_invariants}Invariants}

\subsection{\label{sect:SlidingScale_leadloginv}Leading-Log RGE Invariants}

In this subsection we briefly recall the basic definitions \cite{DeRomeri:2011ie}
for the calculation of the \textit{invariants} \cite{Buckley:2006nv,Hirsch:2008gh,Esteves:2010ff}.
In mSUGRA, since gaugino masses scale with the square of the gauge
couplings, the requirement of GCU fixes the gaugino masses at the
low scale:
\begin{align}
M_{i}\left(m_{SUSY}\right) & =\frac{\alpha_{i}\left(m_{SUSY}\right)}{\alpha_{G}}M_{1/2}\,.\label{eq:SlidingScale_gaugino}
\end{align}
Neglecting the Yukawa and soft trilinear couplings for the soft mass
parameters of the first two sfermions generations, one can write
\begin{alignat}{1}
m_{\widetilde{f}}^{2}-m_{0}^{2} & =\frac{M_{1/2}^{2}}{2\pi\alpha_{G}^{2}}\sum_{R_{j}}\sum_{i=1}^{N}c_{i}^{f,R_{j}}\alpha_{i-}^{R_{j}}\alpha_{i+}^{R_{j}}\left(\alpha_{i-}^{R_{j}}+\alpha_{i+}^{R_{j}}\right)\log\frac{m_{+}^{R_{j}}}{m_{-}^{R_{j}}}\,.\label{eq:SlidingScale_scalar}
\end{alignat}

\begin{table}[tbph]
\begin{centering}
\begin{tabular}{cccccccccccccc}
\toprule 
 &  & \multicolumn{3}{c}{MSSM} &  & \multicolumn{4}{c}{LR} &  & \multicolumn{3}{c}{PS}\tabularnewline
\cmidrule{3-5} \cmidrule{7-10} \cmidrule{12-14} 
 &  & $c_{Y}$ & $c_{L}$ & $c_{3}$ &  & $c_{B-L}$ & $c_{R}$ & $c_{L}$ & $c_{3}$ &  & $c_{R}$ & $c_{L}$ & $c_{4}$\tabularnewline
\midrule
$\widetilde{Q}$ &  & $\frac{1}{30}$ & $\frac{3}{2}$ & $\frac{8}{3}$ &  & $\frac{1}{12}$ & $0$ & $\frac{3}{2}$ & $\frac{8}{3}$ &  & $0$ & $\frac{3}{2}$ & $\frac{15}{4}$\tabularnewline
$\widetilde{U}$ &  & $\frac{8}{15}$ & $0$ & $\frac{8}{3}$ &  & $\frac{1}{12}$ & $\frac{3}{2}$ & $0$ & $\frac{8}{3}$ &  & $\frac{3}{2}$ & $0$ & $\frac{15}{4}$\tabularnewline
$\widetilde{D}$ &  & $\frac{2}{15}$ & $0$ & $\frac{8}{3}$ &  & $\frac{1}{12}$ & $\frac{3}{2}$ & $0$ & $\frac{8}{3}$ &  & $\frac{3}{2}$ & $0$ & $\frac{15}{4}$\tabularnewline
$\widetilde{L}$ &  & $\frac{3}{10}$ & $\frac{3}{2}$ & $0$ &  & $\frac{3}{4}$ & $0$ & $\frac{3}{2}$ & $0$ &  & $0$ & $\frac{3}{2}$ & $\frac{15}{4}$\tabularnewline
$\widetilde{E}$ &  & $\frac{6}{5}$ & $0$ & $0$ &  & $\frac{3}{4}$ & $\frac{3}{2}$ & $0$ & $0$ &  & $\frac{3}{2}$ & $0$ & $\frac{15}{4}$\tabularnewline
\bottomrule
\end{tabular}
\par\end{centering}

\caption{\label{tab:SlidingScale_ci}Values of the $c_{i}^{f,R_{j}}$ coefficients
entering equation \eqref{eq:SlidingScale_scalar}, for $R_{j}=$MSSM,
LR, PS and $f=\widetilde{E},\ab\,\widetilde{L},\ab\,\widetilde{D},\ab\,\widetilde{U},\ab\,\widetilde{Q}$.
Values for the $U(1)_{R}\times U(1)_{B-L}$ regime are not shown,
since equation \eqref{eq:SlidingScale_m2f_U1mixing} should be used
instead. }

\end{table}
Here, the sum over $R_{j}$ runs over the different regimes in the
models under consideration, while the sum over $i$ runs over all
gauge groups in a given regime; $m_{+}^{R_{j}}$ and $m_{-}^{R_{j}}$
are the upper and lower boundaries of the $R_{j}$ regime and $\alpha_{i+}^{R_{j}}$,
$\alpha_{i-}^{R_{j}}$ are the values of the gauge coupling of group
$i$, $\alpha_{i}$, at these scales. As for the coefficients $c_{i}$,
they are twice the quadratic Casimir of the field representations
under each gauge group $i$---see table \eqref{tab:SlidingScale_ci}.
As discussed in chapter \ref{chap:U1_mixing_paper}, in the presence
of multiple $U(1)$ gauge groups the RGEs are different, and this
leads to a generalization of equation \eqref{eq:SlidingScale_scalar}
for the $U(1)$-mixing phase \cite{DeRomeri:2011ie}. Here we just
quote the final result (with a minor correction to the one shown in
this last reference), ignoring the non-$U(1)$ groups:
\begin{alignat}{1}
m_{\widetilde{f}-}^{2}-m_{\widetilde{f}+}^{2} & =\frac{M_{1/2}^{2}}{\pi\alpha_{G}^{2}}\boldsymbol{Q}_{\boldsymbol{f}}^{T}\boldsymbol{A_{-}}\left(\boldsymbol{A_{-}}+\boldsymbol{A_{+}}\right)\boldsymbol{A_{+}Q_{f}}\log\frac{m_{+}}{m_{-}}\,,\label{eq:SlidingScale_m2f_U1mixing}
\end{alignat}
where $m_{+}$ and $m_{-}$ are the boundary scales of the $U(1)$-mixing
regime, and $\boldsymbol{A_{+}}$, $\boldsymbol{A_{-}}$ are the $\boldsymbol{A}$
matrix which generalizes $\alpha$, evaluated in these two limits.
Likewise, $\widetilde{m}_{f+}^{2}$ and $\widetilde{m}_{f-}^{2}$
are the values of the soft mass parameter of the sfermion $\tilde{f}$
at these two energy scales. The equation above is a good approximation
to the result obtained by integration of the following 1-loop RGE
for the soft masses, which assumes unification of gaugino masses and
gauge coupling constants:
\begin{alignat}{1}
\frac{d}{dt}m_{\widetilde{f}}^{2} & =-\frac{4M_{1/2}^{2}}{\alpha_{G}^{2}}\boldsymbol{Q}_{\boldsymbol{f}}^{T}\boldsymbol{A}^{3}\boldsymbol{Q_{f}}\,,\label{eq:SlidingScale_DerivariveOfm2f}
\end{alignat}
where $t=\nicefrac{\log\left(\nicefrac{E}{E_{0}}\right)}{2\pi}$.
Note that in the limit where the $U(1)$-mixing phase extends all
the way up to $m_{G}$, the $\boldsymbol{A}$ matrices measured at
different energy scales will always commute amongst themselves, and
therefore equation \eqref{eq:SlidingScale_m2f_U1mixing} presented
here matches the one in \cite{DeRomeri:2011ie} since both are exact
integrations of \eqref{eq:SlidingScale_DerivariveOfm2f}. However,
if this is not the case, it is expected that there will be a small
discrepancy between the two approximations, which is nevertheless
numerically small and therefore negligible.

From the five soft sfermion mass parameters of the MSSM and one of
the gaugino masses, it is possible to form four different combinations
that, at 1-loop level in the leading-log approximation, do not depend
on the values of $m_{0}$ and $M_{1/2}$, and are therefore called
invariants:
\begin{align}
LE & \equiv\frac{m_{\widetilde{L}}^{2}-m_{\widetilde{E}}^{2}}{M_{1}^{2}}\,,\qquad & QE & \equiv\frac{m_{\widetilde{Q}}^{2}-m_{\widetilde{E}}^{2}}{M_{1}^{2}}\,,\label{eq:SlidingScale_definv1}\\
DL & \equiv\frac{m_{\widetilde{D}}^{2}-m_{\widetilde{L}}^{2}}{M_{1}^{2}}\,,\qquad & QU & \equiv\frac{m_{\widetilde{Q}}^{2}-m_{\widetilde{U}}^{2}}{M_{1}^{2}}\,.\label{eq:SlidingScale_definv2}
\end{align}
These are 4 numbers which carry information on the particle content
and the gauge group of intermediate stages between the low energy
MSSM and full unification, as shown by equations \eqref{eq:SlidingScale_scalar}
and \eqref{eq:SlidingScale_m2f_U1mixing}. We will not discuss in
detail errors in the calculation of these quantities, referring instead
to \cite{DeRomeri:2011ie}, and for classical $SU(5)$ based SUSY
seesaw models to \cite{Hirsch:2008gh,Esteves:2010ff}.

We close this section by asserting that some model variants which
were presented in the previous section will not be testable by measurements
involving invariants at the LHC. According to \cite{Baer:2012vr},
the LHC at $\sqrt{s}=14$ TeV will be able to explore SUSY masses
up to $m_{\widetilde{g}}\sim3.2$ TeV ($3.6$ TeV) for $m_{\widetilde{q}}\approx m_{\widetilde{g}}$
and of $m_{\widetilde{g}}\sim1.8$ TeV ($2.3$ TeV) for $m_{\widetilde{q}}\gg m_{\widetilde{g}}$
with 300 fb$^{-1}$ (3000 fb$^{-1}$). The LEP limit on the chargino,
$m_{\chi^{\pm}}>105$ GeV \cite{Beringer_mod:1900zz}, translates
into a lower bound for $M_{1/2}$, with the precise value depending
on $\Delta b$. For the class-I models with $\Delta b=5$, this leads
to $M_{1/2}\gtrsim1.06$ TeV. One can assume conservatively $m_{0}=0$
GeV and calculate from this lower bound on $M_{1/2}$ a lower limit
on the expected squark masses in the different variants. All variants
with squark masses above the expected reach of the LHC-14 will then
not be testable via measurements of the invariants, and this means
that all single scale models with $\Delta b=5$, for example, will
be untestable.

For completeness we mention that if we take the present LHC limit
on the gluino, $m_{\tilde{g}}\gtrsim1.1$ TeV \cite{ATLAS-CONF-2012-109},
this will translate into a lower limit $M_{1/2}\gtrsim4.31$ TeV for
$\Delta b=5$. We have also checked that models with $\Delta b=4$
can still have squarks with masses testable at LHC, even for the more
recent LHC bound on the gluino mass (see figure \eqref{fig:Introduction_ATLAS-95=000025-confidence}).

\subsection{\label{sub:SlidingScale_invclasses}Classification of invariants}

\noindent For a given model, the invariants defined in equations \eqref{eq:SlidingScale_definv1}--\eqref{eq:SlidingScale_definv2}
differ from the mSUGRA values, and the deviations can be either positive
or negative once new superfields (and/or gauge groups) are added to
the MSSM. The mSUGRA limit is reached in our models when the intermediate
scales are equal to $m_{G}$. However, it should be noted that, in
general, when there are two intermediate scales, the smallest one
(henceforth called $m_{-}$) cannot be pushed all the way up to the
unification scale. Therefore, in those cases, the invariants measured
at the highest possible $m_{-}$ are slightly different from the mSUGRA
invariants.

With this in mind, for each variant of our models, we considered whether
the invariants for $\min\, m_{-}$(=$m_{SUSY}$) are larger or smaller
than for $\max\, m_{-}$, which tends to be within one or two orders
of magnitude of $m_{G}$. With four invariants there are a priori
$2^{4}=16$ possibilities, and in table \eqref{tab:classes} each
of them is assigned a number.

\begin{table}[htb]
\setlength{\tabcolsep}{5pt}

\begin{centering}
\begin{tabular}{ccccccccccccccccc}
\toprule 
Set \# & 1 & 2 & 3 & 4 & 5 & 6 & 7 & 8 & 9 & 10 & 11 & 12 & 13 & 14 & 15 & 16\tabularnewline
\midrule 
$\Delta LE$ & + & + & + & + & + & + & + & + & $-$ & $-$ & $-$ & $-$ & $-$ & $-$ & $-$ & $-$\tabularnewline
$\Delta QE$ & + & + & + & $-$ & + & $-$ & $-$ & $-$ & + & + & + & $-$ & + & $-$ & $-$ & $-$\tabularnewline
$\Delta DL$ & + & + & $-$ & + & $-$ & + & $-$ & $-$ & + & + & $-$ & + & $-$ & + & $-$ & $-$\tabularnewline
$\Delta QU$ & + & $-$ & + & + & $-$ & $-$ & + & $-$ & + & $-$ & + & + & $-$ & $-$ & + & $-$\tabularnewline
\midrule 
Class-I  & {\color{green}\cmark}  & {\color{green}\cmark}  & {\color{red}\xmark}  & {\color{red}\xmark}  & {\color{red}\xmark}  & {\color{red}\xmark}  & {\color{red}\xmark}  & {\color{red}\xmark}  & {\color{red}\xmark}  & {\color{green}\cmark}  & {\color{red}\xmark}  & {\color{red}\xmark}  & {\color{red}\xmark}  & {\color{green}\cmark}  & {\color{red}\xmark}  & {\color{red}\xmark} \tabularnewline
Class-II & {\color{green}\cmark}  & {\color{green}\cmark}  & {\color{green}\cmark}  & {\color{red}\xmark}  & {\color{red}\xmark}  & {\color{green}\cmark}  & {\color{green}\cmark}  & {\color{green}\cmark}  & {\color{red}\xmark}  & {\color{green}\cmark}  & {\color{red}\xmark}  & {\color{red}\xmark}  & {\color{red}\xmark}  & {\color{green}\cmark}  & {\color{red}\xmark}  & {\color{green}\cmark} \tabularnewline
Class-III & {\color{green}\cmark}  & {\color{green}\cmark}  & {\color{green}\cmark} & {\color{red}\xmark}  & {\color{red}\xmark}  & {\color{green}\cmark}  & {\color{green}\cmark}  & {\color{green}\cmark}  &  & {\color{green}\cmark}  &  & {\color{red}\xmark}  & {\color{red}\xmark}  & {\color{green}\cmark}  &  & \tabularnewline
\bottomrule
\end{tabular}
\par\end{centering}

\setlength{\tabcolsep}{6pt}\caption{The 16 different combinations of signs for 4 invariants. We assign
a {}``$+$'' if the corresponding invariant, when the lowest intermediate
scale is set to $m_{SUSY}$, is larger than its value when this scale
is maximized, and {}``$-$'' otherwise. As discussed in the text,
only 9 of the 16 different sign combinations can be realized in the
class-I and class-II models (sets 1, 2, 3, 6, 7, 8, 10, 14 and 16).
In fact, class-I models always fall on sets 1, 2, 10 and 14, and this
can be proven with simple arguments (see text). On the other hand,
there are class-II models will all 9 invariant sign combinations.
Class-III models can conceivably achieve three more sign combinations
(sets 9, 11, 15), but we did not find any such case in a non-exhaustive
search (also, no models in set 16 were found).}

\label{tab:classes} 
\end{table}

However, it is easy to demonstrate that not all of the 16 sets can
be realized in the three classes of models we consider. This can be
understood as follows. If all sfermions have a common mass at the
GUT scale ($m_{0}$), then one can show that
\begin{align}
m_{\widetilde{Q}}^{2}-2m_{\widetilde{U}}^{2}+m_{\widetilde{D}}^{2}-m_{\widetilde{L}}^{2}+m_{\widetilde{E}}^{2} & =0\label{eq:SlidingScale_sumrule}
\end{align}
holds independent of the energy scale at which soft masses are evaluated.
This relation is general, regardless of the combination of intermediate
scales that we may consider. It is a straightforward consequence of
the charge assignments of the Standard Model fermions and can be easily
checked by calculating the Dynkin coefficients of the $Q$, $U$,
$D$, $L$ and $E$ representation in the different regimes. To be
precise, the combination $\left(\textrm{row 1}\right)-2\left(\textrm{row 2}\right)+\left(\textrm{row 3}\right)-\left(\textrm{row 4}\right)+\left(\textrm{row 5}\right)$
of table \eqref{tab:SlidingScale_ci} yields the null row.%
\footnote{Models with a $U(1)_{R}\times U(1)_{B-L}$ stage also obey this relation:
to prove this, we observe that if table \eqref{tab:SlidingScale_ci}
is extended to include 3 more columns, with $q_{R}^{2}$, $q_{X}^{2}$
and $q_{R}q_{X}$ of the different MSSM fields in this regime, we
still have $\left(\textrm{row 1}\right)-2\left(\textrm{row 2}\right)+\left(\textrm{row 3}\right)-\left(\textrm{row 4}\right)+\left(\textrm{row 5}\right)=0$.%
} In terms of the invariants, this relation becomes
\begin{align}
QE & =DL+2QU\,,\label{eq:sumruleinv}
\end{align}
which means that only three of the four invariants are independent.
From equation \eqref{eq:sumruleinv} it is clear that if $\Delta DL$
and $\Delta QU$ are both positive (negative), then $\Delta QE$ must
be also positive (negative). This immediately excludes the sets 4,
5, 12 and 13.

Equation \eqref{eq:SlidingScale_sumrule} is extremely general, in
the sense that any unified model with a combination of stages $\left\{ \textrm{MSSM},\,\textrm{BL},\,\textrm{LR},\,\textrm{PS}\right\} $
will obey it.%
\footnote{Even models with a $SU(5)$ gauge group follow it (assuming the usual
representations assignment of $Q$, $U$, $D$, $L$ and $E$).%
} However, given a restricted set of these stages, one can calculate
other relations among the Dynkin indices of the MSSM sfermions. In
particular, if we ignore the $U(1)$-mixing in the BL regime, and
the MSSM as well, we get one additional relation:
\begin{align}
QU & =LE\,.\label{LRsumrule}
\end{align}
On one hand, the fact that this relation is broken by $U(1)$-mixing
does not appear to be very important here, since the effect is somewhat
small (see figure \eqref{fig:invPS} below); On the other hand, the
MSSM also breaks the relation, but we can account for this since the
field content of the MSSM is known. In particular, the correction
to equation \eqref{LRsumrule} is independent of the variant/$\Delta b$'s
under consideration:
\begin{align}
QU & =LE+f\left(m_{R}\right)\,,\label{LRsumruleApp}
\end{align}
with
\begin{align}
f\left(m_{R}\right) & =\frac{2}{33}\left\{ \left[\frac{33}{10\pi}\alpha_{1}^{MSSM}\log\left(\frac{m_{R}}{m_{SUSY}}\right)-1\right]^{-2}-1\right\} \,.
\end{align}
Here, $\alpha_{1}^{MSSM}$ is the value of $\alpha_{1}$ at $m_{SUSY}$.
It is easy to see that $f(m_{R})$ is always small ($<0.3$), positive
and that it vanishes when $m_{R}\rightarrow m_{SUSY}$. Note that
$m_{R}$ should be seen as the upper energy limit of validity of the
MSSM as an effective field theory; for simplicity we assumed with
this nomenclature that the stage to follows is a $LR$ regime, but
this needs not be the case: in the class-III models it is $m_{B-L}$.

Equation \eqref{LRsumruleApp} can be used to eliminate three more
cases from table \eqref{tab:classes}. Since $f(m_{R})$ is non-negative
and an increasing function of $m_{R}$, it follows that one always
has $\Delta QU\leq\Delta LE$, so it is not possible to have $\Delta LE=-$
and $\Delta QU=+$. This excludes three additional sets from table
\eqref{tab:classes}: 9, 11 and 15, leaving a total of 9 possible
sets. This last statement might conceivably not apply to some models
with $U(1)$-mixing, but it should be noted that for this to happen
the somewhat small mixing effect must break relation \eqref{LRsumruleApp}
with a term which decreases with $m_{R}$ on its left side, such that
$\Delta QU>\Delta LE$. In other words, the small $U(1)$-mixing effect
must dominate over the MSSM's effect which is encoded by the monotonic
increasing function  $f$, and $QU$, $LE$ should be almost constant
with variations of $m_{R}$. See figure \eqref{fig:invBL} below for
two examples where this clearly does not happen. For completeness,
using some simplifications we can write down an approximate expression
which corrects equation \eqref{LRsumruleApp} with this $U(1)$-mixing
effect in class-III models:
\begin{align}
QU & \approx LE+f\left(m_{B-L}\right)-\frac{1}{\pi\sqrt{6}}\frac{\boldsymbol{A}_{RX}^{-1}\left(m_{B-L}\right)}{\boldsymbol{A}_{RR}^{-1}\left(m_{LR}\right)\boldsymbol{A}_{XX}^{-1}\left(m_{LR}\right)}\nonumber \\
 & \quad\quad\times\left[1+2\frac{\boldsymbol{A}_{RR}^{-1}\left(m_{LR}\right)+\boldsymbol{A}_{XX}^{-1}\left(m_{LR}\right)}{\boldsymbol{A}_{RR}^{-1}\left(m_{B-L}\right)+\boldsymbol{A}_{XX}^{-1}\left(m_{B-L}\right)}\right]\log\left(\frac{m_{R}}{m_{B-L}}\right)\,.
\end{align}
The expression $\left[\cdots\right]$ can be taken to be $\approx1$
if the $U(1)$ gauge couplings at $m_{B-L}$ are much weaker than
at $m_{LR}$, and in that case we can see that the magnitude of the
$U(1)$-mixing effect on relation \eqref{LRsumrule} is roughly proportional
to $\boldsymbol{A}_{RX}^{-1}\left(m_{B-L}\right)\log\left(\frac{m_{R}}{m_{B-L}}\right)/\boldsymbol{A}_{RR}^{-1}\left(m_{LR}\right)\boldsymbol{A}_{XX}^{-1}\left(m_{LR}\right)$:
a large running region $\left[m_{B-L},m_{R}\right]$, a large $\boldsymbol{A}_{RX}^{-1}$
at $m_{B-L}$, as well as large coupling constants $\alpha_{R}$ and
$\alpha_{B-L}$ at the matching scale $m_{LR}$ will increase the
effect.

Finally, in class-I models it is possible to eliminate four more sets,
namely all of those with $\Delta DL<0$. It is easy to see with the
help of equation \eqref{eq:SlidingScale_scalar} that this is the
case; in the LR case, the $c_{i}^{L}$ are non-zero for $U(1)_{B-L}$
and $SU(2)_{L}$ with the values $\nicefrac{3}{4}$ and $\nicefrac{3}{2}$,
respectively. Since their sum is smaller than $c_{3}^{D}$ (and $\alpha_{3}$
is larger than the other couplings), $D$ must run faster than $L$
in the LR-regime.

By this reasoning, set 6 seems to be possible in class-I, but is not
realized in our complete scan. However, we found a few examples in
class-II; see below. In fact, it is quite straightforward to understand
why set 6 variants are rare: from equations \eqref{eq:sumruleinv}
and \eqref{LRsumruleApp} we know that $\Delta QE>\Delta DL+2\Delta LE-0.3$
and set 6 requires that $\Delta LE,\,\Delta DL>0$ but with $\Delta QE<0$.
This is possible, but it requires that $0.1\gtrsim\Delta LE,\,\Delta DL>0$,
which is difficult to achieve given that typically $\left|\Delta LE\right|\sim\mathcal{O}\left(10^{0},10^{1}\right)$.
This should also remind us that this classification of the variants
into sets can easily suffer changes if a more accurate calculation
of the invariants is performed. We also note in passing that in the
high-scale seesaw models of type-II \cite{Hirsch:2008gh} and seesaw
type-III \cite{Esteves:2010ff} with running only within the MSSM
group, all invariants run towards larger values, in other words only
set 1 is realized in these cases.

The above discussion serves only as a qualitative classification of
the invariants which are realizable in the different classes of models.
Much of the numerical information contained in these invariants is
therefore ignored by it. To mitigate this issue, in the following
we shall look at some particular variants in each class of models,
and see how the $\Delta b$'s and the intermediate scales affect the
values of the invariants.

\subsection{Invariants in class-I model}

\begin{figure}[h]
\begin{centering}
\begin{tabular}{cc}
\includegraphics[scale=0.76]{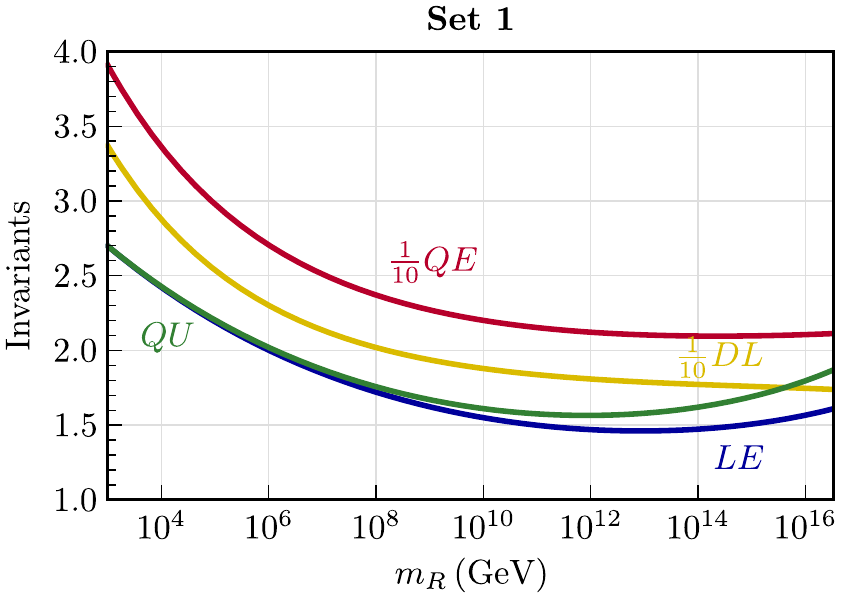}  & \includegraphics[scale=0.76]{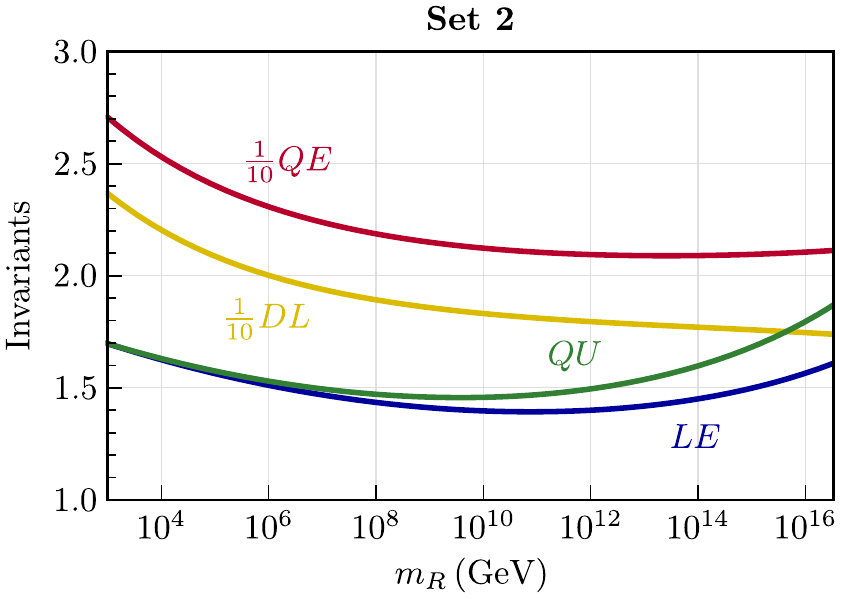}\tabularnewline
\includegraphics[scale=0.76]{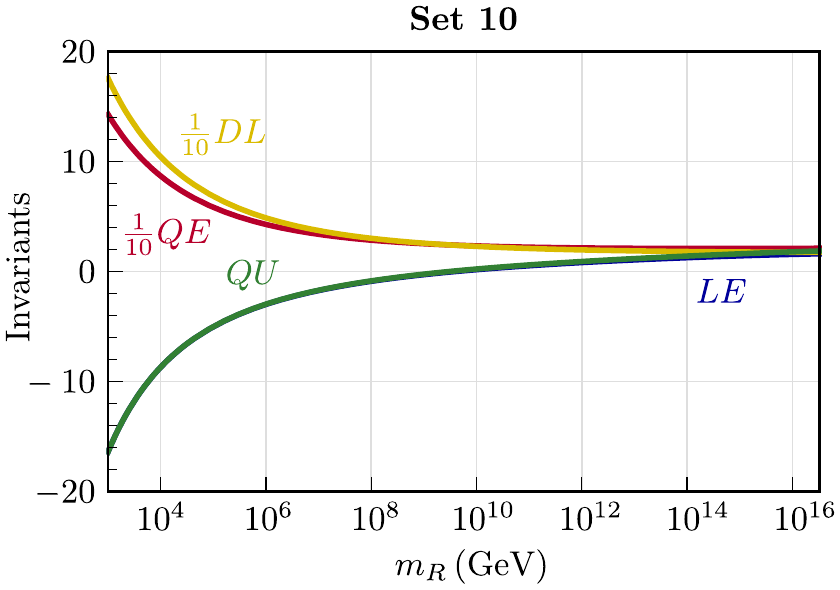}  & \includegraphics[scale=0.76]{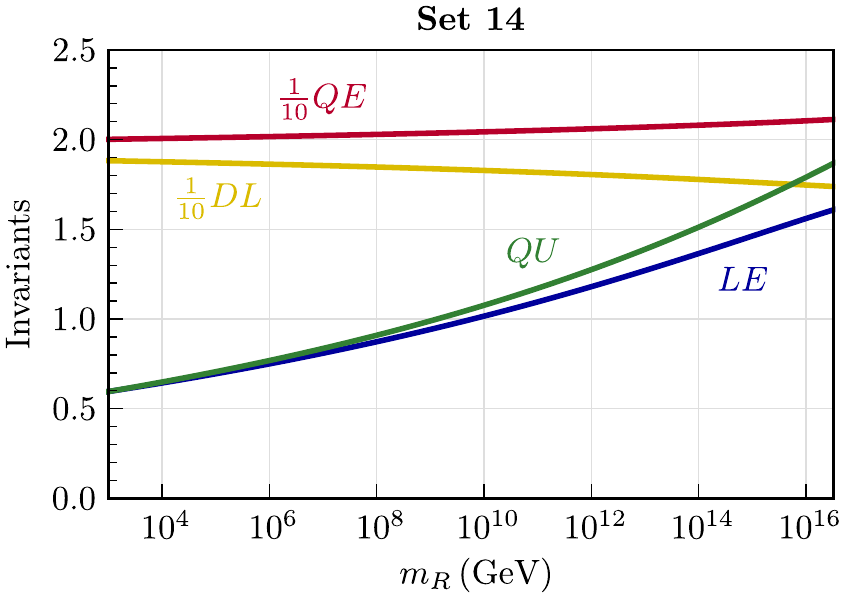} \tabularnewline
\end{tabular}
\par\end{centering}

\caption{\label{fig:invLR}{\small The} $m_{R}$ dependence of the invariants
in class-I models. The values $\Delta b_{i}^{LR}=(\Delta b_{3}^{LR},b_{L}^{LR},\Delta b_{R}^{LR},\Delta b_{BL}^{LR})$
are as follows: ($2,2,9,\nicefrac{1}{2}$) in the set 1 plot, ($1,1,7,1$)
in the set 2 plot, ($4,4,3,29/2$) in the set 10 plot, and ($0,0,2,6$)
in the set 14 plot. For a discussion, see the main text.}
\end{figure}

\noindent Figure \eqref{fig:invLR} shows examples of the $m_{R}$
dependence of the invariants corresponding to the four cases indicated
in table \eqref{tab:classes}: sets 1, 2, 10 and 14. Note that we
have scaled down the invariants $QE$ and $DL$ for practical reasons.
Note also the different scales in the different plots.

In all cases $QU\approx LE$ if the LR regime extends to very low
energies. As explained above, this is a general feature of this class
of models: the separation between the $QU$ and $LE$ is model independent
and thus, experimentally a non-zero measurement of $QU-LE$ allows,
in principle, to determine the scale at which the LR symmetry is broken.

Sets 1 and 2 show a quite similar overall behavior in these examples.
Set 1, however, can also be found in variants of class-I with larger
$\beta$-coefficients, which induce larger quantitative changes with
respect to the mSUGRA values. Note that while it is possible to find
variants within class-I which fall into set 2, due to the similarity
between $QU$ and $LE$ this set can be realized only if both $QU$
and $LE$ are numerically very close to their mSUGRA values. Graphically,
this means that the left endpoint of the $LE$ curve must be higher
than its right endpoint and, at the same time, the opposite must happen
to the $QU$ line. Similarly, since usually $QU$ in equation \eqref{eq:sumruleinv}
is significantly smaller than both $QE$ and $DL$, set 14 typically
implies that $\Delta QE$ and $\Delta DL$, which necessarily have
opposite signs, must be small. Therefore in set 14 $QE$ and $DL$
are close to their mSUGRA values.

In general, when $\Delta b_{3}^{LR}$ is large the invariants vary
strongly with the intermediate scale, as can be seen in the plot shown
for set 10 (figure \eqref{fig:invLR}). The large change is mainly
due to the rapid running of the gaugino masses in these variants,
but also the sfermion spectrum is very {}``deformed'' with respect
to mSUGRA expectations. For example, a negative LE means of course
that left sleptons are lighter than right sleptons, a feature that
can never be found in the {}``pure'' mSUGRA model. Recall that for
solutions with $\Delta b_{3}^{LR}=5$, the value of the squark masses
lies beyond the reach of the LHC.

\subsection{Model class-II}

\begin{figure}[tbph]
\begin{centering}
\begin{tabular}{cc}
\includegraphics[scale=0.76]{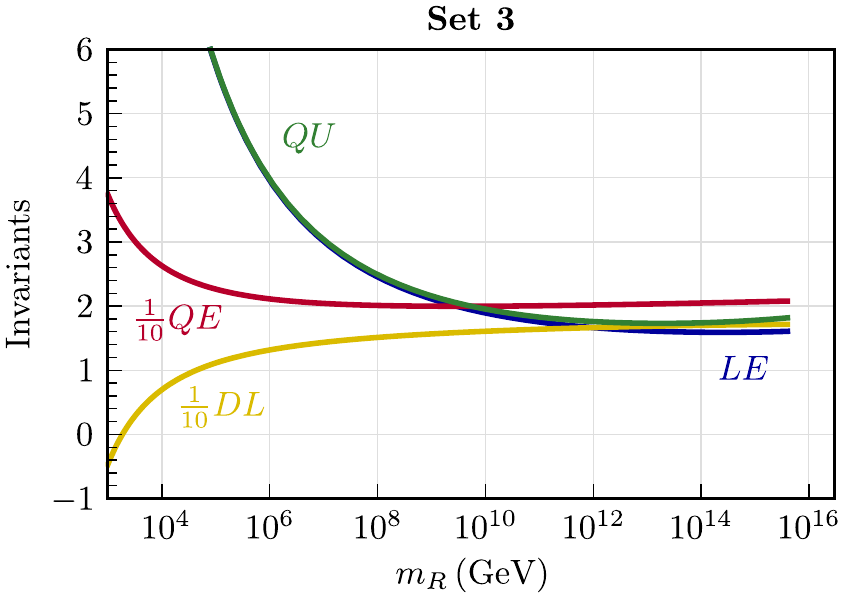}  & \includegraphics[scale=0.76]{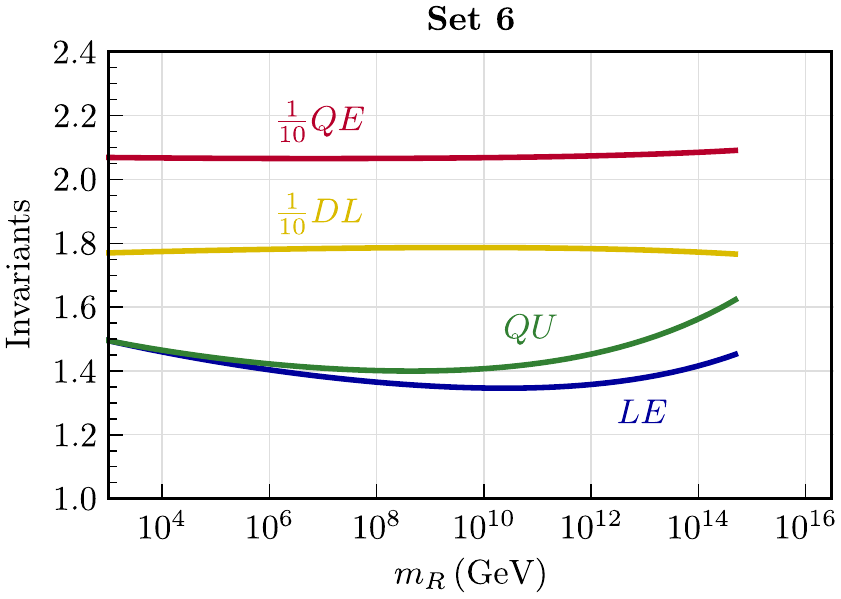} \tabularnewline
\multicolumn{2}{c}{\includegraphics[scale=0.76]{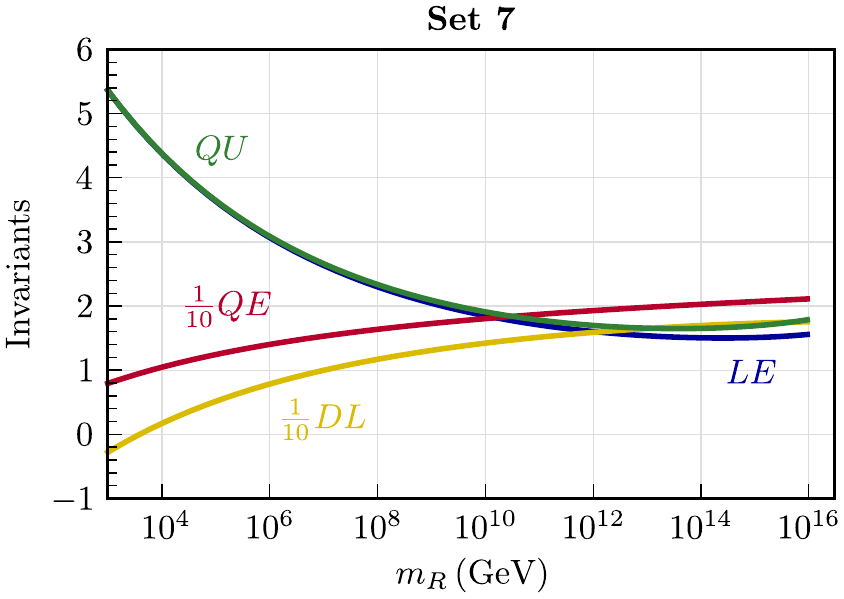} }\tabularnewline
\includegraphics[scale=0.76]{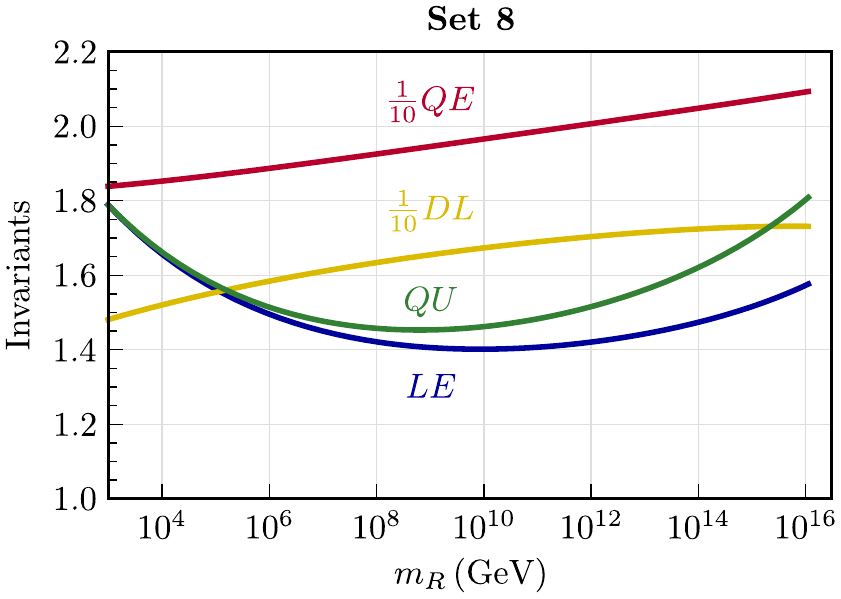}  & \includegraphics[scale=0.76]{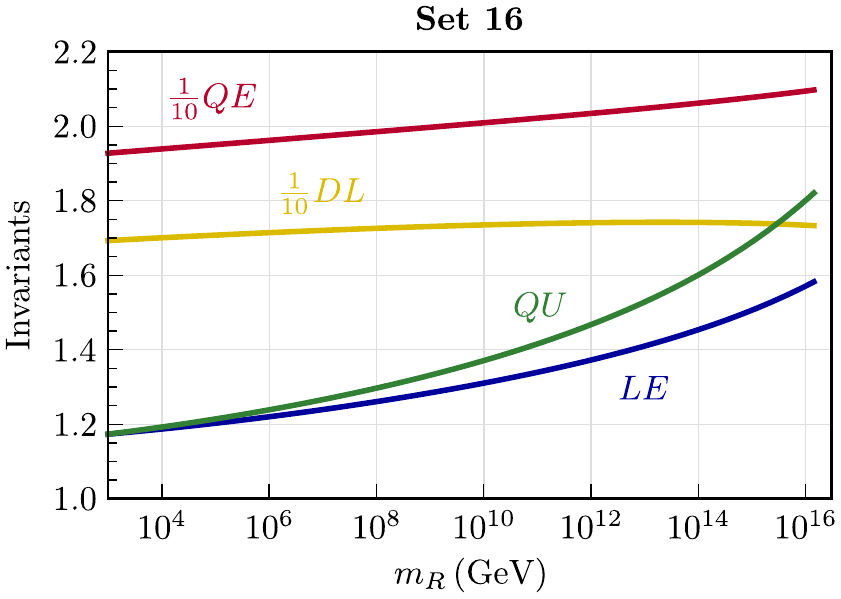} \tabularnewline
\end{tabular}
\par\end{centering}

\caption{{\small \label{fig:invPS}The $m_{R}$ dependence of the invariants
in class-II model. The examples shown correspond to the following
$\Delta b=(\Delta b_{3}^{LR},\ab\Delta b_{L}^{LR},\ab\Delta b_{R}^{LR},\ab\Delta b_{BL}^{LR},\ab\Delta b_{4}^{PS},\ab\Delta b_{L}^{PS},\ab\Delta b_{R}^{PS})$:
($0,\ab1,\ab10,\ab\nicefrac{3}{2},\ab14,\ab9,\ab13$) in the set 3
plot, ($0,\ab0,\ab1,\ab\nicefrac{9}{2},\ab63,\ab60,\ab114$) in the
set 6 plot, ($0,\ab3,\ab12,\ab\nicefrac{3}{2},\ab6,\ab3,\ab15$) in
the set 7 plot, ($0,\ab0,\ab9,\ab\nicefrac{3}{2},\ab11,\ab8,\ab12$)
in the set 8 plot, and ($0,\ab0,\ab7,\ab\nicefrac{3}{2},\ab11,\ab8,\ab10$)
in the set 16 plot.}}
\end{figure}

Figure \eqref{fig:invPS} shows the invariants of class-II models,
corresponding to those cases which are not covered by class-I models.

The example shown in figure \eqref{fig:invPS} for set 3 is similar
to the one of the original prototype model constructed in \cite{DeRomeri:2011ie}.
For set 6 we have found only a few examples, all of them showing invariants
which hardly change with respect to the mSUGRA values, as expected.
The example for set 7 shows that in some variants $QE$ can also decrease
considerably with respect to its mSUGRA value. Set 8 is quantitatively
similar to set 2, and set 16 is numerically similar to set 14. To
distinguish these, highly accurate SUSY mass measurements would be
necessary.

Again we note that larger values of $\Delta b^{LR}$, especially large
$\Delta b_{3}^{LR}$, usually lead to numerically larger changes in
the invariants, making these models in principle easier to test.

\subsection{Model class-III}

\begin{figure}[h]
\begin{centering}
\begin{tabular}{cc}
\includegraphics[scale=0.76]{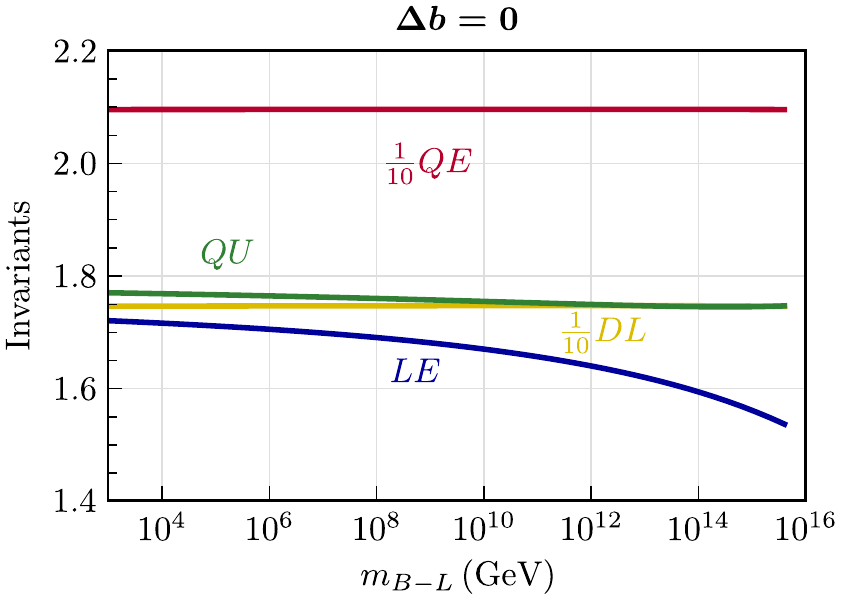}  & \includegraphics[scale=0.76]{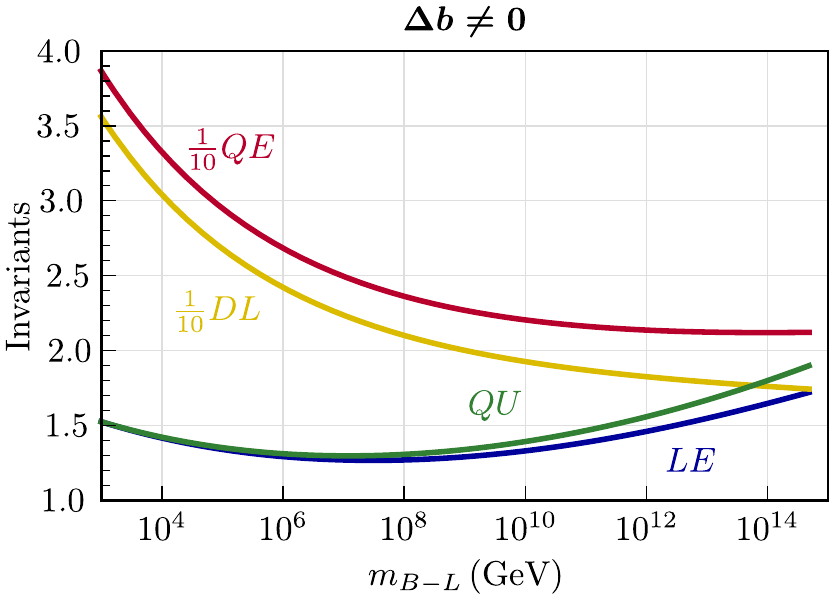} \tabularnewline
\end{tabular}
\par\end{centering}

\caption{{\small \label{fig:invBL}The $m_{B-L}$ dependence of the invariants
in two class-III models. The examples shown correspond to the following
$(\Delta b_{3}^{LR},\ab\Delta b_{L}^{LR},\ab\Delta b_{R}^{LR},\ab\Delta b_{BL}^{LR},\ab\Delta b_{3}^{B-L},\ab\Delta b_{L}^{B-L},\ab\Delta\gamma_{RR},\ab\Delta\gamma_{XR},\ab\Delta\gamma_{XX})$:
$(0,\ab1,\ab3,\ab3,\ab0,\ab0,\ab\nicefrac{1}{2},\ab-\sqrt{\nicefrac{3}{8}},\ab\nicefrac{3}{4})$
in the left plot, and $(2,\ab2,\ab4,\ab8,\ab2,\ab2,\ab\nicefrac{1}{2},\ab-\sqrt{\nicefrac{3}{8}},\ab\nicefrac{11}{4})$
in the one on the right. By comparing the endpoints of the $QU$ and
$LE$ lines we can measure the relative effects of the MSSM and $U(1)$-mixing
in breaking the relation $QU=LE$; with $m_{B-L}$ at its highest,
the models are identical to the MSSM almost up to the unification
scale, so the splitting of the lines $QU$ and $LE$ on the right
of each plot measures the MSSM effect at its maximum. Analogously,
the splitting of the lines $QU$ and $LE$ on the left of each plot
measures the $U(1)$-mixing breaking of the relation $QU=LE$ at its
maximum, without the MSSM's contribution. Clearly, in these two examples
the MSSM effect is bigger.}}
\end{figure}

\noindent Here, the invariants depend on $m_{B-L}$, mildly or strongly
depending on the value of $\Delta b_{3}^{B-L,\, LR}$. For almost
all the solutions with $\Delta b_{3}^{B-L,\, LR}=0$ , the values
of $QU$, $DL$, $QE$ are constant and only $LE$ shows a mild variation
with $m_{B-L}$. This was already pointed out in \cite{DeRomeri:2011ie}.
However, we have found that class-III models can be made with $\Delta b_{3}^{B-L,\, LR}>0$
and, in general, these lead to invariants which are qualitatively
similar to the case of class-I discussed above. In figure \eqref{fig:invBL}
we show two examples of invariants for class-III, one with $\Delta b_{3}^{B-L,\, LR}=0$
and one with $\Delta b_{3}^{B-L,\, LR}=2$.

\subsection{Comparison of classes of models}

\begin{figure}[h]
\begin{centering}
\includegraphics[scale=0.217]{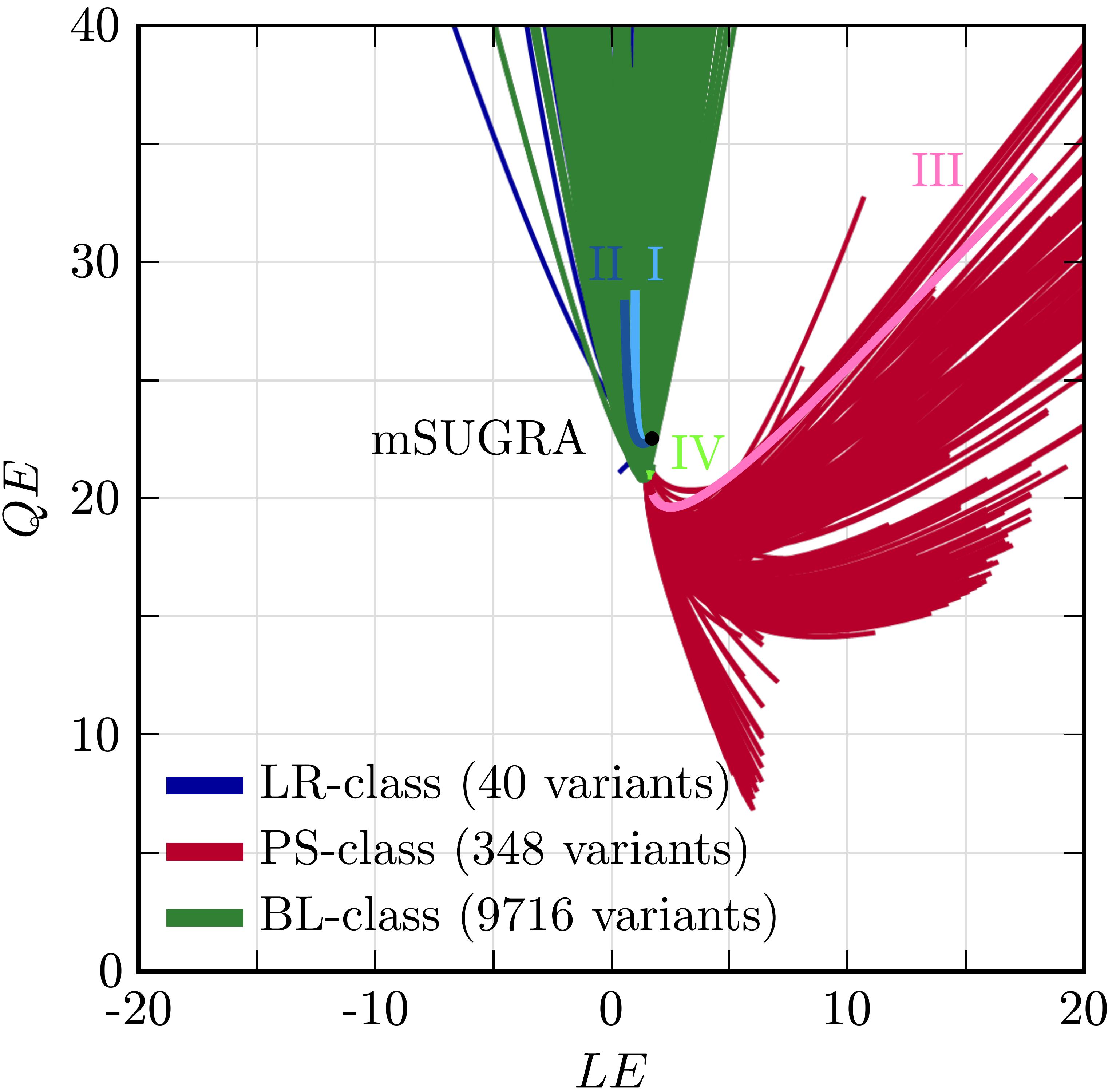}
\par\end{centering}

\caption{\label{fig:LEQE_parametric_plot_RGB}Parametric $\left(LE,QE\right)$
plot for the different variants (see text). The thicker lines labeled
with I, II, III and IV indicate the result for the four prototype
models presented in \cite{DeRomeri:2011ie}.}
\end{figure}

The classification of the variants that we have discussed in subsection
\ref{sub:SlidingScale_invclasses} only takes into account what happens
when the lowest intermediate scale is very low, ${\cal O}(m_{SUSY})$.
When one varies continuously the lowest intermediate scale ($m_{R}$
in class-I and class-II models, or $m_{B-L}$ in class-III models),
each variant draws a line in the 4-dimensional space $\left(LE,QU,DL,QE\right)$.
The dimensionality of such a plot can be lowered if we use the (approximate)
relations between the invariants shown above, namely $QU\approx LE$
and $QE=DL+2QU$. We can then choose two independent ones, for example
$LE$ and $QE$, so that the only non-trivial information between
the 4 invariants is encoded in a $\left(LE,QE\right)$ plot. In this
way, it is possible to simultaneous display the predictions of different
variants. This was done in figure \eqref{fig:LEQE_parametric_plot_RGB},
where LR-, PS- and BL-variants are drawn together. The plot is exhaustive
in the sense that it includes all LR-variants, as well as all PS-
and BL-variants which can have the highest intermediate scale below
$10^{6}$ GeV. In all cases, we required that at unification $\alpha^{-1}$
is larger than $\nicefrac{1}{2}$ when the lowest intermediate scale
is equal to $m_{SUSY}$.

There is a dot in the middle of the figure---the mSUGRA point---which
corresponds to the prediction of mSUGRA models, in the approximation
used. It is expected that every model will draw a line with one end
close to this point. This end-point corresponds to the limit where
the intermediate scales are close to the GUT scale and therefore the
running in the LR, PS and BL phases is small, so the invariants should
be similar to those in mSUGRA models. The general picture is that
lines tend to start (when the lowest intermediate scale is of the
order of $10^{3}$ GeV) outside or at the periphery of the plot, away
from the mSUGRA point and, as the intermediate scales increase, they
converge towards the region of the mSUGRA point, in the middle of
the plot. In fact, note that all the blue lines of LR-class models
do touch this point, because we can slide the LR scale all the way
to $m_{G}$, as mentioned before. But in PS- and BL- models there
are two intermediate scales and often the lowest one cannot be increased
all the way up to $m_{G}$ (either because that would make the highest
intermediate scale bigger than $m_{G}$ or because it would invert
the natural ordering of the two intermediate scales).

It is interesting to note that the BL-class with low $m_{R}$ can
produce the same imprint in the sparticle masses as LR-models. This
is to be expected because with $m_{R}$ close to $m_{B-L}$, the running
in the $U(1)$-mixing phase is small, leading to predictions similar
to LR-models. The equivalent limit for PS-class models is reached
for very high $m_{PS}$, close to the GUT scale (see below). On the
other hand, from figure \eqref{fig:LEQE_parametric_plot_RGB} we can
see that a low $m_{PS}$ actually leads to a very different signal
on the soft sparticle masses. For example, a measurement of $LE\approx10$
and $QE\approx15$, together with compatible values for the other
two invariants ($QU\approx10$ and $DL\approx-5$) would immediately
exclude all classes of models except PS-models, and in addition it
would strongly suggest low PS and LR scales.

\begin{figure}[h]
\begin{centering}
\begin{tabular}{@{}r@{}c}
\includegraphics[scale=0.217]{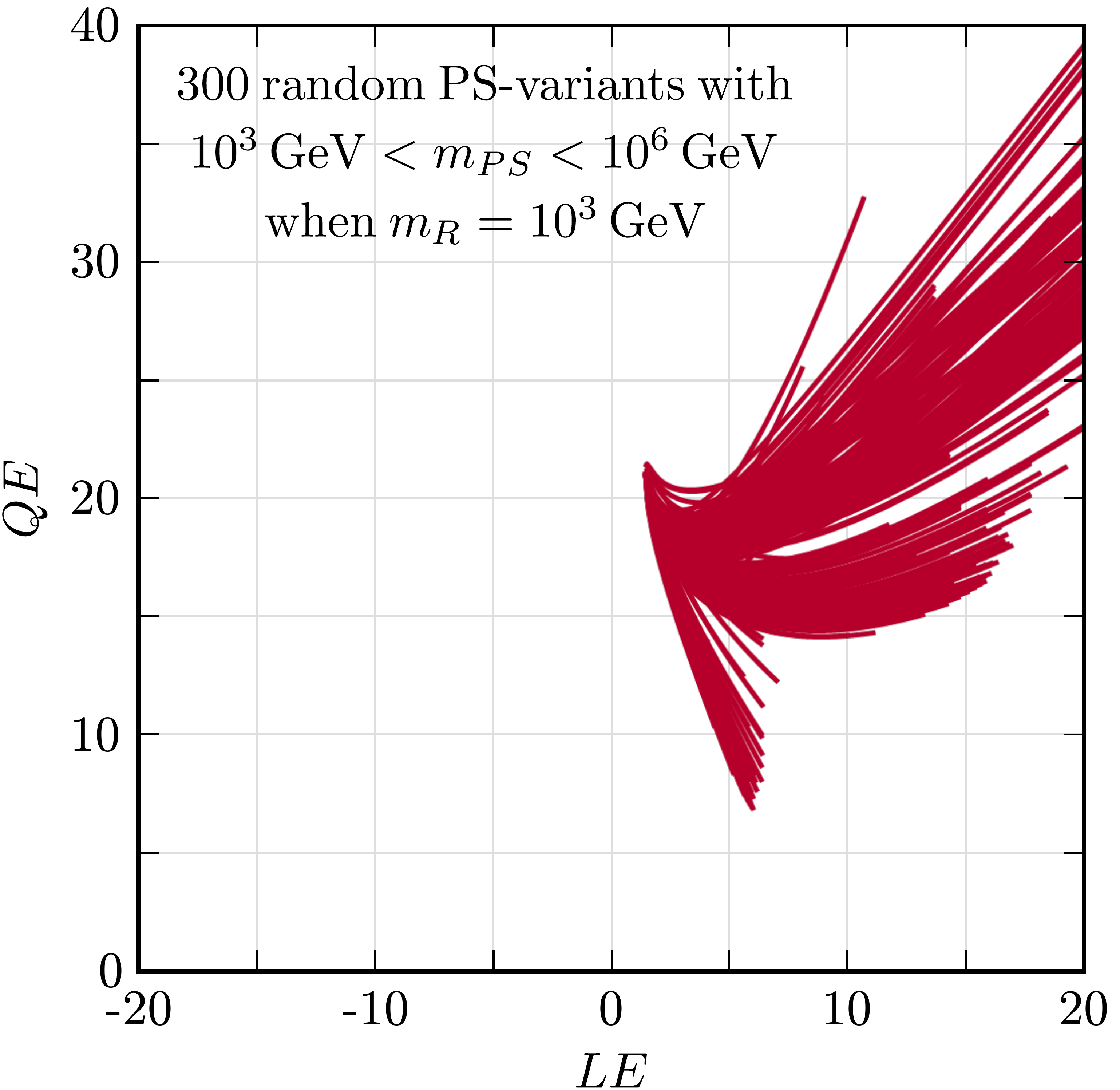} & \includegraphics[scale=0.217]{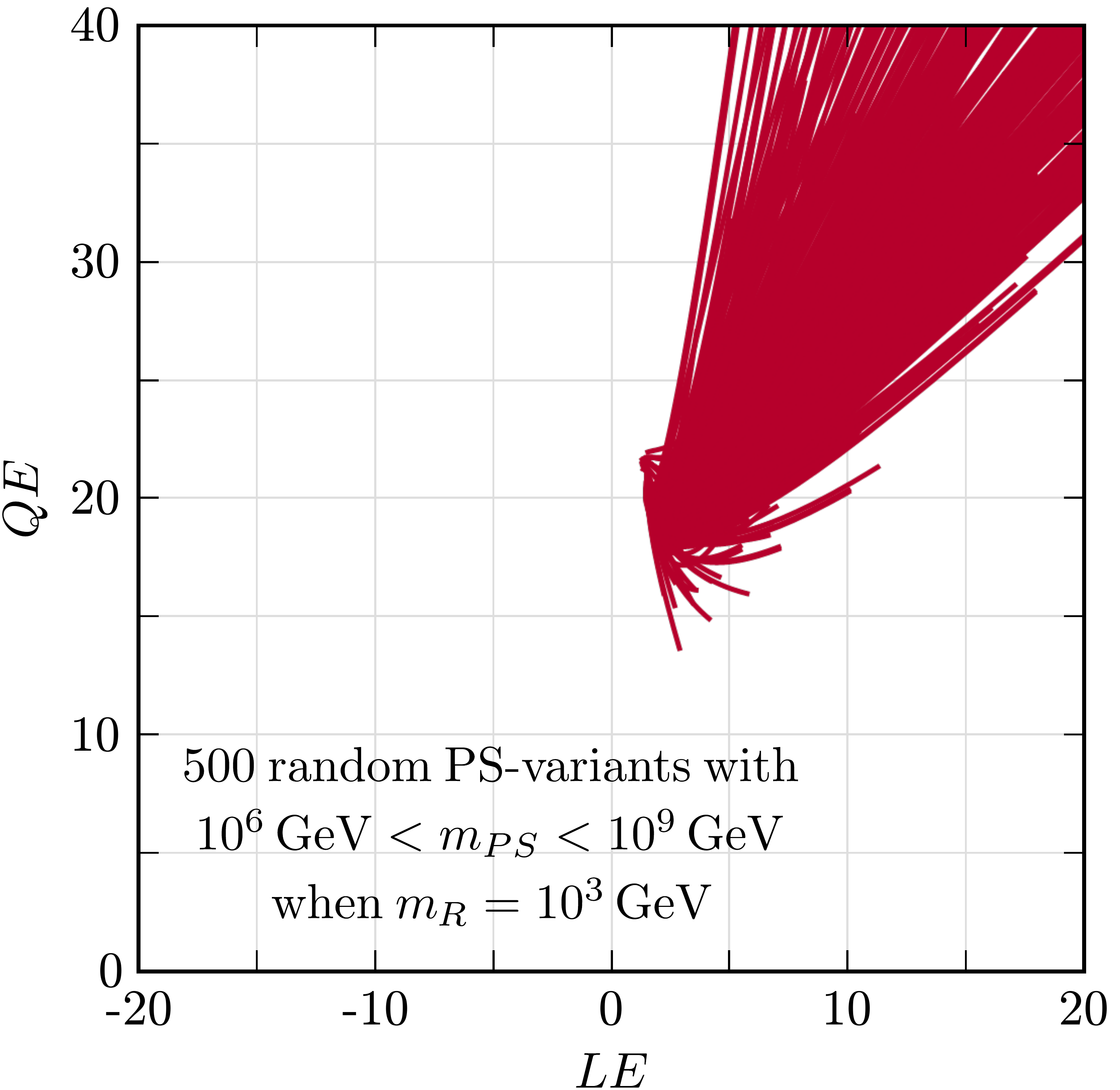}\tabularnewline
\includegraphics[scale=0.217]{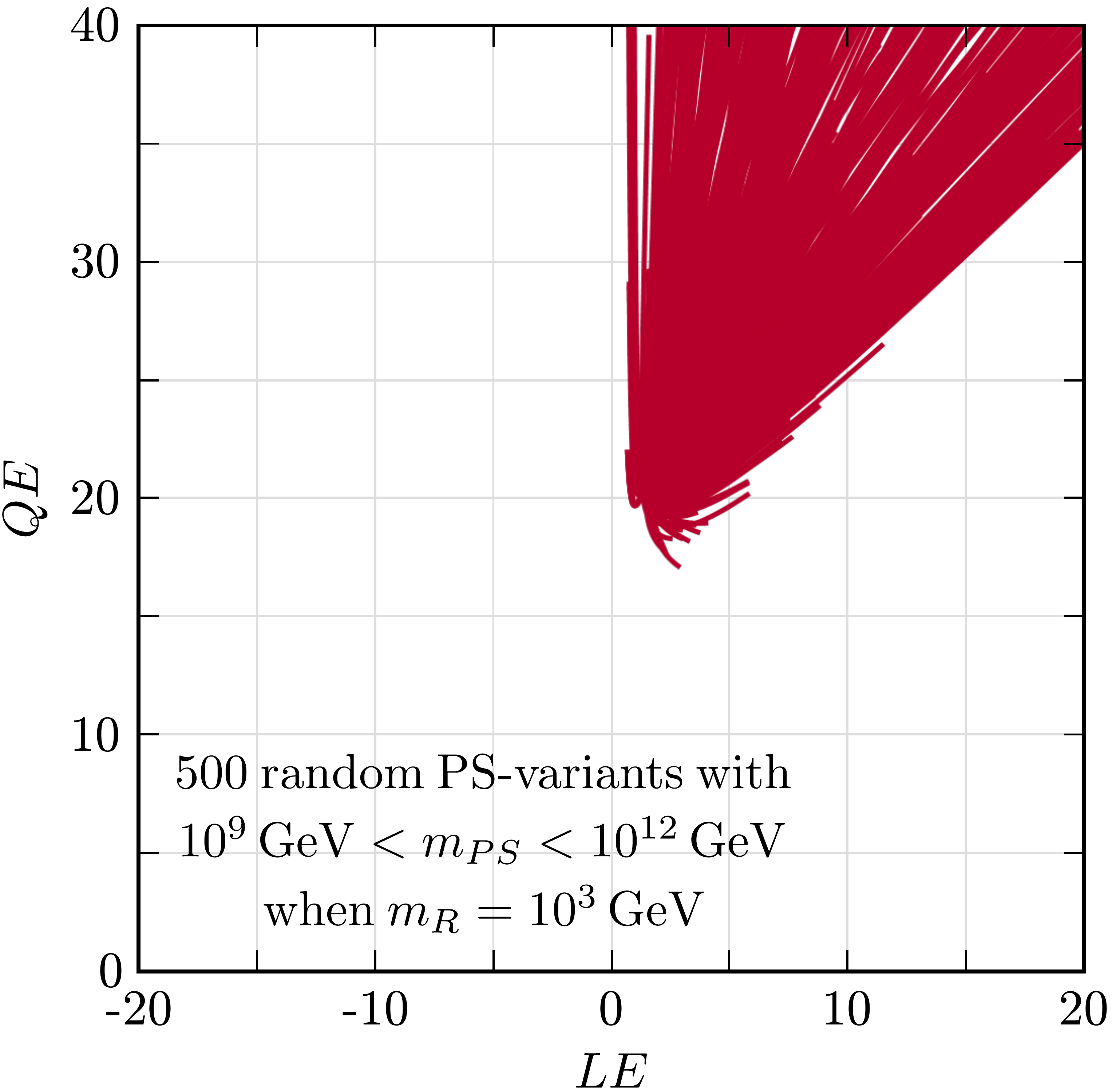} & \includegraphics[scale=0.217]{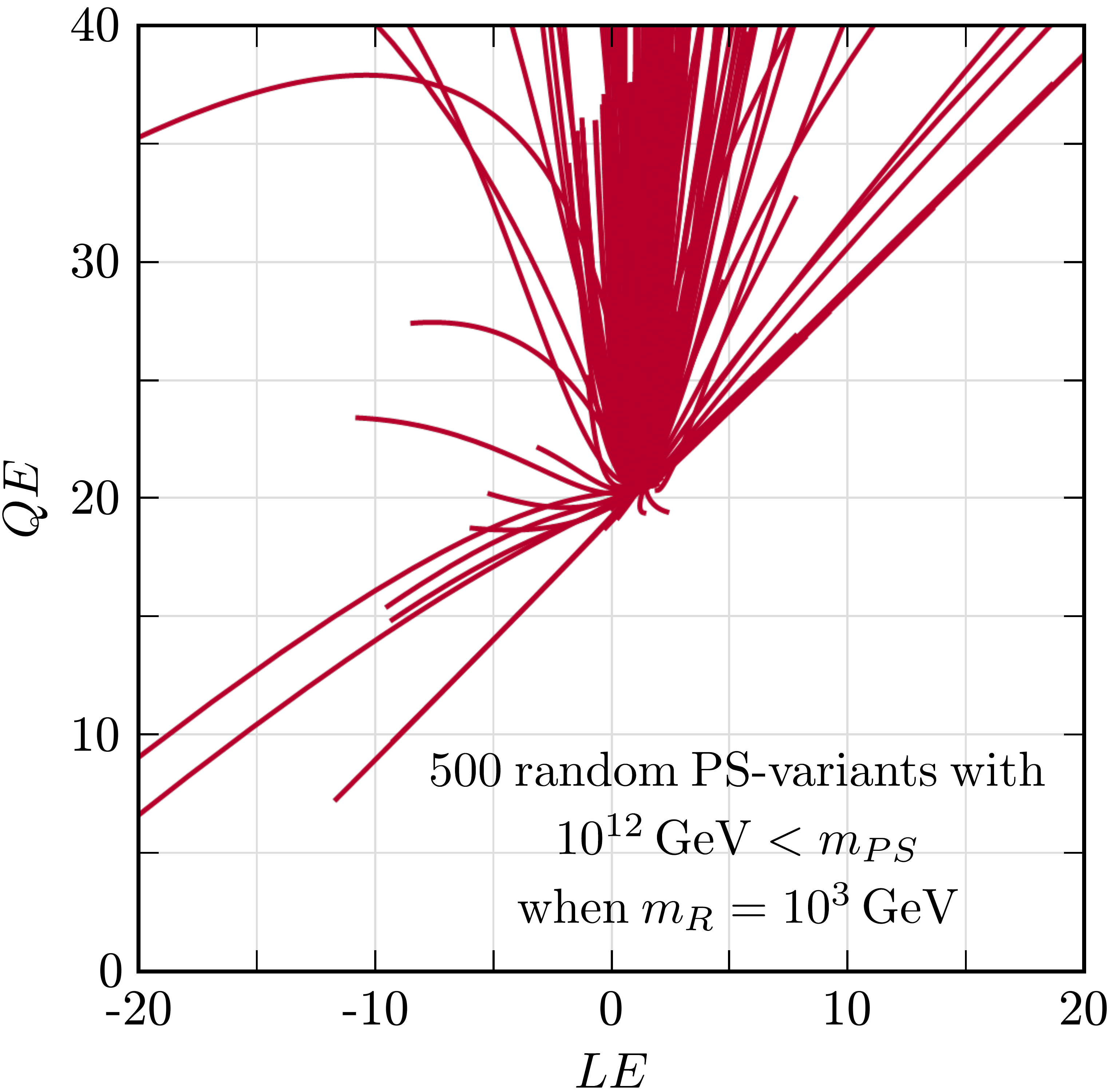}\tabularnewline
\end{tabular}
\par\end{centering}

\caption{\label{fig:LEQE_parametric_plot_PS_only}Parametric $\left(LE,QE\right)$
plots for different PS-variants showing the effect of the PS scale.}
\end{figure}

Figure \eqref{fig:LEQE_parametric_plot_PS_only} illustrates the general
behavior of PS-models as we increase the separation between the $m_{LR}$
and $m_{PS}$ scales. The red region in the $\left(LE,QE\right)$
plot tends to rotate anti-clockwise until it reaches, for very high
$m_{PS}$, the same region of points which is predicted by LR-models.
Curiously, we also see in figure \eqref{fig:LEQE_parametric_plot_PS_only}
that some of these models actually predict different invariant values
from the ones of LR models. What happens in these cases is that since
the PS phase is very short, it is possible to have many active fields
in it which decouple at lower energies. As such, even though the running
is short, the values of the different gauge couplings actually get
very large corrections in this regime, and these are uncommon in other
settings. For example, in this special subclass of PS-models it is
possible for $\alpha_{R}$ to get bigger than $\alpha_{3}$/$\alpha_{4}$
before unification! One can see from figure \eqref{fig:LEQE_parametric_plot_PS_only}
that many (although not all) PS-models can lead to large values of
$LE$. This can happen for both low and high values of $m_{PS}$,
and is a rather particular feature of class-II which is not found
for the other ones.

\section{Summary}

In this chapter, we have discussed $SO(10)$ inspired supersymmetric
models with an extended gauge group near the electroweak scale, consistent
with gauge coupling unification due to a sliding scale mechanism.
We have discussed three different setups, which we call classes of
models. The first and simplest chain breaks $SO(10)$ through a left-right
symmetric stage to the SM group, class-II uses an additional intermediate
Pati-Salam stage, while in class-III we discuss models which break
the LR-symmetric group first into a $U(1)_{R}\times U(1)_{B-L}$ group
before reaching the SM group. We have shown that in each case many
different variants and many configurations (or \textit{proto-models})
for each variant can be constructed.

We have discussed that one can construct sliding models in which an
inverse or linear seesaw is consistent with GCU, as done in earlier
works \cite{Malinsky:2005bi,Dev:2009aw,DeRomeri:2011ie}, as well
as all other known types of seesaws, in principle. We found configurations
for type-I, type-II, type-III seesaw, and even inverse type-III (for
which one example limited to class-II was previously discussed in
\cite{DeRomeri:2011ie}).

Due to the sliding scale property, the different configurations predict
potentially rich and distinctive phenomenology at the LHC, although
by the same reasoning the discovery of any of the additional particles
predicted by the models is not guaranteed. However, even if all the
new particles, including the gauge bosons of the extended gauge group,
lie outside the reach of the LHC, indirect tests of the models are
possible from measurements of SUSY particle masses and couplings.
We have discussed certain combinations of soft parameters, called
\textit{invariants}, and shown that they could be used to gain indirect
information not only on the class of model and its variant realized
in Nature, but also give hints on the scale of beyond-MSSM physics,
which is the energy scale at which the extended gauge group is broken.

We add a few words of caution however. First of all, our analysis
is done completely at the one loop level. It is known from numerical
calculations for seesaw type-II \cite{Hirsch:2008gh} and seesaw type-III
\cite{Esteves:2010ff} that numerically the invariants receive important
shifts at the two loop level. In addition, there are also uncertainties
in the calculation from GUT-scale thresholds and from uncertainties
in the input parameters. For the latter, the most important is most
likely the error on $\alpha_{3}$ \cite{DeRomeri:2011ie}. With the
huge number of models we have considered, taking into account all
of these effects is impractical and, thus, our numerical results should
be considered as approximate. However, should any signs of supersymmetry
be found in the future, improvements in the calculations along these
lines could be easily made, it necessary. More important for the calculation
of the invariants is, of course, the assumption that SUSY breaking
is indeed mSUGRA-like. Tests of the validity of this assumption can
be made also only indirectly. Many of the spectra we find, especially
in class-II models, are actually quite different from standard mSUGRA
expectations and thus pure mSUGRA would give a bad fit to experimental
data, if one of these models is realized in nature. Also, it is important
to keep in mind that by construction our models obey a certain sliding
condition (see the discussion in subsection \ref{sub:SlidingScale_invclasses}),
which means that in principle it is possible that there are other
models that do not satisfy this condition, and yet exhibit the other
interesting features such as low intermediate $B-L$ or $LR$ scales.

So far no signs of supersymmetry have been seen at the LHC, but with
the planned increase of $\sqrt{s}$ for the next run of the accelerator
there is still quite a lot of parameter space to be explored. We note
in this respect that a heavy Higgs with a mass of $m_{h}\sim(125-126)$
GeV, as suggested by the new resonance found by the ATLAS \cite{ATLAS_Higgs_discovery:2012gk}
and CMS \cite{CMS_Higgs_discovery:2012gu} collaborations,%
\footnote{See also the ATLAS \cite{website_ATLAS_Higgs_results} and CMS \cite{website_CMS_Higgs_results}
collaborations' websites for more up-to-date results and analysis.%
} does not necessarily imply a heavy sparticle spectra for the models
studied here. While for a pure MSSM with mSUGRA boundary conditions
it is well-known \cite{Arbey:2011ab,Baer:2011ab,Buchmueller:2011ab,Ellis:2012aa}
that such a hefty Higgs requires multi-TeV scalars,%
\footnote{Multi-TeV scalars are also required if the MSSM with mSUGRA boundary
conditions is extended to include a high-scale seesaw mechanism \cite{Hirsch:2012ti}.%
} all our models have an extended gauge symmetry which means that new
$D$-terms contribute to the Higgs mass \cite{Haber:1986gz,Drees:1987tp},
alleviating the need for large soft SUSY breaking terms, as has been
explicitly shown in \cite{Hirsch:2012kv,Hirsch:2011hg} for one particular
realization of a class-III model \cite{Malinsky:2005bi,DeRomeri:2011ie}.

Finally, many of the configurations (or \textit{proto-models}) which
we have discussed here contain exotic superfields, which might show
up in the LHC. Therefore it might be interesting to do a more detailed
study of the phenomenology of at least some of the models that were
constructed in this chapter.
\cleartooddpage

\chapter{\label{chap:Revisiting_RK}The $\Gamma\left(K\rightarrow e\nu\right)/\Gamma\left(K\rightarrow\mu\nu\right)$
ratio in supersymmetric unified models}

\section{The $R_{K}$ ratio, lepton flavor universality, and lepton flavor
violation}

In chapter \ref{chap:Lepton-flavour-violation} it was mentioned that
LFV has only been observed in the neutral sector, through neutrino
oscillation experiments. In the charged sector there is no evidence
yet that lepton flavor is violated and, from a theoretical point of
view, even the minimally extended Standard Model with massive neutrinos
does not predict it to happen at experimentally detectable rates.
Even so, there is an ongoing effort by different collaborations to
look at such effects in different observables, because in many extensions
of the SM, in particular supersymmetric ones, the flavor of charged
leptons is violated in some processes at rates which are within reach
of present or near future experiments.

In this chapter, following closely reference \cite{Fonseca:2012kr},
we will look at the ratio
\begin{equation}
R_{K}\equiv\frac{\Gamma\left(K^{+}\rightarrow e^{+}\nu\left[\gamma\right]\right)}{\Gamma\left(K^{+}\rightarrow\mu^{+}\nu\left[\gamma\right]\right)}\,,\label{eq:rk:def}
\end{equation}
and see how it relates to cLFV in supersymmetric models, both constrained
and unconstrained, even though we will be particularly interested
in unified models. The B meson decay observables $\textrm{BR}\left(B_{u}\rightarrow\tau\nu\right)$
and $\textrm{BR}\left(B_{s}\rightarrow\mu\mu\right)$, as well as
$\textrm{BR}\left(\tau\rightarrow e\gamma\right)$, depend on some
of the supersymmetric parameters in the same way as $R_{K}$ and therefore
we will take them into consideration in our analysis. On the other
hand, as we shall see latter on, a lightest Higgs with a mass 125
-- 126 GeV does not affect things considerably, even though, as pointed
out previously in this thesis, it does point to a heavy SUSY spectrum.

\begin{figure}[h]
\begin{centering}
\includegraphics[clip,scale=0.8]{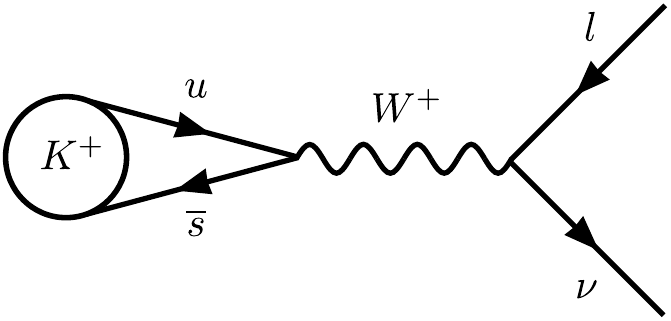}~~~~~~~~~~~\includegraphics[clip,scale=0.8]{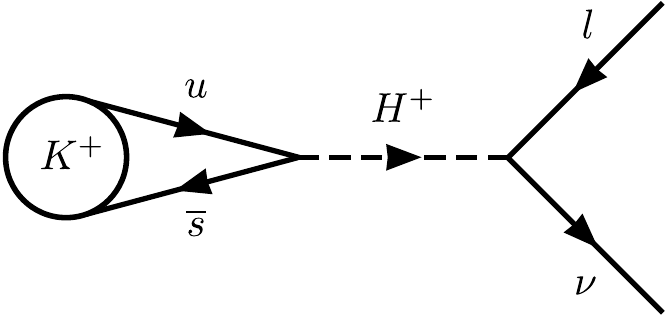}
\par\end{centering}

\caption{\label{fig:RK:SM:Higgs}Tree level contributions to $R_{K}$---through
the $W$ boson, and through a charged Higgs.}
\end{figure}

In the SM, at tree level a charged meson $P^{\pm}$ decays into leptons
through the exchange of a $W$ boson (figure \eqref{fig:RK:SM:Higgs}),
and the decay width is given by

\begin{equation}
\Gamma^{\text{SM}}\left(P^{\pm}\to\ell^{\pm}\nu\right)=\frac{G_{F}^{2}m_{P}m_{\ell}^{2}}{8\pi}\left(1-\frac{m_{\ell}^{2}}{m_{P}^{2}}\right)^{2}f_{P}^{2}|V_{qq^{\prime}}|^{2}\,.\label{eq:SM:Pdecays}
\end{equation}
Here $P$ can be a $\pi,\, K,\, D$ or a $B$ meson, with mass $m_{P}$
and decay constant $f_{P}$, and $G_{F}$ is the Fermi constant, $m_{\ell}$
the lepton mass and $V_{qq^{\prime}}$ the corresponding Cabibbo-Kobayashi-Maskawa
matrix element. This decay width is approximately proportional to
the square of the charged lepton's mass, which makes $R_{K}\sim m_{e}^{2}/m_{\mu}^{2}$
very small, even though the phase space in $K^{+}\rightarrow e^{+}\nu$
is bigger than in $K^{+}\rightarrow\mu^{+}\nu$. The reason for this
is well known: this type of decay, mediated by weak interactions,
is helicity suppressed. This means the following: in its rest frame
the spin 0 meson decays into an almost massless neutrino with left
helicity and consequently a charged anti-lepton with left helicity
as well. However, this last particle is only allowed to have right
chirality by the $W_{\mu}^{+}\overline{\nu}P_{L}\gamma^{\mu}\ell$
interaction, so in the limit where $m_{\ell}\rightarrow0$ the amplitude
of the process vanishes.

As is usually the case, amplitudes of processes involving bound states
of quarks are hampered by hadronic uncertainties in the meson decay
constants. That is the reason why often it is better to work with
ratios, such as $R_{K}$ in equation \eqref{eq:rk:def}, as they are
independent of $f_{P}$ to a very good approximation, and the SM prediction
can then be computed very precisely. Once corrections beyond tree
level are taken into consideration, the SM prediction (inclusive of
internal bremsstrahlung radiation) can be expressed as \cite{Cirigliano:2007xi}
\begin{equation}
R_{K}^{\text{SM}}=\left(\frac{m_{e}}{m_{\mu}}\right)^{2}\left(\frac{m_{K}^{2}-m_{e}^{2}}{m_{K}^{2}-m_{\mu}^{2}}\right)^{2}\left(1+\delta R_{\text{QED}}\right)\,,
\end{equation}
where $\delta R_{\text{QED}}=(-3.60\pm0.04)\%$ is a small electromagnetic
correction accounting for internal bremsstrahlung and structure-dependent
effects. Note that a factor $\nicefrac{g_{e}^{2}}{g_{\mu}^{2}}$ is
implicit in this expression, but since we assume that weak interactions
couple with the same strength to all lepton flavors ($g_{e}=g_{\mu}$),
such an expression is not needed. In any case, it is worth remembering
that this observable also tests the universality of the weak interaction.

The most recent analysis has provided the following value \cite{Cirigliano:2007xi}:
\begin{equation}
R_{K}^{\textrm{SM}}=(2.477\pm0.001)\times10^{-5}\,.\label{eq:Cirigliano:2007xi}
\end{equation}
On the experimental side, the NA62 collaboration has obtained stringent
bounds \cite{Goudzovski:2011tc}: 
\begin{equation}
R_{K}^{\textrm{exp}}=(2.488\pm0.010)\,\times10^{-5}\,,\label{eq:rk:NA62}
\end{equation}
which should be compared with the SM prediction (equation \eqref{eq:Cirigliano:2007xi}).
In order to do so, it is often useful to introduce the following parametrization,
\begin{equation}
R_{K}^{\textrm{exp}}=R_{K}^{\textrm{SM}}\left(1+\Delta r\right)\,,\qquad\Delta r\equiv\nicefrac{R_{K}}{R_{K}^{\textrm{SM}}}-1\,,\label{eq:deltark}
\end{equation}
where $\Delta r$ is a quantity denoting potential contributions arising
from scenarios of new physics. Comparing the theoretical SM prediction
to the current bounds (equations \eqref{eq:Cirigliano:2007xi} and
\eqref{eq:rk:NA62}), one verifies that observation is compatible
with the SM at 1$\sigma$: 
\begin{equation}
\Delta r=\left(4\pm4\right)\times10^{-3}\,.\label{eq:deltarexp}
\end{equation}

Previous analyzes have investigated supersymmetric contributions to
$R_{K}$ in different frameworks, as for instance low-energy SUSY
extensions of the SM (i.e., the unconstrained Minimal Supersymmetric
Standard Model (MSSM)) \cite{Masiero:2005wr,Masiero:2008cb,Girrbach:2012km},
or non-minimal grand unified models (where higher dimensional terms
contribute to fermion masses) \cite{Ellis:2008st}. These studies
have also considered the interplay of $R_{K}$ with other important
low-energy flavor observables, magnetic and electric lepton moments
and potential implications for leptonic CP violation. Distinct computations,
based on an approximate parametrization of flavor violating effects---the
mass insertion approximation (MIA) \cite{Hall:1985dx}---allowed to
establish that SUSY LFV contributions can induce large contributions
to the breaking of lepton universality, as parametrized by $\Delta r$.
The dominant FV contributions are in general associated to charged-Higgs
mediated processes, being enhanced due to non-holomorphic effects---the
so-called {}``HRS'' mechanism \cite{Hall:1993gn}---and require
flavor violation in the $RR$ block of the charged slepton mass matrix.
It is important to notice that these Higgs contributions have been
known to have an impact on numerous observables, and can become especially
relevant for the large $\tan\beta$ regime \cite{Hou:1992sy,Hall:1993gn,Chankowski:1994ds,Babu:1999hn,Carena:1999py,Babu:2002et,Brignole:2003iv,Brignole:2004ah,Arganda:2004bz,Paradisi:2005tk,Paradisi:2006jp,RamseyMusolf:2007yb}.
Also, it has recently been point out \cite{Abada:2012mc} that the
modified $W\ell\nu$ vertex generated in models with sterile neutrinos
can produce a large, measurable change in $R_{K}$.

In the following section, we will therefore explore supersymmetric
contributions to $\Delta r$, and in particular we shall review the
connection between this observable and charged lepton flavor violation.

\section{\label{sec:RK_formulae}$R_{K}$ in supersymmetric models}

In type-II two Higgs doublet models, such as the MSSM, the extended
Higgs sector can play an important role in lepton flavor violating
transitions and decays (see \cite{Hou:1992sy,Hall:1993gn,Chankowski:1994ds,Babu:1999hn,Carena:1999py,Babu:2002et,Brignole:2003iv,Brignole:2004ah,Arganda:2004bz,Paradisi:2005tk,Paradisi:2006jp,RamseyMusolf:2007yb}).
The effects of the additional Higgs are also sizable in meson decays
through a charged Higgs boson, as schematically depicted in figure
\eqref{fig:RK:SM:Higgs}. In particular, for kaons, one finds \cite{Hou:1992sy}
\begin{align}
\Gamma(K^{\pm}\to\ell^{\pm}\nu) & =\Gamma^{\text{SM}}(K^{\pm}\to\ell^{\pm}\nu)\left(1-\tan^{2}\beta\frac{m_{K}^{2}}{m_{H^{+}}^{2}}\frac{m_{s}}{m_{s}+m_{u}}\right)^{2}\,;\label{eq:kaon:gamma:smsusy}
\end{align}
yet, despite this new tree-level contribution, $R_{K}$ is unaffected
as the extra factor does not depend on the (flavored) leptonic part
of the process.

New contributions to $R_{K}$ only emerge at higher order: at one-loop
level, there are box and vertex contributions, wave function renormalization,
which can be both lepton flavor conserving (LFC) and lepton flavor
violating. Flavor conserving contributions arise from loop corrections
to the $W^{\pm}$ propagator, through heavy Higgs exchange (neutral
or charged) as well as from chargino/neutralino-sleptons (in the latter
case stemming from non-universal slepton masses, in other words, a
selectron-smuon mass splitting). As concluded in \cite{Masiero:2005wr},
in the framework of SUSY models where lepton flavor is conserved,
the new contributions to $\Delta r^{\text{SUSY}}$ are too small to
be within experimental reach.

On the other hand, Higgs mediated LFV processes are capable of providing
an important contribution when the kaon decays into a electron plus
a tau-neutrino. For such LFV Higgs couplings to arise, the leptonic
doublet ($L$) must couple to more than one Higgs doublet. However,
at tree level in the MSSM, $L$ can only couple to $H_{d}$, and therefore
such LFV Higgs couplings arise only at loop level, due to the generation
of an effective non-holomorphic coupling between $L$ and $H_{u}^{*}$---the
HRS mechanism \cite{Hall:1993gn}---which is a crucial ingredient
in enhancing the Higgs contributions to LFV observables.

\begin{figure}[h]
\begin{centering}
\includegraphics[clip,scale=0.8]{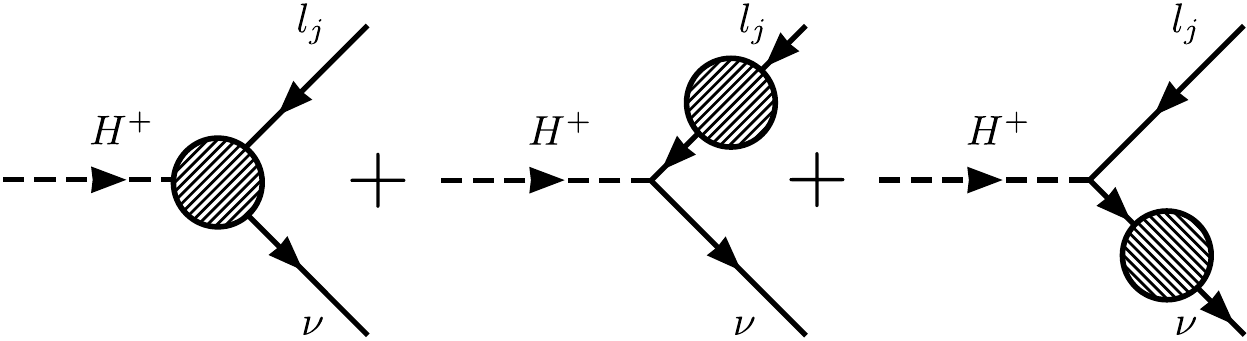}
\par\end{centering}

\caption{\label{fig:RK:Hvertex}Corrections to the $\nu\ell H^{+}$ vertex,
as discussed in the text.}
\end{figure}

From an effective theory approach, the HRS mechanism can be accounted
for by additional terms, corresponding to the higher-order corrections
to the Higgs-neutrino-charged lepton interaction (schematically depicted
in figure \eqref{fig:RK:Hvertex}). At tree-level, the Lagrangian
describing the $\nu\ell H^{\pm}$ interaction is given by
\begin{align}
\mathscr{L}_{0}^{H^{\pm}} & =\overline{\nu}_{L}Y^{\ell\dagger}\ell_{R}H_{d}^{-*}+\textrm{h.c.}\nonumber \\
 & =\left(2^{3/4}G_{F}^{1/2}\right)\tan\beta\,\overline{\nu}_{L}M^{\ell}\ell_{R}H^{+}+\textrm{h.c.}\,,\label{eq:L:nulH0}
\end{align}
with $M^{\ell}=\textrm{diag}\left(m_{e},m_{\mu},m_{\tau}\right)$.
At loop level, two new terms are generated: $\overline{\nu}_{L}\Delta^{+}\ell_{R}H_{u}^{+}-\overline{\ell}_{L}\Delta^{0}\ell_{R}H_{d}^{0}+\textrm{h.c.}$.
The second one, with $\Delta^{0}$, forces a redefinition of the charged
lepton Yukawa couplings, $Y^{\ell\dagger}=\nicefrac{M^{\ell}}{v_{d}}$
$\to$ $Y^{\ell\dagger}\approx\nicefrac{M^{\ell}}{v_{d}}-\Delta^{0}\tan\beta$,
which in turn implies a redefinition of the charged lepton propagator;
the term with $\Delta^{+}$ corrects the Higgs-neutrino-charged lepton
vertex%
\footnote{An extensive discussion on the radiatively induced couplings which
are at the origin of the HRS effect can be found in \cite{Borzumati:1999sp}.%
}. Once these terms are taken into account, the interaction Lagrangian
in equation \eqref{eq:L:nulH0} becomes
\begin{align}
\mathscr{L}^{H^{\pm}} & =\left(2^{3/4}G_{F}^{1/2}\right)\tan\beta\,\overline{\nu}_{L}M^{\ell}\ell_{R}H^{+}\nonumber \\
 & \qquad+\cos\beta\,\overline{\nu}_{L}\left(\Delta^{+}-\Delta^{0}\tan^{2}\beta\right)\ell_{R}H^{+}+\textrm{h.c.}\,.\label{eq:L:nulH}
\end{align}
Since in the $SU(2)_{L}$-preserving limit we have $\Delta^{+}=\Delta^{0}$,
it is reasonable to assume that, after electroweak (EW) symmetry breaking,
both terms remain approximately of the same order of magnitude. Hence,
it is clear that the contribution associated with $\Delta^{0}$ (the
loop contribution to the charged lepton mass term) will be enhanced
by a factor of $\tan^{2}\beta$ when compared to the one associated
with $\Delta^{+}$. This simple discussion elucidates the origin of
the dominant SUSY contribution%
\footnote{\label{fn:RK_footpag5}There are additional corrections to the $\overline{q}q^{\prime}H^{\pm}$
vertex, which are mainly due to a similar modification of the the
quark Yukawa couplings---especially that of the strange quarks. This
amounts to a small multiplicative effect on $\Delta r$ which we will
not discuss here (see \cite{Girrbach:2012km} for details).%
} to $R_{K}$.

To quantify the effect encoded in $\Delta^{+}$ and $\Delta^{0}$,
higher-order effects on the vertex $\overline{\nu}_{L}\, Z^{H}\,\ell_{R}\, H^{+}$
must be considered in a systematic way (see \cite{Bellazzini:2010gn}).
The $Z^{H}$ matrix depends on the following loop-induced quantities:
\begin{itemize}
\item $\eta_{L}^{\ell}$ and $\eta_{L}^{\nu}$ (corrections to the kinetic
terms of $\ell_{L}$ and $\nu_{L}$); 
\item $\eta_{m}^{\ell}$ (correction to the charged lepton mass term); 
\item $\eta^{H}$ (correction to the $\nu\ell H$ vertex). 
\end{itemize}
The expressions for the distinct $\eta$-parameters can be found in
appendix \ref{chap:Renormalization_of_the_NuLH_vertex}. Instead of
$Z^{H}$, which includes both tree and loop level effects, it is more
convenient to use the following combination,
\begin{equation}
-\frac{\tan\beta}{2^{3/4}G_{F}^{1/2}}\left(\frac{m_{K}}{m_{H^{+}}}\right)^{2}\frac{m_{s}}{m_{s}+m_{u}}Z^{H}\left(M^{\ell}\right)^{-1}\equiv\epsilon\mathbb{1}+\Delta\,,\label{eq:epsilondelta}
\end{equation}
where
\begin{align}
\epsilon & =-\tan^{2}\beta\left(\frac{m_{K}}{m_{H^{+}}}\right)^{2}\frac{m_{s}}{m_{s}+m_{u}}\,,\label{eq:epsilon:def}\\
\Delta & =\epsilon\left[\frac{\eta_{L}^{\ell}}{2}-\frac{\eta_{L}^{\nu}}{2}+\left(\frac{\eta^{H}}{2^{3/4}\, G_{F}^{1/2}\tan\beta}-\eta_{m}^{\ell}\right)\left(M^{\ell}\right)^{-1}\right]\,.\label{eq:delta:def}
\end{align}
In the above, $\epsilon$ encodes the tree level Higgs mediated amplitude
(which does not change the SM prediction for $R_{K}$), while $\Delta$,
a matrix in lepton flavor space, encodes the 1-loop effects. From
the simplified approach that led to equation \eqref{eq:L:nulH}, we
expect that the main contribution comes from $\eta_{m}^{\ell}$, which
corrects the charged lepton mass term. This, however, is only true
if the SUSY parameters are such that $\Delta r$ is highly enhanced;
if this is not the case, the remaining $\eta$'s should be taken into
consideration, as we shall do in the numerical calculations shown
in this chapter.

The $\Delta r$ observable is then related to $\epsilon$ and $\Delta$
as follows:
\begin{equation}
\Delta r\equiv\frac{R_{K}}{R_{K}^{\text{SM}}}-1=\frac{\left[\left(\mathbf{1}+\frac{\Delta^{\dagger}}{1+\epsilon}\right)\left(\mathbf{1}+\frac{\Delta}{1+\epsilon}\right)\right]_{ee}}{\left[\left(\mathbf{1}+\frac{\Delta^{\dagger}}{1+\epsilon}\right)\left(\mathbf{1}+\frac{\Delta}{1+\epsilon}\right)\right]_{\mu\mu}}-1\,.\label{eq:deltar:epsilondelta}
\end{equation}
If the slepton mixing is sufficiently large, this expression can be
approximated as
\begin{equation}
\Delta r\approx2\textrm{Re}\left(\Delta_{ee}\right)+\left(\Delta^{\dagger}\Delta\right)_{ee}\,.
\end{equation}
In the above, the first (linear) term on the right hand-side is due
to an interference with the SM process, and is thus lepton flavor
conserving. As shown in \cite{Masiero:2005wr}, this contribution
can be enhanced through both large $RR$ and $LL$ slepton mixing.
On the other hand, the quadratic term $(\Delta^{\dagger}\Delta)_{ee}$
can be augmented mainly through a large LFV contribution from $\Delta_{\tau e}$,
which can only be obtained in the presence of significant $RR$ slepton
mixing.

\subsection{The LFV in the slepton mass matrices as the source of an enhanced
$\Delta r$ }

In order to understand the dependence of $\Delta r$ on the SUSY parameters,
and the origin of the dominant contributions to this observable, an
approximate expression for $\Delta$ is required. Firstly, we remind
that the previous discussion leading to equation \eqref{eq:L:nulH}
suggests that the $\eta_{m}^{\ell}$ term is responsible for the dominant
contributions to $\Delta r$. Thus, in what follows, and for the purpose
of obtaining simple analytical expressions, we shall neglect the contributions
of the other terms (although these are included in the numerical analysis
of section \ref{sec:RK_res}). A fairly simple analytical insight
can be obtained when working in the limit in which the virtual particles
in the loops (sleptons and gauginos) are assumed to have similar masses,
so that their relative mass splittings are small. In this limit, one
can Taylor-expand the loop functions entering $\eta_{m}^{\ell}$ (see
appendix \ref{chap:Renormalization_of_the_NuLH_vertex}); working
to third order in this expansion, and keeping only the terms enhanced
by a factor of $m_{\tau}\,\tan\beta\,\frac{m_{SUSY}}{m_{EW}}$, we
obtain{
\thinmuskip=1mu
\medmuskip=1mu
\thickmuskip=1mu
\begin{align}
\Delta r & \sim\left[1+X\left(1-\frac{9}{10}\frac{\delta}{\overline{m}_{\widetilde{\ell},\chi^{0}}^{2}}\right)\left(m_{\widetilde{L}}^{2}\right)_{e\tau}\right]^{2}-1+X^{2}\left[-\mu^{2}+\delta\left(3-\frac{3}{10}\frac{\mu^{2}+2M_{1}^{2}}{\overline{m}_{\widetilde{\ell},\chi^{0}}^{2}}\right)\right]^{2}\,,\label{eq:RK_deltar:approx:long}
\end{align}
\thinmuskip=3mu
\medmuskip=4.0mu plus 2.0mu minus 4.0mu
\thickmuskip=5.0mu plus 5.0mu
}where $\mu$, $M_{1}$ and $\left(m_{\widetilde{L}}^{2}\right)_{e\tau}$
denote the low-energy values of the Higgs bilinear term, bino soft
breaking mass, and off-diagonal entry of the soft breaking left-handed
slepton mass matrix, respectively (see chapter \ref{chap:The-SM's-shortcomings}).
We have also introduced $\overline{m}_{\widetilde{\ell},\chi^{0}}^{2}=\frac{1}{2}\left(\left\langle m_{\widetilde{\ell}}^{2}\right\rangle +\left\langle m_{\chi^{0}}^{2}\right\rangle \right)$,
the average mass squared of sleptons and neutralinos ($\approx m_{SUSY}^{2}$),
and $\delta=\frac{1}{2}\left(\left\langle m_{\widetilde{\ell}}^{2}\right\rangle -\left\langle m_{\chi^{0}}^{2}\right\rangle \right)$,
the corresponding splitting. The quantity $X$ is given by
\begin{align}
X & \equiv\frac{1}{192\pi^{2}}\, m_{K}^{2}\, g'^{2}\,\mu\, M_{1}\,\frac{\tan^{3}\beta}{m_{H^{+}}^{2}}\,\frac{m_{\tau}}{m_{e}}\,\frac{\left(m_{\widetilde{e}}^{2}\right)_{\tau e}}{(\overline{m}_{\widetilde{\ell},\chi^{0}}^{2})^{3}}\,,\label{eq:deltar:approx:X}
\end{align}
and it illustrates in a transparent (albeit approximate) way the origin
of the terms contributing to the enhancement of $R_{K}$: in addition
to the factor $\tan^{3}\beta/m_{H^{+}}^{2}$ usually associated with
Higgs exchanges, the crucial flavor violating source emerges from
the off-diagonal $(\tau e)$ entry of the right-handed slepton soft
breaking mass matrix.

Using the above analytical approximation, one easily recovers the
results in the literature, usually obtained using the MIA. For instance,
equation (11) of reference \cite{Masiero:2005wr} amounts to
\begin{align}
\Delta r & \sim2X\left(m_{\widetilde{L}}^{2}\right)_{e\tau}+X^{2}\left(m_{\widetilde{L}}^{2}\right)_{e\tau}^{2}+X^{2}\delta^{2}\,,\label{eq:RK_deltar:approx}
\end{align}
which stems from having kept the dominant (crucial) second and third
order contributions in the expansion: $X^{2}\delta^{2}$ and $2X\left(m_{\widetilde{L}}^{2}\right)_{e\tau}+X^{2}\left(m_{\widetilde{L}}^{2}\right)_{e\tau}^{2}$,
respectively.

Regardless of the approximation considered, it is thus clear that
the LFV effects on kaon decays into a $e\nu$ or $\mu\nu$ pair can
be enhanced in the large $\tan\beta$ regime (especially in the presence
of low values of $m_{H^{+}}$), and via a large $RR$ slepton mixing
$\left(m_{\widetilde{e}}^{2}\right)_{\tau e}$. Although the latter
is indeed the privileged source, notice that, as can be seen from
equation \eqref{eq:RK_deltar:approx}, a strong enhancement can be
obtained from sizable flavor violating entries of the left-handed
slepton soft breaking mass, $\left(m_{\widetilde{L}}^{2}\right)_{e\tau}$.
This is in fact a globally flavor conserving effect (which can also
account for negative contributions to $R_{K}$). Previous experimental
measurements of $R_{K}$ appeared to favor values smaller than the
SM theoretical estimation, thus motivating the study of regimes leading
to negative values of $\Delta r$ \cite{Masiero:2005wr}, but these
regimes have now become disfavored in view of the present bounds in
equation \eqref{eq:deltarexp}.

Clearly, these Higgs mediated exchanges, as well as the FV terms at
the origin of the strong enhancement to $R_{K}$, will have an impact
on a number of other low-energy observables, as can be easily inferred
from the structure of equations \eqref{eq:RK_deltar:approx:long}--\eqref{eq:RK_deltar:approx}.
This has been extensively addressed in the literature \cite{Masiero:2005wr,Masiero:2008cb,Ellis:2008st,Girrbach:2012km},
and here we will only briefly discuss the most relevant observables:
electroweak precision data on the anomalous electric and magnetic
moments of the electron, as well as the naturalness of the electron
mass, directly constrain the $\eta_{m}^{\ell}$ corrections (and $\eta_{L}^{\ell}$,
$\eta^{H}$); low-energy cLFV observables, such as $\tau\to\ell\gamma$
and $\tau\to3\ell$ decays are also extremely sensitive probes of
Higgs mediated exchanges, and in the case of $\tau-e$ transitions,
depend on the same flavor violating entries. It has been suggested
that positive and negative values of $\Delta r$ can be of the order
of 1\%, still in agreement with data on the electron's electric dipole
moment and on $\tau\rightarrow\ell\gamma$ \cite{Masiero:2005wr,Masiero:2008cb,Ellis:2008st}.
Finally, other meson decays, such as $B\to\ell\ell$ (and $B\to\ell\nu$),
exhibit a similar dependence on $\tan\beta$, $\tan^{n}\beta/m_{H^{+}}^{4}$
\cite{Parry:2005fp,*Ellis:2007kb} ($n$ ranging from 2 to 6, depending
on the other SUSY parameters), and may also lead to indirect bounds
on $\Delta r$. In particular, the strict bounds on $\textrm{BR}\left(B_{u}\to\tau\nu\right)$
\cite{Amsler:2008zzb} and the recent measurement of $\textrm{BR}\left(B_{s}\to\mu\mu\right)$
\cite{Aaij:2012nna} might severely constrain the allowed regions
in SUSY parameter space for large $\tan\beta$. Although we will come
to this issue in greater detail when discussing the numerical results,
it is clear from the similar nature of the $K^{+}\to\ell\nu$ and
$B_{u}\to\tau\nu$ processes (easily inferred from a generalization
of equation \eqref{eq:kaon:gamma:smsusy}, see for example \cite{Hou:1992sy,Isidori:2006pk})
that light charged Higgs masses, which saturate the bounds on $R_{K}$,
lead to a tension.

Supersymmetric models of neutrino mass generation (such as the SUSY
seesaw) naturally induce sizable cLFV contributions, via radiatively
generated off-diagonal terms in the $LL$ (and to a lesser extent
$LR$) slepton soft breaking mass matrices \cite{Borzumati:1986qx}.
In addition to explaining neutrino masses and mixing, such models
can also easily account for values of $\textrm{BR}\left(\mu\to e\gamma\right)$,
within the reach of the MEG experiment. In view of the recent confirmation
of a large value for the Chooz angle ($\theta_{13}\sim8.8^{\circ}$)
\cite{An:2012eh,*Ahn:2012nd,*Adamson:2012rm} and on the impact it
might have on $\left(m_{\tilde{L}}^{2}\right)_{e\tau}$, in the numerical
analysis of the following section we will also consider different
realizations of the SUSY seesaw (type-I \cite{Minkowski:1977sc,GellMann:1980vs,*Yanagida:1979as,*Glashow:1979nm,*Mohapatra:1979ia},
type-II \cite{Barbieri:1979ag,*Marshak:1980yc,*Cheng:1980qt,*Magg:1980ut,*Lazarides:1980nt,*Schechter:1980gr,*Mohapatra:1980yp},
and inverse \cite{Mohapatra:1986bd}), embedded in the framework of
constrained SUSY models. We will also revisit semi-constrained scenarios
allowing for light values of $m_{H^{+}}$, re-evaluating the predictions
for $R_{K}$ under a full, one loop-computation, and in view of recent
experimental data. Finally, we confront these (semi-)constrained scenarios
with general, low-energy realizations, of the MSSM.

\section{\label{sec:RK_res}Prospects for $R_{K}$: unified vs unconstrained
SUSY models}

In this section we evaluate the SUSY contributions to $R_{K}$, with
the results obtained via the full expressions for $\Delta r$, as
described in section \eqref{sec:RK_formulae} \cite{Fonseca:2012kr}.
These were implemented into the \texttt{SPheno} public code \cite{Porod:2003um,Porod:2011nf},
which was modified to allow the different studies. It is important
to stress that even though some approximations were used (as previously
discussed), the results of the present computation strongly improve
upon those so far reported in the literature (mostly obtained using
the MIA). Although the different contributions cannot be easily disentangled
in a full computation, our results automatically include all one-loop
lepton flavor violating and lepton flavor conserving contributions
(in association with charged Higgs mediation, see footnote \ref{fn:RK_footpag5}).
As mentioned before, we evaluate $R_{K}$ in the framework of constrained
(the cMSSM), semi-constrained (the NUHM) and unconstrained SUSY models
(the general MSSM)---see chapter \ref{chap:The-SM's-shortcomings}
for details on these models. We will also consider the supersymmetrization
of several mechanisms for neutrino mass generation. More specifically,
we have considered the type-I and type-II SUSY seesaw (as detailed
in chapter \ref{chap:Lepton-flavour-violation}). We shall briefly
comment on the inverse SUSY seesaw, and discuss a $LR$ model.

\bigskip{}
In our numerical analysis, we took into account LHC bounds on the
SUSY spectrum \cite{ATLAS:2012ai,*ATLAS:2012ab,*Koay:2012ks,*Aad:2011zj,*Aad:2011ib,*Aad:2011kz,*Aad:2011xm,*Aad:2011xk,*Aad:2011ks,*Aad:2011yf,*daCosta:2011qk,*Aad:2011hh,*Chatrchyan:2011zy,*Chatrchyan:2011qs,*Chatrchyan:2011ek,*Collaboration:2011ida,*Chatrchyan:2011bj,*Chatrchyan:2011ff,*Chatrchyan:2011ah,*Chatrchyan:2011wba,*Chatrchyan:2011bz,*Chatrchyan:2011wc,*Khachatryan:2011tk,*Khachatryan:2010uf},
as well as the constraints from low-energy flavor dedicated experiments
\cite{Amsler:2008zzb}, and neutrino data \cite{Fogli:2011qn,Schwetz:2011zk}.
In particular, concerning lepton flavor violation, we have considered
\cite{Amsler:2008zzb,Adam:2011ch}:
\begin{align}
\text{BR}(\tau\to e\gamma) & <3.3\times10^{-8} & (90\%\text{ C.L.})\,,\label{eq:cLFVbounds1}\\
\text{BR}(\tau\to3\, e) & <2.7\times10^{-8} & (90\%\text{ C.L.})\,,\label{eq:cLFVbounds2}\\
\text{BR}(\mu\to e\gamma) & <2.4\times10^{-12} & (90\%\text{ C.L.})\,,\\
\text{BR}(B_{u}\to\tau\nu) & >9.7\times10^{-5} & (2\,\sigma)\,.\label{eq:cLFVbounds3}
\end{align}
Also relevant are the following B meson bounds from LHCb \cite{Aaij:2012ac}
\begin{align}
\text{BR}(B_{s}\rightarrow\mu\mu) & <4.5\times10^{-9} & (95\%\text{ C.L.})\,,\label{eq:Bbounds1}\\
\text{BR}(B\to\mu\mu) & <1.03\times10^{-9} & (95\%\text{ C.L.})\,.\label{eq:Bbounds2}
\end{align}
It is worth mentioning that, since this analysis was first performed,
the LHCb \cite{Aaij:2012nna} and the MEG collaborations \cite{Adam:2013mnn}
has released new results; in particular there is now evidence for
the decay $B_{s}\rightarrow\mu\mu$ (see chapter \ref{chap:Lepton-flavour-violation}).%
\footnote{The $1\sigma$ upper bound for $\textrm{BR}\left(B_{s}\rightarrow\mu\mu\right)$
obtained recently \cite{Aaij:2012nna} is close to the value in equation
\eqref{eq:Bbounds1}.%
} Nevertheless we find that this does have a significant impact in
our findings.

When addressing models for neutrino mass generation, we take the following
values for the neutrino mixing angles \cite{Schwetz:2011zk} (where
$\theta_{13}$ is already in good agreement with the recent results
from~\cite{An:2012eh,*Ahn:2012nd,*Adamson:2012rm}),
\begin{align}
\sin^{2}\theta_{12} & =0.312_{-0.015}^{+0.017},\quad\sin^{2}\theta_{23}=0.52_{-0.07}^{+0.06},\quad\sin^{2}\theta_{13}\approx0.013_{-0.005}^{+0.007}\,,\label{eq:mixingangles:data}\\[2mm]
\Delta m_{\text{12}}^{2} & =(7.59_{-0.18}^{+0.20})\times10^{-5}\text{ eV}^{2}\,,\quad\Delta m_{\text{13}}^{2}=(2.50_{-0.16}^{+0.09})\times10^{-3}\text{ eV}^{2}\,,\\[2mm]
\end{align}
and all CP violating phases are set to zero.%
\footnote{We will assume that we are in a strictly CP conserving framework,
where all terms are taken to be real. This implies that there will
be no contributions to observables such as electric dipole moments,
or CP asymmetries.%
} See however chapter \ref{chap:Lepton-flavour-violation} for more
up-to-date numbers \cite{Tortola:2012te,Fogli:2012ua}.

\subsection{mSUGRA inspired scenarios: cMSSM and the SUSY seesaw}

We begin by re-evaluating, through a full computation of the one-loop
corrections, the maximal amount of supersymmetric contributions to
$R_{K}$ in constrained SUSY scenarios.

As could be expected from equations \eqref{eq:RK_deltar:approx:long}--\eqref{eq:RK_deltar:approx},
in a strict cMSSM scenario (in agreement with the experimental bounds
above referred to) the SUSY contributions to $R_{K}$ are extremely
small; motivated by the need to accommodate neutrino data, and at
the same time accounting for values of BR($\mu\to e\gamma$) within
MEG reach, we implement type-I and type-II seesaws in mSUGRA-inspired
models. Regarding the heavy-scale mediators, we considered degenerate
right-handed neutrinos, as well as degenerate scalar triplets. We
set the seesaw scale aiming at maximizing the low-energy, non-diagonal
entries of the soft breaking slepton mass matrices, while still in
agreement with the current low-energy bounds (see equations \eqref{eq:cLFVbounds1}--\eqref{eq:Bbounds2}).
In particular, we tried to maximize the $LL$ contributions to $\Delta r$,
i.e., $\left(m_{\widetilde{L}}^{2}\right)_{e\tau}$, and to obtain
BR($\mu\to e\gamma$) within MEG reach (i.e. $10^{-13}\lesssim$ BR($\mu\to e\gamma$)$\lesssim2.4\times10^{-12}$).%
\footnote{Indeed, the more recent bound from MEG is $5.7\times10^{-13}$ \cite{Adam:2013mnn}.%
} However, and due to the fact that both seesaw realizations fail to
account for radiatively induced LFV in the right-handed slepton sector,
one finds values $|\Delta r|\lesssim2\times10^{-8}$. It is worth
emphasizing that if one further requires $m_{h}$ to lie close to
125 GeV, then one is led to regions in mSUGRA parameter space where,
due to the much heavier sparticle masses and typically lower values
of $\tan\beta$, the SUSY contributions to $R_{K}$ become even further
suppressed.

Thus, and even under a full computation of the corrections to the
$\nu\ell H^{+}$ vertex, we nevertheless confirm that, as firstly
put forward in the analyzes of \cite{Masiero:2005wr,Masiero:2008cb}
strictly constrained SUSY and SUSY seesaw models indeed fail to account
for values of $R_{K}$ close to the present limits.

\bigskip{}
Clearly, new sources of flavor violation, associated to the right-handed
sector are required: in what follows, we maintain universality of
soft breaking terms allowing, at the grand unified (GUT) scale, for
a single $\tau-e$ flavor violating entry in $m_{\widetilde{e}}^{2}$.
This approach is somewhat closer to the lines of \cite{Masiero:2005wr,Masiero:2008cb,Ellis:2008st,Girrbach:2012km},
although in our computation we will still conduct a full evaluation
of the distinct contributions to $\Delta r$, and we consider otherwise
universal soft breaking terms. Without invoking a specific framework/scenario
of SUSY breaking that would account for such a pattern, we thus set
\begin{equation}
\delta_{\tau e}^{RR}=\frac{\left(m_{\widetilde{e}}^{2}\right)_{\tau e}}{m_{0}^{2}}\neq0\,.\label{eq:mia:sleptons}
\end{equation}
As discussed above, low-energy constraints on LFV observables (especially
$\tau\rightarrow e\gamma$), severely constrain this entry.

\begin{figure}[tbh]
\centering{}%
\begin{tabular}{cc}
\includegraphics[width=0.42\linewidth]{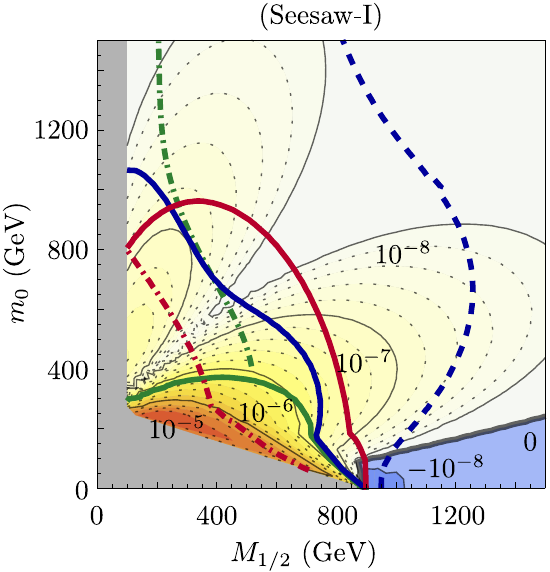}\hspace*{5mm}  & \hspace*{5mm} \includegraphics[width=0.42\linewidth]{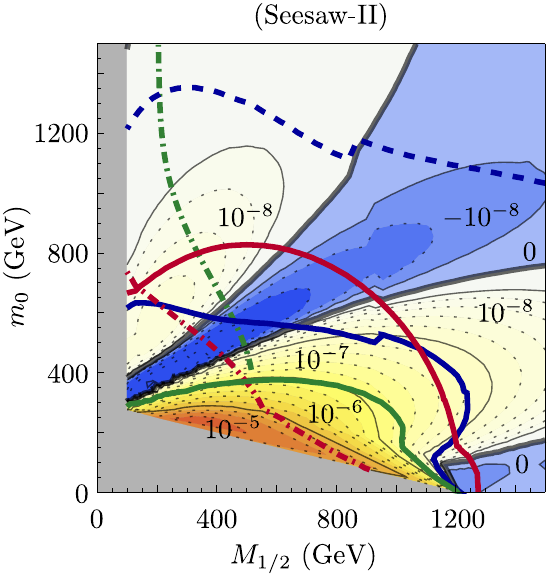} \tabularnewline
\end{tabular}\caption{\label{fig:cMSSM:seesaw}$m_{0}-M_{1/2}$ plane for $\tan\beta=40$
and $A_{0}=-500$ GeV, with $\delta_{\tau e}^{RR}=0.1$. On the left
(right) panel, a type-I (II) SUSY seesaw, considering degenerate heavy
mediators. Contour lines denote values of $\Delta r$ (decreasing
values: positive---in association with an orange-yellow-white color
gradient; negative---blue gradients); solid (gray) regions are excluded
due to the requirement of having the correct EWSB. A green dot-dashed
line corresponds to the present LHC bounds on the cMSSM \cite{website_LHC_Limits_used}.
A full green line delimits the BR($\tau\to e\gamma$) exclusion region,
while full (dot-dashed) red lines correspond to the bounds on BR($B_{s}\rightarrow\mu\mu$)
{[}BR($B_{u}\to\tau\nu$){]}. Finally, the region delimited by blue
lines corresponds to having BR($\mu\to e\gamma$) within MEG reach
(current bound---solid line, future sensitivity---dashed line).}
\end{figure}

In figure \eqref{fig:cMSSM:seesaw}, we present our results for $\Delta r$
scanning the $m_{0}-M_{1/2}$ plane for a regime of large $\tan\beta$.
We have set $\delta_{\tau e}^{RR}=0.1$, $\tan\beta=40$, and taken
$A_{0}=-500$ GeV. The surveys displayed in the panels correspond
to having embedded a type-I (left) or type-II (right) seesaw onto
this near-mSUGRA framework. 

As can be readily seen from figure \eqref{fig:cMSSM:seesaw}, once
the constraints from low-energy observables have been applied, in
the type-I SUSY seesaw, the maximum values for $\Delta r$ are $\mathcal{O}(10^{-7})$,
associated to the region with a lighter SUSY spectra (which is in
turn disfavored by a {}``heavy'' light Higgs). Even for the comparatively
small non-universality, $\delta_{\tau e}^{RR}=0.1$, a considerable
region of the parameter space is excluded due to excessive contributions
to BR($B_{u}\to\tau\nu$) and BR($\tau\to e\gamma$), thus precluding
the possibility of large values of $\Delta r$. In a regime of large
$\tan\beta$, the contributions to BR($B_{s}\rightarrow\mu\mu$) are
also sizable, and LHCb results seem to exclude the regions of the
parameter space where one could still have $\Delta r\sim\mathcal{O}(10^{-6,-7})$.
The excessive SUSY contributions to BR($B_{s}\rightarrow\mu\mu$)
can be somewhat reduced by adjusting $A_{0}$ (in figure \eqref{fig:cMSSM:seesaw}
we used $A_{0}=-500$ GeV) and the values of $\Delta r$ can be slightly
augmented by increasing $\delta_{\tau e}^{RR}$; in the latter case,
the $\tau\to e\gamma$ bound proves to be the most constraining, and
values of $\Delta r$ larger than $\mathcal{O}(10^{-6,-7})$ cannot
be obtained in these constrained SUSY seesaw models.

The situation is somewhat different for the type-II case: first, notice
that a sizable region in the $m_{0}-M_{1/2}$ plane is associated
to negative contributions to $R_{K}$, which are currently disfavored.
In the remaining (allowed) parameter space, the values of $\Delta r$
are slightly smaller than for the type-I case: this is a consequence
of a non trivial interplay between a smaller value for the splitting
$\delta=\frac{1}{2}(\langle{m}_{\widetilde{\ell}}^{2}\rangle-\langle{m}_{\chi^{0}}^{2}\rangle)$
(induced by a lighter spectra), and a lighter charged Higgs boson.

Notice that in both SUSY seesaws it is fairly easy to accommodate
a potential observation of BR($\mu\to e\gamma$) $\sim10^{-13}$ by
MEG, taking for instance $M_{{\rm seesaw}}\sim10^{12}$ GeV for the
type-I and II seesaw mechanisms.

\begin{figure}[tbh]
\centering{}%
\begin{tabular}{cc}
\includegraphics[width=0.45\linewidth]{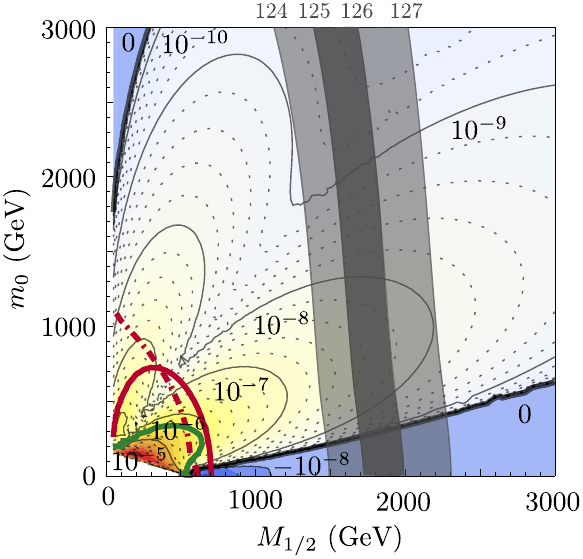}  & \includegraphics[width=0.45\linewidth]{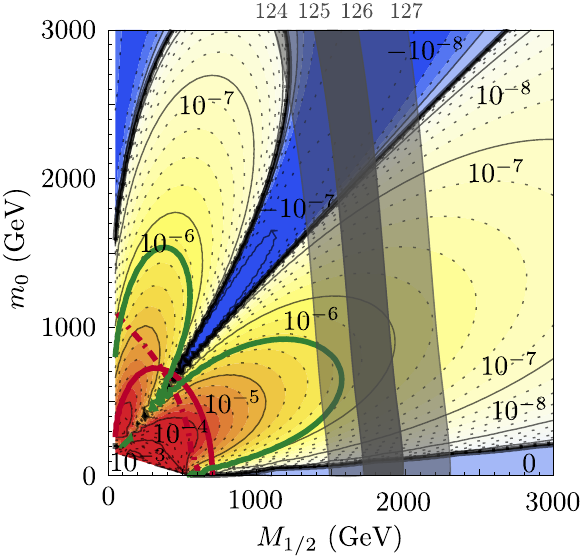} \tabularnewline
\end{tabular}\caption{\label{fig:cMSSM:survey}mSUGRA (type-I seesaw) $m_{0}-M_{1/2}$ plane
for $\tan\beta=40$ and $A_{0}=0$ GeV, with $\delta_{\tau e}^{RR}=0.1$
(left panel) and $\delta_{\tau e}^{RR}=0.7$ (right panel). Contour
lines denote values of $\Delta r$ (decreasing values: positive---in
association with an orange-yellow-white color gradient; negative---blue
gradients). A full green line delimits the BR($\tau\to e\gamma$)
exclusion region, while full (dot-dashed) red lines correspond to
the bounds on BR($B_{s}\rightarrow\mu\mu$) (BR($B_{u}\to\tau\nu$)).
Superimposed are the regions for the Higgs boson mass: the dark band
is for $125\leq m_{h^{0}}\leq126$ (GeV) and the lighter one marks
the region where $124\leq m_{h^{0}}\leq127$ (GeV).}
\end{figure}

In order to conclude this part of the analysis we provide a comprehensive
overview of the constrained MSSM prospects regarding $R_{K}$, presenting
in figure \eqref{fig:cMSSM:survey} a survey of the (type-I seesaw)
mSUGRA parameter space, for two different regimes of $\delta_{\tau e}^{RR}$,
taking all bounds (including the recent ones on $m_{h}$) into account.
The panels of figure \eqref{fig:cMSSM:survey} allow to recover the
information that could be expected from the discussion following figure
\eqref{fig:cMSSM:seesaw}: for fixed values of $A_{0}$ and $\tan\beta$,
increasing $\delta_{\tau e}^{RR}$ indeed allows to augment the SUSY
contributions to $\Delta r$ although, as can be seen from the right-panel,
the constraints from BR($\tau\to e\gamma$) become increasingly harder
to accommodate. Notice that the latter could be avoided by increasing
the SUSY scale (i.e., larger $m_{0}$ and/or $M_{1/2}$). However,
and shown in figure \eqref{fig:cMSSM:survey}, in a constrained SUSY
framework this would lead to heavier charged Higgs masses, and in
turn to suppressed contributions to $\Delta r$.

Although we do not display an analogous plot here, the situation is
very similar for the type-II SUSY seesaw (even though accommodating
$m_{h}\sim125$ GeV is more difficult in these models \cite{Hirsch:2012ti}).

In view of the above discussion, it is clear that even taking into
account all 1-loop corrections to the $\nu\ell H^{+}$ vertex, it
is impossible to saturate $\Delta r$'s current experimental limit
in the framework of constrained SUSY models (and its seesaw extensions
accommodating neutrino data). In this sense, and even though we have
followed a different approach, our results follow the conclusions
of \cite{Ellis:2008st}. We also stress that recent experimental bounds
(both from flavor observables and collider searches) add even more
severe constraints to the maximal possible values of $\Delta r$.

\subsection{mSUGRA inspired scenarios: inverse seesaw and $LR$ models }

We briefly comment here on the prospects of the inverse SUSY seesaw
concerning $R_{K}$: it was pointed out in \cite{Abada:2011hm} that
some flavor violating observables can be enhanced by as much as two
orders of magnitude in a model with the inverse seesaw mechanism.
Within such a framework, right-handed (s)neutrino masses can be relatively
light, and as a consequence these $\nu_{R}$, $\widetilde{\nu}_{R}$
states do not decouple from the theory until the TeV scale, hence
potentially providing important contributions to different low-energy
processes. Nevertheless, the specific contributions to $\Delta r$
are suppressed by a factor $\frac{m_{e}^{2}}{m_{\tau}^{2}}$, with
respect to those discussed above (see equation \eqref{eq:deltar:approx:X}),
so that we do not expect a significant enhancement of SUSY 1-loop
Higgs mediated effects to $R_{K}$ due to the inverse seesaw mechanism.
However, we note that it was shown in \cite{Abada:2012mc} that a
change to the $W\ell\nu$ vertex in such models could potentially
lead to a large $\Delta r$ ($\sim\mathcal{O}\left(1\right)$).

For completeness (and although we do not provide specific details
here), we have considered a specific $LR$ seesaw model \cite{Esteves:2010si}.
In this framework, non-vanishing values of $\delta_{\tau e}^{RR}$
can be dynamically generated. We have numerically verified that typically
one finds $\delta_{\tau e}^{RR}\lesssim0.01$ (we do not dismiss that
larger values might be found, although certainly requiring a considerable
amount of fine-tuning in the input parameters). We have not done a
dedicated $\Delta r$ calculation for this case, but considering that
$\Delta r\propto\left(\delta_{\tau e}^{RR}\right)^{2}$, we also expect
the typical range of $\Delta r$ to be far below the current experimental
sensitivity.

\subsection{mSUGRA inspired scenarios: NUHM}

As can be seen from the approximate expression for $\Delta r$ in
equations \eqref{eq:deltar:approx:X} and \eqref{eq:RK_deltar:approx},
regimes associated with both large $\tan\beta$ and a light charged
Higgs can greatly enhance this observable \cite{Ellis:2008st} ($\Delta r\propto\nicefrac{\tan^{6}\beta}{m_{H^{+}}^{4}}$).
By relaxing the mSUGRA-inspired universality conditions for the Higgs
sector, as occurs in NUHM scenarios (see chapter \ref{chap:The-SM's-shortcomings}),
one can indeed have very low masses for the $H^{+}$ boson at low
energies. This regime corresponds to a narrow strip in parameter space
where $m_{H_{1}}^{2}\approx m_{H_{2}}^{2}$, in particular when both
are close to $-(2.2\:\textrm{TeV})^{2}$. In addition to favoring
electroweak symmetry breaking, since $m_{H^{+}}^{2}\sim\left|m_{H_{1}}^{2}-m_{H_{2}}^{2}\right|$
(even accounting for RG evolution of the parameters down to the weak
scale), it is expected that the charged Higgs can be made very light
with some fine tuning \cite{Ellis:2008st}. In order to explore the
maximal values of $\Delta r$, a small scan was conducted around this
region, where $m_{H^{+}}$ changes very rapidly (see table \eqref{tab:NUHM}).
\begin{table}[tbh]
\centering{}%
\begin{tabular}{cccccc}
\toprule 
 & $m_{0}$ & $M_{1/2}$ & $m_{H_{1}}^{2}$, $m_{H_{2}}^{2}$ & $\tan\beta$ & $\delta_{\tau e}^{RR}$\tabularnewline
 & $\textrm{(GeV)}$ & $\textrm{(GeV)}$ & $\textrm{(Ge\ensuremath{V^{2}})}$ &  & \tabularnewline
\midrule
Min  & 0  & 100  & $-5.2\times10^{6}$  & 40  & 0.1\tabularnewline
Max  & 1500  & 1500  & $-4.6\times10^{6}$  & 40  & 0.7\tabularnewline
\bottomrule
\end{tabular}\caption{\label{tab:NUHM}Range of NUHM parameters leading to the scan of figure
\eqref{fig:NUHM:deltar}. }
\end{table}

\begin{figure}[tbh]

\centering{} %
\begin{tabular}{@{}r@{}c}
\includegraphics[scale=0.77]{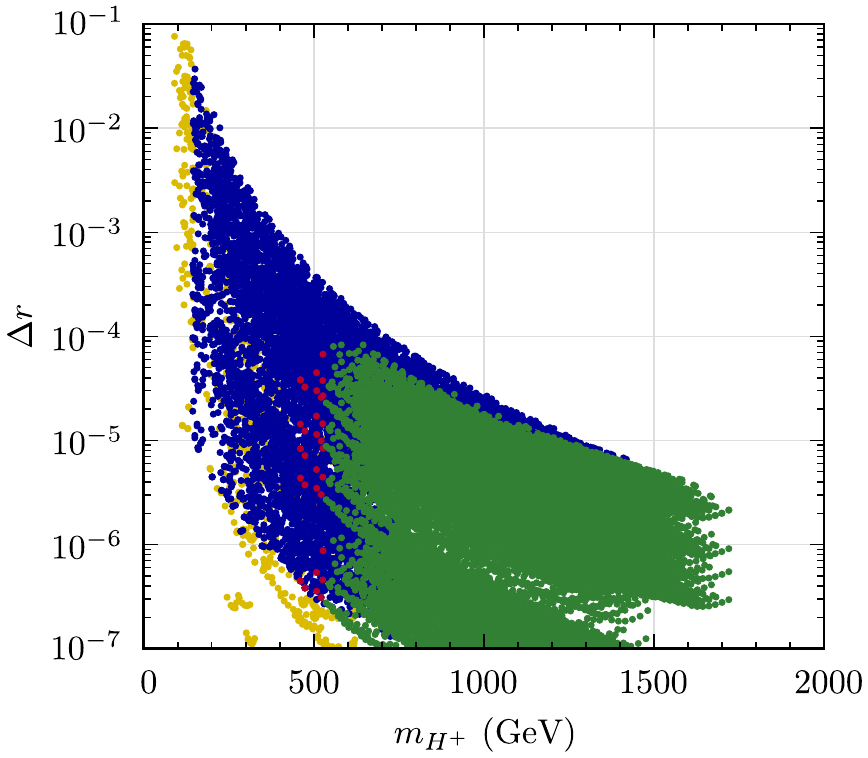}  & \includegraphics[scale=0.77]{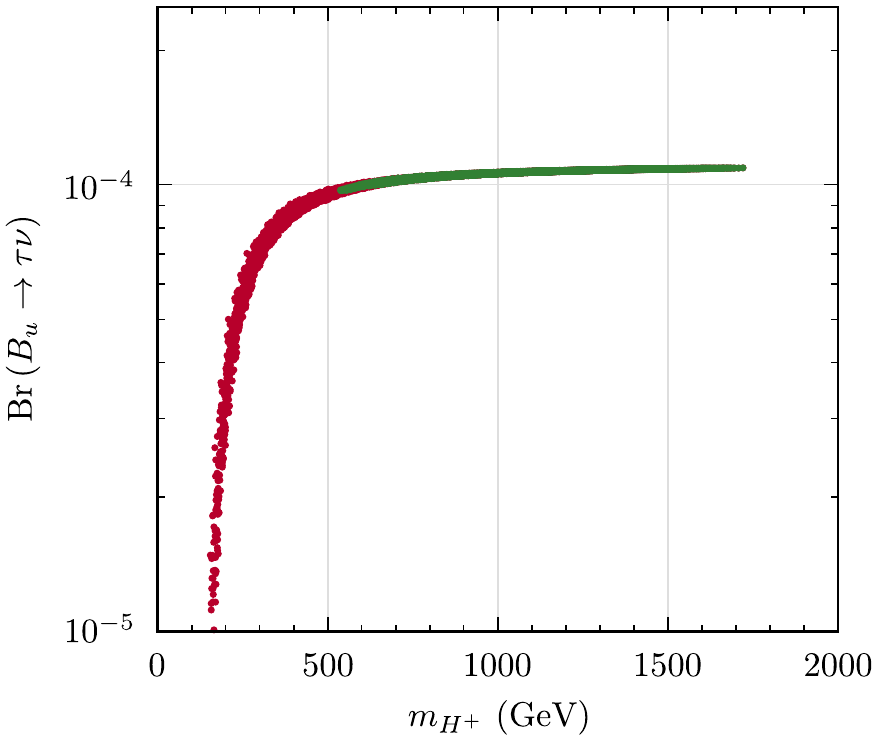} \tabularnewline
\end{tabular}\caption{\label{fig:NUHM:deltar}Left panel: $\Delta r$ as a function of the
charged Higgs mass, $m_{H^{+}}$ (in GeV). Yellow points have been
subject to no cuts, blue points comply with the bounds on the masses
(LEP+LHC), red points satisfy all bounds except BR($B_{u}\to\tau\nu$)
and green points satisfy all bounds. Right panel: BR($B_{u}\to\tau\nu$)
versus $m_{H^{+}}$. Red points satisfy only the bounds on the masses
(LEP+LHC) while green points comply with all bounds.}
\end{figure}

As can be verified from the left-hand panel of figure \eqref{fig:NUHM:deltar},
one could in principle have semi-constrained regimes leading to sizable
values of $R_{K}$, $\mathcal{O}(10^{-2})$. Once all (collider and
low-energy) bounds have been imposed, one has at most $\Delta r\lesssim10^{-4}$
(in association with $m_{H^{+}}\gtrsim500$ GeV). Moreover, it is
interesting to notice that SUSY contributions to BR($B_{u}\to\tau\nu$),
which become non-negligible for lighter $H^{\pm}$, have a negative
interference with those of the SM, lowering the latter branching ratio
to values below the current experimental bound. This can be seen on
the right-hand panel of figure \eqref{fig:NUHM:deltar}. The following
subsection addresses this topic in greater detail.

\subsection{Unconstrained MSSM}

To conclude the numerical discussion, and to allow for a better comparison
between our approach and those usually followed in other analyzes
(for instance \cite{Masiero:2008cb,Girrbach:2012km}), we conduct
a final study of the unconstrained, low-energy MSSM. Thus, and in
what follows, we make no hypothesis concerning the source of lepton
flavor violation, nor on the underlying mechanism of SUSY breaking.
Massive neutrinos are introduced by hand (no assumption being made
on their nature), and although charged interactions do violate lepton
flavor, as parametrized by the PMNS matrix $U$, no sizable contributions
to BR($\mu\to e\gamma$) should be expected, as these would be suppressed
by the light neutrino masses. At low-energies, no constraints (other
than the relevant experimental bounds) are imposed on the SUSY spectrum
(for simplicity, we will assume a common value for all sfermion trilinear
couplings at the low-scale, $A_{i}=A_{0}$). The soft breaking slepton
masses are allowed to be non-diagonal, so that \textit{a priori} a
non-negligible mixing in the slepton sector can occur. In order to
better correlate the source of flavor violation at the origin of $\Delta r$
with the different experimental bounds, we again allow for a single
FV entry in the slepton mass matrices, $\delta_{\tau e}^{RR}\sim0.5$,
setting all other $\delta_{ij}^{XY}$ to zero.

\begin{table}[tbh]
\setlength{\tabcolsep}{4pt}

\begin{centering}
\begin{tabular}{cccccccccccc}
\toprule 
 & $\mu$ & $m_{A}$ & %
\begin{tabular}{@{}c@{}}
$M_{1},$\tabularnewline
$M_{2}$\tabularnewline
\end{tabular} & $M_{3}$ & $A_{0}$ & $m_{\widetilde{L}}$ & $m_{\widetilde{e}}$ & %
\begin{tabular}{@{}c@{}}
$m_{Q},m_{U},$\tabularnewline
$m_{D}$\tabularnewline
\end{tabular} & $\tan\beta$ & $\delta_{\tau e}^{RR}$ & %
\begin{tabular}{@{}c@{}}
other\tabularnewline
$\delta_{ij}^{XY}$\tabularnewline
\end{tabular}\tabularnewline
\midrule
Min & 100  & 50  & 100  & 1100  & -1000  & 100  & 100  & 1200  & 30  & 0.5  & 0\tabularnewline
Max & 3000  & 1500  & 2500  & 2500  & 1000  & 2200  & 2500  & 5000  & 60  & 0.5  & 0\tabularnewline
\bottomrule
\end{tabular}
\par\end{centering}

\setlength{\tabcolsep}{6pt}

\caption{\label{tab:RK_ranges}Range of variation of the unconstrained MSSM
parameters (dimensionful parameters in GeVs). $A_{0}$ denotes the
common value of the low-energy sfermion trilinear couplings. }
\end{table}

In our scan we have varied the input parameters in the ranges collected
in table \eqref{tab:RK_ranges}. We have also applied all relevant
constraints on the low-energy observables, equations \eqref{eq:cLFVbounds1}--\eqref{eq:Bbounds2},
as well as the constraints on the SUSY spectrum \cite{Amsler:2008zzb,ATLAS:2012ai,*ATLAS:2012ab,*Koay:2012ks,*Aad:2011zj,*Aad:2011ib,*Aad:2011kz,*Aad:2011xm,*Aad:2011xk,*Aad:2011ks,*Aad:2011yf,*daCosta:2011qk,*Aad:2011hh,*Chatrchyan:2011zy,*Chatrchyan:2011qs,*Chatrchyan:2011ek,*Collaboration:2011ida,*Chatrchyan:2011bj,*Chatrchyan:2011ff,*Chatrchyan:2011ah,*Chatrchyan:2011wba,*Chatrchyan:2011bz,*Chatrchyan:2011wc,*Khachatryan:2011tk,*Khachatryan:2010uf}.
In particular we have assumed the limits 
\begin{equation}
m_{\widetilde{q}_{L,R}}>1000\text{ GeV}\,,\qquad m_{\widetilde{g}}>1000\text{ GeV}\,,
\end{equation}
which nonetheless can be raised even further without affecting the
$R_{K}$ observable. Concerning the light Higgs boson mass, no constraint
was explicitly imposed, but we note that values close to 125 GeV \cite{ATLAS-CONF-2013-014,CMS-PAS-HIG-13-005},
or even larger, are easily achievable due to the heavy squark masses.

\begin{figure}[tbh]
\centering{}\includegraphics[clip,scale=0.77]{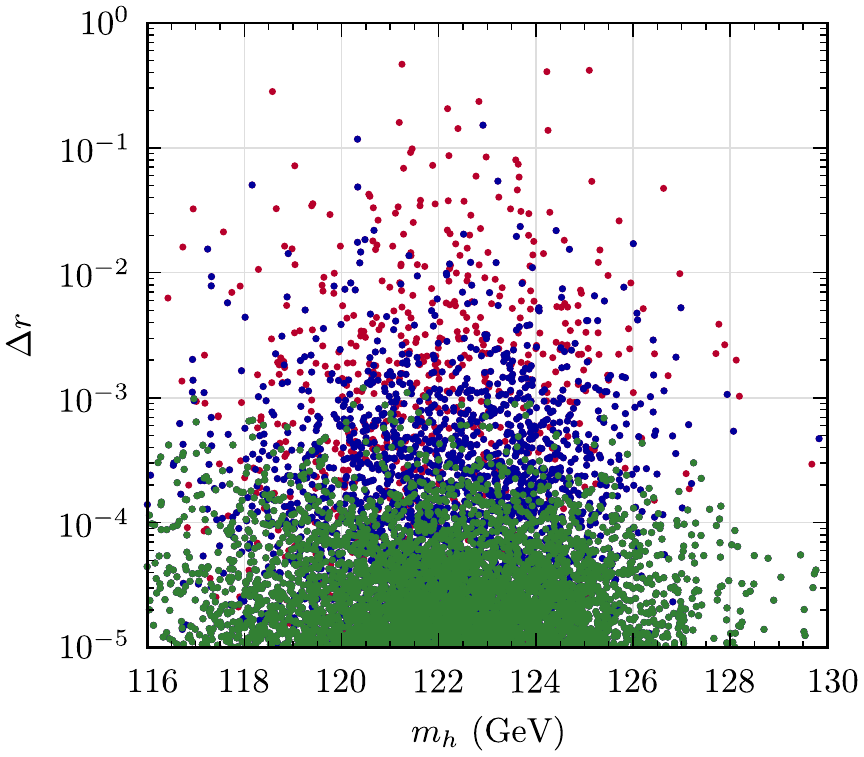}~~\includegraphics[clip,scale=0.77]{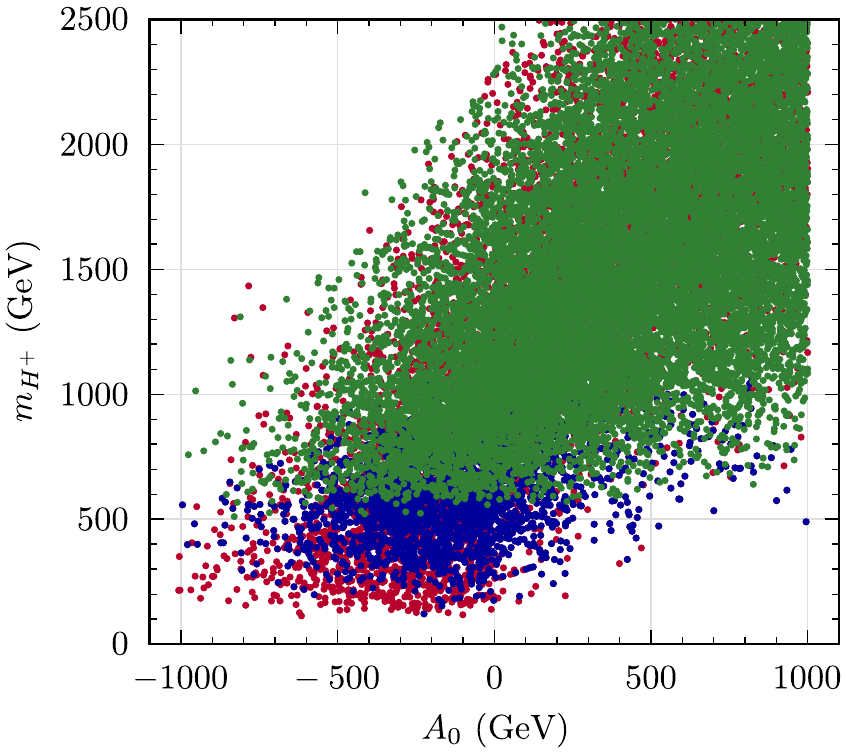}
\caption{\label{fig:RK_DeltaR-mh0}Left panel: $\Delta r$ as a function of
the lightest Higgs boson mass $m_{h}$ (in GeV) for the range of parameters
shown in table \eqref{tab:RK_ranges}. Red points satisfy the bounds
on the spectrum (LEP+LHC), blue points satisfy all bounds except $\textrm{BR}\left(B_{u}\to\tau\nu\right)$
and green points satisfy all bounds. Right panel: $m_{H^{+}}$ versus
$A_{0}$, with the same color code. Both plots were produced by varying
the different input parameters as in table \eqref{tab:RK_ranges}.}
\end{figure}

\begin{figure}[tbh]
\centering{}%
\begin{tabular}{@{}r@{}c}
\includegraphics[clip,scale=0.76]{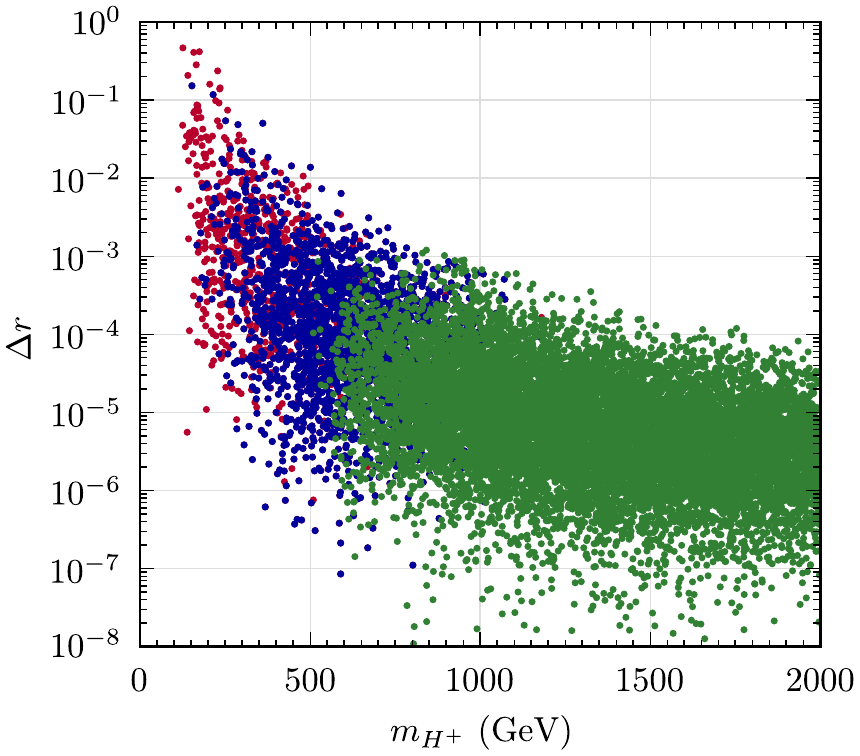} & \includegraphics[clip,scale=0.76]{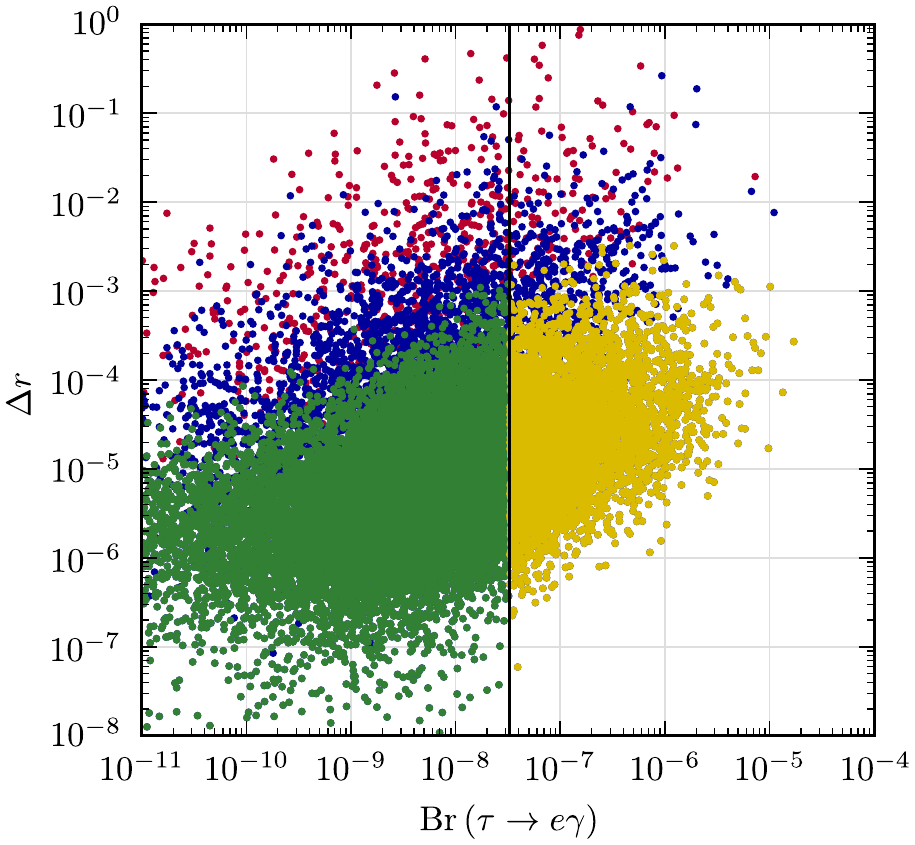}\tabularnewline
\end{tabular} \caption{\label{fig:DeltaRComparisonCuts}Ranges of variation of $\Delta r$
in the unconstrained MSSM as a function of $m_{H^{+}}$ (left panel),
and as a function of BR($\tau\to e\gamma$) (right panel). The different
input parameters were varied as in table \eqref{tab:RK_ranges} (notice
that $\delta_{\tau e}^{RR}=0.5$). On the left panel red points satisfy
the bounds on the masses (LEP+LHC), blue points satisfy all bounds
except $\textrm{BR}\left(B_{u}\to\tau\nu\right)$ and green points
comply with all bounds. Similar color code on the right panel, except
that blue points now comply with all bounds except $\textrm{BR}\left(B_{u}\to\tau\nu\right)$
and BR($\tau\to e\gamma$) while yellow denotes points only failing
the bound on BR($\tau\to e\gamma$).}
\end{figure}

This can be observed from the left panel of figure \eqref{fig:RK_DeltaR-mh0},
where we display the output of the above scan, presenting the values
of $\Delta r$ versus the associated light Higgs boson mass, $m_{h}$.
As expected, no explicit correlation between $m_{h}$ and $\Delta r$
is manifest, nor with the other (relevant) flavor-related low-energy
bounds. For completeness, and to clarify the following discussion,
we present on the right-hand panel of figure \eqref{fig:RK_DeltaR-mh0}
the charged Higgs mass as a function of $A_{0}$, again under a color
scheme denoting the experimental bounds applied in each case. Identical
to what was observed in figure \eqref{fig:NUHM:deltar} (notice that
NUHM models correspond, at low-energies, to a subset of these general
cases), regimes of very light charged Higgs are indeed present, in
association with small to moderate (negative) regimes for $A_{0}$.
Nevertheless, these regimes---which could potentially enhance $\Delta r$---are
likewise excluded due to a conflict with $\textrm{BR}\left(B_{u}\to\tau\nu\right)$.
This can be further confirmed from the left panel of figure \eqref{fig:DeltaRComparisonCuts},
where we display the possible range of variation for $\Delta r$ as
a function of $m_{H^{+}}$, color-coding the different applied bounds.

As can be seen from both panels of figure \eqref{fig:DeltaRComparisonCuts},
values $\Delta r\approx\mathcal{O}(10^{-2}-10^{-1})$ could be obtainable,
in agreement with references \cite{Girrbach:2012km,Masiero:2005wr,Masiero:2008cb,Ellis:2008st}.
However, the situation is substantially altered when one takes into
account the current experimental bounds on $B$ decays ($B_{u}\to\tau\nu$
and $B_{s}\rightarrow\mu\mu$) and $\tau\to e\gamma$. As we can see
from the left panel of figure \eqref{fig:DeltaRComparisonCuts}, once
experimental bounds---other than $B_{u}\to\tau\nu$---are imposed,
one could in principle have $\Delta r^{\text{max}}\approx\mathcal{O}(10^{-2})$;
however, taking into account the limits from BR($B_{u}\to\tau\nu$),
one is now led to $\Delta r\lesssim10^{-3}$.

There are a few comments to be made regarding the impact of the different
low-energy bounds from radiative $\tau$ decays and $B$-physics observables.
Firstly, let us consider the $\tau\to e\gamma$ decay: although directly
depending on $\delta_{\tau e}^{RR}$, its amplitude is (roughly) suppressed
by the fourth power of the average SUSY scale, $m_{SUSY}$. From equations
\eqref{eq:deltar:approx:X} and \eqref{eq:RK_deltar:approx}, $\Delta r$
only depends on the charged Higgs mass: if the latter is assumed to
be an EW scale parameter, $\Delta r$ will be thus independent of
$m_{\text{SUSY}}$ in these unconstrained models. As such, it is possible
to evade the $\tau\to e\gamma$ bound by increasing the soft SUSY
masses, and this can indeed be seen from the right-hand panel of figure
\eqref{fig:DeltaRComparisonCuts}, where a number of blue points are
to the left of the BR($\tau\to e\gamma$) bound line.

Secondly, the $B_{s}\rightarrow\mu\mu$ decay can be a severe constraint
regarding the SUSY contributions to $\Delta r$ in the case of constrained
models (figures \eqref{fig:cMSSM:seesaw} and \eqref{fig:cMSSM:survey}).
We note that $B_{s}\rightarrow\mu\mu$ is approximately proportional
to $A_{0}^{2}$ (see for instance \cite{Isidori:2006pk}) while $\Delta r$
shows no such dependency, thus a regime of small trilinear couplings
easily evades the $B_{s}\rightarrow\mu\mu$ bound.

\noindent Finally, there is the $B_{u}\to\tau\nu$ bound to consider.
Notice that this is a process essentially identical to the charged
kaon decays at the origin of the $R_{K}$ ratio (the only difference
being that the $K^{+}$ meson is to be replaced by a $B_{u}$ and
the $e$/$\mu$ in the decay products by a kinematically allowed $\tau$),
and hence its tree-level decay width can be inferred from equations
\eqref{eq:SM:Pdecays} and \eqref{eq:kaon:gamma:smsusy}. Due to a
negative interference between the SM and the MSSM contributions, given
by the term proportional to $\tan^{2}\beta/m_{H^{\pm}}^{2}$ in equation
\eqref{eq:kaon:gamma:smsusy}, regimes of low $m_{H^{+}}$ lead to
excessively small values of $B_{u}\to\tau\nu$ (below the experimental
bound), effectively setting a lower bound for for $m_{H^{\pm}}^{2}$
(see right panel of figure \eqref{fig:NUHM:deltar}, in relation to
the discussion of NUHM models). In turn, this excludes regimes of
$m_{H^{+}}$ associated to sizable values of $\Delta r$, as is clear
from the comparison of the blue and green regions of the left panel
of figure \eqref{fig:DeltaRComparisonCuts}.

\section{\label{sec:concs}Summary}

In this chapter we have revisited supersymmetric contributions to
$R_{K}=\Gamma\left(K\rightarrow e\nu\right)$$/\Gamma\left(K\rightarrow\mu\nu\right)$,
considering the potential of a broad class of constrained SUSY models
to saturate the current measurement of $R_{K}$ \cite{Fonseca:2012kr}.
We based our analysis in a full computation of the one-loop corrections
to the $\nu\ell H^{+}$ vertex; we have also derived (when possible)
illustrative analytical approximations, which in addition to offering
a more transparent understanding of the role of the different parameters,
also allow to establish a bridge between our results and previous
ones in the literature. Our analysis further revisited the $R_{K}$
observable in the light of new experimental data, arising from flavor
physics as well as from collider searches.

We numerically evaluated the contributions to $R_{K}$ arising in
the context of different minimal supergravity inspired models which
account for observed neutrino data, further considering the possibility
of accommodating a near future observation of a $\mu\to e\gamma$
decay. As expected from the (mostly) $LL$ nature of the radiatively
induced charged lepton flavor violation, type-I and II seesaw mechanisms
implemented in the cMSSM provide minimal contributions to $R_{K}$,
thus implying that such cMSSM SUSY seesaws cannot saturate the present
value for $\Delta r$.

We then considered unified models where the flavor-conserving hypothesis
on the $RR$ slepton sector is relaxed by allowing a non-vanishing
$\delta_{\tau e}^{RR}$. In all models, special attention was given
to experimental constraints, especially four observables which turn
out to play a particularly relevant role: the recent interval for
the lightest neutral Higgs boson mass provided by the CMS and ATLAS
collaborations, BR($B_{s}\rightarrow\mu\mu$), BR($B_{u}\to\tau\nu$)
and BR($\tau\to e\gamma$). These last two exhibit a dependence on
$m_{H^{+}}$ ($B_{u}\to\tau\nu$) and on $\delta_{\tau e}^{RR}$ ($\tau\to e\gamma$)
similar to that of $\Delta r$. SUSY contributions to $\Delta r$
are thus maximized in a regime in which $m_{H^{+}}$ and $\delta_{\tau e}^{RR}$
are such that the experimental limits for $B_{u}\to\tau\nu$ and $\tau\to e\gamma$
are simultaneously saturated; in this regime one must then accommodate
the bounds on other observables, such as $m_{h}$ and BR($B_{s}\rightarrow\mu\mu$).
For a minimal deviation from a pure cMSSM scenario allowing for non-vanishing
values of $\delta_{\tau e}^{RR}$, we can have values for $\Delta r$
at most of the order of $10^{-6}$. In fact, the requirement of having
a Higgs boson mass of 125-126 GeV is much more constraining on the
cMSSM parameter space than, for instance $B_{s}\rightarrow\mu\mu$
(which is sub-dominant, and can be overcome by variations of the trilinear
coupling, $A_{0}$). In order to have $\Delta r\sim\mathcal{O}(10^{-6})$,
one must significantly increase $\delta_{\tau e}^{RR}$ so to marginally
overlap the regions of $m_{h}\sim125$ GeV, while still in agreement
with $\tau\to e\gamma$.

SUSY contributions to $\Delta r$ increase ($\sim10^{-4}$) in models
where the charged Higgs mass can be significantly lowered, as is the
case of NUHM models; larger values are precluded due to $B_{u}\to\tau\nu$
decay constraints.

More general models, as the unconstrained MSSM realized at low-energies,
offer more degrees of freedom, and the possibility to better accommodate/evade
the different experimental constraints. In the unconstrained MSSM,
one can find values of $\Delta r$ one order of magnitude larger,
$\sim10^{-3}$. Again, any further augmentation is precluded due to
incompatibility with the bounds on $B_{u}\to\tau\nu$.

However, $\Delta r\sim\mathcal{O}(10^{-3})$ still remains one order
of magnitude shy of the current experimental sensitivity to $R_{K}$,
and also substantially lower than some of the values previously found
in the literature. As such, if SUSY is indeed discovered, and unless
there is significant progress in the experimental sensitivity to $R_{K}$,
it seems unlikely that the contributions to $R_{K}$ of the SUSY models
studied here will be testable in the near future. On the other hand,
any near-future measurement of $\Delta r$ larger than $\mathcal{O}(10^{-3})$
would unambiguously point towards a scenario different than those
here addressed (mSUGRA-like seesaw, NUHM and the phenomenological
MSSM).

It should be kept in mind that the analysis presented here focused
on the impact of LFV interactions. Should the discrepancy between
the SM and experimental observations turn out to be much smaller than
$10^{-4}$, a more detailed approach and evaluation will then be necessary.
\cleartooddpage
\begin{onehalfspace}

\chapter{\label{chap:Conclusion}Conclusions}
\end{onehalfspace}

In this thesis we have analyzed radiative effects in supersymmetric
models, with particular emphasis on supersymmetric grand unified theories
(SUSY GUTs), based on \cite{Fonseca:2011sy,Fonseca:2011vn,Arbelaez:2013hr,Fonseca:2012kr}.
If one attempts to supersymmetrize the SM in a minimal way, the gauge
coupling constants unify in the resulting model (the MSSM). To study
the evolution of these and other parameters in SUSY models, their
renormalization group equations (RGEs) must be calculated. As pointed
out in this thesis, such equations are known in a generic form for
an arbitrary SUSY model to second order in perturbation theory. However,
application of these generic RGEs to a specific model requires cumbersome
computations involving the fully expanded Lagrangian, which are better
performed by a computer. Otherwise, mistakes may jeopardize the very
precision which is sought by using second order equations. In this
work, following \cite{Fonseca:2011sy}, the \texttt{Susyno} package
for Mathematica was presented: given the gauge group and particle
content, it computes the model Lagrangian and then applies the RGEs
to its parameters. Testing of the program's output showed incorrections
in some of the results available in the literature (since then corrected
by the authors), which emphasizes the importance of automatically
calculating the RGEs of particular models, if true precision is to
be achieved.

While the RGEs for a generic SUSY model were known to two loops or
more, this did not include the important class of models where the
gauge group contains more than a single $U(1)$ factor: for example
the MSSM with an additional $U(1)$, and $SO(10)$ models with an
intermediate $U(1)_{R}\times U(1)_{B-L}$ scale. Indeed, GUTs based
on gauge groups with rank higher than four (such as $SO(10)$ and
$E_{6}$) can have phases where the effective gauge group contains
multiple $U(1)$ factors. In general, this leads to $U(1)$ mixing---gauge
bosons and gauginos of different abelian factor groups can mix---and
there were just some one-loop RGEs for such SUSY models. In \cite{Fonseca:2011vn}
we have extended these results by deriving the two-loop renormalization
group evolution of all parameters. We have also shown that failure
to fully take into account these corrections can lead to errors of
the percent level.

For several $SO(10)$-inspired models, radiative effects on SUSY soft
breaking masses were also analyzed \cite{Arbelaez:2013hr}. In particular,
we studied models where the lowest intermediate scale can be changed
continuously without affecting the unification of the coupling constants.
This sliding condition ensures that these supersymmetric models can
have an extended gauge group near the electroweak scale, which potentially
entails a rich phenomenology at the LHC. By the same reasoning, it
is also conceivable that the intermediate scales of these $SO(10)$-inspired
models are beyond the reach of current direct detection experiments.
If this is the case, it is still possible to indirectly test these
models by measuring the masses of SUSY particles, since radiative
effects imprint on them some of the higher energy behavior of the
theory. In this way, we have showed that it is possible in principle
not only to gain indirect information on the type of intermediate
phases of these SUSY GUT models, but also infer their energy scales.

The above analysis was done at the one-loop level only. Nevertheless,
at the expense of lower precision we were able to analyze a broad
range of $SO(10)$-inspired models. A more precise computation would
require the use of two-loop RGEs, as well as inclusion of threshold
effects. Should any signs of supersymmetry be found in the future,
such improvements in the calculations could easily be carried out.

In some cases, SUSY GUTs also lead to charged lepton flavor violation
(cLFV). The reason why cLFV is so significant is because in the Standard
Model we expect it to be negligible even if we add neutrino masses;
any observation of lepton violating processes, $\ell\rightarrow\ell'\gamma$
for example, would clearly point to the presence of new Physics. These
new Physics could well be SUSY and, in that case, due to the many
new flavored parameters in the soft SUSY breaking Lagrangian, cLFV
could be experimentally detectable. In fact, even assuming that SUSY
is broken in a flavor blind way at some high scale, the SUSY-preserving
neutrino Yukawa couplings are still able to generate cLFV interactions
radiatively. There are many observables affected by this mechanism
of generating cLFV---one of them, the ratio of decay amplitudes $\Gamma\left(K\rightarrow e\nu\right)/\Gamma\left(K\rightarrow\mu\nu\right)\equiv R_{K}$,
was revisited in this thesis (based on \cite{Fonseca:2012kr}). We
studied this ratio in the constrained MSSM with several seesaw realizations
(type I, type II and inverse seesaw), in left-right symmetric models,
and in non-universal Higgs mass models as well. To complete the analysis,
we also considered the prospects of obtaining a large $R_{K}$ in
unconstrained low-energy SUSY models. We took into consideration LEP's
bounds on slepton, chargino and neutralino masses, as well as LHC
bounds on squark, gluino and the lightest Higgs masses. But more importantly
for the saturation of the current experimental sensitivity on $R_{K}$,
we also imposed the $BR\left(B_{s}\rightarrow\mu\mu\right)$, $BR\left(B_{u}\rightarrow\tau\nu\right)$
and $BR\left(\tau\rightarrow e\gamma\right)$ limits. We concluded
that in light of these experimental constraints, SUSY contributions
to the $R_{K}$ observable cannot be as large as previously argued
in the literature and in particular, the effects are (at least) an
order of magnitude lower than the current experimental sensitivity.
Therefore, it seems unlikely that these contributions to $R_{K}$
will be testable in the near future, and if any near-future measurement
of it does detect a discrepancy with the Standard Model value, it
would unambiguously point towards different new Physics.

\medskip{}

We conclude by noting that supersymmetry remains one of the most studied
and attractive candidates to solve some of the Standard Model shortcomings,
particularly in the context of grand unified theories, and radiative
corrections are fundamental in the study of the phenomenology of these
models. Even though no SUSY particles were observed yet at the LHC,
the mass of the recently discovered Higgs boson is within the expected
range if supersymmetry is to be the solution to the hierarchy problem
and also provide a viable dark matter candidate particle. It does
however point to a heavy SUSY spectrum, in the multi TeV range. However,
there is hope that with the restart of the LHC at 13--14 TeV center
of mass energy, as well as with a new generation of low energy experiments
which have started or will start taking data in the near future, that
we may uncover some of the fundamental Physics which lie beyond the
Standard Model.
\cleartooddpage

\appendix
\part{Appendix}\cleartooddpage

\chapter{\label{chap:SM_appendix}Structure of the Standard Model}

This appendix reviews the structure of the Standard Model, which is
a gauge theory based on the group $U\left(1\right)_{Y}\times SU\left(2\right)_{L}\times SU\left(3\right)_{c}$
that is broken down into $U\left(1\right)_{Q}\times SU\left(3\right)_{c}$
by the Higgs mechanism, at low energies.

\section{The gauge theory}

A representation $\Psi$ in table \eqref{tab:Representations-of-the-SM}
of the gauge group $U\left(1\right)_{Y}\times SU\left(2\right)_{L}\times SU\left(3\right)_{c}$
transforms as  
\begin{align}
\Psi & \rightarrow\exp\left[i\left(\alpha T^{Y}+\alpha'^{a}T_{a}^{L}+\alpha''^{b}T_{b}^{c}\right)\right]\Psi
\end{align}
under a local gauge transformation, where $T^{Y}$, $T_{a}^{L}$,
$T_{b}^{c}$ are representation matrices and $\alpha$, $\alpha'^{a}$,
$\alpha''^{b}$ are some real parameters which can be space-time dependent
(see \cite{RomaoValleBook,cheng2000gauge,Djouadi:2005gi} for example).
The Standard Model contains $SU\left(2\right)_{L}$ singlets ($T_{a}^{L}=\boldsymbol{0}$)
and doublets ($T_{a}^{L}=\frac{1}{2}\sigma_{a}$), as well as $SU\left(3\right)_{c}$
singlets ($T_{a}^{c}=\boldsymbol{0}$) and triplets ($T_{a}^{c}=\frac{1}{2}\lambda_{a}$).%
\footnote{In the case of the left-handed quarks $Q=\left(u,d\right)^{T}$, which
are neither singlets of $SU\left(2\right)_{L}$ nor singlets of $SU\left(3\right)_{c}$,
$T_{a}^{SU(2)_{L}}=\frac{1}{2}\sigma_{a}\otimes\mathbb{1}_{3}$ and
$T_{a}^{SU(3)_{c}}=\frac{1}{2}\mathbb{1}_{2}\otimes\lambda_{a}$. %
} Here, $\sigma_{a}$ and $\lambda_{a}$ are the Pauli and Gell-Mann
matrices, respectively. On the other hand, the matrix $T^{Y}$ is
simply given by $Y\mathbb{1}$, where $Y$ is the $U(1)_{Y}$ hypercharge
of the representation. 

The Lagrangian density itself must not change under gauge transformations
and, in order to achieve this, some vector bosons must be introduced:
one $B_{\mu}$ plus three $W_{\mu}^{a}$ (the electroweak gauge bosons),
and eight $G_{\mu}^{a}$ (the gluons). These fields are in the adjoint
representation of the gauge factor groups $U(1)_{Y}$, $SU(2)_{L}$,
and $SU(3)_{C}$, respectively, changing as follows under infinitesimal
transformations:
\begin{align}
B_{\mu} & \rightarrow B_{\mu}-\frac{1}{g'}\partial_{\mu}\alpha\,,\\
W_{\mu}^{a} & \rightarrow W_{\mu}^{a}-\varepsilon_{abc}\alpha'^{b}W_{\mu}^{c}-\frac{1}{g}\partial_{\mu}\alpha'^{a}\,,\\
G_{\mu}^{a} & \rightarrow G_{\mu}^{a}-f_{abc}\alpha''^{b}G_{\mu}^{c}-\frac{1}{g_{s}}\partial_{\mu}\alpha''^{a}\,.
\end{align}
The tensors $\varepsilon_{abc}$ and $f_{abc}$ are structure constants
of $SU(2)_{L}$ and $SU(3)_{c}$:
\begin{align*}
\left[\sigma_{a},\sigma_{b}\right] & =i\varepsilon_{abc}\sigma_{c}\,,\\
\left[\lambda_{a},\lambda_{b}\right] & =if_{abc}\lambda_{c}\,.
\end{align*}

In order to make the Lagrangian density invariant under local gauge
transformations, the partial derivative $\partial_{\mu}$ is replaced
with the covariant one,
\begin{align}
\partial_{\mu} & \rightarrow D_{\mu}=\partial_{\mu}+ig'Y\mathbb{1}B_{\mu}+igT_{a}^{L}W_{\mu}^{a}+ig_{s}T_{b}^{c}G_{\mu}^{b}\,,
\end{align}
and in this way interactions between the gauge bosons and matter/Higgs
fields are generated. For the fermions we obtain
\begin{align}
\mathscr{L}_{\textrm{kin }f}= & \, i\sum_{f}\overline{f}\gamma^{\mu}\partial_{\mu}f-\frac{g_{s}}{2}\sum_{T}\overline{T}\gamma^{\mu}\lambda_{a}TG_{\mu}^{a}-\frac{g}{\sqrt{2}}\sum_{\left(f_{u},f_{d}\right)^{T}}\left(\overline{f}_{u}\gamma^{\mu}P_{L}f_{d}W_{\mu}^{+}+\textrm{h.c.}\right)\nonumber \\
 & -e\sum_{f}q_{f}\overline{f}\gamma^{\mu}fA_{\mu}-\frac{g}{\cos\theta_{W}}\sum_{f}\overline{f}\gamma^{\mu}\left(g_{f}^{V}-g_{f}^{A}\gamma_{5}\right)fZ_{\mu}\,,\label{eq:SM_Lf_kin}
\end{align}
where $f$ are the fermionic field components, $\left(f_{u},f_{d}\right)^{T}$
are the $SU(2)_{L}$ doublets, and $T$ are the $SU(3)_{c}$ triplets.
In preparation to the breaking of the electroweak symmetry, the gauge
bosons $B_{\mu}$ and $W_{\mu}^{3}$ have been rotated to the fields
$A_{\mu}$ (the photon) and $Z_{\mu}$,
\begin{align}
\begin{pmatrix}W_{\mu}^{3}\\
B_{\mu}
\end{pmatrix}= & \begin{pmatrix}\phantom{-}\cos\theta_{W} & \sin\theta_{W}\\
-\sin\theta_{W} & \cos\theta_{W}
\end{pmatrix}\begin{pmatrix}Z_{\mu}\\
A_{\mu}
\end{pmatrix}\,,
\end{align}
where $\theta_{W}$ is the weak mixing angle:
\begin{align}
e & \equiv g\sin\theta_{W}\equiv g'\cos\theta_{W}\,.
\end{align}
On the other hand,
\begin{align}
W_{\mu}^{+} & =\frac{1}{\sqrt{2}}\left(W_{\mu}^{1}-iW_{\mu}^{2}\right)\,.
\end{align}
Lastly, the electric charge $q_{f}$ and the vector/axial-vector coupling
strengths $g_{f}^{V}$/$g_{f}^{A}$ of fermions to the $Z$ boson
can be expressed as a function of the fermion weak isospin $T_{3f}^{L}$
and hypercharge $Y_{f}$:
\begin{align}
q_{f} & =T_{3f}^{L}+Y_{f}\,,\\
g_{f}^{V} & =\left(\frac{1}{2}-\sin^{2}\theta_{W}\right)T_{3f}^{L}-\sin^{2}\theta_{W}Y_{f}\,,\\
g_{f}^{A} & =\frac{1}{2}T_{3f}^{L}\,.
\end{align}

We note as well that in order for the gauge bosons to be dynamical
entities, they need kinetic terms:
\begin{align}
\mathscr{L}_{\textrm{kin }g} & =-\frac{1}{4}B_{\mu\nu}B^{\mu\nu}-\frac{1}{4}W_{\mu\nu}^{a}W_{a}^{\mu\nu}-\frac{1}{4}G_{\mu\nu}^{a}G_{a}^{\mu\nu}\,,
\end{align}
where
\begin{align}
B_{\mu\nu} & \equiv\partial_{\mu}B_{\nu}-\partial_{\nu}B_{\mu}\,,\\
W_{\mu\nu}^{a} & \equiv\partial_{\mu}W_{\nu}^{a}-\partial_{\nu}W_{\mu}^{a}-g\varepsilon_{abc}W_{\mu}^{b}W_{\nu}^{c}\,,\\
G_{\mu\nu}^{a} & \equiv\partial_{\mu}G_{\nu}^{a}-\partial_{\nu}G_{\mu}^{a}-gf_{abc}G_{\mu}^{b}G_{\nu}^{c}\,.
\end{align}

\section{Electroweak symmetry breaking}

Unless the gauge symmetry is broken, fermions will have no mass, as
right- and left-handed matter fields are in different representations
of the $U(1)_{Y}\times SU(2)_{L}$ group. Also, none of the gauge
bosons can be massive. In the Standard Model, the breaking of the
electroweak symmetry is achieved spontaneously through the Higgs mechanism:
the scalar doublet $H$ in table \eqref{tab:Representations-of-the-SM}
acquires a non-vanishing vacuum expectation value $\left\langle H^{\dagger}H\right\rangle $
which is only invariant under one combination of the electroweak generators:
it is usually chosen to be $T^{Q}\equiv T_{3}^{L}+T^{Y}$. Since $SU(3)_{c}$
is also preserved by this vacuum state, we have following breaking
of the gauge group:
\begin{align}
U\left(1\right)_{Y}\times SU\left(2\right)_{L}\times SU\left(3\right)_{c} & \rightarrow U\left(1\right)_{Q}\times SU\left(3\right)_{c}\,.
\end{align}

Considering the Higgs Lagrangian, 
\begin{align}
\mathscr{L}_{H} & =\left(D^{\mu}H\right)^{\dagger}\left(D_{\mu}H\right)-V\left(H\right)\,,\quad V\left(H\right)=\mu^{2}H^{\dagger}H+\lambda\left(H^{\dagger}H\right)^{2}\,,\label{eq:SM_L_H}
\end{align}
one realizes that if the mass squared parameter $\mu^{2}$ is negative
and the quartic coupling $\lambda$ is positive (to ensure vacuum
stability), the potential is minimal when
\begin{align}
\left\langle H^{\dagger}H\right\rangle  & =-\frac{\mu^{2}}{2\lambda}\equiv v^{2}\,.
\end{align}
In a particular gauge---the so-called unitary gauge---one can write
the Higgs doublet as
\begin{align}
H & =\begin{pmatrix}0\\
v+\nicefrac{\phi}{\sqrt{2}}
\end{pmatrix}\,,
\end{align}
where $v$ is just a number, and $\phi$ is a dynamical entity with
a vanishing vacuum expectation value. Substituting this expression
in equation \eqref{eq:SM_L_H} yields
\begin{align}
\mathscr{L}_{H} & =\frac{1}{2}\partial^{\mu}\phi\partial_{\mu}\phi-\frac{1}{2}m_{\phi}^{2}\phi^{2}+\frac{1}{2}m_{Z}^{2}Z^{\mu}Z_{\mu}+m_{W}^{2}W^{+\mu}\left(W_{\mu}^{+}\right)^{*}\nonumber \\
 & -\frac{m_{\phi}^{2}}{2\sqrt{2}v}\phi^{3}-\frac{m_{\phi}^{2}}{16v^{2}}\phi^{4}+\sqrt{2}\frac{m_{W}^{2}}{v}W^{+\mu}\left(W_{\mu}^{+}\right)^{*}\phi+\frac{m_{Z}^{2}}{\sqrt{2}v}Z^{\mu}Z_{\mu}\phi\nonumber \\
 & +\frac{1}{2}\frac{m_{W}^{2}}{v^{2}}W^{+\mu}\left(W_{\mu}^{+}\right)^{*}\phi^{2}+\frac{1}{4}\frac{m_{Z}^{2}}{v^{2}}Z^{\mu}Z_{\mu}\phi^{2}
\end{align}
up to a constant term. The masses in this expression are the following:
\begin{align}
m_{W} & =\frac{gv}{\sqrt{2}}\,,\\
m_{Z} & =\frac{\sqrt{g'^{2}+g^{2}}v}{\sqrt{2}}\,,\\
m_{\phi} & =2\sqrt{\lambda}v\,.
\end{align}
Importantly, the photon $A_{\mu}$ remains massless, in accordance
with stringent experimental bounds \cite{Beringer_mod:1900zz}.

The matter fields acquire mass by interacting with the Higgs doublet:
\begin{align}
\mathscr{L}_{\textrm{Yukawa}} & =-Y_{ij}^{u}\overline{u}_{Ri}Q_{j}\cdot H-Y_{ij}^{d}\overline{d}_{Ri}Q_{j}\cdot H-Y_{ij}^{\ell}\overline{e}_{Ri}L_{j}\cdot H+\textrm{h.c.}\,,
\end{align}
where the flavor indices---$i$ and $j$---are summed over. These
are the Yukawa interactions. After EWSB,
\begin{align}
\mathscr{L}_{\textrm{Yukawa}} & =-m_{ij}^{u}\overline{u}_{Li}u_{Rj}-m_{ij}^{d}\overline{d}_{Li}d_{Rj}-m_{ij}^{\ell}\overline{e}_{Li}e_{Rj}\nonumber \\
 & \phantom{=}-\frac{m_{ij}^{u}}{\sqrt{2}v}\overline{u}_{Li}u_{Rj}\phi-\frac{m_{ij}^{d}}{\sqrt{2}v}\overline{d}_{Li}d_{Rj}\phi-\frac{m_{ij}^{\ell}}{\sqrt{2}v}\overline{e}_{Li}e_{Rj}\phi+\textrm{h.c.}\,,
\end{align}
with
\begin{align}
m_{ij}^{x} & =vY_{ji}^{x*}\,,x=u,d,\ell\,.
\end{align}

There is no theoretical reason for the fermion mass matrices to be
aligned with the charged interactions in equation \eqref{eq:SM_Lf_kin}.
In fact, it has been experimentally established that this is not the
case. Once the weak eigenstates $f_{L/R}$ are rotated to the mass
eigenstates $f_{L/R}^{m}$ by some unitary matrices $U_{L/R}^{f}$,
\begin{align}
\left(f_{L/R}\right)_{i} & \equiv\left(U_{L/R}^{f}\right)_{ij}\left(f_{L/R}^{m}\right)_{j}\,,\quad\textrm{with }U_{L}^{f\dagger}m^{f}U_{R}^{f}=\textrm{diagonal}\,,
\end{align}
the quark charged current ceases to be diagonal:
\begin{align}
\mathscr{L}_{\textrm{kin }f}= & \cdots-\frac{g}{\sqrt{2}}\left(U_{L}^{u\dagger}U_{L}^{d}\right)_{ij}\left(\overline{u}_{Li}^{m}\gamma^{\mu}d_{Lj}^{m}W_{\mu}^{+}+\textrm{h.c.}\right)\,.\label{eq:Appendix_SM_Vckm}
\end{align}
The quark mixing matrix $V\equiv U_{L}^{u\dagger}U_{L}^{d}$ appearing
in this equation is known as the Cabibbo-Kobayashi-Maskawa matrix
\cite{Kobayashi:1973fv}. There is no leptonic analogue in the SM
because, in it, neutrinos do not have mass: since they are degenerate
states, any rotation between the different flavors is physically meaningless
(see chapter \ref{chap:Lepton-flavour-violation}).

For completeness, it should be mentioned that, in order to perform
the quantification of a gauge theory, it is also necessary to provide
a gauge-fixing Lagrangian, $\mathscr{L}_{\textrm{GF}}$, as well as
a ghost Lagrangian $\mathscr{L}_{\textrm{ghost}}$ (see \cite{cheng2000gauge}
and references contained therein). Therefore, the complete Standard
Model Lagrangian density is
\begin{align}
\mathscr{L}_{\textrm{SM}} & =\mathscr{L}_{\textrm{kin }f}+\mathscr{L}_{\textrm{kin }g}+\mathscr{L}_{H}+\mathscr{L}_{\textrm{Yukawa}}+\mathscr{L}_{\textrm{GF}}+\mathscr{L}_{\textrm{ghost}}\,.
\end{align}

\cleartooddpage

\chapter{\label{chap:Implementation-details-of}Implementation of some functions
in the \texttt{Susyno} program}

This appendix discusses how the main functions of the Mathematica
program described in chapter \ref{chap:Susyno} were implemented.
These are almost exclusively group theoretical functions, since coding
the RGEs themselves is a lengthy but simple process. Most of issues
to be discussed below are rather technical, yet they are important
for model building. Since the available literature does not seem to
cover these topics thoroughly, they are discussed here.

Section \ref{sec:Susyno_Building-the-matrices} below analyzes how
the matrices of any representation of any simple Lie group can be
constructed. Quantum mechanics textbooks often discuss how to do this
for the simplest of simple groups, $SU(2)$, but the general case
presents qualitative new features that complicate matters. We note
that with the representation matrices of simple groups, the ones of
products of simple groups (and possibly of abelian groups) are trivial
to obtain, so effectively the method described here can be used to
construct the matrices of any representation used in gauge theories.
In \texttt{Susyno}, this was implemented as the \texttt{RepMatrices}
function. The code is quite efficient, since the 45 matrices with
dimensions $1050\times1050$ of the representation $\left\{ 1,0,0,2,0\right\} $
of $SO(10)$,\\
\\
\texttt{RepMatrices{[}SO10, \{1, 0, 0, 2, 0\}{]}}\\
\\
, are computed in less than one minute in a computer with an Intel
Core i5-2300 CPU (see section \ref{sec:Susyno_running_times}). We
note that in the past some authors have analyzed other methods of
computing the representation matrices of the classical Lie algebras
$SU(n)$, $SO(n)$, $Sp(2n)$ \cite{Gelfand:1950a,Gelfand:1950b,Gould:1981a,Gould:1981b,Molev:1999,Molev:2000}
(for more references, see \cite{Molev:1999}).

Section \ref{sec:Susyno_Invariant-combinations-of} discusses how
to compute the combinations of product of fields which are gauge invariant,
once their transformation/representation matrices are known. For the
$SU(2)$ group, this is equivalent to finding the Clebsch-Gordan coefficients.
Such computation is critical to the construction of the Lagrangian
of a given model. In \texttt{Susyno} this was implemented in the \texttt{Invariants}
function (which is very similar to \texttt{IrrepInProduct}) and performance-wise,
since the method discussed can reasonably be seen as a brute-force
one, it can take many hours once the representations involved have
dimensions above $\thicksim100,200$. The references \cite{Schlosser:1991aa,Koh:1984vs,Anderson:1999em,Anderson:2000ni,Anderson:2001sd,Horst:2010qj}
also discuss the computation of such generalized Clebsch-Gordan coefficients
for some groups/representations.%
\footnote{The \texttt{Clego} program \cite{Horst:2010qj} can only compute tensor
products of non-degenerate or adjoint representations, but the authors
suggest a work around this limitation.%
}

In section \ref{sec:Susyno_Einstein-convention-applied}, without
entering into many details, we discuss briefly the complicated problem
of simplifying the final expressions appearing in the RGEs.

Actually, version 2 of the \texttt{Susyno} program does not compute
an explicit form of the Lagrangian. In other words, the functions
described in sections \ref{sec:Susyno_Building-the-matrices} and
\ref{sec:Susyno_Invariant-combinations-of} are no longer used to
calculate RGEs, but they are still part of the program. The aim of
section \ref{sec:Susyno_A-Lagrangian-less-computation} is to discuss
this Lagrangian-free approach to the computation of the RGEs. With
it, the RGEs can be computed much quicker than with the traditional
way that required building a superpotential and a soft SUSY breaking
Lagrangian for each model. Section \ref{sec:Susyno_running_times}
presents some reference running times.

We emphasize here once more that a detailed description on how to
use these and other functions of the program (see the list in section
\ref{sec:Tests-made}) is provided in the built-in help system of
\texttt{Susyno}.

\section{\label{sec:Susyno_Building-the-matrices}Building the matrices of
an arbitrary representation of a simple group}

In order to build the representation matrices of the algebra of a
group acting on some vector space, a specific basis for both the algebra
and the vector space must be chosen. In particular, once we arrive
at a desired set of generator matrices $\left\{ T_{a}\right\} $,
performing the transformations $T_{a}\rightarrow\sum_{b}\mathcal{O}_{ab}T_{b}$
and/or $T_{a}\rightarrow U^{\dagger}T_{a}U$ for some orthogonal and
unitary matrices $\mathcal{O}$ and $U$, yields another valid set
of generators, which obeys all the normalization requirements often
used in Particle Physics (see subsection \ref{sub:Symmetry_PhysicalConventions}
of the group theory chapter). Because these matrices are basis dependent,
some of the information they contain is not physically relevant. Even
so, the explicit representation matrices are very useful, since many
quantities can be computed from them. We shall then see how these
can be built, for any representation of any simple group.

First, we review how the $SU(2)$ matrices are computed. In preparation
for the general case, we shall use the generators $e,\, f$ and $h$
of equation \eqref{eq:EFH_SU2}. Henceforth denoted by $E,F$ and
$H$, their representation matrices obey the following relations
\begin{align}
\left[E,F\right] & =H\,, & \left[H,E\right] & =2E\,, & \left[H,F\right] & =-2F\,.
\end{align}
This should be compared with the Chevalley-Serre relations \eqref{eq:Serre_Chevalley_relations1}--\eqref{eq:Serre_Chevalley_relations3}
for a generic simple group, taking into account that the Cartan matrix
of $SU(2)$ is simply $\left(2\right)$. We define the state $\left|2m\right\rangle $,
with unit norm ($\left\langle 2m'|2m\right\rangle =\delta_{m,m'}$)
and $m$ a half-integer, to be such that
\begin{align}
H\left|2m\right\rangle  & \equiv\left[2m\right]\left|2m\right\rangle \,,
\end{align}
where $\left[2m\right]=2m$ is presently only a number. Regarding
the notation, notice that there is no sum over $m$ here (summations
will be indicated explicitly). We know that all weights of $SU(2)$
have multiplicity 1, meaning that the eigenspace of $H$ associated
to the eigenvalue $\left[2m\right]$ is always 1-dimensional. Applying
$E$($F$) to $\left|2m\right\rangle $ raises(lowers) $2m$ by two
units:
\begin{align}
H\left(E\left|2m\right\rangle \right) & =\left(EH+2E\right)\left|2m\right\rangle =\left(\left[2m\right]+2\right)\left(E\left|2m\right\rangle \right)\,,\\
H\left(F\left|2m\right\rangle \right) & =\left(FH-2F\right)\left|2m\right\rangle =\left(\left[2m\right]-2\right)\left(F\left|2m\right\rangle \right)\,,
\end{align}
so we can define two new numbers, $\left[2m\right]^{+}$ and $\left[2m\right]_{-}$,
through the following relations:
\begin{align}
E\left|2m\right\rangle  & \equiv\left[2m\right]^{+}\left|2m+2\right\rangle \,,\\
F\left|2m\right\rangle  & \equiv\left[2m\right]_{-}\left|2m-2\right\rangle \,.
\end{align}
Note as well that $E,F$ and $H$ are sparse matrices, with $0$'s
almost everywhere, except for a few entries ($1\times1$ blocks) which
are non-null: 
\begin{gather}
\begin{array}{ccc}
 & \hspace{-11mm}\begin{array}{ccc}
 & 2m\,\,\,\,\end{array}\\
\begin{array}{c}
\\
E=\\
\\
\end{array} & \hspace{-4mm}\begin{pmatrix} & \vdots\,\,\,\,\\
 & \left[2m\right]^{+} & \cdots\\
\,\,
\end{pmatrix} & \hspace{-4mm}\begin{array}{c}
\\
2m+2\quad,\\
\\
\end{array}
\end{array}\begin{array}{ccc}
 & \hspace{-11mm}\begin{array}{ccc}
 & 2m\,\,\,\,\end{array}\\
\begin{array}{c}
\\
\quad F=\\
\\
\end{array} & \hspace{-4mm}\begin{pmatrix} & \vdots\,\,\,\,\\
 & \left[2m\right]_{-} & \cdots\\
\,\,
\end{pmatrix} & \hspace{-4mm}\begin{array}{c}
\\
2m-2\quad,\\
\\
\end{array}
\end{array}\nonumber \\
\begin{array}{ccc}
 & \hspace{-11mm}\begin{array}{ccc}
 & 2m\,\,\,\,\end{array}\\
\begin{array}{c}
\\
H=\\
\\
\end{array} & \hspace{-4mm}\begin{pmatrix} & \vdots\,\,\,\,\\
 & \left[2m\right] & \cdots\\
\,\,
\end{pmatrix} & \hspace{-4mm}\begin{array}{c}
\\
2m\quad.\\
\\
\end{array}
\end{array}\label{eq:Susyno_SU2_matrices}
\end{gather}
The numbers $\left[2m\right]=2m$ are just the weights of the representation;
they are easy to compute and therefore we will assume that they are
known (see for example \cite{Cahn:1985wk}). On the other hand, $\left[2m\right]^{+}$
and $\left[2m\right]_{-}$ are unknown and must be calculated. To
do this, we note that $EF\left|2m\right\rangle $ is equal to $\left[2m\right]_{-}\left[2m-2\right]^{+}\left|2m\right\rangle $
and, commuting $E$ with $F$, it is also the same as $\left(\left[2m\right]^{+}\left[2m+2\right]_{-}+\left[2m\right]\right)\left|2m\right\rangle $;
therefore we obtain the important relation
\begin{align}
\left[2m\right]_{-}\left[2m-2\right]^{+} & =\left[2m\right]^{+}\left[2m+2\right]_{-}+\left[2m\right]\,.\label{eq:Susyno_SU2_recursive_relation}
\end{align}
Assuming that we know $\left[2m\right]^{+}$ and $\left[2m+2\right]_{-}$
for all $m$ above some limit, we can recursively use this relation
to get the remaining values for lower $m$'s. For example, starting
with $2m_{\textrm{max}}$, $\left[2m_{\textrm{max}}\right]^{+}=0$
so
\begin{align}
\left[2m_{\textrm{max}}\right]_{-}\left[2m_{\textrm{max}}-2\right]^{+} & =2m_{\textrm{max}}\,.
\end{align}
Here, we stumble upon an interesting issue: there is an apparent arbitrariness
in building the matrices $E$ and $F$ because, for a given $m$,
the individual values of $\left[2m\right]_{-}$ and $\left[2m-2\right]^{+}$
are undefined; only the combination $\left[2m\right]_{-}\left[2m-2\right]^{+}$
has a definite value. However, note that instead of $E$ and $F$,
we need hermitian generators $T_{1}$ and $T_{2}$ which are complex
linear combinations of $E$ and $F$. The only way that such hermitian
matrices can be built is if $E\propto F^{\dagger}$(see equation \eqref{eq:Susyno_SU2_matrices}),
which implies that for all $m$
\begin{align}
\left[2m\right]_{-} & \propto\left(\left[2m-2\right]^{+}\right)^{*}\,.
\end{align}
Our construction procedure of the representation matrices shows that
it is possible to take the simplest case, $\left[2m\right]_{-}=\left[2m-2\right]^{+}\in\mathbb{R}$,
such that $E=F^{T}$ and $T_{1},T_{2}\propto E+F,i\left(E-F\right)$.
As such,
\begin{align}
\left[2m_{\textrm{max}}\right]_{-} & =\left[2m_{\textrm{max}}-2\right]^{+}=\sqrt{2m_{\textrm{max}}}\,,
\end{align}
and recursive application of equation \eqref{eq:Susyno_SU2_recursive_relation}
yields all the remaining entries $\left[\;\right]_{-}$ and $\left[\;\right]^{+}$
of $E$ and $F$:
\begin{align}
\left[2m_{\textrm{max}}-2\right]_{-}\left[2m_{\textrm{max}}-4\right]^{+} & =\left[2m_{\textrm{max}}\right]_{-}\left[2m_{\textrm{max}}-2\right]^{+}+2m_{\textrm{max}}-2\\
 & \Rightarrow\nonumber \\
\left[2m_{\textrm{max}}-2\right]_{-} & =\left[2m_{\textrm{max}}-4\right]^{+}=\sqrt{4m_{\textrm{max}}-2}\\
 & \;\;\vdots\qquad\qquad\qquad\qquad\qquad\qquad\qquad\qquad\qquad\nonumber 
\end{align}
 In fact, $SU(2)$ is such a simple case that we can write down these
values in a closed form:
\begin{align}
\left[2m\right]_{-} & =\left[2m-2\right]^{+}=\sqrt{\left(s+m\right)\left(s-m+1\right)}\,,
\end{align}
where $s\equiv m_{\textrm{max}}$.

We move now to the general case of an arbitrary simple group of rank
$n$. Our starting point will be the Chevalley-Serre relations \eqref{eq:Serre_Chevalley_relations1}--\eqref{eq:Serre_Chevalley_relations3}
for the representation matrices $E_{i},F_{i},H_{i}$ of the algebra
elements $e_{i},f_{i},h_{i}$. These $3n$ matrices alone do not generate
the whole algebra because there are more raising and lowering matrices
$\widetilde{E}_{x}$ and $\widetilde{F}_{x}$ but, as noted in chapter
\ref{chap:Symmetry}, these can be obtained from the commutators of
the elementary $E_{i}$ and $F_{i}$, $i=1,\cdots,n$. So once the
$3n$ matrices $E_{i},F_{i},H_{i}$ are known, it is trivial to obtain
the remaining generators, and then, as in $SU(2)$, for each pair
of raising and lowering operators $\widetilde{E}_{x}$ and $\widetilde{F}_{x}$,
we can change basis so that all generators are hermitian by considering
instead the combinations $\widetilde{E}_{x}+\widetilde{F}_{x}$ and
$i\left(\widetilde{E}_{x}-\widetilde{F}_{x}\right)$. 

Having dealt with these issues, the only step remaining in order to
get the representation matrices is to actually build the elementary
matrices $E_{i},F_{i},H_{i}$ which obey the relations \eqref{eq:Serre_Chevalley_relations1}--\eqref{eq:Serre_Chevalley_relations3}.%
\footnote{The function \texttt{RepMinimalMatrices} in \texttt{Susyno} returns
just these matrices.%
} We will proceed in analogy to the $SU(2)$ case, and to do so two
complications must be overcome:
\begin{enumerate}
\item As mentioned already, instead of just one set of matrices $\left\{ E,F,H\right\} $,
there are now $n$ such sets $\left\{ E_{i},F_{i},H_{i}\right\} $.
\item The space associated to a given weight is multi-dimensional. This
means that, in general, the quantities $\left[\;\right]_{-}$, $\left[\;\right]^{+}$
and $\left[\;\right]$ in equation \eqref{eq:Susyno_SU2_matrices}
are no longer numbers ($1\times1$ matrices)---instead they are matrices,
which are not even necessarily square.
\end{enumerate}
Recall from chapter \ref{chap:Symmetry} that a state with weight
$\omega$, $\left|\omega\right\rangle $, is such that
\begin{align}
H_{i}\left|\omega\right\rangle  & =\omega\left(i\right)\left|\omega\right\rangle \,,
\end{align}
with $\omega\left(i\right)=\nicefrac{2\left\langle \omega,\alpha_{i}\right\rangle }{\left\langle \alpha_{i},\alpha_{i}\right\rangle }$,%
\footnote{The Dynkin coefficients of a representation, which are used to identify
it, are precisely these $\omega_{\max}\left(i\right)$ numbers (see
chapter \ref{chap:Symmetry}). %
} and $\alpha_{i}$ ($i=1,\cdots,n$) are the simple roots of the algebra.
As mentioned above, in general there is more than a single state with
weight $\omega$, so we must introduce an extra label, $\left|\omega\right\rangle \rightarrow\left|\omega,\lambda\right\rangle $,
such that
\begin{align}
H_{i}\left|\omega,\lambda\right\rangle  & =\omega\left(i\right)\left|\omega,\lambda\right\rangle 
\end{align}
and $\left\langle \lambda',\omega'|\omega,\lambda\right\rangle =\delta_{\omega,\omega'}\delta_{\lambda,\lambda'}$.
Denoting by $n_{\omega}$ the degeneracy of the space associated with
the weight $\omega$, then $\lambda=1,\cdots,n_{\omega}$. However,
we find it more convenient to hide this extra label and instead view
$\left|\omega\right\rangle $ as a $n_{\omega}$-dimensional vector.
Then, it is possible to define a diagonal matrix $\left[\omega\left(i\right)\right]\equiv\omega\left(i\right)\mathbb{1}$
such that
\begin{align}
H_{i}\left|\omega\right\rangle  & =\left[\omega\left(i\right)\right]\left|\omega\right\rangle \,.
\end{align}
Now, according to the Chevalley-Serre relations \eqref{eq:Serre_Chevalley_relations1}--\eqref{eq:Serre_Chevalley_relations3},
applying $E_{j}$($F_{j}$) to $\left|\omega\right\rangle $ raises(lowers)
$\omega$ by the simple root $\alpha_{j}$:
\begin{align}
H_{i}\left(E_{j}\left|\omega\right\rangle \right) & =\left(E_{j}H_{i}+A_{ji}E_{j}\right)\left|\omega\right\rangle =\left[\left(\omega+\alpha_{j}\right)\left(i\right)\right]\left(E_{j}\left|\omega\right\rangle \right)\,,\\
H_{i}\left(F_{j}\left|\omega\right\rangle \right) & =\left(F_{j}H_{i}-A_{ji}F_{j}\right)\left|\omega\right\rangle \,=\left[\left(\omega-\alpha_{j}\right)\left(i\right)\right]\left(F_{j}\left|\omega\right\rangle \right)\,.
\end{align}
As such, we define the matrices $\left[\omega\right]^{i}$ and $\left[\omega\right]_{i}$
through the equations
\begin{align}
E_{i}\left|\omega,\lambda\right\rangle  & \equiv\sum_{\lambda'}\left(\left[\omega\right]^{i}\right)_{\lambda'\lambda}\left|\omega+\alpha_{i},\lambda'\right\rangle \,,\\
F_{i}\left|\omega,\lambda\right\rangle  & \equiv\sum_{\lambda'}\left(\left[\omega\right]_{i}\right)_{\lambda'\lambda}\left|\omega-\alpha_{i},\lambda'\right\rangle \,,
\end{align}
or equivalently\vspace{-2mm}
\begin{align}
E_{i}\left|\omega\right\rangle  & =\left(\left[\omega\right]^{i}\right)^{T}\left|\omega+\alpha_{i}\right\rangle \,,\\
F_{i}\left|\omega\right\rangle  & =\left(\left[\omega\right]_{i}\right)^{T}\left|\omega-\alpha_{i}\right\rangle \,.
\end{align}
The dimensions of $\left[\omega\right]^{i}$ and $\left[\omega\right]_{i}$
are $n_{\omega+\alpha_{i}}\times n_{\omega}$ and $n_{\omega-\alpha_{i}}\times n_{\omega}$,
respectively. Based on this observation, we conclude that the set
of matrices $\left\{ E_{i},F_{i},H_{i}\right\} $ are almost entirely
composed of null blocks, with a few exceptions:
\begin{gather}
\begin{array}{ccc}
 & \hspace{-11mm}\begin{array}{ccc}
 & \omega\,\,\,\,\end{array}\\
\begin{array}{c}
\\
E_{i}=\\
\\
\end{array} & \hspace{-4mm}\begin{pmatrix} & \vdots\,\,\,\,\\
 & \left[\omega\right]^{i} & \cdots\\
\,\,
\end{pmatrix} & \hspace{-4mm}\begin{array}{c}
\\
\omega+\alpha_{i}\quad,\\
\\
\end{array}
\end{array}\,\,\begin{array}{ccc}
 & \hspace{-11mm}\begin{array}{ccc}
 & \omega\,\,\,\,\end{array}\\
\begin{array}{c}
\\
\quad F_{i}=\\
\\
\end{array} & \hspace{-4mm}\begin{pmatrix} & \vdots\,\,\,\,\\
 & \left[\omega\right]_{i} & \cdots\\
\,\,
\end{pmatrix} & \hspace{-4mm}\begin{array}{c}
\\
\omega-\alpha_{i}\\
\\
\end{array}\quad,
\end{array}\nonumber \\
\begin{array}{ccc}
 & \hspace{-11mm}\begin{array}{ccc}
 & \omega\,\,\,\end{array}\\
\begin{array}{c}
\\
H_{i}=\\
\\
\end{array} & \hspace{-4mm}\begin{pmatrix} & \vdots\,\,\,\\
 & \left[\omega\left(i\right)\right] & \cdots\\
\,\,
\end{pmatrix} & \hspace{-4mm}\begin{array}{c}
\\
\omega\quad.\\
\\
\end{array}
\end{array}
\end{gather}
In order to build the diagonal matrices $H_{i}$, we only need the
list of weights $\left\{ \omega\right\} $, and this is easy to obtain
for a given simple Lie algebra (see for example \cite{Cahn:1985wk}).
Also, with Freudenthal's formula, the dimension of the weight space
associated to $\omega$ ($=n_{\omega}$) can be readily calculated,
so we assume that this is known as well. The non-trivial part of the
computation is calculating the ladder operators $E_{i}$ and $F_{i}$,
or equivalently, the matrix blocks $\left[\omega\right]^{i}$ and
$\left[\omega\right]_{i}$ for all weights $\omega$. To proceed we
need the equation analogous to \eqref{eq:Susyno_SU2_recursive_relation}
of $SU(2)$. By applying $E_{i}F_{j}=F_{j}E_{i}+\delta_{ij}H_{j}$
to the vector $\left|\omega\right\rangle $, we get the desired relation:
\begin{align}
\left[\omega-\alpha_{j}\right]^{i}\left[\omega\right]_{j} & =\left[\omega+\alpha_{i}\right]_{j}\left[\omega\right]^{i}+\delta_{ij}\left[\omega\left(i\right)\right]\,.\label{eq:Susyno_Simplegroup_recursive_relation}
\end{align}
If all $\left[\omega+\alpha_{i}\right]_{j}$ and $\left[\omega\right]^{i}$
are known for all $\omega$ above some limit, then this equation can
be used to recursively derive the remaining $\left[\;\right]^{i}$
and $\left[\;\right]_{i}$, with the right-hand side always known
and the left-hand side unknown. For example, starting with the highest
weight $\Lambda\equiv\omega_{\textrm{max}}$ of the representation,
$\left[\Lambda\right]^{i}=0$ since we cannot raise it any further,
and also $n_{\Lambda}=1$ always \cite{Cahn:1985wk}, so $\left[\Lambda\left(j\right)\right]$
are simply the Dynkin indices of the representation $\Lambda_{j}$
(not to be confused with the block matrices $\left[\Lambda\right]_{j}$
of the lowering operators $F_{j}$):
\begin{align}
\left[\Lambda-\alpha_{j}\right]^{i}\left[\Lambda\right]_{j} & =\delta_{ij}\Lambda_{j}\\
 & \;\;\vdots\nonumber 
\end{align}
Assuming that $E_{i}=F_{i}^{\dagger}$ and that these are real matrices,
it follows that $\left[\omega+\alpha_{i}\right]_{i}=\left(\left[\omega\right]^{i}\right)^{T}$.
Using this relation in equation \eqref{eq:Susyno_Simplegroup_recursive_relation},
and also making the shift $\omega\rightarrow\omega+\alpha_{j}$, we
find that 
\begin{align}
\left[\omega\right]^{i}\left(\left[\omega\right]^{j}\right)^{T} & =\left(\left[\omega+\alpha_{i}\right]^{j}\right)^{T}\left[\omega+\alpha_{j}\right]^{i}+\delta_{ij}\left[\left(\omega+\alpha_{i}\right)\left(i\right)\right]\,.\label{eq:Susyno_Simplegroup_recursive_relation_2}
\end{align}
Then, to compute the unknown $\left[\omega\right]^{i}$ on the left-hand
side, it is clear that we must simultaneously solve this equation
for all $\left[\omega\right]^{1},\left[\omega\right]^{2},\cdots,\left[\omega\right]^{n}$,%
\footnote{Note that some of these $\left[\omega\right]^{i}$ might not exist
if $\omega+\alpha_{i}$ is not a weight, or equivalently $\left[\omega\right]^{i}$
is a matrix with size $0\times n_{\omega}$ in these cases. %
} since they all mix together. We can do this by defining a big matrix
\begin{align}
\Omega\left(\omega\right) & \equiv\left(\begin{array}{c}
\left[\omega\right]^{1}\\
\vdots\\
\left[\omega\right]^{i}\\
\vdots
\end{array}\right)
\end{align}
which contains the $\left[\omega\right]^{i}$ ($i=1,\cdots,n$), for
a weight $\omega$. Then, the left-hand side of equation \eqref{eq:Susyno_Simplegroup_recursive_relation_2}
is just the $\left(ij\right)$ block of the matrix $\Omega\left(\omega\right)\Omega^{T}\left(\omega\right)$:
\begin{align}
\Omega\left(\omega\right)\Omega^{T}\left(\omega\right) & =X\left(\omega\right)\,,\label{eq:Susyno_OOX_equation}
\end{align}
where $X\left(\omega\right)$ is a known matrix, because it depends
on weights bigger than $\omega$: its $\left(ij\right)$ block is
$\left(\left[\omega+\alpha_{i}\right]^{j}\right)^{T}\left[\omega+\alpha_{j}\right]^{i}+\delta_{ij}\left[\left(\omega+\alpha_{i}\right)\left(i\right)\right]$.
One must only break this symmetric $X\left(\omega\right)$ into some
matrix $\Omega\left(\omega\right)$ times its transpose and then use
this $\Omega\left(\omega\right)$ to build $X\left(\omega'\right)$
for lower weights $\omega'$, and repeat this process until the lowest
weight is reached. There is always some arbitrariness in this process,
since $\widetilde{\Omega}\left(\omega\right)=\Omega\left(\omega\right)\mathcal{O}$,
for some orthogonal matrix $\mathcal{O}$, also satisfies equation
\eqref{eq:Susyno_OOX_equation}.%
\footnote{It is instructive to compare this general situation with the one encountered
in $SU(2)$, which corresponds to the case where $\Omega\left(\omega\right)$
and $X\left(\omega\right)$ are numbers ($1\times1$ matrices) for
each weight $\omega$.%
} In any case, with this algorithm, all the $\left[\;\right]^{i}$
matrix blocks needed to build the matrices $E_{i}$, $F_{i}$ are
determined and, as previously explained, from there it is easy to
build a complete list of hermitian generators $\left\{ T_{a}\right\} $
of the algebra.

\section{\label{sec:Susyno_Invariant-combinations-of}Invariant combinations
of fields transforming under some representation of a simple group
and generalized Clebsch-Gordan coefficients}

Two $SU(2)$ doublets, $A\equiv\left(A^{+},A^{-}\right)^{T}$ and
$B=\left(A^{+},A^{-}\right)^{T}$, transform as
\begin{align}
A\left(B\right) & \rightarrow\left(\mathbb{1}+i\sum_{a=1}^{3}\varepsilon^{a}\sigma_{a}\right)A\left(B\right)
\end{align}
under an infinitesimal transformation of the group. However, the bilinear
combination $A^{+}B^{-}-A^{-}B^{+}$ of the two fields does not change:
\begin{align}
A^{+}B^{-}-A^{-}B^{+} & \rightarrow A^{+}B^{-}-A^{-}B^{+}\,.\label{eq:Susyno_SU2_invariant}
\end{align}

In order to write the Lagrangian of gauge theories, it is crucial
that all such invariant combinations of the fields are known.%
\footnote{The invariants of a Lie algebra can however be used to refer to something
entirely different \cite{Gruber:0124.17401}.%
} For some combinations of the representations of certain groups ($SU(N)$,
$SO(10)$, ...) there are clever techniques used in Particle Physics
to write down these terms. However, they are usually only applicable
to a few cases, and since \texttt{Susyno} aims at building the Lagrangian
of a model based on any gauge group and with any field content (in
theory at least), these approaches are not adequate. In this section,
our aim is to describe the method used to solve this issue. 

First, we should drop the $+/-$ notation for the components of fields,
in favor of a more general one. As such, we assume that the components
of a field $A$ are labeled as $A^{i}$, such that $A=\sum_{i}A^{i}e_{i}^{(A)}$
where the $e_{i}^{(A)}$ form a basis of the vector space to which
$A$ belongs. Under the gauge symmetry, these basis vectors transform
as
\begin{align}
\delta e_{j}^{(A)} & =i\sum_{a,j'}\varepsilon^{a}\left[T_{a}^{(A)}\right]_{j'j}e_{j'}^{(A)}\,,
\end{align}
where the $T_{a}^{(A)}$ are the algebra generators, in $A$'s representation.
Another perspective is to assume that the $\delta e_{j}^{(A)}$ do
not change ($\delta e_{j}^{(A)}=0$) and instead the components of
the field $A$ transform as follows: 
\begin{align}
\delta A^{j} & =i\sum_{a}\varepsilon^{a}\left[T_{a}^{(A)}A\right]^{j}\,.
\end{align}

Consider now the Kronecker product of two fields $A$ and $B$ given
by
\begin{align}
A\otimes B & =\sum_{i,j}A^{i}B^{j}e_{i}^{(A)}\otimes e_{j}^{(B)}\equiv\sum_{i,j}\left(A\otimes B\right)^{ij}e_{i}^{(A)}\otimes e_{j}^{(B)}\,.
\end{align}
Each component is specified by two indices, $i$ and $j$, but to
view $A\otimes B$ as a vector, we may combine the two indices into
a single one: $\widetilde{k}=1,2,3\cdots,nm$ instead of $\left(i,j\right)=\left(1,1\right),\cdots,\left(1,m\right),\left(2,1\right),\cdots\left(n,m\right)$.
Then, each component of the Kronecker product of $A$ and $B$ transforms
as 
\begin{align}
\delta\left(A\otimes B\right)^{\widetilde{k}} & =i\sum_{a}\varepsilon^{a}\left[\left(T_{a}^{(A)}\otimes\mathbb{1}^{(B)}+\mathbb{1}^{(A)}\otimes T_{a}^{(B)}\right)A\otimes B\right]^{\widetilde{k}}\,.
\end{align}
In other words, $A\otimes B$ transforms as a representation of the
gauge symmetry, with generator matrices given by $T_{a}^{(A\otimes B)}=T_{a}^{(A)}\otimes\mathbb{1}^{(B)}+\mathbb{1}^{(A)}\otimes T_{a}^{(B)}$.
In principle, this representation is reducible and, in fact, if there
is an invariant combination of $A$ and $B$, there must be a trivial
representation in it. Then, consider that 
\begin{align}
\sum_{i,j}\kappa_{ij}A^{i}B^{j} & =\sum_{\widetilde{k}}\kappa_{\widetilde{k}}\left(A\otimes B\right)^{\widetilde{k}}
\end{align}
is such a combination. Gauge invariance is verified if and only if
\begin{align}
\sum_{\widetilde{k}'}\kappa_{\widetilde{k}'}\left[T_{a}^{(A\otimes B)}\right]_{\widetilde{k}'\widetilde{k}} & =0
\end{align}
for all $\widetilde{k}$ and all $a$. Equivalently, all invariant
combinations of the $A$ and $B$ fields are given by vectors $\kappa$
which belong to the nullspace of the matrix:\vspace{-0.8mm}
\begin{gather}
\left(\begin{array}{c}
\left(T_{1}^{(A\otimes B)}\right)^{T}\\
\vdots\\
\left(T_{a}^{(A\otimes B)}\right)^{T}\\
\vdots
\end{array}\right)\,.\label{eq:Susyno_generators_stacking}
\end{gather}
As such, the number of independent invariants is given by the dimension
of this space. We note that this method of finding the invariants
will work for any representation of any group, provided that their
matrices are known. Also, the extension of this method to combinations
involving more than two fields is straightforward.

We now go through the example mentioned at the beginning of this section,
namely $A$ and $B$ are two $SU(2)$ doublets. First, notice that\vspace{-0.8mm}
\begin{align}
\sigma_{a}\otimes\mathbb{1}_{2} & =\left(\begin{array}{cc}
\left(\sigma_{a}\right)_{11}\mathbb{1}_{2} & \left(\sigma_{a}\right)_{12}\mathbb{1}_{2}\\
\left(\sigma_{a}\right)_{21}\mathbb{1}_{2} & \left(\sigma_{a}\right)_{22}\mathbb{1}_{2}
\end{array}\right)\,,\quad\mathbb{1}_{2}\otimes\sigma_{a}=\left(\begin{array}{cc}
\sigma_{a} & 0\\
0 & \sigma_{a}
\end{array}\right)\,.
\end{align}
Then, the components of $A\otimes B=\left(A^{1}B^{1},A^{1}B^{2},A^{2}B^{1},A^{2}B^{2}\right)^{T}$
transform according to the generators
\begin{align}
T_{1}^{(A\otimes B)} & =\left(\begin{array}{cccc}
0 & 1 & 1 & 0\\
1 & 0 & 0 & 1\\
1 & 0 & 0 & 1\\
0 & 1 & 1 & 0
\end{array}\right)\,,\label{eq:Susyno_T1AB_1}\\
T_{2}^{(A\otimes B)} & =\left(\begin{array}{cccc}
0 & -i & -i & 0\\
i & 0 & 0 & -i\\
i & 0 & 0 & -i\\
0 & i & i & 0
\end{array}\right)\,,\\
T_{3}^{(A\otimes B)} & =\left(\begin{array}{cccc}
2 & 0 & 0 & 0\\
0 & 0 & 0 & 0\\
0 & 0 & 0 & 0\\
0 & 0 & 0 & -2
\end{array}\right)\,.\label{eq:Susyno_T1AB_3}
\end{align}
The invariants $\sum_{\widetilde{k}=1}^{4}\kappa_{\widetilde{k}}(A\otimes B)^{\widetilde{k}}=\kappa_{1}A^{1}B^{1}+\kappa_{2}A^{1}B^{2}+\kappa_{3}A^{2}B^{1}+\kappa_{4}A^{2}B^{2}$
are given by the vectors $\kappa=\left(\kappa_{1},\kappa_{2},\kappa_{3},\kappa_{4}\right)^{T}$
which are in the nullspace of the transpose of all three matrices
in equations \eqref{eq:Susyno_T1AB_1}--\eqref{eq:Susyno_T1AB_3}.
One simple way to find them is to stack the transpose of these three
matrices on top of each other in a single $12\times4$ matrix, and
find its nullspace. In this case, we get a 1-dimensional nullspace
generated by the vector $\kappa=\left(0,1,-1,0\right)^{T}$, which
means that the only invariant combination of the fields $A$ and $B$
is $A^{1}B^{2}-A^{2}B^{1}$ (and multiples of it).

As a sightly more elaborate example, consider a third field $C$,
which is a triplet of $SU(2)$. Using as representation matrices of
the generators
\begin{align}
T_{1}^{(C)} & =\frac{1}{\sqrt{2}}\left(\begin{array}{ccc}
0 & 1 & 0\\
1 & 0 & 1\\
0 & 1 & 0
\end{array}\right)\,,\\
T_{2}^{(C)} & =\frac{1}{\sqrt{2}}\left(\begin{array}{ccc}
0 & -i & 0\\
i & 0 & -i\\
0 & i & 0
\end{array}\right)\,,\\
T_{3}^{(C)} & =\left(\begin{array}{ccc}
1 & 0 & 0\\
0 & 0 & 0\\
0 & 0 & -1
\end{array}\right)\,,
\end{align}
the trilinear invariants involving the fields $A$ and $B$ and $C$
are of the form
\begin{align}
\sum_{\widetilde{k}=1}^{12}\kappa_{\widetilde{k}}(A\otimes B\otimes C)^{\widetilde{k}} & =\kappa_{1}A^{1}B^{1}C^{1}+\kappa_{2}A^{1}B^{1}C^{2}+\cdots+\kappa_{12}A^{2}B^{2}C^{3}\,,
\end{align}
where the 12-dimensional vector $\kappa$ must be in the nullspace
of a $36\times12$ matrix made from the $T_{a}^{(A\otimes B\otimes C)}$
generators, which we shall not write down here. Again, it turns out
that a single $\kappa=\left(0,0,\sqrt{2},0,-1,0,0,-1,0,\sqrt{2},0,0\right)$
generates this vector space so, up to a multiplicative factor,
\begin{gather}
\sqrt{2}A^{1}B^{1}C^{3}-A^{1}B^{2}C^{2}-A^{2}B^{1}C^{2}+\sqrt{2}A^{2}B^{2}C^{1}
\end{gather}
is the only trilinear invariant involving two doublets $A$/$B$,
and a triplet $C$. As a final example, if all $A$, $B$ and $C$
were triplets, a similar construction would require finding the nullspace
of a $81\times27$ dimensional matrix; $\varepsilon_{ijk}A^{i}B^{j}C^{k}$
is the only invariant obtained.

A legitimate concern is the time needed to compute invariant combinations
of larger representations with this method. The most demanding part
of the algorithm consists in finding a basis of the nullspace of a
potentially very large matrix: for a simple group with $d$ generators,
and representations of size $n_{A},n_{B},\cdots$ this matrix has
$d\times n_{A}\times n_{B}\times\cdots$ rows and $n_{A}\times n_{B}\times\cdots$
columns. For example, the calculation of the invariant in $\mathbf{120}\otimes\mathbf{120}\otimes\mathbf{54}$
of $SO(10)$ involves a $29160000\times648000$ matrix. Even so, with
a few simple tricks, this computation can be completed in less than
a minute in a modern computer.

Consider the representation matrices $\left\{ E_{i},F_{i},H_{i}\right\} $
obeying the Chevalley-Serre relations obtained in the previous section:
\begin{itemize}
\item The $H_{i}$ are diagonal matrices, so $H_{i}\kappa=0$ implies that
$\kappa_{j}=0$ unless the entries $\left(H_{i}\right)_{jj}$ for
all $i$ are null. This means that we should focus on the subspace
of $A\otimes B\otimes\cdots$ with null weights, which can be readily
calculated.
\item All the remaining generators $\widetilde{E}_{x}$, $\widetilde{F}_{x}$
of a Lie algebra are given by commutators of the Chevalley-Serre elementary
$E_{i}$ or $F_{i}$, so if $E_{i}^{T}\kappa=F_{i}^{T}\kappa=0$ for
all $i$, then it follows that $\widetilde{E}_{x}^{T}\kappa=\widetilde{F}_{x}^{T}\kappa=0$
for all $x$. Therefore, we only need to consider the nullspace of
the matrices $E_{i}^{T}$ and $F_{i}^{T}$.
\item Since $F_{i}=E_{i}^{T}$, the nullspace of $F_{i}^{T}$ is the same
as $E_{i}^{T}F_{i}^{T}$, but $E_{i}^{T}F_{i}^{T}\kappa=\left(F_{i}^{T}E_{i}^{T}-H_{i}\right)\kappa=0$
assuming that $\kappa$ is in the nullspace of both $E_{i}^{T}$ and
$H_{i}$. As such, there is no need to compute the null space of the
$F_{i}^{T}$ matrices.
\end{itemize}
All things considered, for a rank $n$ group, the matrix whose nullspace
we must find has roughly $n\sqrt{n_{A}n_{B}\cdots}$ rows and $\sqrt{n_{A}n_{B}\cdots}$
columns, and most of its entries are null.

To conclude this discussion, we note that finding such invariant combinations
is essentially equivalent to the computation of generalized (i.e.,
non-$SU(2)$) Clebsch-Gordan coefficients. To see this, consider the
following invariant combination of the fields/representations $A$,
$B$ and $C$ of some group:
\begin{align}
\sum_{ijk} & \kappa_{ijk}A^{i}B^{j}C^{k}\,.
\end{align}
Then, a vector $D$ with components $D^{k}\equiv\kappa_{ijk}A^{i}B^{j}$
is in the irreducible representation  conjugate to one of $C$. Therefore,
by fixing $A$ and $B$, and writing all such invariant combinations
for different irreducible representations $C$, we get the explicit
form of the decomposition of $A\otimes B$ in its irreducible parts
(see also \cite{Schlosser:1991aa}).

\section{\label{sec:Susyno_Einstein-convention-applied}Einstein convention
applied to flavor indices}

In this section, we discuss a difficulty in making a program such
as \texttt{Susyno}, arising from the presence of multiple copies of
a representation of the gauge group in a given model. These copies
are usually considered as different flavors of the same field. Flavor
is then an index, which must be carried not only by the fields but
also by the parameters in the Lagrangian which multiply them. An example
would be the Yukawa couplings.

The two-loop RGEs can be quite lengthy, so in most cases, it is preferable
to write an expression such as\\

\noindent \texttt{Yu{[}1,1{]}$^{*}$Yu{[}1,1{]}+Yu{[}1,2{]}$^{*}$Yu{[}1,2{]}+...+Yu{[}3,3{]}$^{*}$Yu{[}3,3{]}}\\
\\
in Einstein notation:\\

\noindent \texttt{Yu{[}i,j{]}$^{*}$Yu{[}i,j{]}}~\\
\texttt{}~\\
This same expression could be expressed as \texttt{Tr{[}Yu$^{\dagger}$Yu{]}},
which is an even more compact notation. However, this last matrix
form cannot handle parameters with more than two flavor indices, so
instead \texttt{Susyno} uses the Einstein summation notation.

A problem then arises due to the presence of dummy indices: we know
that \texttt{Yu{[}i,j{]}$^{*}$Yu{[}i,j{]}} is the same as \texttt{Yu{[}k,i{]}$^{*}$Yu{[}k,i{]}},
but in general, computationally it is far from trivial to establish
the equality of two such terms, particularly when there are parameters
with more than two indices. In order to simplify the final expressions,
this task must nevertheless be carried out. Here, we shall not go
through the algorithm used in the program; we simply note that it
is not perfect, since \texttt{Susyno} can sometimes fail to see that
different forms of the same term are indeed identical.

To complicate matters further, some of these parameters have symmetries:
for instance, some \texttt{P{[}i,j{]}} might be the same as \texttt{P{[}j,i{]}},
which means that \texttt{P{[}i,j{]}$^{*}$P{[}i,j{]}+P{[}i,j{]}$^{*}$P{[}j,i{]}}
equals \texttt{2P{[}i,j{]}$^{*}$P{[}i,j{]}}. In general, a set of
$m$ parameters $P_{f_{1}f_{2}\cdots f_{n}}^{1},\cdots,P_{f_{1}f_{2}\cdots f_{n}}^{m}$
with $n$ flavor indices $f_{i}$, under a permutation $\sigma$ of
these, transforms as
\begin{align}
P_{\sigma\left(f_{1}f_{2}\cdots f_{n}\right)}^{i} & =\sum_{j}S\left(\sigma\right)_{ij}P_{f_{1}f_{2}\cdots f_{n}}^{j}\,,\label{eq:Susyno_Sn_symmetry_of_parameters}
\end{align}
where the matrix $S\left(\sigma\right)$ is a representation of the
symmetric group $S_{n}$. Usually, parameters have at most $n=2$
flavor indices, and the $S_{2}$ group only has two 1-dimensional
irreducible representations: the trivial one, $S\left(\sigma\right)=1$,
and the alternating one, $S\left(\sigma\right)=\textrm{sign}\left(\sigma\right)$.
So
\begin{align}
P_{f_{2}f_{1}} & =\pm P_{f_{1}f_{2}}\,.
\end{align}
In other words, exchanging the flavors of a 2-index parameter results
at most in a sign change. But as shown in equation \eqref{eq:Susyno_Sn_symmetry_of_parameters},
more complex situations can arise---for example, two parameters $P_{f_{1}f_{2}f_{3}}^{1}$
and $P_{f_{1}f_{2}f_{3}}^{2}$ with three flavor indices may be such
that $P_{\sigma\left(f_{1}f_{2}f_{3}\right)}^{1(2)}$ is actually
a linear combination of both $P_{f_{1}f_{2}f_{3}}^{1}$ and $P_{f_{1}f_{2}f_{3}}^{2}$.
In the next section, we return to this topic.

\section{\label{sec:Susyno_A-Lagrangian-less-computation}A Lagrangian-free
computation of the RGEs}

The RGEs of a model can be calculated in two main steps: first the
model's gauge invariant Lagrangian is computed, and then the generic
two-loop RGE formulae of \cite{Martin:1993zk,Yamada:1994id} are applied
to it. Above, we have detailed how the first part can be carried out
for any gauge group and for any field content; the second part is
straightforward, with the exception of the simplification of expressions
discussed in the previous section. Nevertheless, this procedure can
be very time consuming:
\begin{itemize}
\item Calculating the model's Lagrangian requires the computation of all
representation matrices and of all invariant combinations of the fields
(see sections \ref{sec:Susyno_Building-the-matrices} and \ref{sec:Susyno_Invariant-combinations-of});
\item Naive application of the RGEs of a generic SUSY model requires many
summations of symbolic quantities. As an example, one of the terms
in the RGEs of the soft masses is $h_{ilm}h^{jln}Y_{npq}Y^{mpq}$
(see equation \eqref{eq:U1_mixing_example3}) where the tensors $Y$
and $h$ collect all the trilinear couplings in the superpotential
and soft SUSY breaking Lagrangian, respectively. The free indices
($i,j$), as well as the summed/dummy ones ($l,m,m,p,q$), range over
all the field components in a model; in the MSSM, ignoring different
flavors, there are 19 (6 in $\hat{Q}$, $3$ in both $\hat{u}$ and
$\hat{d}$, 2 in $\widehat{L}$, $\widehat{H}_{u}$, $\widehat{H}_{d}$,
and 1 in $\hat{e}$). This means that for each $i,j=1,\cdots,19$
the expression $h_{ilm}h^{jln}Y_{npq}Y^{mpq}$ alone, appearing in
$\left[\beta_{m^{2}}^{(2)}\right]_{i}^{j}$, represents $19^{5}=2476099$
terms. With some ingenuity, the fact that the tensors $Y$ and $h$
are very sparse (most of the entries are null) can be used to reduce
the complexity of this calculation, allowing the MSSM's RGEs to be
computed in a few seconds. Nevertheless, it is clear that the time
needed will scale as some power law $N^{x}$ of the number of field
components $N$ in the model. The expression $h_{ilm}h^{jln}Y_{npq}Y^{mpq}$
suggests that this exponent $x$ is roughly $7$, but due to the increasing
sparseness of the tensors as $N$ increases, the true number is lower.
\end{itemize}
For example, minimalistic $SO(10)$ models contain hundreds of field
components, which makes this approach impractical due to the time
it would take to complete the computations. Given the large representations
involved, there can be memory problems as well.

On the other hand, by inspection of the generic RGEs, it is indisputable
that they do not depend on the explicit basis used for each gauge
representation: a unitary transformation $\Phi_{i}\rightarrow U_{ij}\Phi_{j}$
of all superfields which does not mix different components of different
irreducible representations of the gauge group will not affect the
RGEs. This raises the following question: is it possible to calculate
the basis-independent RGEs of a model, without actually building an
unphysical/basis-dependent Lagrangian? It turns out that such a Lagrangian-free
computation of the RGEs is feasible and also dramatically faster than
the more conventional approach. Therefore, in version 2 of \texttt{Susyno}
this new method was introduced, leading to an extensive rewriting
of the code. Even though there is no longer any need for methods such
as \texttt{RepMatrices} and \texttt{Invariants}, these were kept (and
in some cases extended) since they are useful on their own for other
potential applications. It should be pointed out that the idea of
avoiding altogether the construction of basis dependent quantities,
such as the Lagrangian, is not new---see for example \cite{Cvitanovic:1976am,Cvitanovic:2008zz}.

Let us now analyze how this can be done. We recall that the information
contained in the superpotential and soft SUSY-breaking scalar terms
can be collected in the tensors $Y^{abc},\mu^{ab},L^{a},h^{abc},b^{ab},s^{a}$
and $\left(m^{2}\right)_{b}^{a}$ (see equations \eqref{eq:Introduction_superpotential}
and \eqref{eq:Introduction_Lsoft}). Each of these indices runs over
all fields in the model, and it will prove useful to expand each of
them into 3 sub-indices:
\begin{enumerate}
\item A representation index, denoted by a roman lowercase letter ($a,b,c,\cdots$);
\item A representation-component index, denoted by a roman uppercase letter
($A,B,C,\cdots$);
\item A flavor index, denoted by a Greek lowercase letter ($\alpha,\beta,\xi,\cdots$).
\end{enumerate}
As such,
\begin{eqnarray}
a\overset{\overset{{\scriptscriptstyle \textrm{new}}}{{\scriptscriptstyle \textrm{notation}}}}{\rightarrow}\left(aA\alpha\right) & ;\,\, b\overset{{\scriptscriptstyle \overset{{\scriptscriptstyle \textrm{new}}}{{\scriptscriptstyle \textrm{notation}}}}}{\rightarrow}\left(bB\beta\right) & ;\,\, c\overset{{\scriptscriptstyle \overset{{\scriptscriptstyle \textrm{new}}}{{\scriptscriptstyle \textrm{notation}}}}}{\rightarrow}\left(cC\xi\right)\,.
\end{eqnarray}
For example, the $a$ in $\left(aA\alpha\right)$ could point to the
left-handed leptons representation ($L$), $A=1,2$ would then specify
the doublet's component, and $\alpha=1,2,3$ would stand for one of
the three possible lepton flavors. Consider now a trilinear term in
the superpotential, for example: it involves three superfields, $\Phi_{\left(aA\alpha\right)}\Phi_{\left(bB\beta\right)}\Phi_{\left(cC\xi\right)}$,
whose $A,\, B,\, C$ indices must be contracted in a gauge invariant
way: $\kappa^{ABC}\Phi_{\left(aA\alpha\right)}\Phi_{\left(bB\beta\right)}\Phi_{\left(cC\xi\right)}$.
This tensor $\kappa$ consists of numbers only, and it is specific
to the combination of the representations $a,b$ and $c$, so we shall
use $\kappa\left(abc\right)$ instead of just $\kappa$. However,
the product of three representations may contain more than one independent
gauge invariant combination, so we introduce an extra label $\lambda$
to distinguish them. Finally, we note that the whole expression $\kappa\left(abc\lambda\right)^{ABC}\Phi_{\left(aA\alpha\right)}\Phi_{\left(bB\beta\right)}\Phi_{\left(cC\xi\right)}$
must be multiplied by a parameter $y\left(abc\lambda\right)^{\alpha\beta\xi}$
(such as $Y_{u},Y_{d}$ or $Y_{e}$ in the MSSM) which contains flavor
indices $\alpha,\beta,\xi$. As such, we may write
\begin{align}
\left[W\right]_{\textrm{trilinear part}} & =\frac{1}{6}\sum_{\underset{\underset{A,B,C}{\alpha,\beta,\xi}}{{\scriptscriptstyle a,b,c,\lambda}}}y\left(abc\lambda\right)^{\alpha\beta\xi}\kappa\left(abc\lambda\right)^{ABC}\Phi_{\left(aA\alpha\right)}\Phi_{\left(bB\beta\right)}\Phi_{\left(cC\xi\right)}\,,
\end{align}
or equivalently,
\begin{align}
\left[Y^{abc}\right]_{\textrm{old notation}} & =Y^{\left(aA\alpha\right)\left(bB\beta\right)\left(cC\xi\right)}=\sum_{\lambda}y\left(abc\lambda\right)^{\alpha\beta\xi}\kappa\left(abc\lambda\right)^{ABC}\,.
\end{align}
In this way, we have successfully separated the symbolic part, $y\left(abc\lambda\right)^{\alpha\beta\xi}$,
from the numerical one, $\kappa\left(abc\lambda\right)^{ABC}$. Focusing
on the latter one, we note that
\begin{eqnarray}
\sum_{BC}\kappa\left(abc\lambda\right)^{ABC}\kappa\left(a'bc\lambda'\right)_{A'BC}^{*} & = & f\left(a,\lambda,\lambda'\right)\delta_{a'}^{a}\delta_{A'}^{A}\,.
\end{eqnarray}
The $\delta_{a'}^{a}\delta_{A'}^{A}$ factor is explained by the fact
that the expression on the left must be proportional to the $\kappa\left(aa'\right)^{AA'}$
coefficients of a bilinear term $\kappa\left(aa'\right)^{AA'}\Phi_{\left(aA\right)}\Phi_{\left(a'A'\right)}^{*}$
(omitting the flavor indices and the parameter). As for the $f\left(a,\lambda,\lambda'\right)$
factor in the above equation, we may reasonably orthonormalize the
various invariants in a trilinear combination of the $a$, $b$ and
$c$ representations, such that
\begin{align}
\sum_{ABC}\kappa\left(abc\lambda\right)^{ABC}\kappa\left(abc\lambda'\right)_{ABC}^{*} & \equiv\delta_{\lambda'}^{\lambda}\,,\label{eq:Susyno_kappaABC_normalization}
\end{align}
and consequently
\begin{eqnarray}
\sum_{BC}\kappa\left(abc\lambda\right)^{ABC}\kappa\left(a'bc\lambda'\right)_{A'BC}^{*} & = & \frac{1}{\dim\phi_{a}}\delta_{a'}^{a}\delta_{\lambda'}^{\lambda}\delta_{A'}^{A}\,,\label{eq:Susyno_canoncial_normalization}
\end{eqnarray}
where $\dim\phi_{a}$ is the dimension of the representation $a$
of the gauge group ($A=1,\cdots,\dim\phi_{a}$). Consider now two
generic trilinear tensors, $T^{abc}$ and $U^{abc}$,
\begin{align}
\left[T^{abc}\right]_{\textrm{old notation}} & =\sum_{\lambda}t\left(abc\lambda\right)^{\alpha\beta\xi}\kappa\left(abc\lambda\right)^{ABC}\,,\label{eq:Susyno_T}\\
\left[U^{abc}\right]_{\textrm{old notation}} & =\sum_{\lambda}u\left(abc\lambda\right)^{\alpha\beta\xi}\kappa\left(abc\lambda\right)^{ABC}\,,\label{eq:Susyno_U}
\end{align}
 and two bilinear ones, $D_{b}^{a}$ and ${D'}_{b}^{a}$, which are
diagonal in both representation and representation-component space:
\begin{eqnarray}
\left[D_{b}^{a}\right]_{\textrm{old notation}} & = & d\left(a\right)_{\beta}^{\alpha}\delta_{b}^{a}\delta_{B}^{A}\,,\\
\left[{D'}_{b}^{a}\right]_{\textrm{old notation}} & = & d'\left(a\right)_{\beta}^{\alpha}\delta_{b}^{a}\delta_{B}^{A}\,.
\end{eqnarray}
Assuming that repeated indices are summed over from now on, after
a few simple manipulations of the expressions we conclude that the
generic quantity $U^{amn}D_{n}^{n'}{D'}_{m}^{m'}T_{bmn'}$ is also
diagonal in representation and representation-component space:
\begin{multline}
T^{amn}D_{n}^{n'}{D'}_{m}^{m'}U_{bmn'}\\
=\frac{1}{\dim\phi_{\check{a}}}t\left(\check{a}mn\lambda\right)^{\alpha\mu\nu}d\left(n\right)_{\nu}^{\nu'}d'\left(m\right)_{\mu}^{\mu'}u\left(\check{a}mn\lambda\right)_{\beta\mu'\nu'}^{*}\delta_{b}^{\check{a}}\delta_{B}^{A}\,.\label{eq:Susyno_Lagrangianless_masterequation}
\end{multline}
An inverted hat {}``$\check{\;}$'' was added to the $a$ index
to indicate that it is not to be summed over. This is the most important
equation of this section: we are able to get the result of a multiple-summation
expression without actually performing these sums over the representation-component
indices $A,B,\cdots$. It is not even required to know most of the
group theoretical details such as the gauge group or representations
involved; only $\dim\phi_{a}$ is needed. Substituting the generic
tensors $U,T,D,D'$ by appropriate ones we can compute essentially
all the terms of the two loop RGEs in \cite{Martin:1993zk,Yamada:1994id}.

Consider as an example the two-loop RGEs of the gauge couplings, which
contain a term $Y^{ijk}Y_{ijk}$. Using $T=U=Y$ and ${D}_{b}^{a}={D'}_{b}^{a}=\delta_{b}^{a}$
in the master equation \eqref{eq:Susyno_Lagrangianless_masterequation}
we get 
\begin{align}
\left[Y^{ijk}Y_{ijk}\right]_{\textrm{old notation}} & =y\left(amn\lambda\right)^{\alpha\mu\nu}y\left(amn\lambda\right)_{\alpha\mu\nu}^{*}\,.
\end{align}
To fully appreciate the simplicity of this formula, if the Yukawa
couplings in the MSSM were normalized according to equation \eqref{eq:Susyno_kappaABC_normalization}
we would immediately conclude that $Y^{ijk}Y_{ijk}\overset{*}{=}\textrm{Tr}\left(Y^{u\dagger}Y^{u}+Y^{d\dagger}Y^{d}+Y^{\ell\dagger}Y^{\ell}\right)$.%
\footnote{However, note that usually the normalization in equation \eqref{eq:Susyno_kappaABC_normalization}
is not followed. Nevertheless, adaptation to other normalization schemes
is trivial.%
} As a more complex example, consider the term $Y^{acd}Y_{dmn}Y^{d'mn}Y_{bcd'}$,
which appears in the two-loop RGEs of the anomalous dimensions of
the chiral superfields (equation \eqref{eq:U1_mixing_g5SRCR}). Setting
$D_{d}^{d'}=Y_{dmn}Y^{d'mn}$ and ${D'}_{b}^{a}=\delta_{b}^{a}$ in
equation \eqref{eq:Susyno_Lagrangianless_masterequation} yields
\begin{multline}
Y^{acd}Y_{dmn}Y^{d'mn}Y_{bcd'}\\
=\frac{1}{\dim\phi_{\check{a}}\dim\phi_{d}}y\left(\check{a}cd\lambda\right)^{\alpha\xi\delta}y\left(dmn\lambda'\right)_{\delta\mu\nu}^{*}y\left(dmn\lambda'\right)^{\delta'\mu\nu}y\left(\check{a}cd\lambda\right)_{\beta\xi\delta'}^{*}\delta_{b}^{\check{a}}\delta_{B}^{A}\,.
\end{multline}

For bilinear terms the situation is similar; the entry $\mu^{ab}$
of the $\mu$ tensor is separated as (some $\mu$ parameter) $\times$
(tensor $\kappa$ with representation-component indices):
\begin{align}
\left[\mu^{ab}\right]_{\textrm{old notation}} & =\mu^{\left(aA\alpha\right)\left(bB\beta\right)}=\mu\left(ab\right)^{\alpha\beta}\kappa\left(ab\right)^{AB}\,\,\,\,\textrm{(no sums)}\,.
\end{align}
Note that we do not use here a $\lambda$ label because the product
of two representations contains at most one invariant. Interestingly,
in a model with no singlets, these $\kappa\left(ab\right)^{AB}$ numerical
tensors do not need to be normalized (unlike the trilinear ones $\kappa\left(abc\lambda\right)^{ABC}$---see
equation \eqref{eq:Susyno_kappaABC_normalization}). In other words,
as long as we use consistently the same $\kappa\left(ab\right)^{AB}$
throughout the Lagrangian, its normalization will not affect the RGEs.
As an example of this statement, using $W=\cdots+x\mu\widehat{H}_{u}\cdot\widehat{H}_{d}+\cdots$
and $-\mathscr{L}_{\textrm{soft}}=\cdots+xB\widehat{H}_{u}\cdot\widehat{H}_{d}+\cdots$
in the MSSM for any value of $x$ yields the same RGEs for all the
parameters in the model.

If there are singlets, combinations such as $Y_{amn}b^{mn}$ appearing
in the one-loop RGEs of the tensor $b^{ij}$ require that we relate
$b^{mn}$ with $Y^{amn}$ when representation $a$ is a singlet. The
following convention seems reasonable: bilinear terms involving some
representations $m$ and $n$ are obtained from the trilinear ones
with $m$, $n$ and the singlet representation $\widehat{S}$ by just
deleting the singlet field (and of course using a different parameter
name): for example, from the NMSSM's trilinear term $\left(\textrm{parameter}\right)\widehat{S}\widehat{H}_{u}\cdot\widehat{H}_{d}$
we would write a bilinear one as $\left(\textrm{parameter'}\right)\widehat{H}_{u}\cdot\widehat{H}_{d}$
instead of, for example, $-\left(\textrm{parameter'}\right)\widehat{H}_{u}\cdot\widehat{H}_{d}$
or $2\left(\textrm{parameter'}\right)\widehat{H}_{u}\cdot\widehat{H}_{d}$.
With this convention,
\begin{align}
\left[Y_{amn}b^{mn}\right]_{\textrm{old notation}} & =y\left(\boldsymbol{s}mn\right)_{\alpha\mu\nu}^{*}b\left(mn\right)^{\mu\nu}\delta_{a}^{\boldsymbol{s}}\,,
\end{align}
where $\boldsymbol{s}$ refers to the singlet representation.

For completeness, we note that linear and soft mass terms always have
a trivial representation-component structure,
\begin{align}
\left[L^{a}\right]_{\textrm{old notation}} & =L^{\left(aA\alpha\right)}=l\left(\boldsymbol{s}\right)^{\alpha}\delta_{\boldsymbol{s}}^{a}\,,\\
\left[\left(m^{2}\right)_{b}^{a}\right]_{\textrm{old notation}} & =\left(m^{2}\right)_{\left(bB\beta\right)}^{\left(aA\alpha\right)}=m^{2}\left(\check{a}\right)_{\beta}^{\alpha}\delta_{b}^{\check{a}}\delta_{B}^{A}\,,
\end{align}
so no complications arise from them.\\

In summary, it is possible to completely avoid doing sums over the
components of representations of the gauge group. In fact, the only
group theoretical quantities needed are the following:
\begin{itemize}
\item Dimension of the Lie algebras of each gauge factor group;
\item Quadratic Casimirs and dimensions of the representations under each
gauge factor group;
\item The number of invariants of the gauge group in a given product of
representations.
\end{itemize}
This last piece of information is necessary in order to compute which
are the parameters of a given model. However, as we have already mentioned
in the previous section, this discussion is complicated by the fact
that parameters can have symmetries in their flavor indices. These
symmetries appear only when a parameter multiplies repeated representations,
and its precise nature is inherited from the way these representations
combine to form a singlet. For example, in $SU(2)$, the combination
of $\boldsymbol{3}\otimes\boldsymbol{3}$ is symmetric under a permutation
of the two triplets, so introducing flavor indices $\alpha$ and $\beta$,
the $\mu$ parameter in $\mu^{\alpha\beta}\boldsymbol{3}^{\alpha}\otimes\boldsymbol{3}^{\beta}$
must be symmetric under a permutation of its indices. On the other
hand, $SU(2)$ doublets are pseudo-real representations, which means
that the singlet in $\boldsymbol{2}\otimes\boldsymbol{2}$ changes
sign if the doublets are permuted, so $\mu'$ in $\mu'^{\alpha\beta}\boldsymbol{2}^{\alpha}\otimes\boldsymbol{2}^{\beta}$
must be anti-symmetric under a permutation of its flavor indices.
Crucially, because of this antisymmetry, if there is just one flavor
such a term cannot be formed! Therefore, in order to determine the
parameters in the Lagrangian it is not enough to know if a given product
of representations contains a singlet state; it is also necessary
to known if those combinations are symmetric, antisymmetric or of
mixed symmetry. In general, the product of $n$ copies of the gauge
group $G$ breaks into irreducible representations of $G\times S_{n}$,
where $S_{n}$ is the permutation group of $n$ objects, and these
irreducible representations are sometimes known as \textbf{plethysms}.
If the $G$ invariants are calculated explicitly, as in section \ref{sec:Susyno_Invariant-combinations-of},
then their $S_{n}$ transformation properties can be accessed explicitly.
However, in a Lagrangian-free approach there is no such option. As
such, we have implemented in Mathematica a \texttt{Plethysms} function,%
\footnote{There are also other, more complex functions---\texttt{PermutationSymmetryOfInvariants}
and \texttt{PermutationSymmetryOfTensorProductParts}---which are more
practical and closer to a model builder's needs. Details can be found
on the built-in documentation.%
} following the algorithm in the manual of the \texttt{LiE} program
\cite{LieProgram_website,Leeuwen:1992aa}. Details can be found in
the built-in documentation of \texttt{Susyno} and also in reference
\cite{Leeuwen:1992aa}.\\
\\

Version 2 of \texttt{Susyno} therefore writes the RGEs without computing
basis invariant quantities, such as a Lagrangian, even though all
necessary functions to do so remain in the program. Finally, we must
mention that the canonical normalization assumed in the text above
(equation \eqref{eq:Susyno_kappaABC_normalization} in particular)
does not match the usual one in the MSSM. To make it match, the actual
normalization used by \texttt{Susyno} is
\begin{align}
\sum_{ABC}\kappa\left(abc\lambda\right)^{ABC}\kappa\left(abc\lambda'\right)_{ABC}^{*} & \equiv\sqrt{\dim\phi_{a}\dim\phi_{b}\dim\phi_{c}}\delta_{\lambda'}^{\lambda}\,,
\end{align}
and matching the program's default parameter normalization with the
user's preferred one is easy. In fact, the Lagrangian-free approach
presented here clarifies which are the exact requirements for two
notations to be equivalent, as far as the RGEs of the parameters are
concerned---see subsection \eqref{sub:Susyno_Normalization-of-the}.

\section{\label{sec:Susyno_running_times}Running time}

For reference, this section provides the time needed to run some of
\texttt{Susyno}'s functions in a computer with an Intel Core i5-2300
CPU. In particular, tables \eqref{tab:RunningTime1}, \eqref{tab:RunningTime2}
and \eqref{tab:RunningTime3} contain information on the functions
\texttt{GenerateModel}, \texttt{Invariants} and \texttt{RepMatrices},
respectively.

\begin{table}[h]
\begin{centering}
\setlength{\tabcolsep}{2.9pt}
\par\end{centering}

\begin{centering}
\begin{tabular}{cccccc}
\toprule 
Model & MSSM & RPV MSSM & NMSSM & $SO(10)$ \#1 & $SO(10)$ \#2\tabularnewline
\midrule
Time 1 (s) & 1.1 & 38 & 1.7 & 2.9 & 2.9\tabularnewline
Time 2 (s) & 1.2 & 39 & 1.8 & 3584 & 166\tabularnewline
\bottomrule
\end{tabular}
\par\end{centering}

\begin{centering}
\setlength{\tabcolsep}{6pt}
\par\end{centering}

\caption{\label{tab:RunningTime1}Running time of the command \texttt{GenerateModel}
without ({}``Time 1'') and with ({}``Time 2'') the option \texttt{CalculateEverything->True}
(output was suppressed in both cases). Five models are presented:
MSSM \cite{Martin:1993zk},	R-parity violating MSSM (RPV MSSM) \cite{Allanach:1999mh},	NMSSM
\cite{Ellwanger:2009dp}, and two $SO(10)$ models. Model {}``$SO(10)$
\#1'' contains three $\boldsymbol{16}$ multiplets together with
the representations $\boldsymbol{210}$, $\boldsymbol{126}$, $\overline{\boldsymbol{126}}$,
$\boldsymbol{10}$ (one copy each). In model {}``$SO(10)$ \#2'',
the representation $\boldsymbol{210}$ is replaced by the $\boldsymbol{45}$
plus the $\boldsymbol{54}$. Note that \texttt{GenerateModel} will
always computes the RGEs, regardless of the optional parameters; using
\texttt{CalculateEverything->True} forces the program to calculate
the Lagrangian explicitly (superpotential and soft SUSY breaking Lagrangian).
In this last case, Clebsch-Gordan coefficients must be computed, and
that is why {}``Time 2'' can be significantly larger than {}``Time
1''. Note as well that the RPV MSSM, with or without optional parameters,
takes a substantial amount of time to be computed essentially because
the model contains equal representations, $L$ and $H_{d}$, which
are to be treated differently (in other words, the running time would
be smaller if $H_{d}$ was erased and instead we considered 4 flavors
of $L$).}

\end{table}

\begin{table}[h]
\begin{centering}
\setlength{\tabcolsep}{6pt}
\par\end{centering}

\begin{centering}
\begin{tabular}{cccc}
\toprule 
Input & %
\begin{tabular}{@{}r@{}}
$\boldsymbol{3}\otimes\boldsymbol{6}\otimes\boldsymbol{10}\otimes\boldsymbol{15}\otimes\boldsymbol{15'}\otimes\boldsymbol{42}$\tabularnewline
in $SU(3)$\hspace*{16mm}\tabularnewline
\end{tabular} & %
\begin{tabular}{@{}r@{}}
$\boldsymbol{16}\otimes\boldsymbol{16}\otimes\boldsymbol{\overline{126}}$ \tabularnewline
in $SO(10)$\hspace*{4mm}\tabularnewline
\end{tabular} & %
\begin{tabular}{@{}r@{}}
$\boldsymbol{27}\otimes\boldsymbol{27}\otimes\boldsymbol{27}\otimes\boldsymbol{650}$\tabularnewline
in $E_{6}$\hspace*{12mm}\tabularnewline
\end{tabular}\tabularnewline
\midrule
Time (s) & 1574 & 2.0 & 1230\tabularnewline
\bottomrule
\end{tabular}
\par\end{centering}

\begin{centering}
\setlength{\tabcolsep}{6pt}
\par\end{centering}

\caption{\label{tab:RunningTime2}Running time of the command \texttt{Invariants}
which calculates generalized Clebsch-Gordan coefficients of arbitrary
representations of an arbitrary gauge group. Note that there are fifty
seven $SU(3)$ invariants in $\boldsymbol{3}\otimes\boldsymbol{6}\otimes\boldsymbol{10}\otimes\boldsymbol{15}\otimes\boldsymbol{15'}\otimes\boldsymbol{42}$,
1 $SO(10)$ invariant in $\boldsymbol{16}\otimes\boldsymbol{16}\otimes\boldsymbol{\overline{126}}$,
and 3 $E_{6}$ invariants in $\boldsymbol{27}\otimes\boldsymbol{27}\otimes\boldsymbol{27}\otimes\boldsymbol{650}$.}
\end{table}
\begin{table}[h]
\begin{centering}
\setlength{\tabcolsep}{6pt}
\par\end{centering}

\begin{centering}
\begin{tabular}{cccccc}
\toprule 
Representation & %
\begin{tabular}{@{}r@{}}
$\boldsymbol{100}$\hspace*{4mm}\tabularnewline
in $SU(2)$\tabularnewline
\end{tabular} & %
\begin{tabular}{@{}r@{}}
$\boldsymbol{324}$\hspace*{1mm}\tabularnewline
in $F_{4}$\tabularnewline
\end{tabular} & %
\begin{tabular}{@{}r@{}}
$\boldsymbol{1050}$\hspace*{4mm}\tabularnewline
in $SO(10)$\tabularnewline
\end{tabular} & %
\begin{tabular}{@{}r@{}}
$\boldsymbol{10560}$\hspace*{3mm}\tabularnewline
in $SO(10)$\tabularnewline
\end{tabular} & %
\begin{tabular}{@{}r@{}}
$\boldsymbol{248}$\hspace*{0.6mm}\tabularnewline
in $E_{8}$\tabularnewline
\end{tabular}\tabularnewline
\midrule
Time (s) & 0.2 & 128 & 48 & 5139 & \tabularnewline
\bottomrule
\end{tabular}
\par\end{centering}

\begin{centering}
\setlength{\tabcolsep}{6pt}
\par\end{centering}

\caption{\label{tab:RunningTime3}Running time of the command \texttt{RepMatrices}
which calculates explicitly the matrices of arbitrary representations
of an arbitrary simple gauge group.}
\end{table}

\cleartooddpage

\chapter{\label{chap:Appendix_Two_loop_RGEs}Two-loop RGEs for models with
gauge groups containing at most one $U(1)$}

In this appendix, the two-loop RGEs for a generic softly broken SUSY
model are reproduced from \cite{Martin:1993zk,Yamada:1994id,Jack:1994kd}.
However, these results do not take into account the presence of a
Fayet-Iliopoulos term (see \cite{Jack:1999zs,Jack:2000jr,Jack:2000nm})
nor the presence of non-standard soft supersymmetric breaking terms
$\phi_{i}^{*}\phi_{j}\phi_{k},\;\psi_{i}\psi_{j},\;\psi_{i}\lambda_{a}$---see
\cite{Jack:1999ud,Jack:1999fa,Goodsell:2012fm} for discussions and
analyzes of the RGEs in these cases. With multiple $U(1)$ gauge groups,
the adaptations described in chapter \ref{chap:U1_mixing_paper} must
also be carried out in order to include $U(1)$-mixing effects.

\section{\label{sect-rges-simple}Simple gauge group}

Let us recall that for a general $N=1$ supersymmetric gauge theory
with a generic superpotential 
\begin{equation}
W=\frac{1}{6}Y^{ijk}\Phi_{i}\Phi_{j}\Phi_{k}+\frac{1}{2}\mu^{ij}\Phi_{i}\Phi_{j}+L^{i}\Phi_{i}\,,
\end{equation}
the soft SUSY breaking scalar terms are written in a compact way as
\begin{align}
-\mathscr{L}_{\textrm{soft}}= & \left(\frac{1}{2}M_{a}\lambda^{a}\lambda^{a}+\frac{1}{6}h^{ijk}\phi_{i}\phi_{j}\phi_{k}+\frac{1}{2}b^{ij}\phi_{i}\phi_{j}+s^{i}\phi_{i}+\textrm{h.c.}\right)+\left(m^{2}\right)_{j}^{i}\phi_{i}\phi_{j}^{*}\,,
\end{align}
as mentioned already in chapter \ref{chap:Introduction}. Here we
will follow \cite{Martin:1993zk} and assume that repeated indices
are summed over. The anomalous dimensions of the chiral superfields
are given by 
\begin{align}
\gamma_{i}^{(1)j} & =\frac{1}{2}Y_{ipq}Y^{jpq}-2\delta_{i}^{j}g^{2}C(i)\,,\label{eq:U1_mixing_example1}\\
\gamma_{i}^{(2)j} & =g^{2}Y_{ipq}Y^{jpq}[2C(p)-C(i)]-\frac{1}{2}Y_{imn}Y^{npq}Y_{pqr}Y^{mrj}\nonumber \\
 & +2\delta_{i}^{j}g^{4}[C(i)S(R)+2C(i)^{2}-3C(G)C(i)]\,,\label{eq:U1_mixing_g5SRCR}
\end{align}
and the $\beta$-functions for the gauge couplings are given by 
\begin{align}
\beta_{g}^{(1)} & =g^{3}\left[S(R)-3C(G)\right]\,,\\
\beta_{g}^{(2)} & =g^{5}\left\{ -6[C(G)]^{2}+2C(G)S(R)+4S(R)C(R)\right\} \nonumber \\
 & -g^{3}Y^{ijk}Y_{ijk}C(k)/d(G)\thickspace.
\end{align}
The corresponding RGEs are defined as 
\begin{equation}
\frac{d}{dt}g=\frac{1}{16\pi^{2}}\beta_{g}^{(1)}+\frac{1}{(16\pi^{2})^{2}}\beta_{g}^{(2)}\,.
\end{equation}
We used in this expression \ensuremath{t=\log Q}
, where \ensuremath{Q}
 is the renormalization scale. The $\beta$-functions for the superpotential
parameters can be obtained by using the superfield technique. The
expressions obtained are
\begin{align}
\beta_{Y}^{ijk} & =Y^{ijp}\left[\frac{1}{16\pi^{2}}\gamma_{p}^{(1)k}+\frac{1}{(16\pi^{2})^{2}}\gamma_{p}^{(2)k}\right]+(k\leftrightarrow i)+(k\leftrightarrow j)\,,\\
\beta_{\mu}^{ij} & =\mu^{ip}\left[\frac{1}{16\pi^{2}}\gamma_{p}^{(1)j}+\frac{1}{(16\pi^{2})^{2}}\gamma_{p}^{(2)j}\right]+(j\leftrightarrow i)\,,\\
\beta_{L}^{i} & =L^{p}\left[\frac{1}{16\pi^{2}}\gamma_{p}^{(1)i}+\frac{1}{(16\pi^{2})^{2}}\gamma_{p}^{(2)i}\right]\,.
\end{align}
The expressions for trilinear soft breaking terms are 
\begin{align}
\frac{d}{dt}h^{ijk} & =\frac{1}{16\pi^{2}}\left[\beta_{h}^{(1)}\right]^{ijk}+\frac{1}{(16\pi^{2})^{2}}\left[\beta_{h}^{(2)}\right]^{ijk}\,,
\end{align}
 with 
\begin{align}
\left[\beta_{h}^{(1)}\right]^{ijk} & =\frac{1}{2}h^{ijl}Y_{lmn}Y^{mnk}+Y^{ijl}Y_{lmn}h^{mnk}\nonumber \\
 & -2\left(h^{ijk}-2MY^{ijk}\right)g^{2}C(k)+(k\leftrightarrow i)+(k\leftrightarrow j)\,,\label{eq:U1_mixing_example4}\\
\left[\beta_{h}^{(2)}\right]^{ijk} & =-\frac{1}{2}h^{ijl}Y_{lmn}Y^{npq}Y_{pqr}Y^{mrk}\nonumber \\
 & -Y^{ijl}Y_{lmn}Y^{npq}Y_{pqr}h^{mrk}-Y^{ijl}Y_{lmn}h^{npq}Y_{pqr}Y^{mrk}\nonumber \\
 & +\left(h^{ijl}Y_{lpq}Y^{pqk}+2Y^{ijl}Y_{lpq}h^{pqk}-2MY^{ijl}Y_{lpq}Y^{pqk}\right)g^{2}\left[2C(p)-C(k)\right]\nonumber \\
 & +\left(2h^{ijk}-8MY^{ijk}\right)g^{4}\left[C(k)S(R)+2C(k)^{2}-3C(G)C(k)\right]\nonumber \\
 & +(k\leftrightarrow i)+(k\leftrightarrow j)\,.
\end{align}
For the bilinear soft-breaking parameters, the expressions read 
\begin{align}
\frac{d}{dt}b^{ij} & =\frac{1}{16\pi^{2}}\left[\beta_{b}^{(1)}\right]^{ij}+\frac{1}{(16\pi^{2})^{2}}\left[\beta_{b}^{(2)}\right]^{ij}\,,\label{example2}
\end{align}
with 
\begin{align}
\left[\beta_{b}^{(1)}\right]^{ij} & =\frac{1}{2}b^{il}Y_{lmn}Y^{mnj}+\frac{1}{2}Y^{ijl}Y_{lmn}b^{mn}+\mu^{il}Y_{lmn}h^{mnj}\nonumber \\
 & -2\left(b^{ij}-2M\mu^{ij}\right)g^{2}C(i)+(i\leftrightarrow j)\,,\\
\left[\beta_{b}^{(2)}\right]^{ij} & =-\frac{1}{2}b^{il}Y_{lmn}Y^{pqn}Y_{pqr}Y^{mrj}-\frac{1}{2}Y^{ijl}Y_{lmn}\mu^{mr}Y_{pqr}h^{pqn}\nonumber \\
 & -\mu^{il}Y_{lmn}h^{npq}Y_{pqr}Y^{mrj}-\mu^{il}Y_{lmn}Y^{npq}Y_{pqr}h^{mrj}\nonumber \\
 & -\frac{1}{2}Y^{ijl}Y_{lmn}b^{mr}Y_{pqr}Y^{pqn}+2Y^{ijl}Y_{lpq}\left(b^{pq}-\mu^{pq}M\right)g^{2}C(p)\nonumber \\
 & +\left(b^{il}Y_{lpq}Y^{pqj}+2\mu^{il}Y_{lpq}h^{pqj}-2\mu^{il}Y_{lpq}Y^{pqj}M\right)g^{2}\left[2C(p)-C(i)\right]\nonumber \\
 & +\left(2b^{ij}-8\mu^{ij}M\right)g^{4}\left[C(i)S(R)+2C(i)^{2}-3C(G)C(i)\right]+(i\leftrightarrow j)\,.
\end{align}
The RGEs for the linear soft-breaking parameters are 
\begin{align}
\frac{d}{dt}s^{i} & =\frac{1}{16\pi^{2}}\left[\beta_{s}^{(1)}\right]^{i}+\frac{1}{(16\pi^{2})^{2}}\left[\beta_{s}^{(2)}\right]^{i}\,,
\end{align}
with{\allowdisplaybreaks 
\begin{align}
\left[\beta_{s}^{(1)}\right]^{i} & =\frac{1}{2}Y^{iln}Y_{pln}s^{p}+L^{p}Y_{pln}h^{iln}+\mu^{ik}Y_{kln}b^{ln}+2Y^{ikp}\left(m^{2}\right)_{p}^{l}\mu_{kl}+h^{ikl}b_{kl}\,,\\
\left[\beta_{s}^{(2)}\right]^{i} & =2g^{2}C(l)Y^{ikl}Y_{pkl}s^{p}-\frac{1}{2}Y^{ikq}Y_{qst}Y^{lst}Y_{pkl}s^{p}-4g^{2}C(l)Y_{jnl}\hspace{-1mm}\left(\mu^{nl}M-b^{nl}\right)\hspace{-1mm}\mu^{ij}\nonumber \\
 & -\left[Y^{ikq}Y_{qst}h^{lst}Y_{pkl}+h^{ikq}Y_{qst}Y^{lst}Y_{pkl}\right]L^{p}-4g^{2}C(l)\left(Y^{ikl}M-h^{ikl}\right)Y_{pkl}L^{p}\nonumber \\
 & -\left[Y_{jnq}h^{qst}Y_{lst}\mu^{nl}+Y_{jnq}Y^{qst}Y_{lst}b^{nl}\right]\mu^{ij}+4g^{2}C(l)\left[\vphantom{\left(m^{2}\right)_{p}^{l}}2Y^{ikl}\mu_{kl}|M|^{2}\right.\nonumber \\
 & -\left.Y^{ikl}b_{kl}M-h^{ikl}\mu_{kl}M^{*}+h^{ikl}b_{kl}+Y^{ipl}\left(m^{2}\right)_{p}^{k}\mu_{kl}+Y^{ikp}\left(m^{2}\right)_{p}^{l}\mu_{kl}\right]\nonumber \\
 & -\left[Y^{ikq}Y_{qst}h^{lst}b_{kl}+h^{ikq}Y_{qst}Y^{lst}b_{kl}+h^{ikq}h_{qst}Y^{lst}\mu_{kl}+Y^{ikq}h_{qst}h^{lst}\mu_{kl}\right.\nonumber \\
 & +Y^{ipq}\left(m^{2}\right)_{p}^{k}Y_{qst}Y^{lst}\mu_{kl}+Y^{ikq}Y_{qst}Y^{pst}\left(m^{2}\right)_{p}^{l}\mu_{kl}\nonumber \\
 & +\left.Y^{ikp}\left(m^{2}\right)_{p}^{q}Y_{qst}Y^{lst}\mu_{kl}+2Y^{ikq}Y_{qsp}\left(m^{2}\right)_{t}^{p}Y^{lst}\mu_{kl}\right]\,.
\end{align}
}With these results, the list of the $\beta$-functions for all couplings
is complete. Now, we consider the RGEs for the gaugino masses and
squared masses of scalars. The result for the gaugino masses is 
\begin{align}
\frac{d}{dt}M= & \frac{1}{16\pi^{2}}\beta_{M}^{(1)}+\frac{1}{(16\pi^{2})^{2}}\beta_{M}^{(2)}\,,
\end{align}
with 
\begin{align}
\beta_{M}^{(1)}= & g^{2}\left[2S(R)-6C(G)\right]M\,,\label{eq:U1_mixing_example5}\\
\beta_{M}^{(2)}= & g^{4}\left[-24C(G)^{2}+8C(G)S(R)+16S(R)C(R)\right]M\nonumber \\
 & +2g^{2}\left(h^{ijk}-MY^{ijk}\right)Y_{ijk}C(k)/d(G)\,.
\end{align}
The one- and two-loop RGEs for the scalar mass parameters read 
\begin{align}
\frac{d}{dt}\left(m^{2}\right)_{i}^{j}= & \frac{1}{16\pi^{2}}\left[\beta_{m^{2}}^{(1)}\right]_{i}^{j}+\frac{1}{(16\pi^{2})^{2}}\left[\beta_{m^{2}}^{(2)}\right]_{i}^{j}\,,
\end{align}
with {\allowdisplaybreaks 
\begin{align}
\left[\beta_{m^{2}}^{(1)}\right]_{i}^{j} & =\frac{1}{2}Y_{ipq}Y^{pqn}{(m^{2})}_{n}^{j}+\frac{1}{2}Y^{jpq}Y_{pqn}{(m^{2})}_{i}^{n}+2Y_{ipq}Y^{jpr}{(m^{2})}_{r}^{q}+h_{ipq}h^{jpq}\nonumber \\
 & -8\delta_{i}^{j}\left|M\right|^{2}g^{2}C(i)+2g^{2}{\bf t}_{i}^{Aj}{\rm Tr}[{\bf t}^{A}m^{2}]\,,\\
\left[\beta_{m^{2}}^{(2)}\right]_{i}^{j} & =-\frac{1}{2}{(m^{2})}_{i}^{l}Y_{lmn}Y^{mrj}Y_{pqr}Y^{pqn}-\frac{1}{2}{(m^{2})}_{l}^{j}Y^{lmn}Y_{mri}Y^{pqr}Y_{pqn}\nonumber \\
 & -h_{ilm}Y^{jln}Y_{npq}h^{mpq}-Y_{ilm}Y^{jnm}{(m^{2})}_{n}^{r}Y_{rpq}Y^{lpq}-Y_{ilm}Y^{jnr}{(m^{2})}_{n}^{l}Y_{pqr}Y^{pqm}\nonumber \\
 & -Y_{ilm}Y^{jln}h_{npq}h^{mpq}-2Y_{ilm}Y^{jln}Y_{npq}Y^{mpr}{(m^{2})}_{r}^{q}-h_{ilm}h^{jln}Y_{npq}Y^{mpq}\nonumber \\
 & -Y_{ilm}Y^{jnm}{(m^{2})}_{r}^{l}Y_{npq}Y^{rpq}-Y_{ilm}h^{jln}h_{npq}Y^{mpq}\nonumber \\
 & +\biggl[{(m^{2})}_{i}^{l}Y_{lpq}Y^{jpq}+Y_{ipq}Y^{lpq}{(m^{2})}_{l}^{j}+4Y_{ipq}Y^{jpl}{(m^{2})}_{l}^{q}+2h_{ipq}h^{jpq}\nonumber \\
 & -2h_{ipq}Y^{jpq}M-2Y_{ipq}h^{jpq}M^{*}+4Y_{ipq}Y^{jpq}\left|M\right|^{2}\biggr]g^{2}\left[C(p)+C(q)-C(i)\right]\nonumber \\
 & -2g^{2}{\bf t}_{i}^{Aj}\left({\bf t}^{A}m^{2}\right)_{r}^{l}Y_{lpq}Y^{rpq}+8g^{4}{\bf t}_{i}^{Aj}{\rm Tr}\left[{\bf t}^{A}C(r)m^{2}\right]\nonumber \\
 & +\delta_{i}^{j}g^{4}\left|M\right|^{2}\Big[24C(i)S(R)+48C(i)^{2}-72C(G)C(i)\Big]\nonumber \\
 & +8\delta_{i}^{j}g^{4}C(i)\left\{ {\rm Tr}[S(r)m^{2}]-C(G)\left|M\right|^{2}\right\} \,.\label{eq:U1_mixing_example3}
\end{align}
}Finally, partial expressions for the RGEs for a VEV \ensuremath{v^{i}}
 can be found in \cite{Sperling:2013eva}.\\

\noindent A few comments and clarifications concerning the variables
appearing in the different $\beta$-functions are necessary:
\begin{itemize}
\item $Y_{ijk}=\left(Y^{ijk}\right)^{*}$, $h_{ijk}=\left(h^{ijk}\right)^{*}$,
$\mu_{ij}=\left(\mu^{ij}\right)^{*}$ and $b_{ij}=\left(b^{ij}\right)^{*}$.
\item $d\left(G\right)$ = Dimension of the adjoint representation of group
$G$.
\item $C\left(i\right)$ = Quadratic Casimir invariant of the representation
of the chiral superfield with index $i$.
\item $C\left(G\right)$ = Quadratic Casimir invariant of the adjoint representation
of group $G$.
\item $S\left(R\right)$ = Dynkin index summed over all chiral multiplets.
However $S\left(R\right)C\left(R\right)$ should be interpreted as
the sum of Dynkin indices weighted by the quadratic Casimir invariant.
\item $\boldsymbol{t^{A}}$ = Representation matrices under the gauge group
$G$. Terms with $\boldsymbol{t^{A}}$ are only relevant for $U(1)$
groups.
\item In $\beta_{m^{2}}^{\left(1\right)}$ and $\beta_{m^{2}}^{\left(2\right)}$,
the traces should be understood as traces over all chiral superfields. 
\end{itemize}

\section{Product groups with at most one $U(1)$\label{sect-rges-products}}

To generalize the formulae above to the case of a direct product of
gauge groups, the following substitution rules are needed \cite{Martin:1993zk}.
As long as there is at most one $U(1)$ gauge group, this procedure
will yield the correct results; otherwise the rules must be generalized,
as discussed in section \ref{sect-results}. \allowdisplaybreaks{
\begin{align}
g^{3}C(G) & \rightarrow g_{a}^{3}C(G_{a})\,,\label{eq:ReplacementRule1}\\
g^{3}S(R) & \rightarrow g_{a}^{3}S_{a}(R)\,,\\
g^{5}C(G)^{2} & \rightarrow g_{a}^{5}C(G_{a})^{2}\,,\\
g^{5}C(G)S(R) & \rightarrow g_{a}^{5}C(G_{a})S_{a}(R)\,,\\
g^{5}S(R)C(R) & \rightarrow\sum_{b}g_{a}^{3}g_{b}^{2}S_{a}(R)C_{b}(R)\,,\label{eq:U1_mixing_g5SRCRsubst}\\
g^{3}C(k)/d(G) & \rightarrow g_{a}^{3}C_{a}(k)/d(G_{a})\,,\\
Mg^{2}C(G) & \rightarrow M_{a}g_{a}^{2}C(G_{a})\,,\\
Mg^{2}S(R) & \rightarrow M_{a}g_{a}^{2}S_{a}(R)\,,\\
Mg^{4}C(G)^{2} & \rightarrow M_{a}g_{a}^{4}C(G_{a})^{2}\,,\\
Mg^{4}C(G)S(R) & \rightarrow M_{a}g_{a}^{4}C(G_{a})S_{a}(R)\,,\\
16Mg^{4}S(R)C(R) & \rightarrow8\sum_{b}\left(M_{a}+M_{b}\right)g_{a}^{2}g_{b}^{2}S_{a}(R)C_{b}(R)\,,\\
Mg^{2}C(k)/d(G) & \rightarrow M_{a}g_{a}^{2}C_{a}(k)/d(G_{a})\,,\label{eq:ReplacementRule2}\\
g^{2}C(r) & \rightarrow\sum_{a}g_{a}^{2}C_{a}(r)\,,\label{eq:U1_mixing_example1productX}\\
Mg^{2}C(r) & \rightarrow\sum_{a}M_{a}g_{a}^{2}C_{a}(r)\,,\\
M^{*}g^{2}C(r) & \rightarrow\sum_{a}M_{a}^{*}g_{a}^{2}C_{a}(r)\,,\\
\left|M\right|^{2}g^{2}C(r) & \rightarrow\sum_{a}\left|M_{a}\right|^{2}g_{a}^{2}C_{a}(r)\,,\\
g^{4}C(r)S(R) & \rightarrow\sum_{a}g_{a}^{4}C_{a}(r)S_{a}(R)\,,\\
Mg^{4}C(r)S(R) & \rightarrow\sum_{a}M_{a}g_{a}^{4}C_{a}(r)S_{a}(R)\,,\\
g^{4}C(r)^{2} & \rightarrow\sum_{a}\sum_{b}g_{a}^{2}g_{b}^{2}C_{a}(r)C_{b}(r)\,,\\
Mg^{4}C(r)^{2} & \rightarrow\sum_{a}\sum_{b}M_{a}g_{a}^{2}g_{b}^{2}C_{a}(r)C_{b}(r)\,,\\
g^{4}C(r)C(G) & \rightarrow\sum_{a}g_{a}^{4}C_{a}(r)C(G_{a})\,,\\
Mg^{4}C(r)C(G) & \rightarrow\sum_{a}M_{a}g_{a}^{4}C_{a}(r)C(G_{a})\,,\\
\left|M\right|^{2}g^{4}C(r)C(G) & \rightarrow\sum_{a}\left|M_{a}\right|^{2}g_{a}^{4}C_{a}(r)C(G_{a})\,,\\
48\left|M\right|^{2}g^{4}C(i)^{2} & \rightarrow\sum_{a}\sum_{b}\left(32M_{a}+16M_{b}\right)M_{a}^{*}g_{a}^{2}g_{b}^{2}C_{a}(i)C_{b}(i)\,,\\
\left|M\right|^{2}g^{4}C(i)S(R) & \rightarrow\sum_{a}\left|M_{a}\right|^{2}g_{a}^{4}C_{a}(i)S_{a}(R)\,,\\
g^{2}{\bf t}_{i}^{Aj}\mbox{Tr}\left({\bf t}^{A}m^{2}\right) & \rightarrow\sum_{a}g_{a}^{2}\left({\bf t}_{a}^{A}\right)_{i}^{j}\mbox{Tr}\left({\bf t}_{a}^{A}m^{2}\right)\,,\\
g^{2}{\bf t}_{i}^{Aj}\left({\bf t}^{A}m^{2}\right)_{r}^{l}Y_{lpq}Y^{rpq} & \rightarrow\sum_{a}g_{a}^{2}\left({\bf t}_{a}^{A}\right)_{i}^{j}\left({\bf t}_{a}^{A}m^{2}\right)_{r}^{l}Y_{lpq}Y^{rpq}\,,\\
g^{4}{\bf t}_{i}^{Aj}\mbox{Tr}\left[{\bf t}^{A}C(r)m^{2}\right] & \rightarrow\sum_{a}\sum_{b}g_{a}^{2}g_{b}^{2}\left({\bf t}_{a}^{A}\right)_{i}^{j}\mbox{Tr}\left[{\bf t}_{a}^{A}C_{b}(r)m^{2}\right]\,,\\
g^{4}C(i)\mbox{Tr}\left[S(r)m^{2}\right] & \rightarrow\sum_{a}g_{a}^{4}C_{a}(i)\mbox{Tr}\left[S_{a}(r)m^{2}\right]\,.
\end{align}
}When there is an index $a$ not summed over (equations \eqref{eq:ReplacementRule1}--\eqref{eq:ReplacementRule2}),
this means that the corresponding rule is for the RGEs of a gauge
coupling $g_{a}$ or a gaugino mass $M_{a}$.
\cleartooddpage

\chapter{\label{chap:Matching_conditions}Matching conditions for gauge couplings
and gaugino masses in models with $U(1)$-mixing}

As a complement to chapter \ref{chap:U1_mixing_paper}, in this appendix
we discuss how to proceed with the matching of gauge couplings and
gaugino masses in models with a phase where the gauge group is $U(1)^{n}$
which spontaneously breaks down into $U(1)^{m}$ with $m<n$ \cite{Fonseca:2011vn}.
We shall not deal with non-abelian factors since it is straightforward
to take them into account. Note also that this is a one-loop discussion---at
higher orders the situation becomes more involved, as there are non-trivial
threshold effects to be considered \cite{Weinberg:1980wa,Hall:1980kf},
yielding for example extra factors associated to non-abelian gauge
groups, entering formulas such as equation \eqref{eq:gauge_matching_condition_general}
below---see for instance \cite{Bertolini:2009qj}. The specific shape
of these terms is, however, renormalization scheme dependent.

\section{Gauge couplings}

Here we assume the same setting as the one described in chapter \ref{chap:U1_mixing_paper},
namely that there are scalar fields $\phi_{i}$ (we may ignore their
fermionic partners) with charge vectors $\boldsymbol{Q}_{\boldsymbol{i}}$,
such that the covariant derivative is written as
\begin{alignat}{1}
D_{\mu}\phi_{i} & =\left(\partial_{\mu}-i\boldsymbol{Q}_{\boldsymbol{i}}^{T}\boldsymbol{G}\boldsymbol{A_{\mu}}\right)\phi_{i}\,.\label{eq:covariant_derivative_U1_mixing}
\end{alignat}
Therefore, if the scalar fields acquire a VEV, there will be a gauge
boson mass term
\begin{alignat}{1}
\mathscr{L}_{A} & =\frac{1}{2}\boldsymbol{A}_{\boldsymbol{\mu}}^{T}\boldsymbol{M_{A}^{2}}\boldsymbol{A_{\mu}},\qquad M_{A}^{2}=\sum_{i}2\left|\left\langle \phi_{i}\right\rangle \right|^{2}\boldsymbol{G}^{T}\boldsymbol{Q}_{\boldsymbol{i}}\boldsymbol{Q}_{\boldsymbol{i}}^{T}\boldsymbol{G}\,.
\end{alignat}
Some of the $U(1)$ groups will be broken by these VEVs but in principle
there will be $m$ linearly independent combinations of the $n$ original
$U(1)$'s that remain unbroken. Thus, we should do a $\mathcal{O}_{1}$
rotation in $U(1)$ space such that the first $m$ rotated $U(1)$'s
are the unbroken ones. By doing this, for all $i$ we find new charges
\begin{alignat}{1}
\boldsymbol{Q}'_{i} & \equiv\mathcal{O}_{1}\boldsymbol{Q}_{i}\,,\label{eq:Qprime_definition}
\end{alignat}
and by construction
\begin{alignat}{1}
\forall_{i}\;{\boldsymbol{Q'_{i}}}^{j}\left\langle \phi_{i}\right\rangle  & =0\textrm{ for }j=1,\cdots,m\,,\,\label{eq:matching1}\\
\exists_{i}\;{\boldsymbol{Q'_{i}}}^{j}\left\langle \phi_{i}\right\rangle  & \neq0\textrm{ for }j=m+1,\cdots,n\,.\label{eq:matching2}
\end{alignat}
The gauge boson mass matrix is then of the following form:
\begin{alignat}{1}
\boldsymbol{M_{A}^{2}} & =\boldsymbol{G}^{T}\mathcal{O}_{1}^{T}\begin{pmatrix}\boldsymbol{0} & \boldsymbol{0}\\
\boldsymbol{0} & \boldsymbol{X}
\end{pmatrix}\mathcal{O}_{1}\boldsymbol{G}\,,
\end{alignat}
where $\boldsymbol{X}$ is some non-null $\left(n-m\right)\times\left(n-m\right)$
matrix. $\boldsymbol{M_{A}^{2}}$ will not be block diagonal unless
we also rotate the gauge boson states in such a way that the first
$m$ rotated fields are massless,
\begin{alignat}{1}
\boldsymbol{A'_{\mu}} & \equiv\mathcal{O}_{2}\boldsymbol{A_{\mu}}\,.
\end{alignat}
This gives rise to a new matrix of gauge couplings 
\begin{alignat}{1}
\boldsymbol{G'} & \equiv\mathcal{O}_{1}\boldsymbol{G}\mathcal{O}_{2}^{T}\,,
\end{alignat}
and, in this new basis, we have
\begin{alignat}{1}
{\boldsymbol{M}}_{A'}^{2} & ={\boldsymbol{G'}}^{T}\begin{pmatrix}\boldsymbol{0} & \boldsymbol{0}\\
\boldsymbol{0} & \boldsymbol{X}
\end{pmatrix}\boldsymbol{G'}\equiv\begin{pmatrix}\boldsymbol{0} & \boldsymbol{0}\\
\boldsymbol{0} & \boldsymbol{X'}
\end{pmatrix}\label{eq:GaugeBosonsMassesPrime}
\end{alignat}
for some $\boldsymbol{X'}$. Note that the last equality is true by
construction: the first $m$ gauge bosons are the massless ones, while
the remaining ones are not (an additional rotation might be needed
to bring $\boldsymbol{X'}$ into a diagonal form). Equation \eqref{eq:GaugeBosonsMassesPrime}
will only hold if 
\begin{alignat}{1}
\boldsymbol{G'} & =\begin{pmatrix}\boldsymbol{G'_{SS}} & \boldsymbol{G'_{SB}}\\
\boldsymbol{0} & \boldsymbol{G'_{BB}}
\end{pmatrix}\,.\label{eq:matching3}
\end{alignat}
The interpretation of this $\boldsymbol{0}$ in the $\left(2,1\right)$
block is that the massless gauge bosons will only interact with the
matter fields via the charges ${\boldsymbol{Q}\boldsymbol{'_{i}}}^{j}$
with $j=1,\cdots,m$ as we would expect. To see this we only have
to consider the covariant derivative in the rotated basis, which amounts
to adding primes to equation \eqref{eq:covariant_derivative_U1_mixing}:
\begin{alignat}{1}
D_{\mu}\phi_{i} & =\partial_{\mu}\phi_{i}-i{\boldsymbol{Q}\boldsymbol{'_{i}}}^{T}\boldsymbol{G'}\boldsymbol{A'_{\mu}}\phi_{i}\nonumber \\
 & =\partial_{\mu}\phi_{i}-i\sum_{a,b=1}^{m}{\boldsymbol{Q}\boldsymbol{'}_{\boldsymbol{i}}}^{a}\left(\boldsymbol{G'_{SS}}\right)_{ab}{\boldsymbol{A'_{\mu}}}^{b}\phi_{i}+\left(\textrm{interactions with heavy }{\boldsymbol{A'_{\mu}}}^{b}\textrm{s}\right)\,.
\end{alignat}
Finally, notice that we are still free to rotate the light gauge bosons
${\boldsymbol{A'_{\mu}}}^{1,\cdots,m}$ among themselves, so we cannot
predict completely $\boldsymbol{G'_{SS}}$ since it is basis dependent.
The solution to this problem is nonetheless clear: we should focus
on $\boldsymbol{G'_{SS}}{\boldsymbol{G'_{SS}}}^{T}$ instead. Unfortunately,
the $\left(1,1\right)$ block of $\boldsymbol{G'}{\boldsymbol{G'}}^{T}$
is equal to the combination $\boldsymbol{G'_{SS}}{\boldsymbol{G'_{SS}}}^{T}+\boldsymbol{G'_{SB}}{\boldsymbol{G'_{SB}}}^{T}$
so it cannot be used to extract directly this quantity, but on the
other hand, the $\left(1,1\right)$ block of $\left(\boldsymbol{G'}{\boldsymbol{G'}}^{T}\right)^{-1}$
is suitable for this:
\begin{align}
\left[\left(\mathcal{O}_{1}\boldsymbol{G}\boldsymbol{G}^{T}\mathcal{O}_{1}^{T}\right)^{-1}\right]_{\textrm{block (1,1)}} & =\left[\left(\boldsymbol{G'}{\boldsymbol{G'}}^{T}\right)^{-1}\right]_{\textrm{block (1,1)}}=\left(\boldsymbol{G'_{SS}}{\boldsymbol{G'_{SS}}}^{T}\right)^{-1}\,.\label{eq:gauge_matching_condition_general}
\end{align}
As an example, consider the $U(1)_{R}\times U(1)_{B-L}$ symmetry
of class-III models discussed in chapter \ref{chap:SlidingScale_models},
which breaks down into $U(1)_{Y}$. Before SSB there is a $2\times2$
$U(1)$ gauge couplings matrix
\begin{align}
\boldsymbol{G} & =\begin{pmatrix}g_{RR} & g_{RX}\\
g_{XR} & g_{XX}
\end{pmatrix}\,,
\end{align}
where $X\equiv\sqrt{\nicefrac{3}{5}}\left(B-L\right)$ refers to the
properly normalized $B-L$ abelian group. The hypercharge under $U(1)_{Y}$
is given by the relation $\sqrt{\frac{3}{5}}${[}$U(1)_{R}$ hypercharge{]}
+ $\sqrt{\frac{2}{5}}${[}$U(1)_{B-L}$ hypercharge{]} (see appendix
\ref{chap:Lists_of_superfields_in_LR_models}), so according to equation
\eqref{eq:gauge_matching_condition_general} the gauge coupling matching
condition is
\begin{align}
g_{Y}^{-2} & =\begin{pmatrix}\sqrt{\frac{3}{5}} & \sqrt{\frac{2}{5}}\end{pmatrix}\left(\boldsymbol{G}\boldsymbol{G}^{T}\right)^{-1}\begin{pmatrix}\sqrt{\frac{3}{5}}\\
\sqrt{\frac{2}{5}}
\end{pmatrix}\nonumber \\
 & =\frac{2\left(g_{RR}^{2}+g_{RX}^{2}\right)+3\left(g_{XX}^{2}+g_{XR}^{2}\right)-2\sqrt{6}\left(g_{RR}g_{XR}+g_{XX}g_{RX}\right)}{5\left(g_{RR}g_{XX}-g_{RX}g_{XR}\right)^{2}}\,.\label{eq:gauge_matching_RX}
\end{align}
Only in the limit $g_{RX},\, g_{XR}\rightarrow0$ do we recover the
simplified relation $g_{Y}^{-2}=\nicefrac{3}{5}g_{RR}^{-2}+\nicefrac{2}{5}g_{XX}^{-2}$.

\section{Gaugino masses}

Gauginos interact with the fermionic and scalar components of chiral
superfields, so when the latter acquire VEVs, a mass term is generated,
mixing fermions from chiral and vector superfields. In the rotated,
primed basis we have
\begin{align}
\mathscr{L} & =\begin{pmatrix}\lambda & \psi_{j}\end{pmatrix}\begin{pmatrix}\boldsymbol{M'} & \sqrt{2}\boldsymbol{G}\boldsymbol{'}^{T}\boldsymbol{Q'_{i}}\left\langle \phi_{i}\right\rangle \\
\sqrt{2}{\boldsymbol{Q'_{j}}}^{T}\boldsymbol{G'}\left\langle \phi_{j}\right\rangle  & \frac{\delta^{2}W}{\delta\Phi_{i}\delta\Phi_{j}}
\end{pmatrix}\begin{pmatrix}\lambda\\
\psi_{i}
\end{pmatrix}+\textrm{h.c.}+\cdots\,,
\end{align}
where $\boldsymbol{M'}=\mathcal{O}_{2}\boldsymbol{M}\mathcal{O}_{2}^{T}$
is the rotated gaugino soft mass matrix and $W$ refers to the superpotential.
Now notice that together, equations \eqref{eq:matching1}, \eqref{eq:matching2}
and \eqref{eq:matching3} imply that in block notation $\boldsymbol{G}\boldsymbol{'}^{T}\boldsymbol{Q'_{i}}\left\langle \phi_{i}\right\rangle =\begin{pmatrix}\boldsymbol{0} & \left[\cdots\right]\end{pmatrix}^{T}$.
In other words, the first $m$ rows and columns of the matrix in the
previous equation, which correspond to the $\lambda$'s associated
with the massless gauge bosons, only receive a non-null contribution
from the $\boldsymbol{M'}$ matrix itself. As such, if we define $\boldsymbol{M'}$
in blocks, 
\begin{align}
\boldsymbol{M'} & \equiv\begin{pmatrix}\boldsymbol{M'_{SS}} & \boldsymbol{M'_{SB}}\\
{\boldsymbol{M'}}_{\boldsymbol{SB}}^{T} & \boldsymbol{M'_{BB}}
\end{pmatrix}\,,
\end{align}
the gaugino mass matrix in the broken phase $U(1)^{m}$ is given simply
by the $m\times m$ block $\boldsymbol{M'_{SS}}$. Analogously to
what happens to $\boldsymbol{G}$, here too we have a basis problem:
firstly, we need the rotation matrix $\mathcal{O}_{2}$ in order to
define $\boldsymbol{M'}$; and secondly, we will always retain the
freedom to perform rotations between the massless gauge bosons, which
means that the exact form of the $\left(1,1\right)$ block $\boldsymbol{M'_{SS}}$
of the full gaugino mass matrix $\boldsymbol{M'}$ will always be
basis dependent. Here the sensible solution is to consider a combination
of $\boldsymbol{M'}$ and $\boldsymbol{G'}$ which is $\mathcal{O}_{2}$
invariant and from which it is easy to extract the $\left(1,1\right)$
block $\boldsymbol{M'_{SS}}$. The best solution seems to be the following:
\begin{align}
\left[\mathcal{O}_{1}{\boldsymbol{G}}^{-1\, T}\boldsymbol{M}{\boldsymbol{G}}^{-1}\mathcal{O}_{1}^{T}\right]_{\textrm{block (1,1)}} & =\left[{\boldsymbol{G'}}^{-1\, T}\boldsymbol{M'}{\boldsymbol{G'}}^{-1}\right]_{\textrm{block (1,1)}}\nonumber \\
 & ={\boldsymbol{G'_{SS}}}^{-1\, T}\boldsymbol{M'_{SS}}{\boldsymbol{G'_{SS}}}^{-1}\,.\label{eq:gaugino_matching_condition_general}
\end{align}
We may consider the simple case $U(1)_{R}\times U(1)_{B-L}\rightarrow U(1)_{Y}$
once more. In the $U(1)$-mixing phase, the gaugino mass matrix is
$2\times2$,
\begin{align}
\boldsymbol{M} & =\begin{pmatrix}M_{RR} & M_{RX}\\
M_{RX} & M_{XX}
\end{pmatrix}\,,
\end{align}
and at the matching scale we have
\begin{alignat}{1}
\frac{M_{Y}}{g_{Y}^{2}} & =\frac{1}{5\left(g_{RR}g_{XX}-g_{RX}g_{XR}\right)^{2}}\left[\left(3g_{XX}^{2}+2g_{RX}^{2}-2\sqrt{6}g_{XX}g_{RX}\right)M_{RR}\right.\nonumber \\
 & +\left(2g_{RR}^{2}+3g_{XR}^{2}-2\sqrt{6}g_{RR}g_{XR}\right)M_{XX}\nonumber \\
 & +\left.\left(-4g_{RR}g_{RX}-6g_{XX}g_{XR}+2\sqrt{6}g_{RR}g_{XX}+2\sqrt{6}g_{RX}g_{XR}\right)M_{RX}\right]\,.
\end{alignat}
In the absence of mixing, this expression becomes $\nicefrac{M_{Y}}{g_{Y}^{2}}=\nicefrac{3}{5}\nicefrac{M_{RR}}{g_{RR}^{2}}+\nicefrac{2}{5}\nicefrac{M_{XX}}{g_{XX}^{2}}$
or $M_{Y}=\nicefrac{\left(3g_{XX}^{2}M_{RR}+2g_{RR}^{2}M_{XX}\right)}{\left(3g_{XX}^{2}+2g_{RR}^{2}\right)}$
if we eliminate $g_{Y}$ from the equation.
\cleartooddpage

\chapter{\label{chap:Lists_of_superfields_in_LR_models}Lists of superfields
in Left-Right models}

In chapter \ref{chap:SlidingScale_models} we have considered $SO(10)$-inspired
models which may contain any irreducible representation up to dimension
126 ($\boldsymbol{1}$, $\boldsymbol{10}$, $\boldsymbol{16}$, $\overline{\boldsymbol{16}}$,
$\boldsymbol{45}$, $\boldsymbol{54}$, $\boldsymbol{120}$, $\boldsymbol{126}$,
$\overline{\mathbf{126}}$) \cite{Arbelaez:2013hr}. Once the gauge
group breaks down to $SU(4)\times SU(2)_{L}\times SU(2)_{R}$ or $SU(3)_{C}\times SU(2)_{L}\times SU(2)_{R}\times U(1)_{B-L}$
these $SO(10)$ fields divide into a multitude of different irreducible
representations of these groups. In addition, if $SU(2)_{R}$ is further
broken down further to $U(1)_{R}$, the following branching rules
apply: $\boldsymbol{3}\rightarrow-1,0,+1$; $\boldsymbol{2}\rightarrow\pm\frac{1}{2}$;
$\boldsymbol{1}\rightarrow0$. The Standard Model's hypercharge, in
the canonical normalization, is then equal to the combination $\sqrt{\frac{3}{5}}${[}$U(1)_{R}$
hypercharge{]} + $\sqrt{\frac{2}{5}}${[}$U(1)_{B-L}$ hypercharge{]}.
In tables \eqref{tab:List_of_LR_fields}, \eqref{tab:List_of_PatiSalam_fields}
and \eqref{tab:List_of_LR_fields_U1} we present the list of relevant
fields respecting the conditions above. In these tables we used an
ordered naming of the fields but note that in chapter \ref{chap:SlidingScale_models}
we favor another, less compact, notation where the quantum numbers
under the various groups are indicated explicitly (see for instance
table \eqref{tab:SlidingScale_LR_field_configuration}).

\begin{table}[h]
\begin{centering}
\setlength{\tabcolsep}{2pt}
\par\end{centering}

\begin{centering}
\scalebox{1.0}{
\setlength{\tabcolsep}{3pt}%
\begin{tabular}{c}
\begin{tabular}{ccccccccccccccc}
\toprule 
 & $\Phi_{1}$  & $\Phi_{2}$  & $\Phi_{3}$  & $\Phi_{4}$  & $\Phi_{5}$  & $\Phi_{6}$  & $\Phi_{7}$  & $\Phi_{8}$  & $\Phi_{9}$  & $\Phi_{10}$  & $\Phi_{11}$  & $\Phi_{12}$  & $\Phi_{13}$  & $\Phi_{14}$\tabularnewline
 &  & $\chi$  & $\chi^{c}$  & $\Omega$  & $\Omega^{c}$  & $\Phi$  &  &  & $\delta_{d}$  & $\delta_{u}$  &  &  &  & \tabularnewline
\midrule
 $SU(3)_{C}$  & \textbf{1}  & \textbf{1}  & \textbf{1}  & \textbf{1}  & \textbf{1}  & \textbf{1}  & \textbf{8}  & \textbf{1}  & \textbf{3}  & \textbf{3}  & \textbf{6}  & \textbf{6}  & \textbf{3}  & \textbf{3}\tabularnewline
$SU(2)_{L}$  & \textbf{1}  & \textbf{2}  & \textbf{1}  & \textbf{3}  & \textbf{1}  & \textbf{2}  & \textbf{1}  & \textbf{1}  & \textbf{1}  & \textbf{1}  & \textbf{1}  & \textbf{1}  & \textbf{2}  & \textbf{1}\tabularnewline
$SU(2)_{R}$  & \textbf{1}  & \textbf{1}  & \textbf{2}  & \textbf{1}  & \textbf{3}  & \textbf{2}  & \textbf{1}  & \textbf{1}  & \textbf{1}  & \textbf{1}  & \textbf{1}  & \textbf{1}  & \textbf{1}  & \textbf{2}\tabularnewline
$U(1)_{B-L}$  & 0  & +1  & -1  & 0  & 0  & 0  & 0  & +2  & $-\frac{2}{3}$  & $+\frac{4}{3}$  & $+\frac{2}{3}$  & $-\frac{4}{3}$  & $+\frac{1}{3}$  & $+\frac{1}{3}$\tabularnewline
\midrule 
 {\footnotesize }%
\begin{tabular}{@{}l@{}}
~PS\tabularnewline
origin\tabularnewline
\end{tabular} & {\footnotesize }%
\begin{tabular}{@{}r@{}}
\textbf{\footnotesize $\Psi_{1}$}\tabularnewline
\textbf{\footnotesize $\Psi_{10}$}\tabularnewline
\end{tabular} & \textbf{\footnotesize $\overline{\Psi}_{12}$}{\footnotesize{} } & \textbf{\footnotesize $\Psi_{13}$}{\footnotesize{} } & \textbf{\footnotesize $\Psi_{3}$}{\footnotesize{} } & \textbf{\footnotesize $\Psi_{4}$}{\footnotesize{} } & {\footnotesize }%
\begin{tabular}{@{}r@{}}
\textbf{\footnotesize $\Psi_{2}$}\tabularnewline
\textbf{\footnotesize $\Psi_{7}$}\tabularnewline
\end{tabular} & {\footnotesize }%
\begin{tabular}{@{}r@{}}
\textbf{\footnotesize $\Psi_{10}$}\tabularnewline
\textbf{\footnotesize $\Psi_{11}$}\tabularnewline
\end{tabular} & \textbf{\footnotesize $\overline{\Psi}_{9}$}{\footnotesize{} } & {\footnotesize }%
\begin{tabular}{@{}r@{}}
\textbf{\footnotesize $\Psi_{8}$}\tabularnewline
\textbf{\footnotesize $\Psi_{9}$}\tabularnewline
\end{tabular} & \textbf{\footnotesize $\Psi_{10}$}{\footnotesize{} } & \textbf{\footnotesize $\Psi_{9}$}{\footnotesize{} } & \textbf{\footnotesize $\Psi_{11}$}{\footnotesize{} } & \textbf{\footnotesize $\Psi_{12}$}{\footnotesize{} } & \textbf{\footnotesize $\Psi_{13}$}\tabularnewline
\bottomrule
\end{tabular}\tabularnewline
\tabularnewline
\begin{tabular}{rrrrrrrrrrr}
\toprule 
 & $\Phi_{15}$  & $\Phi_{16}$  & $\Phi_{17}$  & $\Phi_{18}$  & $\Phi_{19}$  & $\Phi_{20}$  & $\Phi_{21}$  & $\Phi_{22}$  & $\Phi_{23}$  & $\Phi_{24}$\tabularnewline
 &  & $\Delta$  & $\Delta^{c}$  &  &  &  &  &  &  & \tabularnewline
\midrule
$SU(3)_{C}$  & \textbf{8}  & \textbf{1}  & \textbf{1}  & \textbf{3}  & \textbf{3}  & \textbf{3}  & \textbf{6}  & \textbf{6}  & \textbf{1}  & \textbf{3}\tabularnewline
$SU(2)_{L}$  & \textbf{2}  & \textbf{3}  & \textbf{1}  & \textbf{2}  & \textbf{3}  & \textbf{1}  & \textbf{3}  & \textbf{1}  & \textbf{3}  & \textbf{2}\tabularnewline
$SU(2)_{R}$  & \textbf{2}  & \textbf{1}  & \textbf{3}  & \textbf{2}  & \textbf{1}  & \textbf{3}  & \textbf{1}  & \textbf{3}  & \textbf{3}  & \textbf{2}\tabularnewline
$U(1)_{B-L}$  & 0  & -2  & -2  & $+\frac{4}{3}$  & $-\frac{2}{3}$  & $-\frac{2}{3}$  & $+\frac{2}{3}$  & $+\frac{2}{3}$  & 0  & $-\frac{2}{3}$\tabularnewline
\midrule 
{\footnotesize }%
\begin{tabular}{@{}l@{}}
~~PS\tabularnewline
origin\tabularnewline
\end{tabular} & \textbf{\footnotesize $\Psi_{7}$}{\footnotesize{} } & \textbf{\footnotesize $\Psi_{16}$}{\footnotesize{} } & \textbf{\footnotesize $\Psi_{17}$}{\footnotesize{} } & \textbf{\footnotesize $\Psi_{7}$}{\footnotesize{} } & {\footnotesize }%
\begin{tabular}{@{}r@{}}
\textbf{\footnotesize $\Psi_{14}$}\tabularnewline
\textbf{\footnotesize $\Psi_{16}$}\tabularnewline
\end{tabular} & {\footnotesize }%
\begin{tabular}{@{}r@{}}
\textbf{\footnotesize $\Psi_{15}$}\tabularnewline
\textbf{\footnotesize $\Psi_{17}$}\tabularnewline
\end{tabular} & \textbf{\footnotesize $\Psi_{16}$}{\footnotesize{} } & \textbf{\footnotesize $\Psi_{17}$}{\footnotesize{} } & \textbf{\footnotesize $\Psi_{5}$}{\footnotesize{} } & \textbf{\footnotesize $\Psi_{6}$}\tabularnewline
\bottomrule
\end{tabular}\tabularnewline
\end{tabular}} 
\par\end{centering}

\setlength{\tabcolsep}{6pt}

\caption{\label{tab:List_of_LR_fields}Naming conventions and transformation
properties of fields in the left-right symmetric regime (excluding
conjugates). The last row exhibits the Pati-Salam regime fields from
which they may originate. The charges under the $U(1)_{B-L}$ group
shown here were multiplied by a factor $\sqrt{\frac{8}{3}}$.}
\end{table}

\begin{table}[h]
\begin{centering}
\scalebox{0.99}{
\setlength{\tabcolsep}{2pt}%
\begin{tabular}{cccccccccccccccccc}
\toprule 
 & $\Psi_{1}$  & $\Psi_{2}$  & $\Psi_{3}$  & $\Psi_{4}$  & $\Psi_{5}$  & $\Psi_{6}$  & $\Psi_{7}$  & $\Psi_{8}$  & $\Psi_{9}$  & $\Psi_{10}$  & $\Psi_{11}$  & $\Psi_{12}$  & $\Psi_{13}$  & $\Psi_{14}$  & $\Psi_{15}$  & $\Psi_{16}$  & $\Psi_{17}$\tabularnewline
\midrule
 $SU(4)$  & \textbf{1}  & \textbf{1}  & \textbf{1}  & \textbf{1}  & \textbf{1}  & \textbf{6}  & \textbf{15}  & \textbf{6}  & \textbf{10}  & \textbf{15}  & \textbf{20'}  & \textbf{4}  & \textbf{4}  & \textbf{6}  & \textbf{6}  & \textbf{10}  & \textbf{10}\tabularnewline
$SU(2)_{L}$  & \textbf{1}  & \textbf{2}  & \textbf{3}  & \textbf{1}  & \textbf{3}  & \textbf{2}  & \textbf{2}  & \textbf{1}  & \textbf{1}  & \textbf{1}  & \textbf{1}  & \textbf{2}  & \textbf{1}  & \textbf{3}  & \textbf{1}  & \textbf{3}  & \textbf{1}\tabularnewline
$SU(2)_{R}$  & \textbf{1}  & \textbf{2}  & \textbf{1}  & \textbf{3}  & \textbf{3}  & \textbf{2}  & \textbf{2}  & \textbf{1}  & \textbf{1}  & \textbf{1}  & \textbf{1}  & \textbf{1}  & \textbf{2}  & \textbf{1}  & \textbf{3}  & \textbf{1}  & \textbf{3}\tabularnewline
\midrule 
 {\footnotesize }%
\begin{tabular}{@{}l@{}}
$SO(10)$ \tabularnewline
Origin\tabularnewline
\end{tabular} & {\footnotesize }%
\begin{tabular}{@{}r@{}}
\textbf{\footnotesize 1}\tabularnewline
\textbf{\footnotesize 54}\tabularnewline
\end{tabular} & {\footnotesize }%
\begin{tabular}{@{}r@{}}
\textbf{\footnotesize 10}\tabularnewline
\textbf{\footnotesize 120}\tabularnewline
\end{tabular} & \textbf{\footnotesize 45}{\footnotesize{} } & \textbf{\footnotesize 45}{\footnotesize{} } & \textbf{\footnotesize 54}{\footnotesize{} } & {\footnotesize }%
\begin{tabular}{@{}r@{}}
\textbf{\footnotesize 45}\tabularnewline
\textbf{\footnotesize 54}\tabularnewline
\end{tabular} & {\footnotesize }%
\begin{tabular}{@{}r@{}}
\textbf{\footnotesize 120}\tabularnewline
\textbf{\footnotesize 126}\tabularnewline
\end{tabular} & {\footnotesize }%
\begin{tabular}{@{}r@{}}
\textbf{\footnotesize 10}\tabularnewline
\textbf{\footnotesize 126}\tabularnewline
\end{tabular} & \textbf{\footnotesize 120}{\footnotesize{} } & \textbf{\footnotesize 45}{\footnotesize{} } & \textbf{\footnotesize 54}{\footnotesize{} } & \textbf{\footnotesize 16}{\footnotesize{} } & \textbf{\footnotesize $\overline{\mathbf{16}}$}{\footnotesize{} } & \textbf{\footnotesize 120}{\footnotesize{} } & \textbf{\footnotesize 120}{\footnotesize{} } & \textbf{\footnotesize 126}{\footnotesize{} } & \textbf{\footnotesize $\overline{\mathbf{126}}$}\tabularnewline
\bottomrule
\end{tabular}} 
\par\end{centering}

\caption{\label{tab:List_of_PatiSalam_fields}Naming conventions and transformation
properties of fields in the Pati-Salam regime (excluding conjugates).
The last row exhibits the $SO(10)$ representations from which they
may originate. }
\end{table}

\begin{table}[h]
\begin{centering}
\scalebox{1.0}{
\setlength{\tabcolsep}{2pt}%
\begin{tabular}{c}
\begin{tabular}{ccccccccccccccccc}
\toprule 
 & $\Phi_{1}^{'}$  & $\Phi_{2}^{'}$  & $\Phi_{3}^{'}$  & $\Phi_{4}^{'}$  & $\Phi_{5}^{'}$  & $\Phi_{6}^{'}$  & $\Phi_{7}^{'}$  & $\Phi_{8}^{'}$  & $\Phi_{9}^{'}$  & $\Phi_{10}^{'}$  & $\Phi_{11}^{'}$  & $\Phi_{12}^{'}$  & $\Phi_{13}^{'}$  & $\Phi_{14}^{'}$  & $\Phi_{15}^{'}$  & $\Phi_{16}^{'}$ \tabularnewline
\midrule
 $SU(3)_{C}$  & \textbf{1}  & \textbf{1}  & \textbf{1}  & 1  & \textbf{1}  & \textbf{1}  & \textbf{1}  & \textbf{8}  & \textbf{1}  & \textbf{3}  & \textbf{3}  & \textbf{6}  & \textbf{6}  & \textbf{3}  & \textbf{3}  & \textbf{3}\tabularnewline
$SU(2)_{L}$  & \textbf{1}  & \textbf{2}  & \textbf{1}  & 1  & \textbf{3}  & \textbf{1}  & \textbf{2}  & \textbf{1}  & \textbf{1}  & \textbf{1}  & \textbf{1}  & \textbf{1}  & \textbf{1}  & \textbf{2}  & \textbf{1}  & \textbf{1}\tabularnewline
$U(1)_{R}$  & 0  & 0  & $-\frac{1}{2}$  & $+\frac{1}{2}$  & 0  & +1  & $+\frac{1}{2}$  & 0  & 0  & 0  & 0  & 0  & 0  & 0  & $-\frac{1}{2}$  & $+\frac{1}{2}$\tabularnewline
$U(1)_{B-L}$  & 0  & +1  & -1  & -1  & 0  & 0  & 0  & 0  & +2  & $-\frac{2}{3}$  & $+\frac{4}{3}$  & $+\frac{2}{3}$  & $-\frac{4}{3}$  & $+\frac{1}{3}$  & $+\frac{1}{3}$  & $+\frac{1}{3}$\tabularnewline
\midrule 
 {\footnotesize }%
\begin{tabular}{@{}l@{}}
~~LR\tabularnewline
origin\tabularnewline
\end{tabular} & %
\begin{tabular}{@{}r@{}}
\textbf{\footnotesize $\Phi_{1}$}\tabularnewline
\textbf{\footnotesize $\Phi_{5}$}{\footnotesize{} }\tabularnewline
\end{tabular} & \textbf{\footnotesize $\Phi_{2}$} & \textbf{\footnotesize $\Phi_{3}$} & \textbf{\footnotesize $\Phi_{3}$}{\footnotesize{} } & %
\begin{tabular}{@{}r@{}}
\textbf{\footnotesize $\Phi_{4}$}\tabularnewline
\textbf{\footnotesize $\Phi_{23}$}{\footnotesize{} }\tabularnewline
\end{tabular} & \textbf{\footnotesize $\Phi_{5}$} & \textbf{\footnotesize $\Phi_{6}$} & \textbf{\footnotesize $\Phi_{7}$} & %
\begin{tabular}{@{}r@{}}
\textbf{\footnotesize $\Phi_{8}$}{\footnotesize{} }\tabularnewline
\textbf{\footnotesize $\bar{\Phi}_{17}$}{\footnotesize{} }\tabularnewline
\end{tabular} & %
\begin{tabular}{@{}r@{}}
\textbf{\footnotesize $\Phi_{9}$}{\footnotesize{} }\tabularnewline
\textbf{\footnotesize $\Phi_{20}$}{\footnotesize{} }\tabularnewline
\end{tabular} & \textbf{\footnotesize $\Phi_{10}$} & %
\begin{tabular}{@{}r@{}}
\textbf{\footnotesize $\Phi_{11}$}\tabularnewline
\textbf{\footnotesize $\Phi_{22}$}{\footnotesize{} }\tabularnewline
\end{tabular} & \textbf{\footnotesize $\Phi_{12}$} & \textbf{\footnotesize $\Phi_{13}$} & \textbf{\footnotesize $\Phi_{14}$}{\footnotesize{} } & \textbf{\footnotesize $\Phi_{14}$}\tabularnewline
\bottomrule
\end{tabular}\tabularnewline
\tabularnewline
\begin{tabular}{cccccccccccccccc}
\toprule 
 & $\Phi_{17}^{'}$  & $\Phi_{18}^{'}$  & $\Phi_{19}^{'}$  & $\Phi_{20}^{'}$  & $\Phi_{21}^{'}$  & $\Phi_{22}^{'}$  & $\Phi_{23}^{'}$  & $\Phi_{24}^{'}$  & $\Phi_{25}^{'}$  & $\Phi_{26}^{'}$  & $\Phi_{27}^{'}$  & $\Phi_{28}^{'}$  & $\Phi_{29}^{'}$  & $\Phi_{30}^{'}$  & $\Phi_{31}^{'}$\tabularnewline
\midrule
 $SU(3)_{C}$  & \textbf{8}  & \textbf{1}  & \textbf{1}  & \textbf{1}  & \textbf{3}  & \textbf{3}  & \textbf{3}  & \textbf{3}  & \textbf{3}  & \textbf{6}  & \textbf{6}  & \textbf{6}  & \textbf{1}  & \textbf{3}  & \textbf{3}\tabularnewline
$SU(2)_{L}$  & \textbf{2}  & \textbf{3}  & \textbf{1}  & \textbf{1}  & \textbf{2}  & \textbf{2}  & \textbf{3}  & \textbf{1}  & \textbf{1}  & \textbf{3}  & \textbf{1}  & \textbf{1}  & \textbf{3}  & \textbf{2}  & \textbf{2}\tabularnewline
$U(1)_{R}$  & $+\frac{1}{2}$  & 0  & -1  & +1  & $-\frac{1}{2}$  & $+\frac{1}{2}$  & 0  & -1  & +1  & 0  & -1  & +1  & +1  & $-\frac{1}{2}$  & $+\frac{1}{2}$\tabularnewline
$U(1)_{B-L}$  & 0  & -2  & -2  & -2  & $+\frac{4}{3}$  & $+\frac{4}{3}$  & $-\frac{2}{3}$  & $-\frac{2}{3}$  & $-\frac{2}{3}$  & $+\frac{2}{3}$  & $+\frac{2}{3}$  & $+\frac{2}{3}$  & 0  & $-\frac{2}{3}$  & $-\frac{2}{3}$\tabularnewline
\midrule 
 {\footnotesize }%
\begin{tabular}{@{}l@{}}
~~LR\tabularnewline
origin\tabularnewline
\end{tabular} & \textbf{\footnotesize $\Phi_{15}$} & \textbf{\footnotesize $\Phi_{16}$}  & \textbf{\footnotesize $\Phi_{17}$} & \textbf{\footnotesize $\Phi_{17}$} & \textbf{\footnotesize $\Phi_{18}$} & \textbf{\footnotesize $\Phi_{18}$} & \textbf{\footnotesize $\Phi_{19}$} & \textbf{\footnotesize $\Phi_{20}$} & \textbf{\footnotesize $\Phi_{20}$} & \textbf{\footnotesize $\Phi_{21}$} & \textbf{\footnotesize $\Phi_{22}$} & \textbf{\footnotesize $\Phi_{22}$} & \textbf{\footnotesize $\Phi_{23}$} & \textbf{\footnotesize $\Phi_{24}$}{\footnotesize{} } & \textbf{\footnotesize $\Phi_{24}$}\tabularnewline
\bottomrule
\end{tabular}\tabularnewline
\end{tabular}} 
\par\end{centering}

\caption{\label{tab:List_of_LR_fields_U1}Naming conventions and transformation
properties of fields in the U(1) mixing regime (excluding conjugates).
The last row exhibits the left-right regime fields from which they
may originate. The charges under the $U(1)_{B-L}$ group shown here
were multiplied by a factor $\sqrt{\frac{8}{3}}$.}
\end{table}

Finally, we point out here that in order for a group $G$ to break
into a subgroup $H\subset G$, there must be one or more fields transforming
non-trivially under $G$ which contain a singlet of $H$ that acquires
a vacuum expectation value. Non-trivial here means that the singlet(s)
of $H$ contained in this(these) field(s) must also break any group
$G'$ such that $H\subset G'\subseteq G$. From this observation alone
we know that certain fields must be present in a fundamental model
if we are to achieve a given breaking sequence---this is shown schematically
in figure \eqref{fig:NecessaryFields} for the groups considered in
chapter \ref{chap:SlidingScale_models}.

\begin{center}
\begin{figure}[tbph]
\begin{centering}
\includegraphics[scale=0.97]{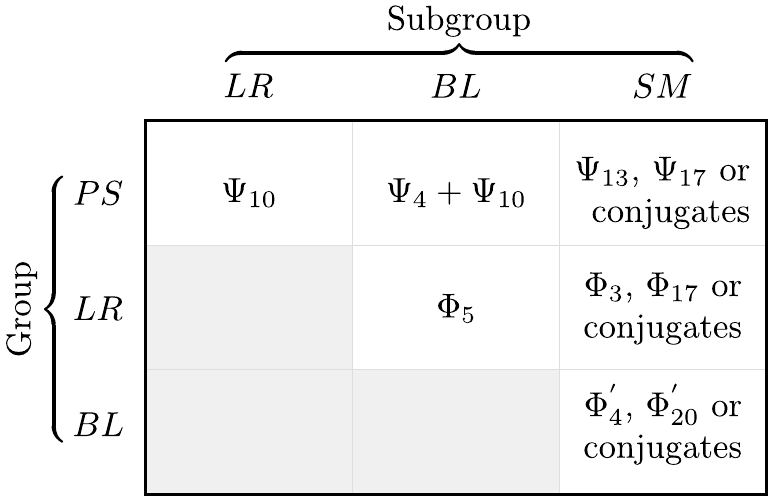}
\par\end{centering}

\caption{\label{fig:NecessaryFields}Specific fields which are needed to break
a group (rows) into a given subgroup (columns).}
\end{figure}

\par\end{center}
\cleartooddpage

\chapter{\label{chap:Renormalization_of_the_NuLH_vertex}Renormalization of
the $\nu\ell H^{+}$ vertex}

In what follows we detail the computation leading to equations \eqref{eq:epsilondelta}--\eqref{eq:delta:def}
of chapter \ref{chap:Revisiting_RK} (see also \citep{Fonseca:2012kr}),
further referring to \citep{Bellazzini:2010gn} for a similar analysis.
As expected, loop effects contribute to both kinetic and mass terms
of charged leptons as well as to the $\nu\ell H^{+}$ vertex:
\begin{align}
\mathscr{L}_{0}^{H^{\pm}} & =i\,\overline{\ell}_{L}\left(\mathbb{1}+\eta_{L}^{\ell}\right)\slashed{\partial}\ell_{L}+i\,\overline{\ell}_{R}\left(\mathbb{1}+\eta_{R}^{\ell}\right)\slashed{\partial}\ell_{R}\nonumber \\[1mm]
 & +i\,\overline{\nu}_{L}\left(\mathbb{1}+\eta_{L}^{\nu}\right)\slashed{\partial}\nu_{L}-\left[\overline{\ell}_{L}\left(M^{\ell0}+\eta_{m}^{\ell}\right)\ell_{R}+\textrm{h.c.}\right]\nonumber \\[1mm]
 & +\left[\overline{\nu}_{L}\left(2^{3/4}G_{F}^{1/2}\,\tan\beta\, M^{\ell0}+\eta^{H}\right)\ell_{R}H^{+}+\textrm{h.c.}\right]\,.
\end{align}
Here $M^{\ell0}$ denotes the bare charged lepton mass and the $\eta$'s
correspond to loop contributions to the various terms. The (new) kinetic
terms can be recast into a canonical form by means of unitary rotations
of the fields ($K_{L}^{\ell}$, $K_{R}^{\ell}$, $K_{L}^{\nu}$),
which are then renormalized by diagonal transformations ($\hat{Z}_{L}^{\ell}$,
$\hat{Z}_{R}^{\ell}$, $\hat{Z}_{L}^{\nu}$):
\begin{align}
\ell_{L}^{\text{old}} & =K_{L}^{\ell}\left(\hat{Z}_{L}^{\ell}\right)^{-\frac{1}{2}}\ell_{L}^{\text{new}}\,, & \hat{Z}_{L}^{\ell} & ={K_{L}^{\ell}}^{\dagger}\left(\mathbb{1}+\eta_{L}^{\ell}\right)K_{L}^{\ell}\,,\\
\ell_{R}^{\text{old}} & =K_{R}^{\ell}\left(\hat{Z}_{R}^{\ell}\right)^{-\frac{1}{2}}\ell_{R}^{\text{new}}\,, & \hat{Z}_{R}^{\ell} & ={K_{R}^{\ell}}^{\dagger}\left(\mathbb{1}+\eta_{R}^{\ell}\right)K_{R}^{\ell}\,,\\
\nu_{L}^{\text{old}} & =K_{L}^{\nu}\left(\hat{Z}_{L}^{\nu}\right)^{-\frac{1}{2}}\nu_{L}^{\text{new}}\,, & \hat{Z}_{L}^{\nu} & ={K_{L}^{\nu}}^{\dagger}\left(\mathbb{1}+\eta_{L}^{\nu}\right)K_{L}^{\nu}\,.
\end{align}
Two unitary rotation matrices ($R_{L}^{\ell}$, $R_{R}^{\ell}$) are
further required to diagonalize the charged lepton mass matrix, and
one finally has
\begin{align}
\ell_{L}^{\text{old}} & =K_{L}^{\ell}\left(\hat{Z}_{L}^{\ell}\right)^{-\frac{1}{2}}\, R_{L}^{\ell}\,\ell_{L}^{\text{new}}\,,\\
\ell_{R}^{\text{old}} & =K_{R}^{\ell}\left(\hat{Z}_{R}^{\ell}\right)^{-\frac{1}{2}}\, R_{R}^{\ell}\,\ell_{R}^{\text{new}}\,,\\
\nu_{L}^{\text{old}} & =K_{L}^{\nu}\left(\hat{Z}_{L}^{\nu}\right)^{-\frac{1}{2}}\, R_{L}^{\ell}\,\nu_{L}^{\text{new}}\,.
\end{align}
In the new basis, the mass terms now read
\begin{align}
\mathscr{L}^{\textrm{mass}} & \equiv-\overline{\ell}_{L}M^{\ell}\ell_{R}+\textrm{h.c.}\nonumber \\
 & =-\overline{\ell}_{L}{R_{L}^{\ell}}^{\dagger}\left[\left(\hat{Z}_{L}^{\ell}\right)^{-\frac{1}{2}}{K_{L}^{\ell}}^{\dagger}\left(M^{\ell0}+\eta_{m}^{\ell}\right)K_{R}^{\ell}\left(\hat{Z}_{R}^{l}\right)^{-\frac{1}{2}}\right]R_{R}^{\ell}\ell_{R}+\textrm{h.c.}\,.
\end{align}
The above equation relates the unknown parameter $M^{\ell0}$ with
the physical mass matrix $M^{\ell}$. Using the latter to rewrite
the $\nu\ell H^{+}$ vertex, one finds
\begin{align}
\mathcal{\mathscr{L}}^{H^{\pm}} & \equiv\overline{\nu}_{L}Z^{H}\ell_{R}H^{+}+\textrm{h.c.}\,,
\end{align}
where
\begin{align}
Z^{H} & =2^{3/4}G_{F}^{1/2}\,\tan\beta\,{R_{L}^{\ell}}^{\dagger}\left(\hat{Z}_{L}^{\nu}\right)^{-\frac{1}{2}}{K_{L}^{\nu}}^{\dagger}\, K_{L}^{\ell}\left(\hat{Z}_{L}^{\ell}\right)^{\frac{1}{2}}\, R_{L}^{\ell}\, M^{\ell}\nonumber \\
 & +{R_{L}^{\ell}}^{\dagger}\left(\hat{Z}_{L}^{\nu}\right)^{-\frac{1}{2}}\,{K_{L}^{\nu}}^{\dagger}\left(-2^{3/4}G_{F}^{1/2}\,\tan\beta\,\eta_{m}^{\ell}+\eta^{H}\right)K_{R}^{\ell}\left(\hat{Z}_{R}^{\ell}\right)^{-\frac{1}{2}}K_{R}^{\ell}\,.
\end{align}
 To one-loop order, this exact expression simplifies to 
\begin{align}
Z^{H}= & 2^{3/4}G_{F}^{1/2}\,\tan\beta\,\left[\left(\mathbb{1}+\frac{\eta_{L}^{\ell}}{2}-\frac{\eta_{L}^{\nu}}{2}\right)M^{\ell}-\eta_{m}^{\ell}\right]+\eta^{H}\,.
\end{align}
 The expressions for the $\eta$'s can be computed from the relevant
Feynman diagrams (assuming zero external momenta):
\begin{align}
-\left(4\pi\right)^{2}\left(\eta_{m}^{\ell}\right)_{ij} & =N_{i\alpha\beta}^{R\left(\ell\right)}N_{j\alpha\beta}^{L\left(\ell\right)*}m_{\chi_{\alpha}^{0}}B_{0}\left(0,m_{\chi_{\alpha}^{0}}^{2},m_{\widetilde{\ell}_{\beta}}^{2}\right)+\nonumber \\
 & +C_{i\alpha\beta}^{R\left(\ell\right)}C_{j\alpha\beta}^{L\left(\ell\right)*}m_{\chi_{\alpha}^{\pm}}B_{0}\left(0,m_{\chi_{\alpha}^{\pm}}^{2},m_{\widetilde{\nu}_{\beta}}^{2}\right)\,,\label{eq:app:etaellm}\\[1mm]
-\left(4\pi\right)^{2}\left(\eta_{R}^{\ell}\right)_{ij} & =N_{i\alpha\beta}^{L\left(\ell\right)}N_{j\alpha\beta}^{L\left(\ell\right)*}B_{1}\left(0,m_{\chi_{\alpha}^{0}}^{2},m_{\widetilde{\ell}_{\beta}}^{2}\right)\nonumber \\
 & +C_{i\alpha\beta}^{L\left(\ell\right)}C_{j\alpha\beta}^{L\left(\ell\right)*}B_{1}\left(0,m_{\chi_{\alpha}^{\pm}}^{2},m_{\widetilde{\nu}_{\beta}}^{2}\right)\,,\label{eq:app:etaellR}\\[1mm]
-\left(4\pi\right)^{2}\left(\eta_{L}^{\ell}\right)_{ij} & =N_{i\alpha\beta}^{R\left(\ell\right)}N_{j\alpha\beta}^{R\left(\ell\right)*}B_{1}\left(0,m_{\chi_{\alpha}^{0}}^{2},m_{\widetilde{\ell}_{\beta}}^{2}\right)\nonumber \\
 & +C_{i\alpha\beta}^{R\left(\ell\right)}C_{j\alpha\beta}^{R\left(\ell\right)*}B_{1}\left(0,m_{\chi_{\alpha}^{\pm}}^{2},m_{\widetilde{\nu}_{\beta}}^{2}\right)\,,\label{eq:app:etaellL}\\[1mm]
-\left(4\pi\right)^{2}\left(\eta_{L}^{\nu}\right)_{ij} & =N_{i\alpha\beta}^{R\left(\nu\right)}N_{j\alpha\beta}^{R\left(\nu\right)*}B_{1}\left(0,m_{\chi_{\alpha}^{0}}^{2},m_{\widetilde{\nu}_{\beta}}^{2}\right)\nonumber \\
 & +C_{i\alpha\beta}^{R\left(\nu\right)}C_{j\alpha\beta}^{R\left(\nu\right)*}B_{1}\left(0,m_{\chi_{\alpha}^{\pm}}^{2},m_{\widetilde{\ell}_{\beta}}^{2}\right)\,,\label{eq:app:etanuL}\\[1mm]
-\left(4\pi\right)^{2}\left(\eta^{H}\right)_{ij} & =C_{i\beta\gamma}^{R\left(\nu\right)}N_{j\alpha\gamma}^{L\left(\ell\right)*}\left[D_{\beta\alpha2}^{L\left(S^{+}\right)*}m_{\chi_{\alpha}^{0}}m_{\chi_{\beta}^{\pm}}C_{0}\left(0,0,0,m_{\chi_{\alpha}^{0}}^{2},m_{\chi_{\beta}^{\pm}}^{2},m_{\widetilde{\ell}_{\gamma}}^{2}\right)\right.\nonumber \\
 & \left.+D_{\beta\alpha2}^{R\left(S^{+}\right)*}dC_{00}\left(0,0,0,m_{\chi_{\alpha}^{0}}^{2},m_{\chi_{\beta}^{\pm}}^{2},m_{\widetilde{\ell}_{\gamma}}^{2}\right)\right]\nonumber \\
 & +N_{i\alpha\gamma}^{R\left(\nu\right)}C_{j\beta\gamma}^{L\left(\ell\right)*}\left[D_{\beta\alpha2}^{L\left(S^{+}\right)*}m_{\chi_{\alpha}^{0}}m_{\chi_{\beta}^{\pm}}C_{0}\left(0,0,0,m_{\chi_{\alpha}^{0}}^{2},m_{\chi_{\beta}^{\pm}}^{2},m_{\widetilde{\nu}_{\gamma}}^{2}\right)\right.\nonumber \\
 & \left.+\, D_{\beta\alpha2}^{R\left(S^{+}\right)*}dC_{00}\left(0,0,0,m_{\chi_{\alpha}^{0}}^{2},m_{\chi_{\beta}^{\pm}}^{2},m_{\widetilde{\nu}_{\gamma}}^{2}\right)\right]\nonumber \\
 & +\, N_{i\alpha\beta}^{R\left(\nu\right)}N_{j\alpha\gamma}^{L\left(\ell\right)*}g_{2\gamma\beta}^{\left(S^{+}\widetilde{\ell}\widetilde{\nu}^{*}\right)}m_{\chi_{\gamma}^{0}}C_{0}\left(0,0,0,m_{\widetilde{\ell}_{\gamma}}^{2},m_{\widetilde{\nu}_{\beta}}^{2},m_{\chi_{\alpha}^{0}}^{2}\right)\,,\label{eq:app:etaH}
\end{align}
with $B_{0,1},\, C_{0},\, C_{0,0}$ denoting the usual loop integral
functions
\begin{align}
B_{0}\left(0,x,y\right) & =\Delta_{\varepsilon}+1-\frac{x\log\frac{x}{\mu^{2}}-y\log\frac{y}{\mu^{2}}}{x-y}\,,\\
B_{1}\left(0,x,y\right) & =-\frac{1}{2}\left[\Delta_{\varepsilon}+\frac{3x-y}{2\left(x-y\right)}-\log\frac{y}{\mu^{2}}+\left(\frac{x}{x-y}\right)^{2}\log\frac{y}{x}\right]\,,\label{eq:app:B1}\\
C_{0}\left(0,0,0,x,y,z\right) & =\frac{xy\log\frac{x}{y}+yz\log\frac{y}{z}+zx\log\frac{z}{x}}{\left(x-y\right)\left(y-z\right)\left(z-x\right)}\,,\\
dC_{00}\left(0,0,0,x,y,z\right) & =\Delta_{\varepsilon}+1\nonumber \\
 & +\frac{x^{2}\left(y-z\right)\log\frac{x}{\mu^{2}}+y^{2}\left(z-x\right)\log\frac{y}{\mu^{2}}+z^{2}\left(x-y\right)\log\frac{z}{\mu^{2}}}{\left(x-y\right)\left(y-z\right)\left(z-x\right)}\,.
\end{align}
Here $d=4-\varepsilon$, $\mu$ is the regularization parameter and
$\Delta_{\varepsilon}=\frac{2}{\varepsilon}-\gamma+\log4\pi$. For
the couplings notation we followed \citep{website_Romao_MSSM}.

The comparison of the above expressions with the corresponding ones
derived in reference \citep{Bellazzini:2010gn}, reveals a fair agreement;
we note nevertheless that the neutralino and chargino masses are absent
from the analogous of equation \eqref{eq:app:etaellm}, and that the
order of the arguments of $B_{1}$ in equations \eqref{eq:app:etaellR},
\eqref{eq:app:etaellL}, \eqref{eq:app:etanuL} appears reversed.
Moreover, we find small discrepancies (which cannot be accounted by
the distinct notations) in the expressions for $\eta_{m}^{\ell}$
and $\eta_{H}$---compare with equations \eqref{eq:app:etaellm} and
\eqref{eq:app:etaH}, respectively.
\cleartooddpage

\bibliographystyle{utphys_mod}
\addcontentsline{toc}{chapter}{\bibname}\bibliography{Bibliography}

\end{document}